\documentclass[12pt,a4paper]{article}  
\usepackage{a41}
\usepackage{cite}
\usepackage{amsmath}
\usepackage{amssymb}
\usepackage{color}

\usepackage[rflt]{floatflt}              
\usepackage{float}

\usepackage[nottoc,notlot,notlof]{tocbibind}

\usepackage{array}

\newcommand{\NS}{\mathrm{NS}}
\newcommand{\ep}{\varepsilon}
\newcommand{\dd}{\mathrm{d}}
\newcommand{\HA}{{\rm H}}
\newcommand{\GeV}{\mathrm{GeV}}
\newcommand{\Li}{{\rm Li}}

\DeclareMathOperator{\arctanh}{arctanh}

\newcommand{\N}{\nonumber} 
\newcommand{\NN}{\nonumber}
\newcommand{\Ctil}{\tilde{C}}
\newcommand{\Ahathat}{\hat{\hat{A}}}

\newcommand*\pFqskip{8mu}
\catcode`,\active
\newcommand*\pFq{\begingroup
        \catcode`\,\active
        \def ,{\mskip\pFqskip\relax}%
        \dopFq
}
\catcode`\,12
\def\dopFq#1#2#3#4#5{%
        {}_{#1}F_{#2}\biggl[\genfrac..{0pt}{}{#3}{#4};#5\biggr]%
        \endgroup
}

\usepackage[english]{babel}
\usepackage[utf8]{inputenc}
\usepackage[T1]{fontenc}
\usepackage{ae}

\usepackage{url}

\usepackage{amsmath, amsthm, amssymb}

\newcommand{\MS}{\overline{\sf MS}}
\newcommand{\MOM}{\sf MOM}

\makeatletter
\renewcommand\section{\@startsection {section}{1}{\z@}%
                                     {-3.5ex \@plus -1ex \@minus -.2ex}%
                                     {2.3ex \@plus.2ex}%
                                     {\normalfont\Large\bfseries\boldmath}}
\renewcommand\subsection{\@startsection{subsection}{2}{\z@}%
                                     {-3.25ex\@plus -1ex \@minus -.2ex}%
                                     {1.5ex \@plus .2ex}%
                                     {\normalfont\large\bfseries\boldmath}}
\renewcommand\subsubsection{\@startsection{subsubsection}{3}{\z@}%
                                     {-3.25ex\@plus -1ex \@minus -.2ex}%
                                     {1.5ex \@plus .2ex}%
                                     {\normalfont\normalsize\bfseries\boldmath}}
\makeatother

\usepackage{graphicx}
\usepackage{color}

\usepackage[unicode=true,bookmarks=true]{hyperref}
\usepackage{bookmark}

\usepackage[a-3a]{pdfx}

\makeatletter
\AtBeginDocument{\let\mathaccentV\AMS@mathaccentV}
\makeatother

\allowdisplaybreaks[4]

\begin{document}

\pagenumbering{Roman}
\setlength{\baselineskip}{0.515cm}

\thispagestyle{empty}
~

\vspace{2.5cm}

\begin{center}

\rule{\textwidth}{1pt}
~

{\huge \bf Massive 2- and 3-loop corrections to }

\vspace{2mm}
{\huge \bf hard scattering processes in QCD}

\rule{\textwidth}{1pt}

\vspace{3cm}
{\LARGE\bf Dissertation}

\vspace{1mm}
{\large \scshape 
zur Erlangung des Doktorgrades an der Fakultät

\vspace{2mm}
für Mathematik, Informatik und Naturwissenschaften

\vspace{2mm}
Fachbereich Physik

\vspace{2mm}
der Universität Hamburg
}

\vspace{2.5cm}
{\large vorgelegt von }

\vspace{0.5cm}
{\LARGE\bf Marco Saragnese}

\vspace{3cm}
{\large Hamburg

\vspace{0.5cm}
2022
}

\end{center}

\clearpage
\thispagestyle{empty}
~
\vspace{5cm}

\begin{table}[h!]
\begin{tabular}{ll}
Gutachter der Dissertation: 					& Prof. Dr. habil. Johannes Bl\"umlein \\[5mm]
												& Prof. Dr. Sven-Olaf Moch \\[1cm]
Zusammensetzung der Prüfungskommission:				& Prof. Dr. habil. Johannes Bl\"umlein  \\[5mm]
													& Prof. Dr. Sven-Olaf Moch \\[5mm]
													& Prof. Dr. Bernd Kniehl \\[5mm]
													& Prof. Dr. Katerina Lipka \\[5mm]
													& Prof. Dr. Gleb Arutyunov \\[1cm]
Vorsitzender der Prüfungskommission:				& Prof. Dr. Bernd Kniehl \\[1cm]
Datum der Disputation:								& 14 Juli 2022 \\[1cm]
Vorsitzender Fach-Promotionsausschusses Physik:		& Prof. Dr. Wolfgang Parak \\[1cm]
Leiter des Fachbereichs Physik:						& Prof. Dr. Günter H. W. Sigl \\[1cm]
Dekan der Fakultät MIN:								& Prof. Dr. Heinrich Graener
\end{tabular}
\end{table}

%
%
%
%

\clearpage
\thispagestyle{empty}
\vspace*{-1cm}
\hspace*{-2.5cm}
\parbox[c][\textheight]{\textwidth}{
\includegraphics[width=209mm,page=1]{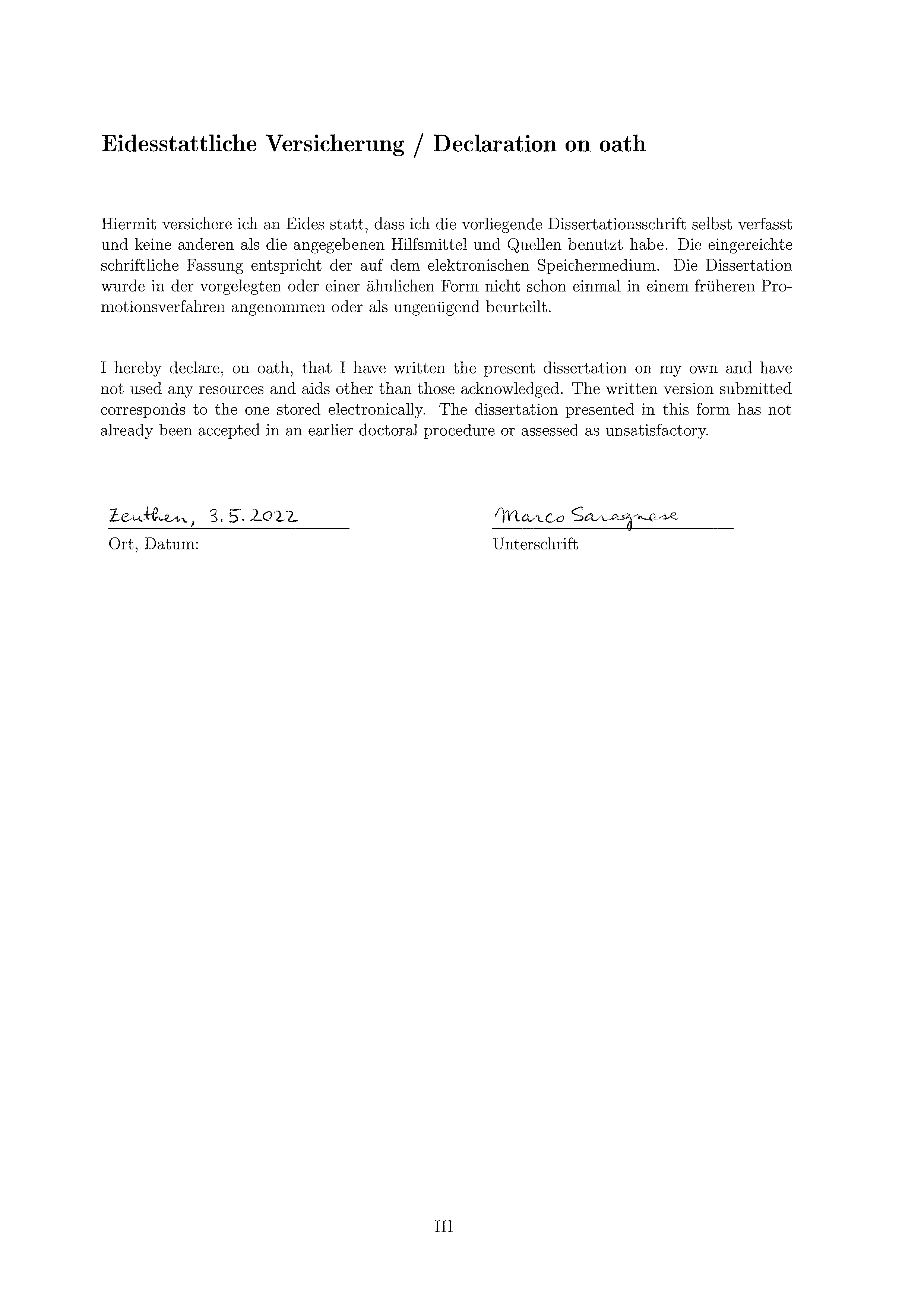}
}

\clearpage
\section*{Abstract}
\pagenumbering{roman}

This thesis deals with calculations of higher-order corrections in perturbative quantum chromodynamics (QCD). The two-mass contributions to the 3-loop, polarized twist-two operator matrix elements (OMEs) $A_{Qq}^{(3),PS}$ and $A_{gg,Q}^{(3)}$ are calculated. The $N$-space result for $A_{gg,Q}^{(3)}$ is obtained analytically as a function of the quark mass ratio, which for $A_{Qq}^{(3),PS}$ is not yet possible. In the $z$-space representation, one obtains for both matrix elements semi-analytical representations in terms of iterated integrals, whereby for reasons of efficiency an additional integral is necessary for some terms.

These universal (process-independent) massive OMEs govern the asymptotic behaviour of the Wilson coefficients in deep-inelastic scattering at large virtualities $Q^2\gg m_{c,b}^2$, with $m_{c,b}$ the charm and bottom quark masses. These corrections are also required to define the variable flavour number scheme. This scheme describes the transition from massive quark corrections to the massless ones for very high momentum scales, which is relevant to the description of collider data.

In the single-mass, polarized case, we derive the logarithmic corrections for the Wilson coefficients of the structure function $g_1$ in the asymptotic region $Q^2\gg m_{c,b}^2$. This is done using the known OMEs and massless Wilson coefficients, using the renormalization group equations.

For the non-singlet structure functions $F_2^{NS}$ and $g_1^{NS}$ we revisit the scheme-invariant evolution operator known for massless quarks and extend it to the massive case with single- and two-mass corrections. In this case, the evolution can effectively be described up to $\mathcal O(a_s^3)$ in the Wilson coefficients, where $a_s=\alpha_s/(4\pi)$ denotes the strong coupling constant. The influence of the hitherto not fully known 4-loop non-singlet anomalous dimension can be described effectively. It turns out that the effect of the theory error in question can be completely controlled. A representation by a Pad\'e approximant proves to be sufficient.

We consider the class of functions of multivariate hypergeometric series and study systems of differential equations obeyed by them. We describe an algorithmic method to solve some classes of such differential systems which delivers a hypergeometric series solution having nested hypergeometric products as summand; we discuss the relationship between these products and Pochhammer symbols. For a number of classical hypergeometric series we derive differential systems and their associated difference equations. We present some examples of series expansions of such functions and of the mathematical objects which arise therein. We also present a Mathematica package which implements algorithms related to the solution of partial linear difference equations, focusing in particular on bounding the degree of the denominator of solutions which are rational functions. These methods are of particular importance when solving multi-leg calculations for Feynman diagrams, but also come into play when hypergeometric methods for multi-loop integrals are used.

We describe a numerical implementation of an $N$-space library for the calculation of scaling violations for  structure functions, which can perform the evolution of parton distribution functions up to NNLO from a parametrization chosen by the user, and encodes massless and massive Wilson coefficients for the structure functions $F_2$ and $g_1$ in the case of photon exchange, and for the structure functions $F_3^{W^+\pm W^-}$ in the case of charged-current exchange. The library contains analytic continuation of the relevant harmonic sums in Mellin-space up to weight 5 and many weight-6 harmonic sums. 
The numerical representation in $x$ space is performed by contour integration around the singularities of the solution of the evolution equations in $N$ space.

\clearpage
\foreignlanguage{german}{
\section*{Zusammenfassung}

\hyphenation{beschrei-ben}

Die vorliegende  Arbeit beschäftigt sich mit Berechnungen von Korrekturen höherer Ordnung in 
der perturbativen Quanten Chromodynamik (QCD). Es werden die 
zweimassigen Beiträge zu den 3-loop massiven, polarisierten twist--2 Operatormatrixelementen (OMEs), 
$A_{Qq}^{(3),PS}$ und $A_{gg,Q}^{(3) }$, berechnet. Das $N$-Raum-Ergebnis für $A_{gg,Q}^{(3)}$ 
erhält man analytisch als Funktion des Massenverhältnisses der schweren Quarks, was f\"ur 
$A_{Qq}^{(3),PS}$ nicht durch iterierte Inegrale m\"oglich ist. In der sog. $z$-Raum Darstellung 
erh\"alt man für beide Matrixelemente
analytische Darstellungen durch iterierte Integrale, wobei aus Effizienzgr\"unden hier\"uber f\"ur 
manche Terme ein weiteres Integral notwendig ist.

Diese universellen (proze\ss{}unabh\"angigen) massiven OMEs bestimmen das asymptotische 
Verhalten der Wilson-Koeffizienten bei tief--inelastischer Streuung f\"ur große Virtualitäten 
$Q^2\gg m_{c,b}^2$. Hier bezeichnen  $m_{c,b}$ die Charm- und
Bottom-Quark-Masse. Diese Korrekturen sind auch erforderlich, um das Variable Flavor Number 
Scheme zu definieren. Dieses Schema beschreibt f\"ur sehr hohe Impulsskalen den \"Ubergang massiver
Quark--Korrekturen in den masslosen Fall, was f\"ur die Beschreibung von Kollider--Daten von 
Bedeutung ist.

Im einmassigen Fall leiten wir die logarithmischen Korrekturen für die 
Wilson-Koeffizienten der Strukturfunktion $g_1$ in der asymptotischen Region $Q^2\gg m_{c,b}^2$ ab.
Dies geschieht unter Verwendung der bekannten OMEs und der masselosen Wilson-Koeffizienten, unter 
Verwendung der Renormierungsgruppengleichungen.

Für die nicht-singulett Strukturfunktionen $F_2^{NS}$ und $g_1^{NS}$ berechnen wir den
schema-invarian\-ten Evolutionsoperator, der f\"ur masslose Quarks bekannt war und erweitern ihn f\"ur 
den massiven Fall mit ein- und zwei-massigen Korrekturen. Hierdurch kann die Evolution in diesem 
Fall effektiv bis zur $O(a_s^3)$ in den Wilsonkoeffizienten beschrieben werden. Hierbei bezeichnet
$a_s = \alpha_s/(4\pi)$ die starke Kopplungskonstante.
Der Einfluss der bisher nicht vollst\"andig bekannten 4--loop nicht-singulett anomalen Dimension
kann effektiv beschrieben werden. Es stellt sich heraus, da\ss{} der Effekt des betreffenden 
Theoriefehlers vollst\"andig kontrolliert werden kann. Eine Darstellung durch eine
Pad\'e-Approximation zeigt sich als ausreichend.

Wir betrachten die Klasse der Funktionen multivariater hypergeometrischer Reihen und untersuchen 
Systeme von Differentialgleichungen und Differenzengleichungen, welche diese beschreiben.
Wir beschreiben ein algorithmisches Verfahren zur Lösung einiger Klassen solcher 
Differentialgleichungssysteme, welche eine hypergeometrische Reihenlösung mit verschachtelten 
hypergeometrischen Produkten als Summanden liefert und  diskutieren die Beziehung zwischen
den Strukturen rationaler Monome aus Pochhammer-Symbolen. Für eine Reihe 
generalisierter klassischer hypergeometrischer Reihen leiten wir Differentialgleichungssysteme
und die zugehörigen Differenzengleichungen her. Wir stellen einige Beispiele für Reihenentwicklungen 
solcher Funktionen und der darin auftretenden mathematischen Objekte vor. Es wird ein 
Mathematica-Paket beschrieben, welches Algorithmen implementiert, die sich auf die Lösung partieller 
linearer Differenzgleichungen beziehen, wobei der Schwerpunkt 
insbesondere auf der Begrenzung des Grades der Nenner von Lösungen liegt, die rationale Funktionen 
sind. Diese Methoden haben besondere Bedeutung bei der L\"osung von sog. multi-leg Berechnungen
bei Feynman Diagrammen, kommen jedoch auch bei Anwendung der hypergeometrischen Methoden f\"ur
multi-loop Diagramme zum Einsatz. 

Wir beschreiben eine numerische Implementierung einer analytischen $N$-Raum Bibliothek zur 
Berechnung von Skalenverletzungen von Strukturfunktionen, welche die Evolution von 
Parton-Verteilungsfunktionen bis zu NNLO aus einer vom Benutzer gewählten Parametrisierung 
durchführen kann und masselose und massive Wilson-Koeffizienten
für das Photon f\"ur die Strukturfunktionen $F_2$ und $g_1$, im Falle des Photonaustausches, und für 
die Strukturfunktion $F_3^{W^+\pm W^-}$ im Falle geladener Str\"ome beschreibt. Die Bibliothek 
enth\"alt im Mellin-Raum analytische Fortsetzungen der relevanten harmonischen Summen bis zum 
Gewicht $w=5$. Die numerische Darstellung im $x$-Raum erfolgt durch eine Kontur-Integration um die
Singularit\"aten der vollkommen analytischen L\"osung der Evolutionsgleichungen im $N$ Raum.

}

\clearpage
\section*{List of publications}

\vspace{1cm}
\noindent
Chapters of this thesis have been published in part in:

\subsection*{ Journal articles }

\noindent
J.~Bl\"umlein and M.~Saragnese,
{\it The $\mathrm N^3\mathrm{LO}$ scheme-invariant QCD evolution of the non-singlet structure functions $F_2^{NS}(x,Q^2)$ and $g_1^{NS}(x,Q^2)$},
Phys. Lett. B \textbf{820} (2021) 136589
[arXiv:2107.01293 [hep-ph]].

\vspace{5mm}
\noindent
J.~Bl\"umlein, A.~De Freitas, M.~Saragnese, C.~Schneider and K.~Sch\"onwald,
{\it Logarithmic contributions to the polarized O($\alpha_s^3$) asymptotic massive Wilson coefficients and operator matrix elements in deeply inelastic scattering},
Phys. Rev. D \textbf{104} (2021) no.3, 034030
[arXiv:2105.09572 [hep-ph]].

\vspace{5mm}
\noindent
J.~Ablinger, J.~Bl\"umlein, A.~De Freitas, A.~Goedicke, M.~Saragnese, C.~Schneider and K.~Sch\"onwald,
{\it The two-mass contribution to the three-loop polarized gluonic operator matrix element $A_{gg,Q}^{(3)}$},
Nucl. Phys. B \textbf{955} (2020) 115059
[arXiv:2004.08916 [hep-ph]].

\vspace{5mm}
\noindent
J.~Ablinger, J.~Bl\"umlein, A.~De Freitas, M.~Saragnese, C.~Schneider and K.~Sch\"onwald,
{\it The three-loop polarized pure singlet operator matrix element with two different masses},
Nucl.~Phys.~B \textbf{952} (2020) 114916
[arXiv:1911.11630 [hep-ph]].

\subsection*{ Proceedings }

\noindent
J.~Ablinger, J.~Bl\"umlein, A.~De Freitas, M.~Saragnese, C.~Schneider and K.~Sch\"onwald,
{\it New 2- and 3-loop heavy flavor corrections to unpolarized and polarized deep-inelastic scattering},
[arXiv:2107.09350 [hep-ph]].

\subsection*{ Preprints }

\noindent
J.~Bl\"umlein, M.~Saragnese and C.~Schneider,
{\it Hypergeometric Structures in Feynman Integrals},
[arXiv:2111.15501 [math-ph]].

\clearpage
\tableofcontents{}

\numberwithin{equation}{section}

\clearpage
\pagenumbering{arabic}
\section{Introduction}

The first experiment in subatomic particle physics was arguably performed by J. Thomson in 1897 \cite{Thomson:1897cm}, see also \cite{Pais:1986nu} for historical accounts of the period.  Thomson was the first to isolate an electron beam (a cathodic ray in the language of the time), speculating that it was composed of charged particles. By measuring their charge to mass ratio he could conclude that those particles were distinct from any known ion, therefore discovering the electron.
Further insights on the structure of matter came from Rutherford's experiment culminating in the establishment of the presence of a positively charged nucleus in 1911 \cite{Geiger:1909,Geiger:1910,Rutherford:1911zz} and therefore discovering the proton. The famous gold-foil experiment is arguably the first scattering experiment in the modern sense. In 1932, Chadwick \cite{Chadwick:1932ma,Chadwick:1932wcf} discovered the neutron.

The notion of a particle-like interpretation of the photon was introduced in the same era, by Planck's theory of blackbody radiation and Einstein's theory of the photoelectric effect, gaining widespread acceptance following Compton's experiments on photon-electron scattering in 1923 \cite{Compton:1923zz}.

Given the known components of the nucleus, it was necessary to deduce the existence of a strong force liable to hold the positively charged particles together. The first theory of the strong force is due to Yukawa in 1934 \cite{Yukawa:1935xg}. Yukawa postulated the existence of a new particle acting as a mediator of the strong force and formulated a prediction of its mass. Yukawa's theory proved inadequate to settle the question of the nature of the strong force completely, as in the following two decades, a large number of new particles and antiparticles, both weakly and strongly interacting, were discovered mainly in the study of cosmic rays.

In the 1960s, the spectra and quantum numbers of the known strongly-interacting particles were classified by Gell-Mann and Ne'eman into multiplets based on a $SU(3)$ symmetry (the ``Eightfold Way'' \cite{GellMann:1962xb}). Based on this model, Gell-Mann successfully predicted the existence of an undiscovered resonance, the $\Omega^-$, its charge, mass and decay rate. It was found as predicted in 1964 \cite{Barnes:1964pd}.
 The success of this model led Gell-Mann \cite{GellMann:1964nj} and Zweig \cite{Zweig:1981pd,Zweig:1964jf} to the interpretation of hadrons as composite objects, mesons and baryons, composed of two or three quarks. At the time, the model contained three flavours of quarks, $u$, $d$, and $s$, explaining the $SU(3)$ symmetry of the  Eightfold Way. A new quantum number, color, was introduced by Greenberg in 1964 \cite{Greenberg:1964pe} to explain the existence of the $\Delta^{++}(uuu)$, $\Delta^-(ddd)$ and $\Omega^-(sss)$ resonances. Without color as a new quantum number, the wavefunction of these resonances would have been totally symmetric, and hence prohibited for Fermions by Pauli's exclusion principle.
 Color was included formally in a Yang-Mills theory by Han and Nambu \cite{Han:1965pf}.

Early experimental evidence for the presence of substructure in the hadrons was obtained by measurements of the anomalous magnetic moment of the proton by Frisch and Stern in 1933 \cite{Frisch:1933} and of the neutron by Alvarez and Bloch in 1939 \cite{Bacher:1933,Alvarez:1940}. Hofstadter's team measured the charge distribution inside the nucleons in the 1950s \cite{Hofstadter:1963} and measured their size to be of the order of $10^{-15}\text{~m}$.

Further evidence for nucleon substructure came from the deep-inelastic scattering experiments performed at SLAC in the 1960s at higher resolution \cite{Panofsky:1968pb,Taylor:1969xi,Bloom:1969kc,Breidenbach:1969kd,Taylor:1991ew,Kendall:1991np,Friedman:1991nq}. In these experiments, based on electron-proton scattering, the measured cross-section was found to be incompatible with the proton being point-like, but was compatible with an internal structure formed by three point-like constituents. The hypothesis of the proton having a uniform charge distribution was definitely excluded.
On the theoretical front, before the deep-inelastic scattering experiments of the late 1960s, the existence of quarks was not universally accepted, because the hypotheses of fractional charge and of confinement were considered too arbitrary; the fact that no quark could be observed directly was for many physicists a reason to accept them only as a book-keeping method. Theoretical research was more focused on the deduction of general properties of the $S$-matrix than on perturbation theory or even field theory. The SLAC experiments were crucial for establishing a baseline of experimental observations that a successful theory of the strong interactions would need to explain. Among these were Bjorken scaling \cite{Bjorken:1968dy}, the observation that the structure functions are approximately independent of the exchanged momentum (it is intimately connected to asymptotic freedom) and the Callan-Gross relation \cite{Callan:1969uq}, a relation between the structure functions $F_L$ and $F_2$ of the proton. Both are successfully explained by Feynman's parton model \cite{Feynman:1969ej,Feynman:1973xc}.

Kinematically, deep-inelastic scattering is described by the exchanged momentum $q^2=-Q^2$ and by $x=Q^2/2p.q$, with $p$ the nucleon momentum (see Section \ref{sec:DIS}). Bjorken scaling refers to the observation that the structure functions, to first approximation, are independent on $Q^2$. The Callan-Gross relation corresponds to the vanishing of the longitudinal structure function, $F_L\ll F_2$. In QCD, these properties are recovered to leading order, and higher-order corrections are calculable in perturbation theory.

Bjorken scaling and the Callan-Gross relation had been predicted in 1969. At the time of Feynman's formulation of the parton model, which provided an explanation of both properties, the existence of quarks started to be commonly accepted in the scientific community. 

Further theoretical developments were the proof of renormalizability of Yang-Mills theory by 't Hooft in 1971 \cite{'tHooft:1971fh}, the formulation of QCD as a $SU(3)$ Yang-Mills theory by Gell-Mann and Fritzsch in 1972 \cite{Fritzsch:1972jv,Fritzsch:1973pi} and the computation of the $\beta$-function by Gross, Wilczek and Politzer in 1973 \cite{Politzer:1973fx,Gross:1973id} establishing the firm theoretical footing for asymptotic freedom.

The predictions of QCD are, to first order in the coupling constant, the same as those of the parton model, but become different at higher order. In particular, QCD predicts a pattern of scaling violations: a specific dependence of the experimental observables on the scale (typically the exchanged virtual momentum).
For example, Bjorken scaling is not valid beyond lowest order, and the dependence of the structure functions on $Q^2$ is calculable. Similarly, the Callan-Gross relation is not valid beyond lowest order. This feature of QCD provides a path to very precise tests of the theory.

The calculation of the theoretical ingredients for this type of analysis started when Gross and Wilczek \cite{Gross:1973ju,Gross:1974cs} and Georgi and Politzer \cite{Georgi:1951sr} first computed the anomalous dimensions of the twist-2 operators to leading order. In the framework of QCD, they account for the anomalous scaling of the parton densities.
The Wilson coefficients, which account for the partonic scattering amplitudes, were first computed in 1978 \cite{Bardeen:1978yd}, and completed in \cite{Furmanski:1981cw} enabling the calculation of the leading order QCD corrections to the parton model.

The scaling of the parton densities was formulated in $x$-space as a set of coupled integro-differential equations in \cite{Altarelli:1977zs,Dokshitzer:1977sg} and, in a fermion-pseudoscalar theory \cite{Drell:1969jm,Drell:1969wd,Drell:1969wb,Drell:1969ca}, in \cite{Gribov:1972ri}. The parton densities and the splitting functions are given in this context an intuitive interpretation related to the partonic content of the hadron and to the amplitude for collinear splitting of the partons. The splitting functions $P_{ij}$ are just Mellin transforms of the already known anomalous dimensions $\gamma_{ij}$,
\begin{equation}
	\gamma_{ij}(N) = -\int_0^1 dx \ x^{N-1} P_{ij}(x) ~.
\end{equation}

Advances in the formulation of theoretical predictions are also due to the factorization technique whereby the infrared and collinear singularities due to initial state partons are absorbed into the bare parton densities to obtain a finite renormalized parton density. The first instance of this idea is due to Politzer in 1977 \cite{Politzer:1977fi}; systematic study on factorization theorems has been performed by a number of authors, among whom Amati, Petronzio and Veneziano \cite{Amati:1978wx,Amati:1978by}, Libby and Sterman, \cite{Libby:1978qf,Libby:1978bx}, in several studies \cite{Mueller:1978xu,Ellis:1978ty,Collins:1981ta,Collins:1985ue,Bodwin:1984hc,Collins:1989gx}; see also the reviews \cite{Sterman:1995fz,Collins:2011zzd}.

Higher order results have been obtained for the deeply inelastic process since the pioneering works of the 1960s.
The Wilson coefficients for the structure functions $F_2$ and $F_L$ and the anomalous dimensions have been computed up to three-loop order in \cite{vanNeerven:1991nn,Zijlstra:1991qc,Zijlstra:1992qd,Larin:1993vu,Larin:1996wd,Retey:2000nq,Blumlein:2004xt,Moch:2004pa,Vogt:2004mw,Moch:2004xu,Vermaseren:2005qc,Blumlein:2009tj,Blumlein:2021enk,Ablinger:2017tan,Ablinger:2014vwa} in massless QCD. For the polarized case, the anomalous dimensions were calculated at LO in \cite{Altarelli:1977zs,Sasaki:1975hk,Ahmed:1976ee}, at NLO in \cite{Mertig:1995ny,Vogelsang:1995vh} and at NNLO in \cite{Moch:2014sna,Moch:2015usa,Behring:2019tus}; a calculation and definition of the M-scheme commonly used for polarized studies appears in \cite{Matiounine:1998re,Behring:2019tus}.
Polarized Wilson coefficients for the structure function $g_1$ were computed in \cite{Bodwin:1989nz,Vogelsang:1990ug}. The Wandzura-Wilczek sum rule was presented in \cite{Wandzura:1977qf}, see also \cite{Blumlein:1996tp,Blumlein:1996vs} for a modern perspective on polarized sum rules.
Higher-order results are also available for structure functions in charged-current exchange in massless QCD \cite{Moch:2008fj}.

The coefficients of the $\beta$-function in QCD are known to NLO~\cite{Caswell:1974gg,Jones:1974mm}, NNLO~\cite{Tarasov:1980au,Larin:1993tp}, N$^3$LO \cite{vanRitbergen:1997va,Czakon:2004bu} and to five-loop order \cite{Baikov:2016tgj,Herzog:2017ohr,Luthe:2017ttg,Chetyrkin:2017bjc}.

The experimental basis for the Standard Model gained another building block in 1974 with the discovery of the $J/\Psi$ resonance \cite{Aubert:1974js,Augustin:1974xw}. The new meson was convincingly interpreted as a $c\bar{c}$ resonance in the quark model, where $c$ denotes the charm quark. The existence of a fourth quark had been conjectured earlier in \cite{Bjorken:1964gz,Glashow:1970gm} in order to preserve a symmetry between leptons and quarks: this fact contributed, together with the discovery of many other charmed particles, to the general acceptance of the quark model. The bottom quark $b$ was identified in 1975 \cite{Herb:1977ek} as a constituent of the $\Upsilon$ meson, and the top quark $t$ was discovered in 1995 \cite{Abe:1995hr,D0:1995jca} at the Tevatron.

In Table \ref{tab:quark_masses} the masses of the six quarks in the $\MS$ scheme are summarized. The $u,d$ and $s$ quarks have a mass which lies in the non-perturbative regime typically characterized by $\Lambda_{QCD}\sim\mathcal O(200\mathrm{~MeV})$. For this reason, in perturbative calculations, they are typically treated as massless. The heavier $c$, $b$ quarks have a mass which is not negligible at the energies probed by scattering experiments and cannot be neglected, while the top quark has a mass much higher still, and is treated as decoupling in most perturbative calculations.

\begin{table}
\begin{centering}
\begin{tabular}{c|r}
$u$ & $2.2_{-0.4}^{+0.5}$ MeV\\[0.15cm]
$d$ & $4.7_{-0.3}^{+0.5}$ MeV\\[0.15cm]
$s$ & $95_{-3}^{+9}$ MeV\\[0.15cm]
$c$ & $1.275_{-0.035}^{+0.025}$ GeV\\[0.15cm]
$b$ & $4.18_{-0.03}^{+0.04}$ GeV\\[0.15cm]
$t$ & $160.0_{-4.3}^{+4.8}$ GeV\\
\end{tabular}
\par\end{centering}
\caption{\small \sf Quark masses quoted in \cite{Tanabashi:2018oca}. The $u$, $d$, $s$ masses are $\MS$ masses at $\mu\approx2$ GeV; the $c$, $b$ masses are the running $\MS$ masses. The $\MS$ $t$ mass is quoted in \cite{Alekhin:2017kpj}.}
\label{tab:quark_masses}
\end{table}

At intermediate energies, the heavy $c$, $b$ quarks are often treated in a variable flavour number scheme, where, depending on the value of $Q^2$, a definite number of quarks is considered light and attributed a parton density. As $Q^2$ increases, when a threshold for quark production is reached, one more quark is treated as light and attributed a parton density, while the other parton densities are redefined according to a set of matching conditions. This type of prescription enables to formulate predictions for a wide range of values of $Q^2$ across the thresholds for the heavy quarks and to limit the occurrence of large logarithms in the perturbative series, which are instead resummed by the renormalization group equation. Various such variable flavour number schemes (VFNS) have been defined in the literature \cite{Collins:1978wz,Aivazis:1993pi,Buza:1996wv,Collins:1998rz,Alekhin:2009ni,Kramer:2000hn,Cacciari:1998it,Forte:2010ta,Thorne:1997ga,Thorne:2006qt,Bonvini:2015pxa}. The matching conditions in the VFNS have been computed in \cite{Ablinger:2017err,Blumlein:2018jfm} for the two-mass case.

Experimentally, the largest kinematic range has been achieved by the HERA experiment at DESY \cite{Abt:1996hi,Ackerstaff:1998av,Derrick:1992nw,Abramowicz:2015mha}, for which the extreme values of $x$ and $Q^2$ covered were $10^{-6}<x<0.65$ and $0.45\mathrm{~GeV^2}<Q^2<50000\mathrm{~GeV^2}$. At $Q^2=10\mathrm{~GeV^2}$, the region up to around $x=10^{-4}$ could be probed \cite{Abramowicz:2015mha}.

Reviews of QCD and deep-inelastic scattering can be found in \cite{Buras:1979yt,Reya:1979zk,Altarelli:1981ax,Field:1989uq,Brock:1993sz,Yndurain,Muta:2010xua,Collins:2011zzd,Blumlein:2012bf}.
Deep-inelastic scattering experiments have been performed with polarized leptons and hadrons since the early experiments at SLAC in the 1970s~\cite{Alguard:1976bm,Baum:1980mh}, and more recently at CERN with the EMC experiment\cite{Ashman:1987hv,Ashman:1989ig}, at SLAC~\cite{Baum:1983ha,Anthony:1993uf,Anthony:1996mw,Abe:1997cx,Abe:1998wq,Anthony:1999rm,Anthony:2000fn}, at CERN~\cite{Adams:1997tq,Adeva:1998vv,Adeva:1999pa,Alexakhin:2006oza,Alekseev:2010hc,Ageev:2005gh}, CLAS~\cite{Zheng:2004ce,Dharmawardane:2006zd} and at the HERMES experiment at DESY~\cite{Ackerstaff:1997ws,Airapetian:2007mh}.
In such experiments, the lepton is longitudinally polarized and the hadron can have longitudinal or transverse polarization. For polarized scattering, a larger number of structure functions than in the unpolarized case appears: in the most general electroweak case, which was studied in~\cite{Blumlein:1998nv}, five unpolarized structure functions $F_i$ and nine polarized ones $g_i$ contribute to the hadronic tensor; in the case of pure photon exchange, however, only $F_{1,2}$ and $g_{1,2}$ appear, see also~\cite{Blumlein:1998nv,Blumlein:1996tp,Lampe:1998eu}.

The contribution due to $g_2$ is suppressed, compared to the contribution due to $g_1$, by a factor of $M^2/s$, with $M$ the mass of the hadron and $s$ the energy in the center-of-mass frame, making it harder to measure experimentally, especially in the kinematic range where higher-twist contributions are less pronounced. Because the theoretical treatment of the polarized structure functions in terms of the operator product expansion closely follows that of the unpolarized case, an opportunity exists for the extension to the structure function $g_1$ of many results already known in the literature.

In particular, the operator matrix elements of twist-two operators are important ingredients in the VFNS, where they enter in the redefinition of the parton densities across a quark mass threshold. Matching relations in the VFNS have been studied at $\mathcal{O}(a_s^2)$ in \cite{Buza:1996wv} and at $\mathcal{O}(a_s^3)$ in \cite{Bierenbaum:2009mv} for the single-mass case, where heavy quarks are decoupled one by one under the assumption $Q^2\gg m_b^2,m_c^2$.

However, because the mass ratio between the charm and the bottom quark is $\eta=m_c^2/m_b^2\sim0.1$, it is desirable for precision applications to build a VFNS with the purpose of matching the high-energy region where both quarks are decoupled and the low-energy region where they are active. To this end, it is necessary to compute operator matrix elements involving both quarks in the loops.
In QCD, operator matrix elements will contain contributions from diagrams with two different massive quarks in the loops starting at $\mathcal{O}(a_s^3)$. This causes the emergence, in their calculation, of classes of special functions and iterated integrals which depend on the ratio of the two masses.
The extension of the VFNS to the decoupling of two quarks has been studied in \cite{Blumlein:2018jfm} and the renormalization of the OMEs in \cite{Buza:1995ie} to $\mathcal{O}(a_s^2)$ and in \cite{Bierenbaum:2009mv,Ablinger:2017err} to $\mathcal{O}(a_s^3)$ for the single-mass and the two-mass case respectively.

The calculation of the matrix elements of twist-two operators has historically been one of the methods by which their anomalous dimensions have been obtained, starting with \cite{Gross:1974cs,Georgi:1951sr}. 
The calculation of the unpolarized single-mass OMEs has been performed to $\mathcal{O}(a_s^2)$ in \cite{Buza:1995ie,Buza:1996wv,Bierenbaum:2007dm,Bierenbaum:2007qe,Bierenbaum:2008yu,Bierenbaum:2009zt} and to $\mathcal{O}(a_s^3)$ in \cite{Blumlein:2006mh,Ablinger:2010ty,Ablinger:2014lka,Ablinger:2014nga,Behring:2014eya,Ablinger:2014vwa,Ablinger:2014uka,Ablinger:2017ptf}, and of the two-mass OMEs in \cite{Ablinger:2011pb,Ablinger:2012qj,Ablinger:2017err,Ablinger:2017xml,Ablinger:2018brx}. The polarized OMEs have been computed in \cite{Buza:1996xr,Bierenbaum:2007zz,Klein:2009ig} and to $\mathcal{O}(a_s^3)$ in \cite{Ablinger:2019etw,Schonwald:2019gmn,Behring:2021asx}. Currently, up to $\mathcal O(a_s^3)$, only the OME $A_{Qg}^{(3)}$ is not fully known, and is likely to fall in the function space of elliptic integrals, at least for some color structures.

From the factorization theorems of QCD, it is possible to obtain the asymptotic form of the Wilson coefficients involving a massive quark, i.e.\ the coefficients of the logarithmic factors $\ln(Q^2/\mu^2)$ and $\ln(Q^2/m^2)$, from the factorization into massive OMEs and massless Wilson coefficients. This method was pioneered in \cite{Buza:1996wv} where the asymptotic two-loop charm contributions to the Wilson coefficients were calculated. Those asymptotic results were found to be in agreement with an analytic calculation in \cite{Blumlein:2016xcy} and with the small-$x$ limit derived in \cite{Catani:1990eg}. To $\mathcal O(a_s^3)$ the logarithmic terms of massive Wilson coefficients have been calculated in \cite{Kawamura:2012cr,Behring:2014eya,Ablinger:2014vwa,Behring:2015zaa}.
For charged-current interactions, the massive Wilson coefficients have been computed with the method of massive OMEs in \cite{Gluck:1997sj,Blumlein:2011zu,Blumlein:2014fqa}, correcting results of \cite{Buza:1997mg}.

The factorization of mass singularities has also found applications in QED: initial-state radiative corrections to the process $e^+ e^-\rightarrow\gamma^*/Z^*$ have been calculated with the method of massive OMEs \cite{Blumlein:2011mi}, which has historically been important to cross-check and to correct a direct computation performed in \cite{Berends:1987ab} and to obtain radiator functions to high orders \cite{Blumlein:2019pqb,Blumlein:2020jrf,Blumlein:2019srk}. Nowadays, these QED results allow a very precise determination of the $Z$-boson width and of the asymmetry in $e^+ e^-$ annihilation \cite{Ablinger:2020qvo,Blumlein:2021jdl}.

This thesis reports on computations which fit into the long-running project of calculating OMEs and heavy-quark corrections to the Wilson coefficients. It is organized as follows:
in Chapter \ref{sec:basic} we review the theoretical basis of deep-inelastic scattering, of the renormalization of the OMEs of twist-two operators, and the definition of the VFNS. We also briefly review the mathematics of nested sums, iterated integrals and of the Mellin transform.
In Chapter \ref{sec:PolDIS}, we describe the calculation of two of the polarized OMEs, namely the contributions to $A_{Qq}^{(3),\text{PS}}$ and $A_{gg,Q}^{(3)}$ with two different quark masses.
For $A_{Qq}^{(3),\text{PS}}$ the result, given in Chapter \ref{sec:AQq3PS}, is obtained by Feynman parametrization and Mellin-Barnes integration. The Mellin-Barnes integrals are calculated by the residue theorem, which turns the integral into a sum, which is then treated using techniques in summation theory. The calculation closely follow that of the unpolarized OME \cite{Ablinger:2019gpu}. Working in the Larin scheme \cite{Larin:1993tq} for the treatment of $\gamma_5$ in dimensional regularization, we computed the OME and compared its poles in the dimensional parameter $\ep$ to the known prediction which is obtainable from the knowledge of OMEs to lower perturbative order and of the renormalization structure of the theory. We could confirm the pole structure of the OME. The constant part in $\ep$ is new and is given in $x$-space in semi-analytic form as iterated integrals over an alphabet which contains root-valued expressions. The $N$-space result is not given, since the respective recurrences are not first-order factorizable and hence the solution falls outside of the function space under consideration. The calculation was published in \cite{Ablinger:2019gpu}.

In Chapter \ref{sec:AggQ3} we calculate the polarized OME $A_{gg,Q}^{(3)}$ in the Larin scheme. The method is the same as for $A_{Qq}^{(3),\text{PS}}$, and closely follows the unpolarized calculation \cite{Ablinger:2018brx}. Here, we could compute the analytic $N$-space result as well, as a function of the ratio $\eta$ of the squared quark  masses. The OME turns out to be expressible in nested harmonic and binomial sums. We also derive the $x$-space result in semi-analytic form using iterated integrals, in a form suitable for numerical evaluations. We also derive a number of identities which re-express many of these iterated integrals into the more familiar harmonic polylogarithms and multiple polylogarithms, which are less cumbersome for numerical evaluation. We review the mathematical objects in which these results are expressed, namely nested sums containing binomial coefficients involving $\eta$ as summands, and iterated integrals whose alphabet contains square roots, at arguments containing~$\eta$. These results were published in \cite{Ablinger:2020snj}.

In Chapter \ref{sec:logWC} we present the asymptotic form, for $Q^2\gg m^2$, of the polarized single-mass Wilson coefficients for the structure function $g_1$ in the Larin scheme. These single-mass Wilson coefficients were obtained by the factorization theorems which state that the massive corrections can be obtained asymptotically as the product of the massive OMEs and the massless Wilson coefficients. The known OMEs and Wilson coefficients allow us to write the logarithmic terms at $\mathcal O(a_s^3)$ with the exclusion of the constant part. These results were published in \cite{Blumlein:2021xlc}.
In Chapter \ref{sec:schemeinv} we describe the the scheme-invariant evolution of the structure functions $F_2^{NS}$ and $g_1^{NS}$ in the asymptotic region accounting for the effects of the $c$ and $b$ quarks to N$^3$LO. These results were published in \cite{Blumlein:2021lmf}.

In Chapter \ref{sec:hyperg} we describe methods to classify some differential systems of hypergeometric type. These systems are obeyed by multivariate hypergeometric series and are in correspondence with difference equations with shifts, obeyed by the summand, involving one variable only. We discuss such a case and describe an algorithm to solve the system of differential equations and recover the summand as a nested hypergeometric product. We discuss how this algorithm works in the classical cases of functions studied by Appell, Horn,  Lauricella, and Exton, as these functions have been used in the physics literature in the context of calculating Feynman integrals. In \cite{Blumlein:2021hbq} we provide a computer algebra package which implements the algorithm and a computer-readable list of these classical functions and of the systems obeyed by them. We also give some examples of how the series expansion of some hypergeometric series can be obtained using the package {\tt Sigma} by Schneider \cite{Schneider:2001,SIG1,SIG2} and of the types of functions arising in those examples. 

The calculation of OMEs to high loop order requires the solution and classification of nested sums, arising for instance from the Mellin-Barnes integration method. These sums, which in physics applications can be very numerous and very complicated, need to be solved, or, in other words, classified in a minimal set of simpler objects, the simplest of which are the harmonic sums. This classification is often done in summation theory by deriving and solving difference equations, through a number of techniques falling under the name of telescoping (see \cite{Petkovsek:1996} for a survey and \cite{Vermaseren:2000we,Moch:2001zr,Moch:2005uc,Bierenbaum:2007qe} for applications to OMEs). The package {\tt Sigma} encodes such techniques for solving univariate difference equations in this context. In principle, the Laporta algorithm \cite{Laporta:2001dd} also gives rise to difference equations. Motivated by the application to physics of univariate difference equations in deep-inelastic scattering, in Chapter~\ref{sec:PLDEsolver} we review the problem of partial linear difference equations in several variables. To date, only a limited number of algorithms are available towards a solution of such equations. We describe a computer algebra implementation of one known approach which targets the solution space of rational functions, possibly containing harmonic sums or Pochhammer symbols in the numerator. The package has been released in \cite{Blumlein:2021hbq}.

In Chapter \ref{sec:program} we describe a numerical library in Fortran which encodes the splitting functions up to NNLO and massless and asymptotic massive Wilson coefficients for the structure functions $F_2$ and $g_1$ for photon exchange, and $F_3^{W^+\pm W^-}$. It also includes Fortran routines for the Wilson coefficients of the polarized and unpolarized Drell-Yan process, as well as for scalar and pseudoscalar Higgs boson production. The library works in $N$-space by encoding the analytic continuation of the Mellin transforms of harmonic polylogarithms, sufficient to evaluate the analytic continuation, through the even or odd moments, of the harmonic sums up to weight 5 and in several cases to weight 6. The library is suitable for the numerical evolution of singlet and non-singlet PDFs given a programmable initial parametrization. It is usable in principle for experimental fits of these structure functions. We study the numerical precision attained over the computation of low integer moments and show the evolution of a test input set of unpolarized and polarized PDFs.

\clearpage
\section{Basic formalism}
\label{sec:basic}

\subsection{Deep-inelastic scattering}
\label{sec:DIS}
\begin{figure}[h]
\centering
\includegraphics[scale=0.8]{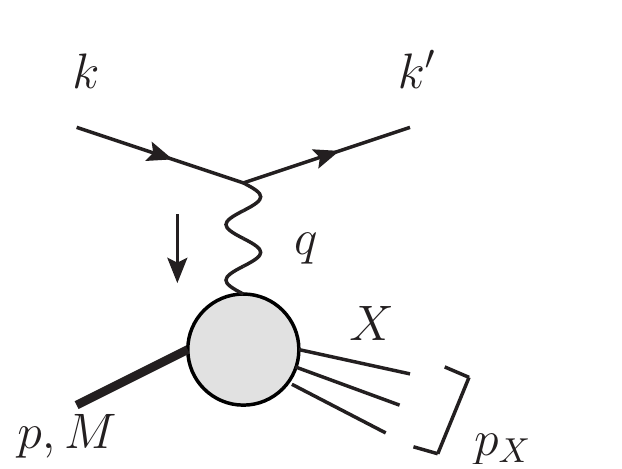}
\caption{\small \sf Kinematics in DIS}
\label{fig:DIS}
\end{figure}

Deep-inelastic scattering (DIS) is the scattering process between a lepton, with momentum $k$, and a hadron, with momentum $p$, in a specific kinematic regime (Figure \ref{fig:DIS}). The lepton interacts with a quark by exchanging an electro-weak boson and a final state is produced. We denote the momentum of the outgoing lepton by $k'$ and the momentum of the final hadronic state by $p_X$.
In many inclusive experiments, only $k'$, but not $p_X$, is measured. A measurement of also $p_X$ has been first possible in the collider experiments H1 and ZEUS at HERA.

We define $q=k-k'$ the momentum of the vector boson. In the reference
frame of the hadron, $p=(M,0)$, $k=(E_{k},\vec{k}).$ Neglecting
the electron mass, we have:
\begin{equation}
q^{2}=(k-k'^{2})=-2E_{k}E'_{k}(1-\cos\theta)=-4E_{k}E'_{k}\sin^{2}\frac{\theta}{2}<0.
\end{equation}
It is customary to use the variables 
\begin{eqnarray}
Q^2   &=& -q^2 > 0, \\
\nu   &=& \frac{p. q}{M}, \\
x     &=& \frac{Q^2}{2p. q}, \\
y     &=& \frac{p.q}{p.k}, \\
W^2   &=& p_X^2
\end{eqnarray}
to define the kinematics of the process.
The variables $x$ and $y$ are called Bjorken variables.
One refers to deep-inelastic scattering if the process occurs in a kinematic region where $Q^2 \gtrsim 4\text{~GeV}^2$ and $W^2 \gtrsim 4\text{~GeV}^2$ \cite{Blumlein:2012bf}, a region where the picture offered by perturbative QCD becomes applicable. Further cuts are typically applied in order to limit the size of power corrections $\mathcal O(M^2/Q^2)$ which would otherwise be visible in the experimental data, but subleading with respect to the logarithmic terms most readily obtainable in QCD.
Target mass corrections, which are one class of such subleading terms, have been discussed in \cite{Nachtmann:1973mr,Georgi:1976ve} and in the context of experimental fits in \cite{Blumlein:2006be,Blumlein:2010rn}. In this thesis, they will be neglected.

Computing the amplitude for the process outlined above, one obtains:
\begin{equation}
i\mathcal{M}=L^{\mu}\frac{-i}{q^{2}}\int dxe^{iq.x}\langle X|J_{\mu}(x)|P\rangle=L^{\mu}\frac{-i}{q^{2}}W_{\mu},
\end{equation}
where $L^{\mu}$ represents the leptonic contribution to the amplitude,
$J^{\mu}$ is the electromagnetic current $J^{\mu}=e_q\bar{q}\gamma^{\mu}q$ in case of pure photon exchange,
and $e_q$ is the quark charge. Squaring the amplitude requires us to
examine the quantities $L^{\mu\nu}$ and $W_{\mu\nu}$. The leptonic
tensor has the form $L^{\mu\nu}\propto\mathrm{Tr}[/\!\!\!k\Gamma^{\mu}/\!\!\!k'\Gamma^{\nu}]$,
$\Gamma^{\mu}$ being the lepton-boson vertex coupling. Because the
hadron is a composite object, such an explicit formula cannot be written
about the hadronic tensor
\begin{align}
W_{\mu\nu}(p,q) & =\sum_{X}(2\pi)^{4}\delta^{4}(p_{X}-p-q)\langle P|J_{\mu}(x)|X\rangle\langle X|J_{\nu}(x)|P\rangle\\
 & =\int d^{4}xe^{iq.x}\langle P|J_{\mu}(x)J_{\nu}(0)|P\rangle.\label{eq:w-tensor}
\end{align}
Nevertheless, its form can be constrained by considering its Lorentz
structure, because, due to the Ward identity and the conservation of the electromagnetic current, $q^{\mu}W_{\mu\nu}=q^{\nu}W_{\mu\nu}=0$.
This restriction allows us to write in general
\begin{eqnarray}
W_{\mu\nu}(p,q) & = & \Bigl(-g_{\mu\nu}+\frac{q_{\mu}q_{\nu}}{q^{2}}\Bigr)F_1(x,Q^2)\nonumber \\
 && + \frac{2x}{Q^2} \Bigl(p_{\mu}-\frac{p.q}{q^2}q_{\mu}\Bigr)\Bigl(p_{\nu}-\frac{p.q}{q^2}q_{\nu}\Bigr)F_2(x,Q^{2}) \nonumber \\
 && + i \varepsilon_{\mu\nu\alpha\beta} \frac{p^\alpha q^\beta}{p.q} F_3(x,Q^2).\label{eq:w-tensor-1}
\end{eqnarray}
The scalar functions $F_i(x,Q^{2})$ are known as \emph{structure functions} and
\begin{equation}
	\ep_{\mu\nu\alpha\beta} = 	\begin{cases} 
								\text{sign}(\sigma) 	& \text{if~} (\mu\nu\alpha\beta) = \sigma(0123), \\
								0						& \text{otherwise},
								\end{cases}
\end{equation}
for $\sigma$ a permutation, is the Levi-Civita tensor.

Formula \eqref{eq:w-tensor-1} is valid if we restrict the analysis to electromagnetic currents. In the case of unpolarized nucleon targets, the term with the Levi-Civita tensor does not contribute, unless weak interactions are considered. In the case of polarized nucleon targets, the hadronic tensor also acquires an antisymmetric part, and the structure functions $g_1$ and $g_2$ are defined by \cite{Blumlein:1996vs,Blumlein:1998nv}
\begin{equation}
	W_{\mu\nu}^A = i \ep_{\mu\nu\lambda\sigma} \Bigl[ \frac{q^\lambda S^\sigma}{p.q} g_1(x,Q^2) + \frac{q^\lambda }{(p.q)^2} (p.q \ S^\sigma - S.q \ p^\sigma) g_2(x,Q^2) \Bigr],
\end{equation}
with $S$ the nucleon spin 4-vector normalized as $S^2=-M^2$.

In the case of pure photon exchange on unpolarized targets, the structure functions can be mapped to the differential cross section of deep-inelastic scattering by \cite{Blumlein:2012bf}
\begin{equation}
	\frac{d^2\sigma^{\gamma,\text{unpol.}}}{dx dy} = \frac{2\pi\alpha^2}{xy Q^2} \Big\{\big[1+(1-y)^2\big] F_2(x,Q^2) -y^2 F_L(x,Q^2) \Big\}
\end{equation}
and therefore can be measured experimentally; they are observables. In the literature, different definitions and normalizations for the structure functions are used; here we follow \cite{Blumlein:2012bf}.
For completeness we repeat, for the case of pure photon exchange on unpolarized targets, the relation between the structure functions and the differential cross-section, \cite{Blumlein:1996vs}
\begin{align}
	\frac{d^2\sigma^{\gamma,\text{pol.}}(\lambda,\pm S_L)}{dx\ dy} &= \pm 2 \pi s \frac{\alpha^2}{Q^4}
\left [
-2\lambda y \left( 2-y-\frac{2 x y M^2}{s} \right) x g_1(x,Q^2) +
8 \lambda \frac{y x^2 M^2}{s} g_2(x,Q^2)
\right],
\label{eq:dsigpolL} \\
	\frac{d^3\sigma^{\gamma,\text{pol.}}(\lambda,\pm S_T)}{dx\ dy\ d\phi} &= \pm \frac{2\alpha^2 s}{Q^4} \sqrt{\frac{M^2}{s}} \sqrt{xy\Bigl(1-y-\frac{xyM^2}{s}\Bigr)} \cos(\theta-\phi) 
\nonumber\\&
	\times \bigl[-2\lambda xy g_1(x,Q^2) -4\lambda x g_2(x,Q^2) \bigr],
\label{eq:dsigpolT}
\end{align}
with $\alpha$ the fine structure constant, $s$ the energy in the center of mass frame, $\lambda$ the helicity of the incoming lepton, $S_{L,T}$ the spin vector of the longitudinally or transversally polarized nucleon, which are 
\begin{align}
	S_L &= (0,0,0,M), \\
	S_T &= M(0,\cos\theta,\sin\theta,0)
\end{align}
in the nucleon rest frame, $\phi$ is the azimuthal angle.

Instead of $F_1$, in the literature it is common to study the longitudinal
structure function
\begin{equation}
F_L(x,Q^2) = F_2(x,Q^2) - 2x F_1(x,Q^2) .
\label{eq:fl}
\end{equation}

By projecting the hadronic tensor \eqref{eq:w-tensor-1} with $g_{\mu\nu}$ and with $p^\mu p^\nu$ and setting $p^2=0$ one can write \cite{Klein:2009ig}
\begin{eqnarray}
	g^{\mu\nu}  W_{\mu\nu}(p,q) &=& \frac{2-D}{2x}   F_2(x,Q^2) + \frac{D-1}{2x} F_L(x,Q^2) , \\
	p^\mu p^\nu W_{\mu\nu}(p,q) &=& \frac{Q^2}{8x^3} F_L(x,Q^2) ,
\end{eqnarray}
with
\begin{equation}
D=4+\ep
\end{equation}
the dimensions of spacetime. These are inverted as
\begin{eqnarray}
	F_2(x,Q^2) &=& \frac{2x}{D-2} \Big[(D-1)\frac{4x^2}{Q^2} p^\mu p^\nu W_{\mu\nu}(p,q) - g^{\mu\nu}W_{\mu\nu}(p,q) \Big]~, \\
	F_L(x,Q^2) &=& \frac{8x^3}{Q^2} p^\mu p^\nu W_{\mu\nu}(p,q)~.
\end{eqnarray}

The process known as deep-inelastic scattering refers to the kinematic
region of $Q^{2}\rightarrow\infty$ as $x$ is kept finite, the Bjorken
limit \cite{Bjorken:1968dy}. In this region it is possible to apply the
methods of perturbative QCD, due to the asymptotic freedom of the
theory.


\subsection{Light-cone dominance}

Consider \cite{Muta:2010xua} the quantity
\begin{equation}
\int d^{4}xe^{iq.x}\langle P|J_{\nu}(0)J_{\mu}(x)|P\rangle\label{eq:current-product-is-zero}
\end{equation}
where the two currents have been interchanged with respect to their order
in $W_{\mu\nu}$. In the physically allowed region, $q^{0}=E-E'>0$.
It can be shown that in this region the quantity (\ref{eq:current-product-is-zero})
is zero. Inserting a complete set of states, one obtains
\begin{equation}
\int d^{4}xe^{iq.x}\langle P|J_{\nu}(0)J_{\mu}(x)|P\rangle=\sum_{X}(2\pi)^{4}\delta^{4}(q-p+p_{X})\langle P|J_{\mu}(x)|X\rangle\langle X|J_{\nu}(x)|P\rangle.
\end{equation}
One can prove that if $q^{0}>0$ then $q-p+p_{X}\neq0$: in the rest
frame of the proton, assuming the equality holds,
\begin{equation}
(q-p)^{2}=p_{X}^{2}\rightarrow q^{2}-2q^{0} M+M^{2}=p_{X}^{2}\rightarrow q^{0}=\frac{1}{2M}(q^{2}+M^{2}-p_{X}^{2}).
\end{equation}
In the physically allowed region, $q^{2}<0$ and $p_{X}^{2}>M^{2}$,
so it is impossible to have $q^{0}>0$.

As a consequence we can rewrite (\ref{eq:w-tensor}) as
\begin{equation}
W_{\mu\nu}=\int d^{4}xe^{iqx}\langle P|[J_{\mu}(x)J_{\nu}(0)]|P\rangle.
\end{equation}
Because of causality, the commutator must vanish for $x^{2}<0$. Additionally,
it can be shown that in the deep-inelastic scattering limit where
$q^{2}\rightarrow-\infty$ and $Q^{2}/2p.q\rightarrow\textrm{constant}$,
the dominant contribution to $W_{\mu\nu}$ is due to the region $0\leq x^{2}<1/Q^{2}$.

This statement is known as \emph{light-cone dominance}: the hadronic
tensor $W_{\mu\nu}$ receives contributions from the product of currents
$J_{\mu}(x)J_{\nu}(0)$ which are dominated by the region $x^{2}\sim0$.

\subsection{The operator product expansion}

The product of two composite operators $O(x_{1})O(x_{2})$ can become
singular in certain limits. Wilson \cite{Wilson:1969zs} considered the
limit $x_{1}\rightarrow x_{2}$ and postulated that such a product, when singular, can be expanded as a linear combination of all other operators $O_{i}$
appearing in the theory which are finite in the limit, with the singular
behaviour encoded in singular coefficient functions $C_{i}(x)$:
\begin{equation}
O(x_{1})O(x_{2})\rightarrow\sum_{i}C_{i}(x)O_{i}(x)\qquad\textrm{as }x_{1}\rightarrow x_{2}.
\end{equation}
For the application to deep inelastic scattering \cite{Symanzik:1971vw,Christ:1972ms,Callan:1972uj},
such an expansion is needed for the light-cone region of the product
of two currents:
\begin{align}
J_{\mu}(x)J_{\nu}(0) & \rightarrow g_{\mu\nu}\left(\frac{\partial}{\partial x}\right)^{2}\sum_{i,n}C_{i,1}^{(n)}(x^{2})x^{\mu_{1}}\cdots x^{\mu_{n}}O_{\mu_{1}\cdots\mu_{n}}^{(i)}(0)\nonumber \\
 & +\frac{1}{x^{2}}\sum_{i,n}C_{i,2}^{(n)}(x^{2})x^{\mu_{1}}\cdots x^{\mu_{n}}O_{\mu\nu\mu_{1}\cdots\mu_{n}}^{(i)}(0)+\cdots\label{eq:ope}
\end{align}

In a massless, free-field theory, it is possible to find the behaviour
of the singular functions $C_{i}(x^{2})$ around $x^{2}=0$ by a power-counting
argument: call $d_{J}$ the mass dimension of the current $J(x)$
and $d_{O}^{(i)}(n)$ that of $O_{\mu_{1}\cdots\mu_{n}}^{(i)}$. Then,
from (\ref{eq:ope}) it follows that the mass dimension of $C_{i}(x^{2})$
is $[C_{i}(x^{2})]=2d_{J}+n-d_{O}^{(i)}(n)$ and
\begin{equation}
C_{i}(x^{2})\sim(x^{2})^{-d_{J}-n/2+d_{O}(n)/2}.
\end{equation}
Thus, the operators corresponding to the minimum value of 
\begin{equation}
\tau_{n}^{(i)}=d_{O}^{(i)}(n)-n,
\end{equation}
a quantity called \emph{twist}, will be dominant in the light-cone expansion. The dominant operators $O_{\mu_{1}\cdots\mu_{n}}^{(i)}$ therefore have twist 2. They are traceless and symmetric, and have definite spin $n$ and dimension $n+2$. The operators can be explicitly written:
\begin{align}
O_{q;\mu_1\cdots\mu_n}^{S}(x) & =i^{n-1} \mathbf S \bigl[ \overline{\psi}(x)\gamma_{\mu_1}D_{\mu_2}\cdots D_{\mu_n}\psi(x) \bigr] - \text{trace terms} \nonumber \\
O_{q;\mu_1\cdots\mu_n}^{NS,(i)}(x) & =i^{n-1} \mathbf S \bigl[ \overline{\psi}(x)\gamma_{\mu_1}D_{\mu_2}\cdots D_{\mu_n}\lambda^{(i)}\psi(x) \bigr] - \text{trace terms} \label{eq:operators}\\
O_{g;\mu_1\cdots\mu_n}^{S}(x) & =i^{n-1} \mathbf {S\,S_p} \bigl[ F_{\mu_1\nu}^a(x) D_{\mu_2}\cdots D_{\mu_{n-1}}F_{\mu_n}^a{}^{\nu}(x) \bigr] - \text{trace terms} \nonumber 
\end{align}
where $\mathbf S$ stands for a symmetrization in the Lorentz indices
\begin{equation}
\mathbf S f_{\mu_{1}\cdots\mu_{n}}=\frac{1}{n!}(f_{\mu_{1}\cdots\mu_{n}}+\textrm{permutations}),
\end{equation}
$\mathbf{S_p}$ is  the trace over $SU(N_c)$ and $\lambda^{(i)}$ are the $SU(N_f)$ generator matrices if we assume the theory to have such a flavour symmetry. In equations (\ref{eq:operators}), the subtraction of trace terms contains factors of $g_{\mu_i\mu_j}$; it is needed in order to make the operators have definite spin. Here, $\psi(x)$ and $F_{\mu\nu}$ denote respectively the quark field and the electromagnetic field strength, and $D_\mu$ is the covariant derivative,
\begin{eqnarray}
D_\mu &=& \partial_\mu-igt^aA_\mu^a,\\
F_{\mu\nu}^a &=&\partial_\mu A_\nu^a-\partial_\nu A_\mu^a +gf^{abc}A_\mu^b A_\nu^c,
\end{eqnarray}
where $t^a$ are the generators of $SU(3)$ in the fundamental representation and $f^{abc}$ are the structure constants.

In the case of polarized scattering, the contributing twist-two operators are \cite{Mertig:1995ny,Sasaki:1975hk}
\begin{align}
	O_{q;\mu_1\cdots\mu_n}^{S,5}(x) & =i^n \mathbf S \bigl[ \overline{\psi}(x)\gamma_5 \gamma_{\mu_1}D_{\mu_2}\cdots D_{\mu_n}\psi(x) \bigr] - \text{trace terms} \nonumber \\
	O_{q;\mu_1\cdots\mu_n}^{NS,(i),5}(x) & =i^n \mathbf S \Bigl[ \overline{\psi}(x) \gamma_5 \gamma_{\mu_1}D_{\mu_2}\cdots D_{\mu_n} \frac{\lambda^{(i)}}{2} \psi(x) \Bigr] - \text{trace terms} \label{eq:pol-operators}\\
O_{g;\mu_1\cdots\mu_n}^{S,5}(x) & =i^n \mathbf {S\,S_p} \Bigl[ \frac{1}{2} \varepsilon^{\mu_1\alpha\beta\gamma} F_{\beta\gamma}^a(x) D_{\mu_2}\cdots D_{\mu_{n-1}}F_{\alpha\mu_n}^a(x) \Bigr] - \text{trace terms} .\nonumber 
\end{align}

\subsection{The forward Compton amplitude}

The optical theorem states that the imaginary part of an amplitude can be related to the amplitude for scattering into all possible final states. The theorem can be applied \cite{Christ:1972ms} in the case of the hadronic tensor to relate it to \emph{forward virtual Compton scattering}, whose amplitude is determined by
\begin{equation}
T_{\mu\nu}=i\int d^{4}x\,e^{iq.x}\langle P|T[j_{\mu}(x)j_{\nu}(0)]|P\rangle.\label{eq:t-mu-nu-definition}
\end{equation}

The theorem states that
\begin{equation}
\frac{1}{2\pi}\textrm{Im}(T_{\mu\nu})=W_{\mu\nu},
\end{equation}
since $\textrm{Im}(T_{\mu\nu})$ is equal to the discontinuity of the tensor in the $q^{0}$ plane. The forward Compton scattering can be more readily computed in a perturbative expansion, and can be related to the hadronic tensor by a dispersion integral: by calling 
\begin{equation}
\omega=\frac{1}{x}=\frac{2p.q}{Q^{2}},
\end{equation}
we have
\begin{equation}
2\int_{0}^{1}dx\,x^{n-2}W_{\mu\nu}=\frac{2}{\pi}\int_{1}^{\infty}\frac{d\omega}{\omega^{n}}\text{Im}(T_{\mu\nu})=\frac{1}{2\pi i}\int_{C}d\omega\frac{T_{\mu\nu}}{\omega^{n}},\label{eq:mellin-w}
\end{equation}
and $C$ is a contour that circles around the branch cuts of $T_{\mu\nu}$ in the $\omega$ plane. Noting the formula
\begin{equation}
\frac{1}{2\pi i}\int_{C}d\omega\,\omega^{m-n}=\delta_{m,n-1}
\end{equation}
it follows that Eq. (\ref{eq:mellin-w}) will pick one power in a series expansion of $T_{\mu\nu}$.

To the forward Compton scattering tensor $T_{\mu\nu}$ one can apply arguments related to an operator product expansion in a way analogous to those applicable to $W_{\mu\nu}$. We report here the results of this operator product expansion as presented in the textbook by Muta \cite{Muta:2010xua}, see also \cite{Moch:1999eb}:
\begin{align}
T[j_{\mu}(x)j_{\nu}(x')]= & (\partial_{\mu}\partial_{\nu}'-g_{\mu\nu}\partial.\partial')O_{L}(x,x')\nonumber \\
 & +(g_{\mu\lambda}\partial_{\rho}\partial_{\nu}'+g_{\rho\nu}\partial_{\mu}\partial_{\lambda}'-g_{\mu\lambda}g_{\rho\nu}\partial.\partial'-g_{\mu\nu}\partial_{\lambda}\partial_{\rho}')O_{2}^{\lambda\rho}(x,x')\\
 & +\textrm{ antisymmetric terms.}\nonumber 
\end{align}
The antisymmetric terms only contribute to polarized scattering processes.
In analogy with (\ref{eq:ope}), the operators have the light-cone expansion
\begin{align}
O_L(x,x') & =\sum_{i,n}C_{L,n}^{(i)}(y^2)y^{\mu_1}\cdots y^{\mu_n}O_{L\mu_1\cdots\mu_n}^{(i)}\Big(\frac{x+x'}{2}\Big)\\
O_2^{\lambda\rho}(x,x') & =\sum_{i,n}C_{2,n}^{(i)}(y^2)y^{\mu_1}\cdots y^{\mu_n}O_{2\mu_1\cdots\mu_n}^{(i)\lambda\rho}\Big(\frac{x+x'}{2}\Big)
\end{align}
where $y=x-x'$, and $i$ runs over the operators in \eqref{eq:operators}. Applying the expansion to (\ref{eq:t-mu-nu-definition}),
one obtains
\begin{equation}
T_{\mu\nu}=2\sum_{i,n}\omega^n\left[\left(g_{\mu\nu}-\frac{q_\mu q_\nu}{q^2}\right)A_{L,n}^{(i)}C_{L,n}^{(i)}(Q^2)+d_{\mu\nu}A_{2,n}^{(i)}C_{2,n}^{(i)}(Q^2)\right],\label{eq:t-tensor-a-c}
\end{equation}
with
\begin{equation}
	d_{\mu\nu} = - g_{\mu\nu} - \frac{p_\mu p_\nu}{(p.q)^2} q^2 + \frac{p_\mu q_\nu+p_\nu q_\mu}{p.q} ,
\end{equation}
\begin{equation}
C_{(2,L),n}^{(i)}(Q^{2})q^{\mu_{1}}\cdots q^{\mu_{n}}/(q^{2})^{n}=i\int d^{4}x\,e^{iq.x}C_{(2,L),n}^{(i)}(x^{2})
\end{equation}
and 
\begin{equation}
A_{L,n}^{(i)}p_{\mu_1}\cdots p_{\mu_n}+\textrm{terms with }g_{\mu_i\mu_j}=\langle P|O_{L\mu_1\cdots\mu_n}^{(i)}(0)|P\rangle,
\end{equation}
with a similar formula for $A_{2,n}^{(i)}$.

From equations \eqref{eq:w-tensor-1}, \eqref{eq:fl},
\eqref{eq:mellin-w}, \eqref{eq:t-tensor-a-c}, and applying the replacement
$n\rightarrow N+1$ to harmonize different conventions present in
the literature, one finally deduces the master formula
\begin{equation}
\int_0^1 dx\,x^{N-1}F_{(2,L)}(x,Q^2) = \sum_i A_{(2,L),N}^{(i)} \Bigl( \frac{p^2}{\mu^2} \Bigr) \, C_{(2,L),N} ^{(i)}\Bigl(\frac{Q^2}{\mu^2}\Bigr) ~.\\
\label{eq:fact-struct-func}
\end{equation}
The operation on the left-hand side is called the Mellin transform,
and can be inverted to give an explicit formula for the structure
functions:
\begin{equation}
F_{(2,L)}(x,Q^{2})=\frac{1}{2\pi i}\int_{c-i\infty}^{c+i\infty}dN\,x^{-N}\sum_iA_{(2,L),N}^{(i)}\Bigl( \frac{p^2}{\mu^2} \Bigr) C_{(2,L),N}^{(i)}\Bigl(\frac{Q^2}{\mu^2}\Bigr).
\label{eq:f-inv-mellin}
\end{equation}

An interesting property with practical consequences for actual computations \cite{Blumlein:1996vs,Politzer:1974fr} is the symmetry
$T_{\mu\nu}(-\omega)=T_{\mu\nu}(\omega)$.
In the unpolarized case, this has the consequence that $C_{(2,L),N}(Q^{2})=0$ for odd $N$. In the case of scattering over a polarized target, instead, $C_{g_1,N}(Q^{2})=0$ for even $N$.

The importance of formula \eqref{eq:f-inv-mellin} lies in the fact that the functions $C_{N}(Q^{2})$, called \emph{Wilson coefficients}, depend only on short-distance physics, and, for this reason, can be computed in perturbative QCD because of the property of asymptotic freedom. Asymptotic freedom refers to the fact that the running coupling constant goes to zero at high energy: this makes the computation of physical quantities possible as the expansion parameter $a_s = \alpha_s/(4\pi)$ is small. By contrast, the quantities $A_{N}$, whose Mellin inverse is related to the parton density through a factorization procedure, depend on long-distance physics: they are the matrix elements between the target hadron state of the operators appearing in the operator product expansion. 

We will discuss later in greater detail how parton densities are defined and also treat possible collinear and mass singularities present in the operator matrix elements. Broadly speaking, in calculations these singularities arise from the initial-state partons, over which the deep-inelastic scattering process is not inclusive. In reality, the initial state partons are confined inside the hadron, and are not themselves asymptotic states. It is therefore reasonable to assume that the initial state infrared and collinear singularities that arise when we compute with initial-state partons are manifestations of the inapplicability of perturbation theory to the long-range nonperturbative regime, and are actually shielded by non-perturbative physics. Therefore, it is sensible to reabsorb them into the unobservable, bare parton density, which encodes the long-distance physics. After this step, the structure function is postulated to be a Mellin convolution of a finite parton density and a finite Wilson coefficient along the lines of \eqref{eq:f-inv-mellin}, and the renormalized parton density becomes scale dependent. The procedure, first applied in \cite{Politzer:1977fi}, is analogous to what happens when renormalizing the coupling constant.

The parton densities cannot be computed perturbatively and must be extracted from experimental data. This can be accomplished by measuring the structure functions experimentally and computing the Wilson coefficients, then applying (\ref{eq:f-inv-mellin}) in a suitable form.

Still, there is predictive value in factorizing the structure function (or its Mellin moments), as one can assume that the $A_{N}$ are universal (see for a discussion \cite{Buras:1979yt,Brock:1993sz}): they will not depend on which scattering process is being studied, and, once they are deduced for a given hadronic target and leptonic probe, they can be used in predictions for processes which involve a different leptonic probe, provided that the virtuality $\mu^2$ does not change sign.

One important remark is the existence of Carlson's theorem \cite{Titchmarsh:1932,Carlson:1914}, which implies that the analytic continuation of the Mellin transform in the complex plane can be derived given the knowledge of the even or odd moments. Extensions and applications of Carlson's theorem to this physical case are derived in \cite{Ablinger:2013cf}.

\subsection{Scaling violation and renormalization group}

Let us describe Eq. \eqref{eq:f-inv-mellin} in greater detail. To the operators \eqref{eq:operators} correspond three parton densities, classifiable according to their symmetry properties under the flavour symmetry. The gluon density $G(N,\mu^2)$ corresponds to the gluon operator. Assuming $N_F$ massless quarks, the singlet combination $\Sigma(N,\mu^2)$ and the non-singlet combination $\Delta_k(N,\mu^2)$ are given by \cite{Zijlstra:1992qd}
\begin{eqnarray}
	\Sigma(N,\mu^2)   &=& \sum_{k=1}^{N_F} [ f_k(N,\mu^2) + f_{\bar k}(N,\mu^2) ] , \label{eq:singletPDF}\\
	\Delta_k(N,\mu^2) &=& f_k(N,\mu^2) + f_{\bar k}(N,\mu^2) - \frac{1}{N_F} \Sigma(N,\mu^2) , \label{eq:NSPDF}
\end{eqnarray}
with $f_k$ and $f_{\bar k}$ the densities of quarks and anti-quarks. The scale $\mu^2$ is the factorization scale, which separates the high-energy and the non-perturbative contributions. In the following, it will be set equal to the renormalization scale, although in principle the two scales can be treated separately. The structure functions then can be written as \cite{Zijlstra:1992qd}
\begin{eqnarray}
	F_{(2,L)}(N_F,N-1,Q^2) &=& \frac{1}{N_F} \sum_{k=1}^{N_F} e_k^2 \Bigl[ \Sigma(N_F,N,\mu^2) C_{q,(2,L)}^S \Bigl(N_F,N,\frac{Q^2}{\mu^2}\Bigr)
\nonumber\\	
								 && +G(N_F,N,\mu^2) C_{g,(2,L)}^S \Bigl(N_F,N,\frac{Q^2}{\mu^2}\Bigr)
\nonumber\\
								 && + N_F \Delta_k(N_F,N,\mu^2) C_{q,(2,L)}^{NS} \Bigl(N_F,N,\frac{Q^2}{\mu^2}\Bigr)
								 \Bigr]~.
\end{eqnarray}
The quarkonic singlet contribution is split into a non-singlet part and a pure-singlet part,
\begin{equation}
	C_{q,i}^S = C_{q,i}^{NS} + C_{q,i}^{PS}, \quad i=2,L.
\end{equation}
The perturbative expansion of the Wilson coefficients is as follows:
\begin{eqnarray}
	C_{g,i}^S \Bigl(N_F,N,\frac{Q^2}{\mu^2}\Bigr) &=& \sum_{k=1}^\infty a_s^k C_{g,i}^{(k),S} \Bigl(N_F,N,\frac{Q^2}{\mu^2}\Bigr) ~, \\
	C_{q,i}^{PS} \Bigl(N_F,N,\frac{Q^2}{\mu^2}\Bigr) &=& \sum_{k=2}^\infty a_s^k C_{q,i}^{(k),PS} \Bigl(N_F,N,\frac{Q^2}{\mu^2}\Bigr) ~, \\
	C_{q,i}^{NS} \Bigl(N_F,N,\frac{Q^2}{\mu^2}\Bigr) &=& \delta_{i,2} + \sum_{k=1}^\infty a_s^k C_{q,i}^{(k),NS} \Bigl(N_F,N,\frac{Q^2}{\mu^2}\Bigr) ~,
\end{eqnarray}
for $i=2,L$. We call
\begin{equation}
	a_s = \frac{\alpha_s}{4\pi} = \Bigl(\frac{g_s}{4\pi}\Bigr)^2.
\end{equation}
the renormalized strong coupling constant.

In order to investigate the behaviour of the Wilson coefficients in the high-$Q^2$ regime, and to formulate predictions for the dependence of the structure functions on $Q^2$, it is necessary to consider the renormalization of the operator matrix elements.
The operators \eqref{eq:operators} are renormalized by \cite{Gross:1974cs}
\begin{eqnarray}
	O_{q;\mu_1\ldots\mu_n}^{NS,(i),\text{bare}} &=& Z^{NS}(\mu^2)   O_{q;\mu_1\ldots\mu_N}^{NS,(i),\text{ren}} , \\[1mm]
	O_{i;\mu_1,\ldots,\mu_N}^{S,\text{bare}}    &=& Z_{ij}^S(\mu^2) O_{j;\mu_1,\ldots,\mu_N}^{S,\text{ren}}, \qquad i,j=q,g .
\end{eqnarray}
The renormalization produces a scale dependence in the renormalized operators, whose anomalous dimension is
\begin{eqnarray}
	\gamma_{qq}^{NS} &=& \mu \bigl(Z^{NS}(\mu^2)\bigr)^{-1} \frac{\partial}{\partial\mu} Z^{NS}(\mu^2) ~,\\
	\gamma_{ij}^S    &=& \mu \bigl(Z_{il}^S(\mu^2)\bigr)^{-1} \frac{\partial}{\partial\mu} Z_{lj}^S(\mu^2) ~.
\end{eqnarray}
Because the structure functions are observables, they must be independent on the renormalization scale $\mu$; therefore their total derivative with respect to $\mu^2$ must vanish,
\begin{equation}
	\mathcal D(\mu^2) F_{(2,L)}(N,Q^2) = 0, 
\label{eq:RGE}
\end{equation}
with
\begin{eqnarray}
	\mathcal D(\mu^2) &=& 	\mu^2 \frac{\partial}{\partial\mu^2} + \beta\bigl(a_s(\mu^2)\bigr) \frac{\partial}{\partial a_s(\mu^2)} -\gamma_m\bigl(a_s(\mu^2)\bigr) m(\mu^2) \frac{\partial}{\partial m(\mu^2)}, \\
	\beta(a_s(\mu^2)) &=& \mu^2 \frac{\partial a_s(\mu^2)}{\partial\mu^2}, \\
	\gamma_m(a_s(\mu^2)) &=& -\frac{\mu^2}{m(\mu^2)} \frac{\partial m(\mu^2)}{\partial\mu^2}.
\end{eqnarray}
Eq.\ \eqref{eq:RGE} is the renormalization group equation \cite{Symanzik:1971vw,Callan:1972uj,Coleman:1985rnk}.
In a massless theory, it follows that \cite{Buras:1979yt,Blumlein:2000wh}
\begin{eqnarray}
	\sum_{i=q,g} \Bigl[\Bigl\{ \mu^2 \frac{\partial}{\partial\mu^2} + \beta(a_s(\mu^2)) \frac{\partial}{\partial a_s(\mu^2)} \Bigr\} \delta_{ij} -\frac{1}{2}\gamma_{ij}^{S,NS} \Bigr] C_{i,(2,L)} \Bigl(\frac{Q^2}{\mu^2},a_s(\mu^2)\Bigr) = 0, \\
	\sum_{j=q,g} \Bigl[\Bigl\{ \mu^2 \frac{\partial}{\partial\mu^2} + \beta(a_s(\mu^2)) \frac{\partial}{\partial a_s(\mu^2)} \Bigr\} \delta_{ij} +\frac{1}{2}\gamma_{ij}^{S,NS} \Bigr] \langle l|\mathcal O_j(\mu^2)|l \rangle = 0.
\end{eqnarray}
(In the non-singlet case, the indices $i,j$ only take the value $q$.) From these equations one can obtain the evolution equations for the parton densities \cite{Gross:1973ju,Gross:1974cs,Georgi:1951sr},
\begin{eqnarray}
	\frac{\partial}{\partial\ln\mu^2}\begin{pmatrix}
\Sigma(N_F,N,\mu^2) \\
G(N_F,N,\mu^2)
\end{pmatrix}
&=&
-\frac{1}{2}
\begin{pmatrix}
\gamma_{qq} & \gamma_{qg} \\
\gamma_{gq} & \gamma_{gg}
\end{pmatrix}
\begin{pmatrix}
\Sigma(N_F,N,\mu^2) \\
G(N_F,N,\mu^2)
\end{pmatrix}
\\
	\frac{\partial}{\partial\ln\mu^2} \Delta_k(N_F,N,\mu^2) &=& -\frac{1}{2} \gamma_{qq}^{NS} \Delta_k(N_F,N,\mu^2)
\end{eqnarray}
and correspondingly
\begin{eqnarray}
	\frac{\partial}{\partial\ln\mu^2}\begin{pmatrix}
C_{q,i}^S (N_F,N,Q^2/\mu^2) \\
C_{g,i}   (N_F,N,Q^2/\mu^2)
\end{pmatrix}
&=&
\frac{1}{2}
\begin{pmatrix}
\gamma_{qq} & \gamma_{gq} \\
\gamma_{qg} & \gamma_{gg}
\end{pmatrix}
\begin{pmatrix}
C_{q,i}^S (N_F,N,Q^2/\mu^2) \\
C_{g,i}   (N_F,N,Q^2/\mu^2)
\end{pmatrix} ,
\\
	\frac{\partial}{\partial\ln\mu^2} C_{q,i}^{NS}(N_F,N,Q^2/\mu^2) &=& \frac{1}{2} \gamma_{qq}^{NS} C_{q,i}^{NS}(N_F,N,Q^2/\mu^2) .
\end{eqnarray}

\subsection{Renormalization in the presence of heavy quarks}

Because of the wide difference in the quark masses, it is natural to divide the quarks into ``light" and ``heavy". Typically, the $u$, $d$, and $s$ quarks are considered light, because they have a mass comparable or smaller than $\Lambda_{QCD}$, which separates perturbative and non-perturbative physics. In perturbative calculations, their mass is neglected.
The masses of the $c$, $b$ and $t$ quarks are outside of the non-perturbative regime, and their treatment in calculations will depend on the scale of the process being considered.
In general, the $t$ quark, because of its extremely high mass, is not considered to produce any effects at the lower energies available for DIS experiments.

The possibility to disregard particles much heavier than the scale under consideration is a consequence of the Appelquist-Carazzone theorem \cite{Appelquist:1974tg}. Its physical meaning is that physics at very high energies should not affect the physics at low energies and should not be discernible only from low-energy phenomena. More precisely, it states that if a low-energy effective theory is constructed by removing the heavy-particle fields, then the effect on the Green's functions that involve only light particles is equivalent to a finite renormalization of the couplings, up to terms suppressed by the heavy mass, $\mathcal{O}(1/M)$. In other words, removing the heavy particles does not produce observable effects. (An exception to this rule applies when the low-energy theory has different symmetries than the high-energy one; in such cases the effect of removing the heavy particles will be visible).

In practice, whether it is appropriate to include the effects of the $c$, $b$ quark, or both, will depend on the observable under consideration and the experimental precision available. However, if calculations are performed in the $\MS$ scheme, the decoupling of heavy particles is not manifest order by order. This is because the $\MS$ renormalization prescription is mass-independent. In dimensional regularization, it prescribes the removal of the $\varepsilon$ poles and of the universal spherical factor
\begin{equation}
S_\varepsilon=\exp\left[\frac{\varepsilon}{2}(\gamma_E-\ln(4\pi))\right]
\end{equation}
for each perturbative order, in $D$ dimensions, with $\gamma_E$ the Euler-Mascheroni constant,
\begin{equation}
	\gamma_E = \lim_{n\rightarrow\infty} \Bigl( \sum_{k=1}^n \frac{1}{k} - \ln(n) \Bigr) ~.
\end{equation}

In the $\MS$ scheme, the decoupling of heavy particles is only manifest after all perturbative orders are summed. However, it is highly desirable in practice to adopt a renormalization scheme which exhibits the decoupling order by order. An example of such a scheme is the CWZ prescription \cite{Collins:1978wz}. 

The CWZ prescription corresponds to a set of subschemes related to each other by matching conditions. If the masses of $n_f$ quarks are
\begin{equation}
m_1, \ldots\, m_{n_\ell}, m_{n_\ell+1},\ldots, m_{n_f}
\end{equation}
and the virtuality of the process is $Q\sim\mu\sim m_{n_\ell}$,
the CWZ prescription is to divide the quarks into $n_\ell$ light or ``active" and $n_h=n_f-n_\ell$ heavy quarks. The light quarks are considered massless and are renormalized in the $\MS$ scheme. Graphs containing heavy quark lines are renormalized using zero-momentum subtraction, also known as BPHZ. Explictly, it is demanded that $\Pi(p^2, m^2)|_{p=0}=0$ for the heavy quarks, where $\Pi(p^2,m^2)$ refers to the contribution from the heavy quarks to the gluon self-energy.

In the CWZ scheme, the decoupling of the heavy quarks is manifest, and the numerical value of the $\beta$-function is the same as in the effective theory with $n_\ell$ quarks renormalized in the $\MS$ scheme.
When implementing a variable flavour number scheme, as a way of resumming the large logarithms that occur, the choice of which subscheme to use is determined by the virtuality $Q$, which should be of the order of $m_{n_\ell}$.

A one-loop calculation in the CWZ scheme can be found in \cite{Qian:1984kf}. Multi-loop calculations can be found in \cite{Buza:1996xr,Buza:1995ie,Klein:2009ig}.


Let us review the renormalization procedure for the massive OMEs developed in \cite{Buza:1995ie,Bierenbaum:2009mv,Klein:2009ig} for the single-mass OMEs; the two-mass case was discussed in \cite{Ablinger:2017err,Schonwald:2019gmn}. This whole section follows the exposition in those References.

The unrenormalized OMEs are first obtained by 2-point functions including self-energies in the external legs, amputated of the external fields. The external legs are kept on-shell, thus avoiding the potential problem of the mixing of non-gauge-invariant operators \cite{Hamberg:1991qt,Hamberg91,Matiounine:1998ky,Collins:1994ee,Blumlein:2021enk,Blumlein:2021ryt,Blumlein:2022ndg}. We call the momentum flowing through them $p$, with $p^2=0$.

The trace terms present in the twist-two operators are projected out by multiplying by an external source
\begin{equation}
	J_{\mu_1\ldots\mu_N} = \Delta_{\mu_1}\cdots\Delta_{\mu_N}~,
\end{equation}
with $\Delta_\mu$ an auxiliary light-like vector, $\Delta^2=0$.

Calling the unrenormalized OMEs, which are denoted by two hats,
\begin{equation}
	\hat{\hat A}_{ij} \Bigl(\frac{m^2}{\mu^2},\varepsilon,N\Bigr) = \langle j | O_i | j \rangle~,
\end{equation}
the general structure of these Green's functions is then \cite{Buza:1995ie,Buza:1996wv}
\begin{eqnarray}
	G_{\mu\nu,l,Q}^{ab} &=& \hat{\hat A}_{lg} \ \delta^{ab} (\Delta.p)^N \Bigl( -g_{\mu\nu}+\frac{\Delta_\mu p_\nu+\Delta_\nu p_\mu}{\Delta.p} \Bigr) ~,\\
	G_{l,Q}^{ij} &=& \hat{\hat A}_{lq} \ \delta^{ij} (\Delta.p)^N /\!\!\!\!\Delta ~,
\end{eqnarray}
depending on whether the external states are gluons or quarks. Here $l$ can indicate a light parton or the heavy quark. One projects out the OMEs through
\begin{eqnarray}
	\hat{\hat A}_{lq} &=& P_q \ G_{l,Q}^{ij} = \frac{\delta^{ij}}{N_c} (\Delta.p)^{-N} \text{Tr}(/\!\!\!p G_{l,Q}^{ij} ) ~,
\end{eqnarray}
for the case of external quarks, with $N_c$ the number of colors. In the case of external gluons, one should distinguish between the two projectors 
\begin{eqnarray}
	P_g^{(1)} &=& -\frac{\delta^{ab}}{N_c^2-1}\frac{1}{D-2} (\Delta.p)^{-N} g_{\mu\nu} ~,\\
	P_g^{(2)} &=& \frac{\delta^{ab}}{N_c^2-1}\frac{1}{D-2} (\Delta.p)^{-N} \Bigl(-g_{\mu\nu} + \frac{\Delta_\mu p_\nu + \Delta_\nu p_\mu}{\Delta.p} \Bigr) \ .
\end{eqnarray}
The projector $P_g^{(2)}$ enforces the transversity of the gluon polarizations and does not require the inclusion of diagrams with external ghosts. Instead, they must be considered if one chooses to perform the calculation using $P_g^{(1)}$. In either case, one has
\begin{equation}
	\hat{\hat A}_{lg} = P_g^{(1,2)} \ G_{\mu\nu,l,Q}^{ab} \ .
\end{equation}
In the case of polarized OMEs, different projectors are used.

The renormalization of the OMEs occurs in four steps: mass renormalization, coupling constant renormalization, operator renormalization and mass factorization. In the first step, the unrenormalized mass $\hat m$ is replaced with the renormalized mass. Considering for illustration the case of one massive quark, the relation reads in the on-mass shell (OMS) scheme \cite{Tarrach:1980up,Nachtmann:1981zg,Gray:1990yh,Broadhurst:1991fy,Fleischer:1998dw}
\begin{equation}
	\hat m = Z_m m = m \bigg[ 1 + \hat a_s \ \Big(\frac{m^2}{\mu^2}\Big)^{\varepsilon/2} \delta m_1 + \hat a_s^2 \ \Big(\frac{m^2}{\mu^2}\Big)^{\varepsilon} \delta m_2 + \mathcal O(\hat a_s^3) \bigg] ~,
\end{equation}
with 
\begin{eqnarray}
    \delta m_1 &=&C_F
                  \left[\frac{6}{\ep}-4+\left(4+\frac{3}{4}\zeta_2\right)\ep
                  \right] \label{delm1}  \\
               &\equiv&  \frac{\delta m_1^{(-1)}}{\ep}
                        +\delta m_1^{(0)}
                        +\delta m_1^{(1)}\ep~, \label{delm1exp} \\
    \delta m_2 &=& C_F
                   \Biggl\{\frac{1}{\ep^2}\Big[18 C_F-22 C_A+8T_F(n_f+N_h)
                    \Big]
                  +\frac{1}{\ep}\Big[-\frac{45}{2}C_F+\frac{91}{2}C_A
\nonumber\\&&                  
                  -14T_F
                   (N_F+N_H)\Big]
                  +C_F\left(\frac{199}{8}-\frac{51}{2}\zeta_2+48\ln(2)\zeta_2
                   -12\zeta_3\right)
\nonumber\\&&                   
                   +C_A\left(-\frac{605}{8}
                  +\frac{5}{2}\zeta_2-24\ln(2)\zeta_2+6\zeta_3\right)
 \N\\ &&
                  +T_F\left[N_F\left(\frac{45}{2}+10\zeta_2\right)+N_H
                  \left(\frac{69}{2}-14\zeta_2\right)\right]\Biggr\}
                  \label{delm2}  \\
               &\equiv&  \frac{\delta m_2^{(-2)}}{\ep^2}
                        +\frac{\delta m_2^{(-1)}}{\ep}
                        +\delta m_2^{(0)}~, \label{delm2exp}
\end{eqnarray}
where $N_F$ is the number of light flavours and $N_H$ that of heavy flavours. 
The constants $\zeta_k$ refer to the Riemann $\zeta$-function at integer values,
\begin{equation}
	\zeta_k = \sum_{n=1}^\infty \frac{1}{n^k}, \qquad k\in\mathbb N, k\ge 2.
\label{eq:zetak}
\end{equation}

After this replacement one can write the mass-renormalized OMEs as \cite{Bierenbaum:2009mv}
\begin{eqnarray}
    \Ahathat_{ij}\Bigl(\frac{m^2}{\mu^2},\ep,N\Bigr) 
                 &=& \delta_{ij}
                 +\hat{a}_s~ 
                   \Ahathat_{ij}^{(1)}\Bigl(\frac{m^2}{\mu^2},\ep,N\Bigr) 
\N\\ &&
                         + \hat{a}_s^2 \left[
                                        \Ahathat^{(2)}_{ij}
                                        \Bigl(\frac{m^2}{\mu^2},\ep,N\Bigr) 
                                      + {\delta m_1} 
                                        \Bigl(\frac{m^2}{\mu^2}\Bigr)^{\ep/2}
                                        m \frac{d}{dm} 
                                                   \Ahathat_{ij}^{(1)}
                                           \Bigl(\frac{m^2}{\mu^2},\ep,N\Bigr) 
                                \right]
\N\\ &&
                         + \hat{a}_s^3 \Biggl[ 
                                         \Ahathat^{(3)}_{ij}
                                           \Bigl(\frac{m^2}{\mu^2},\ep,N\Bigr) 
                                        +{\delta m_1} 
                                         \Bigl(\frac{m^2}{\mu^2}\Bigr)^{\ep/2}
                                         m \frac{d}{dm} 
                                                    \Ahathat_{ij}^{(2)}
                                           \Bigl(\frac{m^2}{\mu^2},\ep,N\Bigr)
\N\\ &&
                                        + {\delta m_2} 
                                          \Bigl(\frac{m^2}{\mu^2}\Bigr)^{\ep}
                                          m \frac{d}{dm} 
                                                    \Ahathat_{ij}^{(1)}
                                           \Bigl(\frac{m^2}{\mu^2},\ep,N\Bigr) 
                                        + \frac{\delta m_1^2}{2}
                                          \Bigl(\frac{m^2}{\mu^2}\Bigr)^{\ep}
                                                   m^2  \frac{d^2}{{dm}~^2}
                                                     \Ahathat_{ij}^{(1)}
                                           \Bigl(\frac{m^2}{\mu^2},\ep,N\Bigr) 
    \Biggr]~. 
\nonumber\\
\label{maren}
\end{eqnarray}
The coupling is renormalized in the $\MS$ scheme as follows:
\begin{eqnarray}
	\hat a_s &=& \Big( Z_g^{\MS}(\ep,N_F) \Big)^2 a_s^{\MS}(\mu^2) \\
		   &=& a_s^{\MS}(\mu^2) \Bigl[ 1 + \delta a^{\MS}_{s, 1}(N_F) 
                                      a^{\MS}_s(\mu^2)
                                 + \delta a^{\MS}_{s, 2}(N_F) 
                                     \big( {a^{\MS}_s}(\mu)\big)^2    
                                      + \mathcal O\Big(\big({a_s^{\MS}}\big)^3\Big) \Bigr]. 
\label{eq:asMS}
\end{eqnarray}
The coefficients in this expansion are related to the $\beta$-function as follows: in dimensional regularization, the renormalization scale $\mu$ is defined through
\begin{eqnarray}
	\hat g_{s,(D)} &=& \mu^{-\ep/2} \hat g_s ~,\\
	\hat g_s^2 &=& (4\pi)^2 \hat a_s ~.
\end{eqnarray}
From the independence of the bare coupling on $\mu$, one derives
\begin{eqnarray}
	0 &=& \frac{d}{d\ln\mu^2} \hat a_{s,(D)} = \frac{d}{d\ln\mu^2} (\mu^{-\ep} \hat a_s) = \frac{d}{d\ln\mu^2} \big[\mu^{-\ep} Z_g(\ep,N_F,\mu^2) a_s(\mu^2) \big]
\end{eqnarray}
from which it follows that
\begin{equation}
	\beta = \frac{\ep}{2} a_s(\mu^2) -2a_s(\mu^2) \frac{d}{d\ln\mu^2} \ln Z_g(\ep,N_F,\mu^2) ~.
\end{equation}
Specializing to the $\MS$ scheme for the coupling constant,
\begin{equation}
	\beta^{\MS}(N_F) = -\beta_0(N_F) \big(a_s^{\MS}\big)^2 -\beta_1(N_F) \big(a_s^{\MS}\big)^3 +\mathcal O\Big(\big(a_s^{\MS}\big)^4\Big)
\end{equation}
and one obtains \cite{Gross:1973id,Politzer:1973fx,Khriplovich:1969aa,Caswell:1974gg,Jones:1974mm}
\begin{eqnarray}
	\delta a^{\MS}_{s, 1}(N_F) &=& \frac{2}{\ep} \beta_0(N_F) \\
	\delta a^{\MS}_{s, 2}(N_F) &=& \frac{4}{\ep^2} \beta_0^2(N_F) + \frac{1}{\ep} \beta_1(N_F)
\end{eqnarray}
with
  \begin{eqnarray}
   \beta_0(N_F)
                 &=& \frac{11}{3} C_A - \frac{4}{3} T_F N_F \label{beta0}~, \\
   \beta_1(N_F)
                 &=& \frac{34}{3} C_A^2 
               - 4 \left(\frac{5}{3} C_A + C_F\right) T_F N_F \label{beta1}~.
  \end{eqnarray}
In order to preserve the condition of having on-shell massless external particles and the decoupling of the massive quarks in the running of the coupling constant, one demands that the gluon self-energy receives no contribution from the heavy quark at zero momentum,
\begin{equation}
	\Pi_{H,BF}(p^2=0,m^2) = 0 ~.
\end{equation}
This condition is enforced in the background field method \cite{Abbott:1980hw,Rebhan:1985yf,Jegerlehner:1998zg,Bierenbaum:2009mv}. In this renormalization scheme the renormalization factor of the coupling constant is defined through
\begin{eqnarray}
	Z_g^{\MOM} (\ep,N_F+1,\mu^2,m^2) &=& \frac{1}{(Z_{A,l}+Z_{A,H})^{1/2}} ~,\\
	Z_{A,l} &=& \big(Z_g^{\MS} (\ep,N_F) \big)^{-2} ~,
\end{eqnarray}
with $Z_{A,(l,H)}$ the contributions to the renormalization factor of the background field due to the light quarks and the heavy quark.

In this $\MOM$ scheme, a formula analogous to \eqref{eq:asMS} holds, but reads \cite{Bierenbaum:2009mv}
\begin{eqnarray}
	\hat a_s &=& \Big( Z_g^{\MOM}(\ep,N_F+1,\mu^2,m^2) \Big)^2 \ a_s^{\MOM}(\mu^2,m^2) \\
		   &=& a_s^{\MOM}(\mu^2,m^2) \Bigl[ 1 + \delta a^{\MOM}_{s, 1} 
                                      a_s^{\MOM}(\mu^2,m^2)
                                 + \delta a^{\MOM}_{s, 2} 
                                     \big( a_s^{\MOM}(\mu)\big)^2    
                                      + \mathcal O\Big(\big({a_s^{\MOM}}\big)^3\Big) \Bigr] ~,
\nonumber\\                                      
\end{eqnarray}
with \cite{Bierenbaum:2009mv}
  \begin{eqnarray}
   \delta a_{s,1}^{\MOM}&=&\Bigl[\frac{2\beta_0(N_F)}{\ep}
                           +\frac{2\beta_{0,Q}}{\ep}f(\ep)
                            \Bigr]~,\label{dela1} \\
   \delta a_{s,2}^{\MOM}&=&\Bigl[\frac{\beta_1(N_F)}{\ep}+
                            \Bigl\{\frac{2\beta_0(N_F)}{\ep}
                              +\frac{2\beta_{0,Q}}{\ep}f(\ep)\Bigr\}^2
                          +\frac{1}{\ep}\Bigl(\frac{m^2}{\mu^2}\Bigr)^{\ep}
                           \Bigl(\beta_{1,Q}+\ep\beta_{1,Q}^{(1)}
                                            +\ep^2\beta_{1,Q}^{(2)}
                           \Bigr)\Bigr] + O(\ep^2)~.\label{dela2}
\nonumber\\
  \end{eqnarray}
  \begin{eqnarray}
   \beta_{0,Q} &=&-\frac{4}{3}T_F~, \label{b0Q} \\
   \beta_{1,Q} &=&- 4 \left(\frac{5}{3} C_A + C_F \right) T_F~, \label{b1Q} \\
   \beta_{1,Q}^{(1)}&=&
                           -\frac{32}{9}T_FC_A
                           +15T_FC_F~, \label{b1Q1} \\
   \beta_{1,Q}^{(2)}&=&
                               -\frac{86}{27}T_FC_A
                               -\frac{31}{4}T_FC_F
                               -\zeta_2\left(\frac{5}{3}T_FC_A
                                        +T_FC_F\right) ~,\label{b1Q2}
  \end{eqnarray}
where
\begin{eqnarray}
	   f(\ep)&\equiv&
                 \Bigl(\frac{m^2}{\mu^2}\Bigr)^{\ep/2}
    \exp \Bigl(\sum_{i=2}^{\infty}\frac{\zeta_i}{i}
                       \Bigl(\frac{\ep}{2}\Bigr)^{i}\Bigr)~. \label{fep}
\end{eqnarray}
From the invariance of the unrenormalized coupling,
\begin{equation}
	\big( Z_g^{\MS}(\ep,N_F+1) \big)^2 \ a_s^{\MS}(\mu^2) = \big( Z_g^{\MOM}(\ep,N_F+1,\mu^2,m^2) \big)^2 \ a_s^{\MOM}(\mu^2)
\end{equation}
one can obtain the relation between $a_s^{\MS}$ and $a_s^{\MOM}$:
  \begin{eqnarray}
   a_s^{\MS}&=&
               a_s^{\MOM}
              +{a_s^{\MOM}}^2\biggl[
                          \delta a^{\MOM}_{s, 1}
                         -\delta a^{\MS}_{s, 1}(N_F+1)
                             \biggr]
              +{a_s^{\MOM}}^{3}\biggl[
                          \delta a^{\MOM}_{s, 2}
                         -\delta a^{\MS}_{s, 2}(N_F+1)
     \N\\ &&
                        -2\delta a^{\MS}_{s, 1}(N_F+1)\Bigl[
                             \delta a^{\MOM}_{s, 1}
                            -\delta a^{\MS}_{s, 1}(n_f+1)
                                                      \Bigr]
                             \biggr]+O({a_s^{\MOM}}^4)~, \label{asmsa}
  \end{eqnarray}
Here, $a_s^{\MS}=a_s^{\MS}(N_F+1)$. Using the MOM scheme for the coupling, one can write the mass- and coupling-renormalized OME, indicated by ${\hat{A}}_{ij}$ \cite{Bierenbaum:2009mv}
  \begin{eqnarray}
   {\hat{A}}_{ij} &=&  \delta_{ij} 
                     + a^{\MOM}_s \Ahathat_{ij}^{(1)}
                     + {a^{\MOM}_s}^2 \left[\Ahathat^{(2)}_{ij}
                     + \delta m_1 \Bigl(\frac{m^2}{\mu^2}\Bigr)^{\ep/2} 
                                 m  \frac{d}{dm} \Ahathat_{ij}^{(1)}
                     + \delta a^{\MOM}_{s,1} \Ahathat_{ij}^{(1)}\right]
\N\\ &&
   + {a^{\MOM}_s}^3 \Biggl[ \Ahathat^{(3)}_{ij}
   + \delta a^{\MOM}_{s,2} \Ahathat_{ij}^{(1)}
   + 2 \delta a^{\MOM}_{s,1} \left( \Ahathat^{(2)}_{ij} 
   +  \delta m_1 \Bigl(\frac{m^2}{\mu^2}\Bigr)^{\ep/2}
      m \frac{d}{dm} \Ahathat_{ij}^{(1)} \right)
\N\\ &&
                + \delta m_1 \Bigl(\frac{m^2}{\mu^2}\Bigr)^{\ep/2}
                           m  \frac{d}{dm} \Ahathat_{ij}^{(2)}
                + \delta m_2 \Bigl(\frac{m^2}{\mu^2}\Bigr)^{\ep}
                          m  \frac{d}{dm} \Ahathat_{ij}^{(1)}
                + \frac{\delta m_1^2}{2} \Bigl(\frac{m^2}{\mu^2}\Bigr)^{\ep}
                m^2  \frac{d^2}{{dm}~^2} \Ahathat_{ij}^{(1)}
    \Biggr]
\N\\&&
    \label{macoren}
  \end{eqnarray}

At this stage, ultraviolet singularities are still present due to the composite operators. In the massless case, they are removed by imposing
\begin{equation}
	A_{ij} \Bigl(\frac{-p^2}{\mu^2},N_F,N \Bigr) = Z_{ij} (\ep, a^{\MS},N_F,N) \ \hat A_{ij} \Bigl(\frac{-p^2}{\mu^2},\ep,N_F,N \Bigr) ~.
\label{eq:opZ}
\end{equation}
The pole structure in the operator $Z$-factors $Z_{ij}$ can be determined by the requirement that 
\begin{eqnarray}
	\gamma_{ij} = \sum_{k=0}^\infty \gamma_{ij}^{(k)} \big(a_s^{\MS}\big)^{k+1} = \mu Z_{il}^{-1}(\mu^2) \frac{\partial}{\partial \mu} Z_{lj}(\mu^2) ~.
\label{eq:opGammas}
\end{eqnarray}
Equations \eqref{eq:opZ} and \eqref{eq:opGammas} have to be specialized to the singlet, non-singlet and pure-singlet cases appropriately, i.e.
   \begin{eqnarray}
    Z_{qq}^{-1}&=&Z_{qq}^{-1, {\sf PS}}+Z_{qq}^{-1, {\sf NS}} 
\label{ZPSNS1}~,\\
    A_{qq}     &=&A_{qq}^{\sf PS}+A_{qq}^{\sf NS} \label{ZPSNS2}~. 
   \end{eqnarray}
From Eq.\ \eqref{eq:opGammas} one determines the pole terms in the $Z$-factors, which can be found written in explicit form in \cite{Bierenbaum:2009mv} and are not repeated here. One obtains the $Z$-factors in the case of $(N_F+1)$ flavours by taking them at $N_F+1$ flavours and applying the scheme transformation between $a_s^{\MS}$ and $a_s^{\MOM}$, i.e.\ the inverse of \eqref{asmsa}. The OMEs are split into a light and a heavy flavour part,
\begin{equation}
	\hat A_{ij}(p^2,m^2,\mu^2,a_s^{\MOM},N_F+1) = \hat A_{ij} \Big(\frac{-p^2}{\mu^2},a_s^{\MS},N_F\Big) + A_{ij}^Q (p^2,m^2,\mu^2,a_s^{\MOM},N_F+1)
\end{equation}
and the operator renormalization of the heavy contribution is \cite{Bierenbaum:2009mv}
  \begin{eqnarray}
   \bar A^Q_{ij}(p^2,m^2,\mu^2,a_s^{\MOM},N_F+1)&=&
               Z^{-1}_{il}(a_s^{\MOM},N_F+1,\mu) 
                  \hat{A}^Q_{ij}(p^2,m^2,\mu^2,a_s^{\MOM},N_F+1)
\N\\ &&
              +Z^{-1}_{il}(a_s^{\MOM},N_F+1,\mu) 
                   \hat{A}_{ij}\Bigl(\frac{-p^2}{\mu^2},a_s^{\MS},N_F\Bigr)
\N\\ &&
              -Z^{-1}_{il}(a_s^{\MS},n_f,\mu)
                   \hat{A}_{ij}\Bigl(\frac{-p^2}{\mu^2},a_s^{\MS},N_F\Bigr)~.
\label{eqXX} 
 \end{eqnarray}
Taking now $p^2=0$, the contribution of the unrenormalized massless OMEs reduces to their tree level value because scaleless loop integrals vanish in dimensional regularization, so they reduce to
\begin{equation}
	\hat A_{ij}(0,a_s^{\MS},N_F) = \delta_{ij}~.
\end{equation}

The remaining collinear singularities in $\bar A^Q_{ij}$ are removed by mass factorization,
\begin{equation}
	A_{ij}^Q = \bar A_{il}^Q \Big(\frac{m^2}{\mu^2},a_s^{\MOM},N_F+1 \Big) \Gamma_{lj}^{-1}~.
\end{equation}
The transition functions $\Gamma_{ij}$ correspond to the inverse of the $Z$-factors in the massless case. We repeat here from \cite{Bierenbaum:2009mv} the formula for the renormalized OMEs which one obtains after these steps:
  \begin{eqnarray}
   && A^Q_{ij}\Bigl(\frac{m^2}{\mu^2},a_s^{\MOM},n_f+1\Bigr)=
\N\\&&\phantom{+}
                a^{\MOM}_s~\Biggl(
                      \hat{A}_{ij}^{(1),Q}\Bigl(\frac{m^2}{\mu^2}\Bigr)
                     +Z^{-1,(1)}_{ij}(n_f+1)
                     -Z^{-1,(1)}_{ij}(n_f)
                           \Biggr)
\N\\&&
           +{a^{\MOM}_s}^2\Biggl( 
                        \hat{A}_{ij}^{(2),Q}\Bigl(\frac{m^2}{\mu^2}\Bigr)
                       +Z^{-1,(2)}_{ij}(n_f+1)
                       -Z^{-1,(2)}_{ij}(n_f)
                       +Z^{-1,(1)}_{ik}(n_f+1)\hat{A}_{kj}^{(1),Q}
                                              \Bigl(\frac{m^2}{\mu^2}\Bigr)
\N\\
&&\phantom{+{a^{\MOM}_s}^2\Biggl(}
                       +\Bigl[ \hat{A}_{il}^{(1),Q}
                               \Bigl(\frac{m^2}{\mu^2}\Bigr)
                              +Z^{-1,(1)}_{il}(n_f+1)
                              -Z^{-1,(1)}_{il}(n_f)
                        \Bigr] 
                             \Gamma^{-1,(1)}_{lj}(n_f)
                         \Biggr)
\N\\ 
&&
          +{a^{\MOM}_s}^3\Biggl( 
                        \hat{A}_{ij}^{(3),Q}\Bigl(\frac{m^2}{\mu^2}\Bigr)
                       +Z^{-1,(3)}_{ij}(n_f+1)
                       -Z^{-1,(3)}_{ij}(n_f)
                       +Z^{-1,(1)}_{ik}(n_f+1)\hat{A}_{kj}^{(2),Q}
                                              \Bigl(\frac{m^2}{\mu^2}\Bigr)
\nonumber\end{eqnarray}\begin{eqnarray}
&&\phantom{+{a^{\MOM}_s}^3\Biggl(}
                       +Z^{-1,(2)}_{ik}(n_f+1)\hat{A}_{kj}^{(1),Q}
                                              \Bigl(\frac{m^2}{\mu^2}\Bigr)
                       +\Bigl[ 
                               \hat{A}_{il}^{(1),Q}
                                 \Bigl(\frac{m^2}{\mu^2}\Bigr)
                              +Z^{-1,(1)}_{il}(n_f+1)
 \N\\ &&\phantom{+{a^{\MOM}_s}^3\Biggl(}
                              -Z^{-1,(1)}_{il}(n_f)
                        \Bigr]
                              \Gamma^{-1,(2)}_{lj}(n_f)
                       +\Bigl[ 
                               \hat{A}_{il}^{(2),Q}
                                 \Bigl(\frac{m^2}{\mu^2}\Bigr)
                              +Z^{-1,(2)}_{il}(n_f+1)
                              -Z^{-1,(2)}_{il}(n_f)
 \N\\ 
&&\phantom{+{a^{\MOM}_s}^3\Biggl(}
                              +Z^{-1,(1)}_{ik}(n_f+1)\hat{A}_{kl}^{(1),Q}
                                              \Bigl(\frac{m^2}{\mu^2}\Bigr)
                        \Bigr]
                              \Gamma^{-1,(1)}_{lj}(n_f)
                        \Biggr)~. \label{GenRen3}
  \end{eqnarray}

By applying \eqref{GenRen3} and \eqref{maren} as well as the pole expansions of the $Z$ and $\Gamma$ factors, the coefficients of the $\ep$-expansion of the OMEs have been predicted in terms of the renormalization constants up to $\mathcal O(a_s^3)$. These predictions are derived by demanding that \eqref{GenRen3} be free of poles in $\ep$.

In analogy with the process described above, the renormalization of the OMEs in the presence of two heavy quarks has been preformed in \cite{Ablinger:2017err,Schonwald:2019gmn}. The main differences to the single-mass case summarized above are due to the fact that the renormalization of one mass will depend on the other; also, two quarks have to be decoupled in the running of the coupling. The procedure is not repeated here and only the relevant results are printed in the next sections for the OMEs respectively under consideration, where we focus on the genuine two-mass contribution to the OMEs, i.e.\ on graphs which contain two different masses.

\subsection{Variable flavour number scheme}
\label{sec:VFNS}

In the so-called fixed-flavour number scheme, heavy quarks are treated as if they were radiatively generated, and are not assigned a parton distribution. Such calculations give rise to logarithms of the type $\ln(Q^2/m^2)$, with $m$ the mass of the heavy quark. In principle, for very large virtualities, these logarithms could become large and ruin the convergence of the perturbative series. In the presently accessible kinematic range at HERA, however, the fixed-flavour number scheme has been shown to be numerically stable \cite{Gluck:1993dpa,Alekhin:2012ig}.

The variable flavour number scheme (VFNS), by contrast, exploits the possibility to factorize the structure functions for $Q^2\gg m^2$ using the massive operator matrix elements and the massless Wilson coefficients, and devises a transition between, e.g., $N_F$ and $N_F+1$ flavours, however, only at very high scales $\mu^2=Q^2$. At the transition scale $\mu^2$, a new PDF is introduced for the heavy quark, which from then on is treated as massless. The transition is designed such that the structure functions are unchanged asymptotically after the new PDF is introduced and the light quark PDFs are adjusted appropriately. The VFNS was discussed in \cite{Collins:1978wz,Aivazis:1993pi,Buza:1996wv,Collins:1998rz,Alekhin:2009ni}; the matching conditions were given at $\mathcal O(a_s^2)$ in \cite{Buza:1996wv}, at $\mathcal O(a_s^3)$ in \cite{Bierenbaum:2009mv} and the VFNS where two quarks are decoupled in \cite{Ablinger:2017err}. This VFNS prescription is also known as the ``zero-mass VFNS".

For concreteness we reproduce here the single-mass VFNS relations given in \cite{Bierenbaum:2009mv}:
\begin{eqnarray}
\label{HPDF1}
f_k(N_F+1,\mu^2,m^2,N) + f_{\overline{k}}(N_F+1,\mu^2,m^2,N)
&=& A_{qq,Q}^{\rm NS}\left(N_F,\frac{\mu^2}{m^2},N\right)
\cdot \bigl[f_k(N_F,\mu^2,N) 
\nonumber\\ && \hspace*{3.8cm}
+ f_{\overline{k}}(N_F,\mu^2,N)\bigr]
\nonumber\\ 
& & \hspace*{-4mm} + \tilde{A}_{qq,Q}^{\rm 
PS}\left(N_F,\frac{\mu^2}{m^2},N\right)
\cdot \Sigma(N_F,\mu^2,N)
\nonumber\\ 
& & \hspace*{-4mm} + \tilde{A}_{qg,Q}\left(N_F,\frac{\mu^2}{m^2},N\right)
\cdot G(N_F,\mu^2,N),
\nonumber\\
\\
\label{fQQB}
f_Q(N_F+1,\mu^2,m^2,N) + f_{\overline{Q}}(N_F+1,\mu^2,m^2,N)
&=&
{A}_{Qq}^{\rm PS}\left(N_F,\frac{\mu^2}{m^2},N\right)
\cdot \Sigma(N_F,\mu^2,N)
\nonumber\\ && \hspace*{-4mm}
+ {A}_{Qg}\left(N_F,\frac{\mu^2}{m^2},N\right)
\cdot G(N_F,\mu^2,N)~.
\nonumber\\
\end{eqnarray}
\begin{eqnarray}
\Sigma(N_F+1,\mu^2,m^2,N) 
&=& \sum_{k+1}^{N_F+1} \big[f_k(N_F+1,\mu^2) + f_{\bar k}(N_F+1,\mu^2) \big]
\nonumber\\
&=& \Biggl[A_{qq,Q}^{\rm NS}\left(N_F, \frac{\mu^2}{m^2},N\right) +
          N_F \tilde{A}_{qq,Q}^{\rm PS}\left(N_F, \frac{\mu^2}{m^2},N\right)
  \nonumber \\ &&
         + {A}_{Qq}^{\rm PS}\left(N_F, \frac{\mu^2}{m^2},N\right)
        \Biggr]
\cdot \Sigma(N_F,\mu^2,N) \nonumber
\\
& & \hspace*{-3mm} + \left[N_F \tilde{A}_{qg,Q}\left(N_F, 
\frac{\mu^2}{m^2},N\right) +
          {A}_{Qg}\left(N_F, \frac{\mu^2}{m^2},N\right) 
\right]
\cdot G(N_F,\mu^2,N) 
\nonumber\\
\\
\Delta_k(N_F+1,\mu^2,m^2,N) &=& 
  f_k(N_F+1,\mu^2,N)
+ f_{\overline{k}}(N_F+1,\mu^2,m^2,N) 
\nonumber\\ &&
- \frac{1}{N_F+1} 
\Sigma(N_F+1,\mu^2,m^2,N)\\
\label{HPDF2}
G(N_F+1,\mu^2,m^2,N) &=& A_{gq,Q}\left(N_F, 
\frac{\mu^2}{m^2},N\right) 
                    \cdot \Sigma(N_F,\mu^2,N)
\nonumber\\
&& 
+ A_{gg,Q}\left(N_F, \frac{\mu^2}{m^2},N\right) 
                    \cdot G(N_F,\mu^2,N)~.
\end{eqnarray}
where $f_{k,\bar k}$ are the light quark and antiquark PDFs, $f_{Q,\bar Q}$ refer to the new PDF of the heavy quark, $G$ is the gluon distribution, $\Sigma$ is the singlet distribution defined in Eq. \eqref{eq:singletPDF}
and $\Delta_k$ the non-singlet distribution, Eq. \eqref{eq:NSPDF}.

The two-mass VFNS derived in \cite{Ablinger:2017err} is also reproduced below, for completeness:
\begin{eqnarray}
\label{eq:VFNS1}
&&
\hspace*{-2.4cm}
      f_k(N_F+2,N,\mu^2,m_1^2,m_2^2) + f_{\overline{k}}(N_F+2,N,\mu^2,m_1^2,m_2^2)= 
\NN\\ &&  \hspace{20mm}
A_{qq,Q}^{\sf NS}\left(N,N_F+2,\frac{m_1^2}{\mu^2},\frac{m_2^2}{\mu^2}\right)
          \cdot\bigl[f_k(N_F,N,\mu^2)+f_{\overline{k}}(N_F,N,\mu^2)\bigr]
\NN\\ &&  \hspace{20mm}
       +\frac{1}{N_F}A_{qq,Q}^{\sf PS}\left(N,N_F+2,\frac{m_1^2}{\mu^2},\frac{m_2^2}{\mu^2}\right)
          \cdot\Sigma(N_F,N,\mu^2)
\NN\\ &&  \hspace{20mm}
       +\frac{1}{N_F}A_{qg,Q}\left(N,N_F+2,\frac{m_1^2}{\mu^2},\frac{m_2^2}{\mu^2}\right)
          \cdot G(N_F,N,\mu^2), \label{HPDF12M} \\
     && 
\hspace*{-2.4cm}\label{fQQB2M}
        f_Q(N_F+2,N,\mu^2,m^2_1,m_2^2) + f_{\overline{Q}}(N_F+2,N,\mu^2,m^2_1,m_2^2)=
\NN\\ &&  \hspace{20mm}
        A_{Qq}^{\sf PS}\left(N,N_F+2,\frac{m_1^2}{\mu^2},
\frac{m_2^2}{\mu^2},
\right)
          \cdot \Sigma(N_F,N,\mu^2)
\NN\\ &&  \hspace{20mm}
       +A_{Qg}\left(N,N_F+2,\frac{m_1^2}{\mu^2},\frac{m_2^2}{\mu^2}\right)
          \cdot G(N_F,N,\mu^2)~.   
    \end{eqnarray}
In this case, the flavor singlet, non-singlet and gluon densities for $(N_F+2)$ flavors are given by
    \begin{eqnarray}
     \Sigma(N_F+2,N,\mu^2,m^2_1,m_2^2) 
      &=& \Biggl[
             A_{qq,Q}^{\sf NS}\left(N,N_F+2,\frac{m_1^2}{\mu^2},\frac{m_2^2}{\mu^2}\right)
            +A_{qq,Q}^{\sf PS}\left(N,N_F+2,\frac{m_1^2}{\mu^2},\frac{m_2^2}{\mu^2}\right)
\NN \\ && \hspace*{-20mm}
            +A_{Qq}^{\sf PS}\left(N,N_F+2,\frac{m_1^2}{\mu^2},\frac{m_2^2}{\mu^2}\right)
          \Biggr]  \cdot \Sigma(N_F,N,\mu^2)
\NN \\ && \hspace*{-23mm} 
          +\left[
                A_{qg,Q}\left(N,N_F+2,\frac{m_1^2}{\mu^2},\frac{m_2^2}{\mu^2}\right)
               +A_{Qg}\left(N,N_F+2,\frac{m_1^2}{\mu^2},\frac{m_2^2}{\mu^2}\right)
          \right]   \cdot G(N_F,N,\mu^2)~,
\NN\\ \\
     \Delta_k(N_F+2,N,\mu^2,m_1^2,m_2^2)
      &=& f_k(N_F+2,N,\mu^2,m_1^2,m_2^2)+f_{\overline{k}}(N_F+2,N,\mu^2,m_1^2,m_2^2)
\NN\\ &&
         -\frac{1}{N_F+2}\Sigma(N_F+2,N,\mu^2,m_1^2,m_2^2)~, \\
     \label{HPDF22M}
     G(N_F+2,N,\mu^2,m^2_1, m^2_2) 
      &=& A_{gq,Q}\left(N,N_F+2,\frac{m_1^2}{\mu^2},\frac{m_2^2}{\mu^2}\right) 
                    \cdot \Sigma(N_F,N,\mu^2)
\NN\\ && 
         +A_{gg,Q}\left(N,N_F+2,\frac{m_1^2}{\mu^2},\frac{m_2^2}{\mu^2}\right) 
                    \cdot G(N_F,N,\mu^2)~.
\end{eqnarray}
One can observe how the two-mass contributions to the OME enter the definitions of the parton densities. Due to the presence of diagrams with two different quark masses, decoupling the charm and bottom quarks one at a time becomes theoretically an ill-defined procedure at higher order. This is one main motivation for the computation of the two-mass contributions to OMEs.

\subsection{Mathematical methods}

\subsubsection{The Mellin transform}

The Mellin transform is a central mathematical operation in the study of deep-inelastic scattering, because the Bjorken variable $x$ and the variable $N$ are conjugate to each other with respect to it. In Eq.\ \eqref{eq:f-inv-mellin} we saw an example of this fact.

The Mellin transform \cite{mellin1896,mellin1092,paris-kaminski} of a function $f(x)$ is defined as
\begin{equation}
\mathbf{M}\left[f\right](N)=F(N)=\int_{0}^1 x^{N-1}f(x)dx ,
\end{equation}
whenever the integral exists.
In our physical applications, poles of $\mathbf{M}\left[f\right](N)$ arise along the negative real axis, and the rightmost pole will be located at $N=1$.

In Table \ref{tab:Commonly-appearing-Mellin}, a few common cases of Mellin transforms are summarized.
\begin{table}[h!]
\begin{centering}
\begin{tabular}{|c|c|}
\hline 
$f(x)$ & $\mathbf M [f(x)](N)=F(N)$\tabularnewline
\hline 
\hline 
$x^a$ & $\frac{1}{N+a}$ \tabularnewline
\hline 
$(1+x)^{-a}$ & $\frac{\Gamma(s)\Gamma(a-s)}{\Gamma(a)}$\tabularnewline
\hline 
$\delta(1-x)$ & $1$\tabularnewline
\hline 
$\ln^a x$ & $(-1)^a N^{-a-1} a!$ \tabularnewline
\hline 
$\ln^a f(x)$ & $F^{(a)}(N)$ \tabularnewline
\hline
$f^{(a)}(x)$ & $\frac{\Gamma(a+1-N)}{\Gamma(1-N)}F(N-a)$ \tabularnewline
\hline
\end{tabular}
\par\end{centering}
\caption{\label{tab:Commonly-appearing-Mellin}Some elementary properties of the Mellin transform.}
\end{table}

From the Mellin transform of a function it is possible to recover the original function using the inverse Mellin transform
\begin{equation}
f(x)=\frac{1}{2\pi i}\int_{c-i\infty}^{c+i\infty}x^{-N}F(N)dN .
\end{equation}

The Mellin convolution of two functions is defined as
\begin{equation}
\left[f\otimes g\right](x)=\int_0^\infty dx_1 \int_0^\infty dx_2 \, \delta(x-x_1 x_2)f(x_1)g(x_2).
\end{equation}
A property of the Mellin convolution is:
\begin{equation}
\mathbf{M}\left[f\otimes g\right](s)=\left(\mathbf{M}\left[f\right](s)\right)\left(\mathbf{M}\left[g\right](s)\right),
\end{equation}
or, in other words, the Mellin convolution becomes an ordinary product in Mellin space.

The functions of relevance to physics which we are concerned with, such as parton distribution functions, are defined in the interval $[0,1]$, and vanish outside of this interval.

\subsubsection{Nested sums}

From the earliest computations of anomalous dimensions in QCD, sums have appeared in physical quantities \cite{Gross:1974cs,Georgi:1951sr}, even though a more systematic study of these objects in the context of high-energy physics appeared only much later \cite{Vermaseren:1998uu,Blumlein:1998if}.

The first class of sums to be studied systematically \cite{Vermaseren:1998uu,Blumlein:1998if}, see also \cite{Moch:2001zr} is that of the harmonic sums, defined recursively as
\begin{eqnarray}
	S_{n_1,\ldots,n_k}(N) &=& \sum_{i=1}^{N} \frac{(\text{sign}(n_1))^i}{i^{|n_1|}} S_{n_2,\ldots,n_k}(i), \qquad n_i \in \mathbb Z \backslash\{0\}, \label{eq:defHSums1}\\
	S_\emptyset &=& 1. \label{eq:defHSums2}
\end{eqnarray}
The quantity $w = \sum|n_i|$ is called the {\it weight} of the sum, and $k$ is called the {\it depth}. For any given weight there are $2\cdot 3w-1$ different harmonic sums.

Generalizations of these objects which have been encountered in QCD computations have the form \cite{Ablinger:2013cf,Moch:2001zr,Ablinger:2013jta}
\begin{equation}
	S_{n_1,\ldots,n_k}(x_1,\ldots,x_k;N) = \sum_{i=1}^N \frac{x_1^i}{i^{|n_1|}} S_{n_2,\ldots,n_k}(x_2,\ldots,x_k;i), \qquad x_i\neq 0 .
\label{eq:genSsums}
\end{equation}
%
A further generalization is the class of cyclotomic sums \cite{Ablinger:2011te}, defined as
\begin{equation}
	S_{\{a_1,b_1,c_1\},\ldots,\{a_k,b_k,c_k\}}(x_1,\ldots,x_k;N) = \sum_{i=1}^N \frac{x_1^i}{(a_1 i+b_1)^{c_1}} S_{\{a_2,b_2,c_2\},\ldots,\{a_k,b_k,c_k\}}(x_2,\ldots,x_k;i) .
\end{equation}
%
In three-loop massive calculations, in addition, more complex sums, both finite and infinite, involving binomial summands have been found to contribute, \cite{Fleischer:1998nb,Davydychev:2003mv,Weinzierl:2004bn,Ablinger:2014bra,Ablinger:2015tua}. Examples of such summands are
\begin{eqnarray}
	&&\binom{2i}{i}^r \frac{x^i}{i}, \qquad r=\pm 1,  \\
	&& \frac{4^i}{i \binom{2i}{i}} \Bigl(\frac{\eta}{\eta-1}\Bigr)^i, \qquad 0<\eta<1,
\end{eqnarray}
where in our applications $\eta$ is the ratio of the squares of two quark masses.
No specific symbol has been used for this type of sums in the literature and they are typically written explicitly.

Nested sums obey algebraic and structural properties \cite{Blumlein:2009ta,Blumlein:2009fz,Blumlein:2003gb}, most importantly the algebraic quasi-shuffle algebra \cite{Hoffman:2000}. This algebra can be derived from relations of the type
\begin{equation}
	\Bigl(\sum_{i=1}^N a_i\Bigr) \Bigl(\sum_{i=1}^N b_i\Bigr) = \sum_{i=1}^N a_i \sum_{j=1}^i b_j + \sum_{i=1}^N b_i \sum_{j=1}^i a_j - \sum_{i=1}^N a_i b_i .
\end{equation}
Such relations allow to reduce the sums to a smaller set of ``basis'' sums which are algebraically independent.

The classes of sums described above have been an object of interest for mathematicians, and Mathematica packages exist to perform algebraic reductions and inverse Mellin transforms. We have extensively used {\tt HarmonicSums} \cite{HARMONICSUMS} in this thesis.

\subsubsection{Iterated integrals}

Iterated integrals are a powerful way to represent many classes of functions which arise naturally in the computation of Feynman integrals. They naturally appear, for example, when the method of differential equations is used to compute the Feynman integrals \cite{Henn:2013pwa,Blumlein:2018cms}, as well as in other areas \cite{Ablinger:2013cf,Moch:2001zr}. In this thesis, they occur in the $x$-space representation of OMEs, as the result of inverse Mellin transforms.

Iterated integrals are functions defined as
\begin{equation}
G\left(\left\{f_1(\tau),f_2(\tau),\cdots,f_n(\tau)\right\},z\right)
=\int_0^z  d\tau_1~f_1(\tau_1)  
G\left(\left\{f_2(\tau),\cdots,f_n(\tau)\right\},\tau_1\right),
\label{eq:Gfunctions}
\end{equation}
with
\begin{equation}
 G\Biggl(\Biggl\{\underbrace{\frac{1}{\tau},\frac{1}{\tau},
  \cdots,\frac{1}{\tau}}_{\text{n times}}\Biggr\},z\Biggr)
\equiv
\frac{1}{n!} \ln^n(z)~.
\label{eq:Gfunctions2}
\end{equation}
The set $\{f_i\}$ of functions appearing in a given physical problem is called the {\it alphabet}.

One example of iterated integrals studied in the context of high-energy physics is that of the harmonic polylogarithms or HPLs \cite{Remiddi:1999ew,Poincare:1884,Lappo:1953,Chen:1961}, which can be considered a special case of iterated integrals. They are defined as
\begin{eqnarray}
\HA_{b,\vec{a}}(x) &=& \int_0^x dy f_b(y) \HA_{\vec{a}}(y),~~~\HA_\emptyset = 1,~a_i, b~\in~\{0,1,-1\}~,
\label{eq:HPLdef1}
\end{eqnarray}
where
\begin{eqnarray}
f_0(x) = \frac{1}{x},~~~f_1(x) = \frac{1}{1-x},~~~f_{-1}(x) = \frac{1}{1+x}~,
\label{eq:HPLdef2}
\end{eqnarray}
and
\begin{equation}
\HA_{\underbrace{\text{\scriptsize 0},\ldots,\text{\scriptsize 0}}_{\text{\scriptsize n times}}}(x)
=
\frac{1}{n!} \ln^n(x)~.
\label{eq:HPLdef3}
\end{equation}
The number of indices in $H_{\vec a}(x)$ is called the \textit{weight} of the HPL. A numerical library for the evaluation of HPLs up to weight 5 has been published in \cite{Gehrmann:2001pz} and to weight 8 in \cite{Ablinger:2018sat}.

Harmonic polylogarithms are closely related to the Mellin transform of harmonic sums. For example, defining the ``+''-distribution as
\begin{equation}
	\int_0^1 dx \, f(x) \bigl(g(x)\bigr)_+ = \int_0^1 dx \, \bigl(f(x)-f(1)\bigr) g(x)
\end{equation}
one has
\begin{equation}
	S_1(N) = \mathbf M\Big[\Bigl(\frac{1}{x-1}\Bigr)_+\Big] .
\end{equation}
By repeated integration-by-parts, one can obtain the Mellin transform of HPLs in terms of harmonic sums (possibly evaluated at infinity) provided that appropriate regularizations such as the +-distribution are used in the Mellin transform.

Another special class is that of cyclotomic multiple polylogarithms \cite{Ablinger:2011te}, whose alphabet is
\begin{equation}
	\Bigl\{\frac{1}{x}\Bigr\}\cup\Big\{ \frac{x^b}{\Phi_n(x)} \mid n \in \mathbb N_+,0\leq b\leq\varphi(k) \Big\}
\end{equation}
\begin{equation}
	\Phi_n(x) = \prod_{\substack{1\le k\le n\\\text{gcd}(k,n)=1}} \Bigl(x-e^{2i\pi \frac{k}{n}} \Bigr)
\end{equation}
where $\varphi(n)$ is Euler's totient function. The polynomials $\Phi_n(x)$ are called cyclotomic polynomials. Cyclotomic harmonic polylogarithms are related through the Mellin transform to the cyclotomic harmonic sums.

Iterated integrals also obey shuffle algebras \cite{Blumlein:2003gb} due to the identity
\begin{equation}
	\Bigl(\int_0^x dy \, f(y) \Bigr) \Bigl(\int_0^x dy \, g(y) \Bigr) = \int_0^x dy \, f(y) \int_0^y dz \, g(z) + \int_0^x dy \, g(y) \int_0^y dz \, f(z) ,
\end{equation}
which allow to reduce the iterated integrals to a basis. Many algorithms pertaining to the symbolic manipulation of iterated integrals are available in {\tt HarmonicSums}.

\subsubsection{Mellin-Barnes integration}

The Feynman parametrization is one of the standard ways to compute Feynman integrals. It consists in the repeated application to Feynman integrals of the identity
\begin{equation}
	\frac{1}{A_1^{\nu_1}\cdots A_n^{\nu_n}} = \frac{\Gamma(\sum_{i=1}^n \nu_i)}{\Gamma(\nu_1)\cdots\Gamma(\nu_n)} \int_0^1 dx_1\cdots\int_0^1 dx_n \, \frac{x_1^{\nu_1-1}\cdots x_n^{\nu_n-1}}{(x_1 A_1+\cdots+x_n A_n)^{x_1+\cdots+x_n}} \,\delta\Big(1-\sum_{i=1}^n \nu_i\Big),
\end{equation}
which is valid for $A_i>0$, $\text{Re}(\nu_i)>0$. This identity is one of the methods which can be used to turn the integrals over the loop momenta into integrals over scalar Feynman parameters $x_i$. The evaluation of the integrals over $x_i$ is then the source of complexity in the Feynman representation, and the integrals will in general evaluate to special functions, in many cases yet unknown. One way to approach the integrals in the Feynman parametrization, which has been used in the calculations in this thesis, in part is through Mellin-Barnes integrals.

The Mellin-Barnes formula \cite{MB1a,MB1b,MB2,MB3} 
\begin{equation}
	\frac{1}{(A+B)^\lambda} = \frac{1}{2\pi i \Gamma(\lambda)} \int_{-i\infty}^{i\infty} d\sigma \,\Gamma(\sigma+\lambda)\Gamma(-\sigma) \frac{A^\sigma}{B^{\lambda+\sigma}}
\label{eq:mbFormula}
\end{equation}
is one important tool for the evaluation of Feynman integrals; early applications can be found in \cite{Bergere:1973fq,Ussyukina:1975}, see also \cite{Smirnov:2006ry}.
In \eqref{eq:mbFormula}, the integration contour must separate the infinite sets of ascending and descending poles of the $\Gamma$-functions, and must otherwise stretch in the direction of the imaginary axis. 
This formula is used to disentangle polynomials appearing from the Feynman parametrization. In general, the contour can be quite involved, and can necessitate the separate calculation of one or more residues, particularly if the integral is divergent in the dimensional parameter $\ep$. In any case, symbolic algebra packages exist to perform the analytic continuation and $\ep$-expansion of Mellin-Barnes integrals as well as for numerical evaluation. In this thesis we made use of {\tt MB} and {\tt MBResolve} \cite{Czakon:2005rk,Smirnov:2009up}.

The evaluation of Mellin-Barnes integrals with these packages is not possible in general if the integrand depends on the further symbolic parameter $N$, as is the case in the calculation of OMEs. For this application, the packages were therefore used only to compute moments. In the general case, one can apply the residue theorem to evaluate the integral into a (nested) infinite sum, and turn to algorithms in summation theory from the package {\tt Sigma} \cite{Schneider:2001,SIG1,SIG2}.

\clearpage
\section{Polarized deep-inelastic scattering}
\label{sec:PolDIS}
%
%

In the following, we present the calculation of the two-mass contributions to $A_{Qq}^{(3),\mathrm{PS}}$ and to $A_{gg,Q}^{(3)}$, which were performed in \cite{Ablinger:2019gpu} and in \cite{Ablinger:2020snj} respectively.

\subsection{The two-mass contribution to the polarized operator matrix element \texorpdfstring{$A_{Qq}^{(3),\mathrm{PS}}$}{AQq(3),PS}}
\label{sec:AQq3PS}
The calculation presented here for the polarized $A_{Qq}^{(3),\mathrm{PS}}$ closely mirrors that of the corresponding unpolarized OME, which was performed in \cite{Ablinger:2017xml}.

The Feynman rule used for the operator insertion, for the quark-quark-gluon vertex, is taken from \cite{Mertig:1995ny}: one has
\begin{equation}
O_a^\mu(p,q)=-g T_a \Delta^\mu/\!\!\!p\gamma_5\sum_{i=0}^{N-2}(\Delta.p)^{N-i-2}(-\Delta.q)^i,
\end{equation}
where $\Delta^\mu$ is a light-like vector, $\Delta^2=0$ and $p$, $q$ are incoming quark momenta.

When using dimensional regularization in polarized physics, a choice must be made for the analytic continuation from four to $D$ dimensions of chiral quantities, such as $\gamma_5$, which are intrinsically four-dimensional. This calculation has been performed in the Larin scheme \cite{Larin:1993tq}, which is the definition
\begin{equation}
\gamma_5 = \frac{i}{24}\gamma_\mu\gamma_\nu\gamma_\rho\gamma_\delta\varepsilon^{\mu\nu\rho\delta}.
\label{eq:gamma5Larin}
\end{equation}
The contraction of two Levi-Civita tensors is then
\begin{equation}
\varepsilon_{\alpha\beta\gamma\delta}\varepsilon^{\mu\nu\rho\sigma}=\left|\begin{array}{cccc}
g_{\alpha\mu} & g_{\alpha\nu} & g_{\alpha\rho} & g_{\alpha\sigma}\\
g_{\beta\mu} & g_{\beta\nu} & g_{\beta\rho} & g_{\beta\sigma}\\
g_{\gamma\mu} & g_{\gamma\nu} & g_{\gamma\rho} & g_{\gamma\sigma}\\
g_{\delta\mu} & g_{\delta\nu} & g_{\delta\rho} & g_{\delta\sigma}
\end{array}\right| ~.
\label{eq:LeviCivita}
\end{equation}
Other scheme choices have been made in the literature; in particular, the anomalous dimensions have been computed in the M-scheme, which was first defined in \cite{Matiounine:1998re} and is obtained from the Larin scheme through a finite renormalization.

In order to compare with the literature, care must be taken to adopt a consistent scheme choice: see \cite{Behring:2019tus} for the relationship between the anomalous dimensions in the two schemes.

The pole structure for the OME can be derived from the renormalization structure; see \cite{Ablinger:2017err} for this particular case. The predicted pole structure, which was used to check the calculation, is presented in Eq. \eqref{Ahhhqq3PSQ}:
\begin{eqnarray}
\hat{\hat{\tilde{A}}}_{Qq}^{(3),\rm{PS, tm}} &=&
\frac{8}{3 \ep^3} \gamma_{gq}^{(0)} \hat{\gamma}_{qg}^{(0)} \beta_{0,Q}
+\frac{1}{\ep^2}\biggl[
2 \gamma_{gq}^{(0)} \hat{\gamma}_{qg}^{(0)} \beta_{0,Q} \left(L_1+L_2\right)
+\frac{1}{6} \hat{\gamma}_{qg}^{(0)} \hat{\gamma}_{gq}^{(1)}
-\frac{4}{3} \beta_{0,Q} \hat{\gamma}_{qq}^{\rm{PS},(1)}
\biggr] 
\nonumber\\&&
+\frac{1}{\ep}
\biggl[
\gamma_{gq}^{(0)} \hat{\gamma}_{qg}^{(0)} \beta_{0,Q} \left(L_1^2+L_1 L_2+L_2^2\right)
+\biggl\{\frac{1}{8} \hat{\gamma}_{qg}^{(0)} \hat{\gamma}_{gq}^{(1)}
- \beta_{0,Q} \hat{\gamma}_{qq}^{\rm{PS},(1)}
\biggr\} \left(L_2+L_1\right)
\nonumber\\&&
+\frac{1}{3} \hat{\tilde{\gamma}}_{qq}^{(2),\rm{PS}}
-8 a_{Qq}^{(2),\rm{PS}} \beta_{0,Q}
+ \hat{\gamma}_{qg}^{(0)} a_{gq}^{(2)}
\biggr] 
+\tilde{a}_{Qq}^{(3),\rm{PS}}\left(m_1^2,m_2^2,\mu^2\right)
                   \label{Ahhhqq3PSQ},
\end{eqnarray}
where
\begin{eqnarray}
\hat{\gamma}_{ij} &=& \gamma_{ij}(N_F+2) - \gamma_{ij}(N_F),
\\
\hat{\tilde{\gamma}}_{ij} &=& \frac{\gamma_{ij}(N_F+2)}{N_F+2}  - \frac{\gamma_{ij}(N_F)}{N_F} ,
\end{eqnarray}
and the notation $a_{ij}$, $\bar{a}_{ij}$ denote the respective $\mathcal{O}(\ep^0)$, $\mathcal{O}(\ep)$ terms of the OMEs, while $\zeta_k$ is the Riemann $\zeta$-function at integer values, Eq. \eqref{eq:zetak}.
The quantity $\tilde{a}_{Qq}^{(3),\rm{PS}}\left(m_1^2,m_2^2,\mu^2\right)$ is the object of this calculation.

\begin{figure}[ht]
\begin{center}
\begin{minipage}[c]{0.19\linewidth}
     \includegraphics[width=1\textwidth]{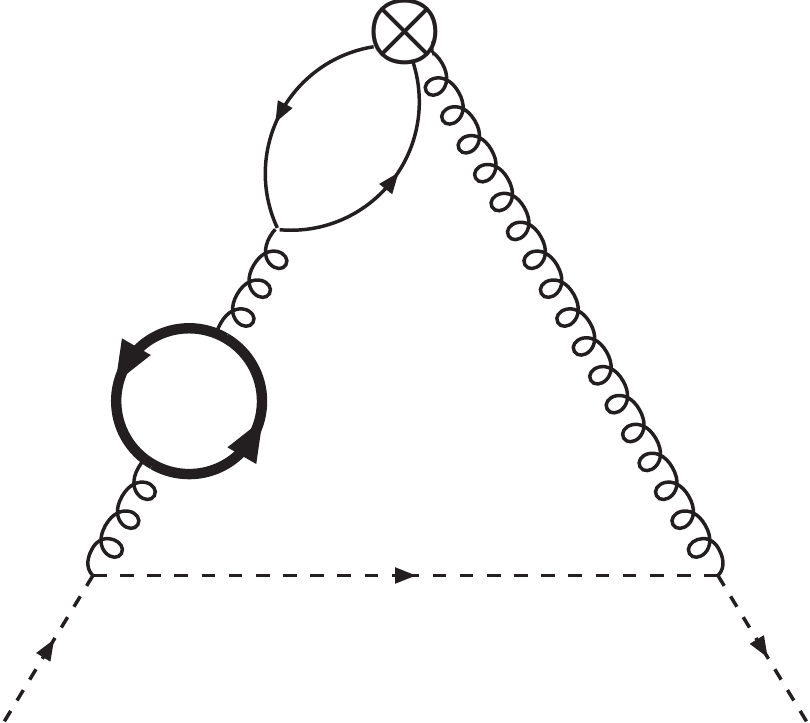}
\vspace*{-11mm}
\begin{center}
{\footnotesize (1)}
\end{center}
\end{minipage}
\hspace*{1mm}
\begin{minipage}[c]{0.19\linewidth}
     \includegraphics[width=1\textwidth]{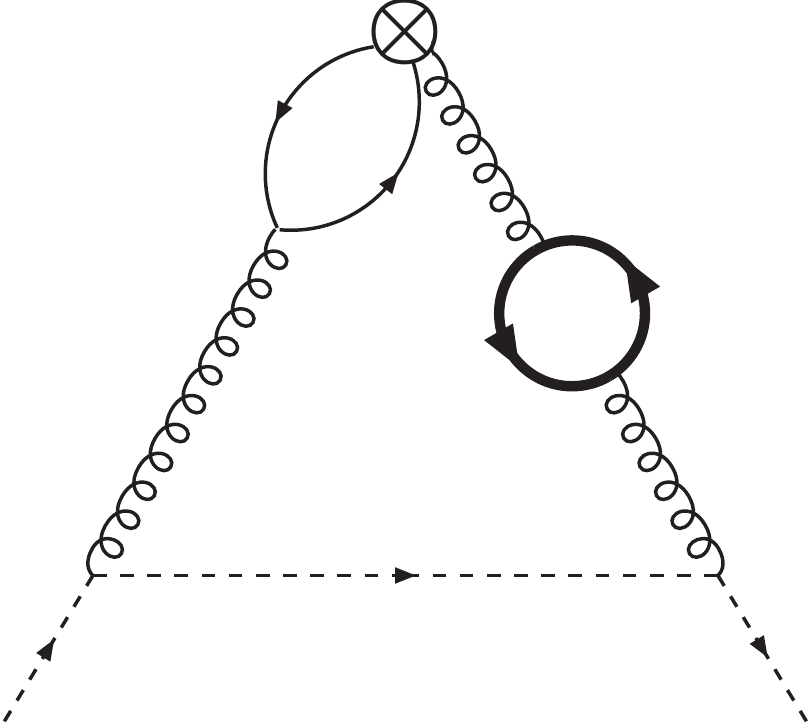}
\vspace*{-11mm}
\begin{center}
{\footnotesize (2)}
\end{center}
\end{minipage}
\hspace*{1mm}
\begin{minipage}[c]{0.19\linewidth}
     \includegraphics[width=1\textwidth]{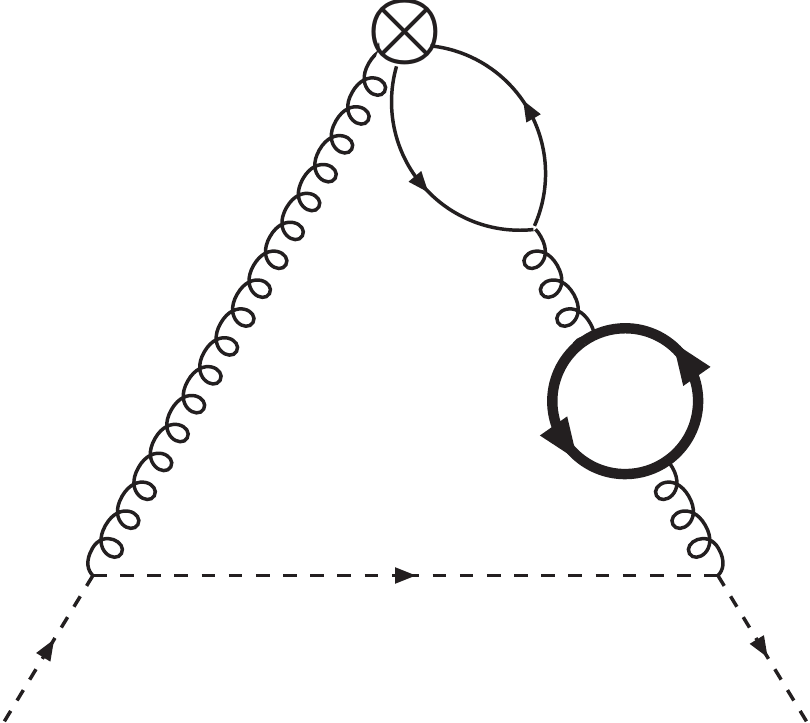}
\vspace*{-11mm}
\begin{center}
{\footnotesize (3)}
\end{center}
\end{minipage}
\hspace*{1mm}
\begin{minipage}[c]{0.19\linewidth}
     \includegraphics[width=1\textwidth]{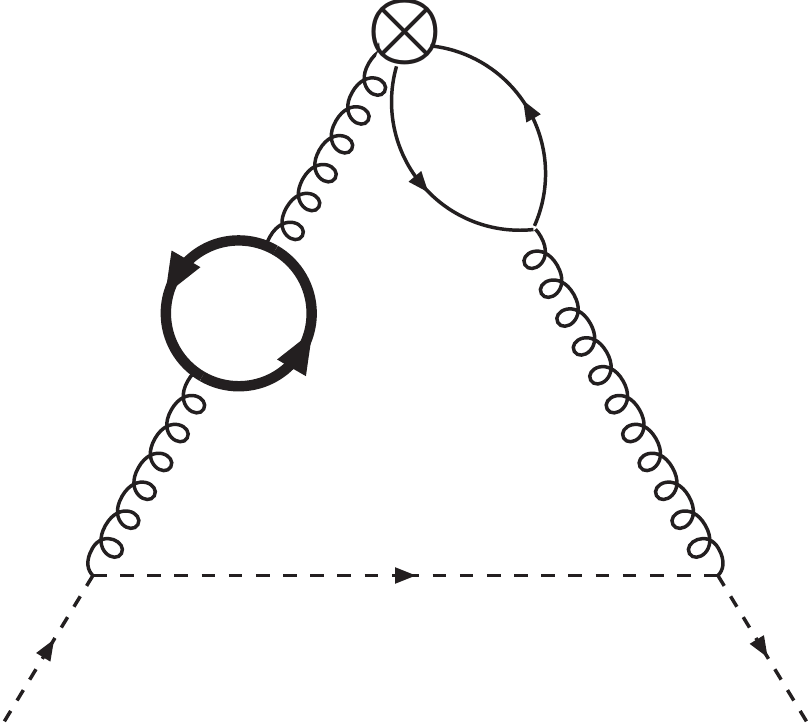}
\vspace*{-11mm}
\begin{center}
{\footnotesize (4)}
\end{center}
\end{minipage}

\vspace*{5mm}

\begin{minipage}[c]{0.19\linewidth}
     \includegraphics[width=1\textwidth]{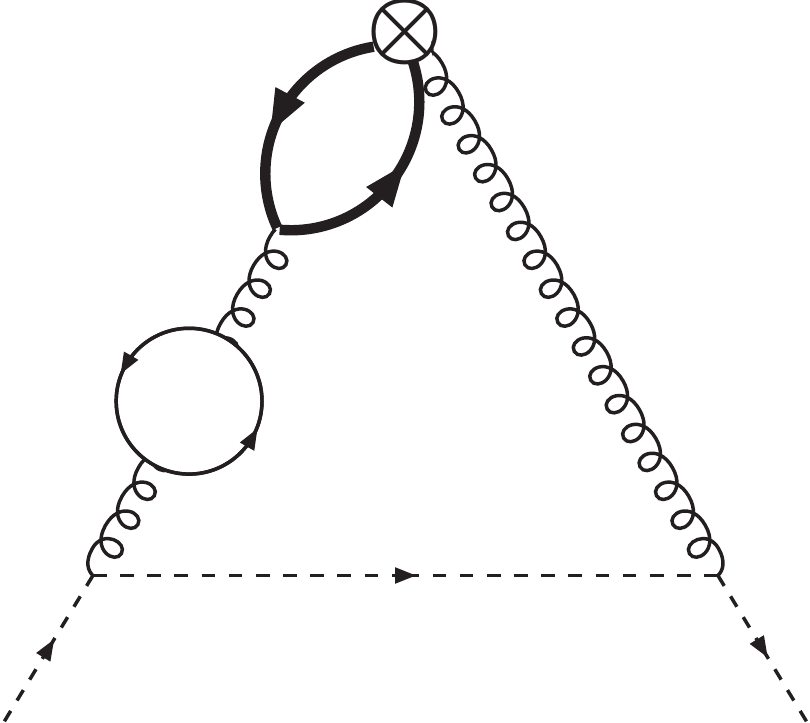}
\vspace*{-11mm}
\begin{center}
{\footnotesize (5)}
\end{center}
\end{minipage}
\hspace*{1mm}
\begin{minipage}[c]{0.19\linewidth}
     \includegraphics[width=1\textwidth]{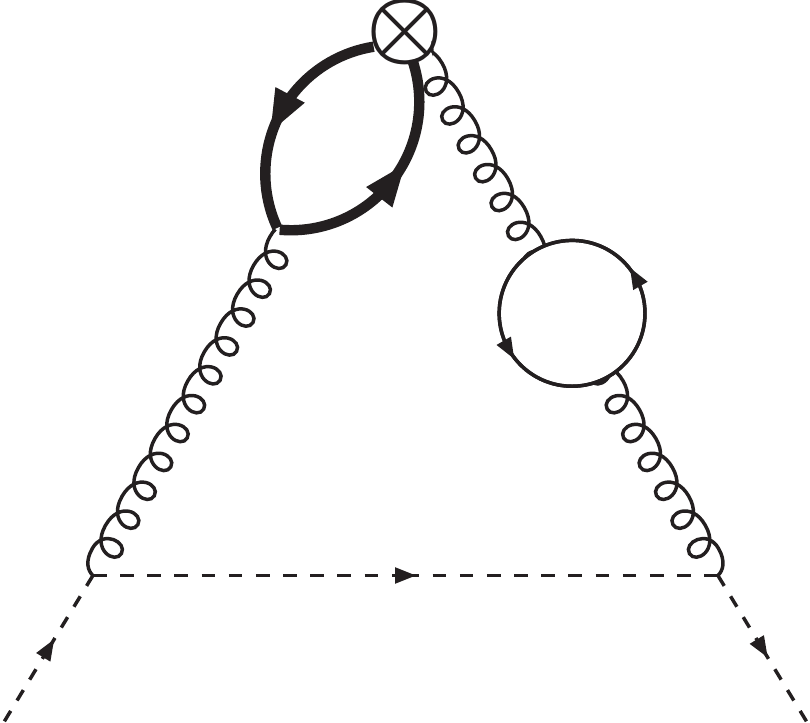}
\vspace*{-11mm}
\begin{center}
{\footnotesize (6)}
\end{center}
\end{minipage}
\hspace*{1mm}
\begin{minipage}[c]{0.19\linewidth}
     \includegraphics[width=1\textwidth]{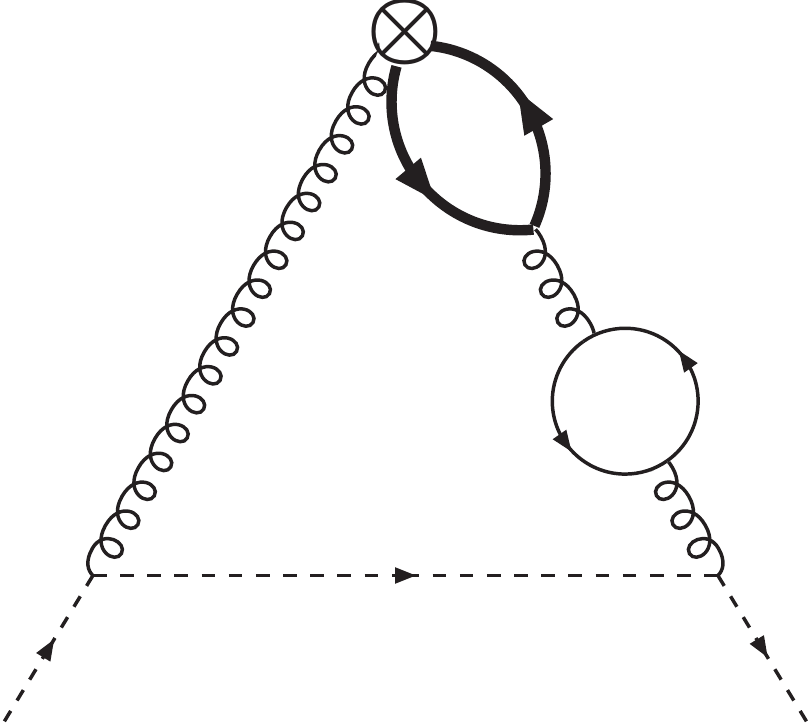}
\vspace*{-11mm}
\begin{center}
{\footnotesize (7)}
\end{center}
\end{minipage}
\hspace*{1mm}
\begin{minipage}[c]{0.19\linewidth}
     \includegraphics[width=1\textwidth]{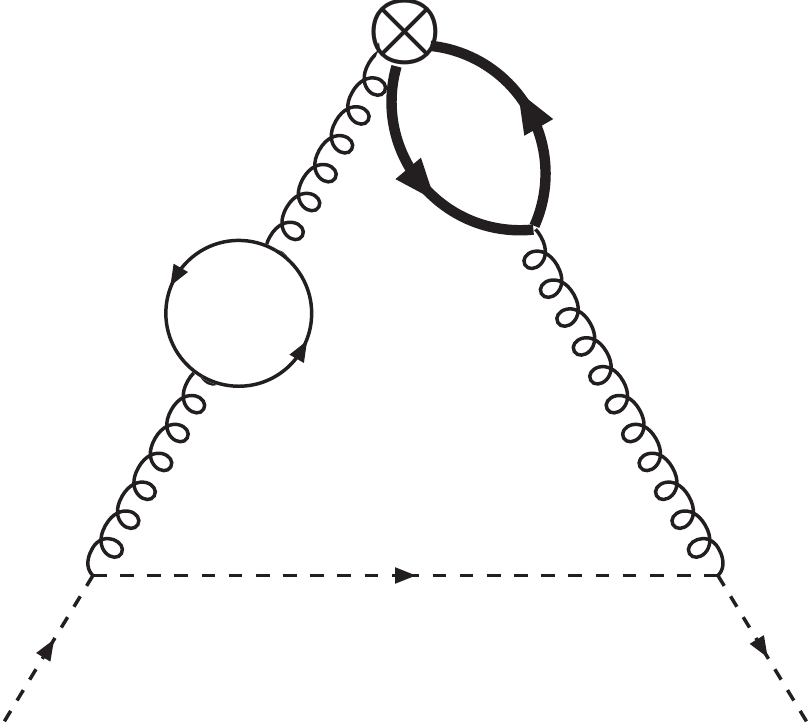}
\vspace*{-11mm}
\begin{center}
{\footnotesize (8)}
\end{center}
\end{minipage}

\vspace*{5mm}

\begin{minipage}[c]{0.19\linewidth}
     \includegraphics[width=1\textwidth]{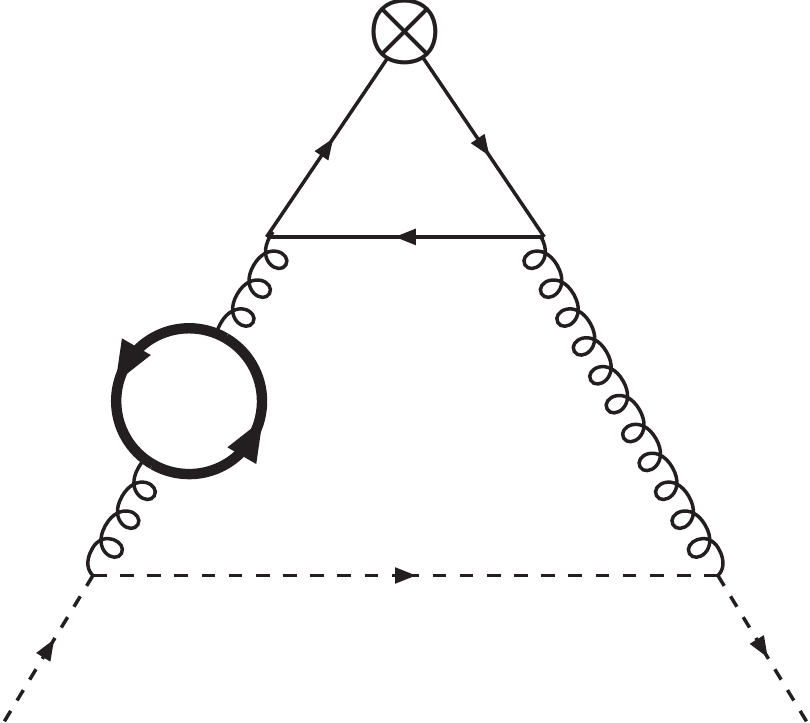}
\vspace*{-11mm}
\begin{center}
{\footnotesize (9)}
\end{center}
\end{minipage}
\hspace*{1mm}
\begin{minipage}[c]{0.19\linewidth}
     \includegraphics[width=1\textwidth]{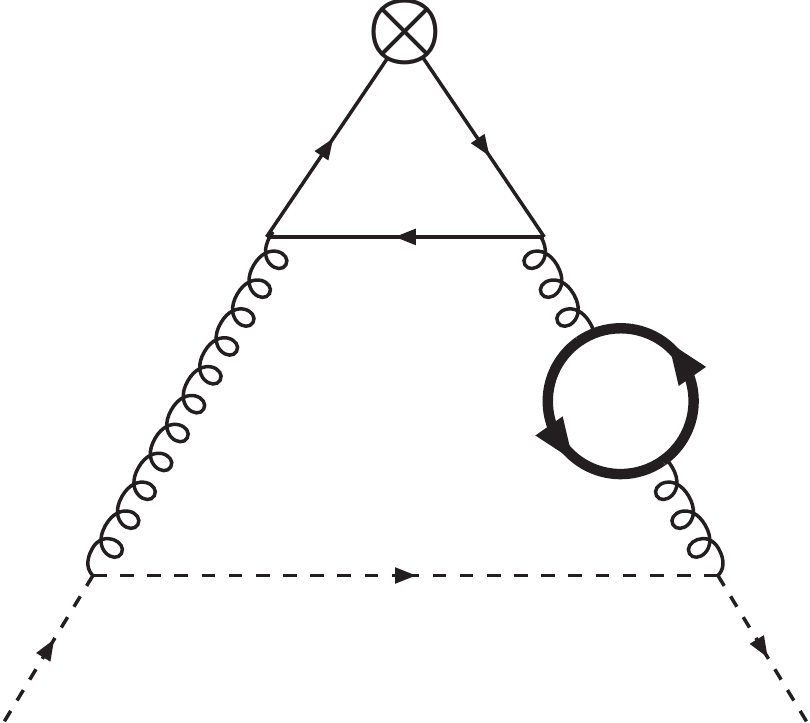}
\vspace*{-11mm}
\begin{center}
{\footnotesize (10)}
\end{center}
\end{minipage}
\hspace*{1mm}
\begin{minipage}[c]{0.19\linewidth}
     \includegraphics[width=1\textwidth]{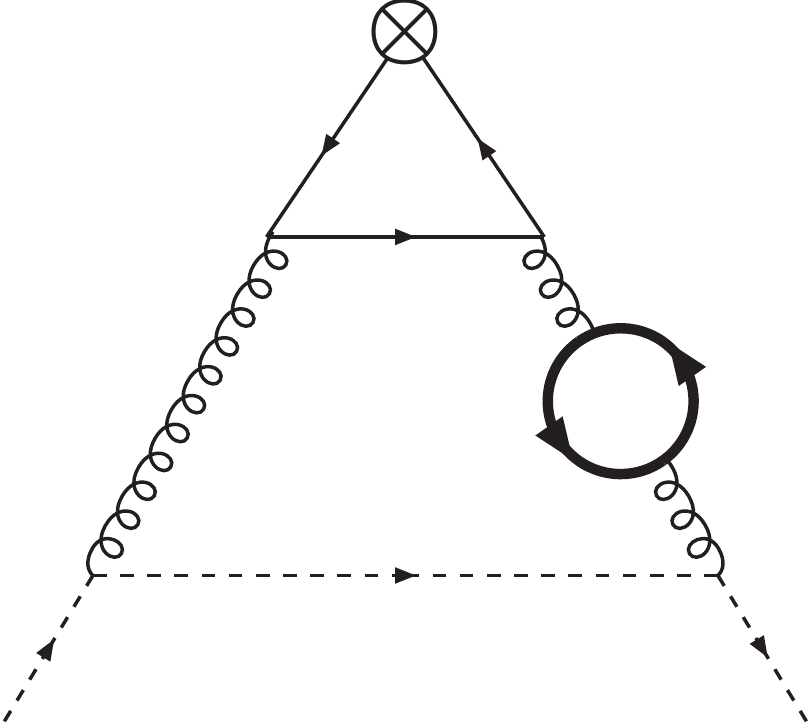}
\vspace*{-11mm}
\begin{center}
{\footnotesize (11)}
\end{center}
\end{minipage}
\hspace*{1mm}
\begin{minipage}[c]{0.19\linewidth}
     \includegraphics[width=1\textwidth]{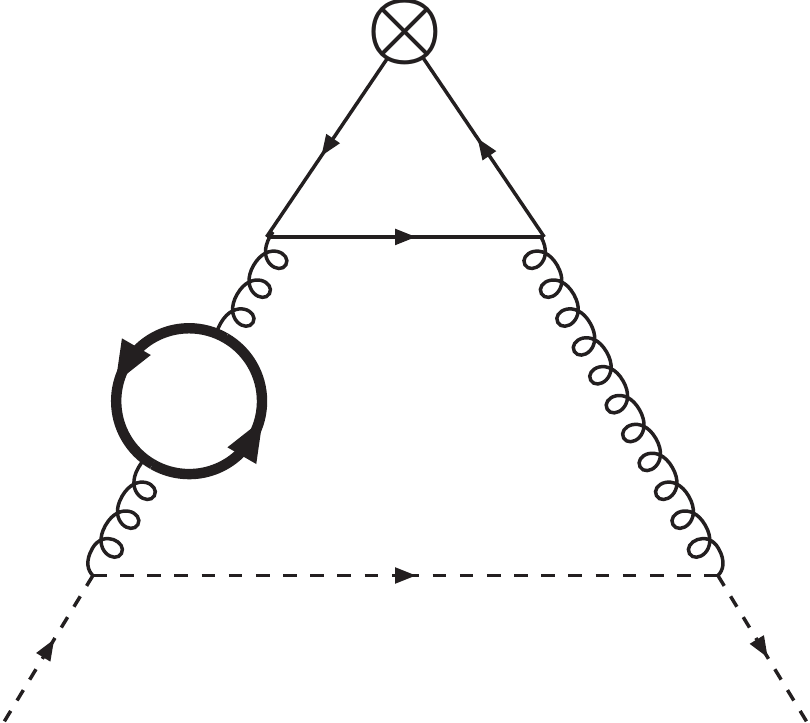}
\vspace*{-11mm}
\begin{center}
{\footnotesize (12)}
\end{center}
\end{minipage}

\vspace*{5mm}

\begin{minipage}[c]{0.19\linewidth}
     \includegraphics[width=1\textwidth]{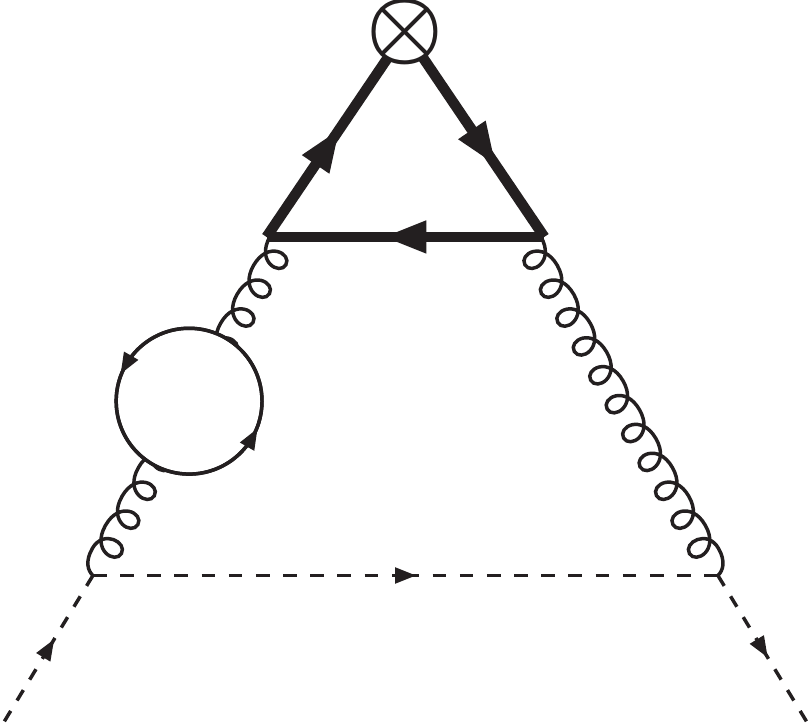}
\vspace*{-11mm}
\begin{center}
{\footnotesize (13)}
\end{center}
\end{minipage}
\hspace*{1mm}
\begin{minipage}[c]{0.19\linewidth}
     \includegraphics[width=1\textwidth]{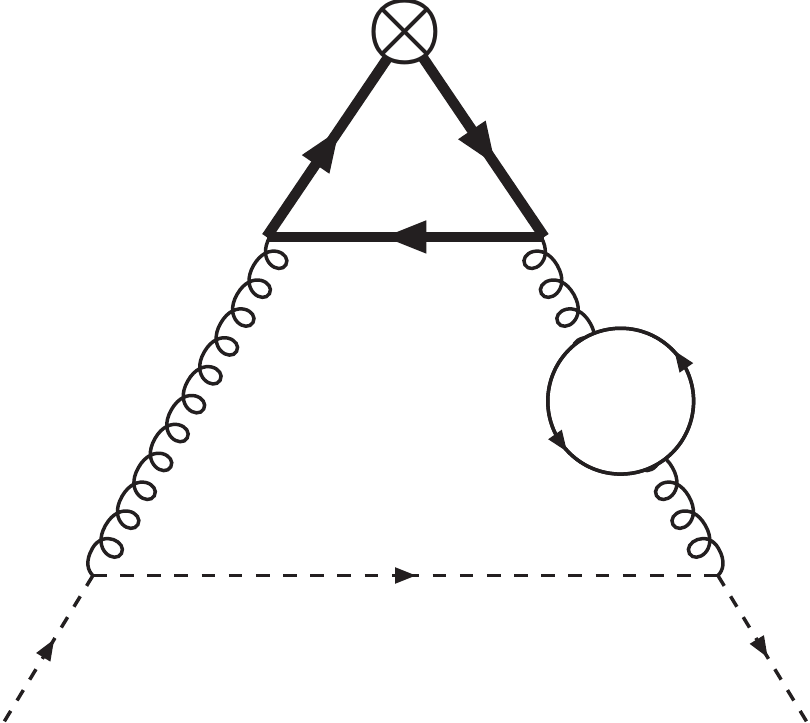}
\vspace*{-11mm}
\begin{center}
{\footnotesize (14)}
\end{center}
\end{minipage}
\hspace*{1mm}
\begin{minipage}[c]{0.19\linewidth}
     \includegraphics[width=1\textwidth]{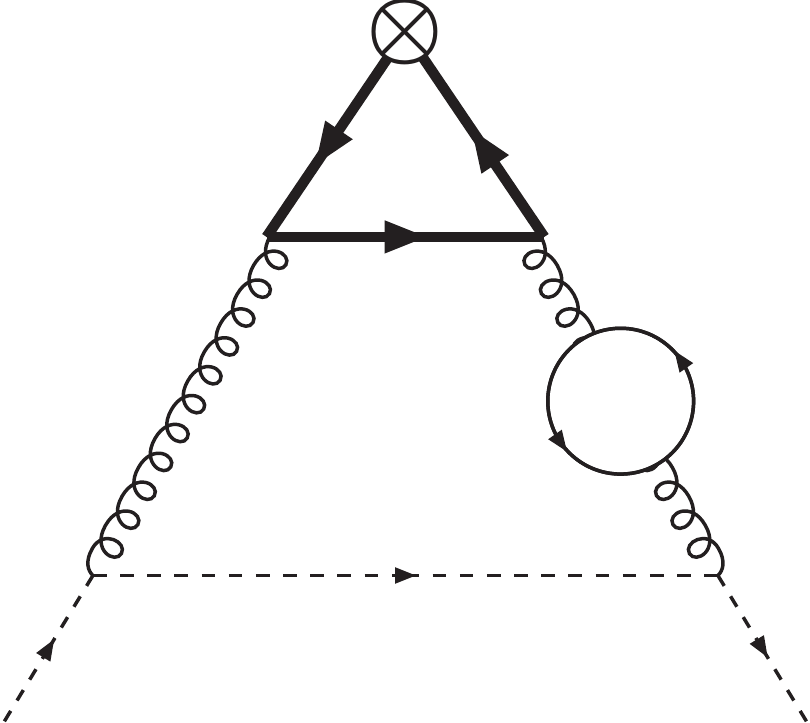}
\vspace*{-11mm}
\begin{center}
{\footnotesize (15)}
\end{center}
\end{minipage}
\hspace*{1mm}
\begin{minipage}[c]{0.19\linewidth}
     \includegraphics[width=1\textwidth]{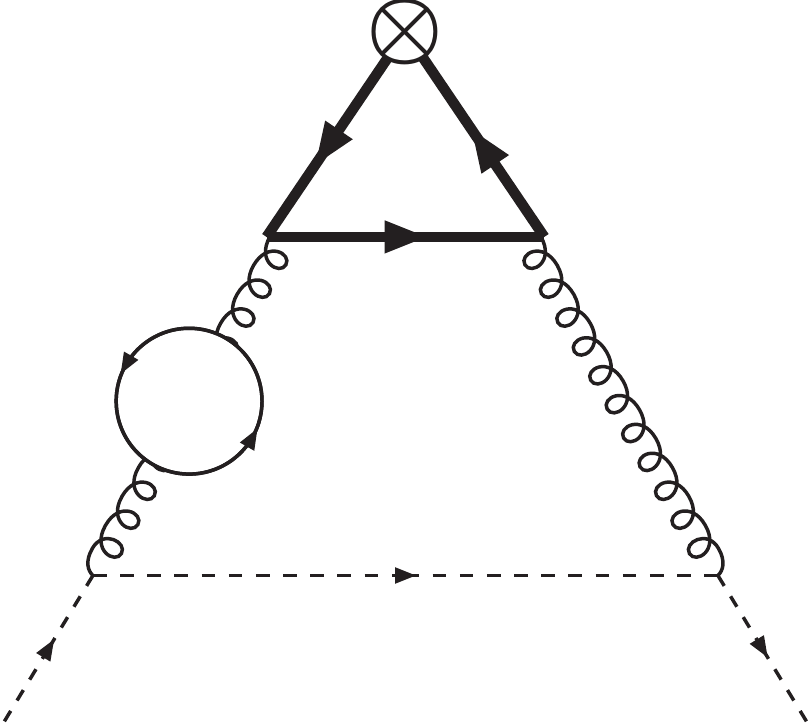}
\vspace*{-11mm}
\begin{center}
{\footnotesize (16)}
\end{center}
\end{minipage}
\caption{\sf \small The diagrams for the two-mass contributions to $\tilde{A}_{Qq}^{(3), \rm PS}$. The dashed arrow line represents the external
massless quarks, while the thick solid arrow line represents a quark of mass $m_1$, and the thin arrow line a quark of mass $m_2$. We assume $m_1 > m_2$.}
\label{PSdiagrams}
\end{center}
\end{figure}

The contributing diagrams are shown in Figure \ref{PSdiagrams}. The unrenormalized OME is obtained from their sum by applying the projector \cite{Behring:2019tus,Schonwald:2019gmn}
\begin{eqnarray}
\label{eq:PqNEW}
P_q \hat{G}_l^{ij} = - \delta_{ij} \frac{i (\Delta.p)^{-N-1}}{4 N_c (D-2)(D-3)} \ep_{\mu \nu p \Delta} {\rm tr} \left[p
\hspace*{-2mm}
\slash
\gamma^\mu \gamma^\nu \hat{G}_l^{ij}\right].
\end{eqnarray}
The numerator algebra was performed in {\tt Form} \cite{Vermaseren:2000nd} and in {\tt Mathematica}. 
Of the contributing diagrams, the nonzero ones are 9--12 and 13--16, which are related by a symmetry under the exchange of the two heavy quarks. One obtains for the unrenormalized OME
\begin{eqnarray}
A_{Qq}^{(3), {\rm PS, tm}}(N) &=&
2 \left[1 + (-1)^{N-1} \right] D_9(m_1,m_2,N) + 2 \left[1 + (-1)^{N-1} \right] D_9(m_2,m_1,N),
\label{eq:D9}
\end{eqnarray}
where we define the variable
\begin{equation}
\eta = \frac{m_2^2}{m_1^2}<1.
\end{equation}
The calculation made use of the following Mellin-Barnes representation for the massive bubbles of Figure \ref{PSbubbles}:

\begin{figure}[ht]
\begin{center}
\begin{minipage}[c]{0.21\linewidth}
     \includegraphics[width=1\textwidth]{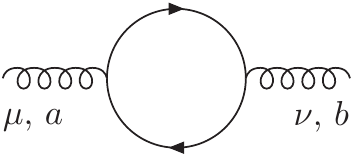}
\vspace*{-7mm}
\begin{center}
{\footnotesize ($a_1$)}
\end{center}
\end{minipage}
\hspace*{11mm}
\hspace*{11mm}
\begin{minipage}[c]{0.21\linewidth}
     \includegraphics[width=1\textwidth]{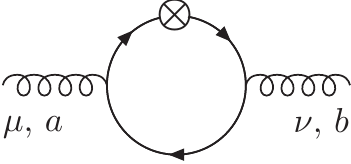}
\vspace*{-7mm}
\begin{center}
{\footnotesize ($b_2$)}
\end{center}
\end{minipage}
\caption{\sf \small Massive bubbles appearing in the Feynman diagrams shown in Figure \ref{PSdiagrams}.}
\label{PSbubbles}
\end{center}
\end{figure}

\begin{eqnarray}
I^{\mu\nu,ab}_{a_1}(k) &=&
-\frac{8iT_F g_s^2}{(4\pi)^{D/2}}\delta_{ab}(k^2g^{\mu\nu}-k^\mu k^\nu)\int_0^1 
dx\frac{\Gamma(2-D/2)(x(1-x))^{D/2-1}}{\left(-k^2+\frac{m^2}{x(1-x)}\right)^{2-D/2}},\\
I^{\mu\nu,ab}_{b_2}(k) &=&
\alpha_s T_F i e^{-\gamma_E \varepsilon/2}(k\cdot\Delta)^{N-1} (\mu^2)^{-\varepsilon/2}S_\varepsilon \epsilon^{\Delta k 
\mu 
\nu}\int_0^1 dx\, x^{N+D/2-1}(1-x)^{D/2-1}\nonumber\\
&&
\times\Bigg\{\left(-k^2+\frac{m^2}{x(1-x)}\right)^{-2+D/2}2\Gamma(2-D/2)\left[(D-6)x^{-2}+(D+2N)x^{-1}\right]\\\nonumber&&
+\left(-k^2+\frac{m^2}{x(1-x)}\right)^{-3+D/2}4\Gamma(3-D/2)(1-x)^{-1}\Big[m^2(x^{-3}+x^{-2})\nonumber\\
&&
+(-k^2)(1-x^{-1})\Big]\Bigg\},
\end{eqnarray}

After the Feynman parametrization and the Mellin-Barnes decomposition are performed, one obtains for Diagram 9 the representation

\begin{eqnarray}
D_{9}(m_{1},m_{2},N) & = & C_F T_F^2 \alpha_s^3 S_\varepsilon^3 \frac{16}{2+\varepsilon}\Bigg\{ 4(2-\varepsilon)J_{1}-8\eta 
J_{2}-8(N+3)J_{3}+8J_{4}+8\bigg(2+\frac{\varepsilon}{2}+N\bigg)
\nonumber\\ 
& &
\times J_{5}
-8J_{6}-(\varepsilon-2)^{2}J_{7}+2(2-\varepsilon)\eta J_{8}+2(2-\varepsilon)(3+N)J_{9}-2(2-\varepsilon)J_{10}
\nonumber\\
 &  & -2(2-\varepsilon)\bigg(2+\frac{\varepsilon}{2}+N\bigg)J_{11}+2(2-\varepsilon)J_{12}-8\eta J_{13}+2(2-\varepsilon)\eta 
J_{14}\Bigg\},
\end{eqnarray}
with
\begin{eqnarray}
J_{1} & = & \left(\frac{m_{1}^{2}}{\mu^{2}}\right)^{\frac{3}{2}\varepsilon}\frac{\Gamma(N)}{\Gamma\big(1+\frac{\varepsilon 
}{2}+N\big)}\int_{0}^{1}dx\,(1-x)^{\frac{\varepsilon}{2}}x^{-1+\frac{\varepsilon }{2}+N}B_{1}\left(\frac{\eta}{x(1-x)}\right),\\
J_{2} & = & \left(\frac{m_{1}^{2}}{\mu^{2}}\right)^{\frac{3}{2}\varepsilon}\frac{\Gamma(N)}{\Gamma\big(1+\frac{\varepsilon 
}{2}+N\big)}\int_{0}^{1}dx\,(1-x)^{\frac{\varepsilon}{2}}x^{-1+\frac{\varepsilon }{2}+N}B_{3}\left(\frac{\eta}{x(1-x)}\right),\\
J_{3} & = & \left(\frac{m_{1}^{2}}{\mu^{2}}\right)^{\frac{3}{2}\varepsilon}\frac{\Gamma(N)}{\Gamma\big(1+\frac{\varepsilon 
}{2}+N\big)}\int_{0}^{1}dx\,(1-x)^{\frac{\varepsilon}{2}}x^{\frac{\varepsilon }{2}+N}B_{1}\left(\frac{\eta}{x(1-x)}\right),\\
J_{4} & = & \left(\frac{m_{1}^{2}}{\mu^{2}}\right)^{\frac{3}{2}\varepsilon}\frac{\Gamma(N)}{\Gamma\big(1+\frac{\varepsilon 
}{2}+N\big)}\int_{0}^{1}dx\,(1-x)^{\frac{\varepsilon}{2}}x^{\frac{\varepsilon }{2}+N}B_{2}\left(\frac{\eta}{x(1-x)}\right),\\
J_{5} & = & \left(\frac{m_{1}^{2}}{\mu^{2}}\right)^{\frac{3}{2}\varepsilon}\frac{\Gamma(N)}{\Gamma\big(1+\frac{\varepsilon 
}{2}+N\big)}\int_{0}^{1}dx\,(1-x)^{\frac{\varepsilon}{2}}x^{1+\frac{\varepsilon }{2}+N}B_{1}\left(\frac{\eta}{x(1-x)}\right),\\
J_{6} & = & \left(\frac{m_{1}^{2}}{\mu^{2}}\right)^{\frac{3}{2}\varepsilon}\frac{\Gamma(N)}{\Gamma\big(1+\frac{\varepsilon 
}{2}+N\big)}\int_{0}^{1}dx\,(1-x)^{\frac{\varepsilon}{2}}x^{1+\frac{\varepsilon }{2}+N}B_{2}\left(\frac{\eta}{x(1-x)}\right),\\
J_{7} & = & \left(\frac{m_{1}^{2}}{\mu^{2}}\right)^{\frac{3}{2}\varepsilon}\frac{\Gamma(N+1)}{\Gamma\big(2+\frac{\varepsilon 
}{2}+N\big)}\int_{0}^{1}dx\,(1-x)^{\frac{\varepsilon}{2}}x^{-1+\frac{\varepsilon }{2}+N}B_{1}\left(\frac{\eta}{x(1-x)}\right),\\
J_{8} & = & \left(\frac{m_{1}^{2}}{\mu^{2}}\right)^{\frac{3}{2}\varepsilon}\frac{\Gamma(N+1)}{\Gamma\big(2+\frac{\varepsilon 
}{2}+N\big)}\int_{0}^{1}dx\,(1-x)^{\frac{\varepsilon}{2}}x^{-1+\frac{\varepsilon }{2}+N}B_{3}\left(\frac{\eta}{x(1-x)}\right),\\
J_{9} & = & \left(\frac{m_{1}^{2}}{\mu^{2}}\right)^{\frac{3}{2}\varepsilon}\frac{\Gamma(N+1)}{\Gamma\big(2+\frac{\varepsilon 
}{2}+N\big)}\int_{0}^{1}dx\,(1-x)^{\frac{\varepsilon}{2}}x^{\frac{\varepsilon }{2}+N}B_{1}\left(\frac{\eta}{x(1-x)}\right),\\
J_{10} & = & \left(\frac{m_{1}^{2}}{\mu^{2}}\right)^{\frac{3}{2}\varepsilon}\frac{\Gamma(N+1)}{\Gamma\big(2+\frac{\varepsilon 
}{2}+N\big)}\int_{0}^{1}dx\,(1-x)^{\frac{\varepsilon}{2}}x^{\frac{\varepsilon }{2}+N}B_{2}\left(\frac{\eta}{x(1-x)}\right),\\
J_{11} & = & \left(\frac{m_{1}^{2}}{\mu^{2}}\right)^{\frac{3}{2}\varepsilon}\frac{\Gamma(N+1)}{\Gamma\big(2+\frac{\varepsilon 
}{2}+N\big)}\int_{0}^{1}dx\,(1-x)^{\frac{\varepsilon}{2}}x^{1+\frac{\varepsilon }{2}+N}B_{1}\left(\frac{\eta}{x(1-x)}\right),\\
J_{12} & = & \left(\frac{m_{1}^{2}}{\mu^{2}}\right)^{\frac{3}{2}\varepsilon}\frac{\Gamma(N+1)}{\Gamma\big(2+\frac{\varepsilon 
}{2}+N\big)}\int_{0}^{1}dx\,(1-x)^{\frac{\varepsilon}{2}}x^{1+\frac{\varepsilon }{2}+N}B_{2}\left(\frac{\eta}{x(1-x)}\right),\\
J_{13} & = & \left(\frac{m_{1}^{2}}{\mu^{2}}\right)^{\frac{3}{2}\varepsilon}\frac{\Gamma(N)}{\Gamma\big(1+\frac{\varepsilon 
}{2}+N\big)}\int_{0}^{1}dx\,(1-x)^{\frac{\varepsilon}{2}}x^{-2+\frac{\varepsilon }{2}+N}B_{3}\left(\frac{\eta}{x(1-x)}\right),\\
J_{14} & = & \left(\frac{m_{1}^{2}}{\mu^{2}}\right)^{\frac{3}{2}\varepsilon}\frac{\Gamma(N+1)}{\Gamma\big(2+\frac{\varepsilon 
}{2}+N\big)}\int_{0}^{1}dx\,(1-x)^{\frac{\varepsilon}{2}}x^{-2+\frac{\varepsilon }{2}+N}B_{3}\left(\frac{\eta}{x(1-x)}\right).
\end{eqnarray}
The functions $B_i$ are given by
\begin{eqnarray}
B_1(\xi) &=& \frac{1}{2 \pi i} \int_{-i \infty}^{i \infty} d\sigma \, \xi^{\sigma} \,
\Gamma(-\sigma) \Gamma(-\sigma+\varepsilon) \Gamma\left(\sigma-\frac{3 \varepsilon}{2}\right) \Gamma\left(\sigma 
-\frac{\varepsilon}{2}\right)
\frac{\Gamma^2(\sigma+2-\varepsilon)}{\Gamma(2 \sigma+4-2 \varepsilon)},
\label{B1}
\\
B_2(\xi) &=& \frac{1}{2 \pi i} \int_{-i \infty}^{i \infty} d\sigma \, \xi^{\sigma} \,
\Gamma(-\sigma) \Gamma(-\sigma+\varepsilon) \Gamma\left(\sigma-\frac{3 \varepsilon}{2}\right) 
\Gamma\left(\sigma+1-\frac{\varepsilon}{2}\right)
\frac{\Gamma^2(\sigma+2-\varepsilon)}{\Gamma(2 \sigma+4-2 \varepsilon)},
\nonumber \\   
\label{B2}
\\
B_3(\xi) &=& \frac{1}{2 \pi i} \int_{-i \infty}^{i \infty} d\sigma \, \xi^{\sigma} \,
\Gamma(-\sigma) \Gamma(-\sigma-1+\varepsilon) \Gamma\left(\sigma+1-\frac{3 \varepsilon}{2}\right) 
\Gamma\left(\sigma+1-\frac{\varepsilon}{2}\right)
\nonumber \\ && \phantom{\frac{1}{2 \pi i} \int_{-i \infty}^{i \infty} d\sigma} \times
\frac{\Gamma^2(\sigma+3-\varepsilon)}{\Gamma(2 \sigma+6-2 \varepsilon)}.
\label{B3}
\end{eqnarray}
where we take 
\begin{equation}
\xi = \frac{1}{\eta x(1-x)}
\end{equation}
for diagram 9.
It was possible to reduce this representation to the same class of functions as in the unpolarized calculation, as is expected from the fact that the two cases differ only in numerator structures. The calculation was then performed in $x$ space by taking residues of the functions $B_i$ in $\sigma$ and expanding in $\ep$. Because the functions $B_i$ are the same as those appearing in the unpolarized calculation, it was possible to refer to their $\mathcal{O}(\ep^0)$ behaviour as computed in \cite{Ablinger:2017xml}, where this expansion was performed with the packages {\tt MB} \cite{Czakon:2005rk}, {\tt MBresolve} \cite{Smirnov:2009up}, {\tt Sigma}~\cite{Schneider:2001,SIG1,SIG2}, {\tt HarmonicSums}~\cite{HARMONICSUMS,Ablinger:2011te,Ablinger:2013cf}, {\tt EvaluateMultiSums} and {\tt SumProduction} \cite{EMSSP}. In this way, the integrals appearing in $B_i$ are evaluated in terms of sums involving harmonic sums.

The convergence of the integrals depends on the value of $\xi$. This implies that the interval $x\in [0,1]$ needs to be split into three intervals
\begin{eqnarray}   
[0,\eta_-],~~~[\eta_-,\eta_+],~~~[\eta_+,1],~~~\text{with}~~\eta_\pm = \frac{1}{2}\left(1 \pm \sqrt{1-\eta}\right),
\end{eqnarray}   
where the contours appearing in the functions $B_i$ are closed to the right (for the second region) or to the left (for the first and the third), giving rise to two different functional forms for the residue sums.
The constant parts in $\ep$, $B_i^{(0)}$, depend on harmonic sums \cite{Vermaseren:1998uu,Blumlein:1998if}, defined in Eqs.\ \eqref{eq:defHSums1} and \eqref{eq:defHSums2}.

Then, the prefactors appearing in $J_i$ can be expanded in $\ep$, giving rise, after partial fractioning, to denominators of the type
\begin{equation}
\frac{1}{N+l}, \quad {\rm with} \quad l \in \{0,1\},
\end{equation}
which can be absorbed inside an integral using the relation
\begin{eqnarray}
\frac{1}{N+l} \int_a^b dx \, x^{N-1} f(x) &=&
\frac{b^{N+l}}{N+l} \int_a^b dy \frac{f(y)}{y^{l+1}} - \int_a^b dx \, x^{N+l-1} \int_a^x dy \frac{f(y)}{y^{l+1}}
\label{absorbN1}
\\ &=&
\frac{a^{N+l}}{N+l} \int_a^b dy \frac{f(y)}{y^{l+1}} + \int_a^b dx \, x^{N+l-1} \int_x^b dy \frac{f(y)}{y^{l+1}}.
\label{absorbN2}
\end{eqnarray}   
The purpose of this method is to ultimately obtain $\tilde{a}_{Qq}^{(3),\rm{PS}}$ in $x$-space, by leaving one of the Feynman parameters unintegrated.

The final result is expressed in terms of generalized iterated integrals, as defined in Eqs. \eqref{eq:Gfunctions}, \eqref{eq:Gfunctions2}
and harmonic polylogarithms \cite{Remiddi:1999ew} which can be considered a special case of iterated integrals and are defined as in \eqref{eq:HPLdef1}, \eqref{eq:HPLdef2}, \eqref{eq:HPLdef3}.

In this case, the letters appearing in the iterated integrals are 
\begin{equation}
\frac{1}{\tau}, \quad \sqrt{4-\tau} \sqrt{\tau}, \quad \frac{\sqrt{1-4 \tau}}{\tau}.
\end{equation}

\subsubsection{The \texorpdfstring{$x$}{x}-space result}
We obtain the following expression for the $O(\ep^0)$ term of the unrenormalized 3-loop two-mass pure singlet operator matrix element: 
\begin{eqnarray}
\tilde{a}_{Qq}^{(3), \rm PS}(x) &=& C_F T_F^2 \Biggl\{
R_0(m_1,m_2,x) +\big(\theta(\eta_--x)+\theta(x-\eta_+)\big) x \, g_0(\eta,x)
\nonumber \\ &&
+\theta(\eta_+-x) \theta(x-\eta_-) \biggl[x \, f_0(\eta,x)
-\int_{\eta_-}^x dy \left(f_1(\eta,y)+\frac{x}{y} f_3(\eta,y)\right)\biggr]
\nonumber \\ &&
+\theta(\eta_--x) \int_x^{\eta_-} dy \left(g_1(\eta,y)+\frac{x}{y} g_3(\eta,y)\right)
\nonumber \\ &&
-\theta(x-\eta_+) \int_{\eta_+}^x dy \left(g_1(\eta,y)+\frac{x}{y} g_3(\eta,y)\right)
\nonumber \\ &&
+x \, h_0(\eta,x) +\int_x^1 dy \left(h_1(\eta,y)+\frac{x}{y} h_3(\eta,y)\right)
\nonumber \\ &&
+\theta(\eta_+-x) \int_{\eta_-}^{\eta_+} dy \left(f_1(\eta,y)+ \frac{x}{y} f_3(\eta,y)\right)
\nonumber \\ &&
+\int_{\eta_+}^1 dy \left(g_1(\eta,y)+\frac{x}{y} g_3(\eta,y)\right)
\Biggr\}.
\label{aQq}
\end{eqnarray}
Here we follow the notation used in Ref.~\cite{Ablinger:2017xml}. Compared to that notation, in the present case no functions carrying the index 2 occur.
The functions $g_i(\eta,x)$ in Eq.~(\ref{aQq}) shall not be confounded with polarized structure functions, also often 
denoted
by $g_i$. Here $\theta(z)$ denotes the Heaviside function
\begin{eqnarray}
\theta(z) = \left\{\begin{array}{ll} 1 &~~z \geq 0\\
                                     0 &~~z < 0
\end{array} \right.
\label{eq:Heaviside}
\end{eqnarray}
by which we divide the interval $x\in [0,1]$ as described earlier. We define for convenience
\begin{equation}
u=\frac{x(1-x)}{\eta},~~~~
v=\frac{\eta}{x(1-x)}
\end{equation}
and
\begin{equation}
L_1 = \ln\left(\frac{m_1^2}{\mu^2}\right), \quad L_2 = \ln\left(\frac{m_2^2}{\mu^2}\right),
\label{L1L2}
\end{equation}
with $\mu$ the renormalization scale. If in the following expressions the harmonic polylogarithms $H_{\vec{a}}$ are given without argument it is understood that their argument is $x$.
The functions appearing in Eq.~(\ref{aQq}) are given by

\begin{eqnarray}
R_0(m_1,m_2,x)&=&32
        \bigg( L_1^3 +L_1  L_2  (L_1  +L_2  ) + L_2^3 \bigg)
\bigg[5 (-1+x) -2 (1+x) \HA_0 \bigg]\nonumber\\
& & +128 L_1  L_2  \bigg[
        (x+1) \bigg(\frac{2}{3}\HA_{0,1}-\frac{10}{9}\HA_0-\frac{2}{3}\zeta_2\bigg)
        +(x-1)\bigg(\frac{10}{9}-\frac{5}{3}\HA_1\bigg)
\bigg]\nonumber\\
&&+32 \big(
         L_1^2
        + L_2^2
\big)
\bigg[(x+1)\bigg(
		\frac{2}{3}  \HA_{0,1}
		+\HA_0^2
		-\frac{2}{3} \zeta_2
		\bigg)
		+(x-1)\bigg(
		\frac{1}{9}
        -\frac{5}{3} \HA_1
        \bigg)\nonumber\\
&&        +\frac{1}{9} (17-37 x) \HA_0
\bigg]\nonumber\\
&&+64 (L_1 +L_2 ) \bigg[
        (1+x) \bigg(
        \Big(
                2 \HA_{0,1}
                -\frac{8 \zeta_2}{3}
        \Big) \HA_0
        -\frac{2}{9} \HA_0^3
        -\frac{10}{3} \HA_{0,0,1}\nonumber\\
&&        -\frac{4}{3} \HA_{0,1,1}
        +\frac{14}{3} \zeta_3
\bigg)
+(x-1) \bigg(
        \frac{442}{27}
        +\frac{5}{3} \HA_1^2
        -\frac{5}{9} \HA_1 \Big(
                1+9 \HA_0\Big)
\bigg)\nonumber\\
&&-\frac{2}{27} (56+137 x) \HA_0
+\frac{1}{9} (-5+4 x) \HA_0^2
+\frac{2}{9} (-17+28 x) \HA_{0,1}\nonumber\\
&&+\frac{2}{9} (-28+17 x) \zeta_2
\bigg]
+\frac{64}{1215} \Bigg[(1+x) \bigg(
        \Big(
                3240 \HA_{0,0,1}
                +1620 \HA_{0,1,1}
        \Big) \HA_ 0\nonumber\\
&&        +\Big(
                -1620 \HA_{0,1}
                +945 \zeta_2
        \Big) \HA_ 0^2
        +90 \HA_ 0^4
        -1080 \HA_{0,0,0,1}\nonumber\\
&&        -2700 \HA_{0,0,1,1}
        +540 \HA_{0,1,1,1}
        +1296 \zeta_2^2
\bigg)+
(-1+x) \bigg(
        20 (437+54 x)
        +\nonumber\\
&&        \Big(
                1080 \HA_ 0
                +4050 \HA_ 0^2
                +2025 \zeta_2
        \Big) \HA_ 1
        -225 \HA_ 1^3\nonumber\\
&&        -45 \HA_ 1^2 \Big(
                11+45 \HA_ 0\Big)
\bigg)
+\Big(
        -10 \big(
                -842+1111 x+81 x^2\big)\nonumber\\
&&        -540 (-7+11 x) \HA_{0,1}
        -45 (-53+73 x) \zeta_2
        -4860 (1+x) \zeta_3
\Big) \HA_ 0\nonumber\\
&&+165 (19+37 x) \HA_ 0^2
-30 (-19+8 x) \HA_ 0^3
+30 (-1+x) (157+27 x) \HA_ 1\nonumber\\
&&+\Big(
        -30 (61+169 x)
        -810 (1+x) \zeta_2
\Big) \HA_{0,1}
+180 (-11+25 x) \HA_{0,0,1}\nonumber\\
&&+180 (-14+13 x) \HA_{0,1,1}
+15 (131+329 x) \zeta_2
+90 (-55+29 x) \zeta_3
\Bigg],
\\
g_0(\eta,x)&=&
-\frac{32 (1-x)}{9} \Bigg[
        -\frac{16 (-1+x) x}{\eta }
        +18 \bigg(
                -\frac{2 (\eta 
                -4 (-1+x) x
                )^2}{9 \eta ^2}
                +\frac{1}{3} \zeta_2
        \bigg) \HA_0\big(
                u\big)
\nonumber\\
&&        +5 \HA_0^2\big(
                u\big)
        +2 \bigg(
                -1
                +\frac{(\eta 
                -4 (1-x) x
                )^{3/2}}{\eta ^{3/2}}
        \bigg) \zeta_2
        +\HA_0^3\big(
                u\big)
\Bigg]\nonumber\\
&&-\frac{64 (1-x)}{9} \Bigg[
        \bigg(
                2 \frac{(\eta 
                -4 (1-x) x
                )^{3/2}}{\eta ^{3/2}}
                -3 \zeta_2
        \bigg) G\left(
                \left\{\frac{\sqrt{1-4 \tau }}{\tau }\right\},u\right)\Bigg]\nonumber\\
&&+\frac{64 (1-x)}{3} \left[ 
        G^2\left(
                \left\{\frac{\sqrt{1-4 \tau }}{\tau }\right\},u\right)
+
        G\left(
                \left\{\frac{\sqrt{1-4 \tau }}{\tau },\frac{\sqrt{1-4 \tau \
}}{\tau },\frac{1}{\tau }\right\},u\right)\right]\nonumber\\
&&
-\frac{64 (1-x)}{9} 
        G\left(\left\{
                \frac{\sqrt{1-4 \tau }}{\tau },\frac{1}{\tau \
}\right\},u\right)
\frac{(\eta 
-4 (1-x) x
)^{3/2}}{\eta ^{3/2}},\\
g_1(\eta,x)&=&\frac{64}{27 \eta ^2 x} \bigg[
        -6 (1-x) \HA_ 0\big(
                u\big) P_ 2
        -8 \eta  (-1+x) (1+x) (7 \eta 
        +24 (1-x) x
        )\nonumber\\
&&        +3 \eta ^2 (-1+x) (-5+13 x) \HA_ 0^2\big(
                u\big)
        -3 \eta ^2 (1-x) (-1+2 x) \HA_ 0^3\big(
                u\big)\nonumber\\
&&        -(6 (1-x)) \bigg(
                (1+x) \eta ^{3/2}
                -4 \eta  (1+x) \sqrt{\eta 
                -4 (1-x) x
                }
                +2 (-1+x) x (1+10 x) \nonumber\\
&&           \times     \sqrt{\eta 
                -4 (1-x) x
                }
        \bigg) \sqrt{\eta } \zeta_2
\bigg]
+\frac{128 (-1+x)}{9 x} \bigg[
        \Big(
                -4 \frac{\sqrt{\eta 
                -4 (1-x) x
                }}{\eta ^{3/2}} P_ 1\nonumber\\
&&                -3 (-1+2 x) \zeta_2
        \Big) G\left(
                \left\{\frac{\sqrt{1-4 \tau }}{\tau }\right\},u\right)\bigg]\nonumber\\
&&-\frac{128 (-1+x) (-1+2 x)}{3 x} 
        G^2\left(
                \left\{\frac{\sqrt{1-4 \tau }}{\tau }\right\},u\right)\nonumber\\
&&-\frac{128 (-1+x) (-1+2 x)}{3 x} 
       G\left(
                \left\{\frac{\sqrt{1-4 \tau }}{\tau },\frac{\sqrt{1-4 \tau \
}}{\tau },\frac{1}{\tau }\right\},u\right)\nonumber\\
&&-\frac{256 (-1+x) P_ 1}{9 x} 
        G\left(
                \left\{\frac{\sqrt{1-4 \tau }}{\tau },\frac{1}{\tau \
}\right\},u\right) 
\frac{\sqrt{\eta 
-4 (1-x) x
}
}{\eta ^{3/2}},\\
g_3(\eta,x)&=&-\frac{32}{27 \eta ^2 x} \bigg[
        8 \eta  (1-x) P_4
        -6 (1-x) \HA_0\big(
                u\big) P_5
        +3 \eta ^2 (-1+x) (-5+8 x)\nonumber\\
&&         \times \HA_0^2\big(
                u\big)
        +3 \eta ^2 (-1+x)^2 \HA_0^3\big(
                u\big)
        -(6 (1-x)) \bigg(
                (1+2 x) \eta ^{3/2}\nonumber\\
&&                -\eta  (4+5 x) \sqrt{\eta 
                -4 (1-x) x
                }
                +2 (-1+x) x (1+8 x) \sqrt{\eta 
                -4 (1-x) x
                }
        \bigg) \sqrt{\eta } \zeta_2
\bigg]\nonumber\\
&&-\frac{64 (1-x)}{9 x} \bigg[
        \bigg(
                2 \frac{\sqrt{\eta 
                -4 (1-x) x
                }}{\eta ^{3/2}} P_3
                +3 (-1+x) \zeta_2
        \bigg) G\left(
                \left\{\frac{\sqrt{1-4 \tau }}{\tau }\right\},u\right)\bigg]\nonumber\\
&&+\frac{64 (-1+x)^2}{3 x} 
        G^2\left(
                \left\{\frac{\sqrt{1-4 \tau }}{\tau }\right\},u\right)\nonumber\\
&&+\frac{64 (-1+x)^2}{3 x} 
        G\left(
                \left\{\frac{\sqrt{1-4 \tau }}{\tau },\frac{\sqrt{1-4 \tau \
}}{\tau },\frac{1}{\tau }\right\},u\right)\nonumber\\
&&-\frac{64 (1-x) P_3}{9 x} 
        G\left(
                \left\{\frac{\sqrt{1-4 \tau }}{\tau },\frac{1}{\tau \
}\right\},u\right) \frac{\sqrt{\eta 
-4 (1-x) x
}}{\eta ^{3/2}},
\end{eqnarray}
with the polynomials
\begin{eqnarray}
P_1&=&2 \eta  (x+1)-10 x^3+9 x^2+x,\\
P_2&=&3 \eta ^2 (2 x \zeta_2+x-\zeta_2+1)+8 \eta  x \left(10 x^2-9 \
x-1\right)-16 (1-x)^2 x^2 (10 x+1),\\
P_3&=&\eta  (5 x+4)+2 x \left(-8 x^2+7 x+1\right),\\
P_4&=&7 \eta  (x+1)+6 x \left(-5 x^2+x+4\right),\\
P_5&=&\eta ^2 (x (3 \zeta_2+5)-3 \zeta_2+3)+8 \eta  x \left(8 x^2-7 \
x-1\right)-16 (x-1)^2 x^2 (8 x+1),
\end{eqnarray}
and
\begin{eqnarray}
f_0(\eta,x)&=&\bigg[
        -
        \frac{16 (1-x)}{3} 
                G\left(
                        \left\{\frac{1}{\tau },\sqrt{4-\tau } \sqrt{\tau \
}\right\},v\right)
        -\frac{4 P_6}{9 (-1+x) x^2} \Big[
                -1
                +2 \HA_0\big(
                        v\big)\Big] \nonumber\\
&&                       \times \frac{(-\eta 
        +4 (1-x) x
        )^{3/2}}{\eta ^{3/2}}
\bigg] G\left(
        \left\{\sqrt{4-\tau } \sqrt{\tau }\right\},v\right)\nonumber\\
&&+\frac{4 (1-x)}{3} \bigg[
        \Big[
                -1+2 \HA_0\big(
                        v\big)\Big] G^2\left(
                \left\{\sqrt{4-\tau } \sqrt{\tau }\right\},v\right)\bigg]\nonumber\\
&&+\frac{16 (1-x)}{3} 
        G\left(
                \left\{\frac{1}{\tau },\sqrt{4-\tau } \sqrt{\tau },\sqrt{4-\tau } \sqrt{\tau }\right\},v\right)
+\frac{1}{18} \bigg[
        -1536 (1-x)\nonumber\\
&&        -\frac{9 \eta ^4}{(-1+x)^3 x^4}
        -\frac{80 \eta ^3}{(-1+x)^2 x^3}
        -\frac{104 \eta ^2}{(-1+x) x^2}
        +\frac{576 \eta }{x}
        +\frac{4  P_7}{(-1+x)^3 x^4}\HA_0\big(
                v\big)\nonumber\\
&&        -320 (1-x) \HA_0^2\big(
                v\big)
        +64 (1-x) \HA_0^3\big(
                v\big)
        -\Big(128 (1-x)\Big) \Big(
                5
               -3 \HA_0\big(
                        v\big)\Big) \zeta_2\nonumber\\
&&        +768 (1-x) \zeta_3
\bigg]
-\frac{8 P_6}
{9 (1-x) x^2} 
        G\left(
                \left\{\frac{1}{\tau },\sqrt{4-\tau } \sqrt{\tau 
}\right\},v\right) \nonumber\\
&&\times\frac{(-\eta 
+4 (1-x) x
)^{3/2}}{\eta ^{3/2}},\\
f_1(\eta,x)&=&\frac{1}{27 x^5} \Bigg[
        -\frac{1}{(1-x)^3} \bigg[
                6912 \eta  (-1+x)^3 x^4
                -27 \eta ^4 (-1+2 x)
                -240 \eta ^3 (-1+x) x \nonumber\\
&&                (-1+2 x)
                +512 (-1+x)^4 x^4 (-16+11 x)
                +12 \HA_ 0\big(
                        v\big) P_ 9\nonumber\\
&&                -24 \eta ^2 (-1+x)^2 x^2 (-25+2 x)
                +192 (-1+x)^4 x^4 (-5+13 x) \HA_ 0^2\big(
                        v\big)\nonumber\\
&&                -192 (-1+x)^4 x^4 (-1+2 x) \HA_ 0^3\big(
                        v\big)
        \bigg]
        +384 (1-x) x^4 \bigg(
                5
                -13 x\nonumber\\
&&                +3 (-1+2 x) \HA_ 0\big(
                        v\big)
        \bigg) \zeta_2
        -2304 (-1+x) x^4 (-1+2 x) \zeta_3
\Bigg]
+\nonumber\\
&&\bigg[
        \frac{32 (-1+x) (-1+2 x)}{3 x} 
                G\left(
                        \left\{\frac{1}{\tau },\sqrt{4-\tau } \sqrt{\tau 
}\right\},v\right)\nonumber\\
&&        -\frac{8 P_ 8}{9 (1-x) x^3} \bigg(
                -1+2 \HA_ 0\big(
                        v\big)\bigg) \frac{\sqrt{-
\eta 
        +4 (1-x) x
        }}{\eta ^{3/2}}
\bigg] \nonumber\\
&& \times G\left(
        \left\{\sqrt{4-\tau } \sqrt{\tau }\right\},v\right)
-\frac{8 (-1+x) (-1+2 x)}{3 x} \bigg[
        \bigg(
                -1+2 \HA_ 0\big(
                        v\big)\bigg) \nonumber\\
&& \times G^2\left(
                \left\{\sqrt{4-\tau } \sqrt{\tau }\right\},v\right)\bigg]
+\frac{32 (1-x) (-1+2 x)}
{3 x} \nonumber\\
&& 
        \times G\left(
                \left\{\frac{1}{\tau },\sqrt{4-\tau } \sqrt{\tau },\sqrt{4-\tau } \sqrt{\tau }\right\},v\right)
-\frac{16 P_ 8}{9 (-1+x) x^3} \nonumber\\
&&        \times G\left(
                \left\{\frac{1}{\tau },\sqrt{4-\tau } \sqrt{\tau 
}\right\},v\right) \frac{\sqrt{-\eta 
+4 (1-x) x
}}{\eta ^{3/2}},\\
f_3(\eta,x)&=&\frac{1}{54 (-1+x)^2 x^5} \bigg[
        27 \eta ^4
        -240 \eta ^3 (1-x) x
        -5184 \eta  (-1+x)^2 x^4
        -1024 (-8+x) \nonumber\\
&&        \times (-1+x)^3 x^4
        -12 \HA_ 0\big(
                v\big) P_{10}
        -24 \eta ^2 (-1+x) x^2 (25+11 x)\nonumber\\
&&        -192 (-1+x)^3 x^4 (-5+8 x) \HA_ 0^2\big(
                v\big)
        +192 (-1+x)^4 x^4 \HA_ 0^3\big(
                v\big)\nonumber\\
&&        +384 (-1+x)^3 x^4 \bigg(
                5
                -8 x
                -3 (1-x) \HA_ 0\big(
                        v\big)
        \bigg) \zeta_2
        +2304 (-1+x)^4 x^4 \zeta_3
\bigg]\nonumber\\
&&+\bigg[
        -\frac{16 (-1+x)^2}{3 x}
                G\left(
                        \left\{\frac{1}{\tau },\sqrt{4-\tau } \sqrt{\tau 
}\right\},v\right)
        -\frac{4 P_{11}}{9 x^3} \bigg(
                -1\nonumber\\
&&                +2 \HA_ 0\big(
                        v\big)\bigg) \frac{\sqrt{-
\eta 
        +4 (1-x) x
        }}{\eta ^{3/2}}
\bigg] G\left(
        \left\{\sqrt{4-\tau } \sqrt{\tau }\right\},v\right)\nonumber\\
&&+\frac{4 (-1+x)^2}{3 x} 
\bigg[
        \bigg(
                -1+2 \HA_ 0\big(
                        v\big)\bigg) G\left(
                \left\{\sqrt{4-\tau } \sqrt{\tau }\right\},v\right)^2\bigg] \nonumber\\
&&+\frac{16 (-1+x)^2}{3 x} 
        G\left(
                \left\{\frac{1}{\tau },\sqrt{4-\tau } \sqrt{\tau },\sqrt{4-\tau } \sqrt{\tau }
                \right\},v\right)
+\frac{8 P_{11}}{9 x^3} \nonumber\\
&& \times       G\left(
                \left\{\frac{1}{\tau },\sqrt{4-\tau } \sqrt{\tau 
}\right\},v\right) \frac{\sqrt{-\eta 
+4 (1-x) x
}}{\eta ^{3/2}},
\end{eqnarray}
with
\begin{eqnarray}
P_6    &=&  3 \eta ^2+6 \eta  (1-x) x+4 (x-1)^2 x^2,\\
P_7    &=&  3 \eta ^4-24 \eta ^3 (1-x) x+20 \eta ^2 (x-1)^2 x^2-160 \eta  \
            (x-1)^3   x^3-128 (x-1)^4 x^4,\\
P_8    &=&  \eta ^3 (6 x-3)+6 \eta ^2 x \left(2 x^2-3 x+1\right)-8 \eta  \
            (x-1)^2 x^2 (8 x-1)\nonumber\\
       & &  +8 (x-1)^3 x^3 (10 x+1),\\
P_9    &=&  \eta ^4 (6 x-3)+24 \eta ^3 x \left(2 x^2-3 x+1\right)-4 \eta ^2 (x-1)^2 x^2 (2 x+11)\nonumber\\
       & &  -64 \eta  (x-1)^3 x^3 (8 x-1)+96 (x-1)^4 x^4 (x+4),\\
P_{10} &=&  3 \eta ^4-24 \eta ^3 (1-x) x-4 \eta ^2 x^2 \left(7 x^2+4 x-11\right)-32 \eta  (x-1)^2 x^3 (11 x-2)\nonumber\\
       & &  +32 (x-1)^3 x^4 (7 x+12),\\
P_{11} &=&  3 \eta ^3-6 \eta ^2 (1-x) x-4 \eta  x^2 \left(11 x^2-13 x+2\right)+8 (1-x)^2 x^3 (8 x+1).
\end{eqnarray}
The functions $h_i$ are defined as follows 
\begin{eqnarray}
h_i(\eta,x) = g_i\left(\frac{1}{\eta},x\right), \quad i=0,1,3.
\end{eqnarray}
\begin{figure}[H]\centering
\includegraphics[width=0.7\textwidth]{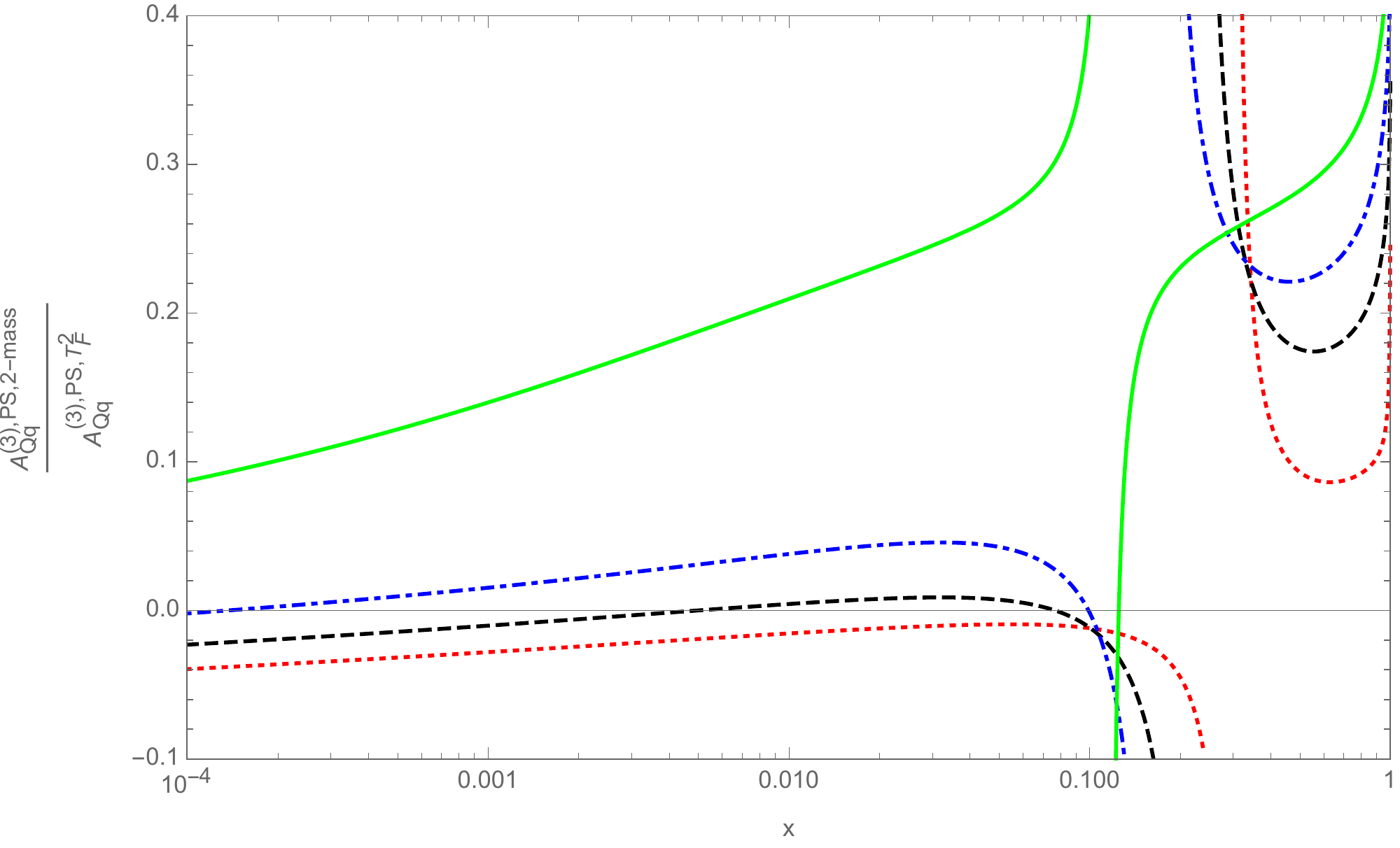}
\caption[]{\label{PSFIGnum} \sf
The ratio of the 2-mass contributions to the massive OME $A_{Qq}^{{\sf PS},(3)}$ to all contributions to
$A_{Qq}^{{\sf PS},(3)}$
of $O(T_F^2)$ as a function of $x$ and $\mu^2$.
Dotted line (red): $\mu^2 = 30~\GeV^2$.
Dashed line (black): $\mu^2 = 50~\GeV^2$.
Dash-dotted line (blue): $\mu^2 = 100~\GeV^2$.
Full line (green): $\mu^2 = 1000~\GeV^2$. Here the on-shell heavy quark masses $m_c = 1.59~\GeV$ and $m_b = 4.78~\GeV$
\cite{Alekhin:2012vu,Agashe:2014kda} have been used, from \cite{Ablinger:2019gpu}.
}
\end{figure}
Figure \ref{PSFIGnum} shows the ratio of the two-mass contribution to the complete $O(T_F^2C_F)$ term. The two-mass correction grows relatively larger with $\mu^2$ and can reach the order of 40\% of the total $O(T_F^2C_F)$ contribution.

The result is difficult to obtain in Mellin-$N$ space and we refrain from this. The main reason for this is the appearance of the Heaviside functions in \eqref{eq:Heaviside}. In general one will obtain recursions not factorizing in first order.

\subsection{The two-mass contribution to the polarized matrix element \texorpdfstring{$A_{gg,Q}^{(3)}$}{Agg,Q(3)}}
\label{sec:AggQ3}
\begin{figure}[ht]
\begin{center}
\begin{minipage}[c]{0.20\linewidth}
  \includegraphics[width=1\textwidth]{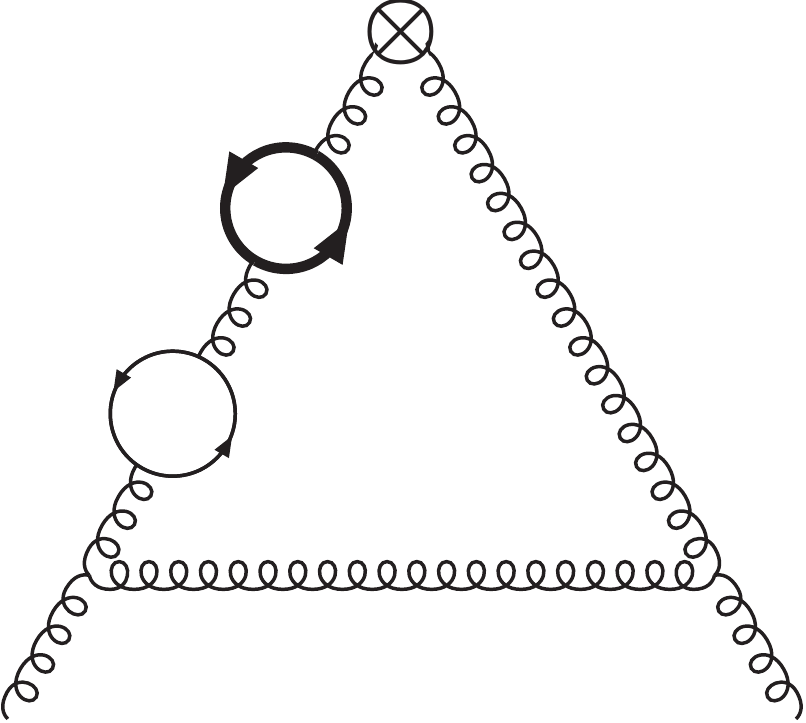}
\vspace*{-11mm}
\begin{center}
{\footnotesize (1)}
\end{center}
\end{minipage}
\hspace*{1mm}
\begin{minipage}[c]{0.20\linewidth}
  \includegraphics[width=1\textwidth]{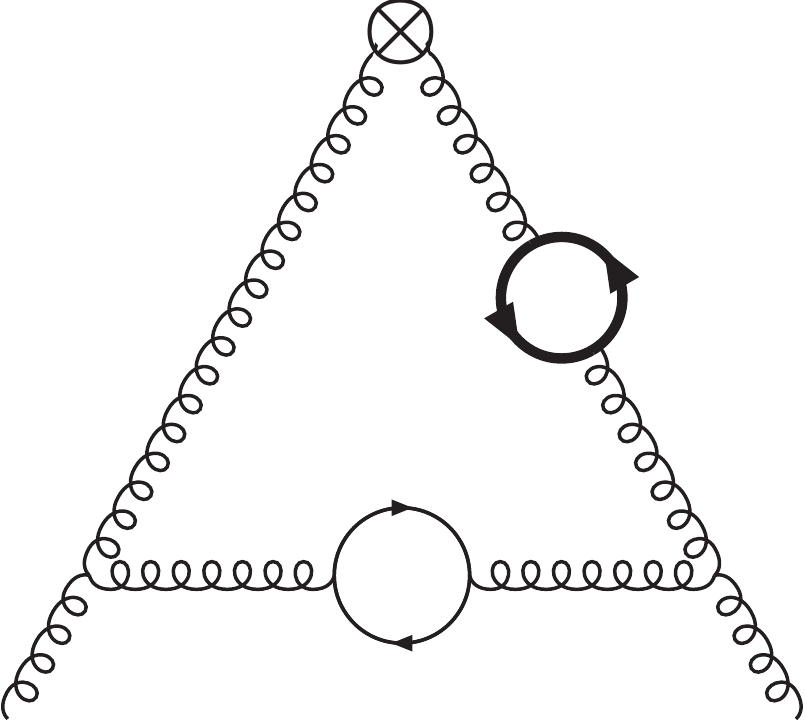}
\vspace*{-11mm}
\begin{center}
{\footnotesize (2)}
\end{center}
\end{minipage}
\hspace*{1mm}
\begin{minipage}[c]{0.20\linewidth}
  \includegraphics[width=1\textwidth]{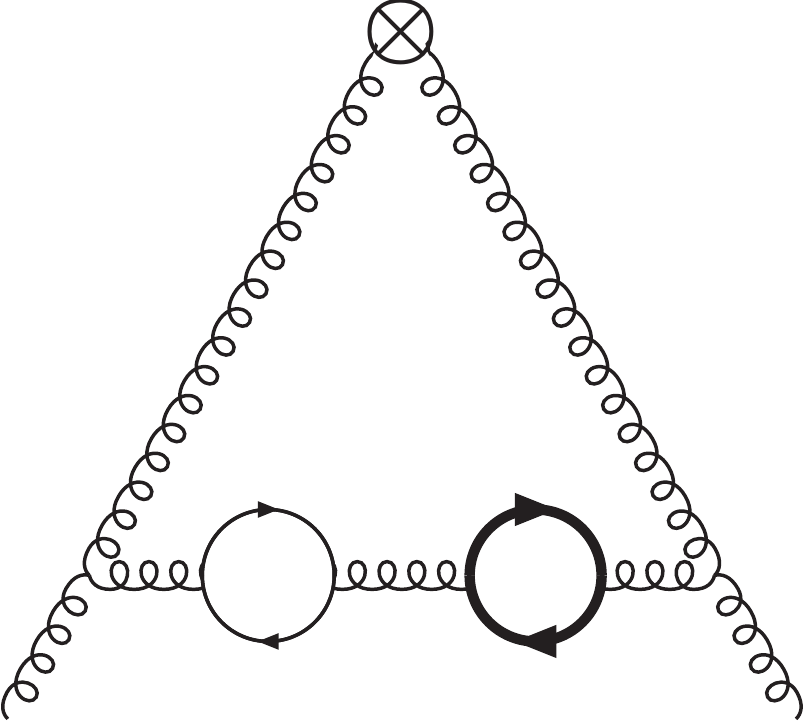}
\vspace*{-11mm}
\begin{center}
{\footnotesize (3)}
\end{center}
\end{minipage}
\hspace*{1mm}
\begin{minipage}[c]{0.20\linewidth}
  \includegraphics[width=1\textwidth]{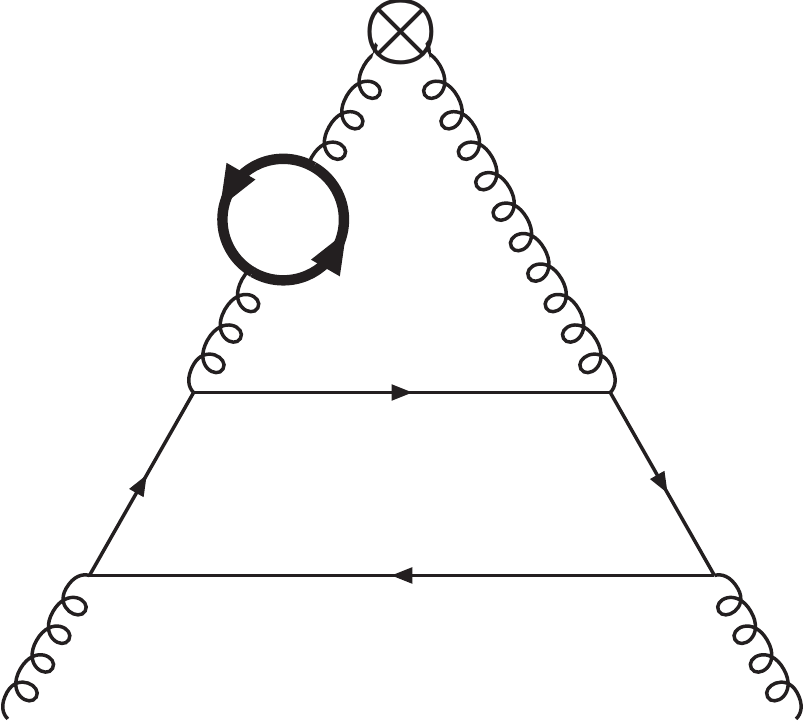}
\vspace*{-11mm}
\begin{center}
{\footnotesize (4)}
\end{center}
\end{minipage}

\vspace*{5mm}

\begin{minipage}[c]{0.20\linewidth}
  \includegraphics[width=1\textwidth]{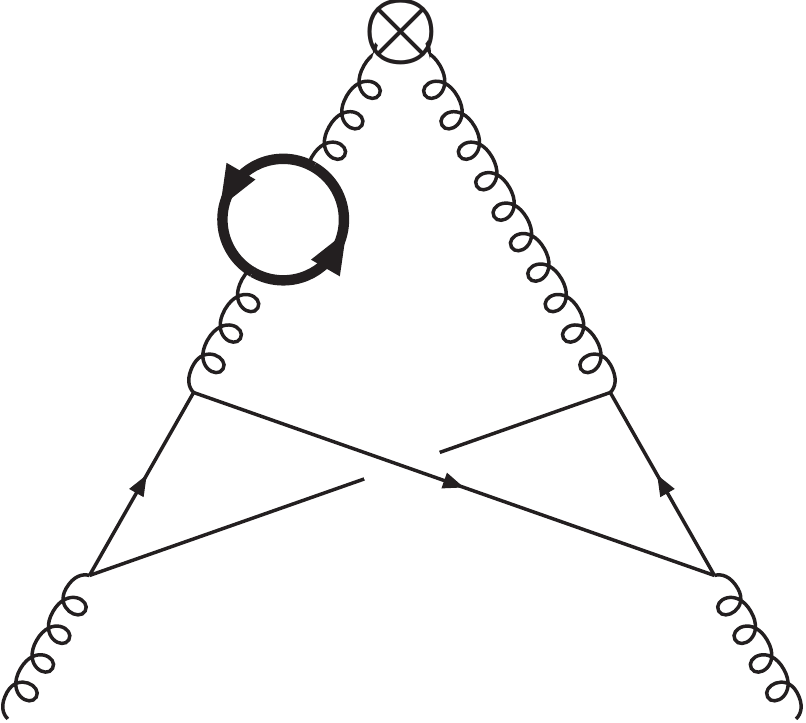}
\vspace*{-11mm}
\begin{center}
{\footnotesize (5)}
\end{center}
\end{minipage}
\hspace*{1mm}
\begin{minipage}[c]{0.20\linewidth}
  \includegraphics[width=1\textwidth]{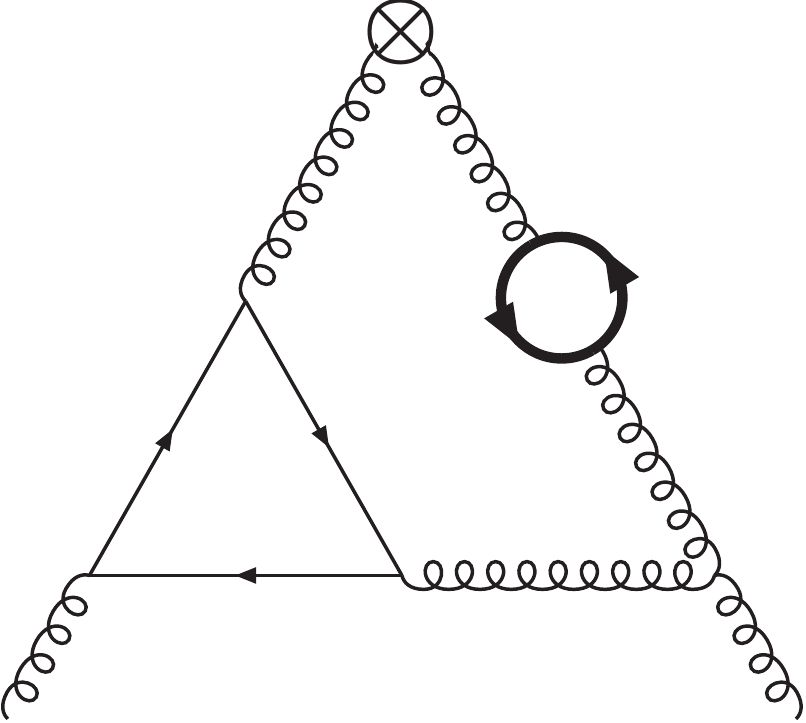}
\vspace*{-11mm}
\begin{center}
{\footnotesize (6)}
\end{center}
\end{minipage}
\hspace*{1mm}
\begin{minipage}[c]{0.20\linewidth}
  \includegraphics[width=1\textwidth]{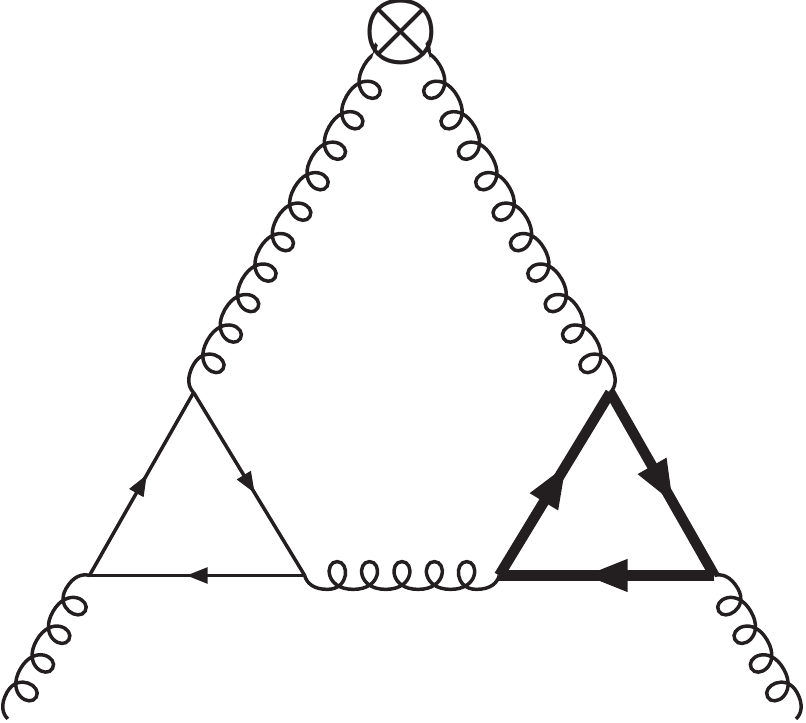}
\vspace*{-11mm}
\begin{center}
{\footnotesize (7)}
\end{center}
\end{minipage}
\hspace*{1mm}
\begin{minipage}[c]{0.20\linewidth}
  \includegraphics[width=1\textwidth]{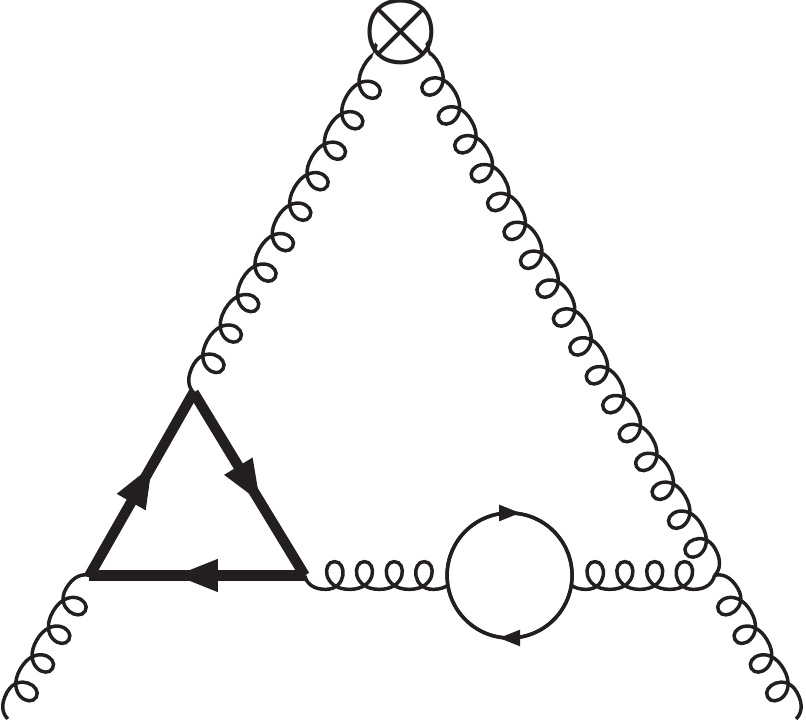}
\vspace*{-11mm}
\begin{center}
{\footnotesize (8)}
\end{center}
\end{minipage}

\vspace*{5mm}

\begin{minipage}[c]{0.20\linewidth}
  \includegraphics[width=1\textwidth]{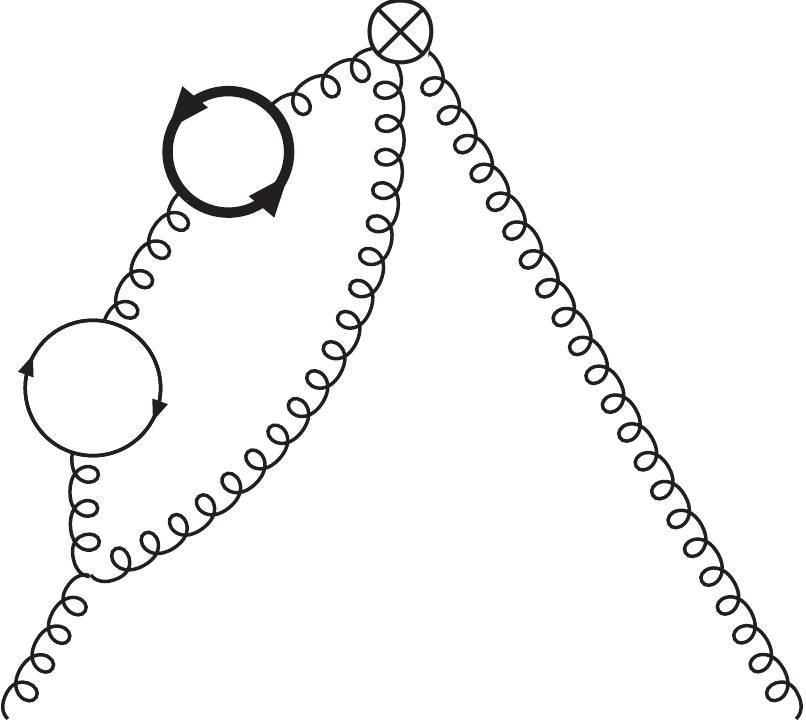}
\vspace*{-11mm}
\begin{center}
{\footnotesize (9)}
\end{center}
\end{minipage}
\hspace*{1mm}
\begin{minipage}[c]{0.20\linewidth}
  \includegraphics[width=1\textwidth]{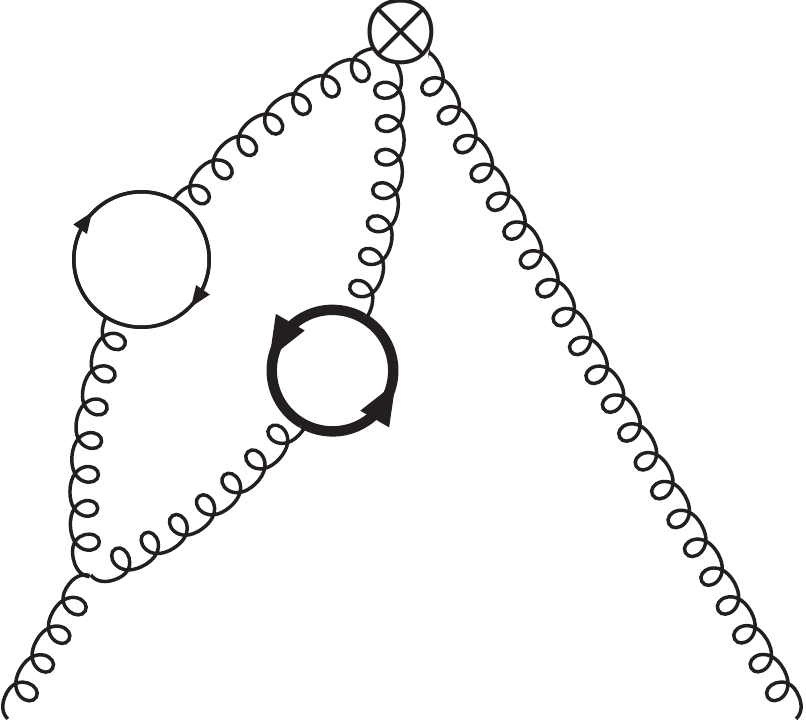}
\vspace*{-11mm}
\begin{center}
{\footnotesize (10)}
\end{center}
\end{minipage}
\hspace*{1mm}
\begin{minipage}[c]{0.20\linewidth}
  \includegraphics[width=1\textwidth]{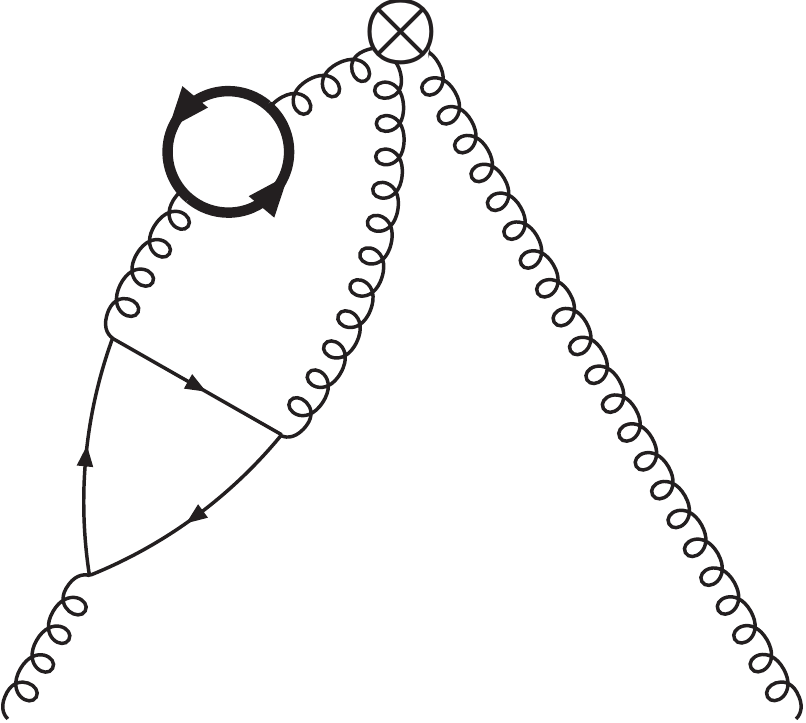}
\vspace*{-11mm}
\begin{center}
{\footnotesize (11)}
\end{center}
\end{minipage}
\hspace*{1mm}
\caption{\small \sf The 11 different topologies for $\tilde{A}_{gg,Q}^{(3)}$. Curly lines: gluons; thin arrow 
lines: lighter massive quark; thick arrow lines: heavier massive quark; the symbol $\otimes$ represents the corresponding 
local operator insertion, cf.~\cite{Mertig:1995ny} for the related Feynman rules.
\label{fig:GGdiagrams}}
\end{center}
\end{figure}

The calculation of the OME $A_{gg,Q}^{(3)}$ proceeds from the diagrams of Figure \ref{fig:GGdiagrams}.
The renormalization of the OME is performed in \cite{Ablinger:2017err} and provides a check by predicting the pole terms of the full calculation.

We perform the calculation in $N$-space following the methods of the unpolarized calculation \cite{Ablinger:2018brx}, which will be described in what follows. Then we obtain the $z$-space result by an inverse Mellin transform.
The Feynman rules applied to the operator insertion are found in \cite{Mertig:1995ny}, and to the sum of the Feynman diagrams we apply the projector \cite{Bierenbaum:2009mv,Klein:2009ig}
\begin{equation}\label{eq:projector}
A_{gg,Q}=\frac{\delta^{ab}}{N_c^2-1}
                    \frac{1}{(D-2)(D-3)}
                    (\Delta\cdot p)^{-N-1}\ep^{\mu\nu\rho\sigma}
                    \Delta \hat{G}^{ab}_{Q,\mu\nu}
                    \Delta_{\rho}p_{\sigma}\,.
\end{equation}
We adopt the Larin scheme. The Feynman parametrization used for each diagram is chosen such that, after performing the Dirac algebra with {\tt FORM} \cite{Vermaseren:2000nd}, numerator structures are not cancelled against denominators. At the price of having to deal with more complicated denominator structures, this allows us to reduce their number, which makes it easier to manipulate these terms.

We use the Feynman parametrization
\begin{eqnarray}
    \Pi^{\mu\nu}_{ab} ( k ) = -i \frac{ 8 T_F g^2}{(4\pi)^{D/2}} \delta_{ab} ( k^2 g^{\mu\nu} - k^\mu k^\nu ) 
\int\limits_{0}^{1} \text{d}x \frac{ \Gamma(2-D/2) \left( x(1-x) \right)^{D/2-1} }
{ \left( -k^2 + \frac{m^2}{x(1-x)} \right)^{2-D/2} }
\end{eqnarray}
for massive quark bubbles. After the Feynman parametrization is obtained for the whole diagram, the numerator algebra is performed through the identities
\begin{eqnarray}
 \int \frac{\dd^D k}{(2\pi)^D}  k_{\mu_1} k_{\mu_2} f(k^2) 
&=&  \frac{g_{\mu_1\mu_2}}{D} \int \frac{\dd^D k}{(2\pi)^D}  k^2 f(k^2), \\
\int \frac{\dd^D k}{(2\pi)^D}  k_{\mu_1} k_{\mu_2} k_{\mu_3} k_{\mu_4} f(k^2)  
&=& \frac{S_{\mu_1\mu_2\mu_3\mu_4}}{D(D+2)}  \int \frac{\dd^D k}{(2\pi)^D} (k^2)^2 f(k^2),
\nonumber\\
\\
\int \frac{\dd^D k}{(2\pi)^D}  k_{\mu_1} k_{\mu_2} k_{\mu_3} k_{\mu_4} k_{\mu_5} k_{\mu_6} f(k^2)  
&=& \frac{S_{\mu_1\mu_2\mu_3\mu_4\mu_5\mu_6}}{D(D+2)(D+4)} 
\int \frac{\dd^D k}{(2\pi)^D}  (k^2)^3 f(k^2),
\end{eqnarray}
with the symmetric tensors
\begin{eqnarray}
S_{\mu_1\mu_2\mu_3\mu_4} &=& g_{\mu_1\mu_2}g_{\mu_3\mu_4} + g_{\mu_1\mu_3}g_{\mu_2\mu_4} 
+ g_{\mu_1\mu_4} g_{\mu_2\mu_3} 
\\
S_{\mu_1\mu_2\mu_3\mu_4\mu_5\mu_6} &=& 
g_{\mu_1\mu_2} 
\left[
  g_{\mu_3\mu_4} g_{\mu_5\mu_6} 
+ g_{\mu_3\mu_5} g_{\mu_4\mu_6}
+ g_{\mu_3\mu_6} g_{\mu_4\mu_5}\right]
\nonumber\\
&&
+g_{\mu_1\mu_3} 
\left[
  g_{\mu_2\mu_4} g_{\mu_5\mu_6} 
+ g_{\mu_2\mu_5} g_{\mu_4\mu_6}
+ g_{\mu_2\mu_6} g_{\mu_4\mu_5}\right]
\nonumber\\
&&
+g_{\mu_1\mu_4} 
\left[
  g_{\mu_2\mu_3} g_{\mu_5\mu_6} 
+ g_{\mu_2\mu_5} g_{\mu_3\mu_6}
+ g_{\mu_2\mu_6} g_{\mu_3\mu_5}\right]
\nonumber\\
&&
+g_{\mu_1\mu_5} 
\left[
  g_{\mu_2\mu_3} g_{\mu_4\mu_6} 
+ g_{\mu_2\mu_4} g_{\mu_3\mu_6}
+ g_{\mu_2\mu_6} g_{\mu_3\mu_4}\right]
\nonumber\\
&&
+g_{\mu_1\mu_6} 
\left[
  g_{\mu_2\mu_3} g_{\mu_4\mu_5} 
+ g_{\mu_2\mu_4} g_{\mu_3\mu_5}
+ g_{\mu_2\mu_5} g_{\mu_3\mu_4}\right].
\end{eqnarray}
The scalar integrals can then be performed using the relation
\begin{eqnarray}
\int \frac{\dd^D k}{(2\pi)^D} \frac{ (k^2)^m }{ ( k^2 + R^2 )^n } 
&=& \frac{1}{(4\pi)^{D/2}} \frac{\Gamma(n-m-D/2)}{\Gamma(n)} \frac{\Gamma(m+D/2)}{\Gamma(D/2)} 
\left( R^2 \right)^{m-n+D/2} .
\end{eqnarray}
After the integrals over the loop momenta are performed, only integrals over the Feynman parameters are left. They always appear in the form
\begin{equation}\label{eq:feynIntegrals}
    \prod\limits_{i=1}^{j} \int\limits_0^1 \mathrm{d} x_i \ x_i^{a_i} (1-x_i)^{b_i} \ R_0^N \ \left[ R_1 
\ m_1^2 + R_2 \ m_2^2 \right]^{-s},
\end{equation} 
where $R_0$ is a polynomial in the Feynman parameters $x_i$, and $R_1$ and $R_2$ are rational functions 
in $x_i$.
At this point the polynomial $R_0$ can be treated by applying the binomial theorem (multiple times if necessary)
\begin{equation}
    (A + B)^N = \sum\limits_{i=0}^{N} \binom{N}{i} A^i B^{N-i}
\label{eq:I2}
\end{equation}
while the factor $\left[ R_1 \ m_1^2 + R_2 \ m_2^2 \right]^{-s}$ is treated via a Mellin-Barnes decomposition \eqref{eq:mbFormula}.

The integrals over the Feynman parameters are then turned into infinite sums using the residue theorem. This sum representation is treated analytically by the algorithms of {\tt Sigma}, {\tt HarmonicSums}, {\tt EvaluateMultiSums} and {\tt SumProduction}, which reduce these nested sums into classes of hypergeometric sums which are shown, algorithmically, to be independent.

The target function space is, for the $N$-space result, that of harmonic sums, defined in Eqs. \eqref{eq:defHSums1} and \eqref{eq:defHSums2}, generalized harmonic sums \eqref{eq:genSsums},
cyclotomic sums~\cite{Ablinger:2011te} and binomial sums~\cite{Ablinger:2014bra}. Harmonic polylogarithms~\cite{Remiddi:1999ew} will also appear in the result.

The calculation was checked by computing fixed $N$-moments using {\tt MB} and {\tt MBResolve} and comparing the results with those obtained from the packages {\tt Q2E/EXP}~\cite{Harlander:1997zb,Seidensticker:1999bb}. In $N$-space, the result appears to exhibit spurious poles for $N=1/2,~3/2$. It has been verified analytically that the expression is actually regular at those points.

\subsubsection{The \texorpdfstring{$N$}{N}-space solution}
We obtain, for the constant part in $\ep$ of the $N$-space $\tilde{A}_{gg,Q}^{(3)}$,

%
where the argument of the harmonic polylogarithms is implied to be $\eta$ where omitted, and that of the harmonic sums to be $N$ where omitted.

The polynomials $P_i$ are:
\begin{eqnarray}
P_1&=&-27
-36 N^2
-36 N (-2+\eta )
+54 \eta 
+5 \eta ^2
\\
P_2&=&5
-2 \big(
        -11-18 N+18 N^2\big) \eta 
+5 \eta ^2
\\
P_3&=&372 \eta 
+N^2 \big(
        5-102 \eta +5 \eta ^2\big)
+N \big(
        5-66 \eta +5 \eta ^2\big)
\\
P_4&=&-5
+18 (-3+2 N) \eta 
+9 (-3+2 N) (-1+2 N) \eta ^2
\\
P_5&=&-80 \big(
        5+22 \eta +5 \eta ^2\big)
+3 N^2 \big(
        71-46 \eta +71 \eta ^2\big)
+3 N \big(
        167+18 \eta +167 \eta ^2\big)
\\
P_6&=&-80 \big(
        5+22 \eta +5 \eta ^2\big)
+3 N^2 \big(
        111+64 \eta +111 \eta ^2\big)
+3 N \big(
        207+128 \eta +207 \eta ^2\big)
\\
P_7&=&-(2+N) \Big[
        27
        +36 N^2
        +36 N (-2+\eta )
        -54 \eta 
        -5 \eta ^2
\Big]\\
P_8&=&24
-48 N
-18 N^3
+5 N^2 (9+\eta )
\\
P_9&=&(2+N) (-4+3 N) (-2+3 N) (1+\eta )\\
P_{10}&=&-24 \eta 
+48 N \eta 
+18 N^3 \eta 
-5 N^2 (1+9 \eta )
\\
P_{11}&=&9
+160 \eta 
+50 N^2 \eta 
+140 N^3 \eta 
+9 \eta ^2
-9 N \big(
        7+10 \eta +7 \eta ^2\big)
\\
P_{12}&=&(2+N) \Big[
        11
        -54 \eta 
        +11 \eta ^2
        +18 N^2 \big(
                1+\eta ^2\big)
        -36 N \big(
                1-\eta +\eta ^2\big)
\Big]\\
P_{13}&=&(2+N) \Big[
        -5
        +18 (-3+2 N) \eta 
        +9 (-3+2 N) (-1+2 N) \eta ^2
\Big]\\
P_{14}&=&168-536 N-407 N^2-278 N^3-167 N^4\\
P_{15}&=&-(2+N) \Big[
        -24
        +48 N
        +18 N^3
        -5 N^2 (9+\eta )
\Big]\\
P_{16}&=&-96+370 N+277 N^2+12 N^3+9 N^4\\
P_{17}&=&48-27 N+263 N^2+634 N^3+344 N^4\\
P_{18}&=&-1632+7072 N+8611 N^2+3078 N^3+1539 N^4\\
P_{19}&=&-20 N^4 (10+81 \eta )
-4 N^3 \big(
        41+855 \eta +135 \eta ^2\big)
+N \big(
        41+855 \eta -765 \eta ^2-87 \eta ^3\big)
\nonumber\\&&        
-3 \big(
        17+60 \eta -135 \eta ^2+2 \eta ^3\big)
+N^2 \big(
        254+1125 \eta +180 \eta ^2+45 \eta ^3\big)
\\
P_{20}&=&50 N^2 \eta 
-660 N^3 \eta 
-440 N^4 \eta 
-3 \big(
        3+160 \eta +3 \eta ^2\big)
+9 N \big(
        7+30 \eta +7 \eta ^2\big)
\\
P_{21}&=&(2+N) \Big[
        -24 \eta 
        +48 N \eta 
        +18 N^3 \eta 
        -5 N^2 (1+9 \eta )
\Big]\\
P_{22}&=&60 N^2 (1+N)^2 \eta\\
P_{23}&=&\big(
        3-21 N+25 N^2+50 N^3+25 N^4\big) (-1+\eta ) (1+\eta )\\
P_{24}&=&-36 N^3 (-1+\eta ) \eta 
+36 N^4 \eta ^2
+N (1+15 \eta ) (-5+21 \eta )
-4 \big(
        5+27 \eta ^2\big)
\nonumber\\&&
-N^2 \big(
        5+18 \eta +189 \eta ^2\big)
\\
P_{25}&=&384 \eta 
-1664 N \eta 
+18 N^3 \big(
        4-93 \eta +4 \eta ^2\big)
+9 N^4 \big(
        4-93 \eta +4 \eta ^2\big)
\nonumber\\&&        
+N^2 \big(
        36-2501 \eta +36 \eta ^2\big)
\\
P_{26}&=&-48 \big(
        1+160 \eta +\eta ^2\big)
-16 N \big(
        4-1405 \eta +4 \eta ^2\big)
+3 N^4 \big(
        71+134 \eta +71 \eta ^2\big)
\nonumber\\&&        
+N^2 \big(
        101+15234 \eta +101 \eta ^2\big)
+2 N^3 \big(
        357+418 \eta +357 \eta ^2\big)
\\
P_{27}&=&56 N^4 (1+\eta ) \big(
        1+\eta ^2\big)
-6 (1+\eta ) \big(
        5+86 \eta +5 \eta ^2\big)
-N (1+\eta ) \big(
        41+390 \eta +41 \eta ^2\big)
\nonumber\\&&        
+4 N^3 (1+\eta ) \big(
        41+405 \eta +41 \eta ^2\big)
+2 N^2 (1+\eta ) \big(
        53+972 \eta +53 \eta ^2\big)
\\
P_{28}&=&3 \big(
        17+62 \eta -270 \eta ^2+62 \eta ^3+17 \eta ^4\big)
+20 N^4 \big(
        10+81 \eta +81 \eta ^3+10 \eta ^4\big)
\nonumber\\&&        
-N \big(
        41+768 \eta -1530 \eta ^2+768 \eta ^3+41 \eta ^4\big)
\nonumber\\&&        
+4 N^3 \big(
        41+855 \eta +270 \eta ^2+855 \eta ^3+41 \eta ^4\big)
\nonumber\\&&        
-2 N^2 \big(
        127+585 \eta +180 \eta ^2+585 \eta ^3+127 \eta ^4\big)
\\
P_{29}&=&144-51 N-585 N^2+190 N^3+36 N^4+56 N^5\\
P_{30}&=&-800 N^6
-8 N^5 (169+270 \eta )
+4 N^4 \big(
        599-645 \eta +30 \eta ^2\big)
-2 N^3 \big(
        -1565-3810 \eta 
\nonumber\\&&
        +225 \eta ^2+6 \eta ^3\big)
+3 N^2 \big(
        -355+2255 \eta +335 \eta ^2+53 \eta ^3\big)
\nonumber\\&&        
+3 \big(
        43+705 \eta +405 \eta ^2+175 \eta ^3\big)
-2 N \big(
        349+4530 \eta +855 \eta ^2+342 \eta ^3\big)
\\
P_{31}&=&-400 N^6
+24 \eta  (80+3 \eta )
-4 N^5 (119+128 \eta )
+N^3 \big(
        847+2314 \eta -513 \eta ^2\big)
\nonumber\\&&        
+N \big(
        129-4882 \eta -171 \eta ^2\big)
-4 N^4 \big(
        -359+172 \eta +3 \eta ^2\big)
\nonumber\\&&        
+4 N^2 \big(
        -239+572 \eta +189 \eta ^2\big)
\\
P_{32}&=&-280 N^6
-100 N^5 (11+81 \eta )
+N^3 \big(
        13967+41445 \eta -1755 \eta ^2-2325 \eta ^3\big)
\nonumber\\&&        
+27 \big(
        -61-160 \eta -45 \eta ^2+2 \eta ^3\big)
-90 N^4 \big(
        -65-72 \eta +27 \eta ^2+20 \eta ^3\big)
\nonumber\\&&        
+N^2 \big(
        5143+15660 \eta -405 \eta ^2+174 \eta ^3\big)
\nonumber\\&&        
+3 N \big(
        -1141-3285 \eta +1125 \eta ^2+377 \eta ^3\big)
\\
P_{33}&=&2 \big(
        -42-29 N+68 N^2+47 N^3+88 N^4+99 N^5+33 N^6\big)\\
P_{34}&=&528-224 N+2008 N^2+7149 N^3+4239 N^4+279 N^5+93 N^6\\
P_{35}&=&400 N^6
-24 \eta  (80+3 \eta )
+4 N^5 (119+128 \eta )
+4 N^4 \big(
        -359+172 \eta +3 \eta ^2\big)
\nonumber\\&&        
+N \big(
        -129+4882 \eta +171 \eta ^2\big)
-4 N^2 \big(
        -239+572 \eta +189 \eta ^2\big)
\nonumber\\&&        
+N^3 \big(
        -847-2314 \eta +513 \eta ^2\big)
\\
P_{36}&=&-1088-800 N+1664 N^2+1081 N^3+1675 N^4+1899 N^5+633 N^6\\
P_{37}&=&800 N^6
+8 N^5 (169+270 \eta )
-4 N^4 \big(
        599-645 \eta +30 \eta ^2\big)
+2 N^3 \big(
        -1565-3810 \eta
\nonumber\\&&
         +225 \eta ^2+6 \eta ^3\big)
-3 N^2 \big(
        -355+2255 \eta +335 \eta ^2+53 \eta ^3\big)
\nonumber\\&&        
-3 \big(
        43+705 \eta +405 \eta ^2+175 \eta ^3\big)
+2 N \big(
        349+4530 \eta +855 \eta ^2+342 \eta ^3\big)
\\
P_{38}&=&-400 N^6 \eta ^2
+24 (3+80 \eta )
-4 N^5 \eta  (128+119 \eta )
+N \big(
        -171-4882 \eta +129 \eta ^2\big)
\nonumber\\&&        
-4 N^2 \big(
        -189-572 \eta +239 \eta ^2\big)
+4 N^4 \big(
        -3-172 \eta +359 \eta ^2\big)
\nonumber\\&&        
+N^3 \big(
        -513+2314 \eta +847 \eta ^2\big)
\\
P_{39}&=&-512 \eta 
-256 N \eta 
+1024 N^2 \eta 
+3 N^3 \big(
        5+282 \eta +5 \eta ^2\big)
+9 N^5 \big(
        5+282 \eta +5 \eta ^2\big)
\nonumber\\&&        
+3 N^6 \big(
        5+282 \eta +5 \eta ^2\big)
+N^4 \big(
        45+2282 \eta +45 \eta ^2\big)
\\
P_{40}&=&400 N^6 \eta ^2
-24 (3+80 \eta )
+4 N^5 \eta  (128+119 \eta )
+N^3 \big(
        513-2314 \eta -847 \eta ^2\big)
\nonumber\\&&        
+N \big(
        171+4882 \eta -129 \eta ^2\big)
+4 N^2 \big(
        -189-572 \eta +239 \eta ^2\big)
\nonumber\\&&        
-4 N^4 \big(
        -3-172 \eta +359 \eta ^2\big)
\\
P_{41}&=&400 N^6 \big(
        1-\eta +\eta ^2\big)
-3 \big(
        109+446 \eta +109 \eta ^2\big)
+3 N^2 \big(
        151-1446 \eta +151 \eta ^2\big)
\nonumber\\&&        
+4 N^5 \big(
        169+101 \eta +169 \eta ^2\big)
-2 N^4 \big(
        599-1214 \eta +599 \eta ^2\big)
\nonumber\\&&        
+N \big(
        691+4694 \eta +691 \eta ^2\big)
-N^3 \big(
        1559+2026 \eta +1559 \eta ^2\big)
\\
P_{42}&=&-800 N^6 \eta ^3
-8 N^5 \eta ^2 (270+169 \eta )
+4 N^4 \eta  \big(
        30-645 \eta +599 \eta ^2\big)
\nonumber\\&&        
+3 N^2 \big(
        53+335 \eta +2255 \eta ^2-355 \eta ^3\big)
+3 \big(
        175+405 \eta +705 \eta ^2+43 \eta ^3\big)
\nonumber\\&&        
-2 N \big(
        342+855 \eta +4530 \eta ^2+349 \eta ^3\big)
+2 N^3 \big(
        -6-225 \eta +3810 \eta ^2+1565 \eta ^3\big)
\\
P_{43}&=&-280 N^6 \eta ^3
-100 N^5 \eta ^2 (81+11 \eta )
-27 \big(
        -2+45 \eta +160 \eta ^2+61 \eta ^3\big)
\nonumber\\&&        
+90 N^4 \big(
        -20-27 \eta +72 \eta ^2+65 \eta ^3\big)
-3 N \big(
        -377-1125 \eta +3285 \eta ^2+1141 \eta ^3\big)
\nonumber\\&&        
+N^2 \big(
        174-405 \eta +15660 \eta ^2+5143 \eta ^3\big)
\nonumber\\&&        
+N^3 \big(
        -2325-1755 \eta +41445 \eta ^2+13967 \eta ^3\big)
\\
P_{44}&=&800 N^6 \eta ^3
+8 N^5 \eta ^2 (270+169 \eta )
-4 N^4 \eta  \big(
        30-645 \eta +599 \eta ^2\big)
\nonumber\\&&        
-3 \big(
        175+405 \eta +705 \eta ^2+43 \eta ^3\big)
+2 N \big(
        342+855 \eta +4530 \eta ^2+349 \eta ^3\big)
\nonumber\\&&        
+3 N^2 \big(
        -53-335 \eta -2255 \eta ^2+355 \eta ^3\big)
-2 N^3 \big(
        -6-225 \eta +3810 \eta ^2+1565 \eta ^3\big)
\\
P_{45}&=&-800 N^6 \big(
        1+\eta ^4\big)
+3 \big(
        43+880 \eta +810 \eta ^2+880 \eta ^3+43 \eta ^4\big)
\nonumber\\&&        
+N^3 \big(
        3130+7608 \eta -900 \eta ^2+7608 \eta ^3+3130 \eta ^4\big)
-8 N^5 \big(
        169+270 \eta +270 \eta ^3+169 \eta ^4\big)
\nonumber\\&&        
-2 N \big(
        349+4872 \eta +1710 \eta ^2+4872 \eta ^3+349 \eta ^4\big)
\nonumber\\&&        
-3 N^2 \big(
        355-2308 \eta -670 \eta ^2-2308 \eta ^3+355 \eta ^4\big)
\nonumber\\&&        
+4 N^4 \big(
        599-645 \eta +60 \eta ^2-645 \eta ^3+599 \eta ^4\big)
\\
P_{46}&=&-800 N^6 (-1+\eta ) (1+\eta ) \big(
        1+\eta ^2\big)
-698 N (-1+\eta ) (1+\eta ) \big(
        1+12 \eta +\eta ^2\big)
\nonumber\\&&        
+3 (-1+\eta ) (1+\eta ) \big(
        43+530 \eta +43 \eta ^2\big)
-8 N^5 (-1+\eta ) (1+\eta ) \big(
        169+270 \eta +169 \eta ^2\big)
\nonumber\\&&        
-3 N^2 (-1+\eta ) (1+\eta ) \big(
        355-2202 \eta +355 \eta ^2\big)
\nonumber\\&&        
+4 N^4 (-1+\eta ) (1+\eta ) \big(
        599-645 \eta +599 \eta ^2\big)
\nonumber\\&&        
+2 N^3 (-1+\eta ) (1+\eta ) \big(
        1565+3816 \eta +1565 \eta ^2\big)
\\
P_{47}&=&800 N^6 \big(
        1+\eta ^4\big)
-3 \big(
        43+880 \eta +810 \eta ^2+880 \eta ^3+43 \eta ^4\big)
\nonumber\\&&        
+N \big(
        698+9744 \eta +3420 \eta ^2+9744 \eta ^3+698 \eta ^4\big)
+8 N^5 \big(
        169+270 \eta +270 \eta ^3+169 \eta ^4\big)
\nonumber\\&&        
+3 N^2 \big(
        355-2308 \eta -670 \eta ^2-2308 \eta ^3+355 \eta ^4\big)
\nonumber\\&&        
-4 N^4 \big(
        599-645 \eta +60 \eta ^2-645 \eta ^3+599 \eta ^4\big)
\nonumber\\&&        
-2 N^3 \big(
        1565+3804 \eta -450 \eta ^2+3804 \eta ^3+1565 \eta ^4\big)
\\
P_{48}&=&-560 N^7
-40 N^6 (62+405 \eta )
+20 N^5 \big(
        -190-2187 \eta +729 \eta ^2\big)
\nonumber\\&&        
+N^2 \big(
        1417+12690 \eta -5535 \eta ^2-240 \eta ^3\big)
-9 \big(
        17+60 \eta -135 \eta ^2+2 \eta ^3\big)
\nonumber\\&&        
+10 N^4 \big(
        -260-3807 \eta +2430 \eta ^2+15 \eta ^3\big)
-N^3 \big(
        107+9045 \eta -4455 \eta ^2+75 \eta ^3\big)
\nonumber\\&&        
-3 N \big(
        -91-1665 \eta +1575 \eta ^2+137 \eta ^3\big)
\\
P_{49}&=&-(2+N) \Big[
        -216 \eta 
        -144 N \eta 
        -696 N^2 \eta 
        +148 N^6 \eta 
        -30 N^4 \big(
                9-16 \eta +9 \eta ^2\big)
\nonumber\\&&                
        -9 N^5 \big(
                15-38 \eta +15 \eta ^2\big)
        -N^3 \big(
                135+1022 \eta +135 \eta ^2\big)
\Big]\\
P_{50}&=&225
+1630 \eta 
+3456 \eta ^2
+2466 \eta ^3
+415 \eta ^4
\nonumber\\&&
+N \big(
        930+5372 \eta +6912 \eta ^2+2820 \eta ^3+350 \eta ^4\big)
\nonumber\\&&        
-152 N^6 \eta  (1+\eta ) \big(
        5+22 \eta +5 \eta ^2\big)
-16 N^7 \eta  (1+\eta ) \big(
        5+22 \eta +5 \eta ^2\big)
\nonumber\\&&        
-192 N^3 \big(
        -5+3 \eta +135 \eta ^2+157 \eta ^3+30 \eta ^4\big)
\nonumber\\&&        
-72 N^4 \big(
        -5+53 \eta +405 \eta ^2+427 \eta ^3+80 \eta ^4\big)
\nonumber\\&&        
-12 N^5 \big(
        -5+218 \eta +1296 \eta ^2+1318 \eta ^3+245 \eta ^4\big)
-6 N^2 \big(
        -225-830 \eta 
\nonumber\\&&        
        +864 \eta ^2+1854 \eta ^3+385 \eta ^4\big)
\\
P_{51}&=&396+690 N+518 N^2+240 N^3-289 N^4-432 N^5+494 N^6+588 
N^7+147 N^8\\
P_{52}&=&560 N^8
-216 (-1+\eta ) \eta  (80+3 \eta )
+40 N^7 (34+405 \eta )
+N^4 \big(
        44539+73899 \eta
\nonumber\\&&        
         +1629 \eta ^2-7215 \eta ^3\big)
+60 N^6 \big(
        -310-561 \eta +81 \eta ^2+60 \eta ^3\big)
\nonumber\\&&        
+9 N \big(
        -420-6451 \eta +4306 \eta ^2+189 \eta ^3\big)
-2 N^5 \big(
        7334+31887 \eta -414 \eta ^2+375 \eta ^3\big)
\nonumber\\&&        
-9 N^2 \big(
        1731-919 \eta +137 \eta ^2+391 \eta ^3\big)
+3 N^3 \big(
        9966+26631 \eta -11136 \eta ^2+959 \eta ^3\big)
\\
P_{53}&=&-1080 N^2 (1+N)^2 \eta ^2 \big(
        207-456 N+89 N^2-56 N^3+92 N^4\big)\\
P_{54}&=&-540 N (1+N)^2 \eta ^2 \big(
        144-51 N-585 N^2+190 N^3+36 N^4+56 N^5\big)\\
P_{55}&=&N^2 (1+N) \Big[
        288 N^5 \eta ^2
        -36 N^4 \eta  (-8+3 \eta )
        -160 \big(
                1+3 \eta ^2\big)
        -4 N^3 \big(
                -5-9 \eta +438 \eta ^2\big)
\nonumber\\&&                
        +N \big(
                -275-270 \eta +801 \eta ^2\big)
        +N^2 \big(
                25-522 \eta +1485 \eta ^2\big)
\Big]\\
P_{56}&=&560 N^8 \eta ^3
+40 N^7 \eta ^2 (405+34 \eta )
+216 (-1+\eta ) (3+80 \eta )
-60 N^6 \big(
        -60
\nonumber\\&&        
        -81 \eta +561 \eta ^2+310 \eta ^3\big)
-9 N \big(
        -189-4306 \eta +6451 \eta ^2+420 \eta ^3\big)
\nonumber\\&&        
-9 N^2 \big(
        391+137 \eta -919 \eta ^2+1731 \eta ^3\big)
-2 N^5 \big(
        375-414 \eta +31887 \eta ^2+7334 \eta ^3\big)
\nonumber\\&&        
+3 N^3 \big(
        959-11136 \eta +26631 \eta ^2+9966 \eta ^3\big)
\nonumber\\&&        
+N^4 \big(
        -7215+1629 \eta +73899 \eta ^2+44539 \eta ^3\big)
\\
P_{57}&=&-51840 \eta ^2
+103680 N \eta ^2
+80 N^8 \eta  \big(
        405-10412 \eta +405 \eta ^2\big)
\nonumber\\&&        
-4 N^7 \big(
        -2700-20331 \eta +165688 \eta ^2+1539 \eta ^3\big)
\nonumber\\&&        
-36 N^6 \big(
        204+378 \eta -49156 \eta ^2+1863 \eta ^3\big)
\nonumber\\&&        
-9 N^2 \big(
        459+5184 \eta +20987 \eta ^2+3618 \eta ^3\big)
\nonumber\\&&        
-2 N^5 \big(
        13500+171477 \eta -942766 \eta ^2+3807 \eta ^3\big)
\nonumber\\&&        
-3 N^3 \big(
        -2025-24489 \eta +185329 \eta ^2+4077 \eta ^3\big)
\nonumber\\&&        
+N^4 \big(
        18360+94851 \eta +70490 \eta ^2+58239 \eta ^3\big)
\\
P_{58}&=&N \Big[
        2240 N^9 (-1+\eta ) (1+\eta ) \big(
                1+\eta ^2\big)
        +N^4 \big(
                6264+156159 \eta +63990 \eta ^2
\nonumber\\&&                
                -251679 \eta ^3-13850 
\eta ^4\big)
        +N^3 \big(
                -6175-104082 \eta +24435 \eta ^2+51822 \eta ^3+3020 
\eta ^4\big)
\nonumber\\&&
        +N^2 \big(
                -525-58899 \eta -7155 \eta ^2+87081 \eta ^3+4230 \eta 
^4\big)
        +27 \big(
                -17-96 \eta +135 \eta ^2
\nonumber\\&&                
                +34 \eta ^3\big)
        +160 N^8 \big(
                -48-405 \eta +405 \eta ^3+55 \eta ^4\big)
\nonumber\\&&                
        +8 N^7 \big(
                -450-13887 \eta +17937 \eta ^3+860 \eta ^4\big)
\nonumber\\&&                
        +9 N \big(
                25-482 \eta -1845 \eta ^2+2134 \eta ^3+100 \eta ^4\big)
\nonumber\\&&                
        -10 N^5 \big(
                -2214-30615 \eta +486 \eta ^2+36153 \eta ^3+2834 \eta 
^4\big)
\nonumber\\&&
        -4 N^6 \big(
                -3660-21279 \eta +7290 \eta ^2+11559 \eta ^3+3620 
\eta ^4\big)
\Big]\\
P_{59}&=&36288 \eta 
-19872 N \eta 
-220032 N^2 \eta 
-252192 N^3 \eta 
+N^5 \big(
        -61155+394298 \eta -61155 \eta ^2\big)
\nonumber\\&&        
+N^6 \big(
        -17415+597938 \eta -17415 \eta ^2\big)
-320 N^4 \big(
        81+526 \eta +81 \eta ^2\big)
\nonumber\\&&        
+36 N^{11} \big(
        405-3766 \eta +405 \eta ^2\big)
+12 N^{12} \big(
        405-3766 \eta +405 \eta ^2\big)
\nonumber\\&&        
-5 N^9 \big(
        3483-56554 \eta +3483 \eta ^2\big)
+N^{10} \big(
        3645+44314 \eta +3645 \eta ^2\big)
\nonumber\\&&        
+10 N^7 \big(
        4455-2386 \eta +4455 \eta ^2\big)
+2 N^8 \big(
        7695-64514 \eta +7695 \eta ^2\big)
\\
P_{60}&=&1080 N^3 (1+N)^2 (2+N)^4 \eta ^3 \big(
        576+1173 N-2988 N^2-4835 N^3+2674 N^4+572 N^5
\nonumber\\&&
        +248 N^6\big)\\
P_{61}&=&-5806080 \eta ^3
-7464960 N \eta ^3
+39628800 N^2 \eta ^3
\nonumber\\&&
+N^4 \big(
        273375-2809566 \eta +5935680 \eta ^2+522717086 \eta 
^3+2240865 \eta ^4-4790016 \eta ^5\big)
\nonumber\\&&
+864 N^{15} \eta  \big(
        400+3639 \eta +589 \eta ^2+3639 \eta ^3+400 \eta ^4\big)
\nonumber\\&&        
-768 N^3 \eta  \big(
        1377+4293 \eta -244034 \eta ^2+4293 \eta ^3+1377 \eta ^4\big)
\nonumber\\&&        
+432 N^{14} \eta  \big(
        7856+80679 \eta +16689 \eta ^2+80679 \eta ^3+7856 \eta 
^4\big)
\nonumber\\&&
-16 N^{11} \eta  \big(
        450738-6962355 \eta -48003782 \eta ^2-6819795 \eta ^3+483138 
\eta ^4\big)
\nonumber\\&&
+8 N^{13} \eta  \big(
        1570752+18898191 \eta +7445585 \eta ^2+18898191 \eta 
^3+1570752 \eta ^4\big)
\nonumber\\&&
+4 N^{12} \eta  \big(
        4566888+73693719 \eta +72807653 \eta ^2+73622439 \eta 
^3+4550688 \eta ^4\big)
\nonumber\\&&
+4 N^5 \big(
        236925+975942 \eta +27088992 \eta ^2+117199930 \eta 
^3+25711587 \eta ^4-325728 \eta ^5\big)
\nonumber\\&&
-12 N^8 \big(
        28350+912078 \eta +59340789 \eta ^2+250807007 \eta 
^3+58332339 \eta ^4+564588 \eta ^5\big)
\nonumber\\&&
-36 N^9 \big(
        6075+1870314 \eta +30831858 \eta ^2+19135996 \eta ^3+30932703 
\eta ^4+1867884 \eta ^5\big)
\nonumber\\&&
-8 N^{10} \big(
        6075+7288137 \eta +73658484 \eta ^2-103483664 \eta 
^3+74427579 \eta ^4+7437582 \eta ^5\big)
\nonumber\\&&
+4 N^7 \big(
        18225+8623746 \eta +9768924 \eta ^2-816997534 \eta 
^3+15310539 \eta ^4+9807156 \eta ^5\big)
\nonumber\\&&
+2 N^6 \big(
        443475+12987162 \eta +132854256 \eta ^2-575608138 \eta 
^3+137257821 \eta ^4
\nonumber\\&&
+12137472 \eta ^5\big).
\end{eqnarray}


\subsubsection{The result in momentum fraction $z$-space}
From the $N$-space result, the Mellin inversion has been performed using the algorithms encoded in {\tt HarmonicSums}. The result, which we reproduce here from~\cite{Ablinger:2020snj}, is reported as a one-dimensional integral, in a form amenable to numerical evaluation. It depends on iterated integrals of the type \eqref{eq:Gfunctions} on an alphabet of root-valued letters. The iterated integrals are the same as those appearing in the unpolarized calculation, and can be found in Appendix~D of \cite{Ablinger:2018brx}.

In what follows, the argument of the functions $G_i$ is implied to be $z$ in the formulas for $\tilde{a}_{gg,Q}^{(3),+} (z)$ and $\tilde{a}_{gg,Q}^{(3),\text{reg}}(z)$, and it is implied to be $y$ in the functions $\Phi_i$ which follow. Such arguments are omitted in the interest of brevity. Defining for the inverse Mellin transform of $\tilde{a}_{gg,Q}^{(3)}$

\begin{eqnarray}
\tilde{a}_{gg,Q}^{(3)} (N) &=& \tilde{a}_{gg,Q}^{(3),\delta}  + \int_{0}^{1} \text{d} z \ \left( z^{N-1} - 1 \right) \ \tilde{a}_{gg,Q}^{(3),+} (z)
 + \int_{0}^{1} \text{d} z \ z^{N-1} \ \tilde{a}_{gg,Q}^{(3),\text{reg}} (z) ~,
\end{eqnarray}
we find


with the polynomials
\begin{eqnarray}
Q_1&=&-405
-405 \eta 
+10412 \eta ^2
-405 \eta ^3
-405 \eta ^4
+405 z (-1+\eta )^2 \big(
        1+\eta +\eta ^2\big)
\\
Q_2&=&-z (-1+\eta )^2 \big(
        1+\eta ^4\big)
-2 \eta  \big(
        1+\eta ^4\big)
+z^2 (-1+\eta )^2 (1+\eta )^2 \big(
        1-\eta +\eta ^2\big)
\\
Q_3&=&2 \eta  \big(
        1-\eta +\eta ^2\big)
+z^3 (-1+\eta )^2 \big(
        1+\eta +\eta ^2\big)
+z \big(
        1-6 \eta +6 \eta ^2-6 \eta ^3+\eta ^4\big)
\nonumber\\&&        
-z^2 (-1+\eta )^2 \big(
        2-\eta +2 \eta ^2\big)
\\
Q_4&=&30 \eta 
+88 \eta ^3
+30 \eta ^5
+z \big(
        15-60 \eta +103 \eta ^2-176 \eta ^3+103 \eta ^4-60 \eta ^5+15 
\eta ^6\big)
\nonumber\\&&
+15 z^3 (-1+\eta )^2 (1+\eta )^2 \big(
        1-\eta +\eta ^2\big)
\nonumber\\&&        
-z^2 (-1+\eta )^2 \big(
        30+15 \eta +88 \eta ^2+15 \eta ^3+30 \eta ^4\big)
\\
Q_5&=&5
+22 \eta 
+5 \eta ^2
+2 z \big(
        1-10 \eta +\eta ^2\big)
\\
Q_6&=&45
+302 \eta 
+45 \eta ^2
+z \big(
        27-10 \eta +27 \eta ^2\big)
\\
Q_7&=&135
-3436 \eta 
+135 \eta ^2
+z \big(
        81+596 \eta +81 \eta ^2\big)
\\
Q_8&=&z^2 \big(
        -287+62 \eta -287 \eta ^2\big)
+120 z^4 \big(
        1-10 \eta +\eta ^2\big)
-7 \big(
        1-\eta +\eta ^2\big)
\nonumber\\&&        
+240 z^3 \big(
        1+8 \eta +\eta ^2\big)
-6 z \big(
        11+64 \eta +11 \eta ^2\big)
\\
Q_9&=&16 \big(
        73+90 \eta +163 \eta ^2+90 \eta ^3+73 \eta ^4\big)\\
Q_{10}&=&20 (1+\eta )^2 \big(
        73+17 \eta +73 \eta ^2\big)\\
Q_{11}&=&1
-70 \eta 
+\eta ^2
+8 z \big(
        1+40 \eta +\eta ^2\big)
\\
Q_{12}&=&29
+2540 \eta 
+29 \eta ^2
+2 z \big(
        16-1523 \eta +16 \eta ^2\big)
\\
Q_{13}&=&9
+400 \eta 
+9 \eta ^2
+4 z \big(
        9+100 \eta +9 \eta ^2\big)
\\
Q_{14}&=&109
+446 \eta 
+109 \eta ^2
+64 z \big(
        1+14 \eta +\eta ^2\big)
\\
Q_{15}&=&9
+160 \eta 
+9 \eta ^2
+8 z \big(
        9+20 \eta +9 \eta ^2\big)
\\
Q_{16}&=&-81
-410 \eta 
-81 \eta ^2
+z \big(
        81+550 \eta +81 \eta ^2\big)
\\
Q_{17}&=&-81
-820 \eta 
-81 \eta ^2
+3 z \big(
        27+260 \eta +27 \eta ^2\big)
\\
Q_{18}&=&59
+226 \eta 
+59 \eta ^2
+4 z \big(
        59+346 \eta +59 \eta ^2\big)
\\
Q_{19}&=&9
-11246 \eta 
+9 \eta ^2
+8 z \big(
        261+1568 \eta +261 \eta ^2\big)
\\
Q_{20}&=&-5 \big(
        17739+24192 \eta +397052 \eta ^2+24192 \eta ^3+17739 \eta 
^4\big)
\nonumber\\&&
+z \big(
        88695+567 \eta +2287160 \eta ^2+567 \eta ^3+88695 \eta 
^4\big)
\\
Q_{21}&=&-49 \eta  \Big[
        -z (-1+\eta )^2
        +z^2 (-1+\eta )^2
        -\eta 
\Big]\\
Q_{22}&=&147 \eta  \big(
        1+\eta ^2
\big)
\Big[-z (-1+\eta )^2
        +z^2 (-1+\eta )^2
        -\eta 
\Big]\\
Q_{23}&=&5 \eta  \big(
        807+574 \eta +807 \eta ^2\big)
+z \big(
        3285-6985 \eta -7249 \eta ^2-6985 \eta ^3+3285 \eta ^4\big)
\nonumber\\&&        
+3 z^3 (-1+\eta )^2 \big(
        1095+2693 \eta +1095 \eta ^2\big)
\nonumber\\&&        
-z^2 (-1+\eta )^2 \big(
        6570+11699 \eta +6570 \eta ^2\big)
\\
Q_{24}&=&\eta  \big(
        2421+792 \eta -33824 \eta ^2+792 \eta ^3+2421 \eta ^4\big)
\nonumber\\&&        
+z \big(
        1971-5121 \eta -40574 \eta ^2+67548 \eta ^3-40574 \eta 
^4-5121 \eta ^5+1971 \eta ^6\big)
\nonumber\\&&
+z^3 (-1+\eta )^2 \big(
        1971+7137 \eta +1684 \eta ^2+7137 \eta ^3+1971 \eta ^4\big)
\nonumber\\&&        
-z^2 (-1+\eta )^2 \big(
        3942+8379 \eta -32140 \eta ^2+8379 \eta ^3+3942 \eta ^4\big)
\\
Q_{25}&=&z \big(
        9855-4695 \eta -27488 \eta ^2-123060 \eta ^3-27488 \eta 
^4-4695 \eta ^5+9855 \eta ^6\big)
\nonumber\\&&
+5 \eta  \big(
        2421+4974 \eta +8324 \eta ^2+4974 \eta ^3+2421 \eta ^4\big)
\nonumber\\&&        
+z^3 (-1+\eta )^2 \big(
        9855+29223 \eta +89560 \eta ^2+29223 \eta ^3+9855 \eta 
^4\big)
\nonumber\\&&
-z^2 (-1+\eta )^2 \big(
        19710+56343 \eta +131180 \eta ^2+56343 \eta ^3+19710 \eta 
^4\big)
\\
Q_{26}&=&49 \big(
        1-\eta +\eta ^2\big)
+3840 z^5 \big(
        1+14 \eta +\eta ^2\big)
+60 z^4 \big(
        13-898 \eta +13 \eta ^2\big)
\nonumber\\&&        
-58 z^3 \big(
        173+1102 \eta +173 \eta ^2\big)
+z \big(
        1463+412 \eta +1463 \eta ^2\big)
\nonumber\\&&        
+z^2 \big(
        7187+64438 \eta +7187 \eta ^2\big)
\end{eqnarray}

The functions $\Phi_1,\ldots,\Phi_8$, which appear as arguments of a further integral, are:


The polynomials $R_i$ are:
\begin{eqnarray}
R_1&=&-20 (1+\eta )^2 \big(
        1-10 \eta +\eta ^2\big)\\
R_2&=&-16 \big(
        1-9 \eta -8 \eta ^2-9 \eta ^3+\eta ^4\big)\\
R_3&=&\eta  \big(
        729-862 \eta +729 \eta ^2\big)
+27 y^2 (-1+\eta )^2 \big(
        1+46 \eta +\eta ^2\big)
\nonumber\\&&        
-27 y \big(
        1+70 \eta -126 \eta ^2+70 \eta ^3+\eta ^4\big)
\\
R_4&=&18 \eta ^2
+2 y \eta  \big(
        13-50 \eta +13 \eta ^2\big)
+y^3 (-1+\eta )^2 \big(
        1+46 \eta +\eta ^2\big)
\nonumber\\&&        
-y^2 (-1+\eta )^2 \big(
        1+74 \eta +\eta ^2\big)
\\
R_5&=&2 \eta ^2 \big(
        81-34 \eta +81 \eta ^2\big)
\nonumber\\&&        
+y^2 \big(
        9-576 \eta +1391 \eta ^2-1360 \eta ^3+1391 \eta ^4-576 \eta 
^5+9 \eta ^6\big)
\nonumber\\&&
+9 y^4 (-1+\eta )^2 (1+\eta )^2 \big(
        1-10 \eta +\eta ^2\big)
-18 y^3 (-1+\eta )^2 \big(
        1-21 \eta -21 \eta ^3+\eta ^4\big)
\nonumber\\&&        
+2 y \eta  \big(
        126-385 \eta +302 \eta ^2-385 \eta ^3+126 \eta ^4\big)
\\
R_6&=&2 \eta ^2 \big(
        81-62 \eta +81 \eta ^2\big)
\nonumber\\&&        
+y^2 \big(
        9-576 \eta +1447 \eta ^2-1472 \eta ^3+1447 \eta ^4-576 \eta 
^5+9 \eta ^6\big)
\nonumber\\&&
+9 y^4 (-1+\eta )^2 (1+\eta )^2 \big(
        1-10 \eta +\eta ^2\big)
-18 y^3 (-1+\eta )^2 \big(
        1-21 \eta -21 \eta ^3+\eta ^4\big)
\nonumber\\&&        
+2 y \eta  \big(
        126-413 \eta +358 \eta ^2-413 \eta ^3+126 \eta ^4\big)
\\
R_7&=&-20 (1+\eta )^2 \big(
        5+22 \eta +5 \eta ^2\big)\\
R_8&=&-16 \big(
        5+27 \eta +32 \eta ^2+27 \eta ^3+5 \eta ^4\big)\\
R_9&=&2 \eta  \big(
        11-70 \eta +11 \eta ^2\big)
+y^2 (-1+\eta )^2 \big(
        5+86 \eta +5 \eta ^2\big)
\nonumber\\&&        
-y (-1+\eta )^2 \big(
        5+118 \eta +5 \eta ^2\big)
\\
R_{10}&=&\eta  \big(
        243-790 \eta +243 \eta ^2\big)
+9 y^2 (-1+\eta )^2 \big(
        5+86 \eta +5 \eta ^2\big)
\nonumber\\&&        
-9 y \big(
        5+98 \eta -270 \eta ^2+98 \eta ^3+5 \eta ^4\big)
\\
R_{11}&=&y \Big[
        -280 \eta ^3
        +y^2 \big(
                15-96 \eta +757 \eta ^2-1736 \eta ^3+757 \eta ^4-96 
\eta ^5+15 \eta ^6\big)
\nonumber\\&&
        +3 y^4 (-1+\eta )^2 (1+\eta )^2 \big(
                5+22 \eta +5 \eta ^2\big)
\nonumber\\&&                
        -6 y^3 (-1+\eta )^2 \big(
                5+21 \eta +108 \eta ^2+21 \eta ^3+5 \eta ^4\big)
\nonumber\\&&                
        +4 y \eta  \big(
                24-79 \eta +254 \eta ^2-79 \eta ^3+24 \eta ^4\big)
\Big]\\
R_{12}&=&y \Big[
        -368 \eta ^3
        +y^2 \big(
                15-96 \eta +845 \eta ^2-1912 \eta ^3+845 \eta ^4-96 
\eta ^5+15 \eta ^6\big)
\nonumber\\&&
        +3 y^4 (-1+\eta )^2 (1+\eta )^2 \big(
                5+22 \eta +5 \eta ^2\big)
\nonumber\\&&                
        -6 y^3 (-1+\eta )^2 \big(
                5+21 \eta +108 \eta ^2+21 \eta ^3+5 \eta ^4\big)
\nonumber\\&&                
        +4 y \eta  \big(
                24-101 \eta +298 \eta ^2-101 \eta ^3+24 \eta ^4\big)
\Big]\\
R_{13}&=&-20 (1+\eta )^2 \big(
        29-74 \eta +29 \eta ^2\big)\\
R_{14}&=&16 \big(
        -29+45 \eta +16 \eta ^2+45 \eta ^3-29 \eta ^4\big)\\
R_{15}&=&-55 \eta  \big(
        243-382 \eta +243 \eta ^2\big)
+y \big(
        783+23949 \eta -54320 \eta ^2+23949 \eta ^3+783 \eta ^4\big)
\\
R_{16}&=&(1+\eta ) \Big[
        270 \eta ^2
        +2 y \eta  \big(
                233-982 \eta +233 \eta ^2\big)
        +y^3 (-1+\eta )^2 \big(
                29+974 \eta +29 \eta ^2\big)
\nonumber\\&&                
        -y^2 (-1+\eta )^2 \big(
                29+1498 \eta +29 \eta ^2\big)
\Big]\\
R_{17}&=&-2 \eta ^2 \big(
        1215-2686 \eta +1215 \eta ^2\big)
\nonumber\\&&        
-9 y^2 (-1+\eta )^2 \big(
        29-482 \eta +810 \eta ^2-482 \eta ^3+29 \eta ^4\big)
\nonumber\\&&        
+y^3 (-1+\eta )^2 \big(
        261-144 \eta -1610 \eta ^2-144 \eta ^3+261 \eta ^4\big)
\nonumber\\&&        
-4 y \eta  \big(
        1179-3476 \eta +4250 \eta ^2-3476 \eta ^3+1179 \eta ^4\big)
\\
R_{18}&=&-2 \eta ^2 \big(
        1215-2486 \eta +1215 \eta ^2\big)
\nonumber\\&&        
-9 y^2 (-1+\eta )^2 \big(
        29-482 \eta +810 \eta ^2-482 \eta ^3+29 \eta ^4\big)
\nonumber\\&&        
+y^3 (-1+\eta )^2 \big(
        261-144 \eta -1210 \eta ^2-144 \eta ^3+261 \eta ^4\big)
\nonumber\\&&        
-4 y \eta  \big(
        1179-3376 \eta +4150 \eta ^2-3376 \eta ^3+1179 \eta ^4\big)
\\
R_{19}&=&-20 (1+\eta )^2 \big(
        55+26 \eta +55 \eta ^2\big)\\
R_{20}&=&-16 \big(
        55+81 \eta +136 \eta ^2+81 \eta ^3+55 \eta ^4\big)\\
R_{21}&=&\eta  \big(
        -2187+4202 \eta -2187 \eta ^2\big)
+3 y (-1+\eta )^2 \big(
        55+1594 \eta +55 \eta ^2\big)
\\
R_{22}&=&(1+\eta ) \Big[
        324 \eta ^2
        +2 y \eta  \big(
                337-1526 \eta +337 \eta ^2\big)
        +y^3 (-1+\eta )^2 \big(
                55+1594 \eta +55 \eta ^2\big)
\nonumber\\&&                
        -y^2 (-1+\eta )^2 \big(
                55+2378 \eta +55 \eta ^2\big)
\Big]\\
R_{23}&=&-108 (-3+\eta ) \eta ^2 (-1+3 \eta )
+y^3 (-1+\eta )^2 (1+\eta )^2 \big(
        55+26 \eta +55 \eta ^2\big)
\nonumber\\&&        
-y^2 (-1+\eta )^2 \big(
        55-538 \eta +1942 \eta ^2-538 \eta ^3+55 \eta ^4\big)
\nonumber\\&&        
-4 y \eta  \big(
        196-573 \eta +890 \eta ^2-573 \eta ^3+196 \eta ^4\big)
\\
R_{24}&=&-4 \eta ^2 \big(
        81-250 \eta +81 \eta ^2\big)
+y^3 (-1+\eta )^2 (1+\eta )^2 \big(
        55+26 \eta +55 \eta ^2\big)
\nonumber\\&&        
-y^2 (-1+\eta )^2 \big(
        55-538 \eta +1862 \eta ^2-538 \eta ^3+55 \eta ^4\big)
\nonumber\\&&        
-4 y \eta  \big(
        196-553 \eta +850 \eta ^2-553 \eta ^3+196 \eta ^4\big)
\\
R_{25}&=&\eta 
+\eta ^3
+y \eta  \big(
        1+\eta ^2\big)
+y^2 (-1+\eta )^2 \big(
        1+\eta +\eta ^2\big)
\\
R_{26}&=&y (1+\eta ) \Big[
        -y (-1+\eta )^2 (1+\eta )^2
        +y^2 (-1+\eta )^2 \big(
                1+\eta +\eta ^2\big)
\nonumber\\&&                
        -\eta  \big(
                1+\eta +\eta ^2\big)
\Big]\\
R_{27}&=&2 \eta ^3
-y \eta  (1+\eta )^2 \big(
        1-\eta +\eta ^2\big)
+y^3 (-1+\eta )^2 (1+\eta )^2 \big(
        1-\eta +\eta ^2\big)
\nonumber\\&&
-y^2 (-1+\eta )^2 \big(
        1+2 \eta +2 \eta ^3+\eta ^4\big)
\\
R_{28}&=&-128 \big(
        1+15 \eta +16 \eta ^2+15 \eta ^3+\eta ^4\big)\\
R_{29}&=&-32 (1+\eta )^2 \big(
        1+14 \eta +\eta ^2\big)\\
R_{30}&=&189
+810 y^2 (-1+\eta )^4
-4518 \eta 
-2458 \eta ^2
-4518 \eta ^3
+189 \eta ^4
\nonumber\\&&
-27 y \big(
        37-308 \eta +30 \eta ^2-308 \eta ^3+37 \eta ^4\big)
\\
R_{31}&=&90 y^4 (-1+\eta )^4
-2 \eta  \big(
        7-158 \eta +7 \eta ^2\big)
\nonumber\\&&        
+y \big(
        -21+634 \eta -1418 \eta ^2+634 \eta ^3-21 \eta ^4\big)
\nonumber\\&&        
+4 y^2 (-1+\eta )^2 \big(
        33-296 \eta +33 \eta ^2\big)
-3 y^3 (-1+\eta )^2 \big(
        67-262 \eta +67 \eta ^2\big)
\\
R_{32}&=&y \big(
        63-1488 \eta +5429 \eta ^2+5816 \eta ^3+5429 \eta ^4-1488 
\eta ^5+63 \eta ^6\big)
\nonumber\\&&
+144 y^4 (-1+\eta )^2 (1+\eta )^2 \big(
        1+14 \eta +\eta ^2\big)
\nonumber\\&&        
-45 y^3 (-1+\eta )^2 \big(
        5+140 \eta +222 \eta ^2+140 \eta ^3+5 \eta ^4\big)
\nonumber\\&&        
+2 \eta  \big(
        21-474 \eta -470 \eta ^2-474 \eta ^3+21 \eta ^4\big)
\nonumber\\&&        
+2 y^2 \big(
        9+2640 \eta -3217 \eta ^2-3472 \eta ^3-3217 \eta ^4+2640 \eta 
^5+9 \eta ^6\big)
\\
R_{33}&=&y \big(
        63-1488 \eta +4789 \eta ^2+7096 \eta ^3+4789 \eta ^4-1488 
\eta ^5+63 \eta ^6\big)
\nonumber\\&&
+144 y^4 (-1+\eta )^2 (1+\eta )^2 \big(
        1+14 \eta +\eta ^2\big)
\nonumber\\&&        
-45 y^3 (-1+\eta )^2 \big(
        5+140 \eta +222 \eta ^2+140 \eta ^3+5 \eta ^4\big)
\nonumber\\&&        
+2 \eta  \big(
        21-474 \eta -790 \eta ^2-474 \eta ^3+21 \eta ^4\big)
\nonumber\\&&        
+2 y^2 \big(
        9+2640 \eta -2897 \eta ^2-4112 \eta ^3-2897 \eta ^4+2640 \eta 
^5+9 \eta ^6\big)
\\
R_{34}&=&-10 (1+\eta )^2 \big(
        109+446 \eta +109 \eta ^2\big)\\
R_{35}&=&-8 \big(
        109+555 \eta +664 \eta ^2+555 \eta ^3+109 \eta ^4\big)\\
R_{36}&=&-387
-5958 \eta 
-10102 \eta ^2
-5958 \eta ^3
-387 \eta ^4
\nonumber\\&&
+108 y^2 (-1+\eta )^2 \big(
        11-14 \eta +11 \eta ^2\big)
\nonumber\\&&        
-9 y \big(
        89-1312 \eta -210 \eta ^2-1312 \eta ^3+89 \eta ^4\big)
\\
R_{37}&=&43
+794 \eta 
-1530 \eta ^2
+794 \eta ^3
+43 \eta ^4
+12 y^3 (-1+\eta )^2 \big(
        11-14 \eta +11 \eta ^2\big)
\nonumber\\&&        
-y^2 (-1+\eta )^2 \big(
        221-866 \eta +221 \eta ^2\big)
\nonumber\\&&        
+2 y \big(
        23-835 \eta +1576 \eta ^2-835 \eta ^3+23 \eta ^4\big)
\\
R_{38}&=&y \Big[
        5120 \eta ^3
        +y \big(
                129+1332 \eta -4129 \eta ^2-18568 \eta ^3-4129 \eta 
^4+1332 \eta ^5+129 \eta ^6\big)
\nonumber\\&&
        -6 y^4 (-1+\eta )^2 (1+\eta )^2 \big(
                109+446 \eta +109 \eta ^2\big)
\nonumber\\&&                
        +3 y^3 (-1+\eta )^2 \big(
                479+3536 \eta +5250 \eta ^2+3536 \eta ^3+479 \eta 
^4\big)
\nonumber\\&&
        -2 y^2 \big(
                456+3195 \eta -3680 \eta ^2-7910 \eta ^3-3680 \eta 
^4+3195 \eta ^5+456 \eta ^6\big)
\Big]\\
R_{39}&=&y \Big[
        2560 \eta ^3
        +y \big(
                129+1332 \eta -6689 \eta ^2-13448 \eta ^3-6689 \eta 
^4+1332 \eta ^5+129 \eta ^6\big)
\nonumber\\&&
        -6 y^4 (-1+\eta )^2 (1+\eta )^2 \big(
                109+446 \eta +109 \eta ^2\big)
\nonumber\\&&                
        +3 y^3 (-1+\eta )^2 \big(
                479+3536 \eta +5250 \eta ^2+3536 \eta ^3+479 \eta 
^4\big)
\nonumber\\&&
        -2 y^2 \big(
                456+3195 \eta -4960 \eta ^2-5350 \eta ^3-4960 \eta 
^4+3195 \eta ^5+456 \eta ^6\big)
\Big]\\
R_{40}&=&-160 (1+\eta )^2 \big(
        17+883 \eta +17 \eta ^2\big)\\
R_{41}&=&-128 \big(
        17+900 \eta +917 \eta ^2+900 \eta ^3+17 \eta ^4\big)\\
R_{42}&=&-5 \big(
        1557-21798 \eta +3602 \eta ^2-21798 \eta ^3+1557 \eta ^4\big)
\nonumber\\&&        
+2 y \big(
        9009-53793 \eta +49720 \eta ^2-53793 \eta ^3+9009 \eta ^4\big)
\\
R_{43}&=&(1+\eta ) \Big[
        10 \eta  \big(
                53-690 \eta +53 \eta ^2\big)
\nonumber\\&&                
        +y \big(
                865-15172 \eta +30678 \eta ^2-15172 \eta ^3+865 \eta 
^4\big)
\nonumber\\&&
        +2 y^3 (-1+\eta )^2 \big(
                1001-3034 \eta +1001 \eta ^2\big)
\nonumber\\&&                
        -y^2 (-1+\eta )^2 \big(
                2867-17786 \eta +2867 \eta ^2\big)
\Big]\\
R_{44}&=&2 \eta  \big(
        -265+3450 \eta +2022 \eta ^2+3450 \eta ^3-265 \eta ^4\big)
\nonumber\\&&        
+8 y^3 (-1+\eta )^2 \big(
        34+1969 \eta +4900 \eta ^2+1969 \eta ^3+34 \eta ^4\big)
\nonumber\\&&        
+y^2 (-1+\eta )^2 \big(
        593-27584 \eta -34050 \eta ^2-27584 \eta ^3+593 \eta ^4\big)
\nonumber\\&&        
-y \big(
        865-12898 \eta +35567 \eta ^2+24180 \eta ^3+35567 \eta 
^4-12898 \eta ^5+865 \eta ^6\big)
\\
R_{45}&=&2 \eta  \big(
        -795+10350 \eta +10666 \eta ^2+10350 \eta ^3-795 \eta ^4\big)
\nonumber\\&&        
+8 y^3 (-1+\eta )^2 \big(
        102+5907 \eta +13550 \eta ^2+5907 \eta ^3+102 \eta ^4\big)
\nonumber\\&&        
+3 y^2 (-1+\eta )^2 \big(
        593-27584 \eta -34050 \eta ^2-27584 \eta ^3+593 \eta ^4\big)
\nonumber\\&&        
-y \big(
        2595-38694 \eta +97501 \eta ^2+81740 \eta ^3+97501 \eta 
^4-38694 \eta ^5+2595 \eta ^6\big)
\\
R_{46}&=&-20 (1+\eta )^2 \big(
        1127+6478 \eta +1127 \eta ^2\big)\\
R_{47}&=&-16 \big(
        1127+7605 \eta +8732 \eta ^2+7605 \eta ^3+1127 \eta ^4\big)\\
R_{48}&=&1305
+16902 \eta 
+1666 \eta ^2
+16902 \eta ^3
+1305 \eta ^4
\nonumber\\&&
+3 y (-1+\eta )^2 \big(
        257-4018 \eta +257 \eta ^2\big)
\\
R_{49}&=&(1+\eta ) \Big[
        -3 \eta  \big(
                43+1046 \eta +43 \eta ^2\big)
        +2 y^2 (-1+\eta )^2 \big(
                89+4957 \eta +89 \eta ^2\big)
\nonumber\\&&                
        +y^3 (-1+\eta )^2 \big(
                257-4018 \eta +257 \eta ^2\big)
\nonumber\\&&                
        -y \big(
                435+5633 \eta -15640 \eta ^2+5633 \eta ^3+435 \eta ^4
\big)
\Big]\\
R_{50}&=&y \big(
        -435-4249 \eta +17803 \eta ^2+8690 \eta ^3+17803 \eta ^4-4249 
\eta ^5-435 \eta ^6\big)
\nonumber\\&&
-y^3 (-1+\eta )^2 (1+\eta )^2 \big(
        1127+6478 \eta +1127 \eta ^2\big)
\nonumber\\&&        
-\eta \big(
        129+3138 \eta +3074 \eta ^2+3138 \eta ^3+129 \eta ^4\big)
\nonumber\\&&        
+2 y^2 (-1+\eta )^2 \big(
        781+7358 \eta +5990 \eta ^2+7358 \eta ^3+781 \eta ^4\big)
\\
R_{51}&=&y \big(
        -435-4249 \eta +14523 \eta ^2+15250 \eta ^3+14523 \eta 
^4-4249 \eta ^5-435 \eta ^6\big)
\nonumber\\&&
-y^3 (-1+\eta )^2 (1+\eta )^2 \big(
        1127+6478 \eta +1127 \eta ^2\big)
\nonumber\\&&        
-3 \eta \big(
        43+1046 \eta +2118 \eta ^2+1046 \eta ^3+43 \eta ^4\big)
\nonumber\\&&        
+2 y^2 (-1+\eta )^2 \big(
        781+7358 \eta +7630 \eta ^2+7358 \eta ^3+781 \eta ^4\big)
\\
R_{52}&=&\eta 
+\eta ^3
+y \eta  \big(
        1+\eta ^2\big)
+y^2 (-1+\eta )^2 \big(
        1+\eta +\eta ^2\big)
\\
R_{53}&=&y (1+\eta ) \Big[
        -y (-1+\eta )^2 (1+\eta )^2
        +y^2 (-1+\eta )^2 \big(
                1+\eta +\eta ^2\big)
\nonumber\\&&                
        -\eta  \big(
                1+\eta +\eta ^2\big)
\Big]\\
R_{54}&=&2 \eta ^3
-y \eta  (1+\eta )^2 \big(
        1-\eta +\eta ^2\big)
+y^3 (-1+\eta )^2 (1+\eta )^2 \big(
        1-\eta +\eta ^2\big)
\nonumber\\&&        
-y^2 (-1+\eta )^2 \big(
        1+2 \eta +2 \eta ^3+\eta ^4\big).
\end{eqnarray}
%
In the Appendices of~\cite{Ablinger:2020snj}, a set of relations and evaluations of iterated integrals can be found, which were used for this calculation.  We present further relations in Appendices \ref{sec:A} and \ref{sec:appB}.

\subsubsection{Numerical results}
\begin{figure}[H]
\begin{center}
\includegraphics[width=.7\textwidth]{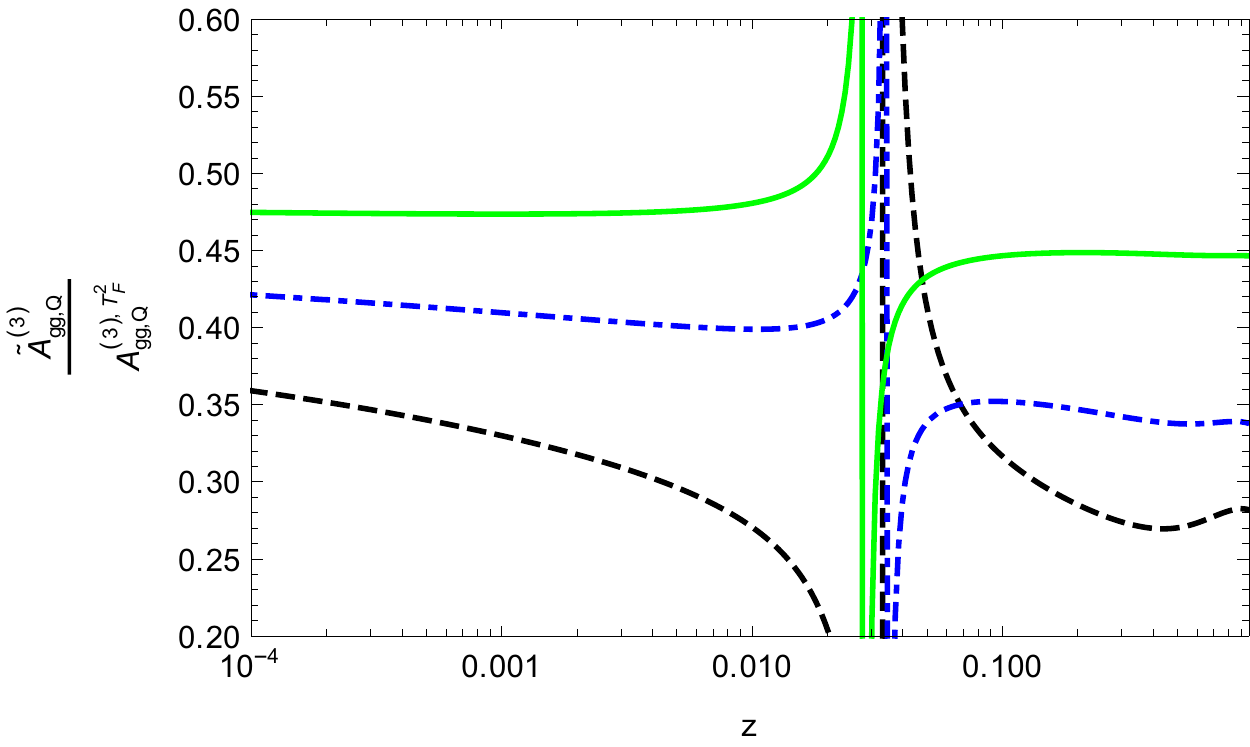}
\caption{\sf%
The ratio of the two mass contribution to the total
contribution at $O(T_F^2)$ for the
polarized massive OME $A_{gg,Q}^{(3)}$ as a function of the momentum
fraction $z$ and the virtuality $\mu^2$.
Dashed line: $\mu^2 = 50~\GeV^2$;
Dash--dotted line: $\mu^2 = 100~\GeV^2$;
Full line: $\mu^2 = 1000~\GeV^2$.
For the values of $m_c$ and $m_b$ we refer to the on--shell heavy 
quark masses
$m_c = 1.59~\GeV$ and $m_b = 4.78~\GeV$ \cite{Alekhin:2012vu,Agashe:2014kda}, from \cite{Ablinger:2020snj}.}
\label{fig:RAT}
\end{center}
\end{figure}

A numerical evaluation of the size of this correction compared to the total $\mathcal{O}(T_F^2)$ is shown in Figure~\ref{fig:RAT}, depicting the ratio between the two-mass contribution and the total contribution to $O(T_F^2)$. Their appreciable size of $O(0.4)$ makes it necessary to include them in numerical studies.

\subsubsection{Summary}

We computed the two-mass contributions to the polarized OMEs $A_{gg,Q}^{(3)}$ and $A_{Qq}^{(3),PS}$ in the Larin scheme at $\mathcal O(a_s^3)$ in semi-analytic form in $z$-space and, for $A_{gg,Q}^{(3)}$, analytically in $N$-space. We show that their numerical contribution is of the same order of magnitude as the single-mass contribution and conclude that two-mass effects should be taken into account in the variable flavour number scheme.

\clearpage
\section[\texorpdfstring{The logarithmic single-mass contributions to the polarized asymptotic $\mathcal{O}(a_s^3)$ Wilson coefficients in deeply inelastic scattering}{The logarithmic single-mass contributions to the polarized asymptotic O(aₛ³) Wilson coefficients in deeply inelastic scattering}]{The logarithmic single-mass contributions to the polarized asymptotic $\mathcal{O}(a_s^3)$ Wilson coefficients in deeply inelastic scattering}
\label{sec:logWC}

Heavy flavour contributions to the structure functions in deep-inelastic scattering need to be taken into account in precision studies of DIS. They arise when quark masses are not neglected in the DIS process. These heavy flavour contributions can be attributed to the Wilson coefficients
\begin{eqnarray}
\mathcal{C}_i^{S,PS,NS}\Big(N,N_F+1,\frac{Q^2}{\mu^2},\frac{m^2}{\mu^2}\Big)
&=&C_i^{S,PS,NS}\Big(N,N_F,\frac{Q^2}{\mu^2} \Big) + H_i^{S,PS}(N,N_F+1,\frac{Q^2}{\mu^2},\frac{m^2}{\mu^2}\Big) 
\nonumber\\&&
+ L_i^{S,PS,NS}(N,N_F+1,\frac{Q^2}{\mu^2},\frac{m^2}{\mu^2}\Big) ~,
\end{eqnarray}
where the symbol $N_F+1$ refers to $N_F$ massless flavours and one additional massive flavour \cite{Buza:1995ie,Bierenbaum:2009mv}. In this notation, the Wilson coefficients $H_i$ identify the contributions where the off-shell photon couples to a heavy quark, and $L_i$ to those where it couples to a light quark. The light flavour Wilson coefficients are denoted by $C_i$.

In the asymptotic limit $Q^2\gg m^2$, where $m$ is the mass of the heavy quark, the asymptotic form of the Wilson coefficients can be obtained through the factorization formula
\begin{eqnarray}
\mathcal{C}_j^{S,PS,NS,asymp} \Big(N,N_F+1,\frac{Q^2}{\mu^2},\frac{m^2}{\mu^2}\Big)
&=& 
\nonumber\\&&
\hspace{-5cm}
\sum_i A_{ij}^{S,PS,NS}\Big(N,N_F+1,\frac{m^2}{\mu^2}\Big) C_i^{S,PS,NS}\Big(N,N_F+1,\frac{Q^2}{\mu^2}\Big) +\mathcal{O}\Big(\frac{m^2}{Q^2}\Big)
\end{eqnarray}
with $A_{i,j}^{S,PS,NS}$ the matrix elements of twist-2 operators between partonic states, which contain the contributions of heavy quarks in the loops:
\begin{equation}
A_{i,j}^{S,PS,NS}\Big(N,N_F+1,\frac{m^2}{\mu^2}\Big) = \langle j| O_i^{S,NS} |j \rangle, \qquad j=q,g.
\end{equation}
Both the light Wilson coefficients $C_i$ and the OMEs $A_{ij}$ can be calculated perturbatively in a series in $a_s$,
\begin{eqnarray}
C_i\Big(N,\frac{Q^2}{\mu^2}\Big) &=& \sum_{k=0}^\infty a_s^k \ C_i^{(k)}\Big(N,\frac{Q^2}{\mu^2}\Big) ~,\\
A_{ij}^{S,PS,NS}\Big(N,\frac{m^2}{\mu^2}\Big) &=& \sum_{k=0}^\infty a_s^k \ A_{ij}^{(k),S,PS,NS}\Big(N,\frac{m^2}{\mu^2}\Big) ~.
\end{eqnarray}
After the renormalization procedure is carried out, in our case in the $\MS$ scheme for the strong coupling constant and the on-mass-shell scheme for the mass of the heavy quark, the light Wilson coefficients and the OMEs take the form 
\begin{eqnarray}
A_{ij}^{(k)} &=& \sum_{\ell=0}^k A_{ij}^{(k,\ell)} \ln^\ell\frac{m^2}{\mu^2} ~,\\
C_i^{(k)} &=& \sum_{\ell=0}^k C_i^{(k,\ell)} \ln^\ell\frac{Q^2}{\mu^2} ~.
\label{eq:logexpansion}
\end{eqnarray}
Treating $\gamma_5$, we work in the Larin scheme.

With the definitions
\begin{eqnarray}
\tilde f(N_f) &\equiv & \frac{f(N_F)}{N_F} ~,\\
\hat f(N_F) &\equiv & f(N_F+1)-f(N_F) ~,
\end{eqnarray}
the heavy quark Wilson coefficients take the asymptotic form \cite{Bierenbaum:2009mv}
\begin{eqnarray}
     \label{eqWIL1}
     L_{q,g_1}^{\sf NS,Q}(N_F+1) &=& 
     a_s^2 \Bigl[A_{qq,Q}^{(2), {\sf NS}}(N_F+1) +
     \hat{C}^{(2), {\sf NS}}_{q,g_1}(N_F)\Bigr]
     \N\\
     &+&
     a_s^3 \Bigl[A_{qq,Q}^{(3), {\sf NS}}(N_F+1)
     +  A_{qq,Q}^{(2), {\sf NS}}(N_F+1) C_{q,g_1}^{(1), {\sf NS}}(N_F+1)
       \N \\
     && \hspace*{5mm}
     + \hat{C}^{(3), {\sf NS}}_{q,g_1}(N_F)\Bigr]~,  \\
      \label{eqWIL2}
      L_{q,g_1}^{\sf PS}(N_F+1) &=& 
     a_s^3 \Bigl[~A_{qq,Q}^{(3), {\sf PS}}(N_F+1)
     +  A_{gq,Q}^{(2)}(N_F+1)~~N_F\Ctil_{g,g_1}^{(1)}(N_F+1) \N \\
     && \hspace*{5mm}
     + N_F \hat{\Ctil}^{(3), {\sf PS}}_{q,g_1}(N_F)\Bigr]~,
     \\
     \label{eqWIL3}
      L_{g,g_1}^{\sf S}(N_F+1) &=& 
     a_s^2 A_{gg,Q}^{(1)}(N_F+1)N_F \Ctil_{g,g_1}^{(1)}(N_F+1)
     \N\\ &+&
      a_s^3 \Bigl[~A_{qg,Q}^{(3)}(N_F+1) 
     +  A_{gg,Q}^{(1)}(N_F+1)~~N_F\Ctil_{g,g_1}^{(2)}(N_F+1)
     \N\\ && \hspace*{5mm}
     +  A_{gg,Q}^{(2)}(N_F+1)~~N_F\Ctil_{g,g_1}^{(1)}(N_F+1)
     \N\\ && \hspace*{5mm}
     +  ~A^{(1)}_{Qg}(N_F+1)~~N_F\Ctil_{q,g_1}^{(2), {\sf PS}}(N_F+1)
     + N_F \hat{\Ctil}^{(3)}_{g,g_1}(N_F)\Bigr]~,
 \\ \N \\
\label{eq:WILPS}
     H_{q,g_1}^{\sf PS}(N_F+1)
     &=& a_s^2 \Bigl[~A_{Qq}^{(2), {\sf PS}}(N_F+1)
     +~\Ctil_{q,g_1}^{(2), {\sf PS}}(N_F+1)\Bigr]
     \\
     &+& a_s^3 \Bigl[~A_{Qq}^{(3), {\sf PS}}(N_F+1)
     +~\Ctil_{q,g_1}^{(3), {\sf PS}}(N_F+1) \N\\
 && 
     + A_{gq,Q}^{(2)}(N_F+1)~\Ctil_{g,g_1}^{(1)}(N_F+1) 
     + A_{Qq}^{(2), {\sf PS}}(N_F+1)~C_{q,g_1}^{(1), {\sf NS}}(N_F+1) 
        \Bigr]~,       \label{eqWIL4}
         \N\\ 
\label{eq:WILS}
     H_{g,g_1}^{\sf S}(N_F+1) &=& a_s \Bigl[~A_{Qg}^{(1)}(N_F+1)
     +~\Ctil^{(1)}_{g,g_1}(N_F+1) \Bigr] \N\\
     &+& a_s^2 \Bigl[~A_{Qg}^{(2)}(N_F+1)
     +~A_{Qg}^{(1)}(N_F+1)~C^{(1), {\sf NS}}_{q,g_1}(N_F+1)\N\\ && 
     \hspace*{5mm}
     +~A_{gg,Q}^{(1)}(N_F+1)~\Ctil^{(1)}_{g,g_1}(N_F+1) 
     +~\Ctil^{(2)}_{g,g_1}(N_F+1) \Bigr]
     \N\\ &+&
     a_s^3 \Bigl[~A_{Qg}^{(3)}(N_F+1)
     +~A_{Qg}^{(2)}(N_F+1)~C^{(1), {\sf NS}}_{q,g_1}(N_F+1)
     \N\\ &&
     \hspace*{5mm}
     +~A_{gg,Q}^{(2)}(N_F+1)~\Ctil^{(1)}_{g,g_1}(N_F+1)
     \N\\ && \hspace*{5mm}
     +~A_{Qg}^{(1)}(N_F+1)\Bigl\{
     C^{(2), {\sf NS}}_{q,g_1}(N_F+1)
     +~\Ctil^{(2), {\sf PS}}_{q,g_1}(N_F+1)\Bigr\}
     \N\\ && \hspace*{5mm}
     +~A_{gg,Q}^{(1)}(N_F+1)~\Ctil^{(2)}_{g,g_1}(N_F+1)
     +~\Ctil^{(3)}_{g,g_1}(N_F+1) \Bigr] ~.
\label{eqWIL5}
\end{eqnarray}

In the unpolarized case, the asymptotic form of the heavy flavour Wilson coefficients has been computed with this method to $\mathcal O(a^3)$ in \cite{Kawamura:2012cr,Behring:2014eya} and in the case of the structure function $g_1$ the non-singlet Wilson coefficient $L_{q,g_1}^{NS}$ was calculated in \cite{Behring:2015zaa}; see also \cite{Behring:2015roa}.

We briefly review how the logarithmic terms in the massless Wilson coefficients are obtained following \cite{vanNeerven:2000uj}. From the renormalization group equation
\begin{equation}
\sum_i\Big[\frac{\partial}{\partial \ln\mu^2} +\beta(a_s) \frac{\partial}{\partial a_s} -\gamma_{ij} \Big] C_i\Big(a_s,\frac{Q^2}{\mu^2}\Big) = 0
\label{eq:rge}
\end{equation}
and
\begin{equation}
\beta(a_s) = - \sum_{k=0}^\infty a_s^{k+2} \beta_k , 
\end{equation}
inserting the ansatz \eqref{eq:logexpansion} into \eqref{eq:rge} one obtains the explicit form of $C_i^{(k,\ell)}$ for $\ell>0$ in terms of lower-order quantities. For instance, one can write for the non-singlet Wilson coefficient in $N$-space
\begin{eqnarray}
	C_q^{NS(0,0)} &=& 1 \\
	C_q^{NS(1,1)} &=& -C_q^{NS(0, 0)} \gamma_{qq}^{NS(0)} \\
	C_q^{NS(2,2)} &=& \frac{1}{2} \big(-\beta_0 C_q^{NS(1, 1)} - c_q^{NS(1, 1)} \gamma_{qq}^{NS(0)} \big) \\
	C_q^{NS(2,1)} &=& -\beta_0 C_q^{NS(1, 0)} - C_q^{NS(1, 0)} \gamma_{qq}^{NS(0)} - C_q^{NS(0, 0)} \gamma_{qq}^{NS(1)}.
\end{eqnarray}
Formulas for $C_g^{S(k,\ell)}$ and $C_q^{S(k,\ell)}$ can be derived in a very similar way.

In the case of the structure function $g_1$, the massless Wilson coefficients are not currently known to 3-loop accuracy. In the formulas reported in Appendix \ref{sec:polOMEs} and Appendix \ref{sec:polWCs}, the terms $C_i^{(3)}$ are all left symbolic. However, they cannot affect the logarithmic terms $\ln(Q^2/m^2)$.

The form of the renormalized massive OMEs has been derived in \cite{Bierenbaum:2009mv,Klein:2009ig}. In \cite{Blumlein:2021xlc} we have collected the explicit form of the renormalized OMEs in the Larin scheme, in terms of harmonic sums \cite{Vermaseren:1998uu,Blumlein:1998if,Blumlein:2003gb,Blumlein:2009fz}. These formulas, which we do not repeat here due to their length, have been collected using the Mathematica packages {\tt HarmonicSums} and {\tt Sigma}.

The explicit results on the logarithmic corrections are reported in Appendix \ref{sec:polOMEs} and Appendix~\ref{sec:polWCs}.

\clearpage
\thispagestyle{empty}
~

\clearpage
\section[\texorpdfstring{N$^3$LO scheme-invariant evolution of the non-singlet structure functions $F_2^{NS}$ and $g_1^{NS}$}{N³LO scheme-invariant evolution of the non-singlet structure functions F₂,NS and g₁,NS}]%
{N$^3$LO scheme-invariant evolution of the non-singlet structure functions $F_2^{NS}$ and $g_1^{NS}$}
\label{sec:schemeinv}

The measurement of DIS structure functions provides an ideal framework for the measurement of the strong coupling constant $a_s(M_Z)$, which is known today with a precision of $\mathcal O(1\%)$. Fits of the experimental data are typically performed by parametrizing the non-perturbative parton distribution functions through an appropriate functional form with free parameters, and by building the theoretical prediction for the structure functions through a convolution with the DIS Wilson coefficients. A global fit then delivers the parameters and the measurement of $a_s(M_Z)$ by an error minimization procedure, along with the respective uncertainties, see e.g.\ \cite{Blumlein:2006be} for a study of $F_2$ and \cite{Blumlein:2010rn} for a study of $g_1$.

In the non-singlet case, it is also possible to directly compare the structure functions at two virtualities, considering an experimentally determined input at a starting scale $Q_0^2$ and fitting data at $Q^2>Q_0^2$. In the massless case, the relationship between the structure function at two different virtualities only depends on one parameter, namely $a_s$, and is scheme-invariant, i.e.\ independent on the factorization scheme which defines the factorization of the structure function into the Wilson coefficients and the parton distribution functions \cite{Furmanski:1981cw,Blumlein:2000wh}, see also \cite{Blumlein:2005rq} for the explicit treatment of the singlet case. In principle, this makes it possible to perform a one-parameter fit for $a_s$, potentially reducing the uncertainty on its measurement.

This type of framework could be extended by considering massive quarks, by importing precision measurements of the masses $m_b$, $m_c$ from other sources. In \cite{Blumlein:2021lmf}, we completed the formalism of \cite{Blumlein:2000wh} by also considering the asymptotic heavy Wilson coefficients, and perform a numerical study of their effects. An implementation of the formalism has been developed in a numerical code which performs the evolution in $N$-space and obtains the structure functions in momentum fraction space by performing the inverse Mellin transform through a numerical integration in the complex $N$-plane. The evolution of the PDFs in $N$-space has been considered before in \cite{Vogt:2004ns}; the analyses of \cite{Blumlein:2006be,Blumlein:2010rn} have also been performed in $N$-space.

In an analysis on experimental data, it would be important to apply the cuts $W^2>15\text{~GeV}^2$, $Q^2>10\text{~GeV}^2$, with $W$ the invariant mass of the hadrons in the final state, in order to exclude higher-twist effects from the experimental sample and ensure that the description given by the asymptotic Wilson coefficients is sufficiently accurate.

\subsection{Flavour decomposition}
The non-singlet flavour distributions can be written as \cite{Vogt:2004ns}
\begin{equation}
	v_{k^2-1}^\pm = \sum_{l=1}^k (q_l\pm \bar q_l) - k(q_k\pm \bar q_k) ,
\end{equation} 
with $q_i$ the quark distributions. For three light flavours, they are
\begin{eqnarray}
	v_0^\pm &=& 0 \\
	v_3^\pm &=& (u\pm\bar u) - (d\pm\bar d) \\
	v_8^\pm &=& (u\pm\bar u) + (d\pm\bar d) - 2(s\pm\bar s)
\end{eqnarray}
and in general
\begin{eqnarray}
	q_i+\bar q_i &=& \frac{1}{N_F} \Sigma - \frac{1}{i}v_{i^2-1}^+ + \sum_{l=i+1}^{N_F} \frac{1}{l(l-1)} v_{l^2-1}^+, \\
	\Sigma &=& \sum_{l=1}^{N_F}(q_l+\bar q_l) ,
\end{eqnarray}
and the nucleon structure functions are given by
\begin{eqnarray}
	F_2^p &=& x \Big[ \frac{2}{9} \Sigma + \frac{1}{6} v_3^+ + \frac{1}{18} v_8^+ \Big] \\
	F_2^d &=& \frac{1}{2} [ F_2^p + P_2^n ] = x \Big[ \frac{2}{9} \Sigma + \frac{1}{18} v_8^+ \Big]
\end{eqnarray}
with a similar relation for $g_1^{p,d}$. A projection on the singlet distribution would require charged current structure functions \cite{Blumlein:2012bf}
\begin{equation}
	\frac{1}{2}[W_2^{p,+} + W_2^{p,-}] = x \Sigma
\end{equation}
with the index $\pm$ indicating the exchange of a $W^+$ or $W^-$ boson. We will consider the flavour non-singlet combinations
\begin{eqnarray}
	F_2^{NS}   &=& F_2^p - F_2^d = \frac{1}{6} x C_q^{NS,+} \otimes v_3^+ , \\
	x g_1^{NS} &=& \frac{1}{6} x \Delta C_q^{NS,+} \otimes \Delta v_3^+ .
\end{eqnarray}

\subsection{The non-singlet evolution}
We derive a relation, valid in $N$-space,
\begin{equation}
	F^{NS}(Q^2) = E_{NS}(Q^2,Q_0^2) F_{NS}(Q_0^2),
\end{equation}
where $F^{NS}$ refers to $F_2^{NS}$ or to $g_1^{NS}$, solving the evolution equation
\begin{equation}
\label{eq:nsEvolution}
\frac{d}{dt}\ln\left[F^{\NS}(Q^{2})\right]=\frac{d}{dt}\ln\left[C^{\NS}(Q^{2})\right]+\frac{d}{dt}\ln\left[q^{\NS}(Q^{2})\right] .
\end{equation}
The Wilson coefficient is given by
\begin{equation}
\label{eq:WilCoeff}
C(Q^{2})=1+\sum_{k=1}^{\infty}a_{s}^{k}(Q^{2}) C_k, \quad C_k = c_k + h_k(L_c,L_b).
\end{equation}
Here $c_k$ denote the expansion coefficients of the massless Wilson coefficients and $h_k$ of the massive Wilson coefficient, with 
\begin{eqnarray}
	L_c=\ln\frac{Q^2}{m_c^2}, \quad L_b=\ln\frac{Q^2}{m_b^2} .
\end{eqnarray}
In the non-singlet case the heavy flavor corrections contribute from $\mathcal{O}(a_s^2)$ onward. One has
\begin{eqnarray}
h_{2}&=&\hat{h}_{2}(L_{c})+\hat{h}_{2}(L_{b})
\\
h_{3}&=&\hat{h}_{3}(L_{c})+\hat{h}_{3}(L_{b})+\hat{\hat{h}}_{3}(L_{c},L_{b})
\end{eqnarray}
where $\hat{h}$ denote the single mass and $\hat{\hat{h}}$ the double mass contributions.

One may rewrite the differential operator
\begin{equation}
	\frac{d}{d \ln(Q^2)} = \frac{d a_s(Q^2)}{d\ln(Q^2)} \frac{d}{d a_s(Q^2)}
\end{equation}
with
\begin{equation}
	\frac{d a_s}{d\ln(Q^2)} = -\sum_{k=0}^\infty \beta_k a_s^{k+2} .
\label{eq:AS}
\end{equation}

The evolution equation for the non-singlet quark density is 
\begin{equation}
\label{eq:AP}
\frac{d}{dt}\ln\left[q^{\NS}(Q^{2})\right]= \sum_{k=0}^\infty P_{k}a^{k+1}(Q^{2}) ,
\end{equation}
where $\beta_k$ are the expansion coefficients of the QCD-$\beta$ function and $P_{k,qq}^{NS}\equiv P_k^{NS}$ are the splitting functions. The anomalous dimensions are related to the splitting functions by\footnote{Our normalizations are such that a factor of two has to be applied to those given in \cite{Moch:2004pa,Vogt:2004mw}}
\begin{equation}
\gamma_{ij,(k)}(N)=-\int_{0}^{1}dxx^{N-1}P_{ij,(k)}(x) ,
\end{equation}
where $\gamma_{ij,(k)}(N)$ are the expansion coefficients of the non-singlet anomalous dimensions.

One obtains to $N^3LO$
\begin{eqnarray}
E_{NS}(Q^2,Q_0^2) & =&\left(\frac{a}{ a_0 }\right)^{-\frac{ P_0 }{2\beta_0 }}
\Biggl\{
1
+\frac{a- a_0 }{2\beta_0 ^2} \biggl\{
\Bigl[1 + a^2  C_2(Q^2) - a_0 ^2  C_2(Q_0^2) \Bigr] \bigl(2 \beta_0 ^2  C_1 -\beta_0   P_1 +\beta_1  P_0 \bigr)
\nonumber\\&&    
-\frac{\bigl(a^2- a_0 ^2\bigr)}{4 \beta_0 ^3} \bigl(2\beta_0 ^2  C_1 -\beta_0   P_1 +\beta_1   P_0 \bigr) \Bigl[2\beta_0 ^3 C_1 ^2+\beta_0 ^2 P_2 - \beta_0 \beta_1 P_1 + \bigl(\beta_1 ^2-\beta_0  \beta_2 \bigr) P_0 \Bigr]
\nonumber\\&&
+\frac{\bigl(a^2+a  a_0 + a_0 ^2\bigr)}{3 \beta_0 ^2} \Bigl[2\beta_0 ^4  C_1 ^3-\beta_0 ^3  P_3 +\beta_0 ^2 \beta_1 P_2 + \bigl(\beta_0 ^2 \beta_2 -\beta_0  \beta_1 ^2\bigr) P_1 
\nonumber\\&&
+ \bigl(\beta_0 ^2 \beta_3 -2 \beta_0  \beta_1  \beta_2 +\beta_1 ^3\bigr) P_0 \Bigr]
+\frac{a- a_0 }{4 \beta_0 ^2} \bigl(2\beta_0 ^2  C_1 -\beta_0   P_1 +\beta_1   P_0 \bigr)^2
\nonumber\\&&
+\frac{(a- a_0 )^2}{24 \beta_0 ^4} \bigl(2\beta_0 ^2  C_1 -\beta_0   P_1 +\beta_1   P_0 \bigr)^3
-\frac{a+ a_0 }{2 \beta_0 } \Bigl[2\beta_0 ^3  C_1 ^2+\beta_0 ^2  P_2 -\beta_0  \beta_1 P_1 
\nonumber\\&&
+ P_0  \bigl(\beta_1 ^2-\beta_0  \beta_2 \bigr)
 \Bigr]
\biggr\}
+a^2  C_2(Q^2) - a_0 ^2  C_2(Q_0^2) 
- C_1  \Bigl[a^3  C_2(Q^2) - a_0 ^3  C_2(Q_0^2) \Bigr]
\nonumber\\&&
+ a^3 C_3(Q^2) - a_0 ^3  C_3(Q_0^2) 
\Biggr\} .
\label{eq:ENS}
\end{eqnarray}
Here we used the notation $a=a_s(Q^2)$, $a_0=a_s(Q_0^2)$ and $P_i=P_{qq,(i)}^{NS}$, considered in $N$-space.

The heavy quark contributions to the Wilson coefficients are given by \cite{Blumlein:2016xcy,Ablinger:2014vwa,Behring:2015zaa,Bierenbaum:2009mv}
\begin{eqnarray}
\label{H2}
\hat{h}_2^{(Q)} &=& - \frac{\beta_{0,Q}}{4} P_{qq,(0)} \ln^2\left(\frac{Q^2}{m^2}\right)
                   + \frac{1}{2} \hat{P}_{qq,(1)}^{\rm NS} \ln\left(\frac{Q^2}{m^2}\right)
                   + a_{qq}^{(2), \rm NS} + \frac{\beta_{0,Q}}{4} \zeta_2 P_{qq,(0)} + \hat{C_q}^{(2),\rm NS}
\\
\label{H3}
\hat{h}_3^{(Q)} &=& -\frac{1}{6} P_{qq,(0)} \beta_{0,Q} \left(\beta_0 + 2 \beta_{0,Q}\right) 
\ln^3\left(\frac{Q^2}{m^2}\right)
+ \frac{1}{4} \Biggl[-2 P_{qq,(1)}^{\rm NS} \beta_{0,Q} + 2 \hat{P}_{qq,(1)}^{\rm NS} \left(\beta_0 + 
\beta_{0,Q}\right) 
\nonumber\\ && 
- \beta_{1,Q} P_{qq,(0)} \Biggr] \ln^2\left(\frac{Q^2}{m^2}\right)
- \frac{1}{2} \Biggl[- \hat{P}_{qq,(2)}^{\rm NS} - \left(4 a_{qq,Q}^{(2),\rm NS} + \zeta_2 \beta_{0,Q} 
P_{qq,(0)}\right)  \left(\beta_0 +  \beta_{0,Q}\right) 
\nonumber\\ && 
- P_{qq,(0)} \beta_{1,Q}^{(1)}\Biggr] 
\ln\left(\frac{Q^2}{m^2}\right)
+ 4 \bar{a}_{qq,Q}^{(2),\rm NS}(\beta_0+\beta_{0,Q}) + P_{qq,(0)} \beta_{1,Q}^{(2)} 
+ \frac{1}{6} P_{qq,(0)} \beta_0 \beta_{0,Q} \zeta_3 
\nonumber\\ && 
+ \frac{1}{4} P_{qq,(1)}^{\rm NS} \beta_{0,Q} \zeta_2 
- 2 \delta m_1^{(1)} \beta_{0,Q} P_{qq,(0)} - \delta m_1^{(0)} \hat{P}_{qq,(1)}^{\rm NS} + 2 \delta m_1^{(-1)} 
a_{qq,Q}^{(2),\rm NS} + a_{qq,Q}^{(3),\rm NS} 
\nonumber\\ && + \Biggl[- \frac{\beta_{0,Q}}{4} P_{qq,(0)} \ln^2\left(\frac{Q^2}{m^2}\right)
                   + \frac{1}{2} \hat{P}_{qq,(1)}^{\rm NS} \ln\left(\frac{Q^2}{m^2}\right)
                   + a_{qq}^{(2), \rm NS} + \frac{\beta_{0,Q}}{4} \zeta_2 P_{qq,(0)}\Biggr] C_q^{(1),\rm NS}
\nonumber\\ &&
                   + \hat{C}_q^{(3),\rm NS}.  
\end{eqnarray}
and the two-mass contribution by \cite{Ablinger:2017err}
\begin{eqnarray}
\label{H23}
\hat{\hat{h}}_3^{\rm NS} &=& P_{qq}^{(0)} \beta_{0,Q}^2\left[\frac{2}{3} \left(L_c^3 + L_b^3\right) + \frac{1}{2} 
\left(L_c^2 L_b +L_c L_b^2 \right)\right] - \beta_{0,Q} \hat{P}_{qq}^{(1),\rm NS} \left(L_c^2 + L_b^2\right)
-\left[4 a_{qq,Q}^{(2),\rm NS} \beta_{0,Q} \right. 
\nonumber\\ && \left.
- \frac{1}{2} \beta_{0,Q}^2 P_{qq}^{(0)} \zeta_2\right] (L_c + L_b)
+ 8 \bar{a}_{qq,Q}^{(2),\rm NS} \beta_{0,Q} + \tilde{a}^{(3),\rm NS}_{qq,Q}(m_c,m_b,Q^2).
\end{eqnarray}
The two-mass term is the same in the unpolarized and polarized case. We employed the definition
\begin{eqnarray}
\hat{f}(x,N_F) = f(x,N_F+1) - f(x,N_F).
\end{eqnarray}

The perturbative solution for $a_s(Q^2)$ is given in the $\overline{\sf MS}$ scheme by \cite{Chetyrkin:1997sg}
\begin{eqnarray}
a_s(Q^2) &=& \frac{1}{\beta_0 L} 
- \frac{\beta_1}{\beta_0^3 L^2} \ln(L)
+ \frac{1}{\beta_0^3 L^3} \Biggl[\frac{\beta_1^2}{\beta_0^2 }(\ln^2(L) - \ln(L) -1) + \frac{\beta_2}{\beta_0}\Biggr]
\nonumber\\ &&
+ \frac{1}{\beta_0^4 L^4} \Biggl[\frac{\beta_1^3}{\beta_0^3 }\Biggl(-\ln^3(L) + \frac{5}{2} \ln^2(L) + 2 \ln(L) - 
\frac{1}{2}\Biggr) - 3 \frac{\beta_1 \beta_2}{\beta_0^2} \ln(L) + \frac{\beta_3}{2 \beta_0}\Biggr],
\label{eq:asSol}
\end{eqnarray}
with $L = \ln(Q^2/\Lambda_{\rm QCD}^2)$. Here the integration constant for solving \eqref{eq:AS} is chosen by 
$(\beta_1/\beta_0^2) \ln(\beta_0)$ \cite{Bardeen:1978yd}. The expansion coefficients of the 
$\beta$-function to N$^3$LO were calculated in \cite{Gross:1973id,Politzer:1973fx,Caswell:1974gg,Jones:1974mm,Tarasov:1980au,Larin:1993tp,vanRitbergen:1997va,Czakon:2004bu}.
The flavor matching conditions were given in \cite{Chetyrkin:1997sg}. 
The expansion coefficients of the renormalized mass were given in  \cite{Gray:1990yh,Broadhurst:1991fy}.
The constant and $O(\ep)$ parts of the massive unrenormalized operator matrix elements at $O(a_s^k)$
are denoted by $a_{ij}^{(k)}$ and $\bar{a}_{ij}^{(k)}$, respectively, cf.~\cite{Buza:1995ie,Bierenbaum:2007qe,Bierenbaum:2008yu,POL21}.

In the numerical evaluation shown below, an approximate form is used for the three-loop massless Wilson coefficients. For all other Wilson coefficients, the analytic Mellin-space representations are used. After reduction to a basis of independent harmonic sums \cite{Blumlein:2003gb}, the objects depend on 32 harmonic sums \cite{Vermaseren:1998uu,Blumlein:1998if} up to weight 6; with weight-6 sums appearing only in the 3-loop Wilson coefficient.

The harmonic sums in $N$ space are calculated in the complex plane after representing them using the Mellin transforms of harmonic polylogarithms, and computing their asymptotic expansions \cite{Blumlein:1998if,Blumlein:2009fz,Blumlein:2009ta}. Together with exact step relations in $N$, an accurate numerical evaluation of the harmonic sums becomes possible in the complex plane. The analytic continuation of harmonic sums needed for the anomalous dimensions up to 3 loops has first been performed in \cite{Blumlein:2005jg}.

In the case of $F_2^{NS}$, the relevant splitting functions are $P_{qq,(k)}^{NS,+}$, whereas for $g_1^{NS}$ they are $P_{qq,(k)}^{NS,-}$. The massless Wilson coefficients have been calculated in \cite{Bardeen:1978yd,vanNeerven:1991nn,Zijlstra:1991qc,Zijlstra:1992qd,Vermaseren:2005qc} and \cite{Vogelsang:1990ug,Wandzura:1977qf} respectively. For the four-loop splitting functions, which are not currently known in analytic form, we employ in numerical illustrations below the Pad\'e approximant 
\begin{eqnarray}
\label{eq:PADE}
P_{qq}^{3, \pm, \rm NS}(N) \approx \frac{P_{qq}^{2, \pm, \rm NS}(N)^2}{P_{qq}^{1, \pm, \rm NS}(N)} .
\end{eqnarray}

Low moments of the four-loop splitting functions have been calculated in \cite{Baikov:2006ai,Baikov:2015tea,Velizhanin:2014fua,Ruijl:2016pkm,Davies:2016jie,Moch:2017uml}. A comparison of these exact moments to the Pad\'e approximant is shown in Table \ref{TAB1}.

A previous analysis \cite{Blumlein:2006be} has showed that a $100\%$ error on $P_{qq,(3)}^{NS}$ would determine an error of 2 MeV on $\Lambda_{QCD}$, well below the experimental error currently of $\delta\Lambda_{QCD}=26\text{~MeV}$.

The leading small-$x$ terms of $P_3^{NS,+}$ and $P_2^{NS,-}$ have been studied in \cite{Kirschner:1983di,Blumlein:1995jp,Bartels:1995iu}; the leading large-$N_F$ behaviour of the splitting functions has also been given in \cite{Gracey:1994nn}. 

\begin{table}[H]\centering
\renewcommand{\arraystretch}{1.3}
\begin{tabular}{|r|l|r|l|}
\hline
\multicolumn{1}{|c}{$N$} & 
\multicolumn{1}{|c}{$\delta \gamma^{+,\rm NS}$} & 
\multicolumn{1}{|c}{$N$} & 
\multicolumn{1}{|c|}{$\delta \gamma^{-,\rm NS}$} \\
\hline
 2 & 0.208822541 &  1 &  0.0          \\
 4 & 0.123728742 &  3 &  0.147102092  \\
 6 & 0.087155544 &  5 &  0.101634935  \\
 8 & 0.064949195 &  7 &  0.074593595  \\
10 & 0.049680399 &  9 &  0.056598595  \\
12 & 0.038394815 & 11 &  0.043633919  \\
14 & 0.029638565 & 13 &  0.033767853  \\
16 & 0.022602035 & 15 &  0.025956941  \\
\hline
\end{tabular}
\caption[]{
\label{TAB1}
\sf The relative error comparing the exact moments of the four--loop anomalous dimensions, $\gamma^{(3),\pm, \rm 
NS}$, with the Pad\'e approximation (\ref{eq:PADE}).}
\renewcommand{\arraystretch}{1.0}
\end{table}

\subsection{Numerical results}

In our numerical illustration we employ the values of the charm and bottom quarks of $m_c = 1.59\text{~GeV}$ \cite{Alekhin:2012vu} and $m_b = 4.78\text{~GeV}$ \cite{Zyla:2020zbs}.

The input structure function in the unpolarized case is built from the non-singlet parton distribution \cite{Blumlein:2006be}
\begin{eqnarray}
xq^{\rm NS}(x,Q_0^2) &=& 
\frac{1}{3} \biggl[
        0.262 \ x^{0.298} (1-x)^{4.032} (1+6.042 \sqrt{x} +35.49 x)
\nonumber\\&&        
           ~~~~~-1.085 \ x^{0.5} (1-x)^{5.921} (1-3.618 \sqrt{x} +16.41 x) \biggr] 
\end{eqnarray} 
at $Q_0^2=4\text{~GeV}^2$; for the polarized case we use a fit of the structure function of \cite{Blumlein:2010rn} at $Q_0^2 = 10~\text{GeV}^2$.

We also employ for the purpose of illustration the quantity
\begin{eqnarray}
F_2^{\rm h}(N,Q^2) = \left[E_{\rm NS} - \left. E_{\rm NS}\right|_{\rm h=0}\right] F_2(N,Q_0^2).
\end{eqnarray}

\begin{figure}[H]
        \centering
        \includegraphics[width=0.49\textwidth]{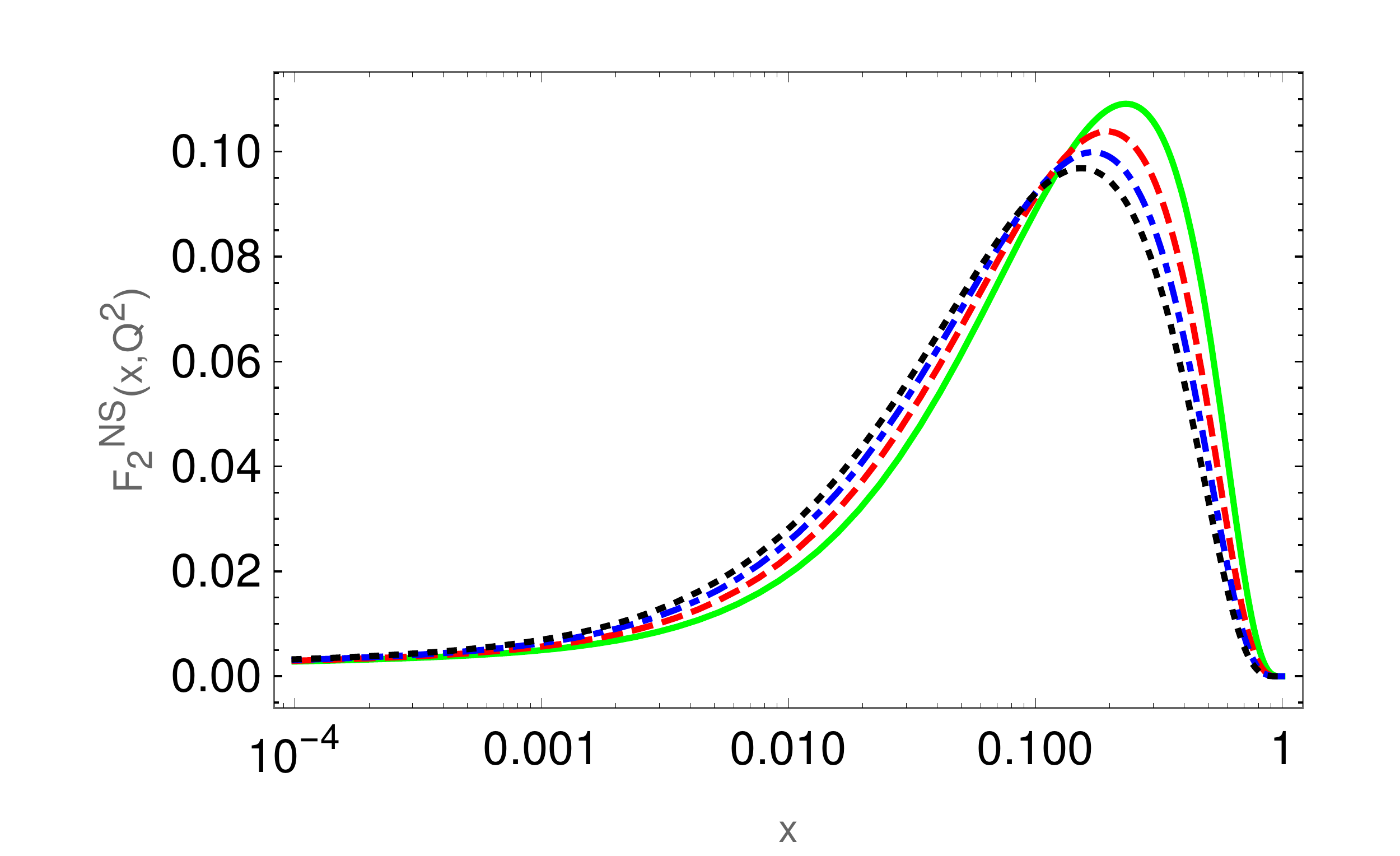}
        \includegraphics[width=0.49\textwidth]{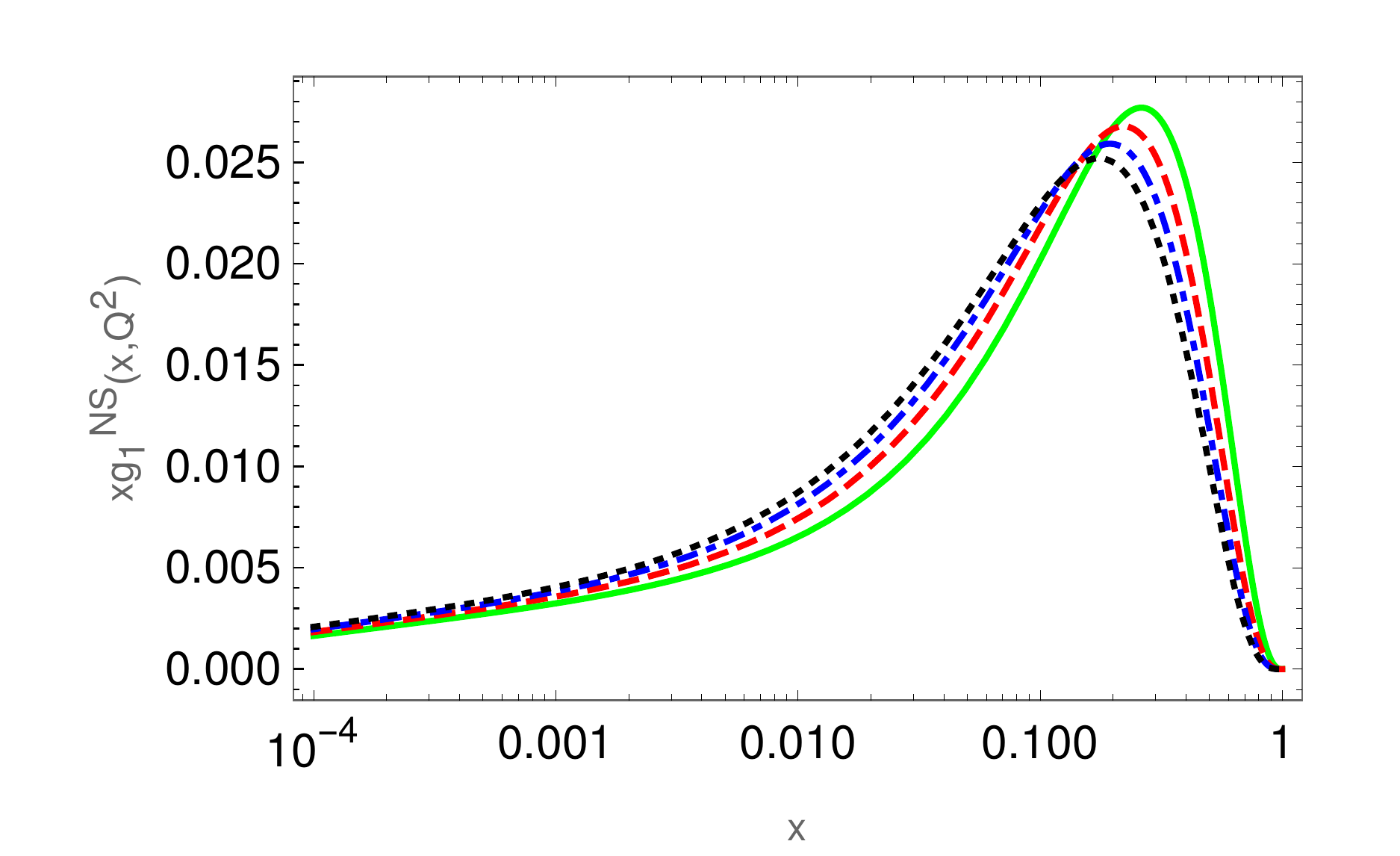}
        \caption{\sf Left:~The structure function $F_2^{\rm NS}$ at N$^3$LO. Right:~The structure function 
$xg_1^{\rm NS}$ at N$^3$LO. Full lines: $Q^2 = 10~\GeV^2$; dashed lines: $100~\GeV^2$;
dash-dotted lines: $1000~\GeV^2$; 
dotted lines: $10000~\GeV^2$.}
\label{fig1}
\end{figure}

\begin{figure}[H]
        \centering
        \includegraphics[width=0.49\textwidth]{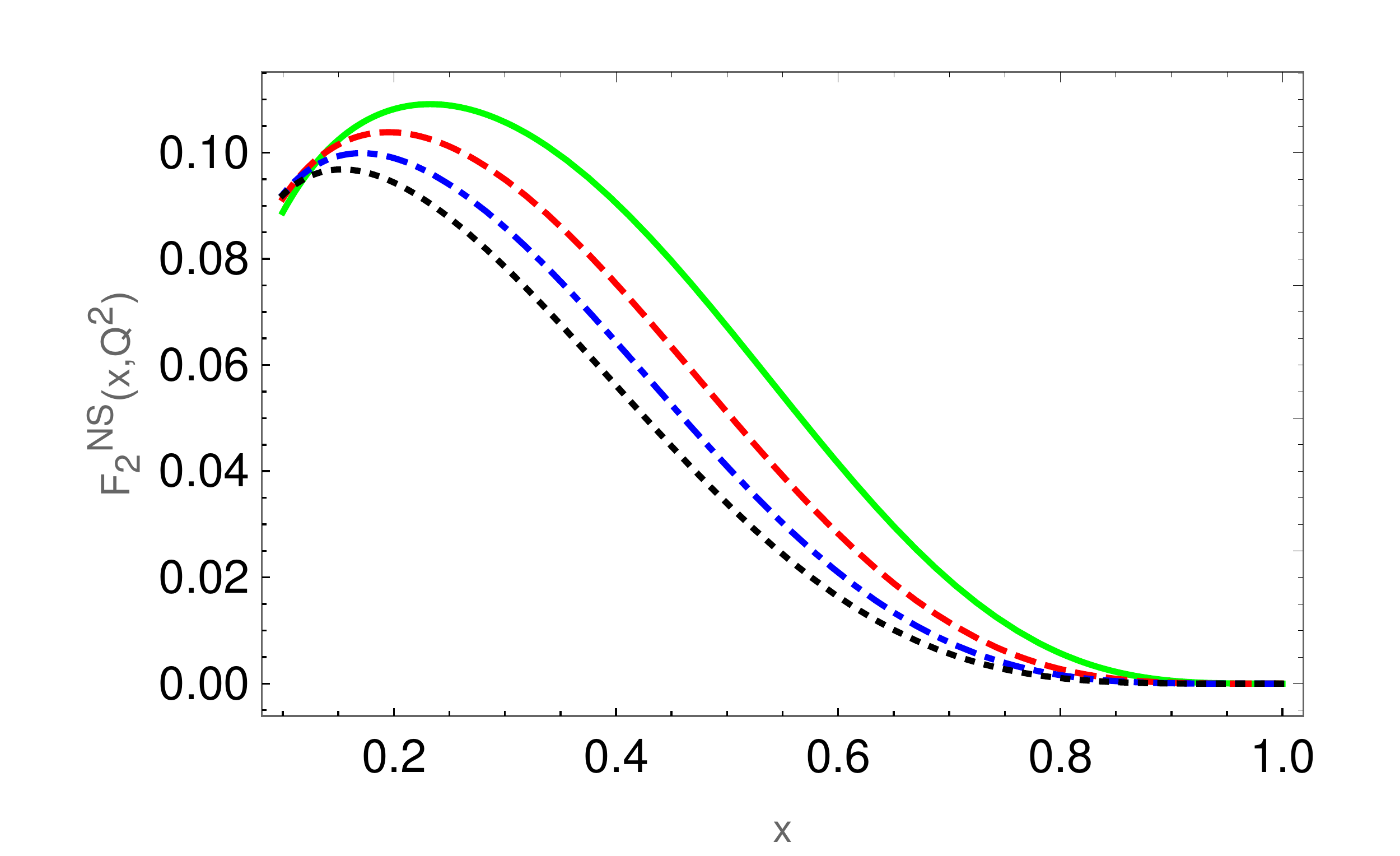}
        \includegraphics[width=0.49\textwidth]{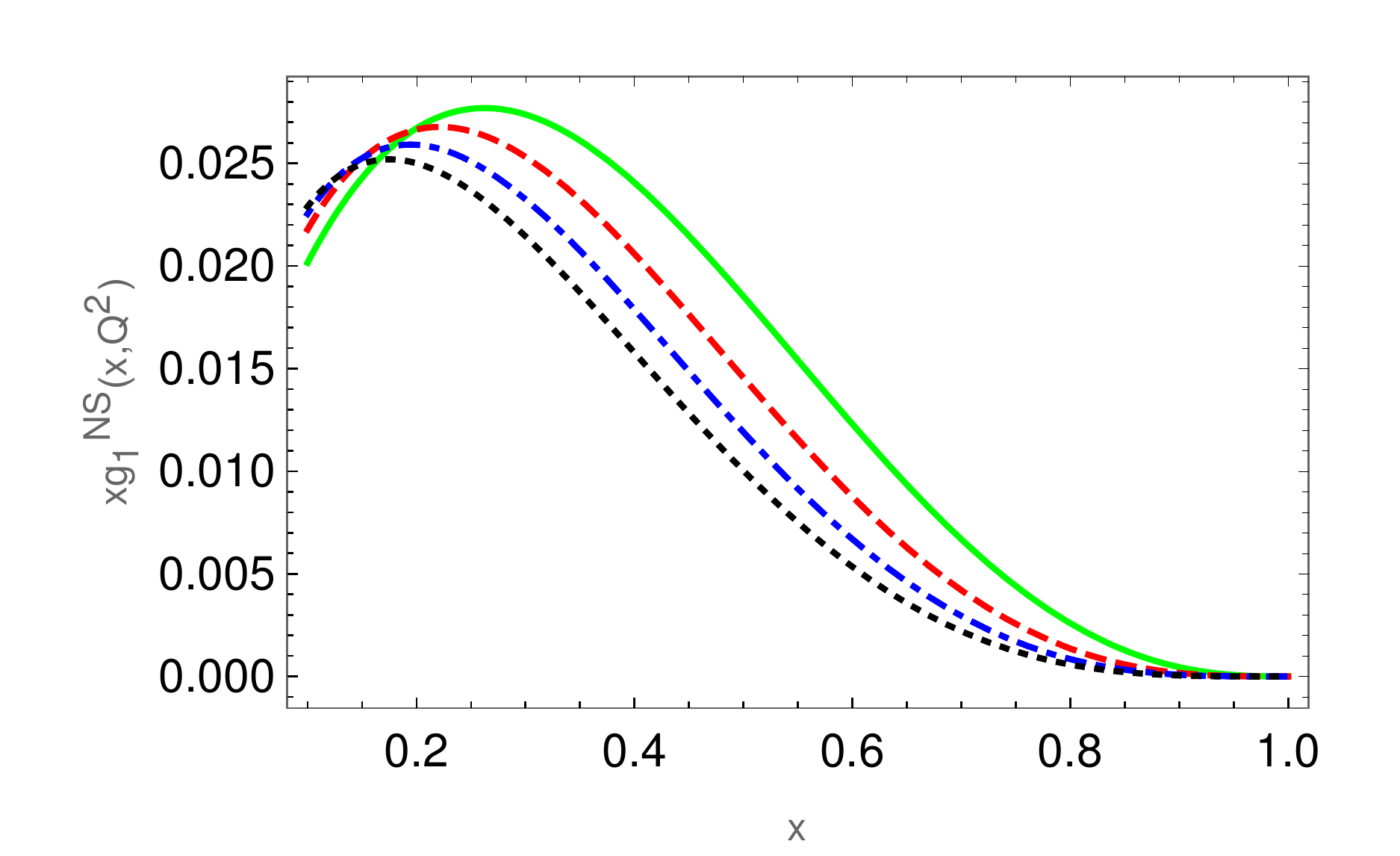}
        \caption{\sf The same as Figure~1 but with expanded large $x$ region.}
\label{fig2}
\end{figure}

\begin{figure}[H]
        \centering
        \includegraphics[width=0.49\textwidth]{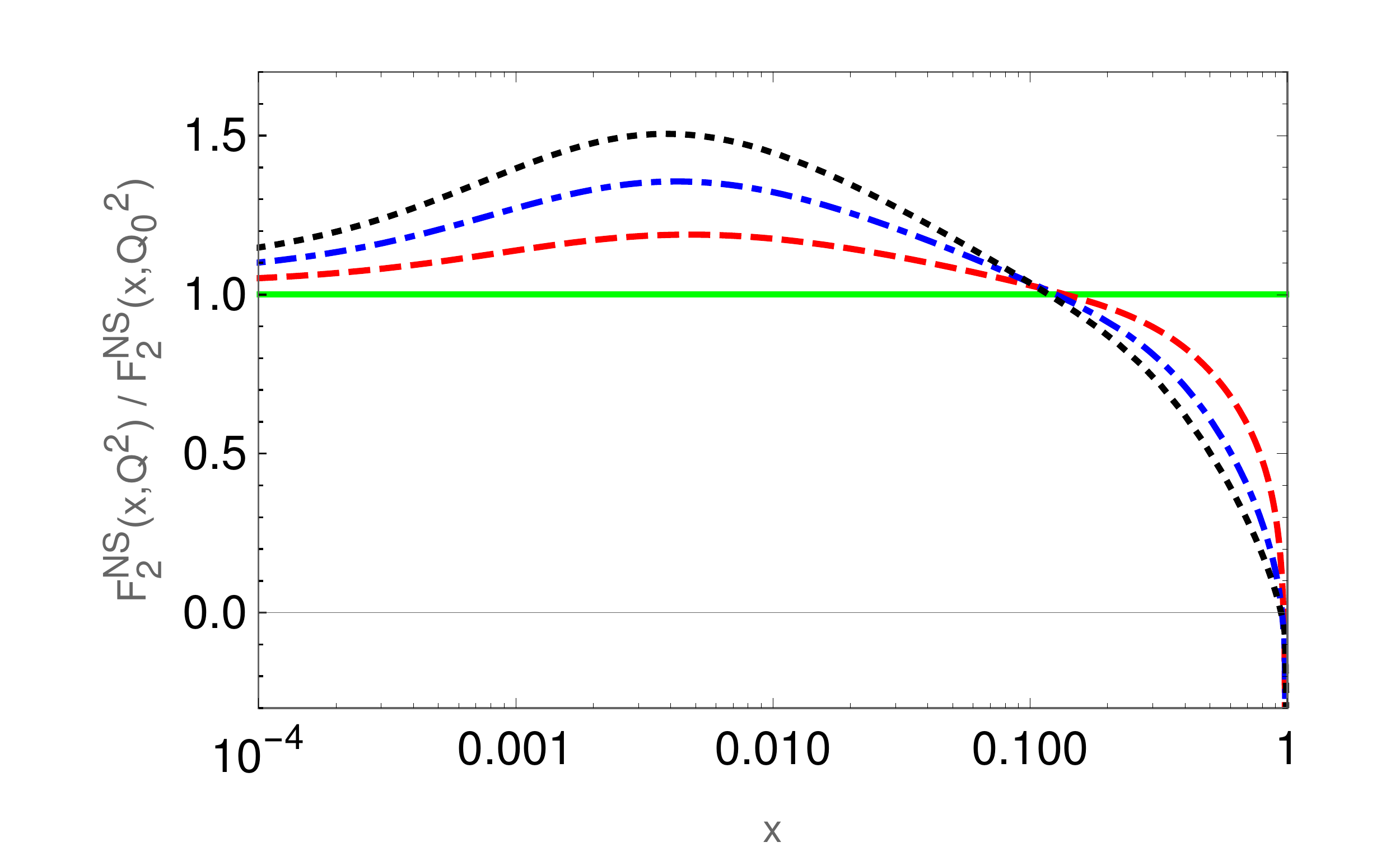}
        \includegraphics[width=0.49\textwidth]{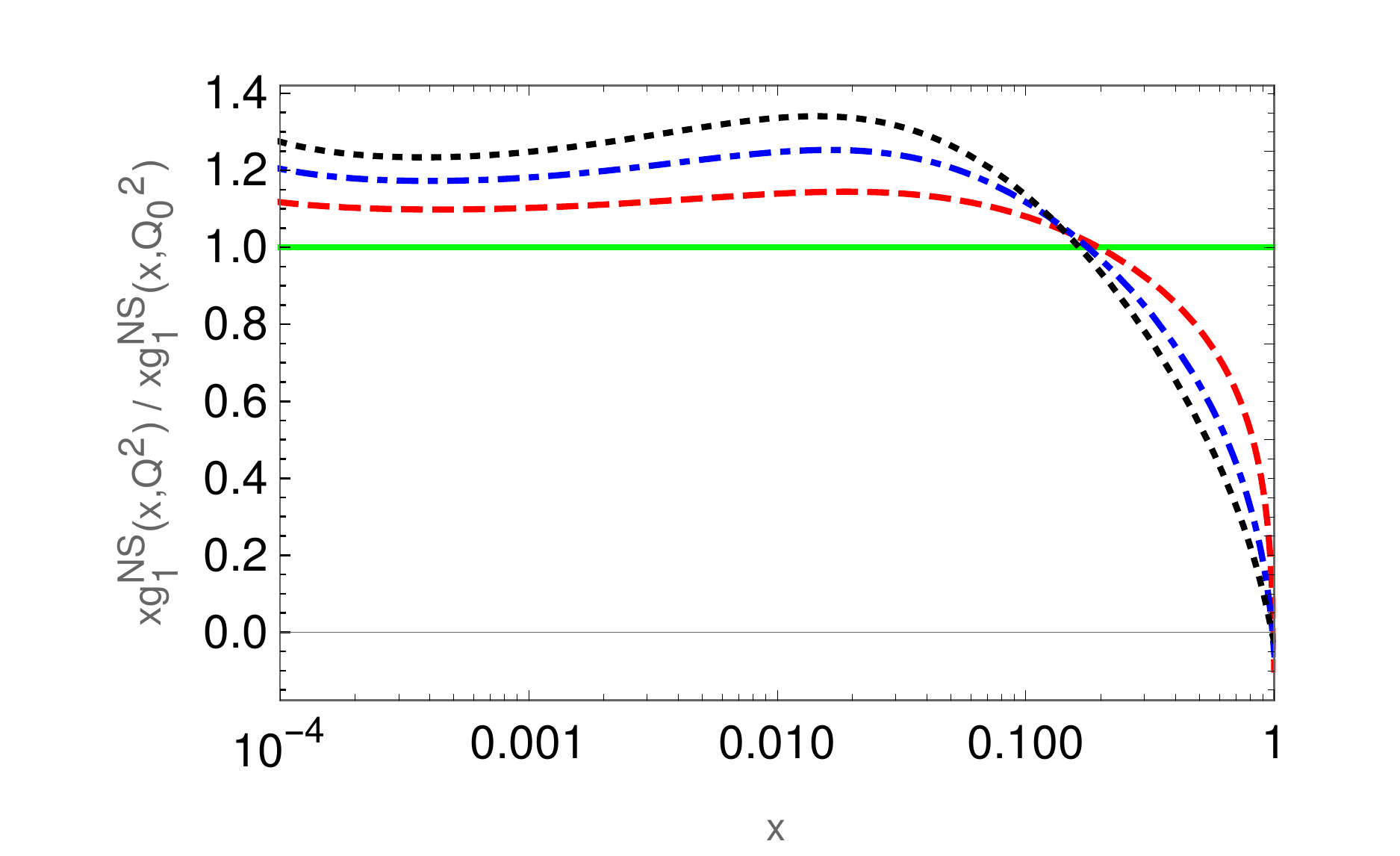}
        \caption{\sf Left:~The relative contribution of $F_2^{\rm NS}$ in the evolution from $Q^2 = 10~\GeV^2$ to $10000~\GeV^2$.
        Right:~The same for the structure function $xg_1^{\rm NS}$.
\label{fig3}}
\end{figure}

\begin{figure}[H]
        \centering
        \includegraphics[width=0.49\textwidth]{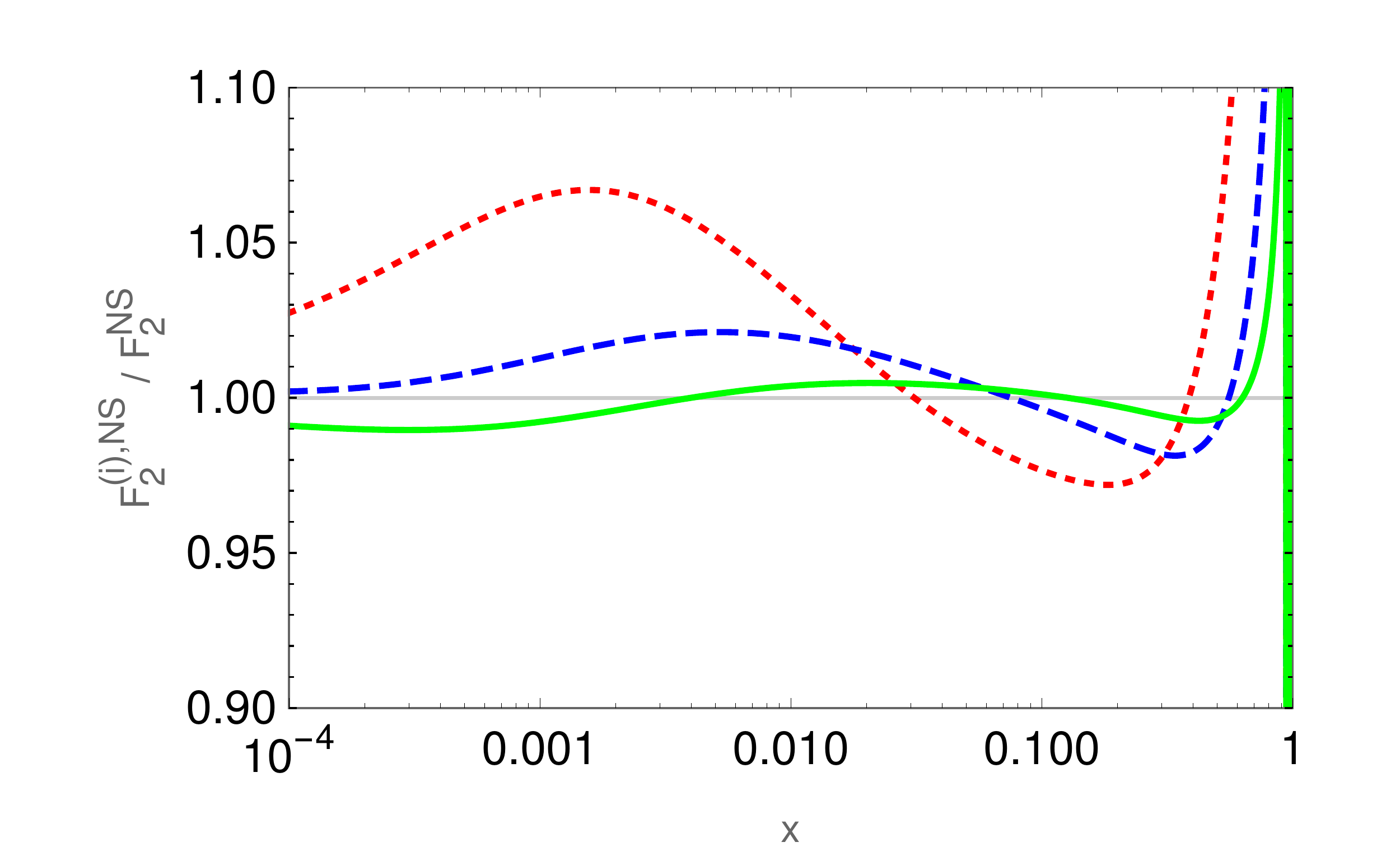}
        \includegraphics[width=0.49\textwidth]{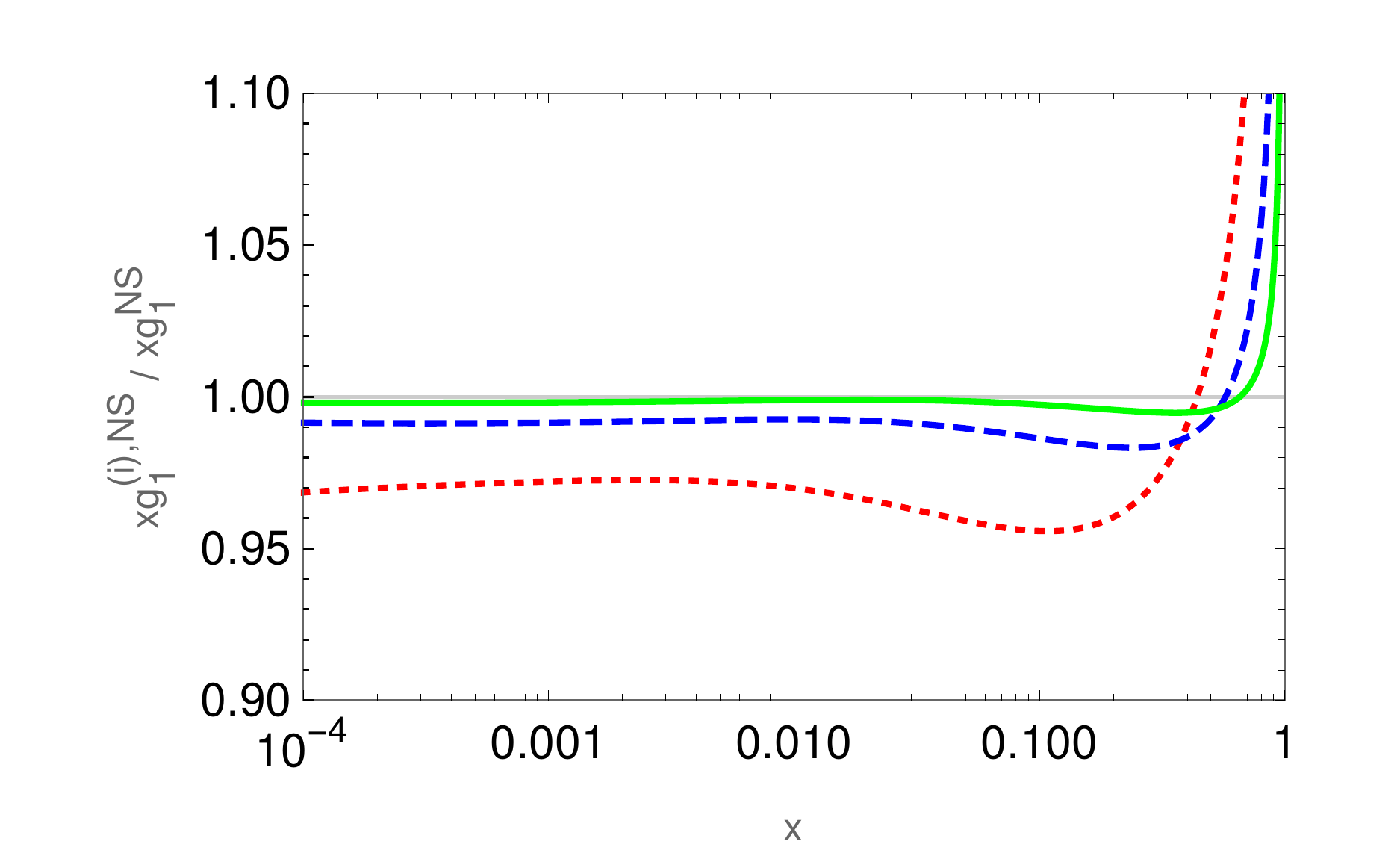}
        \caption{\sf Left:~The relative contributions from LO (dotted lines), NLO (dashed lines) and NNLO (full lines)
to the structure function
$F_2^{\rm NS}$ at N$^3$LO at $Q^2 = 100~\GeV^2$ as an example. Right:~The same for the structure function $xg_1^{\rm NS}$.
\label{fig4}}
\end{figure}

\begin{figure}[H]
        \centering
        \includegraphics[width=0.49\textwidth]{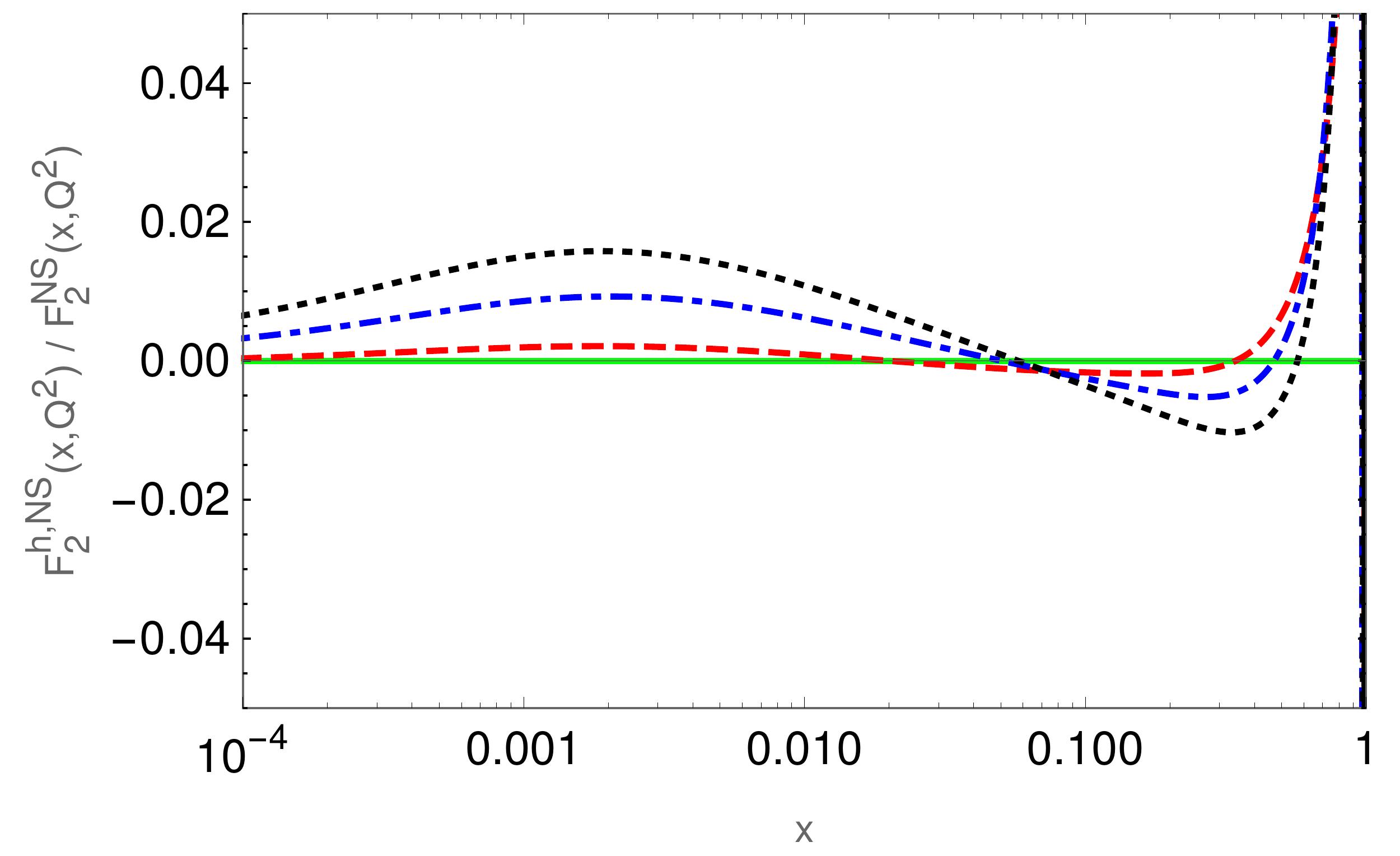}
        \includegraphics[width=0.49\textwidth]{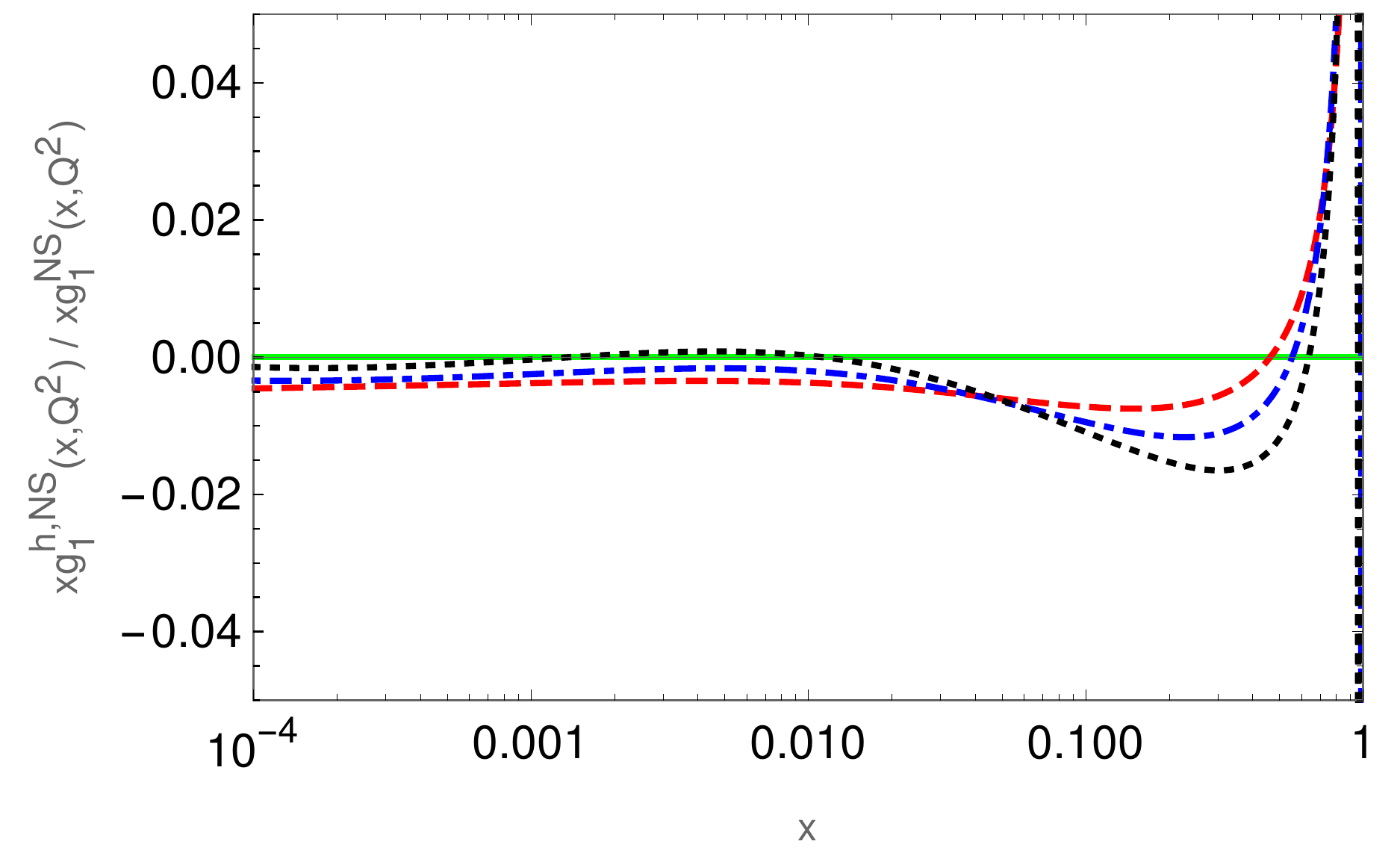}
        \caption{\sf Left:~The relative contribution of the heavy flavor contributions due 
to $c$ and $b$ quarks to the structure function $F_2^{\rm NS}$ at N$^3$LO;
dashed lines: $100~\GeV^2$;
dashed-dotted lines: $1000~\GeV^2$;
dotted lines: $10000~\GeV^2$. Right:~The same
for the structure function $xg_1^{\rm NS}$ at N$^3$LO.} 
\label{fig5}
\end{figure}

\begin{figure}[H]
        \centering
        \includegraphics[width=0.49\textwidth]{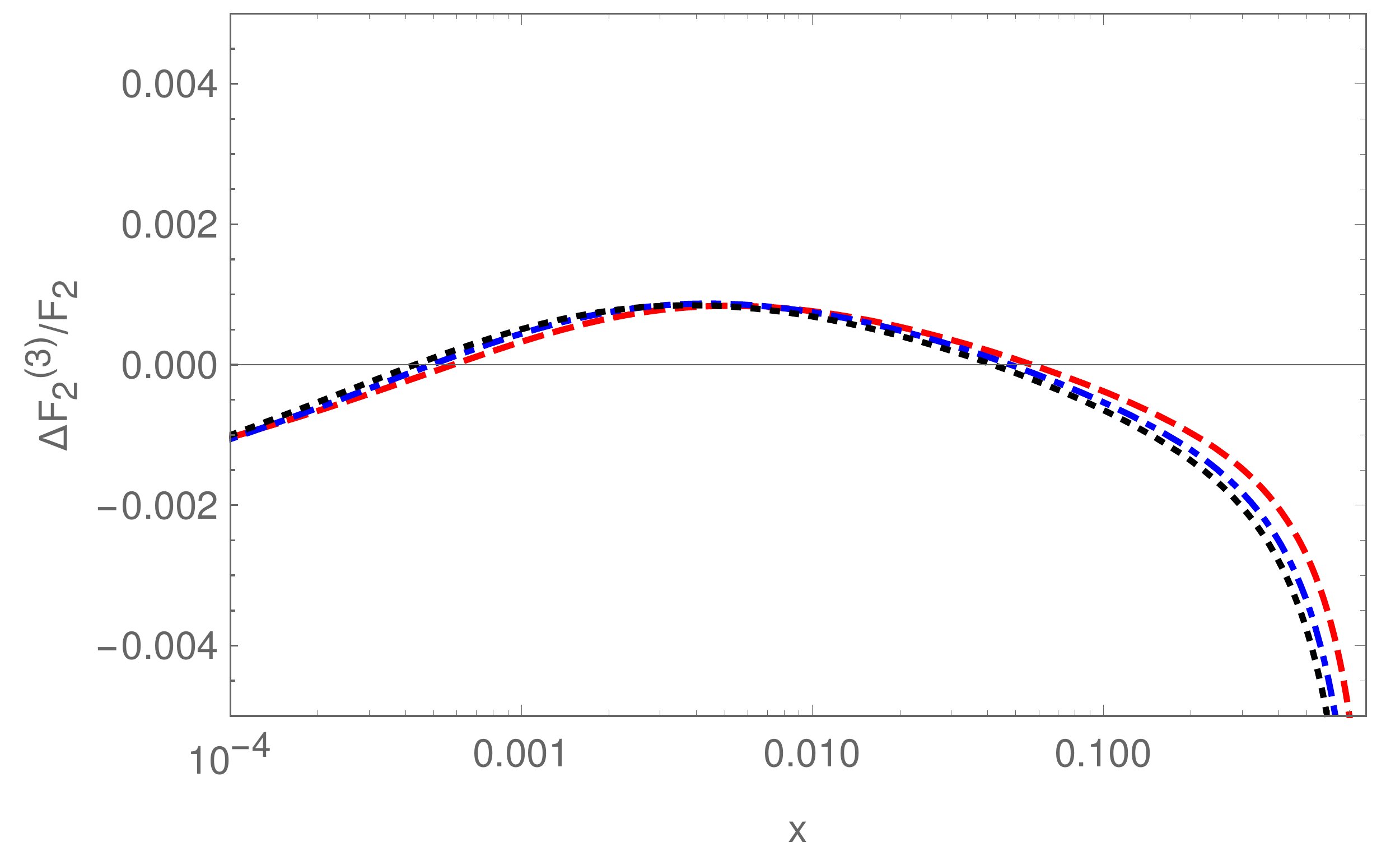}
        \includegraphics[width=0.49\textwidth]{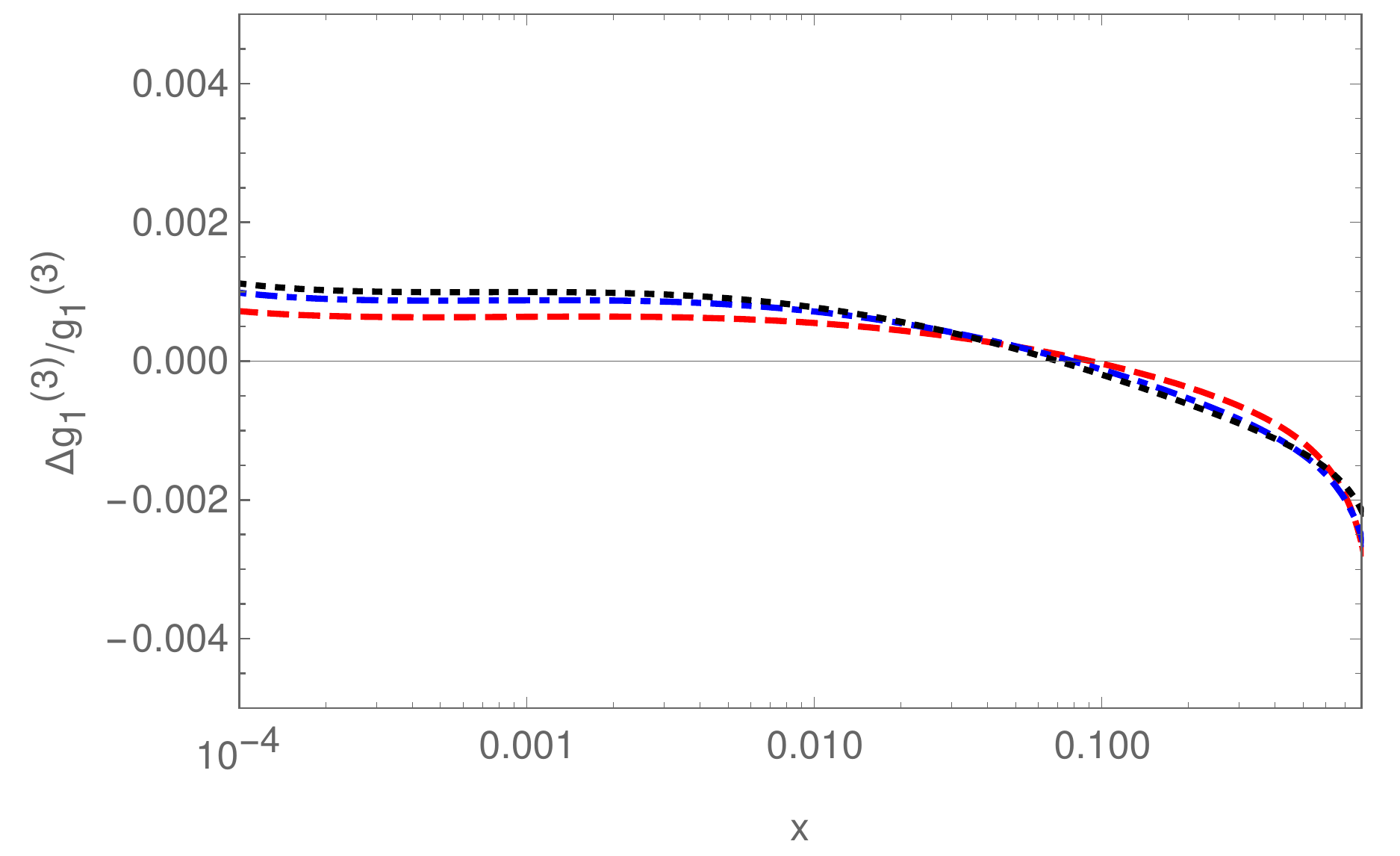}
        \caption{\sf The effect of the variation of $P_{qq}^{3, \pm, \rm NS}$ around the value in Eq.~(\ref{eq:PADE}) by $\pm 
100\%$. Dashed lines: $Q^2 = 100~\GeV^2$, Dash-dotted lines: $Q^2 = 1000~\GeV^2$; Dotted lines: $Q^2 = 10000~\GeV^2$.
Left: $F_2^{\rm NS}$; Right: $xg_1^{\rm NS}$.}
\label{fig6}
\end{figure}

In Fig. \ref{fig1} we show the scheme-invariant evolution of the non-singlet structure functions $F_2^{NS}$ and $xg_1^{NS}$, in the kinematic region $Q^2\in [10, 10^4]\text{~GeV}^2$. In Fig. \ref{fig2} we expand the representation for the region of larger values of $x$. 
In Fig. \ref{fig3} we illustrate the relative effect of the scale evolution in $Q^2$ both for $F_2^{NS}$ and $xg_1^{NS}$ comparing to the starting scale $Q_0^2$.
In Fig.~\ref{fig4} we show the ratio of the results obtained at leading order (LO), next-to-leading order (NLO), and next-to-next-to leading order (NNLO) to the N$^3$LO results at $Q^2 = 100~\GeV^2$.
In Fig.~\ref{fig5} we illustrate the relative size of the heavy flavor parts for the same region in $Q^2$ in the unpolarized and polarized cases.  In the important region $x \leq 0.4$ the heavy flavor corrections reach the size of $\sim 1\%$.
In Fig.~\ref{fig6} we illustrate the effect of the half difference if putting $P_{qq}^{3, \pm, \rm NS} = 2 {P_{qq}^{2, \pm, \rm NS}}^2/ P_{qq}^{2, \pm, \rm NS}$ and $P_{qq}^{3, \pm, \rm NS} = 0$ for both $F_2^{\rm NS}$ and $xg_1^{\rm NS}$. This rescaled correction is in the sub--percent range. Moreover, the impact on $\Lambda_{\rm QCD}$ comes from the slope in $Q^2$ which is seen to be rather small.


\clearpage
\thispagestyle{empty}
~

\clearpage
\section{Hypergeometric functions and differential systems}
\label{sec:hyperg}

Hypergeometric functions are known to occur in the computation of Feynman integrals \cite{Hamberg91,Davydychev:2003mv,Bierenbaum:2007qe,Schlosser:2013hbz,Kalmykov:2020cqz}. In the mathematical literature, hypergeometric functions of various kinds have been studied by a number of authors \cite{HYPKLEIN,HYPBAILEY,SLATER1,APPELL1,APPELL2,KAMPE1,KAMPE2,BORNG,HORN,EXTON72a,EXTON1,EXTON2,SRIKARL,Lauricella:1893,Saran:1954,Saran:1955,ERDELYI}.

A major methodological advance in the calculation of Feynman diagrams was the development of the method of integration by parts \cite{Lagrange:1760,Gauss:1813,Green:1828,Ostrogradski:1831,Chetyrkin:1981qh,Laporta:2001dd,CRUSHER,Studerus:2009ye,vonManteuffel:2012np}, used together with systems of differential equations obeyed by sets of Feynman integrals \cite{Kotikov:1990kg,Bern:1992em,Remiddi:1997ny,Gehrmann:1999as,Ablinger:2015tua,Kotikov:2010gf,Henn:2013pwa,Ablinger:2018zwz}. Systems obtained as the result of IBP reduction can be decoupled \cite{Zuercher:94,BCP13} using methods available in the package OreSys \cite{ORESYS}. In the case of partial differential equations, there exists the method of Janet bases \cite{JANET}.

Calculations in which hypergeometric functions have played a role are among others \cite{Boos:1990rg,Fleischer:2003rm,Watanabe:2013ova,Bluemlein:2017rbi,Phan:2018cnz} at one loop, multileg level, and \cite{Anastasiou:1999ui,Anastasiou:1999cx,Bauberger:1994nk,Ablinger:2012qm} at higher loop order.

Other authors \cite{delaCruz:2019skx,Klausen:2019hrg} have emphasized the relationship between Feynman integrals and Gelfand-Kapranov-Zelevinsky hypergeometric systems \cite{Gelfand:1989,Gelfand:1990,Gelfand:1992,Saito:2000}, which have solutions in terms of hypergeometric functions.

In \cite{Blumlein:2021hbq} we systematized the study of a class of differential systems obeyed by certain classes of hypergeometric functions and investigated in some concrete examples how the series expansion of hypergeometric functions with respect to a parameter is obtained in the context of difference ring methods \cite{Karr:81,Schneider:01,Schneider:07d,Schneider:10b,Schneider:10c,Schneider:15a,Schneider:08c,Schneider:2021,LinearSolver} implemented in the package {\tt Sigma} \cite{Schneider:2001,SIG1,SIG2} and {\tt EvaluateMultiSums} \cite{EMSSP}.

We follow here the exposition of \cite{Blumlein:2021hbq}, in which we present our package {\tt HypSeries} which employs strategies useful for the classifications of systems of differential equations having hypergeometric solutions.

\subsection{Differential equations}
\label{sec:delist}

Hypergeometric functions satisfy differential equations and have integral representations. For instance, the function $_2F_1$ has the property 
\begin{eqnarray}
\pFq{2}{1}{a_1,a_2}{b_1}{z} = \frac{\Gamma(b_1)}{\Gamma(a_1) \Gamma(b_1-a_1)} \int_0^1 dx x^{a_1-1} (1-x)^{b_1-a_1-1}
(1-zx)^{-a_2}
\end{eqnarray}
and satisfies the differential equation
\begin{eqnarray}
x(1-x) \frac{d^2}{dx^2} + (c -(a+b+1)x) \frac{d}{dx} - ab ~.
\label{eq:D2F1g}
\end{eqnarray}
For the function $_{3}F_2$ one obtains
\begin{eqnarray}
x^2(1-x) \frac{d^3}{dx^3} + x (A_{{2}}+B_{{2}} x) \frac{d^2}{dx^2} 
+ (A_{{1}}+B_{{1}} x) \frac{d}{dx} + C,
\label{eq:D3F2}
\end{eqnarray}
with $A_{{2}} = b_1+b_2 +1, B_{{2}} = -(3 + a_1 + a_2 +a_3), A_{{1}} = b_1b_2, 
B_{{1}} = -(a_2 a_1 + a_3 a_1 + a_2 a_3 + a_1 +a_2 +a_3 +1), C = - a_1 a_2 a_3$,
while $_{p+1}F_p$ can be represented by the iterative integral
\begin{eqnarray}
\int_0^1 dx x^{a-1} (1-x)^{b-1} 
\pFq{p}{q}{a_1,...,a_p}{b_1,...b_q}{x z} = \frac{\Gamma(a) \Gamma(b)}{\Gamma(a+b)} \pFq{p+1}{q+1}
{a_1,...,a_p,a}{b_1,...b_q,a+b}{z} 
\end{eqnarray}
and satisfies the differential equation
\begin{eqnarray}
x^p (1-x) \frac{d^{p+1}}{dx^{p+1}} + \sum_{k = 1}^{p} x^{k-1}(A_k + B_k x) \frac{d^k}{dx^k} + C 
\end{eqnarray}
since it is annihilated by the differential operator
\begin{eqnarray}
x \frac{d}{dx} \Biggl( x \frac{d}{dx} + b_1 -1 \Biggr) ... \Biggl( x \frac{d}{dx} + b_q -1 \Biggr)
- x \Biggl( x \frac{d}{dx} + a_1  \Biggr) ... \Biggl( x \frac{d}{dx} + a_p \Biggr).
\label{eq:PFQ1}
\end{eqnarray}
The products of the differential operators in (\ref{eq:PFQ1}) $\vartheta = x (d/dx) \equiv x \partial_x$, 
can be written in the following form
\begin{eqnarray}
\vartheta &=& x \partial_x
\\
\vartheta^2 &=& x \partial_x + x^2 \partial_x^2
\\
\vartheta^3 &=& x \partial_x + 3 x^{{2}} \partial_x^2 + x^3 \partial_x^3
\\
\vartheta^4 &=& x \partial_x + 7 x^{{2}} \partial_x^2 + 6 x^3 \partial_x^3+ x^4 \partial_x^4
\\
\vartheta^5 &=& x \partial_x + 15 x^{{2}} \partial_x^2 + 25 x^3 \partial_x^3+ 10 x^4 \partial_x^4 + x^5 
\partial_x^5,~~\text{etc.}
\end{eqnarray}

One can parametrize the equations obeyed by the Horn hypergeometric functions \cite{APPELL1,APPELL2,KAMPE1,KAMPE2,HORN} $F_1$ to $F_4$, $G_1$ to $G_3$, and $H_1$ to $H_7$, including the Appell functions \cite{APPELL1,APPELL2} as follows:
\begin{eqnarray}
a
+ (b x+c) \partial_x
+x (d+e x) \partial_x^2
+f y \partial_y
+(g y+h x y) \partial_{x,y}^2
+j y^2 \partial_y^2 &=& 0
\label{APP1}
\\
a_1
+ (b_1 y+ c_1) \partial_y
+y (d_1+e_1
   y) \partial_y^2
+f_1 x \partial_x
+ (g_1  x+ h_1 x y) \partial_{x,y}^2
+j_1 x^2 \partial_x^2 &=& 0,
\label{APP2}
\end{eqnarray}
with the example of the Appell $F_1$ function
\begin{eqnarray}
F_1 &:& x(1-x)\partial_x^2 +y(1-x) \partial_{x,y}^2 + (A+B x)\partial_x + C y\partial_y + D 
\\
F_1 &:& y(1-y)\partial_y^2 +x(1-y) \partial_{x,y}^2 + (A +B' y)\partial_y + C' x\partial_x + D'. 
\end{eqnarray}

In physical applications, the functions $S_1$ and $S_2$ appeared in \cite{Anastasiou:1999ui,Anastasiou:1999cx}. They are annihilated by the differential operators
\begin{eqnarray}
S_1 &:& 
a+(c+b x) \partial_x +x (d+e x) \partial_x^2 +x^2 (l+p x) \partial_x^3+f y \partial_y +x 
(q+r x) y \partial_x^2 \partial_y +j y^2 \partial_y^2 
\nonumber\\ &&
+s x y^2 \partial_x \partial_y^2 +(g y+h x y) 
\partial_{x,y}^2
\\
 &:& 
a_1+f_1 x \partial_x + j_1 x^2 \partial_x^2 +(c_1+b_1 y) \partial_y+y (d_1+e_1 y) \partial_y^2 +(g_1 x+h_1 x 
y) \partial_{x,y}^2
\\
S_2 &:&
a+(c+b x) \partial_x +x (d+e x) \partial_x^2 +x^2 (l+p x) \partial_x^3 +f y \partial_y +x 
(q+r x) y \partial_x^2 \partial_y
+j y^2 \partial_y^2
\nonumber\\ &&
+s x y^2 \partial_x \partial_y^2 +(g y+h x y) 
\partial_{x,y}^2
\\
 &:&
a_1+f_1 x \partial_x +(c_1+b_1 y) \partial_y +j_1 x^2 \partial_x^2 \partial_y +y (d_1+e_1 y) 
\partial_y^2+q_1 x y \partial_x \partial_y^2 +p_1 y^2 \partial_y^3 
\nonumber\\ &&
+(g_1 x+h_1 x y) \partial_{x,y}^2.
\end{eqnarray}
For the Kamp\'e de F\'eriet function 
\begin{eqnarray}
	F^{p;q;k}_{l;m;n} \left[ \begin{array}{cc}
	(a_p) ; (b_q) ; (c_k) & \\
	& x, y \\
	(\alpha_l) ; (\beta_m) ; (\gamma_n) &
	\end{array}
	\right]
	&=& \sum_{r,s=0}^\infty \, \frac{\prod_{j=1}^p (a_j)_{r+s} 
        \prod_{j=1}^q (b_j)_r \prod_{j=1}^k (c_{{j}})_s}{\prod_{j=1}^l (\alpha_j)_{r+s} \prod_{j=1}^m (\beta_j)_r \prod_{j=1}^n 
       (\gamma_j)_s } \frac{x^r}{r!} \frac{y^s}{s!} \\
	&=& \sum_{r,s=0}^\infty f[r,s] x^r y^s
\end{eqnarray}
one obtains the following annihilating differential operators \cite{KAMPE1,KAMPE2}
\begin{eqnarray}
	\prod_{j=1}^p (x \partial_x + y \partial_y +a_j) \prod_{j=1}^q (x \partial_x +b_j) -{\partial_x} \prod_{j=1}^l (x \partial_x + y \partial_y -1+\alpha_j) \prod_{j=1}^m ({x \partial_x} -1+\beta_j) = 0,
\\
	\prod_{j=1}^p (x \partial_x +y \partial_y +a_j) \prod_{j=1}^k (y \partial_y +c_j) -{\partial_y} \prod_{j=1}^l (x 
\partial_x + y \partial_y -1+\alpha_j) \prod_{j=1}^n (y \partial_y -1+\gamma_j) = 0.
\end{eqnarray}

The differential operators for the triple hypergeometric series \cite{SRIKARL} read
{
\begin{eqnarray}
D_{3,1} &=& A+(B_0+B_1 x) \partial_x +x (E_0+E_1 x) \partial_x^2 + C_1 y \partial_y +{F_1} y^2 \partial_y^2 +(H_0+H_1 x) y 
\partial_{x,y}^2
\nonumber\\ &&
+D_1 z \partial_z 
+G_1 z^2 \partial_z^2 +(L_0+L_1 x) z \partial_{x,z}^2 + S_1 y z \partial_{y,z}^2
\\
D_{3,2} &=& A'+{B_1'} x \partial_x +E_1' x^2 \partial_x^2 +(C_0'+C_1' y) \partial_y +y (F_0'+F_1' y) 
\partial_y^2+x 
(H_2'+H_1' y) \partial_{x,y}^2
\nonumber\\ &&
+ D_1' z \partial_z 
+ G_1' z^2 \partial_z^2 + L_1' x z \partial_{x,z}^2
+(S_0'+S_1' y) z \partial_{y,z}^2
\\
D_{3,3} &=& 
A''+ B_1'' x \partial_x + {E''_1} x^2 \partial_x^2 + C_1'' y \partial_y + {F_1''} y^2 \partial_y^2
+ H_1'' x y \partial_{x,y}^2 + (D''_0 + D_1'' z) \partial_z 
\nonumber\\ &&
+ z (G_0'' + G_1'' z) \partial_z^2
+x {(L_2'' + L''_1 z)} \partial_{x,z}^2 + y (S_2''+ S_1'' z) \partial_{y,z}^2 .
\end{eqnarray}
}

The differential operators for the quadruple versions are given by
{
\begin{eqnarray}
D_{4,1} &=&
A + E_1 t \partial_t +L_1 t^2 \partial_t^2 +(B_0+B_1 x)\partial_x +x (F_0+F_1 x) \partial_x^2+t 
(P_0+P_1 x) \partial_{t,x}^2 +C_1 y \partial_y
\nonumber\\ &&
+G_1 y^2 \partial_y^2 +R_1 t y \partial_{t,y}^2 + (M_0+M_1 x) y 
\partial_{x,y}^2 + D_1 z \partial_z + H_1 z^2 \partial_z^2 + S_1 t z \partial_{t,z}^2 
\nonumber\\ &&
+(N_0+N_1 x) z 
\partial_{x,z}^2 +Q_1 y z \partial_{y,z}^2
\\ 
D_{4,2} &=& 
A'+E_1' t \partial_t + L_1' t^2 \partial_t^2 + B_1' x \partial_x +F_1' x^2 \partial_x^2 + P_1' t x 
\partial_{t,x}^2+(C_0'+ C_1' y) \partial_y
+y (G_0' + G_1' y) \partial_y^2 
\nonumber\\ &&
+ t (R_0'+R_1' y) \partial_{t,y}^2 + x (M_2' + M_1' y) 
\partial_{x,y}^2 + D_1' z \partial_z +H_1' z^2 \partial_z^2 +S_1' t z \partial_{t,z}^2 + N_1' x z 
\partial_{x,z}^2 
\nonumber\\ &&
+ (Q_0'+Q_1' y) z \partial_{y,z}^2
\\
D_{4,3} &=& 
A''+E_1'' t \partial_t + L_1'' t^2 \partial_t^2 + B_1'' x \partial_x + F_1'' x^2 \partial_x^2
+ P_1'' t x \partial_{t,x}^2  + C_1'' y \partial_y + G_1'' y^2 \partial_y^2 + R_1'' t y \partial_{t,y}^2 
\nonumber\\ &&
+ M_1'' x 
y \partial_{x,y}^2
+(D_0'' + D_1'' z) \partial_z + z (H_0''+H_1'' z) \partial_z^2 + t (S_0''+S_1'' z) \partial_{t,z}^2+x 
(N''_2 + N_1'' z) \partial_{x,z}^2 
\nonumber\\ &&
+ y (Q_2''+Q_1'' z) \partial_{y,z}^2
\\
D_{4,4} &=& 
A''' + (E_0''' + E_1''' t) \partial_t + t (L_0''' + L_1''' t) \partial_t^2 + B_1''' x \partial_x
+ F_1''' x^2 \partial_x^2 + (P_2''' + P_1''' t) x \partial_{t,x}^2 
\nonumber\\ &&
+ C_1''' y \partial_y + G_1''' y^2 
\partial_y^2 + (R_2''' + R_1''' t) y \partial_{t,y}^2 + M_1''' x y \partial_{x,y}^2 + D_1''' z \partial_z + H_1''' z^2 
\partial_z^2 
\nonumber\\ &&
+ (S_2'''+ S_1''' t) z \partial_{t,z}^2 + N_1''' x z \partial_{x,z}^2 + Q_1''' y z \partial_{y,z}^2.
\end{eqnarray}
}
They cover the functions $K_i,~~i = 1 ... 21$ of Refs.~\cite{EXTON72a,EXTON1}.

\subsection{Recursions} 
\label{sec:reclist} 

The formal power series ansatz
\begin{eqnarray}
\sum_{k_1, ..., k_n = 0}^\infty f[k_1, ..., k_n] x_1^{k_1} ... x_n^{k_n}
\label{eq:TAYLOR}
\end{eqnarray}
with $f[k_1, ..., k_n]$ hypergeometric induces a recurrence relation on the expansion coefficients, deriving from the differential equations of Section~\ref{sec:delist}. (In addition, hypergeometric functions obey contiguous relations in their parameters \cite{Gauss:1876,Kalmykov:2020cqz,Paule:2021gaw}).

We list here the recursions corresponding to the differential systems from the previous Section:
\begin{eqnarray}
_2F_1 &:& 
(C + n (1 - n + B_1)) f[n] + (1 + n) (n + A_1) f[1 + n] = 0
\label{eq:R2F1}
\\
_3F_2 &:& 
{
\big( n B_1 + (n-1)n B_2 +C-(n-2) (n-1) n \big) f[n]
    }
\nonumber\\ &&
{
+ \big( (n+1) A_1 + n (n+1) A_2 +  (n-1)n(n+1)\big) f[n+1] = 0
}    
\label{eq:R3F2}
\\
_{p+1}F_p &:& 
{
\Biggl[\frac{C}{n!} - \frac{1}{(n-p-1)!} + \sum_{k=1}^p \frac{B_k}{(n-k)!} \Biggr] f[n]
}
\nonumber\\ && 
{
+ (n+1) \Biggl[ \frac{1}{(n-p)!} +  \sum_{k=1}^p \frac{A_k}{(n-k+1)!} \Biggr] f[n+1] = 0.
}
\end{eqnarray}

In the two--variable cases the expansion coefficients of the  Horn--type functions obey
\begin{eqnarray}
&&\Big[a+b m+e (m-1) m+n \big(f+h m+j (n-1)\big)\Big] f[m,n]
\nonumber\\&&\qquad\qquad\qquad\qquad\qquad\qquad\qquad\qquad
+(1+m) (c+d m+g n) f[1+m,n]=0,
\\
&&\Big[a_1+f_1 m+j_1 (m-1) m+n \big(b_1+h_1 m+e_1 (n-1)\big)\Big] f[m,n]
\nonumber\\&&\qquad\qquad\qquad\qquad\qquad\qquad\qquad\qquad
+(1+n) (c_1+g_1 m+d_1 n) f[m,1+n]=0,
\end{eqnarray}
and for the $S_1$-functions one has
\begin{eqnarray}
&&\Big[a+b m+e (m-1) m+f n+h m n+j (n-1) n+(m-2) (m-1) m p+(m-1) m n r
\nonumber\\&&
+m (n-1) n s\Big] f[m,n]
+(1+m) \Big[c+g n+m \big(d+l (m-1)+n q\big)\Big] f[1+m,n]=0,
\\
&&\Big[a_1+f_1 m+j_1 (m-1) m+n \big(b_1+h_1 m+e_1 (n-1)\big)\Big] f[m,n]
\nonumber\\&&
+(1+n) (c_1+g_1 m+d_1 n) f[m,1+n]=0,
\end{eqnarray}
as, likewise, for the 
$S_2$-functions
\begin{eqnarray}
&&\Big[a+b m+e (m-1) m+f n+h m n+j (n-1) n+(m-2) (m-1) m p+(m-1) m n r+
\nonumber\\&&
m (n-1) n s\Big] f[m,n]+(1+m) \Big[c+g n+m \big(d+l (-1+m)+n q\big)\Big] f[1+m,n]=0,
\\
&&\Big[a_1+f_1 m+n \big(b_1+h_1 m+e_1 {(n-1)}\big)\Big] f[m,n]+(1+n) \Big[c_1+n (d_1+(-1+n) p_1)
\nonumber\\&&
+m \big(g_1+j_1 (m-1)+n q_1\big)\Big] f[m,1+n]=0.
\end{eqnarray}
For the expansion coefficients $f[r,s]$ of the Kamp\'e de F\'eriet functions the recurrences read
\begin{eqnarray}
	\prod_{j=1}^p (r+s+a_j) \prod_{j=1}^q (r+b_j) f[r,s] - (r+1) \prod_{j=1}^l (r+s+\alpha_j) \prod_{j=1}^m (r+\beta_j) 
f[r+1,s] = 0,
\\
	\prod_{j=1}^p (r+s+a_j) \prod_{j=1}^k (s+c_j) f[r,s] - (s+1) \prod_{j=1}^l (r+s+\alpha_j) \prod_{j=1}^n (s+\gamma_j) 
f[r,s+1] =0.
\end{eqnarray}

The coefficients in the 3-variable cases obey
\begin{eqnarray}
&&\Big[A+ B_1  m+ E_1  (m-1) m+ C_1  n+ H_1  m n+ F_1  (n-1) n+ D_1  p+ L_1  m p+ G_1  (p-1) p
\nonumber\\&&
+n p  S_1 \Big] f[m,n,p]+(1+m) ( B_0 + E_0  m+ H_0  n+ L_0  p) f[1+m,n,p]=0,
\\
&&\Big[ A' + B'_1  m+ E'_1  (m-1) m+ C'_1  n+ H'_1  m n+ F'_1  (n-1) n+ D'_1  p+ L'_1  m p+ G'_1  (p-1) p
\nonumber\\&&
+n p  S'_1 \Big] f[m,n,p]+(1+n) ( C'_0 + H'_2  m+ F'_0  n+p  S'_0 ) f[m,1+n,p]=0,
\\
&&\Big[ A'' + B''_1  m+ E''_1  (m-1) m+ C''_1  n+ H''_1  m n+ F''_1  (n-1) n+ D''_1  p+ L''_1  m p+ G''_1  (p-1) p
\nonumber\\&&
+n p  S''_1 \Big] f[m,n,p]+(1+p) ( D''_0 + L''_2  m+ G''_0  p+n  S''_2 ) f[m,n,1+p]=0.
\end{eqnarray}

For the 4-variable systems one has
\begin{eqnarray}
&&\Big[A+ B_1  m+ F_1  (m-1) m+ C_1  n+m  M_1  n+ G_1  (n-1) n+ D_1  p+m  N_1  p+ H_1  (p-1) p
\nonumber\\&&
+ E_1  q+m  P_1  q+ L_1  (q-1) q+n p  Q_1 +n q  R_1 +p q  S_1 \Big] f[m,n,p,q]+(1+m) ( B_0 + F_0  m
\nonumber\\&&
+ M_0  n+ N_0  p+ P_0  q) f[1+m,n,p,q]=0,
\\[2mm]
&&\Big[ A' + B'_1  m+ F'_1  (m-1) m+ C'_1  n+m  M'_1  n+ G'_1  (n-1) n+ D'_1  p+m  N'_1  p+ H'_1  (p-1) p
\nonumber\\&&
+ E'_1  q+m  P'_1  q+ L'_1  (q-1) q+n p  Q'_1 +n q  R'_1 +p q  S'_1 \Big] f[m,n,p,q]+(1+n) ( C'_0 +m  M'_2 
\nonumber\\&&
+ G'_0  n+p  Q'_0 +q  R'_0 ) f[m,1+n,p,q]=0,
\\[2mm]
&&\Big[ A'' + B''_1  m+ F''_1  (m-1) m+ C''_1  n+m  M''_1  n+ G''_1  (n-1) n+ D''_1  p+m  N''_1  p+ H''_1  (p-1) p
\nonumber\\&&
+ E''_1  q+m  P''_1  q+ L''_1  (q-1) q+n p  Q''_1 +n q  R''_1 +p q  S''_1 \Big] f[m,n,p,q]+(1+p) ( D''_0 +m  N''_2 
\nonumber\\&&
+ H''_0  p+n  Q''_2 +q  S''_0 ) f[m,n,1+p,q]=0,
\\[2mm]
&&\Big[ A''' + B'''_1  m+ F'''_1  (m-1) m+ C'''_1  n+m  M'''_1  n+ G'''_1  (n-1) n+ D'''_1  p+m  N'''_1  p
\nonumber\\&&
+ H'''_1  (p-1) p+ E'''_1  q+m  P'''_1  q+ L'''_1  (q-1) q+n p  Q'''_1 +n q  R'''_1 +p q  S'''_1 \Big] f[m,n,p,q]
\nonumber\\&&
+(1+q) ( E'''_0 +m  P'''_2 + L'''_0  q+n  R'''_2 +p  S'''_2 ) f[m,n,p,1+q]=0.
\end{eqnarray}

\subsection{The Solution of the Recursions} 
\label{sec:recsol}
Given a field $\mathbb K$ with characteristic zero, a multiple hypergeometric series is a function
\begin{equation}
	f(x_1,\ldots,x_r) = \sum_{n_i\geq0} A(n_1,\ldots,n_r) x_1^{n_1}\cdots x_r^{n_r}
\label{eq:hyp-series}	
\end{equation}
where $A:\mathbb N^r\to \mathbb K$, is hypergeometric, i.e. 
\begin{equation}
s_i\,A(n_1,\ldots,n_i,\ldots,n_r) = t_i\,A(n_1,\ldots,n_i+1,\ldots,n_r) , \qquad i=1,\ldots,r
\label{eq:hyp-ratio}
\end{equation}
for polynomials $s_i,t_i\in \mathbb K[n_1,\ldots,n_r]$ being coprime.

It was remarked in the previous sections how hypergeometric series satisfy differential equations which can be mapped into difference equations for the coefficients of the Taylor expansion. In general, not many algorithms exist to solve such a multivariate system of difference equations having the target solution space being that of hypergeometric functions; we refer the reader to Section~\ref{sec:PLDEsolver} for an exposition of one direction more targeted toward the class of rational functions.

However, if a system of differential equations induces a system of recurrences of the form \eqref{eq:hyp-ratio}, then a method to investigate its solutions is possible using Sigma. Let us concentrate on a system of linear differential equations of the form
\begin{equation}
\Big[
s_i \Bigl(x_1\frac{\partial}{\partial x_i}, \ldots, x_i \frac{\partial}{\partial x_i}, \ldots, x_r \frac{\partial}{\partial x_r} \Bigr) 
- \frac{1}{x_i} t_i \Bigl(x_1 \frac{\partial}{\partial x_1}, \ldots, x_i\frac{\partial}{\partial x_i}-1, \ldots, x_r \frac{\partial}{\partial x_r} \Bigr) 
\Big] f(x_1,\ldots,x_r)
= 0.
\label{eq:DEsystem}
\end{equation}
One has the property
\begin{equation}
x_i \frac{\partial}{\partial x_i} x_1^{n_1}\cdots x_r^{n_r} = n_i x_1^{n_1}\cdots x_r^{n_r}
\end{equation}
which implies that for a polynomial $p(n_1,\ldots,n_r)$ we have
\begin{equation}
p \Big(x_1\frac{\partial}{\partial x_1}, \ldots, x_i \frac{\partial}{\partial x_i}, \ldots, x_r \frac{\partial}{\partial x_r} \Bigr) x_1^{n_1}\cdots x_r^{n_r} = p(n_1,\ldots,n_r) x_1^{n_1}\cdots x_r^{n_r} .
\end{equation}
Thus
\begin{align}
\label{eq:t-term}
&\Big[ s_i \Bigl(x_1\frac{\partial}{\partial x_1}, \ldots, x_i \frac{\partial}{\partial x_i}, \ldots, x_r \frac{\partial}{\partial x_r} \Bigr) \Big] f(x_1,\ldots,x_r)
\nonumber\\
&\qquad \qquad \qquad \qquad \qquad = \sum_{n_i\ge 0} s_i(n_1,\ldots,n_i,\ldots,n_r)  A(n_1,\ldots,n_i,\ldots,n_r) x_1^{n_1}\cdots x_r^{n_r}	\\ 
&\Big[ t_i \Bigl(x_1 \frac{\partial}{\partial x_1}, \ldots, x_i\frac{\partial}{\partial x_i}-1, \ldots, x_r \frac{\partial}{\partial x_r} \Bigr) \Big] f(x_1,\ldots,x_r)
\nonumber\\
&\qquad \qquad \qquad \qquad \qquad = \sum_{n_i\ge 0} t_i(n_1,\ldots,n_i-1,\ldots,n_r)  A(n_1,\ldots,n_i,\ldots,n_r) x_1^{n_1}\cdots x_r^{n_r} 
\end{align}
and therefore, dividing the second equation by $x_i$ from the left,
\begin{align}
\label{eq:s-term}
&\Big[ \frac{1}{x_i} t_i \Bigl(x_1 \frac{\partial}{\partial x_1}, \ldots, x_i\frac{\partial}{\partial x_i}-1, \ldots, x_r \frac{\partial}{\partial x_r} \Bigr) \Big] f(x_1,\ldots,x_r)
\nonumber\\
& \qquad \qquad \qquad \qquad \qquad = \sum_{n_i\ge 0} 
t_i(n_1,\ldots,n_i-1,\ldots,n_r)  A(n_1,\ldots,n_i,\ldots,n_r) 
x_1^{n_1}\cdots x_i^{n_i-1} \cdots x_r^{n_r}. 
\end{align}
The coefficient of the term $x_1^{n_1}\cdots x_i^{n_i}\cdots x_r^{n_r}$ in 
\eqref{eq:t-term} and in 
\eqref{eq:s-term} is respectively
\begin{equation}
s_i(n_1,\ldots,n_i,\ldots,n_r) A(n_1,\ldots,n_i,\ldots,n_r)
\end{equation}
and
\begin{equation}
t_i(n_1,\ldots,n_i,\ldots,n_r) A(n_1,\ldots,n_i+1,\ldots,n_r).
\end{equation}
This shows, due to \eqref{eq:hyp-ratio}, that~\eqref{eq:DEsystem} holds.

For example, for the case of the Gauss hypergeometric function
\begin{equation}
_2F_1(a,b;c;x) = \sum_{n\ge 0} \frac{(a)_n (b)_n}{(c)_n n!}x^n
\end{equation}
one has
\begin{eqnarray}
A(n) &=& \frac{(a)_n (b)_n}{(c)_n n!} \\
s(n) &=& (a+n)(b+n) \\
t(n) &=& (n+1)(c+n) 
\end{eqnarray}
and the differential equation obeyed by $_2F_1(a,b;c;x)$ is, from \eqref{eq:DEsystem},
\begin{equation}
\Big[ \Big(a+ x \frac{\partial}{\partial x} \Big) \Big(b+ x \frac{\partial}{\partial x} \Big) - \frac{1}{x} \Big(x \frac{\partial}{\partial x}\Big)	\Big( x \frac{\partial}{\partial x} -1 +c \Big) \Big] {}_2F_1(a,b;c;x) = 0.
\end{equation}

To summarize, a system of equations of the type \eqref{eq:DEsystem} induces first-order recurrences, which can be studied using the methods of Sigma as described next.

\subsection{An algorithm for hypergeometric products}
\label{sec:solveProd}

We consider a system \eqref{eq:hyp-ratio} and look for a hypergeometric solution $A$. The univariate case is well-known and treatable under the methods of Sigma. To be explicit, for $r=1$, the polynomials $s_1,t_1$ will be nonzero for all $k>\lambda_1,k,\lambda_1\in \mathbb N$. Then, using the hypergeometric property, one writes
\begin{equation}\label{Equ:Prod1}
\begin{split}
A(n_1)&=\frac{s_1(n_1-1)}{t_1(n_1-1)}A(n_1-1)\\
&=\frac{s_1(n_1-1)}{t_1(n_1-1)}\frac{s_1(n_1-2)}{t_1(n_1-2)}A(n_1-2)=\dots=\left(\prod_{k=\lambda_1+1}^{n_1}\frac{s_1(k-1)}{t_1(k-1)}\right)A(\lambda_1)
\end{split}
\end{equation}
and a formula for $A$ is obtained as a hypergeometric product $\prod_{k=\lambda_1+1}^{n_1}\frac{s_1(k-1)}{t_1(k-1)}$. For the multivariate case, we seek a generalization and a solution in the form of a nested product.

One calls a sequence non-trivial if its zero points vanish on a polynomial in $\mathbb K[n_1,\ldots,n_r]$, \cite{AP:02}. For such a sequence, the system \eqref{eq:hyp-ratio} implies a compatibility condition: calling
\begin{equation}
	R_i = \frac{s_i}{t_i} \in K(n_1,\ldots,n_r)
\end{equation}
then for there to be solutions it must also be that~\cite[Prop~4]{AP:02}
\begin{equation}
\frac{R_i(n_1,\dots,n_j+1,\dots,n_r)}{R_i(n_1,\dots,n_j,\dots,n_r)}=\frac{R_j(n_1,\dots,n_i+1,\dots,n_r)}{R_j(n_1,\dots,n_i,\dots,n_r)}.
\label{Equ:CompatibilityProp}
\end{equation}

Relevant to this context, the Ore-Sato theorem \cite{Ore:1929,Sato:1990,OSGeneral} states what a general form of the hypergeometric solution, if it exists, must have; it is a product of factorials and hypergeometric terms. Here we specialize to the additional assumption that $s_i,t_i\neq0$ for all $(n_1,\ldots,n_r)\in \mathbb N^r$ with $n_i>\lambda_i$. (In general no algorithm exists \cite{Hilbert10} to find the $\lambda_i$, although it is often possible in practically occurring cases).

Under these assumptions, one can write
\begin{equation}
A(n_1,\ldots,n_i+1,\ldots,n_r) = R_i(n_1,\dots,n_r)\,A(n_1,\ldots,n_i,\ldots,n_r) , \qquad i=1,\ldots,r
\end{equation}
and 
\begin{equation}
A(n_1,\dots,n_r)=c\,A(\lambda_1,\dots,\lambda_r)
\end{equation}
for any $(n_1,\dots,n_r)\in\mathbb N^r$ with $n_i\geq\lambda_i$ and some $c\in\mathbb K\setminus\{0\}$.

One can derive under these assumptions a formula for $A$ as the nested product
\begin{equation}
c\,\left(\prod_{k=\lambda_1}^{n_1}h_1(k,n_2,\dots,n_r)\right)\left(\prod_{k=\lambda_2}^{n_2}h_2(k,n_3\dots,n_r)\right)\dots \left(\prod_{k=\lambda_r}^{n_r}h_r(k)\right)
\label{Equ:FinalProdForm}
\end{equation}
with $c=A(\lambda_1,\dots,\lambda_r)\in\mathbb K\setminus\{0\}$ in a recursive manner. The case $r=1$ is treated in \eqref{Equ:Prod1}. Otherwise one can write 
\begin{equation}
	A(n_1,\dots,n_r)=A(\lambda_1,n_2,\dots,n_r)\prod_{k=\lambda_1+1}^{n_1}h_1(k,n_2,\dots,n_r)
\end{equation}
with 
\begin{equation}
	h_1(k,n_2,\dots,n_r)=R_1(k-1,n_2,\dots,n_r)=\frac{s_1(k-1,n_2,\dots,n_r)}{t_1(k-1,n_2,\dots,n_r)}
\end{equation}
Then one considers
\begin{equation}
	A'(n_2,\dots,n_r):=A(\lambda_1,n_2,\dots,n_r)
\end{equation}
which satisfies
\begin{equation}
A'(n_2,\ldots,n_i+1,\ldots,n_r) = R_i(\lambda_1,n_2,\dots,n_r)\,A'(n_2,\ldots,n_i,\ldots,n_r) , \qquad i=2,\ldots,r
\end{equation}
with $R_i(\lambda_1,n_2,\dots,n_r)\in\mathbb K[n_2,\dots,n_r]$ and obtains 
\begin{equation}
A'(n_2,\dots,n_r)=c\,\left(\prod_{k=\lambda_1}^{n_2}h_2(k,n_3\dots,n_r)\right)
\dots \left(\prod_{k=\lambda_r}^{n_r}h_r(k)\right),
\end{equation}
with 
$c=A'(\lambda_2,\dots,\lambda_r)=A(\lambda_1,\lambda_2,\dots,\lambda_r)\in\mathbb K\setminus\{0\}$ 
and $h_i(x,n_{i+1},\dots,n_r)\in\mathbb K(x,n_{i},\dots,n_r)$ with $2\leq i\leq r$. 

Proceeding in the same way by recursion one obtains \eqref{Equ:FinalProdForm}. We show now the results of the application of the algorithm in some examples.

\subsubsection{Examples}
\label{sec:exProdExp}

We consider the differential equation
(\ref{eq:D2F1g}) which leads to the recurrence (\ref{eq:R2F1}) for the expansion coefficient $f[n]$. The recursion is 
of order one and is solved for $f[n] = 0$. {\tt Sigma} obtains the product solution
\begin{eqnarray}
f[n] = 
\frac{\prod_{i_1=1}^n \big(
        2
        +B_1
        -C
        -3 i_1
        -B_1 i_1
        +i_1^2
\big)}{n! (A_1)_n}
\equiv \frac{\prod_{i_1=1}^n \big[-C +B_1(1-i_1) + (1-i_1)(2-i_1)
\big]}{n! (A_1)_n},
\label{eq:T1}
\nonumber\\
\end{eqnarray}
%
One can factorize the product in (\ref{eq:T1}) in terms of Pochhammer symbols by      
\begin{eqnarray}
f[n] = \frac{(\alpha_1)_n (\alpha_2)_n}{(A_1)_n n!},
\end{eqnarray}
with 
\begin{eqnarray}
\alpha_{1(2)} = -\frac{1}{2}(1+B_1) \mp  \frac{1}{2} \sqrt{(1+B_1)^2 + 4 C}.
\end{eqnarray}
By replacing $A_1,B_1$ and $C$ by
\begin{eqnarray}
C \rightarrow -a b,~~A_1 \rightarrow c,~~B_1 \rightarrow -1 - a - b
\end{eqnarray}
one obtains
\begin{eqnarray}
f[n] = \frac{(a)_n (b)_n}{(c)_n n!}.
\end{eqnarray}

In the case of the hypergeometric function $_3F_2$, the differential equation \eqref{eq:D3F2} implies the recurrence (\ref{eq:R3F2}) for $f[n]$ with $f[n]=1$, which has the solution
\begin{eqnarray}
f[n] = \frac{\prod_{i_1=1}^n 
[-C
+B_1 \big(
        1-i_1\big)
-B_2 \big(
        2-i_1
\big)
{ \big(1-i_1\big) }
-\big(
        3-i_1
\big)
\big(2-i_1
\big)
\big(1-i_1\big)]}
{n!~\prod_{i_1=1}^n
[A_1
-A_2 \big(
        1-i_1\big)
+\big(
        2-i_1
\big)
\big(1-i_1\big)]}.
\label{eq:3F2a}
\end{eqnarray}
Eq.~(\ref{eq:3F2a}) can be rewritten in terms of radicals by
\begin{eqnarray}
f[n] = \frac{
(\alpha_1)_n
(\alpha_2)_n
(\alpha_3)_n}{ {n!} \big(
        -\frac{1}{2}
        +\frac{A_2 }{2}
        -\frac{z_5}{2}
\big)_n \big(
        -\frac{1}{2}
        +\frac{A_2 }{2}
        +\frac{z_5}{2}
\big)_n},
\end{eqnarray}
with
\begin{eqnarray}
\alpha_1 &=& 
        1
        -\frac{z_4}{3}
        +\frac{\sqrt[3]{z_1
        +z_2}
        }{6 \sqrt[3]{2}}
        -\frac{i \sqrt[3]{z_1
        +z_2
        }}{2 \sqrt[3]{2} \sqrt{3}}
        +\frac{\sqrt[3]{-2} z_3}{3 \sqrt[3]{z_1
        +z_2
        }}
\\
\alpha_2 &=&
        1
        -\frac{z_4}{3}
        -\frac{\sqrt[3]{z_1
        +z_2
        }}{3 \sqrt[3]{2}}
        -\frac{\sqrt[3]{2} z_3}{3 \sqrt[3]{z_1
        +z_2
        }}
\\
\alpha_3 &=& 
        1
        -\frac{z_4}{3}
        +\frac{\sqrt[3]{z_1
        +z_2
        }}{6 \sqrt[3]{2}}
        +\frac{i \sqrt[3]{z_1
        +z_2
        }}{2 \sqrt[3]{2} \sqrt{3}}
        -\frac{(-1)^{2/3} \sqrt[3]{2} z_3}{3 \sqrt[3]{z_1
        +z_2
        }}
\\
z_1 &=&  27 C + (3 + B_2) (9 B_1 + B_2 (3 + 2 B_2))
\\
z_2 &=&  \sqrt{-4 (3 + 3 B_1 + 
    B_2 (3 + B_2))^3 + (27 C + (3 + B_2) (9 B_1 + 
      B_2 (3 + 2 B_2)))^2}
\\
z_3 &=&  3 + 3 B_1 + B_2 (3 + B_2)
\\
z_4 &=&  6 + B_2
\\
z_5 &=&  \sqrt{-4 A_1 + (A_2-1)^2}.
\end{eqnarray}

After performing the replacements
\begin{eqnarray} A_{{2}} &\rightarrow& b_1+b_2 +1,~~ B_{{2}} \rightarrow -(3 + a_1 + a_2 +a_3),~~A_{{1}} \rightarrow b_1 b_2, 
\nonumber\\ 
B_{{1}} &\rightarrow& -(a_2 a_1 + a_3 a_1 + a_2 a_3 + a_1 +
a_2 +a_3 +1),~~C \rightarrow - a_1 a_2 a_3
\label{eq:VIETA}
\end{eqnarray}
one obtains
\begin{eqnarray} 
f[n] = \frac{(a_1)_n (a_2)_n (a_3)_n}{(b_1)_n (b_2)_n n!}.
\end{eqnarray}
%
The replacements are related to Vieta's theorem \cite{VIETA} for the roots $r_i$ of the algebraic equation
\begin{eqnarray} 
x^n + \sum_{k = 1}^{n} a_{n-k} x^{n-k} = 0,
\end{eqnarray}
which obey
\begin{eqnarray} 
-a_{n-1} &=& r_1 + ... + r_n
\nonumber\\
 a_{n-2} &=& r_1 (r_2 + ... + r_n) + r_2(r_3+ ... r_n) + ... r_{n-1} r_n 
\nonumber\\
&\vdots& 
\nonumber\\
(-1)^n a_0 &=& r_1 ... r_n.
\end{eqnarray}
For the function $_{p+1}F_p$ the product solution is 
\begin{eqnarray} 
f[n] = 
\prod_{i=1}^n \frac{\frac{1}{(i-p-2)!} - \sum_{k=1}^p \frac{B_k}{(i-k-1)!} - \frac{C}{(i-1)!}}
{\frac{i}{(i-p-1)!} + \sum_{k=1}^p \frac{A_k i}{(i-k)!} } .
\end{eqnarray}

In the multivariate case, we list the product solutions for the Horn--type functions, from Eqs. \eqref{APP1}, \eqref{APP2}, 
\begin{eqnarray}
f_{\rm H}[m,n] &=& 
\Biggl[
        \prod_{i_1=1}^m \frac{-a
        +b
        -2 e
        -f n
        +h n
        +j n
        -j n^2
        -b i_1
        +3 e i_1
        -h n i_1
        -e i_1^2
        }{\big(
                c
                -d
                +g n
                +d i_1
        \big) i_1}\Biggr]
\nonumber\\&& \times        
        \Biggl[\prod_{i_1=1}^n \frac{-a_1
+b_1
-2 e_1
-b_1 i_1
+3 e_1 i_1
-e_1 i_1^2
}{\big(
        c_1
        -d_1
        +d_1 i_1
\big) i_1}\Biggr]
\label{eq:HORN2}
\end{eqnarray}
and for the functions $S_1$ and $S_2$ one has
\begin{eqnarray}
f_{\rm S_1}[m,n] &=& 
\Biggl[
        \prod_{i_1=1}^n \frac{- a_1 
        + b_1 
        -2  e_1 
        - b_1  i_1
        +3  e_1  i_1
        - e_1  i_1^2
        }{\big(
                 c_1 
                - d_1 
                + d_1  i_1
        \big) i_1}\Biggr] 
\nonumber\\&& \times    
\Biggl[   
        \prod_{i_1=1}^m 
\bigg(        
        \frac{1}{\big(
        c
        -d
        +2 l
        +g n
        -n q
        +d i_1
        -3 l i_1
        +n q i_1
        +l i_1^2
\big) i_1}	(-a
+b
\nonumber\\&&
-2 e
-f n
+h n
+j n
-j n^2
+6 p
-2 n r
-n s
+n^2 s
-b i_1
\nonumber\\&&
+3 e i_1
-h n i_1
-11 p i_1
+3 n r i_1
+n s i_1
-n^2 s i_1
-e i_1^2
+6 p i_1^2
-n r i_1^2
\nonumber\\&&
-p i_1^3
)
\bigg)
\bigg]
\\
f_{\rm S_2}[m,n] &=& 
\bigg(
        \prod_{i_1=1}^n \frac{- a_1 
        + b_1 
        -2  e_1 
        - b_1  i_1
        +3  e_1  i_1
        - e_1  i_1^2
        }{\big(
                 c_1 
                -d_1
                +2  p_1 
                +d_1 i_1
                -3  p_1  i_1
                + p_1  i_1^2
        \big) i_1}\bigg)
\nonumber\\&& \times     
\bigg[   
         \prod_{i_1=1}^m 
\bigg(         
         \frac{1}{\big(
        c
        -d
        +2 l
        +g n
        -n q
        +d i_1
        -3 l i_1
        +n q i_1
        +l i_1^2
\big) i_1}
(-a
+b
-2 e
-f n
\nonumber\\&&
+h n
+j n
-j n^2
+6 p
-2 n r
-n s
+n^2 s
-b i_1
+3 e i_1
-h n i_1
-11 p i_1
+3 n r i_1
\nonumber\\&&
+n s i_1
-n^2 s i_1
-e i_1^2
+6 p i_1^2
-n r i_1^2
-p i_1^3
)
\bigg)\Bigr].
\end{eqnarray}
Eq.~(\ref{eq:HORN2}) can be rewritten as
\begin{eqnarray} 
f[n,m] &=& \frac{\displaystyle (-1)^{m+n}} 
{\displaystyle m! n! \Biggl(
        \frac{c_1}{d_1}\Biggr)_n \Biggl(
        \frac{c}{d}
        +\frac{g n}{d}
\Biggr)_m}
\Biggl(
        \frac{e}{d}\Biggr)^m \Biggl(
        \frac{e_1}{d_1}\Biggr)^n \Biggl(
        -\frac{1}{2}
        +\frac{b_1}{2 e_1}
        -\frac{r_1}{2 e_1}
\Biggr)_n 
\nonumber\\ && \times
\Biggl(
        -\frac{1}{2}
        +\frac{b_1}{2 e_1}
        +\frac{r_1}{2 e_1}
\Biggr)_n \Biggl(
        -\frac{1}{2}
        +\frac{b}{2 e}
        +\frac{h n}{2 e}
        -\frac{r_2}{2 e}
\Biggr)_m \Biggl(
        -\frac{1}{2}
        +\frac{b}{2 e}
        +\frac{h n}{2 e}
        +\frac{r_2}{2 e}
\Biggr)_m,
\nonumber\\
\label{eq:FMN}
\end{eqnarray}
with
\begin{eqnarray}\label{Equ:AlgebraicExt}
r_1 &=& \sqrt{(b_1 - e_1)^2 - 4 a_1 e_1}
\\
r_2 &=& 
{
\sqrt{(b - 3 e + h n)^2 - 4 e (a - b + 2 e + f n - h n - j n + j n^2)}.
}
\end{eqnarray}
They can be factorized into the usual Pochhammer representation if appropriate replacements are obeyed.

In the tri--variate cases one obtains
\begin{eqnarray}
	f[m,n,p] &=&
\bigg(
        \prod_{i_1=1}^m \frac{-A
        + B_1
        -2  E_1
        - B_1 i_1
        +3  E_1 i_1
        - E_1 i_1^2
        }{\big(
                B_0 
                -E_0 
                +E_0  i_1
        \big) i_1}
\bigg)
\nonumber\\&& \times
\bigg[\prod_{i_1=1}^n 
\bigg(
\frac{1}{\big(
                 C'_0 
                - F'_0 
                + H'_2  m
                + F'_0  i_1
        \big) i_1}
		(- A' 
        + C'_1 
        -2  F'_1 
        - B'_1  m
\nonumber\\&&        
        + E'_1  m
        + H'_1  m
        - E'_1  m^2
        - C'_1  i_1
        +3  F'_1  i_1
        - H'_1  m i_1
        - F'_1  i_1^2
        )  
\bigg)              
        \bigg]
\nonumber\\&& \times   
\bigg[     
         \prod_{i_1=1}^p 
\bigg(         
         \frac{1}{\big(
        D''_0 
        - G''_0 
        + L''_2  m
        +n  S''_2 
        + G''_0  i_1
\big) i_1}
(-A'' 
+ D''_1 
-2  G''_1 
- B''_1  m
+ E''_1  m
\nonumber\\&&
+ L''_1  m
- E''_1  m^2
- C''_1  n
+ F''_1  n
- H''_1  m n
- F''_1  n^2
+n  S''_1 
- D''_1  i_1
+3  G''_1  i_1
- L''_1  m i_1
\nonumber\\&&
-n  S''_1  i_1
- G''_1  i_1^2
)
\bigg)
\bigg].
\end{eqnarray}

Finally, in the four--variable case the product solution reads 
\begin{eqnarray}
f[m,n,p,q] &=&	\bigg(
        \prod_{i_1=1}^m 
        \frac{-A
        +B_1
        -2 F_1
        -B_1 i_1
        +3 F_1 i_1
        -F_1 i_1^2
        }{\big(
                B_0
                -F_0
                +F_0 i_1
        \big) i_1}
\bigg)
\nonumber\\&&\times
\bigg[
\prod_{i_1=1}^n
\frac{1}{\big(
                C'_0
                -G'_0
                +m M'_2
                +G'_0 i_1
        \big) i_1}
        (-A'
        +C'_1
        -2 G'_1
        -B'_1 m
        +F'_1 m
\nonumber\\&&        
        -F'_1 m^2
        +m M'_1
        -C'_1 i_1
        +3 G'_1 i_1
        -m M'_1 i_1
        -G'_1 i_1^2
        )
\bigg]
\nonumber\\&&\times
\bigg[
\prod_{i_1=1}^p 
\frac{1}{\big(
                D''_0
                -H''_0
                +m N''_2
                +n Q''_2
                +H''_0 i_1
        \big) i_1}
(-A''
        +D''_1
        -2 H''_1
        -B''_1 m
        +F''_1 m
\nonumber\\&&        
        -F''_1 m^2
        -C''_1 n
        +G''_1 n
        -m M''_1 n
        -G''_1 n^2
        +m N''_1
        +n Q''_1
        -D''_1 i_1
        +3 H''_1 i_1
\nonumber\\&&        
        -m N''_1 i_1
        -n Q''_1 i_1
        -H''_1 i_1^2
        )        
        \bigg]
\nonumber\\&&\times
\bigg[
         \prod_{i_1=1}^q \frac{1}{\big(
        E'''_0
        -L'''_0
        +m P'''_2
        +n R'''_2
        +p S'''_2
        +L'''_0 i_1
\big) i_1}
(-A'''
+E'''_1
-2 L'''_1
-B'''_1 m
\nonumber\\&& 
+F'''_1 m
-F'''_1 m^2
-C'''_1 n
+G'''_1 n
-m M'''_1 n
-G'''_1 n^2
-D'''_1 p
+H'''_1 p
-m N'''_1 p
\nonumber\\&& 
-H'''_1 p^2
+m P'''_1
-n p Q'''_1
+n R'''_1
+p S'''_1
-E'''_1 i_1
+3 L'''_1 i_1
-m P'''_1 i_1
-n R'''_1 i_1
\nonumber\\&& 
-p S'''_1 i_1
-L'''_1 i_1^2
)
\bigg].
\end{eqnarray}

\subsection{Computing the expansion in $\ep$}
\label{sec:epExpansion}

Algorithms for the series expansion of hypergeometric series would be of interest for their applicability to physics. One may try to obtain the $\ep$-expansion of the hypergeometric series as nested sums using the difference ring methods of {\tt EvaluateMultiSums}, but this is not always possible. An alternative is to try to obtain the series expansion of the summand and to sum the terms of the expansion. For an explanation of an algorithms which can work on nested products of the type discussed in Section \ref{sec:solveProd}, we refer to \cite[Sec. 5]{Blumlein:2021hbq}. For these methods to be applicable, the convergence region of the hypergeometric function needs to be understood, as well as the conditions under which the summation quantifier commutes with the differential operator in $\ep$.

Here we reproduce from \cite{Blumlein:2021hbq} two examples in which these methods are employed to obtain the $\ep$ expansion of hypergeometric functions.

\subsubsection{Example 1}

Consider for example the system of equations
\begin{eqnarray}
	\Bigg[ (x-1) y \partial_{x,y}^2+ \Big[x \Big(2 \varepsilon +\frac{7}{2}\Big)-\varepsilon +1\Big]\partial_x + (x-1) x \partial_x^2
&\nonumber\\
	+y (2 \varepsilon +1) \partial_y +\frac{3}{2} (2 \varepsilon +1) \Bigg] f(x,y) &= 0, \\
	\Bigg[ x (y-1) \partial_{x,y}^2 +x (4-\varepsilon ) \partial_x+ \Big[y \Big(\frac{13}{2}-\varepsilon \Big)-\varepsilon +1\Big]\partial_y 
	&\nonumber\\
	+(y-1) y \partial_y^2+\frac{3 (4-\varepsilon )}{2}
	\Bigg]f(x,y) &= 0,
\end{eqnarray}
for which we search for a solution of the form~\eqref{eq:hyp-series} with $r=2$ where $x_1=x$ and $x_2=y$. 
Computing a first--order recurrence system of $A(n_1,n_2)=A(m,n)$ and solving it by the method presented in Section~\ref{sec:solveProd} provides the solution
\begin{equation}
	f(x,y) = \sum_{m,n= 0}^\infty A(m,n) = \sum_{m,n= 0}^\infty \frac{x^m y^n \big(
        \frac{3}{2}\big)_{m
+n
} (4-\varepsilon )_n (1+2 \varepsilon )_m}{m! n! (-1+\varepsilon )_{m
+n
}}.
\label{eq:example-summand}
\end{equation}
A series expansion of the summand $A(m,n)$ in \eqref{eq:example-summand} up to $\mathcal O(\varepsilon^0)$ gives
\begin{eqnarray}
A(m,n)&=&	-\frac{1}{6} \frac{x^m y^n (3+n)! \big(
        \frac{3}{2}\big)_{m
+n
}}{n! (-2
+m
+n
)! \varepsilon}
+
\frac{1}{36} \bigg[
        -\frac{1}{(1+n) (2+n) (3+n) (m
        +n
        ) (-1
        +m
        +n
        )} 
\nonumber\\&& \times        
        \big(
                -36
                -30 n
                +17 n^2
                +97 n^3
                +79 n^4
                +17 n^5
                +m^2 \big(
                        36+115 n+84 n^2+17 n^3\big)
\nonumber\\&&                
                +m \big(
                        36+89 n
                        +218 n^2
                        +163 n^3+34 n^4\big)
        \big)
        -12 S_1({m})
        +6 S_1({n})
        +6 S_1({m+n})
\bigg] 
\nonumber\\&& \times
\frac{x^m y^n (3+n)! \big(
        \frac{3}{2}\big)_{m
+n
}}{n! (-2
+m
+n
)!}
+\mathcal O(\varepsilon).
\end{eqnarray}
A series expansion of \eqref{eq:example-summand} in the region $0<x<\sqrt{y}$, $0<y<\frac{1}{2}$,
\begin{equation}
	f(x,y) = \frac{1}{\varepsilon} f_{-1}(x,y) + f_0(x,y) +\mathcal O(\varepsilon)
\end{equation}
 is possible using {\tt EvaluateMultiSum} and results in an expression involving the sums
\begin{eqnarray}
	R_0 &=& \sum_{i=1}^{\infty } \frac{x^i \big(\frac{3}{2}\big)_i}{i!} = -1+\frac{1}{(1-x)^{3/2}}
\\
	R_1 &=& \sum_{i=1}^{\infty } \frac{y^i \big(\frac{3}{2}\big)_i}{i!} = -1+\frac{1}{(1-y)^{3/2}}
\end{eqnarray}
at $\mathcal O(\varepsilon^{-1})$. The function $f_{-1}(x,y)$ reads
\begin{eqnarray}
	f_{-1}(x,y) &=& -\frac{15 x^6 }{4 (x-y)^4(1-x)^{7/2}}
-\frac{15 y^3}{64 (x-y)^4 (1-y)^{13/2}} \big[
        y^3 \big(
                160+80 y-10 y^2+y^3\big)
\nonumber\\&&                                
        -x y^2 \big(
                576+176 y
                -64 y^2+5 y^3\big)
        +x^3 \big(
                -320+120 y-36 y^2+5 y^3\big)
\nonumber\\&&
        +3 x^2 y \big(
                240+8 y-22 y^2+5 y^3\big)
\big] .
\end{eqnarray}

In addition, one encounters at $\mathcal O(\varepsilon^0)$ the sums 
\begin{eqnarray}
R_2 &=& \sum_{i=1}^{\infty } \frac{x^i \big(
        \frac{3}{2}\big)_i}{(1+2 i)^2 i!} = -1 + \frac{1}{\sqrt{x}} \arcsin\big( \sqrt{x}\big)
\\ 
R_3 &=& \sum_{i=1}^{\infty } \frac{y^i \big(
        \frac{3}{2}\big)_i}{i i!} = -2
+2 \ln(2)
+2 \frac{1}{\sqrt{1-y}}
-2 \HA_{-1}\big(
        \sqrt{1-y}\big)
\\ 
R_4 &=& \sum_{i_1=1}^{\infty } \frac{x^{i_1} \big(\frac{3}{2}\big)_{i_1}
}{i_1!} 
        \sum_{i_2=1}^{i_1} \frac{1}{1+2 i_2}
= \frac{1}{2} \frac{\HA_1(x)}{(1-x)^{3/2}}
\\ 
R_5 &=& \sum_{i_1=1}^{\infty } \frac{x^{i_1} 
\big(\frac{3}{2}\big)_{i_1}}{i_1!} 
        \sum_{i_2=1}^{i_1} \frac{y^{-i_2} i_2!}{\big(
                \frac{3}{2}\big)_{i_2}}
\nonumber\\&=&
\frac{y}{(1-x) (y-x)}
-\frac{1}{(1-x)^{3/2}}
+\frac{y}{2(1-x)^{3/2} \sqrt{1-y}} \Big[
        i \pi 
        -\HA_0\big( \sqrt{1-y} -\sqrt{1-x} \big)
\nonumber\\&&
        -2 \HA_{-1}\big( \sqrt{1-y}\big)
        +\HA_0(y)
        +\HA_0\big( \sqrt{1-y}+\sqrt{1-x} \big)
\Big] 
\\ 
R_6 &=& \sum_{i_1=1}^{\infty } \frac{x^{i_1} y^{i_1}
\big(\big(
                \frac{3}{2}\big)_{i_1}\big)^2}{\big(
        i_1!\big)^2}  
        \sum_{i_2=1}^{i_1} \frac{y^{-i_2} i_2!}{\big(
                \frac{3}{2}\big)_{i_2}}
\nonumber\\&=&
\frac{1}{2}\int_0^1 dt \bigg[
\frac{-1+t}{\pi  (-1
+t
+y
)} \Big[
        \frac{1}{(1
        -(1-t) x
        )^2}
         \big[
                4 E(x -t x)
                -2 [1
                -(1-t) x
                ] K(x-t x)
        \big]
\nonumber\\&&
        -\frac{4 E(x y)
        +2 (-1
        +x y
        ) K(x y)
        }{(-1
        +x y
        )^2}
\Big] \frac{1}{\sqrt{t}}
\bigg]
\\ 
R_7 &=& \sum_{i_1=1}^{\infty } \frac{x^{i_1} y^{i_1} 
\big(\big(
                \frac{3}{2}\big)_{i_1}\big)^2}{\big(
        i_1!\big)^2 \big(
        1+2 i_1\big)^2} 
        \sum_{i_2=1}^{i_1} \frac{y^{-i_2} i_2!}{\big(
                \frac{3}{2}\big)_{i_2}}
=
{ \frac{1}{\pi} } \int_0^1 dt \frac{t-1}{\sqrt{t}(t+y-1) } \big[ K(x(1-t))
-K(x y) \big]
\nonumber\\&&
\\ 
R_8 &=& \sum_{i_1=1}^{\infty } \frac{y^{i_1} 
\big(\frac{3}{2}\big)_{i_1}}{i_1! \big(
        1+2 i_1\big)} 
        \sum_{i_2=1}^{i_1} \frac{x^{i_2} y^{-i_2}}{i_2}
=
-2 \frac{\HA_0\big(
        \sqrt{1-x}+\sqrt{1-y}\big)}{\sqrt{1-y}}
+2 \frac{\HA_{-1}\big(
        \sqrt{1-y}\big)}{\sqrt{1-y}}
\\ 
R_9 &=& \sum_{i_1=1}^{\infty } \frac{x^{i_1} 
\big(\frac{3}{2}\big)_{i_1}}{i_1!} \big(
        \sum_{i_2=1}^{i_1} \frac{y^{-i_2} i_2!}{\big(
                \frac{3}{2}\big)_{i_2}}
\big)
\big(
        \sum_{i_2=1}^{i_1} \frac{y^{i_2} \big(
                \frac{3}{2}\big)_{i_2}}{i_2!}
\big)
\nonumber\\&=&
\frac{1}{(1-y)^{5/2}} \Big[ \left(\frac{1}{(1-x)^{3/2}}-1\right) \left((1-y)^{3/2}-1\right) 
\biggl(\sqrt{1-y} y
   \Big(\HA_{-1}\big(\sqrt{1-y}\big)
\nonumber\\&&   
   -\frac{\HA_0(y)}{2}+\frac{i \pi }{2}\Big)-y+1\biggr) \Big]
+\sum_{i_ 1=1}^{\infty } \bigg\{
        \frac{1}{\pi  
(1-y)^2 \sqrt{1- y} \Gamma \big(
                1+i_ 1\big) \Gamma \big(
                2+i_ 1\big)} 
\nonumber\\&& \times
                \bigg[ 4 x^{i_ 1} y^{1+i_ 1} \Big[
                1
                -y
                +\sqrt{1-y} y \Big(
                        \frac{i \pi }{2}
                        -\frac{1}{2} \HA_ 0(y)
                        +\HA_ {-1}\big(
                                \sqrt{1-y}\big)
                \Big)
        \Big] \Gamma \big(
                \frac{3}{2}+i_ 1\big) 
\nonumber\\&& \times                
                \Gamma \big(
                \frac{5}{2}+i_ 1\big)
                 \,_ 2F_ 1\Big(
                -\frac{1}{2},1+i_ 1;2+i_ 1;y\Big) \bigg]
        -\frac{1}{(-1+y) 
\sqrt{\pi -\pi  y} \Gamma \big(
                1+i_ 1\big)}
\nonumber\\&& \times                                
                 \Big[2 x^{i_ 1} \Gamma \big(
                \frac{3}{2}+i_ 1\big) 
                \,_ 2F_ 1\Big(
                -\frac{1}{2},1+i_ 1;2+i_ 1;y\Big) 
                \,_ 2F_ 1\Big(
                1,2+i_ 1;\frac{5}{2}+i_ 1;\frac{1}{y}\Big) \Big]
\nonumber\\&&
        -\frac{1}{(-1+y) \sqrt{\pi -\pi  y} \big(
                3+2 i_ 1\big)}           
                \Big[ 2 \sqrt{\pi } x^{i_ 1} y^{-1-i_ 1} \big(
                -1+(1
                -y
                )^{3/2}\big)
\nonumber\\&& \times                
                 \,_ 2F_ 1\Big(
                1,2+i
                _ 1;\frac{5}{2}+i_ 1;\frac{1}{y}
        \Big)
\big(1+i_ 1\big) \Big] 
\bigg\}
\\ 
R_{10} &=& \sum_{i_1=1}^{\infty } \frac{y^{i_1} \big(
        \frac{3}{2}\big)_{i_1} S_1\big({i_1}\big) i_1}{i_1!} 
=
-3 \ln(2) y \frac{1}{(1-y)^{5/2}}
+\frac{3}{2} y \frac{\HA_1(y)}{(1-y)^{5/2}}
\nonumber\\&&
+3 y \frac{\HA_{-1}\big(
        \sqrt{1-y}\big)}{(1-y)^{5/2}}
+\Big[
        1
        +y \big(
                3-2 \sqrt{1
                -y
                }\big)
        -\sqrt{1-y}
\Big] \frac{1}{(1-y)^{5/2}},
\end{eqnarray}
as well as the combination
\begin{eqnarray}
R_{11} &=&\big( 1-\big(1-x\big)^{3/2} \big)
        \sum_{i_1=1}^{\infty } \frac{y^{i_1} 
\big(\frac{3}{2}\big)_{i_1}}{i_1! \big(
                1+2 i_1\big)}
                \sum_{i_2=1}^{i_1} \frac{y^{-i_2} i_2!}{\big(
                        \frac{3}{2}\big)_{i_2}}
\nonumber\\&&
        -\big(
                1-x\big)^{3/2} 
        \sum_{i_1=1}^{\infty } \frac{y^{i_1} 
\big(\frac{3}{2}\big)_{i_1}}{i_1! \big(
                1+2 i_1\big)}\Big(
                \sum_{i_2=1}^{i_1} \frac{y^{-i_2} i_2!}{\big(
                        \frac{3}{2}\big)_{i_2}}
        \Big)
\Big(
                \sum_{i_2=1}^{i_1} \frac{x^{i_2} \big(
                        \frac{3}{2}\big)_{i_2}}{i_2!}
        \Big)
\nonumber\\&=&
\frac{1}{4} \bigg\{
        \Big[
                2
                -2 \sqrt{1-x}
                +2 x \sqrt{1-x}
                -3 x F_1\Big(
                        \frac{5}{2};\frac{1}{2},1;2;x y,x
                \Big)
\big(1-x\big)^{3/2}
        \Big]
\big[i \pi  y
\nonumber\\&&
                +2 \sqrt{1-y}                
                -y \HA_0(y)
                +2 y \HA_{-1} \big(\sqrt{1-y}\big)
        \big]\bigg\} \frac{1}{\sqrt{1-y}}
\nonumber\\&&
-
\sum_{i_1=1}^{\infty } \Big[ \frac{x^{1+ i_1} \big(
        1-x\big)^{3/2} \Gamma \big(
        \frac{1}{2}+i_1\big)  }{\sqrt{\pi } y \Gamma \big(
        1+i_1\big)}     
        \,_2F_1\Big(
        1,2+i_1;\frac{5}{2}+i_1;\frac{1}{y}\Big)
\nonumber\\&& \times
        \,_2F_1\Big(
        1,\frac{5}{2}+i_1;2+i_1;x\Big) \Big].
\end{eqnarray}
The harmonic polylogarithms \cite{Remiddi:1999ew} are defined by Eqs. \eqref{eq:HPLdef1}-\eqref{eq:HPLdef3}.

One can further employ the relations
\begin{eqnarray}
_2F_1\Big(\frac{3}{2},\frac{3}{2};1,z\Big) &=& \frac{2(z-1)K(z)+4E(z)}{\pi(z-1)^2}
\\
K(z) &=& \int_0^1 \frac{1}{\sqrt{(1-t^2)(1-z t^2)}} dt = \frac{\pi}{2} \,_2F_1\Bigr(\frac{1}{2},\frac{1}{2};1;z  \Bigl)
\\
E(z) &=& \int_0^1 \frac{\sqrt{1-z t^2}}{\sqrt{1-t^2}} dt = \frac{\pi}{2} \,_2F_1\Bigl(-\frac{1}{2},\frac{1}{2},1,z  \Bigr).
\end{eqnarray}
The function $f_0(x,y)$ reads
{
\begin{eqnarray}
f_0(x,y) &=& 
\frac{5 R_{10} y^2}{16 (1-y)^4 (x
-y
)^4} \Bigl[
        y^3 \big(
                160+80 y-10 y^2+y^3\big)
        -x y^2 \big(
                576+176 y-64 y^2+5 y^3\big) 
\nonumber\\&&
        +x^3 \big(
                -320+120 y-36 y^2+5 y^3\big)
        +3 x^2 y \big(
                240+8 y-22 y^2+5 y^3\big)
\Bigr]
\nonumber\\&&
-\frac{15 R_8 y^3}{32 (1-y)^6 (x
-y
)^4} \Bigl[
        y^3 \big(
                160+80 y-10 y^2+y^3\big)
        -x y^2 \big(
                576+176 y-64 y^2+5 y^3\big)
\nonumber\\&&
        +x^3 \big(
                -320+120 y-36 y^2+5 y^3\big)
        +3 x^2 y \big(
                240+8 y-22 y^2+5 y^3\big)
\Bigr]
\nonumber\\&&
+\frac{1}{128 (1-x)^2 (1-y)^6 y (x
-y
)^4} \big(
        -960 x^7 (-1+y)^7
        +y^5 \big(
                128-1344 y+1536 y^2
\nonumber\\&&                
                -4240 y^3+4110 y^4-223 y^5+33 
y^6\big)
        +8 x^6 y \big(
                4-24 y+60 y^2-4880 y^3+1860 y^4
\nonumber\\&&                
                -564 y^5+79 y^6\big)
        +x^5 y \big(
                -832+2752 y-1920 y^2+76160 y^3+69280 y^4-8772 
y^5
\nonumber\\&&
+2427 y^6-495 y^7\big)
        -x y^4 \big(
                512-4544 y+3264 y^2-34480 y^3+4240 y^4+3363 y^5
\nonumber\\&&                
                -141 
y^6+66 y^7\big)
        +x^3 y^2 \big(
                -512+384 y+11008 y^2+83680 y^3+169980 y^4+6287 
y^5
\nonumber\\&&
+5931 y^6+747 y^7-305 y^8\big)
        +x^2 y^3 \big(
                768-4736 y-2272 y^2-84288 y^3-56570 y^4
\nonumber\\&&                
                +11627 
y^5-3549 y^6+387 y^7+33 y^8\big)
        +x^4 y \big(
                128+1984 y-9792 y^2-29440 y^3
\nonumber\\&&                
                -180320 y^4-63768 
y^5+10536 y^6-8583 y^7+2055 y^8\big)
\big)
\nonumber\\&&
+R_1
 \biggl[
        -
        \frac{1}{128 (1-x)^2 (1-y)^5 y (x
        -y
        )^5} \Bigl[
                960 x^7 (-1+y)^6
                -y^6 \big(
                        128+4800 y-4640 y^2
\nonumber\\&&                        
                        +4480 y^3-270 y^4+33 
y^5\big)
                +2 x y^5 \big(
                        320+10944 y-4016 y^2+4664 y^3+1749 y^4
\nonumber\\&&                        
                        -101 
y^5+33 y^6\big)
                +x^6 y \big(
                        -448+1920 y-12800 y^2+5440 y^3+1920 y^4-628 
y^5+65 y^6\big)
\nonumber\\&&
                -x^2 y^4 \big(
                        1280+38080 y+20128 y^2-6304 y^3+18546 
y^4-4204 y^5+406 y^6+33 y^7\big)
\nonumber\\&&
                +2 x^3 y^3 \big(
                        640+15040 y+32080 y^2-9896 y^3+11559 y^4-3791 
y^5-491 y^6+169 y^7\big)
\nonumber\\&&
                -5 x^4 y^2 \big(
                        128+1856 y+11712 y^2+1856 y^3-2076 y^4+1727 
y^5-2058 y^6+448 y^7\big)
\nonumber\\&&
                +2 x^5 y \big(
                        64+320 y+8000 y^2+15360 y^3-12080 y^4+5164 
y^5-4170 y^6+935 y^7\big)
        \Bigr]
\nonumber\\&&
        -\frac{15 R_5 x^6 (1-y)^2}{2 (1-x)^2 y (x
        -y
        )^4}
\biggr]
+R_0 \biggl[
        \frac{1}{16 (1-x)^2 (1-y)^6 y (x
        -y
        )^5} \Bigl[
                -120 x^8 (-1+y)^7
\nonumber\\&&
                +y^6 \big(
                        -32+384 y+2988 y^2+140 y^3-15 y^4\big)
                +5 x y^5 \big(
                        32-352 y-2676 y^2-1772 y^3
\nonumber\\&&                        
                        -95 y^4+12 y^5\big)
                +5 x^2 y^4 \big(
                        -64+608 y+4688 y^2+7516 y^3+1723 y^4+100 
y^5-18 y^6\big)
\nonumber\\&&
                +x^6 y \big(
                        20-624 y+3324 y^2+8040 y^3+18380 y^4-6720 
y^5+2164 y^6-329 y^7\big)
\nonumber\\&&
                +5 x^3 y^3 \big(
                        64-448 y-4056 y^2-12396 y^3-6809 y^4-612 
y^5-10 y^6+12 y^7\big)
\nonumber\\&&
                -5 x^4 y^2 \big(
                        32-64 y-1888 y^2-9120 y^3-11664 y^4-1436 
y^5-158 y^6+40 y^7+3 y^8\big)
\nonumber\\&&
                +x^5 y \big(
                        32+416 y-2848 y^2-12000 y^3-46800 y^4-11880 
y^5+430 y^6-200 y^7+85 y^8\big)
\nonumber\\&&
                +x^7 \big(
                        -60+304 y-444 y^2-360 y^3-2780 y^4-1080 
y^5+1536 y^6-701 y^7+120 y^8\big)
        \Bigr]
\nonumber\\&&
        +\frac{15 R_3
         x^6}{4 (1-x)^2 (x
        -y
        )^4}
        -\frac{15 R_1 x^6 (1-y)}{2 (1-x)^2 y (x
        -y
        )^4}
\biggr]
+\frac{15 R_9 x^6 (1-y)^2}{2 (1-x)^2 y (x
-y
)^4}
\nonumber\\&&
+\frac{15 R_6 x^6 (1-y) (-1
+(2+x) y
)}{4 (1-x)^2 y (x
-y
)^4}
+\frac{15 R_3 x^6}{4 (1-x)^2 (x
-y
)^4}
+\frac{15 R_2 x^6 (1-y)}{4 (1-x)^2 y (x
-y
)^4}
\nonumber\\&&
+\frac{15 R_4 x^6 (1-y)}{2 (1-x)^2 y (x
-y
)^4}
-\frac{15 R_7 x^6 (1-y)}{4 (1-x)^2 y (x
-y
)^4}
-\frac{15 R_{11} x^6 (1-y) }{2 y (x
-y)^4 (1-x)^{7/2}}.
\end{eqnarray}

The sums $R_i$ could be treated using the methods of \cite{Ablinger:2014bra} which are encoded in {\tt HarmonicSums}.

\subsubsection{Example 2}
\label{sec:ex2}

Consider for example the system of equations
\begin{eqnarray}
	1+\varepsilon +(2-x+\varepsilon ) \partial_x+2 x (1+x) 
\partial_x^2=0\\
	2-\varepsilon +(1-2 y+2 \varepsilon ) \partial_y+y (3+y) \partial_y^2=0 ~.
\end{eqnarray}
We can write its solution as
\begin{equation}
	\mathcal F(x,y) = \sum_{x,y\ge 0}A(m,n) x^m y^n
\end{equation}
with
\begin{equation}
	A(m,n) = \bigg(
        \prod_{i_1=1}^m \frac{-6
        -\varepsilon 
        +7 i_1
        -2 i_1^2
        }{\big(
                \varepsilon 
                +2 i_1
        \big) i_1}\bigg) \prod_{i_1=1}^n \frac{-6
+\varepsilon 
+5 i_1
-i_1^2
}{\big(
        -2
        +2 \varepsilon 
        +3 i_1
\big) i_1}.
\end{equation}
The quantity $A(m,n)$ can also be expressed as
\begin{eqnarray}
	A(m,n) &=& \frac{(-1)^m  \big(
        -\frac{3}{4}-\frac{1}{4} \sqrt{1
        -8 \varepsilon 
        }\big)_m \big(
        \frac{1}{4} \big(
                -3+\sqrt{1
                -8 \varepsilon 
                }\big)\big)_m }{\big(
        1+\frac{\varepsilon }{2}\big)_m  \Gamma (1+m) }
\nonumber\\&&\times
\frac{ (-1)^n 3^{-n} \big(
        -\frac{3}{2}-\frac{1}{2} \sqrt{1
        +4 \varepsilon 
        }\big)_n \big(
        \frac{1}{2} \big(
                -3+\sqrt{1
                +4 \varepsilon 
                }\big)\big)_n}{\big(
        \frac{1}{3}+\frac{2 \varepsilon }{3}\big)_n \Gamma (1+n)}
\end{eqnarray}
and $\mathcal F(x,y)$ can be rewritten as 
\begin{eqnarray}
	\mathcal F(x,y) &=& \Big( \sum_{m\ge 0} x^m f_1(m,\varepsilon) \Big) \Big( \sum_{n\ge 0}y^n f_2(n,\varepsilon) \Big).
	\\
	&=& F_1(x,\varepsilon) F_2(y,\varepsilon).
\end{eqnarray}
Expanding $F_1$ and $F_2$ in a series in $\varepsilon$ using {\tt EvaluateMultiSums}, one can write an expression containing infinite (nested) sums. These are rewritten as iterated integrals following \cite{Ablinger:2014bra}. Two of the sums are written in semi-analytic form as definite integrals by writing part of the summand as the Mellin transform of a function. For example, we encounter the sum
\begin{equation}
	s_1=\sum_{i=1}^{\infty } \frac{(-1)^i x^i \big(
        -\frac{3}{2}+i\big)! \big(
        \sum_{j=1}^i \frac{1}{1+2 j}\big) S_1({i})}{i i!}.
\end{equation}
By isolating the term $i=1$ and applying the Legendre duplication formula
\begin{equation}
	\Gamma\Big(z+\frac{1}{2}\Big) = \sqrt{\pi} \frac{\Gamma(2z)}{2^{2z-1} \Gamma(z)}
\end{equation}
and the identity 
\begin{equation}
	\Gamma(2z) = \frac{1}{2} \binom{2z}{z} \Gamma(z) \Gamma(z+1)
\end{equation}
we write
\begin{eqnarray}
	s_1 &=& -\frac{1}{3} x \sqrt{\pi }
+
\sum_{i=1}^{\infty } \frac{(-1)^{1+i} 2^{-2 i} \sqrt{\pi } x^{1+i} 
\binom{2 i}{i}}{(1+i)^3 (3+2 i)}
+
\sum_{i=1}^{\infty } \frac{(-1)^{1+i} 2^{-2 i} \sqrt{\pi } x^{1+i} 
\binom{2 i}{i} 
\sum_{j=1}^i \frac{1}{1+2 j}}{(1+i)^3}
\nonumber\\&&
+
\sum_{i=1}^{\infty } \frac{(-1)^{1+i} 2^{-2 i} \sqrt{\pi } x^{1+i} 
\binom{2 i}{i} S_1({i})}{(1+i)^2 (3+2 i)}
\nonumber\\&&
+
\sum_{i=1}^{\infty } \frac{(-1)^{1+i} 2^{-2 i} \sqrt{\pi } x^{1+i} 
\binom{2 i}{i} \big(
        \sum_{j=1}^i \frac{1}{1+2 j}\big) S_1({i})}{(1+i)^2}.
\end{eqnarray}
The first three sums are treated following \cite{Ablinger:2014bra}. The fourth sum can be written as
\begin{eqnarray}
	t_1 &=& \sum_{i=1}^{\infty } \frac{(-1)^{1+i} 2^{-2 i} \sqrt{\pi } x^{1+i} 
\binom{2 i}{i} \big(
        \sum_{j=1}^i \frac{1}{1+2 j}\big) S_1({i})}{(1+i)^2}
\nonumber\\&=&
        \sum_{i=1}^\infty \frac{(-1)^{1+i}2^{-2i}\sqrt{\pi} {x^{1+i}} \binom{2i}{i}}{(1+i)^2} \int_0^1 dz \bigg\{ (z^i-1) \frac{1}{2 (-1+z)} \bigg[
        -2
        +2 z
        +\big(
                1+\sqrt{z}\big) G\Big(
                \frac{\sqrt{\tau }}{1-\tau };z\Big)
\nonumber\\&&                
        +2 \sqrt{z} \big(1-\ln (2)\big)
\bigg] \bigg\}
\nonumber\\&=&
\int_0^1 dz \bigg\{ { \frac{1}{2(z-1)} } \bigg[
        -2
        +2 z
        +\big(
                1+\sqrt{z}\big) G\Big(
                \frac{\sqrt{\tau }}{1-\tau };z\Big)    
        +2 \sqrt{z} \big(1-\ln (2)\big)
\bigg]
\nonumber\\&& \times
\sum_{i=1}^\infty \frac{\big(
        -1+z^i\big) (-1)^{1+i} 2^{-2 i} \sqrt{\pi } x^{1+i} \binom{2 i}{i}}{(1+i)^2}
\bigg\}
\nonumber\\&=&
\int_0^1 dz \bigg\{ { \frac{1}{2(z-1)} }
\bigg[
        -2
        +2 z
        +\big(
                1+\sqrt{z}\big) G\Big(
                \frac{\sqrt{\tau }}{1-\tau };z\Big)    
        +2 \sqrt{z} \big(1-\ln (2)\big)
\bigg]
\bigg[
-\frac{4}{z} \Big[
        -1
        +z
\nonumber\\&&        
        +\sqrt{1
        +x z
        }
        -z \sqrt{1+x}
        +z \HA_0\Big(
                \frac{1}{2} \big(
                        1+\sqrt{1+x}\big)\Big)
        -\HA_0\big(
                \frac{1}{2} \Big(
                        1+\sqrt{1+x z}\big)\Big)
\Big] \sqrt{\pi }
\bigg]
\bigg\}.
\end{eqnarray}
The $\varepsilon$ expansion of $F_1(x,\varepsilon)$ and $F_2(x,\varepsilon)$ then can be written by
\begin{eqnarray}
	F_1(x,\varepsilon) &=& 1
-
\frac{x}{2}
+\varepsilon  \biggl\{
        -1
        +\sqrt{1+x}
        +\frac{1}{4} x \big(
                -9+4 \sqrt{1
                +x
                }\big)
        +\frac{1}{2} (-2+x) \HA_0(x)
\nonumber\\&&        
        +\frac{1}{2} (2-x)  G_{3}(x)
\biggr\}
+\varepsilon ^2 \Biggl\{
        \frac{1}{8} \Bigl[
                20 \big(
                        -1+\sqrt{1
                        +x
                        }\big)
                +x \big(
                        -33
                        +4 x
                        +20 \sqrt{1+x}
                \big)
        \Bigr]
\nonumber\\&&
        +(-2+x) \Bigl(
                -\frac{1}{4} \HA_0(x)^2
                +\frac{1}{4} \HA_{0,-1}(x)
                +\frac{ G_{11}(x)}{4}
                -\frac{ G_{12}(x)}{4}
                +\frac{ G_{5}(x)^2}{8}
        \Bigr)
\nonumber\\&&        
        +\frac{1}{2} (2-x) \bigl( G_{8}(x)
        + G_{9}(x)
        \bigr)
        +\frac{1}{4} (-4+13 x) \HA_0(x)
        +\frac{1}{2} (1+x)^{3/2} \HA_{-1}(x)
\nonumber\\&&        
        +\biggl[
                1
                -\frac{13 x}{4}
                +\frac{1}{2} (-2+x) \HA_0(x)
        \biggr]  G_{3}(x)
        +\biggl[
                \frac{1}{2} (1+x)^{3/2}
                +\frac{1}{4} (2-x) \HA_0(x)
        \biggr]  G_{5}(x)
\Biggr\}
\nonumber\\&&
+\varepsilon ^3 \Biggl\{
        \frac{ G_{5}(x)^2}{16} \Bigl[
                x \big(
                        17-4 \sqrt{1
                        +x
                        }\big)
                -4 \big(
                        3+\sqrt{1
                        +x
                        }\big)
        \Bigr]
        + G_{11}(x) \Bigl[
                \frac{1}{8} \big(
                        x \big(
                                13-6 \sqrt{1
                                +x
                                }\big)
\nonumber\\&&                                
                        -6 \sqrt{1+x}
                \big)
                +\frac{1}{8} (-2+x) \HA_0(x)
        \Bigr]
        + G_{12}(x) \biggl[
                \frac{1}{8} \Bigl[
                        2 \big(
                                6+\sqrt{1
                                +x
                                }\big)
                        +x \big(
                                -17+2 \sqrt{1
                                +x
                                }\big)
                \Bigr]
\nonumber\\&&                
                +\frac{3}{8} (-2+x) \HA_0(x)
        \biggr]
        + G_{3}(x) \biggl[
                \frac{9}{2}
                -\frac{105 x}{8}
                +\frac{1}{4} (-14+17 x) \HA_0(x)
                +\frac{1}{4} (2-x) \HA_0(x)^2
        \biggr]
\nonumber\\&&        
        + G_{5}(x) \Biggl[
                (-2+x) \Bigl(
                        -\frac{1}{16} \HA_0(x)^2
                        +\frac{1}{8} \HA_{0,-1}(x)
                        +\frac{3  G_{11}(x)}{8}
                        -\frac{ G_{12}(x)}{8}
                \Bigr)
\nonumber\\&&                
                +\frac{1}{4} \biggl[
                        -6 x
                        -2 x^2
                        +(1+x) (5+16 x) \sqrt{1+x}
                \biggr]
                +\frac{1}{8} \biggl[
                        x \big(
                                -13+6 \sqrt{1
                                +x
                                }\big)
\nonumber\\&&                                
                        +6 \sqrt{1+x}
                \biggr] \HA_0(x)
                -\frac{1}{4} (1+x)^{3/2} \HA_{-1}(x)
        \Biggr]
        +(-2+x) \big(
                \frac{1}{12} \HA_0(x)^3
                -\frac{3}{8} \HA_{0,0,-1}(x)
\nonumber\\&&                
                -\frac{1}{8} \HA_{0,-1,-1}(x)
                -\frac{ G_{18}(x)}{8}
                -\frac{5  G_{19}(x)}{8}
                -\frac{3  G_{20}(x)}{8}
                +\frac{ G_{21}(x)}{8}
                -\frac{3  G_{22}(x)}{4}
                +\frac{ G_{23}(x)}{4}
\nonumber\\&&                
                -\frac{1}{24}  G_{5}(x)^3
        \big)
        +\frac{1}{2} (2-x) ( G_{14}(x)
        + G_{15}(x)
        + G_{16}(x)
        + G_{17}(x)
        )
\nonumber\\&&        
        +\frac{1}{240} \Biggl\{
                -60 (-2+x) 
                        \int_0^1 dz \biggl\{
                                -\frac{1}{(-1+z) z}2 \sqrt{\pi } \Bigl[
                                        -2
                                        +2 z
                                        +\big(
                                                1+\sqrt{z}\big) G_1(z)
\nonumber\\&&
                                        -2 \sqrt{z} \bigl(-1+\ln (2)\bigr)
                                \Bigr]
\biggl[-1
                                        +z
                                        -\sqrt{1+x} z
                                        +\sqrt{1
                                        +x z
                                        }
                                        +z \HA_0\Big(
                                                \frac{1}{2} \big(
                                                        1+\sqrt{1+x}\big)\Big)
\nonumber\\&&                                                        
                                        -\HA_0\Big(
                                                \frac{1}{2} \big(
                                                        1+\sqrt{1+x 
z}\big)\Big)
                                \biggr] \biggr\}
                +180 x 
                        \int_0^1 dz \biggl\{
                                -
                                \frac{1}{4 \sqrt{1
                                +x z
                                }}\sqrt{\pi } x \big(
                                        -1+\sqrt{1
                                        +x z
                                        }
                                \big)
\biggl[4
\nonumber\\&&
                                        +2 \sqrt{z} \Bigl[
                                                -8
                                                +2 z^{3/2}
                                                -3 \sqrt{z} \big(
                                                        2
                                                        +\zeta_2
                                                \big)
                                                +z \big(6-4 \ln (2)\big)
                                                +8 \ln (2)
                                        \Bigr]
                                        +4 \HA_0(z)
\nonumber\\&&
                                        +4 \HA_1(z)
                                        +2 \Bigl[
                                                -3
                                                -4 \sqrt{z}
                                                +3 z
                                                +2 z^{3/2}
                                                +(-1+z) \HA_0(z)
                                                +(-1+z) \HA_1(z)
                                                +2 \ln (2)
\nonumber\\&&
                                                -2 z \ln (2)
                                        \Bigr]  G_{1}(z)
                                        +(-1+z)  G_{1}(z)^2
                                        -2 (-1+z)  G_{6}(z)
                                        -2 (-1+z)  G_{7}(z)
                                        +6 \zeta_2
                                \biggr]\biggr\}
\nonumber\\&&                                
                +\biggl[
                        x \Bigl[
                                -2215
                                +4 x \big(
                                        435
                                        +96 x
                                        -500 \sqrt{1+x}
                                \big)
                                +60 \sqrt{1+x}
                        \Bigr]
\nonumber\\&&                        
                        +2060 \big(
                                -1+\sqrt{1
                                +x
                                }\big)
                \biggr] \sqrt{\pi }
        \Biggr\} \frac{1}{\sqrt{\pi }}
        +\frac{3}{8} (-12+35 x) \HA_0(x)
        +\frac{1}{8} (14-17 x) \HA_0(x)^2
\nonumber\\&&        
        +\frac{1}{4} (11-16 x) (1+x)^{3/2} \HA_{-1}(x)
        +\frac{5}{8} (1+x)^{3/2} \HA_{-1}(x)^2
        +\frac{1}{8} \Bigl[
                x \big(
                        17-2 \sqrt{1
                        +x
                        }\big)
\nonumber\\&&                        
                -2 \big(
                        6+\sqrt{1
                        +x
                        }\big)
        \Bigr] \HA_{0,-1}(x)
        +8 \sqrt{x}  G_{10}(x)
        +8 \sqrt{x}  G_{13}(x)
        +\frac{ G_{2}(x)}{2}
\nonumber\\&&        
        +\Bigl[
                2 (17-4 x) \sqrt{x}
                -8 \sqrt{x}  G_{5}(x)
        \Bigr]  G_{4}(x)
        +2 (-9+4 x) \sqrt{x}  G_{4}(x)
        +\biggl[
                \frac{7}{2}
                -\frac{17 x}{4}
\nonumber\\&&                
                +\frac{1}{2} (-2+x) \HA_0(x)
        \biggr]  G_{8}(x)
        +\biggl[
                -\frac{7}{2} (-1+x)
                +\frac{1}{2} (-2+x) \HA_0(x)
        \biggr]  G_{9}(x)
\Biggr\}
\nonumber\\&&
+\mathcal{O}(\varepsilon^4),
\end{eqnarray}
\begin{eqnarray}
	F_2(y,\varepsilon) &=& 1
-2 y
-
\frac{1}{4} (-20+y) y \varepsilon 
+\varepsilon ^2 \biggl\{
        -\frac{1}{48} y \big(
                480-765 y-56 y^2+64 y^3+12 y^4\big)
\nonumber\\&&
        +\frac{1}{4} (-9+4 y) y^{2/3} (3+y)^{4/3}  G_{26}(y)
        +(1-2 y)  G_{30}(y)
\biggr\}
+\varepsilon ^3 \Biggl\{
        \frac{1}{192} y \big(
                3840-21453 y
\nonumber\\&&
-1672 y^2+1638 y^3+280 y^4-6 y^5\big)
        + G_{26}(y) \biggl[
                \frac{1}{16} \big(
                        243-108 y+2 y^2\big) y^{2/3} (3+y)^{4/3}
\nonumber\\&&
                +\frac{1}{6} (9-4 y) y^{2/3} (3+y)^{4/3} \HA_0(y)
                +\frac{1}{6} (-9+4 y) y^{2/3} (3+y)^{4/3} \HA_{-3}(y)
\nonumber\\&&
                +\frac{1}{270} \big(
                        -1215+108 y+4 y^2\big) y^{2/3}  G_{24}(y)
                +\frac{7}{5} (-1+2 y)  G_{25}(y)
                +\frac{2}{3} (-1+2 y)  G_{28}(y)
\nonumber\\&&
                -\frac{2}{3} (-1+2 y)  G_{29}(y)
        \biggr]
        +\big(
                -1215+108 y+4 y^2
        \big)
\biggl[-\frac{1}{270} y^{2/3}  G_{27}(y)
                -\frac{1}{270} y^{2/3}  G_{33}(y)
        \biggr]
\nonumber\\&&
        +(-1+2 y) \Bigl(
                -\frac{7  G_{34}(y)}{5}
                +\frac{2  G_{35}(y)}{3}
        \Bigr)
        +\biggl[
                \big(
                        \frac{2}{3}-\frac{4 y}{3}\big)  G_{31}(y)
                +\frac{2}{3} (-1+2 y)  G_{32}(y)
        \biggr]  G_{25}(y)
\nonumber\\&&
        +\frac{1}{20} \big(
                -52+204 y-5 y^2\big)  G_{30}(y)
        +\frac{1}{6} (-9+4 y) y^{2/3}
         (3+y)^{4/3}  G_{31}(y)
\nonumber\\&&
        -\frac{1}{6} (3+y) (-9+4 y) y^{2/3} \sqrt[3]{3+y}  G_{32}(y)
        -\frac{2}{3} (-1+2 y)  G_{36}(y)
\Biggr\}+\mathcal{O}(\varepsilon^4).
\end{eqnarray}
By multiplying $F_1(x)$ and $F_2(y)$ one obtains the series expansion of $\mathcal F(x,y)$, with $0<x<1, 0<y<1$.
The functions $G_i$ are iterated integrals \cite{Ablinger:2014bra} defined as in Eq. \eqref{eq:Gfunctions}, and they are listed in Appendix \ref{sec:ex2GL}.

\subsection{Example 3}

In \cite{Blumlein:2021hbq} we also considered as an example the system of two differential equations in two variables implied by the following differential operators,
\begin{eqnarray}
&&	-\alpha  \beta -x^6 \partial_x^6-\beta  x^5 \partial_x^5-5 x^5 y \partial_x^5 \partial_y-15 x^5 \partial_x^5-10 \beta  x^4 \partial_x^4-10 x^4 y^2 \partial_x^4 \partial_y^2-5 \beta  x^4 y \partial_x^4 \partial_y
\nonumber\\&&	
	-60 x^4 y \partial_x^4 \partial_y-66 x^4 \partial_x^4-26 \beta  x^3 \partial_x^3-10 x^3 y^3 \partial_x^3 \partial_y^3-10 \beta  x^3 y^2 \partial_x^3 \partial_y^2-90 x^3 y^2 \partial_x^3 \partial_y^2
\nonumber\\&&	
	-40 \beta x^3 y \partial_x^3 \partial_y-198 x^3 y \partial_x^3
   \partial_y-96 x^3 \partial_x^3-18 \beta  x^2 \partial_x^2-5 x^2 y^4 \partial_x^2 \partial_y^4-10 \beta  x^2 y^3 \partial_x^2 \partial_y^3
\nonumber\\&&   
   -60 x^2 y^3 \partial_x^2 \partial_y^3-60 \beta  x^2 y^2 \partial_x^2 \partial_y^2-198 x^2 y^2
   \partial_x^2 \partial_y^2-78 \beta  x^2 y \partial_x^2 \partial_y-192 x^2 y \partial_x^2 \partial_y
\nonumber\\&&   
   -38 x^2 \partial_x^2-\alpha  x \partial_x-2 \beta  x \partial_x+\gamma  \partial_x-x y^5 \partial_x \partial_y^5-5 \beta  x y^4 \partial_x
   \partial_y^4-15 x y^4 \partial_x \partial_y^4
\nonumber\\&&   
   -40 \beta  x y^3 \partial_x \partial_y^3-66 x y^3 \partial_x \partial_y^3-78 \beta  x y^2 \partial_x \partial_y^2-96 x y^2 \partial_x \partial_y^2-36 \beta  x y \partial_{x y}-38 x y \partial_{x y}
\nonumber\\&&   
   +y
   \partial_{x y}+x \partial_x^2-2 x \partial_x-\beta  y^5 \partial_y^5-10 \beta  y^4 \partial_y^4-26 \beta  y^3 \partial_y^3-18 \beta  y^2 \partial_y^2-2 \beta  y \partial_y,
\\  [1cm]
&& 
-\alpha  \beta_1 -\beta_1  x^5 \partial_x^5-x^5 y \partial_x^5 \partial_y-10 \beta_1  x^4 \partial_x^4-5 x^4 y^2 \partial_x^4 \partial_y^2-5 \beta_1  x^4 y \partial_x^4
   \partial_y-15 x^4 y \partial_x^4 \partial_y
\nonumber\\&&   
   -26 \beta_1  x^3 \partial_x^3-10 x^3 y^3 \partial_x^3 \partial_y^3-10 \beta_1  x^3 y^2 \partial_x^3 \partial_y^2-60 x^3 y^2 \partial_x^3 \partial_y^2-40 {\beta_1} x^3 y \partial_x^3 \partial_y
\nonumber\\&&   
   -66 x^3 y \partial_x^3 \partial_y-18 \beta_1  x^2 \partial_x^2-10 x^2 y^4 \partial_x^2 \partial_y^4-10 \beta_1  x^2 y^3 \partial_x^2 \partial_y^3-90 x^2 y^3 \partial_x^2
   \partial_y^3
\nonumber\\&&   
   -60 \beta_1  x^2 y^2 \partial_x^2 \partial_y^2-198 x^2 y^2 \partial_x^2 \partial_y^2-78 \beta_1  x^2 y \partial_x^2 \partial_y-96 x^2 y \partial_x^2 \partial_y-2 \beta_1  x \partial_x
\nonumber\\&&   
   -5 x
   y^5 \partial_x \partial_y^5-5 \beta_1  x y^4 \partial_x \partial_y^4-60 x y^4 \partial_x \partial_y^4-40 \beta_1  x y^3 \partial_x \partial_y^3-198 x y^3 \partial_x \partial_y^3
\nonumber\\&&   
   -78 \beta_1  x y^2 \partial_x
   \partial_y^2-192 x y^2 \partial_x \partial_y^2-36 \beta_1  x y \partial_{x y}+x \partial_{x y}-38 x y \partial_{x y}-y^6 \partial_y^6-\beta_1  y^5 \partial_y^5
\nonumber\\&&   
   -15 y^5 \partial_y^5-10 \beta_1 
   y^4 \partial_y^4-66 y^4 \partial_y^4-26 \beta_1  y^3 \partial_y^3-96 y^3 \partial_y^3-18 \beta_1  y^2 \partial_y^2-38 y^2 \partial_y^2-\alpha  y \partial_y
\nonumber\\&&   
   -2 \beta_1  y \partial_y+\gamma
    \partial_y+y \partial_y^2-2 y \partial_y.
\end{eqnarray}

Assuming a hypergeometric solution
\begin{equation}
	\mathcal F(x,y) = \sum_{m,n\ge 0} f(m,n) x^m y^n,
\end{equation}
the coefficients $f(m,n)$ must obey
\begin{eqnarray}
	(m+1) (\gamma +m+n) f(m+1,n)-(\beta +m) \left(\alpha +(m+n)^5+(m+n)^3\right) f(m,n)=0,\\
	(n+1) (\gamma +m+n) f(m,n+1)-(\beta_1 +n) \left(\alpha +(m+n)^5+(m+n)^3\right) f(m,n)=0.
\end{eqnarray}
Solving these two equations with the help of {\tt Sigma}, one obtains
\begin{eqnarray}
	f(m,n) &=& \bigg(
        \prod_{i_1=1}^n \frac{\big(
                -1
                +{\beta_1}
                +i_1
        \big)
\big(-2
                +\alpha 
                +8 i_1
                -13 i_1^2
                +11 i_1^3
                -5 i_1^4
                +i_1^5
        \big)}{\big(
                -1
                +\gamma 
                +i_1
        \big) i_1}\bigg) 
\nonumber\\&& \times        
        \prod_{i_1=1}^m \frac{\big(
        -1
        +\beta 
        +i_1
\big)
}{\big(
        -1
        +n
        +\gamma 
        +i_1
\big) i_1}
\big(-2
        +8 n
        -13 n^2
        +11 n^3
        -5 n^4
        +n^5
        +\alpha 
        +8 i_1
        -26 n i_1
\nonumber\\&&        
        +33 n^2 i_1
        -20 n^3 i_1
        +5 n^4 i_1
        -13 i_1^2
        +33 n i_1^2
        -30 n^2 i_1^2
        +10 n^3 i_1^2
        +11 i_1^3
        -20 n i_1^3
\nonumber\\&&        
        +10 n^2 i_1^3
        -5 i_1^4
        +5 n i_1^4
        +i_1^5
\big) ~.
\end{eqnarray}
This quantity cannot be analytically expressed as a product of Pochhammer symbols due to the high degree of the polynomials appearing.

\subsection{A brief descriptions of the commands of {\tt HypSeries}} 
\label{sec:C}
Techniques for the solution and classification of the hypergeometric systems \eqref{eq:DEsystem} are implemented in the Mathematica package {\tt HypSeries}, which is attached to the paper \cite{Blumlein:2021hbq}.
The package requires Sigma, EvaluateMultiSums and HarmonicSums \cite{Schneider:2001,SIG1,SIG2,HARMONICSUMS,EMSSP} to be loaded.

The commands 
\[ 
{\tt solveDE1},~~
{\tt solveDE2},~~ 
{\tt solveDE3},~~ 
{\tt solveDE4}
\]
check whether the corresponding set of one to four equations in as many variables has hypergeometric solutions by consulting internal lists of cases, i.e.\ those discussed in Sec. \ref{sec:delist}. The syntax is e.g.
\[{\tt {solveDE4}[\{eq1==0,eq2==0,eq3==0,eq4==0\},\{x,y,z,t\},\{m,n,p,q\}]}.\]

More general solutions are possible by using the command {\tt DEProductSolution}. One has to provide the required $n$ differential equations in the list 
\[ {\tt sys = \{eq1==0,...,eqn==0\}}.\] 
Then 
\[
{\tt DEProductSolution[sys,\{x,y,...\},\{m,n,...\}]}
\]
returns the respective expansion coefficient {\tt f[m,n,p,q]}. Here the tools described in Section~\ref{sec:solveProd} are utilized.

Conversely, from a Pochhammer ratio {\tt A = f[m,n,p,q]} the command
\[
{\tt {findDE}[A,\{x,y,...\},\{m,n,...\}]}
\]
returns the system of differential equations obeyed by
\[
f(x,y,...) = \sum_{m,n,...\ge =0}^\infty f[m,n,...] \, x^m y^n \cdots
\]
Given a differential equation {\tt eq} in $n$ variables, the command {\tt findRE} 
\[
{\tt findRE[eq==0,\{x,y,\ldots\}, f[m,n,\ldots]]\} }
\]
returns a corresponding recurrence for {\tt f[m,n,...]}. The last two commands implement the techniques presented in the beginning of Section~\ref{sec:recsol}.

The convergence conditions for a number of two- and three-variate hypergeometric functions can be accessed from internal tables using the commands {\tt findCond2} and  
{\tt findCond3}. One first has to determine the corresponding function label {\tt fcn}
via {\tt classifier2}, {\tt classifier3}, as e.g.
\[
\tt classifier3[f[m,n,p], \{x,y,z\}, \{m,n,p\}]
\]
returning {\tt fcn}. Then 
\[
 {\tt findCond3[fcn,\{x,y,z\}}]           
\]
returns the convergence conditions, which are in some cases given in implicit form and are taken from the tabulation in \cite{SRIKARL}.

The function {\tt CheckDE[sol,eq]} provides a way to check if an expression satisfies a differential equation in a series expansion. It returns a sum which, in that case, should be of higher order in $\ep$ if evaluated.

In \cite{Blumlein:2021hbq} a notebook is attached where the usage of the package is presented along with the examples from the previous Sections, together with Mathematica files containing the definitions of the hypergeometric series treated by the package, their convergence conditions, and the translation table to their differential systems.

\clearpage
\thispagestyle{empty}
~

\clearpage
\section{Partial difference equations with rational coefficients}
\label{sec:PLDEsolver}

We examine the problem of linear partial difference equations in several variables, with the solution space being that of rational functions, possibly containing harmonic sums or Pochhammer symbols in the numerator. The corresponding problem in one variable is widely studied, and algorithms exist to find not only rational solutions but also hypergeometric solutions in a wide class of cases \cite{Abramov:1971,Abramov:1989,Schneider:2004,Schneider:2005,Schneider:2005b,Abramov:2021}; these algorithms are implemented in the package {\tt Sigma}. It is well-known how univariate difference equations arise frequently in the calculation of Feynman diagrams, for example, in calculating the master integrals in many applications in QCD, one can reduce the problem to a difference equation, which, for OMEs, is in the Mellin variable $N$, see e.g. \cite{Vermaseren:2000we,Moch:2005uc,Bierenbaum:2007qe}.

In the multivariate case, there are fewer known algorithms that deal with difference equations than in the univariate case. We implemented in a Mathematica package called {\tt SolvePartialLDE} \cite{Blumlein:2021hbq} the algorithms of \cite{Kauers:2010,Kauers:2011} and describe here how we complemented them with flexible heuristic methods that may be useful, potentially, in future applications to Feynman integrals. As we will see in greater detail in the following sections, the basic idea is to constrain the denominator of the solution or at least parts of the denominator, in a way that we will describe more precisely. Once the denominator is constrained, the problem of finding the numerator can be reduced to that of a linear system of equations, and our program can accept an ansatz provided by a user for what type of object may appear in the numerator, chosen in the space of harmonic sums and/or Pochhammer symbols. The package can also solve the difference equation in a series expansion in one parameter, which in applications would be the dimensional parameter $\ep$, and can factor out a hypergeometric factor chosen by the user as we will elucidate in the following sections. The exposition in this chapter follows \cite{Blumlein:2021hbq}.

\subsection{Description of the basic problem}
A partial linear difference equation (PLDE) is an equation for an unknown function $y(n_1,\ldots,n_r)\in \mathbb K(n_1,\ldots,n_r)$, here a rational function in $r$ variables. We define the \textit{shift operators} 
 $N_{\mathbf s}$ with respect to the \textit{shift} $\mathbf s=(s_1,\ldots,s_r)\in \mathbb Z^r$ as:

\begin{equation}
	N_\mathbf s y(n_1,\ldots,n_R) = y(n_1+s_1,\ldots,n_r+s_r).
\end{equation}
A PLDE is then an equation of the type
\begin{equation}
	\sum_{\mathbf s \in S} a_\mathbf s N_\mathbf s y = f
\label{eq:PLDE}	
\end{equation}
with $a_{\mathbf s}$ and $f$ polynomials in $n_1,\ldots,n_r$ and $S$ is a finite subset of $\mathbb Z^r$ called the \textit{shift set} or \textit{structure set}. Because \eqref{eq:PLDE} is linear, its general solution is the sum of a particular solution and of a linear combination of solutions of the homogeneous equation with $f=0$.

An example of the type of equation under consideration is:
\begin{equation}
-(1 + k + n^2) y(n, k) + (4 + k + 2 n + n^2) y(1 + n, 2 + k) = 0.
\end{equation}
It has the shift set $S=\{(0,0), (1,2)\}$ and its coefficients are
\begin{eqnarray}
a_{(0,0)} &=& -(1+k+n^2),\\
a_{(1,2)} &=& (4 + k + 2 n + n^2).
\end{eqnarray}

For the purpose of the algorithms dealt with in our package, one must distinguish between \textit{periodic} and \textit{aperiodic} polynomials \cite{Kauers:2010,Kauers:2011}. A polynomial $p$ is periodic if there exist infinitely many shifts, mapping $p$ into $p'$, such that $\text{gcd}(p,p')\neq 1$. A polynomial is aperiodic if it is not periodic.
For example, with respect to the variables $\{n,k\}$, the polynomial $(n + k + 2)$ is periodic and the polynomial $(n^2 + k + 6)$ is aperiodic. Any polynomial can be factorized into a periodic and an aperiodic part. Given a PLDE, there are algorithms which constrain the periodic and the aperiodic part of the denominator of the solution. As we will describe further, the aperiodic part can always be constrained, but this is not guaranteed for the periodic part. In the following we describe our implementation choices in our Mathematica package \cite{Blumlein:2021hbq}.

\subsection{Denominator bounds}
\label{sec:denBounds}
The algorithms in \cite{Kauers:2010,Kauers:2011} aim at formulating a \textit{denominator bound} for the solution of \eqref{eq:PLDE}. A denominator bound is a polynomial $d$ such that for any solution $y=\frac{n}{p}$ of \eqref{eq:PLDE} it must be $p|d$. As observed already in the univariate case \cite{Abramov:1971,Abramov:1989}, obtaining a denominator bound is valuable for the following reason: a naive way to solve the PLDE is by formulating an ansatz for the solution, i.e.\ a rational function in the variables $n_1,\ldots,n_k$ with undetermined coefficients $c_\mathbf k$,
\begin{equation}
y(n_1,\ldots,n_r)=\frac{\sum\limits_{\mathbf k} c_\mathbf k \prod\limits_i n_i^{k_i} }{\sum\limits_{\mathbf k'} c_{\mathbf k'} \prod\limits_i n_i^{k'_i}}.
\label{eq:rational-ansatz}
\end{equation}
By plugging this ansatz in \eqref{eq:PLDE} and clearing denominators one obtains a set of constraints on the $c_\mathbf k$ and $c_{\mathbf k'}$ by imposing the equality of every monomial in the variables $n_i$ on both sides of the
equation. However, these equations are, in general, non-linear.

However, if the denominator bound can be obtained, then only an ansatz for the numerator is required and the  equations for the unknown coefficients are linear. The ansatz for the numerator can then be made to involve harmonic sums \cite{Vermaseren:1998uu,Blumlein:1998if} and/or Pochhammer symbols.

If we write the solution to a partial linear difference equation as $y=\frac{n}{uv}$ with $u$ aperiodic and $v$ periodic, it is always possible to calculate a bound $d_a$ for the aperiodic part $u$ of the denominator. We refer to \cite{Kauers:2010} for a description of how the aperiodic denominator bound is calculated. 

For the periodic part $v$ it is not always possible to obtain a complete denominator bound for a PLDE. This is illustrated for example by the equation
\begin{equation}
y(n+1,k) - y(n,k+1) = 0,
\label{eq:noDenBound}
\end{equation}
which is satisfied by $\frac{1}{(n+k)^\alpha}$ for any $\alpha \in \mathbb N$. Clearly, no polynomial can be a denominator bound for Eq. \eqref{eq:noDenBound}.
This example shows that it is impossible to formulate in all cases a complete denominator bound, because arbitrary powers of periodic factors can appear in the denominator of the solution of some equations. However, it is often possible to formulate a partial bound for $v$ and to deduce some properties of its unknown factors. (A partial bound is a bound for some, but not all, the periodic factors). Specifically, the algorithm in \cite{Kauers:2011} works by successively examining the periodic factors of $a_\mathbf p$ with $\mathbf p$ a ``corner point'' of the shift set of the equation (see \cite{Kauers:2011} for a definition). It guarantees that $v$ has the property
\begin{equation}
	v \mid (d_p \cdot v_{\text{semi-known}} \cdot v_{\text{unknown}})
\end{equation}
where $d_p$ is an explicitly given polynomial obtained through the algorithm, $v_{\text{semi-known}}$ is a polynomial whose factors are to be taken from $\{N_\mathbf s (p^m) | \mathbf s \in \mathbb Z^r, m \in \mathbb Z\}$, where $p\in P$ and $P$ is a subset of the periodic factors of the coefficients of the corner points, which is identified by the algorithm; and $v_{\text{unknown}}$ is a polynomial such that $\text{spread}(v_{\text{unknown}}) = V$ and $V$ is a lattice in $\mathbb Z^r$ identified by the algorithm. The spread of a polynomial $u$ is defined by
\begin{equation}
	\text{spread}(u) = \{\mathbf s\in \mathbb Z^r \mid \text{gcd}(u,N_\mathbf s u) \neq 1 \}.
\label{eq:spread}
\end{equation}
In our implementation, if $P\neq\varnothing$ or $V\neq\varnothing$, then corresponding pieces of information are printed in order to guide the user to the formulation of an ansatz for $v_{\text{semi-known}}$ and $v_{\text{unknown}}$. Once the user has decided on an ansatz $d_\text{user}$ for the missing factors in the denominator, it can be included in the search when looking for the numerator of the solution through the option {\tt InsertDenFactor $\rightarrow$} $d_{\text{user}}$ of our package.

\subsection{Determination of the numerator}
\label{sec:num}

Once the denominator has been constrained, possibly including an ansatz chosen by the user, one may look for the numerator of the solution. It has been shown in \cite{AP:12} based on \cite{Hilbert10} that in general the problem is unsolvable, and no algorithm exists that can find all solutions. Still, one can find some solutions by taking a polynomial ansatz $\text{num}(c_i)$ and substituting the rational function
\begin{equation}\label{Equ:PLDEAnsatz}
y=\frac{\text{num}(c_i)}{d_a d_p d_{\text{user}}}
\end{equation}
into the equation \eqref{eq:PLDE}. After clearing denominators, one can formulate a linear system for the unknown coefficients $c_i$ such that the equation is satisfied.

In our experience, the determination of the $c_i$ is more computationally demanding than the determination of the denominator bound; for this reason we propose the following strategy, which is implemented in our package as released in \cite{Blumlein:2021hbq}. When the PLDE does not contain any symbolic parameters (such as the dimensional regulator $\ep$, or ratios of invariants) other than the shift variables, one may obtain constraints on the undetermined $c_i$ simply by plugging, sufficiently many times, random numerical values for the shift variables. This allows to quickly obtain a linear system for the $c_i$ without resorting to more expensive symbolic comparisons of the coefficients of many monomials and to Gaussian elimination in a potentially large and highly redundant system.

If there are symbols present, instead, one may consider performing a first pass with the symbols replaced by random numbers, with the purpose of identifying and removing redundant constraints. Then, after removing the redundant equations for the $c_i$, the system can be solved in a stepwise manner, i.e.\ considering one at a time the constraints produced by one monomial, and plugging the result in the rest of the equation. This is what our package does when the function {\tt SolvePLDE} is called.

It is certainly possible that the use of random numbers to generate constraints can cause the system to generate two (or more) equations for the $c_i$ which are not independent. The probability of such an occurrence can be made arbitrarily small by choosing a sufficiently large range over which the random numbers are chosen. In any event, the consequence of an unfortunate draw of random numbers can only cause the software to output more functions misidentified as solutions when in fact they are not; it cannot cause the software to miss any solutions. By explicitly checking the result, one can guard against this remote possibility, at the expense of additional computation time. The user should be aware of this aspect when calling the function {\tt SolvePLDE}.

In the following we elaborate further enhancements in order to extend the solution space from the rational function case to more general classes of functions. Besides the examples below, further examples for each aspect can be found in the Mathematica notebooks attached to the paper \cite{Blumlein:2021hbq}.

\subsubsection{Treatment of a hypergeometric prefactor}
\label{sec:hypergeometric-prefactor}

Given a PLDE \eqref{eq:PLDE}
\begin{equation}
	\sum_{\mathbf s \in S} a_\mathbf s N_\mathbf s y = f
\label{eq:PLDE1}	
\end{equation}
one may want to identify solutions of the form
\begin{equation}
y' = r y
\label{eq:hypFactor}
\end{equation}
with $r=r(n_1,\ldots,n_r)$ a hypergeometric function of its arguments, i.e.\ a function such that the ratio
\begin{equation}
\frac{N_{\mathbf e_i} r}{r} = \frac{r(n_1,\ldots,n_i+1,\ldots n_r)}{r(n_1,\ldots,n_i,\ldots,n_r)} 
\end{equation}
is for all $i$ a rational function of the variables $n_i$. Examples of 
hypergeometric functions are Pochhammer symbols, factorials, 
$\Gamma$-functions, binomial symbols, and obviously rational functions and polynomials.

The demand \eqref{eq:hypFactor} implies that $y$ must satisfy another PLDE,
\begin{equation}
\sum_{\mathbf s \in S} a'_\mathbf s N_\mathbf s y' = f',
\label{eq:PLDE2}
\end{equation}

The transformation from \eqref{eq:PLDE1} to \eqref{eq:PLDE2} is useful whenever it is possible to formulate an ansatz for $r$. Once some specific form can be postulated for $r$, the equation \eqref{eq:PLDE1} is obtained by substitution and by exploiting the hypergeometric property.
Consider for example the equation
\begin{eqnarray}
&&
(1 + k) (\varepsilon + k) (1 + k + n^2) \, y(n, k) - 2 k (2 + k + n^2) \, y(n, 1 + k) 
\nonumber\\ &&
+ (1 + k) (\varepsilon + k) (2 + k + 2 n + n^2) \, y(1 + n, k) = 0.
\label{eq:example_5}
\end{eqnarray}
We assume that its solution is 
\begin{equation}
y(n,k) = (\varepsilon)_k \, y'(n,k)
\end{equation}
with $y'$ a rational function of $n$ and $k$ and $(\varepsilon)_k$ the Pochhammer symbol
\begin{equation}
(\varepsilon)_k = \varepsilon(\varepsilon+1)\cdots(\varepsilon+k-1).
\end{equation}
Then one derives a difference equation for $y'$, namely
\begin{eqnarray}
&& (\varepsilon + k) \big[(1 + k^2 + n^2 + k (2 + n^2)) y'(n, k) - 
2 k (2 + k + n^2) y'(n, 1 + k)
\nonumber\\&&
+ (2 + k^2 + 2 n + n^2 + k (3 + 2 n + n^2)) y'(1 + n, k)\big] = 0 .
\end{eqnarray}
We can now solve the new equation, obtaining
\begin{equation}
y'(n,k)=\frac{k}{1 + k + n^2}.
\end{equation}
From this we conclude that the solution of \eqref{eq:example_5} is 
\begin{equation}
y(n,k) = (\varepsilon)_k \, \frac{k}{1 + k + n^2} C,
\end{equation}
for some constant $C\in\mathbb K(\varepsilon)$.

\subsubsection{Finding solutions in terms of nested sums}
\label{sec:findingNestedSums}

In physical applications, it is known that the solutions of difference equations occurring in actual problems contain (cyclotomic) harmonic sums \cite{Vermaseren:1998uu,Blumlein:1998if,Ablinger:2011te} or their generalizations. The algorithm we described can be adapted to look for these objects in the numerator of the solution by modifying the ansatz \eqref{Equ:PLDEAnsatz} to be formed by a linear combination, with unknown coefficients, of a polynomial expression in a finite list of harmonic sums having coefficients in $\mathbb K[n_1,\ldots,n_r]$. The list of harmonic sums must be shift-stable, i.e.\ a shift in any of the variables must not introduce new harmonic sums not already included in the list, and they also should be linearly independent. These properties can be guaranteed by the quasi-shuffle algebra that the harmonic sums satisfy or by difference ring methods \cite{Blumlein:2003gb,Ablinger:2015lpa,Schneider:2021}.

After plugging the ansatz into the equation and clearing denominators, the nested sums at shifted arguments can be rewritten using relations of the type
\begin{equation}
\label{Equ:SumShiftRules}
S_1(n+i) = \frac{1}{n+i} + S_1(n+i-1)
\end{equation}
and similarly for all other nested sums, until only unshifted nested sums appear. Then one applies a coefficient comparison on the power products in the harmonic sums and the shift variables. One notices how the number of unknowns $c_i$ and of equations rises very fast with the degree of the polynomial chosen for the ansatz and with the number of harmonic sums under consideration; the homomorphic image techniques described in the beginning of Section~\ref{sec:num} are instrumental to perform these calculations in reasonable time.

This heuristic method provides in many cases the desired solution. For instance, consider the equation from \cite{Blumlein:2021hbq}
\begin{eqnarray}
	&&(-k-1) \left(k+n^2+2 n+1\right) f(n,k)
	+k \left(k+n^2+2 n+2\right) f(n,k+1)
\nonumber\\&&		
	+2 (k+1) \left(k+n^2+4 n+4\right) f(n+1,k)
	-2 k \left(k+n^2+4 n+5\right) f(n+1,k+1)
\nonumber\\&&	
	-(k+1) \left(k+n^2+6 n+9\right) f(n+2,k)
	+k \left(k+n^2+6 n+10\right) f(n+2,k+1)=0,
\nonumber\\
\end{eqnarray}
Looking for solutions of the form described, with a numerator of degree up to 2, containing the harmonic sums $S_1(n),S_1(k),S_{2,1}(n)$ the algorithm finds the denominator
\begin{equation}
	d_p=1 + k + 2 n + n^2
\end{equation}
and the corresponding numerators of the solutions of the homogeneous equation:
\begin{eqnarray}
	1, k, k^2, n, k n, S_1(k), k S_1(k), n S_1(k), S_1(k)^2,k S_1(n), k S_{2,1}(n).
\end{eqnarray}

\subsubsection{Matching the solution to initial values}
\label{sec:initVal}

Our package allows to look for solutions that conform to a given set of initial conditions. This general solution is found by building a linear combination with undetermined coefficients of the solutions of the homogeneous equation, plus a particular solution of the equation. Next, the initial values are plugged in, and a system of equations is obtained. In the case that the system contains symbolic parameters other than the shift variables, the undetermined coefficients to be searched are not just numbers. In that case, the coefficients of the linear combination are taken to be general rational functions in the parameters up to some chosen degree. The combination of the solutions will be of particular importance for the next subsection.

\subsubsection{Finding the solution in a series expansion}
\label{sec:expansion}

In many applications it is desirable to obtain the Laurent series expansion of the solution of a difference equation. This may be easier to achieve than the derivation of a complete solution, because, at each order in the expansion, it is possible to derive a difference equation where the expansion parameter is absent, therefore the linear system to find the coefficients $c_i$ can potentially be solved much more quickly. The procedure, described in the following, generalizes the univariate case given in~\cite{Blumlein:2010zv}. It assumes that the initial values of the solution in its $\ep$-expansion are known. 

Consider the case where \eqref{eq:PLDE} contains a parameter $\varepsilon$ in the coefficients:
\begin{equation}
\sum_{\mathbf s \in S} a_\mathbf s(n_i,\varepsilon) N_\mathbf s y(n_i) = f(n_i,\varepsilon) ,
\label{eq:PLDE-eps}
\end{equation}
where the coefficients $a_{\mathbf s}(n_i,\varepsilon)$ are polynomials in the shift variables and in the parameter $\varepsilon$. We search for a solution of \eqref{eq:PLDE-eps} which has, around $\varepsilon=0$, a Laurent expansion starting from the power $\varepsilon^{-\ell}$ of the parameter, with $\ell$ known,
\begin{equation}
	y_\varepsilon(n_i) = \varepsilon^{-\ell} y_{-\ell}(n_i) + \cdots + y_0(n_i) + \varepsilon y_1(n_i) + \cdots + \varepsilon^c y_c(n_i) ,
\label{eq:y_eps}	
\end{equation}
and we assume that the right-hand side of the equation can be expanded in a series in $\varepsilon$ as
\begin{equation}
	f =  \varepsilon^{-\ell} f_{-\ell}(n_i) + \varepsilon^{-\ell+1} f_{-\ell+1} (n_i) + \cdots .
\label{eq:f_eps}	
\end{equation}
We assume also that the $a_{\mathbf s}(n_i,\varepsilon=0)$ are not all zero, so that an overall power of $\varepsilon$, if present in the equation, has been factored out.
Then, one may proceed by inserting \eqref{eq:y_eps} and \eqref{eq:f_eps} into \eqref{eq:PLDE-eps} and doing a coefficient comparison of the $\varepsilon^{-\ell}$ terms, obtaining
\begin{equation}
	\sum_{\mathbf s \in S} a_\mathbf s(n_i,\varepsilon=0) N_\mathbf s y_{-\ell}(n_i) = f_{-\ell}(n_i).
\label{eq:PLDE_eps-l}
\end{equation}
Equation \eqref{eq:PLDE_eps-l} is now free of $\varepsilon$, which facilitates the task of finding a solution and reduces the computational time required. If \eqref{eq:PLDE_eps-l} can be uniquely solved for $y_{-\ell}$ and the solution matched to initial values, one can move to the next higher power in $\varepsilon$ by plugging the solution into \eqref{eq:y_eps}. In this new equation one does a coefficient comparison of the next power in $\varepsilon$ and solves for $y_{\ell+1}$. The process is repeated as many times as needed until all the terms of interest in the Laurent expansion are obtained.

For instance, consider the equation
\begin{eqnarray}
&&\big[3 (k+1) (n+1)+4 (n+1)+1\big] (4 k n^2 \varepsilon ^3+5 n \varepsilon +6 \varepsilon ^2+1) f(n+1,k+1)
\nonumber\\&&
-(3 k n+4 n+1) \big[4 (k+1) (n+1)^2 \varepsilon ^3+5 (n+1)
   \varepsilon +6 \varepsilon ^2+1\big] f(n,k) = 0 .
\label{eq:example-series}   
\end{eqnarray}
Together with a list of 25 initial values, our procedure to compute the expansion encounters at order 
$\varepsilon^{-2},\varepsilon^{-1},\varepsilon^0$ the equations
\begin{eqnarray}
	(-3 k n-4 n-1) f(n,k)+(3 k n+3 k+7 n+8) f(n+1,k+1) &=& \tau , 
\end{eqnarray}
with $\tau = 0, 5, 0$, respectively,
which are free of $\varepsilon$. The series solution of \eqref{eq:example-series} is found to be
\begin{equation}
	f(n,k) = \frac{1}{\varepsilon ^2 (3 k n+4 n+1)}+\frac{5 n}{\varepsilon  (3 k n+4 n+1)}+\frac{6}{3 k n+4 n+1} + \mathcal O(\varepsilon).
\end{equation}

\subsubsection{A brief descriptions of the commands of {\tt solvePartialLDE}}
\label{sec:solvePartialLDE_commands}

\vspace*{1mm}
\noindent
The {\tt Mathematica} package {\tt SolvePartialLDE.m} implements the 
aforementioned 
algorithms for solving partial linear difference equations. It requires {\tt Sigma} and {\tt HarmonicSums} to be loaded. Additionally, the software {\tt Singular} \cite{DGPS} must be installed, and made available by the interface~\cite{SingularInterface} to Mathematica. The installation path of Singular can be set using the command appropriate for the user's system, e.g.
\\[2mm]
{\tt <\,<Singular.m\\
SingularCommand = "{\it (path to)}/Singular-4.1.3-x86\_64-Linux/bin/Singular"
}.
\\[2mm]

The functions available are
\begin{itemize}
	\item $\mathtt {spread[p, q, \{n, k, ...\} (, \{eps,...\} ) ]}$: this function calculates the spread, Eq.\ \eqref{eq:spread}, of the polynomials $p$ and $q$, in the variables $n,k,\ldots$. The symbols in the optional list are treated as an extension to the field over which the polynomials are defined. If the polynomials $p$ and $q$ contain symbolic parameters other than $n,k,\ldots$, such as for instance the dimensional regulator $\varepsilon$, they must be declared in the second list.
	\item $\mathtt {dispersion[p, q, \{n, k, ...\} (, \{eps,...\} ) ]}$: this function calculates the dispersion (it is the maximum of the spread) of the polynomials $p$ and $q$ in the variables $n,k,\ldots$. The second optional list has the same function as in the function {\tt spread}.
	\item $\mathtt {SolvePLDE[eq==rhs,f[n,k,...] ,(options) ] }$. This command solves the linear partial difference equation. It has the following available options:
	\begin{itemize}
		\item {\tt UseObject $\rightarrow$} {\em list of Harmonic sums and/or Pochhammer symbols}
			\\ Allows to define a list of harmonic sums and Pochhammer symbols to be searched in the numerator of the solution.

		\item {\tt PLDEdegBound $\rightarrow$} {\em d}
			\\ Allows to choose the total degree $d$ of the ansatz for the numerator of the solution. Defaults to 0.
		\item {\tt InsertDenFactor $\rightarrow$} {\em factors}
			\\ In the case the periodic denominator bound was 
not 
complete, the user may force the search to include {\em factors} in the denominator.
		\item {\tt PLDESymbols $\rightarrow$} {\em list}
			\\ Any symbols appearing other than the shift variables must be declared in {\em list}.	
		\item {\tt InitialValues $\rightarrow$} {\em list}
			\\ A list of initial values in the form $\{\{var_1\rightarrow val_1, var_2\rightarrow val_2,\ldots, initial value\},\ldots\}$
		\item {\tt SymbolDegree $\rightarrow$} {\em d}
			\\ When initial conditions are provided, a linear combination of the homogeneous solutions is built, having as coefficients rational functions in the symbols. This option sets the maximum total degree of the numerator and denominator of those rational functions.
	\end{itemize}
	\item $\mathtt { SolveExpand[eq==rhs,f[n,k,...], PLDEExpandIn\rightarrow\{\varepsilon,\ell_{min},\ell_{max}\} , }$
	\\
	$ \mathtt{ InitialValues\rightarrow\{\ldots\} ,(options) ] }$ : this command solves the PLDE in a series expansion in the parameter $\varepsilon$, as described in Section \ref{sec:expansion}. The options are the same as for {\tt SolvePLDE}.
	\item $\mathtt { expandHypergPref[eq==rhs, f[n,k,...], fac] }$. This command derives a new equation whose solution has the hypergeometric factor {\tt fac} removed, as described in Section \ref{sec:hypergeometric-prefactor}.
\end{itemize}

\clearpage
\thispagestyle{empty}
~

\clearpage
\section{A numerical library for DIS structure functions}
\label{sec:program}

We present a Fortran library \cite{library} of use for numerically calculating the structure functions $F_2(x,Q^2)$, $g_1(x,Q^2)$ for electromagnetic current exchange and $F_3^{W^+ \pm W^-}(x,Q^2)$ for charged-current exchange in the asymptotic regime of large exchanged momentum $Q^2$, by convolution in $N$-space of parametrized parton distribution functions (PDFs) with Wilson coefficients. The PDFs can be evolved by the library from a parametrization in $N$-space at an initial scale $Q_0^2$ specifiable by the user, and are evolved in the fixed-flavour-number scheme by employing the splitting functions which are known up to $\mathcal O(a_s^3)$ \cite{Moch:2004pa,Vogt:2004mw,Moch:2014sna}. The library accepts as inputs the combinations $u\pm\bar u$, $d\pm\bar d$ and $s\pm\bar s$ in the unpolarized case and $\Delta u+\Delta\bar u$, $\Delta d+\Delta\bar d$, $\Delta s+\Delta\bar s$ in the polarized case. It also encodes the Wilson coefficients for the above-mentioned deep-inelastic processes and for the Drell-Yan process and Higgs production, which have been calculated in \cite{Blumlein:2005im}. In the case of $F_2(x,Q^2)$, the known two-mass contributions of the Wilson coefficients are included up to $\mathcal O(a_s^3)$ \cite{Ablinger:2017err}, whereas for $g_1(x,Q^2)$ and $F_3^{W^+ \pm W^-}(x,Q^2)$ single-mass contributions are considered \cite{Behring:2015zaa,Blumlein:2021xlc,Gluck:1997sj,Blumlein:2011zu,Blumlein:2014fqa,Behring:2015roa}.

The applicability of these asymptotic representations is due to the factorization theorems for QCD \cite{Collins:1989gx,Buza:1996wv}, and any intervening corrections due to heavy quark effects to the Wilson coefficients are suppressed as $\mathcal O(m_{c,b}^2/Q^2)$ in the region $Q^2\gg m_{b,c}^2$. Furthermore, the asymptotic values of the structure functions are used as an ingredient in many definitions of variable flavour number schemes \cite{Buza:1996wv,Alekhin:2009ni}.

The $N$-space evaluation is obtained by the asymptotic expansion of the harmonic sums in the limit $|N|\to\infty$, see \cite{Blumlein:1998if,Blumlein:2009fz,Blumlein:2009ta}, together with recursion relations. In this way, it is possible to express the physical quantities in the complex $N$ plane. We show what accuracy is attainable in the evaluation of moments of the Wilson coefficients and perform the evolution of a set of test PDFs with our library.

\subsection{The structure functions $\boldmath{F_2}$ and $\boldmath{F_L}$}

In the fixed flavour number scheme with $N_F=3$, the structure functions $F_{2,L}(x,Q^2)$ are written as the sum of purely massless and massive contributions as follows \cite{Ablinger:2017err}:
\begin{eqnarray}
F_i(x,Q^2) = F_i^{\rm massless}(x,Q^2) + F_i^{\rm heavy}(x,Q^2)~, \quad i=2,L.
\label{eq:Fi}
\end{eqnarray}
The massless part can be written as
\begin{eqnarray}
 \frac{1}{x} F_i^{\rm massless}(x,Q^2)&=& \sum_{q} e_q^2 \Biggl\{\frac{1}{N_F} \Biggl[
 \Sigma(x,\mu^2) \otimes C_{i,Q}^{\rm S}\left(x,\frac{Q^2}{\mu^2}\right)
 +G\left(x,\mu^2\right) \otimes C_{i,g}\left(x,\frac{Q^2}{\mu^2}\right)\Biggr]
 \NN\\&&
 + \Delta_{q}(x,\mu^2) \otimes C_{i,q}^{\rm NS}\left(x,\frac{Q^2}{\mu^2}\right) \Biggr\}~,
 \quad i=2,L~,
\label{eq:F2L_Factorization}
\end{eqnarray}
with $\Sigma$ and $\Delta_k$ the flavor singlet and non-singlet distributions given by Eqs. \eqref{eq:singletPDF} and \eqref{eq:NSPDF}, and $G$ denoting the gluon density. 

In our code, for $F_2$, we consider the contributions of both the $c$ and the $b$ quark in the asymptotic region $Q^2\gg m_{c,b}^2$, limited to the OMEs which are fully known in $N$-space, as specified below.
The heavy quark part is then given by
\begin{eqnarray}
\label{eq:F2heavy}
    \frac{1}{x}   F_{(2,L)}^{\rm  heavy}(x,N_F\!\!\!&+&\!\!\!2,Q^2,m_1^2,m_2^2) =\NN\\
       &&\sum_{k=1}^{N_F}e_k^2\Biggl\{
                   L_{q,(2,L)}^{\sf NS}\left(x,N_F+2,\frac{Q^2}{\mu^2}
                                                ,\frac{m^2_1}{\mu^2}
                                                ,\frac{m^2_2}{\mu^2}
\right)
                \otimes
                   \Bigl[f_k(x,\mu^2,N_F)+f_{\overline{k}}(x,\mu^2,N_F)\Bigr]
\NN\\ &&\hspace{14mm}
               +\frac{1}{N_F}L_{q,(2,L)}^{\sf PS}\left(x,N_F+2,\frac{Q^2}{\mu^2}
                                                ,\frac{m^2_1}{\mu^2}
                                                ,\frac{m^2_2}{\mu^2}
\right)
                \otimes
                   \Sigma(x,\mu^2,N_F)
\NN\\ &&\hspace{14mm}
               +\frac{1}{N_F}L_{g,(2,L)}^{\sf S}\left(x,N_F+2,\frac{Q^2}{\mu^2}
                                                 ,\frac{m^2_1}{\mu^2}
                                                 ,\frac{m^2_2}{\mu^2}
\right)
                \otimes
                   G(x,\mu^2,N_F)
                             \Biggr\}
\NN\\ && \hspace{7mm}
                   +\tilde{\tilde{H}}_{q,(2,L)}^{\sf PS}\left(x,N_F+2,\frac{Q^2}{\mu^2}
                                        ,\frac{m^2_1}{\mu^2}
                                        ,\frac{m^2_2}{\mu^2}
\right)
                \otimes
                   \Sigma(x,\mu^2,N_F)
\NN\\ &&\hspace{7mm}
                  +\tilde{\tilde{H}}_{g,(2,L)}^{\sf S}\left(x,N_F+2,\frac{Q^2}{\mu^2}
                                           ,\frac{m^2_1}{\mu^2}
                                           ,\frac{m^2_2}{\mu^2}
\right)
                \otimes
                   G(x,\mu^2,N_F)~.
\end{eqnarray}

The massive Wilson coefficients in their asymptotic form are given in \cite{Ablinger:2017err} and read:
\begin{eqnarray}
     \label{eqWIL12M}
     L_{q,(2,L)}^{\sf NS}(N_F+2) &=& 
     a_s^2 \Bigl[A_{qq,Q}^{(2), {\sf NS}}(N_F+2)~\delta_2 +
     \hat{C}^{(2), {\sf NS}}_{q,(2,L)}(N_F)\Bigr]
     \NN\\
     &+&
     a_s^3 \Bigl[A_{qq,Q}^{(3), {\sf NS}}(N_F+2)~\delta_2
     +  A_{qq,Q}^{(2), {\sf NS}}(N_F+2) C_{q,(2,L)}^{(1), {\sf NS}}(N_F+2)
       \NN \\
     && \hspace*{5mm}
     + \hat{C}^{(3), {\sf NS}}_{q,(2,L)}(N_F)\Bigr]~,  \\
      \label{eqWIL22M}
      L_{q,(2,L)}^{\sf PS}(N_F+2) &=& 
     a_s^3 \Bigl[~A_{qq,Q}^{(3), {\sf PS}}(N_F+2)~\delta_2
     +  A_{gq,Q}^{(2)}(N_F+2) N_F\Ctil_{g,(2,L)}^{(1)}(N_F+2) \NN \\
     && \hspace*{5mm}
     + N_F \hat{\Ctil}^{(3), {\sf PS}}_{q,(2,L)}(N_F)\Bigr]~,
     \\
     \label{eqWIL32M}
      L_{g,(2,L)}^{\sf S}(N_F+2) &=& 
     a_s^2 A_{gg,Q}^{(1)}(N_F+2)N_F \Ctil_{g,(2,L)}^{(1)}(N_F+2)
     \NN\\ &+&
      a_s^3 \Bigl[~A_{qg,Q}^{(3)}(N_F+2)~\delta_2 
     +  A_{gg,Q}^{(1)}(N_F+2) N_F\Ctil_{g,(2,L)}^{(2)}(N_F+2)
     \NN\\ && \hspace*{5mm}
     +  A_{gg,Q}^{(2)}(N_F+2) N_F\Ctil_{g,(2,L)}^{(1)}(N_F+2)
     \NN\\ && \hspace*{5mm}
     +  ~A^{(1)}_{Qg}(N_F+2) N_F\Ctil_{q,(2,L)}^{(2), {\sf PS}}(N_F+2)
     + N_F \hat{\Ctil}^{(3)}_{g,(2,L)}(N_F)\Bigr]~,
 \\ \NN \\
\label{eq:WILPS2M}
     \tilde{\tilde{H}}_{q,(2,L)}^{\sf PS}(N_F+2)
     &=& \sum_{i=1}^2 e_{Q_i}^2 a_s^2 \Bigl[~A_{Qq}^{(2), {\sf PS}}(N_F+2,m_i^2)~\delta_2
     +~\Ctil_{q,(2,L)}^{(2), {\sf PS}}(N_F+2)\Bigr]
     \\
     &+& a_s^3 \Bigl[~\tilde{\tilde{A}}_{Qq}^{(3), {\sf PS}}(N_F+2)~\delta_2
     + \sum_{i=1}^2 e_{Q_i}^2 \Bigl[~\Ctil_{q,(2,L)}^{(3), {\sf PS}}(N_F+2) \NN\\
 && 
     + A_{gq,Q}^{(2)}(N_F+2)~\Ctil_{g,(2,L)}^{(1)}(N_F+2) 
     \NN\\&&
     + A_{Qq}^{(2), {\sf PS}}(N_F+2)~C_{q,(2,L)}^{(1), {\sf NS}}(N_F+2) 
        \Bigr] \Bigr]~,       \label{eqWIL42M}
         \NN\\ 
\label{eq:WILS2M}
     \tilde{\tilde{H}}_{g,(2,L)}^{\sf S}(N_F+2) &=&   \sum_{i=1}^2 e_{Q_i}^2\Bigl[
a_s \Bigl[~A_{Qg}^{(1)}(N_F+2)~\delta_2
+~\Ctil^{(1)}_{g,(2,L)}(N_F+2) \Bigr] \NN\\
     &+& a_s^2 \Bigl[~A_{Qg}^{(2)}(N_F+2)~\delta_2
     +~A_{Qg}^{(1)}(N_F+2)~C^{(1), {\sf NS}}_{q,(2,L)}(N_F+2)\NN\\ && 
     \hspace*{5mm}
     +~A_{gg,Q}^{(1)}(N_F+2)~\Ctil^{(1)}_{g,(2,L)}(N_F+2) 
     +~\Ctil^{(2)}_{g,(2,L)}(N_F+2) \Bigr]\Bigr]
     \NN\\ &+&
     a_s^3 \Bigl[~\tilde{\tilde{A}}_{Qg}^{(3)}(N_F+2)~\delta_2
     + \sum_{i=1}^2 e_{Q_i}^2 \Bigl[A_{Qg}^{(2)}(N_F+2)~C^{(1), {\sf NS}}_{q,(2,L)}(N_F+2)
     \NN\\ &&
     \hspace*{5mm}
     +~A_{gg,Q}^{(2)}(N_F+2)~\Ctil^{(1)}_{g,(2,L)}(N_F+2)
     \NN\\ && \hspace*{5mm}
     +~A_{Qg}^{(1)}(N_F+2)\Bigl\{
     C^{(2), {\sf NS}}_{q,(2,L)}(N_F+2)
     +~\Ctil^{(2), {\sf PS}}_{q,(2,L)}(N_F+2)\Bigr\}
     \NN\\ && \hspace*{5mm}
     +~A_{gg,Q}^{(1)}(N_F+2)~\Ctil^{(2)}_{g,(2,L)}(N_F       +2)
     +~\Ctil^{(3)}_{g,(2,L)}(N_F+2) \Bigr]\Bigr]~.
\label{eqWIL52M}
\end{eqnarray}
Here the symbol $\delta_2$ takes the values
\begin{eqnarray}
\delta_2 = \left\{ \begin{array}{lll} 1 & \text{for} & F_2 \\
                                      0 & \text{for} & F_L. 
\end{array} \right.
\end{eqnarray}
In this notation, we use
\begin{eqnarray}
	\tilde{f}(x)&=&\frac{f(x)}{x}~, \label{tildenotation}
	\\
	\hat{f}(x)&=&f(x+2)-f(x)~. \label{hatnotation}
\end{eqnarray}
The OMEs $A_{ij}(N_F+2)$ have the structure
\begin{equation}
	A_{ij}(N_F+2) = A_{ij}(m_1) + A_{ij}(m_2) + \bar A_{ij}(m_1,m_2) + \bar A_{ij}(m_2,m_1)
\end{equation}
in which $\bar A_{ij}(m_1,m_2)$ denotes the part for which the current couples to the quark of mass $m_1$. The double tilde in $\tilde{\tilde{H}}_{q,(2,L)}$ and $\tilde{\tilde{H}}_{g,(2,L)}^{\sf S}$ refers to the charge weighting in the third-order OMEs: 
\begin{equation}
\tilde{\tilde{A}}_{ij}^{(3)} = e_{Q_1}^2 A_{ij}^{(3)}(m_1) + e_{Q_2}^2 A_{ij}^{(3)}(m_2) 
+   e_{Q_1}^2 \bar{A}_{ij}^{(3)}(m_1,m_2)
+   e_{Q_2}^2 \bar{A}_{ij}^{(3)}(m_2,m_1)~.
\end{equation}

We implement in a numerical program the evolution of the structure function $F_2(x,Q^2)$ according to Eq. \eqref{eq:Fi}, \eqref{eq:F2L_Factorization}, \eqref{eq:F2heavy}, by performing in $N$-space the evolution of the PDFs and pairing them to the Wilson coefficients by multiplication in $N$-space. As a last step, by a numerical contour integral, $F_2(x,Q^2)$ is obtained.

At present, some of the quantities in Eqs. \eqref{eqWIL12M}-\eqref{eqWIL52M} are unknown in full analytic form. They are: $a_{Qg}^{(3)}$ and the two-mass asymmetrical OMEs $\bar A_{Qg}^{(3)}$ and $\bar A_{Qq}^{(3),{\sf PS}}$.
For this reason, $\bar A_{Qq}^{(3),{\sf PS}}$ and $\tilde{\tilde A}_{Qg}^{(3)}$ are not implemented in our code. For $a_{Qq}^{(3),{\sf PS}}$, which depends on generalized harmonic sums, we employ an approximate representation.

For $a_{Qg}^{(3)}$, approximate representations exist, based on partial information obtained from interpolations of fixed moments and from the known leading-logarithmic contributions to the heavy Wilson coefficients in the small-$x$ limit and their double-log contributions in the large-$x$ limit \cite{Kawamura:2012cr}. The approximation in \cite{Kawamura:2012cr} was further refined in \cite{Alekhin:2017kpj}; it is not included at present in our code.

As for the massless Wilson coefficients, pertaining the approximate parametrizations of the three-loop Wilson coefficients, there appear to be discrepancies between Eq.\ (4.13) in the text of \cite{Vermaseren:2005qc} and the attachments to the paper. In particular, an inverse Mellin transform of the formula presented in the attachments would suggest that a term in the 4th line in the text needs to be corrected to read $+932.089L_0$ i.e.\ a sign flip. A second inconsistency is in the attachment to the paper where the $N$-space approximation is given. There, it appears that the flavour constant $fl_{11}^g$ should be multiplied by $N_F^2$ instead of $N_F$. We suggest these changes so that the formulas become the Mellin transforms of each other and agree numerically with the moments presented in \cite{Retey:2000nq}: with the definitions
\begin{equation}
	x_1=1-x, \qquad L_0 = \ln(x), \qquad L_1 = \ln(x_1), \qquad fl_{11}^g = \frac{\langle e\rangle^2}{\langle e^2\rangle}
\end{equation}
where $\langle e^i\rangle$ refers to the average of the $i$-th power of the  quark charges, the correct formula appears to be:
\begin{eqnarray}
c_{2,g}^{(3)}(x) &\cong& \textcolor{blue}{N_F} \Big\{\frac{966}{81} L_1^5 - \frac{1871}{18} L_1^4 
    + 89.31 L_1^3 + 979.2 L_1^2 - 2405 L_1 + 
    1372 x_1 L_1^4 
\nonumber\\&&    
    - 15729 - 310510 x + 331570 x^2 - 244150 x L_0^2 - 253.3 x L_0^5 
\nonumber\\&&       
    + L_0 L_1 (138230 - 237010 L_0) - 11860 L_0 - 
    700.8 L_0^2 - 1440 L_0^3 
\nonumber\\&&       
    + \frac{4961}{162} L_0^4 - \frac{134}{9} L_0^5 - 
    x^{-1} (6362.54 \textcolor{red}{+} 932.089 L_0) + 0.625 \delta(x_1) \Big\} 
\nonumber\\&&    
    + 
 \textcolor{blue}{N_F^2} \Big\{\frac{131}{81} L_1^4 - 14.72 L_1^3 + 3.607 L_1^2 - 226.1 L_1 + 4.762 - 
    190 x - 818.4 x^2 
\nonumber\\&&      
    - 4019 x L_0^2 - L_0 L_1 (791.5 + 4646 L_0) + 
    739.0 L_0 + 418.0 L_0^2 + 104.3 L_0^3 
\nonumber\\&&      
    + \frac{809}{81} L_0^4 + \frac{12}{9} L_0^5 + 
    84.423 x^{-1} \Big\} 
\nonumber\\&&      
    + 
 \textcolor{blue}{fl_{11}^g N_F^2 } \Big\{3.211 L_1^2 + 19.04 x L_1 + 0.623 x_1 L_1^3 - 64.47 x + 
    121.6 x^2 - 45.82 x^3 
\nonumber\\&&      
    - x L_0 L_1 (31.68 + 37.24 L_0) + 
    11.27 x^2 L_0^3 - 82.40 x L_0 - 16.08 x L_0^2 
\nonumber\\&&      
    + \frac{520}{81} x L_0^3 + 
    \frac{20}{27} x L_0^4 \Big\} ~.
\label{eq:approxC2g3}
\end{eqnarray}
The difference to the text of \cite{Vermaseren:2005qc} is marked in red. The Mellin transform of \eqref{eq:approxC2g3} is
\begin{eqnarray}
c_{2,g}^{(3)}(N) &\cong&
\textcolor{blue}{N_F} \bigg\{
        \frac{1}{27000 (N-1)^2 N^6 (1+N)^6 (2+N)} \big(
                96480000+473848000 N+1224036000 N^2
\nonumber\\&&
                +2069001600 
N^3+1862075200 N^4-699200400 N^5-39517220484 N^6
\nonumber\\&&
-70794783706 
N^7+37122825948 N^8+130572396885 N^9+45633145850 N^{10}
\nonumber\\&&
-51332369409 
N^{11}-42336233496 N^{12}-9514319157 N^{13}-27750330 N^{14}
\nonumber\\&&
+16875 
N^{15}\big)
        +\Big[
                \frac{1}{N^3 (1+N)^4}(474020+2034310 N+3399445 N^2+2702152 
N^3
\nonumber\\&&
+1041370 N^4+147850 N^5+2405 N^6)
                -\frac{3 \big(
                        8931+566662 N+8931 N^2\big) S_2}{100 N (1+N)^2}
\nonumber\\&&                        
                -\frac{1610 S_2^2}{9 N}
                -\frac{4 (-22825+1871 N) S_3}{9 N (1+N)}
                -\frac{3220 S_4}{9 N}
        \Big] S_1
        +\Big[
                \frac{48}{5 N (1+N)^3} \big(
                        102-1409 N
\nonumber\\&&
                        +306 N^2+102 N^3\big)
                +\frac{(22825-1871 N) S_2}{3 N (1+N)}
                -\frac{6440 S_3}{27 N}
        \Big] S_1^2
\nonumber\\&&        
        +\Big(
                \frac{-8931-566662 N-8931 N^2}{100 N (1+N)^2}
                -\frac{3220 S_2}{27 N}
        \Big) S_1^3
        +\frac{(22825-1871 N) S_1^4}{18 N (1+N)}
        -\frac{322 S_1^5}{27 N}
\nonumber\\&&        
        +\Big[
                \frac{2 \big(
                        1185050+3903173 N+4558059 N^2+2229119 
N^3+348023 N^4\big)}{5 N^2 (1+N)^3}
                -\frac{6440 S_3}{27 N}
        \Big] S_2
\nonumber\\&&        
        +\frac{(22825-1871 N) S_2^2}{6 N (1+N)}
        +
        \frac{\big(
                23692069+46835338 N+23692069 N^2\big) S_3}{50 N 
(1+N)^2}
\nonumber\\&&
        +\frac{(22825-1871 N) S_4}{3 N (1+N)}
        -\frac{2576 S_5}{9 N}
        -\frac{10 (47402+13823 N) \zeta_2}{N^2}
        -\frac{474020 \zeta_3}{N}
\bigg\}
\nonumber\\&&
+\textcolor{blue}{N_F^2} \bigg\{
        \frac{1}{27000 (N-1) N^6 (1+N)^3 (2+N)} \big(
                8640000+8656000 N+10073200 N^2
\nonumber\\&&                
                +9115000 N^3-30552800 
N^4+8345452 N^5+539404372 N^6-148164201 N^7
\nonumber\\&&
-258366915 N^8-67625199 
N^9-24818805 N^{10}\big)
        +\Big(
                \frac{92920-7915 N+2261 N^2}{10 N^3}
\nonumber\\&&                
                +\frac{1104 S_2}{25 N}
                +\frac{1048 S_3}{81 N}
        \Big) S_1
        +\Big(
                \frac{3607}{1000 N}
                +\frac{262 S_2}{27 N}
        \Big) S_1^2
        +\frac{368 S_1^3}{25 N}
        +\frac{131 S_1^4}{81 N}
\nonumber\\&&
        +\frac{(9292000-787893 N) S_2}{1000 N^2}
        +\frac{131 S_2^2}{27 N}
        +\frac{233036 S_3}{25 N}
        +\frac{262 S_4}{27 N}
        +\frac{(-18584+1583 N) \zeta_2}{2 N^2}
\nonumber\\&&
        -\frac{9292 \zeta_3}{N}
\bigg\}
\nonumber\\&&
+\textcolor{blue}{fl_{11}^g} \textcolor{blue}{N_F^{\textcolor{red}{2}}} \bigg\{
        \frac{1}{13500 (1+N)^5 (2+N)^4 (3+N)} \big(
                97917774+305888240 N
\nonumber\\&&                
                +418317438 N^2                
                +369372616 N^3+267827242 N^4+162867261 N^5+71110703 N^6
\nonumber\\&&                
                +19153530 N^7
				+2741175 N^8+152685 N^9\big)
        +\Big(
                \frac{13749-34880 N-9520 N^2}{500 (1+N)^3}
\nonumber\\&&                
                -\frac{1869 S_2}{1000 N (1+N)}
        \Big) S_1        
        +\frac{\big(
                3211+8291 N+3211 N^2\big) S_1^2}{1000 N (1+N)^2}
        -\frac{623 S_1^3}{1000 N (1+N)}
\nonumber\\&&
        +\frac{\big(
                3211+51091 N-28469 N^2\big) S_2}{1000 N (1+N)^2}
        +\frac{7 (-89+5320 N) S_3}{500 N (1+N)}
        +\frac{2 (-535+396 N) \zeta_2}{25 (1+N)^2}
\nonumber\\&&        
        -\frac{1862 \zeta_3}{25 (1+N)}
\bigg\} .
\label{eq:eq:approxC2g3N}
\end{eqnarray}
The difference to the attachment of \cite{Vermaseren:2005qc} is marked in red.

For the structure function $F_L$ for photon exchange, the charm-quark contributions to the Wilson coefficients have been calculated to $\mathcal O(a^2)$ in \cite{Buza:1995ie,Blumlein:2016xcy} with full mass dependence and in the asymptotic approximation, and to $\mathcal O(a^3)$ in \cite{Behring:2014eya}. The massless Wilson coefficients are known from \cite{Zijlstra:1992qd,Moch:2004xu,Vermaseren:2005qc}. It is known that the asymptotic approximation to the charm contribution at $\mathcal O(a^2)$ differs significantly from the analytic expression for virtualities below $Q^2\sim 1000\text{~GeV}^2$. For this reason, in our numerical library, only the massless Wilson coefficient for $F_L$ are coded. At $\mathcal O(a^3)$, our program uses the approximate formulas of \cite{Moch:2004xu}. The structure function $F_L(x,Q^2)$ is coded according to Eq. \eqref{eq:F2L_Factorization}.


\subsection{The structure function $g_1$ }

In the case of $g_1(x,Q^2)$, we consider the contributions to the Wilson coefficients due to one heavy quark. Their explicit form can be found in \cite{Behring:2015zaa,Blumlein:2021xlc}, and reads:
\begin{eqnarray} \frac{1}{x}g_1(x,Q^2) &=& 
      \sum_{k=1}^{N_F}e_k^2\Biggl\{ 
                   L_{q,g_1}^{\sf NS}\left(x,N_F+1,\frac{Q^2}{\mu^2}
                                                ,\frac{m^2}{\mu^2}\right)
                \otimes 
                   \Bigl[\Delta f_k(x,\mu^2,N_F)+\Delta f_{\overline{k}}(x,\mu^2,N_F)\Bigr]
\N\\ &&\hspace{14mm}
               +\frac{1}{N_F}L_{q,g_1}^{\sf PS}\left(x,N_F+1,\frac{Q^2}{\mu^2}
                                                ,\frac{m^2}{\mu^2}\right) 
                \otimes 
                   \Delta \Sigma(x,\mu^2,N_F)
\N\\ &&\hspace{14mm}
               +\frac{1}{N_F}L_{g,g_1}^{\sf S}\left(x,N_F+1,\frac{Q^2}{\mu^2}
                                                 ,\frac{m^2}{\mu^2}\right)
                \otimes 
                   \Delta G(x,\mu^2,N_F) 
                             \Biggr\}
\N\\
       &+&e_Q^2\Biggl[
                   H_{q,g_1}^{\sf PS}\left(x,N_F+1,\frac{Q^2}{\mu^2}
                                        ,\frac{m^2}{\mu^2}\right) 
                \otimes 
                   \Delta \Sigma(x,\mu^2,N_F)
\N\\ &&\hspace{7mm}
                  +H_{g,g_1}^{\sf S}\left(x,N_F+1,\frac{Q^2}{\mu^2}
                                           ,\frac{m^2}{\mu^2}\right)
                \otimes 
                   \Delta G(x,\mu^2,N_F)
                                  \Biggr]~.
\end{eqnarray}
In our numerical program, the polarized splitting functions and Wilson coefficients are coded. They can be obtained from \cite{Zijlstra:1993sh,Moch:2014sna} and from \cite{Behring:2019tus}.


\subsection{The structure function $xF_3^{W^+ - W^-}$ }

For charged-current DIS processes, the structure functions $F_i^{W^\pm}$ are defined by \cite{Behring:2016hpa}
\begin{eqnarray}
\frac{d \sigma^{\nu(\bar{\nu})}}{dx dy} 
&=& \frac{G_F^2 s}{4 \pi} \frac{M_W^4}{(M_W^2 + Q^2)^2} 
\\ && \times
\Biggl\{
\left(1 + (1-y)^2\right) F_2^{W^\pm}(x,Q^2)
- y^2 F_L^{W^\pm}(x,Q^2)
\pm \left(1 - (1-y)^2\right) xF_3^{W^\pm}(x,Q^2)\Biggr\}
\nonumber\\
\frac{d \sigma^{l(\bar{l})}}{dx dy} 
&=& \frac{G_F^2 s }{4 \pi} \frac{M_W^4}{(M_W^2 + Q^2)^2} 
\\ && \times
\Biggl\{
\left(1 + (1-y)^2\right) F_2^{W^\mp}(x,Q^2)
- y^2 F_L^{W^\mp}(x,Q^2)
\pm \left(1 - (1-y)^2\right) xF_3^{W^\mp}(x,Q^2)\Biggr\}
\nonumber
\end{eqnarray}
in terms of the differential cross sections, with $y=q.P/l.P$, $x=Q^2/ys$, $s=(l+P)^2$, $G_F$ the Fermi constant, $M_W$ the $W$-boson mass and $l,P$ are the momenta of the incoming lepton and proton. The combinations
\begin{eqnarray}
xF_3^{W^+ \mp W^-}(x,Q^2) = xF_3^{W^+}(x,Q^2) \pm xF_3^{W^-}(x,Q^2)
\label{eq:F3pm}
\end{eqnarray}
are usually considered because they exist for odd or even moments respectively \cite{Bardeen:1978yd}.

Here we consider the combination $xF_3^{W^+ - W^-}(x,Q^2)$, for which the asymptotic charm contributions to the Wilson coefficients have been calculated in \cite{Blumlein:2014fqa} to $\mathcal O(a^2)$ correcting the results in \cite{Buza:1997mg}, and in \cite{Behring:2015roa} to $\mathcal O(a^3)$, and the exact $\mathcal O(a)$ in \cite{Blumlein:2011zu}.

One can write from the factorization theorems \cite{Behring:2015roa}
\begin{eqnarray}
F_3^{W^+ - W^-}(x,Q^2) &=& 
\Bigl[
  |V_{du}|^2 (d - \overline{d}) 
+ |V_{su}|^2 (s - \overline{s}) 
+ V_{u}      (u - \overline{u})\Biggr] \otimes \Biggl[
C_{q,3}^{W^+ - W^-, \rm NS} +  
L_{q,3}^{W^+ - W^-, \rm NS} \Biggr] \nonumber\\
&&   
+ \Bigl[
  |V_{dc}|^2 (d - \overline{d}) 
+ |V_{sc}|^2 (s - \overline{s}) \Biggr] \otimes
H_{q,3}^{W^+ - W^-, \rm NS}~.
\label{eq:F3WMW}
\end{eqnarray}
Notice how there is no dependence on the gluon density. The coefficients $V_{ij}$ are those of the CKM matrix, whose values are \cite{ParticleDataGroup:2020ssz}
\begin{eqnarray}
	|V_{du}| &=& 0.97370 \\
	|V_{su}| &=& 0.2245  \\
	|V_{dc}| &=& 0.221   \\
	|V_{sc}| &=& 0.987
\end{eqnarray}
and
\begin{equation}
	V_u = |V_{du}|^2 + |V_{su}|^2.
\end{equation}
%
The asymptotic expressions of the heavy quark Wilson coefficients take the form \cite{Behring:2015roa}
\begin{eqnarray}
L_{q,3}^{W^+ - W^-,\rm NS}(N_F+1) &=& a_s^2 A_{qq,Q}^{(2),\rm NS} + \hat{C}_{q,3}^{(2),W^+ - W^-, \rm NS}(N_F) + 
a_s^3 \Biggl[A_{qq,Q}^{(3),\rm NS}  
\nonumber\\  &&
+ A_{qq,Q}^{(2),\rm NS} C_{q,3}^{(1),W^+ - W^-, \rm NS}(N_F+1) + \hat{C}_{q,3}^{(3),W^+ - W^-, \rm NS}(N_F)\Biggr]~, 
\label{eq:WIL1}
\\ 
H_{q,3}^{W^+ - W^-,\rm NS}(N_F+1) &=& 
L_{q,3}^{W^+ - W^-,\rm NS}(N_F+1) + {C}_{q,3}^{W^+ - W^-, \rm 
NS}(N_F),
\label{eq:H3}
\end{eqnarray}
with the notation
\begin{equation}
	\hat f(N_F) = f(N_F+1)-f(N_F).
\end{equation}
In our program, we calculate $xF_3^{W^+ - W^-}(x,Q^2)$ numerically in $N$-space using Eq. \eqref{eq:F3WMW} by pairing the valence PDFs obtained by evolving an input parametrization and the logarithmic contributions to the Wilson coefficients computed for $\mu^2=Q^2$ where $\mu^2$ denotes the factorization and renormalization scales.

\subsection{The structure function $xF_3^{W^+ + W^-}$ }

For the structure function $xF_3^{W^+ + W^-}$, defined as in \eqref{eq:F3pm}, the asymptotic factorization derived in \cite{Blumlein:2014fqa} reads:
\begin{align}
\label{eq:3flOddComb3}
F_3^{W^+ + W^-}
  = &
  \phantom{+}
  \left(
  \vert V_{du} \vert^2
  (d + \bar{d})
  + 
  \vert V_{su} \vert^2
  (s + \bar{s})
  -
  V_u
  (u + \bar{u})
  \right)
  (
   C_{3,q}^{W^++W^-,\text{NS}} 
  +L_{3,q}^{W^++W^-,\text{NS}}
  )
\nonumber\\ &
  + 
  \left(
  \vert V_{dc} \vert^2
  (d + \bar{d})
  + 
  \vert V_{sc} \vert^2
  (s + \bar{s})
  \right)
  H_{3,q}^{W^++W^-,\text{NS}}
\nonumber\\ &
  +
  2V_c
  \left[
  H_{3,q}^{W,\text{PS}}
  \Sigma
  +
  H_{3,g}^{W}
  G
  \right]
.
\end{align}
Explicit formulas for the asymptotic Wilson coefficients are given in \cite{Blumlein:2014fqa} in $N$-space, obtained from the factorization theorem. These formulas are encoded in our numerical program.

\subsection{Drell-Yan process}

The Drell-Yan process refers to the inclusive lepton pair production from two hadrons,
\begin{equation}
	H_1 + H_2 \to \ell_1+\ell_2+X,
\end{equation}
with the invariant mass of the lepton pair denoted by $Q^2$ and the CM energy of the hadrons by $s$.
From the mass factorization theorems \cite{Collins:1985ue,Collins:1988ig,Bodwin:1984hc}, the hadronic structure function can be written as
\begin{equation}
	W^{DY}(x,Q^2) = \sum_{i,j=q,\bar q,g} \int_0^1 dx_1\int_0^1 dx_2 \int_0^1 dz \ \delta(x-x_1 x_2 z) \ f_{i/H_1}(x_1,\mu^2) \ f_{j/H_2}(x_2,\mu^2) \ \Delta_{ij}^{DY} \Bigl(z,\frac{Q^2}{\mu^2}\Bigr) ,
	\label{eq:sigmaDY}
\end{equation}
where $f_{i,j}$ are the parton distribution functions, and $\mu^2$ is the mass factorization and renormalization scale, here set to be equal, and
\begin{equation}
	x=\frac{Q^2}{s} .
\end{equation}
The Drell-Yan structure function is related to the differential cross-section by
\begin{equation}
	\frac{1}{x}\frac{d\sigma^{DY}(x,Q^2)}{dQ^2} = \sigma^V(Q^2) \ W^{DY}(x,Q^2) ,
\end{equation}
and $\sigma^V(Q^2)$ is the point-like cross-section, which depends on which of the standard-model bosons $V$ is  exchanged. In the case of photon exchange, one has \cite{Hamberg:1990np}
\begin{equation}
	\sigma^\gamma(Q^2) = \frac{16\pi^2 a_s^2}{3 N_c Q^4} .
\end{equation}

The Wilson coefficients, $\Delta_{ij}^{DY}$, can be calculated in perturbation theory and have been computed at order $a_s$ in \cite{Altarelli:1978id,Humpert:1979qk,Humpert:1979hb,Humpert:1980uv} and at order $a_s^2$ in \cite{Hamberg:1990np}, see also \cite{Blumlein:2019srk}. Recently, the order $a_s^3$ has been discussed in \cite{Duhr:2021vwj}; the calculation limited to photon exchange only was presented in \cite{Duhr:2020seh} and for charged-current exchange in \cite{Duhr:2020sdp}.

After a Mellin transform, Eq. \eqref{eq:sigmaDY} can be written as:
\begin{equation}
	\mathbf M[W^{DY}](N,Q^2) = \mathbf M[f_i](N,\mu^2) \ \mathbf M[f_j](N,\mu^2) \ \mathbf M[\Delta_{ij}^{DY}]\Bigl(N,\frac{Q^2}{\mu^2}\Bigr) .
\end{equation}
In \cite{Blumlein:2005im}, the analytic expressions for the Mellin transforms of the Wilson coefficients $\Delta_{ij}^{DY}$ have been given. In $N$-space, these Wilson coefficients are written in terms of harmonic sums. We have included the implementation of the quantities $\Delta_{ij}^{DY}(N)$ in our numerical library and evaluated some of the lower moments in Table \ref{tab:mellinDY}. We adopt the same notation as \cite{Hamberg:1990np}, namely
\begin{equation}
	\Delta_{ij} = \Delta_{ij}^{(0)} + a_s \Delta_{ij}^{(1)} + a_s \Delta_{ij}^{(2)} .
\end{equation}
At lowest order, the only relevant process is $q+\bar q\to V$, corresponding to 
\begin{equation}
	\Delta_{q\bar q}^{(0)}(N)=1.
\end{equation}
At order $a_s$, one has the one-loop correction to this process, and the processes $q+\bar q\to V+g$ and $q(\bar q)+g\to V+q(\bar q)$ giving rise to $\Delta_{q\bar q}^{(1)}$ and $\Delta_{qg}^{(1)}$. At order $a_s^2$ one additionally finds the $qq$ and $gg$ processes, and up to three particles are present in the final state.  In \cite{Hamberg:1990np}, the diagrams for the process $q\bar q\to V+q+\bar q$ have been classified into six groups labeled from $A$ to $F$, and the possible interference terms between them has been calculated. In their notation, for example, $\Delta_{q\bar q,A\bar C}^{(2)}$ refers to the interference $AC^\dag+CA^\dag$. At order $a_s^2$, the Wilson coefficients fall into five types: (i) the $q\bar q$ non-singlet, itself divided into a collinearly singular part
\begin{equation}
	\Delta_{q\bar q}^{(2),NS} = \Delta_{q\bar q}^{(2),S+V} + \Delta_{q\bar q}^{(2),C_A} + \Delta_{q\bar q}^{(2),C_F} + \Delta_{q\bar q,A\bar A}^{(2)} + 2 \Delta_{q\bar q,A\bar C}^{(2)} + \beta_0 \Delta_{q\bar q}^{(1)} \ln\Bigl(\frac{\mu_R^2}{\mu^2}\Bigr) ,
\end{equation}
where the notation $S+V$ refers to pieces obtainable in a soft gluon approximation, and a part free of mass singularities, denoted by the terms $\Delta_{q\bar q,B\bar B}^{(2)}$, $\Delta_{q\bar q,B\bar C}^{(2)}$ and, contributing only for $V\neq\gamma$, $\Delta_{q\bar q,A\bar B}^{(2)}$; (ii) the $q(\bar q)g$
\begin{equation}
	\Delta_{qg}^{(2)} = \Delta_{\bar qg}^{(2)} = \Delta_{qg}^{(2),C_A} + \Delta_{qg}^{(2),C_F} + \beta_0 \Delta_{qg}^{(1)} \ln\Bigl(\frac{\mu_R^2}{\mu^2}\Bigr)
\end{equation}
(iii) the $qq$ singlet and non-identical quark (composed of the contributions $\Delta_{q\bar q,C\bar C}^{(2)}$ and $\Delta_{q\bar q,C\bar D}^{(2)}$), (iv) the identical $qq$ (formed by the contributions $\Delta_{qq,C\bar E}^{(2)}$ and $\Delta_{qq,C\bar F}^{(2)}$); (v) the $gg$ contribution
\begin{equation}
	\Delta_{gg}^{(2)} = \Delta_{gg}^{(2),C_A} + \Delta_{gg}^{(2),C_F} .
\end{equation}

The same formalism applies to the case of longitudinally polarized hadrons. In the paper \cite{Blumlein:2005im}, drawing from the $\mathcal O(a_s^2)$ results of \cite{Ravindran:2003gi,Ravindran:2002na}, the polarized Wilson coefficients $\delta\Delta_{ij}^{DY}$ have been given analytically in $N$-space. A numerical implementation is included in our numerical library and an evaluation of the first Mellin moments is in Table \ref{tab:mellinDYpol}.

\subsection{Higgs boson production}

In \cite{Ravindran:2003um}, the production of Higgs bosons from a hadronic collision was studied. The total cross-section for the process
\begin{equation}
	H_1+H_2\to B+X ,
\end{equation}
inclusive over unobserved hadrons $X$, where $B=(H,A)$ denotes a scalar or a pseudoscalar Higgs boson, is given by
\begin{equation}
	\sigma_{tot,B}(x,m^2) = \frac{\pi G_B^2}{8(N_c^2-1)} \sum_{i,j=q,\bar q,g} \int_x^1 dx_1 \int_{x/x_1}^1 dx_2 \ f_{i/H_1}(x_1,\mu^2) \ f_{j/H_2}(x_2,\mu^2) \ \Delta_{ij,B}\Bigl(\frac{x}{x_1 x_2},\frac{m^2}{\mu^2}\Bigr) .
\end{equation}
Here, $m$ is the mass of the Higgs boson, $\mu^2$ is the factorization and renormalization scale. In \cite{Ravindran:2003um}, the calculation was performed in the limit of large top-quark mass, $m_t\to\infty$. In this model, the effective coupling constant $G_B$ is defined through
\begin{eqnarray}
	G_B &=& -2^{5/4} a_s(\mu_R^2) \ G_F^{1/2} \tau_B \ F_B(\tau_B) \ \mathcal C_B\Bigl(a_s(\mu_R^2),\frac{\mu_R^2}{m_t^2}\Bigr),\\
	\tau &=& \frac{4m_t^2}{m^2} ,\\
	F_H(\tau) &=& 1+(1-\tau)f(\tau), \quad F_A(\tau) = f(\tau) \cot\beta , \\[1mm]
	f(\tau) &=& \begin{cases}
				\arcsin^2 \frac{1}{\sqrt\tau} & \text{if~} \tau\ge 1, \\
				-\frac{1}{4}\Bigl( \ln\frac{1-\sqrt{1-\tau}}{1+\sqrt{1-\tau}} +\pi i\Bigr) & \text{if~} \tau<1,\\
			  \end{cases}
\end{eqnarray}
with $G_F$ the Fermi constant and $\cot\beta$ the mixing angle in the two-Higgs-doublet model. The coefficients $\mathcal C_B$ are computable in a series in $a_s$ and can be found in \cite{Ravindran:2003um}, where the authors also calculated the Wilson coefficients $\Delta_{ij,H}$ and $\Delta_{ij,A-H}=\Delta_{ij,A}-\Delta_{ij,H}$.

In lowest order, Higgs production proceeds via the gluon fusion process $gg\to B$ through a top quark triangle loop, corresponding to $\Delta_{gg}^{(0)}$. At NLO, the possible partonic reactions are $g+g\to g+B$, $g+q(\bar q)\to q(\bar q)+B$ and $q+\bar q\to g+B$, from which the Wilson coefficients $\Delta_{gg}^{(1)}$, $\Delta_{qg}^{(1)}$ and $\Delta_{q\bar q}^{(1)}$ are obtained. At NNLO, the notation employed is
\begin{eqnarray}
	\Delta_{gg}^{(2)} &=& C_A^2 \Delta_{gg}^{(2),C_A^2} + C_A T_F N_F \Delta_{gg}^{(2),C_A T_F N_F} + C_FT_FN_F \Delta_{gg}^{(2),C_F T_F N_F}, \\
	\Delta_{qg}^{(2)} &=& C_F^2 \Delta_{qg}^{(2),C_F^2} + C_A C_F \Delta_{qg}^{(2),C_AC_F} + C_F T_F N_F \Delta_{qg}^{(2),C_F T_F N_F}, \\
	\Delta_{q_1q_2}^{(2)} &=& C_F^2 \Delta_{q_1q_2}^{(2),C_F^2}, \\
	\Delta_{qq}^{(2)} &=& C_A C_F^2 \Delta_{qq}^{(2),C_A C_F^2} + C_F^3 \Delta_{qq}^{(2),C_F^3} + C_F^2 \Delta_{qq}^{(2),C_F^2}.
\end{eqnarray}

In the paper \cite{Blumlein:2005im} the Mellin transform of the Wilson coefficients was calculated analytically; these quantities are included in our numerical library. An evaluation of the lowest moments can be found in Tables \ref{tab:mellinDYpol} and \ref{tab:mellinPSH}.

We remark that inclusive Higgs production has been studied at N$^3$LO in \cite{Anastasiou:2014vaa,Li:2014afw,Mistlberger:2018etf} and an $x$-space program has been released in \cite{Dulat:2018rbf}.

\subsection{Notation and conventions}

We follow the conventions in \cite{Klein:2009ig} for the normalization of the anomalous dimensions. The parton densities satisfy\footnote{As in \cite{Klein:2009ig} Eq. (2.115)}
\begin{equation}\label{eq:AP2}
\frac{\partial}{\partial\ln Q^{2}}\left(\begin{array}{c}
\Sigma(x,Q^{2})\\
G(x,Q^{2})
\end{array}\right)=-\frac{1}{2}\left(\begin{array}{cc}
\gamma_{qq} & \gamma_{qg}\\ 
\gamma_{gq} & \gamma_{gg}
\end{array}\right)\left(\begin{array}{c}
\Sigma(x,Q^{2})\\
G(x,Q^{2})
\end{array}\right),
\end{equation}
and we will use the notations $a=a_s(Q^2)$ and $a_0=a_s(Q_0^2)$.
The anomalous dimensions are expanded in a series in $a_s$ as follows:
\begin{equation}
\gamma_{ij}=a\gamma_{ij}^{(0)}+a^{2}\gamma_{ij}^{(1)}+a^{3}\gamma_{ij}^{(2)},
\end{equation}
while the running of $a$ is given by
\begin{equation}\label{eq:aRun}
\frac{\partial a}{\partial\ln Q^{2}}=-\beta_{0}a^{2}-\beta_{1}a^{3}-\beta_{2}a^{4}.
\end{equation}
We further define\footnote{As in \cite{Klein:2009ig} Eq. (2.117)}
\begin{equation}
\gamma_{ij}=-\mathbf{M}[P_{ij}].
\end{equation}
We systematically follow \cite{Klein:2009ig} in all conventions. The running of the coupling is performed by Eq. \eqref{eq:asSol}.
\footnote{
The definition of the anomalous dimensions in \cite{Klein:2009ig} differs from the one adopted in \cite{Moch:2004pa,Vogt:2004mw}.
To recover the anomalous dimensions of \cite{Klein:2009ig}, it is sufficient to replace $N_F\rightarrow 2 T_F N_F$ from those of \cite{Moch:2004pa,Vogt:2004mw} and multiply by two.
In the convention of \cite{Moch:2004pa,Vogt:2004mw}, the Altarelli-Parisi equation \eqref{eq:AP2} loses the factor one-half.
}

\subsection{Evolution of the singlet PDFs}

The singlet PDFs satisfy the equation \cite{Blumlein:1997em,Blumlein:2000wh}
\begin{eqnarray}
\frac{\partial}{\partial a}\left(\begin{array}{c}
\Sigma(N,a)\\
G(N,a)
\end{array}\right) &=& \frac{1}{2(a^2 \beta_0 +a^3 \beta_1 +a^4\beta_2 +\cdots)}
\left(\begin{array}{cc}
\gamma_{qq}^S & \gamma_{qg}\\
\gamma_{gq}   & \gamma_{gg}
\end{array}\right)
\left(\begin{array}{c}
\Sigma(N,a)\\
G(N,a)
\end{array}\right)\\
&=& -\frac{1}{a} \left[\mathbf{R}_{0}+\sum_{k=1}^{\infty}a^{k}\mathbf{R}_{k}\right]\left(\begin{array}{c}
\Sigma(N,a)\\
G(N,a)
\end{array}\right)\label{eq:APt}
\end{eqnarray}
where
\begin{eqnarray}
\gamma_{jk}        &=& \sum_{i=0}^\infty a^{i+1} \gamma_{jk}^{(i)} \\
\gamma_{qq}^{S(i)} &=& \gamma_{qq}^{(+),(i)} + \gamma_{qq}^{PS,(i)} 
\end{eqnarray}
and
\begin{eqnarray}
\boldsymbol{\gamma}^{(i)} &=& \left(\begin{array}{cc}
\gamma_{qq}^{(i)} & \gamma_{qg}^{(i)}\\
\gamma_{gq}^{(i)} & \gamma_{gg}^{(i)}
\end{array}\right)
= - \mathbf P^{(i)},\\
\mathbf R_0 &=& \frac{1}{2\beta_0}\mathbf P^{(0)},\\
\mathbf R_k &=& \frac{1}{2\beta_0}\mathbf P^{(k)}-\sum_{i=1}^{k}\frac{\beta_i}{\beta_0}\mathbf R_{k-i} .
\end{eqnarray}
Its perturbative solution can be given as a series expansion around the lowest order solution $\mathbf L$, as in \cite{Furmanski:1981cw,Blumlein:1997em}:
\begin{eqnarray}
\left(\begin{array}{c}
\Sigma(N,a)\\
G(N,a)
\end{array}\right) 
=
\left[ \mathbf 1 + \sum_{k=1}^\infty a^k \mathbf U_k \right] \mathbf L(a,a_0)
\left[ \mathbf 1 + \sum_{k=1}^\infty a_0^k \mathbf U_k \right]^{-1}
\left(\begin{array}{c}
\Sigma(N,a_0)\\
G(N,a_0)
\end{array}\right)
\end{eqnarray}
with
\begin{eqnarray}
\mathbf L(a,a_0) &=& \Big(\frac{a}{a_0} \Big)^{-\mathbf R_0}\\
&=& \mathbf e_- \Big(\frac{a}{a_0} \Big)^{-r_-} + \mathbf{e}_+ \Big(\frac{a}{a_0} \Big)^{-r_+} 
\label{eq:Lmatrix}
\end{eqnarray}
where $r_\pm$ are the eigenvalues of $\mathbf R_0$,
\begin{equation}
r_\pm = \frac{1}{4\beta_0} [ P_{qq}^{(0)} + P_{gg}^{(0)} \pm \sqrt{ (P_{qq}^{(0)}-P_{gg}^{(0)})^2 + 4 P_{gq}^{(0)} P_{qg}^{(0)} } ]
\label{eq:eigenvalues}
\end{equation}
and $\mathbf{e}_\pm$ are the projectors
\begin{equation}
\mathbf e_\pm = \frac{1}{r_\pm - r_\mp } [ \mathbf R_0 - r_\mp \mathbf 1 ]
\label{eq:L-projectors} 
\end{equation}
such that
\begin{equation}
\mathbf R_0 = r_- \mathbf{e}_- + r_+ \mathbf{e}_+.
\label{eq:R0}
\end{equation}
The matrices $\mathbf U_i$ are calculated by
\begin{eqnarray}
\left[ \mathbf U_1,\mathbf R_0 \right] &=& \mathbf R_1 + \mathbf U_1, \nonumber \\
\left[ \mathbf U_2,\mathbf R_0 \right] &=& \mathbf R_2 + \mathbf R_1 \mathbf U_1 + 2 \mathbf U_2, \nonumber \\
\vdots \nonumber\\
\left[ \mathbf U_k,\mathbf R_0 \right] &=& \mathbf R_k + \sum_{i=1}^{k-1} \mathbf R_{k-i} \mathbf U_i + k \mathbf U_k = \mathbf{\widetilde R}_k + k\mathbf U_k,
\label{eq:U-commutators}
\end{eqnarray}
which implies
\begin{equation}
\mathbf U_k = - \frac{1}{k} \left[ 
                  \mathbf e_- \mathbf{\widetilde R}_k \mathbf e_- 
                 +\mathbf e_+ \mathbf{\widetilde R}_k \mathbf e_+ 
                  \right]
               +\frac{\mathbf e_+ \mathbf{\widetilde R}_k \mathbf e_-}{r_- - r_+ -k}
               +\frac{\mathbf e_- \mathbf{\widetilde R}_k \mathbf e_+}{r_+ - r_- -k}.
\label{eq:Uk}
\end{equation}
To $\mathcal O(a^2)$, the perturbative solution of \eqref{eq:APt} then reads
\begin{eqnarray}
\left(\begin{array}{c}
\Sigma(N,a)\\
G(N,a)
\end{array}\right) &=& \Big[ \mathbf L + a \mathbf U_1 \mathbf L -a_0 \mathbf L \mathbf U_1 + a^2 \mathbf U_2 \mathbf L 
\nonumber\\
&& - a a_0 \mathbf U_1 \mathbf L \mathbf U_1 + a_0^2 \mathbf L (\mathbf U_1^2 - \mathbf U_2) \Big]
\left(\begin{array}{c}
\Sigma(N,a_0)\\
G(N,a_0)
\end{array}\right) ~.
\label{eq:solAP}
\end{eqnarray}

\subsection{Evolution of the non-singlet PDFs}

Analogous to \eqref{eq:APt}, one can write for the non-singlet case \cite{Blumlein:1997em}
\begin{eqnarray}
	\frac{\partial}{\partial a} q^{NS}(N,a) &=& \frac{1}{2(a^2\beta_0+a^3\beta_1+a^4\beta_2)} \gamma_{qq}^{NS} q^{NS}(N,a) = -\frac{1}{a}\Big[ R_0^{NS} + a^k R_k^{NS} \Big] q^{NS}(N,a), \label{eq:solAPNS}
\nonumber \\&&\\
	\gamma_{qq}^{NS} &=& \sum_{k=1}^\infty a^{k+1} \gamma_{qq}^{(k)NS} ,
\end{eqnarray}
with
\begin{eqnarray}
	R_0^{NS} &=& \frac{1}{2\beta_0}P_{qq}^{(0)NS} \label{eq:R0NS} ~,\\
	R_k^{NS} &=& \frac{1}{2\beta_0} P_{qq}^{(k)NS} - \sum_{i=1}^k \frac{\beta_i}{\beta_0} R_{k-i}^{NS} ~.
\end{eqnarray}
A solution in a series expansion in $a,a_0$ can be written similarly to the non-singlet case, with matrix relations reducing to scalar relations, as follows:
\begin{eqnarray}
	q^{NS}(N,a) &=& (1 + a U_1 + a^2 U_2) L (1+a_0 U_1 + a_0^2 U_2 )^{-1} q^{NS}(N,a_0) \nonumber\\
				&=& L \Big[1 + (a-a_0) U_1 + a^2 U_2 - a a_0 U_1^2 + a_0^2(U_1^2-U_2) \Big] q^{NS}(N,a_0) \nonumber\\
				&=& L \Big[1 + (a-a_0) U_1 + (a-a_0)^2 \frac{U_1^2}{2} +(a^2-a_0^2) \Big(U_2 - \frac{U_1^2}{2}\Big) \Big] q^{NS}(N,a_0)
\label{eq:qNS}				
\end{eqnarray}
with $L$ the lowest-order solution
\begin{equation}
	L= \Big( \frac{a}{a_0} \Big)^{-R_0^{NS}} 
\end{equation}
and the quantities $U_i$ determined by
\begin{eqnarray}
	0 &=& R_1^{NS} + U_1 , \\
	0 &=& R_2^{NS} + R_1^{NS} U_1 + 2 U_2 \\
	  &\vdots& \nonumber\\
	0 &=& R_k^{NS} + \sum_{i=1}^{k-1} R_{k-i}^{NS} U_i + k U_k
\end{eqnarray}
or explicitly
\begin{eqnarray}
	U_1 &=& \frac{1}{2\beta_0}(-P_{qq}^{(1)NS} + \frac{\beta_1}{\beta_0}P_{qq}^{(0)NS}) \label{eq:U1NS}\\
	U_2 &=& \frac{1}{2 \beta_0} \Big[ \frac{\beta_2 P_{qq}^{(0)NS}}{2 \beta_0}
			+\frac{1}{4 \beta _0}\Big(P_{qq}^{(1)NS}-\frac{\beta_1 P_{qq}^{(0)NS}}{\beta_0}\Big)^2
			+\frac{\beta_1}{2\beta_0} \Big(P_{qq}^{(1)NS} - \frac{\beta_1 P_{qq}^{(0)NS}}{\beta _0} \Big)
\nonumber\\&&
			-\frac{P_{qq}^{(2)NS}}{2} \Big] ~.
\label{eq:U2NS}
\end{eqnarray}

\subsection{Implementation}

The evolution program works in $N$-space by analytically continuing the relevant harmonic sums through their asymptotic representation and recursion properties \cite{Ablinger:2013cf,Blumlein:2009fz}.

The PDFs are decomposed in $SU(N_F=3)$
\begin{eqnarray}
	v_3^+(N,Q_0^2) &=& (u+\bar u)(N,Q_0^2) - (d+\bar d)(N,Q_0^2) , \label{eq:v3P}\\
	v_8^+(N,Q_0^2) &=& (u+\bar u)(N,Q_0^2) + (d+\bar d)(N,Q_0^2) -2 (s+\bar s)(N,Q_0^2) , \\
	\Sigma(N,Q_0^2) &=& (u+\bar u)(N,Q_0^2) + (d+\bar d)(N,Q_0^2) + (s+\bar s)(N,Q_0^2) , \label{eq:sigmaPDF}
\end{eqnarray}
for the case of $F_2$; the same decomposition is applied to the polarized PDFs for the calculation of $g_1$.

For $xF_3^{W^+ - W^-}$, the analogous decomposition 
\begin{eqnarray}
	v_3^-(N,Q_0^2) &=& (u-\bar u)(N,Q_0^2) - (d-\bar d)(N,Q_0^2) , \label{eq:v3M}\\
	v_8^-(N,Q_0^2) &=& (u-\bar u)(N,Q_0^2) + (d-\bar d)(N,Q_0^2) -2 (s-\bar s)(N,Q_0^2) , \\
	q^V  (N,Q_0^2) &=& (u-\bar u)(N,Q_0^2) + (d-\bar d)(N,Q_0^2) + (s-\bar s)(N,Q_0^2)
\end{eqnarray}
is used. Furthermore, one majorly has $s=\bar s$.
The non-singlet distributions $v_3^\pm$, $v_8^\pm$ are evolved to the virtuality $Q^2$ by applying \eqref{eq:qNS}, and the singlet distribution $\Sigma$ by applying \eqref{eq:solAP}. Equations \eqref{eq:v3P}-\eqref{eq:sigmaPDF} are then inverted to obtain the quark and antiquark PDFs $(q+\bar q)$, and similarly for the valence distributions.

Next, the structure functions $F_2(N,Q^2)$ and $g_1(x,Q^2)$ are formed from \eqref{eq:Fi}, \eqref{eq:F2L_Factorization} and \eqref{eq:F2heavy} and their polarized counterpart, and are Mellin-inverted to $x$-space by integrating numerically over a contour. The inverse Mellin transform of the function $F_i(N,Q^2)$ is performed by
\begin{equation}
F_i(x,Q^2) = \frac{1}{\pi} \mathrm{Im} \left[ \int_C dN\, x^{-N}F_i(N,Q^2) \right] ,
\end{equation}
with the contour defined by
\begin{equation}
N = c_0 + t e^{i\phi}, \quad c_0 = 1.5,\quad 0<t<10^3, \quad \phi=\frac{3\pi}{4}.
\end{equation}
This contour is subdivided into 20 segments, logarithmically spaced, and each of the 20 integrals is evaluated by a Gaussian quadrature with 32 points.

\subsubsection{Analytic continuation}

The harmonic sums need to be analytically continued to complex values of $N$, which is done by the asymptotic expansion and recursion relations \cite{Ablinger:2013cf,Blumlein:2009fz}. To make a concrete example, one obtains for $S_1$ the asymptotic expansion
\begin{eqnarray}
	S_1(N) &=& \ln N +\gamma_E +\frac{1}{2 N} -\frac{1}{12 N^2} +\frac{1}{120 N^4} -\frac{1}{252 N^6} +\frac{1}{240 N^8} -\frac{1}{132 N^{10}} +\frac{691}{32760 N^{12}}
\nonumber\\&&
	 -\frac{1}{12 N^{14}} +\frac{3617}{8160 N^{16}} -\frac{43867}{14364 N^{18}} + \frac{174611}{6600 N^{20}} +\mathcal O\big(\frac{1}{N^{21}}\big)
\end{eqnarray}
valid for $N\to\infty$, which, together with the repeated application of the recurrence
\begin{equation}
	S_1(N) = -\frac{1}{N+1} + S_1(N+1), \qquad N\in \mathbb C, \ N\notin \mathbb Z_- \cup\{0\} ,
\end{equation}
allows to compute the analytic continuation with high accuracy in the complex plane.

Let us first present a list of harmonic sums which is sufficient to encode the anomalous dimensions up to three loops and the Wilson coefficients up to two loops. 
The non-alternating harmonic sums encountered are
\begin{equation}
	S_1, S_2, S_3, S_4, S_5, S_6, S_{2,1}, S_{3,1}, S_{4,1}, S_{2,3}, S_{2,1,1}, S_{2,2,1}, S_{3,1,1}, S_{2,1,1,1} .
\end{equation}
They are coded by asymptotic expansion for $|N|>15$ and by applying recursions inside the disk $|N|<15$.
The alternating harmonic sums up to weight 5 which we encounter are
\begin{equation}
\begin{gathered}
	S_{-1}, S_{-2}, S_{-3}, S_{-4}, S_{-5}, S_{2,-1}, S_{-2,1}, S_{-3,1}, S_{-2,2}, S_{-3,2}, S_{2,-3}, S_{-2,3}, S_{-4,1}, \\
	S_{-2,1,1}, S_{-2,2,1}, S_{2,1,-2}, S_{-3,1,1}, S_{-2,1,-2}, S_{-2,1,1,1} .
\end{gathered}
\end{equation}
In order to code their analytic continuation, we employ the functions
\begin{eqnarray}
	\beta(N)	&=& \frac{1}{2} \biggl[ \psi \biggl( \frac{N+1}{2} \biggr) - \psi \biggl( \frac{N}{2} \biggr) \biggr] \\
	\beta^{(k)}(N)&=& \frac{d^k}{dN^k}\beta(N) \\
	f_1(N) 		&=& \mathbf M \biggl[ \frac{\HA_{0,1}		}{x+1} \biggr] (N) \\
	f_2(N)		&=& \mathbf M \biggl[ \frac{\HA_{0,-1}		}{x+1} \biggr] (N) \\
	f_3(N) 		&=& \mathbf M \biggl[ \frac{\HA_{0,0,1}		}{x+1} \biggr] (N) \\
	f_4(N) 		&=& \mathbf M \biggl[ \frac{\HA_{0,1,0}		}{x+1} \biggr] (N) \\
	f_5(N) 		&=& \mathbf M \biggl[ \frac{\HA_{0,0,1,0}	}{x+1} \biggr] (N) \\
	f_6(N) 		&=& \mathbf M \biggl[ \frac{\HA_{0,-1,0,0}	}{x+1} \biggr] (N) \\
	f_7(N) 		&=& \mathbf M \biggl[ \frac{\HA_{0,1,1}		}{x+1} \biggr] (N) \\
	f_8(N) 		&=& \mathbf M \biggl[ \frac{\HA_{0,1,0,1}	}{x+1} \biggr] (N) \\
	f_9(N) 		&=& \mathbf M \biggl[ \frac{\HA_{0,-1,-1,0}	}{x+1} \biggr] (N) \\
	f_{10}(N)	&=& \mathbf M \biggl[ \frac{\HA_{0,1,1,1}	}{x+1} \biggr] (N) \\
	f_{11}(N)	&=& \mathbf M \biggl[ \frac{\HA_{0,0,1,1}	}{x+1} \biggr] (N) \\
	f_{12}(N)	&=& \mathbf M \biggl[ \frac{\HA_{0,0,0,1}	}{x+1} \biggr] (N) \\
	f_{13}(N)	&=& \mathbf M \biggl[ \frac{\HA_{0,1,0,0}	}{x+1} \biggr] (N) \\
	f_{14}(N)	&=& \frac{3}{8} \zeta_2 \zeta_3 + \mathbf M \biggl[ \frac{4 \HA_{0,-1}^2 -8 \HA_0 \HA_{0,-1,-1} - 4 \zeta_2 \HA_{0,-1} + \zeta_2^2 + \zeta_3 \HA_0}{8(1-x)} \biggr] 
\end{eqnarray}
There exist relations \cite{Blumlein:2003gb} between these functions, namely:
\begin{eqnarray}
	f_4(N) 		&=& \mathbf M\biggl[ \frac{\HA_0 \HA_{0,1}}{x+1} \biggr] -2 f_3(N) , \\
	f_5(N) 		&=& \mathbf M\biggl[ \frac{\HA_0 \HA_{0,0,1}}{x+1} \biggr] -3 f_{12}(N) , \\
	f_8(N)		&=& \mathbf M\biggl[ \frac{\HA_{0,1}^2}{x+1} \biggr] -2 f_{11}(N) , \\
	f_{13}(N) 	&=& \mathbf M\biggl[ \frac{\HA_0^2 \HA_{0,1}}{2(x+1)} -2 \frac{\HA_0 \HA_{0,0,1}}{x+1} \biggr] +3 f_{12}(N) .
\end{eqnarray}
The asymptotic expansion of these functions can be obtained similarly to the case non-alternating harmonic sums, for example:
\begin{eqnarray}
f_7(N) &=&
\frac{1}{4 N^3}
+\frac{5}{16 N^4}
-\frac{7}{48 N^5}
-\frac{553}{576 N^6}
-\frac{449}{2880 N^7}
+\frac{14143}{2880 N^8}
+\frac{13523}{3360 N^9}
-\frac{48812441}{1209600 N^{10}}
\nonumber\\&&
-\frac{76577261}{1209600 N^{11}}
+\frac{7416007}{15120 N^{12}}
+\frac{2236826303}{1900800 N^{13}}
-\frac{2317056701681}{279417600 N^{14}}
-\frac{4517188480391}{165110400 N^{15}}
\nonumber\\&&
+\frac{1074395008571}{5765760 N^{16}}
+\frac{1441529428321447}{1816214400 N^{17}}
-\frac{413219201857699}{76876800 N^{18}}
-\frac{777809511672210671}{27445017600 N^{19}}
\nonumber\\&&
+\frac{13589624465891861}{70171920 N^{20}}
+(\ln(N) + \gamma_E) \biggl[
-\frac{1}{3 N^4}
-\frac{11}{24 N^5}
+\frac{149}{240 N^6}
\nonumber\\&&
+\frac{469}{240 N^7}
-\frac{661}{252 N^8}
-\frac{67379}{5040 N^9}
+\frac{9179}{480 N^{10}}
+\frac{1393813}{10080 N^{11}}
-\frac{7033}{33 N^{12}}
-\frac{5001819}{2464 N^{13}}
\nonumber\\&&
+\frac{220711619}{65520 N^{14}}
+\frac{19348413013}{480480 N^{15}}
-\frac{860107}{12 N^{16}}
-\frac{499342522543}{480480 N^{17}}
+\frac{32237449303}{16320 N^{18}}
\nonumber\\&&
+\frac{553305879870769}{16336320 N^{19}}
-\frac{491600492471}{7182 N^{20}}
\biggr]
+ (\ln(N) + \gamma_E)^2 \biggl[
-\frac{1}{4 N^2}
-\frac{1}{4 N^3}
+\frac{1}{4 N^5}
\nonumber\\&&
-\frac{3}{4 N^7}
+\frac{17}{4 N^9}
-\frac{155}{4 N^{11}}
+\frac{2073}{4 N^{13}}
-\frac{38227}{4 N^{15}}
+\frac{929569}{4 N^{17}}
-\frac{28820619}{4 N^{19}}
\biggr]
\nonumber\\&&
+\biggl[
-\frac{1}{4 N^2}
-\frac{1}{4 N^3}
+\frac{1}{4 N^5}
-\frac{3}{4 N^7}
+\frac{17}{4 N^9}
-\frac{155}{4 N^{11}}
+\frac{2073}{4 N^{13}}
-\frac{38227}{4 N^{15}}
+\frac{929569}{4 N^{17}}
\nonumber\\&&
-\frac{28820619}{4 N^{19}}
\biggr] \zeta_2
+\biggl[
\frac{1}{2 N}
+\frac{1}{4 N^2}
-\frac{1}{8 N^4}
+\frac{1}{4 N^6}
-\frac{17}{16 N^8}
+\frac{31}{4 N^{10}}
-\frac{691}{8 N^{12}}
+\frac{5461}{4 N^{14}}
\nonumber\\&&
-\frac{929569}{32 N^{16}}
+\frac{3202291}{4 N^{18}}
-\frac{221930581}{8 N^{20}}
\biggr] \zeta_3  +\mathcal O\Bigl(\frac{1}{N^{21}}\bigr)\\
f_{14} &=&
\frac{1}{16} \zeta_2 \zeta_3
+\frac{5}{8} \zeta_5
-\frac{1}{16 N^4}
+\frac{1}{8 N^6}
-\frac{1}{16 N^7}
-\frac{13}{32 N^8}
+\frac{1}{2 N^9}
+\frac{33}{16 N^{10}}
-\frac{73}{16 N^{11}}
\nonumber\\&&
-\frac{1517}{96 N^{12}}
+\frac{109}{2 N^{13}}
+\frac{527}{3 N^{14}}
-\frac{13821}{16 N^{15}}
-\frac{261131}{96 N^{16}}
+\frac{17899}{N^{17}}
+\frac{2709997}{48 N^{18}}
-\frac{7590505}{16 N^{19}}
\nonumber\\&&
-\frac{729036941}{480 N^{20}}
+\mathcal O\Bigl(\frac{1}{N^{21}}\Bigr)
\end{eqnarray}
One also has the relations
\begin{eqnarray}
	S_{-1}(N) 	&=& (-1)^N \beta(N + 1) - \ln(2) \\
	S_{-2}(N) 	&=& (-1)^{N + 1} \beta^{(1)}(N + 1) - \frac{\zeta_2}{2} \\
	S_{-3}(N) 	&=& (-1)^N \frac{\beta^{(2)}(N + 1)}{2} - \frac{3}{4} \zeta_3 \\
	S_{-4}(N) 	&=& (-1)^{N + 1} \frac{\beta^{(3)}(N + 1)}{6} - \frac{7}{20} \zeta_2^2 \\
	S_{-5}(N) 	&=& (-1)^N \frac{\beta^{(4)}(N + 1)}{24} - \frac{15}{16} \zeta_5 \\
	S_{2,-1}(N)	&=& \Bigl( -S_2({N}) +S_{-2}({N}) -\frac{\zeta_2}{2} \Bigr) \ln(2) 
					-\frac{1}{2} S_{-1}({N}) \zeta_2
					+\frac{1}{4} \zeta_3
\nonumber\\&&
					+(-1)^N f_2(N+1) \\
	S_{-2,1}(N)	&=& \ln(2) \zeta_2 +S_{-1}({N}) \zeta_2 -\frac{5}{8} \zeta_3 +(-1)^{1+N} f_1(N+1) \\
	S_{-3,1}(N)	&=& 2 \Li_4\Bigl(\frac{1}{2}\Bigr) +\frac{\ln(2)^4}{12} 
					-\frac{1}{2} \ln(2)^2\zeta_2 +S_{-2}({N}) \zeta_2 -\frac{3}{5} \zeta_2^2 
					+\frac{3}{4} \ln(2) \zeta_3 -S_{-1}({N}) \zeta_3 
\nonumber\\&&
					+(-1)^N f_3(N+1)\\
	S_{-2,2}(N)	&=& -4 \Li_4\Bigl(\frac{1}{2}\Bigr) -\frac{\ln(2)^4}{6} +\ln(2)^2 \zeta_2 
					+\frac{51}{40} \zeta_2^2 -\frac{3}{2} \ln(2) \zeta_3 
					+2 S_{-1}({N}) \zeta_3 
\nonumber\\&&					
					+(-1)^N f_4(N+1)\\
	S_{-3,2}(N)	&=& -\frac{6}{5} \ln(2) \zeta_2^2 -\frac{6}{5} S_{-1}({N}) \zeta_2^2
					+2 S_{-2}({N}) \zeta_3 +\frac{3}{8} \zeta_2 \zeta_3 +\frac{11}{32} \zeta_5
\nonumber\\&&
					+(-1)^{1+N} f_5(N+1) \\
	S_{2,-3}(N)	&=& -\frac{21}{20} \ln(2) \zeta_2^2 -\frac{21}{20} S_{-1}({N}) \zeta_2^2
					-\frac{3}{4} S_2({N}) \zeta_3 +\frac{3}{4} S_{-2}({N}) \zeta_3
					+\zeta_2 \zeta_3 -\frac{41}{32} \zeta_5 
\nonumber\\&&					
					+(-1)^N f_6(N+1)\\
	S_{-2,3}(N)	&=& \frac{6}{5} \ln(2) \zeta_2^2 +\frac{6}{5} S_{-1}({N}) \zeta_2^2
					-\frac{3}{4} \zeta_2 \zeta_3 +\frac{21}{32} \zeta_5 +(-1)^{1+N} f_{13}(N+1)\\
	S_{-4,1}(N)	&=& S_{-3}({N}) \zeta_2 +\frac{2}{5} \ln(2) \zeta_2^2
					+\frac{2}{5} S_{-1}({N}) \zeta_2^2 -S_{-2}({N}) \zeta_3
					+\frac{3}{4} \zeta_2 \zeta_3 -\frac{59}{32} \zeta_5
\nonumber\\&&
					+(-1)^{1+N} f_{12}(N+1)\\
	S_{-2,1,1}(N)&=& -\Li_4\Bigl(\frac{1}{2}\Bigr) -\frac{\ln(2)^4}{24} +\frac{1}{4} \ln(2)^2 \zeta_2
					+\frac{1}{8} \zeta_2^2 +\frac{1}{8} \ln(2) \zeta_3
					+S_{-1}({N}) \zeta_3 
\nonumber\\&&					
					+(-1)^{1+N} f_7(N+1)\\
	S_{-2,2,1}(N)&=& 4 \Li_5\Bigl(\frac{1}{2}\Bigr) +4 \Li_4\Bigl(\frac{1}{2}\Bigr) \ln(2) 
					+\frac{2 \ln(2)^5}{15}
					-\frac{2}{3} \ln(2)^3 \zeta_2 +S_{-2,1}({N}) \zeta_2
\nonumber\\&&					
					-\frac{3}{10} \ln(2) \zeta_2^2 -\frac{3}{10} S_{-1}({N}) \zeta_2^2
					+\frac{7}{4} \ln(2)^2 \zeta_3 -\frac{9}{8} \zeta_2 \zeta_3
					-\frac{89}{64} \zeta_5 
\nonumber\\&&					
					+(-1)^N f_8(N+1)\\ 
	S_{2,1,-2}(N)&=& \frac{1}{2} \ln(2) S_2({N}) \zeta_2
					-\frac{1}{2} \ln(2) S_{-2}({N}) \zeta_2
					-\frac{1}{2} S_{2,1}({N}) \zeta_2
					+\frac{1}{2} S_{2,-1}({N}) \zeta_2
\nonumber\\&&
					+\frac{1}{8} \ln(2) \zeta_2^2
					+\frac{1}{8} S_{-1}({N}) \zeta_2^2
					-\frac{1}{8} S_2({N}) \zeta_3
					+\frac{1}{8} S_{-2}({N}) \zeta_3
					+\frac{11}{8} \zeta_2 \zeta_3
					-\frac{177}{64} \zeta_5
\nonumber\\&&
					+(-1)^N f_9(N+1)\\ 
	S_{-3,1,1}(N)&=& -2 \Li_5\Bigl(\frac{1}{2}\Bigr)
					-2 \Li_4\Bigl(\frac{1}{2}\Bigr) \ln(2)
					-\frac{\ln(2)^5}{15}
					+\frac{1}{3} \ln(2)^3 \zeta_2
					-\frac{1}{10} \ln(2) \zeta_2^2
\nonumber\\&&
					-\frac{1}{10} S_{-1}({N}) \zeta_2^2
					-\frac{7}{8} \ln(2)^2 \zeta_3
					+S_{-2}({N}) \zeta_3
					+\frac{7}{8} \zeta_2 \zeta_3
					+\frac{15}{32} \zeta_5
\nonumber\\&&
					+(-1)^N f_{11}(N+1)\\ 
	S_{-2,1,-2}(N)&=& \frac{1}{8} \bigl( -4 S_{-2,1}({N}) \zeta_2 -S_{-2}({N}) \zeta_3
					-3 \zeta_2 \zeta_3
					\bigr)
					+f_{14}(N+1) \\
	S_{-2,1,1,1}(N)&=& \Li_5\Bigl(\frac{1}{2}\Bigr)
					+\Li_4\Bigl(\frac{1}{2}\Bigr) \ln(2)
					+\frac{\ln(2)^5}{30}
					-\frac{1}{6} \ln(2)^3 \zeta_2
					+\frac{2}{5} \ln(2) \zeta_2^2
\nonumber\\&&
					+\frac{2}{5} S_{-1}({N}) \zeta_2^2
					+\frac{7}{16} \ln(2)^2 \zeta_3
					-\frac{7}{16} \zeta_2 \zeta_3
					-\frac{27}{32} \zeta_5
					+(-1)^{1+N} f_{10}(N+1)
\end{eqnarray}

The asymptotic expansions have been obtained using {\tt HarmonicSums}, \cite{HARMONICSUMS}. The {\tt Fortran} routines have been created using {\tt Form} \cite{Vermaseren:2000nd}. Historically, the analytic continuation of these harmonic sums has first been given in an accurate numerical representation in \cite{Blumlein:2005jg}.

We also produced Fortran routines for the analytic continuation of the weight-6 harmonic sums which contribute to the Wilson coefficient $c_{2,q}^{(3)NS}$ and a code to calculate the analytic continuation of the coefficient itself. The relevant sums are
\begin{equation}
\begin{gathered}
	S_{-6},S_{-5,1},S_{-4,-2},S_{-4,2},S_{-3,3},S_{-2,-3},S_{4,-2},S_{4,2},S_{5,1},S_{-4,1,1},S_{-3,-2,1},
S_{-3,1,-2},\\	S_{-3,2,1},S_{-2,2,-2},S_{-2,2,2},S_{-2,3,1},S_{2,-3,1},S_{2,3,1},S_{3,1,-2},S_{3,2,1},S_{4,1,1},S_{-3,1,1,1},\\
S_{-2,-2,1,1},S_{-2,1,1,2},S_{-2,2,1,1},S_{2,-2,1,1},S_{2,2,1,1},S_{3,1,1,1},S_{-2,1,1,1,1},S_{2,1,1,1,1}
\end{gathered}
\end{equation}
and their analytic continuation is obtained in the same way as for the weight-5 sums, i.e.\ by rewriting them as Mellin transforms of HPLs, and through asymptotic expansion for $N\to\infty$ and by recursion relations. In addition to the constants found for the weight-5 sums, one encounters the constants 
\begin{equation}
	s_6 \equiv S_{-5,-1}(\infty)
\end{equation}
and $\Li_6(\frac{1}{2})$. For example,
\begin{eqnarray}
	S_{2,-2,1,1}(N) &=& 4 \Li_6\Big(\frac{1}{2}\Big)
+4 \Li_5\Big(\frac{1}{2}\Big) \ln(2)
+
\frac{\ln(2)^6}{18}
+\frac{\ln(2)^4}{24 N^2}
+s_6
+\Li_4\Big(\frac{1}{2}\Big) \Big(
        2 \ln(2)^2
        +\frac{1}{N^2}
\Big)
\nonumber\\&&
+(-1)^N \biggl\{
        -\frac{\Li_4\big(\frac{1}{2}\big)}{N^2}
        -\frac{\ln(2)^4}{24 N^2}
        +\Big(
                \frac{\ln(2)^2}{4 N^2}
                -\frac{11 \zeta_3}{16 N}
        \Big) \zeta_2
        +\frac{\zeta_2^2}{8 N^2}
        +\frac{\ln(2) \zeta_3}{8 N^2}
\nonumber\\&&        
        +\frac{41 \zeta_5}{32 N}
        +\mathrm M\Bigl[\frac{H_{0,-1,0,1,1}(x)}{1+x}\Bigr](N)
\biggr\}
+\Bigl[
        -\Li_4\Big(\frac{1}{2}\Big)
        -\frac{\ln(2)^4}{24}
        +\frac{1}{4} \ln(2)^2 \zeta_2
\nonumber\\&&        
        +\frac{1}{8} \zeta_2^2
        +\frac{1}{8} \ln(2) \zeta_3
\Bigr] S_2(N)
+\Bigl(
        -\frac{\zeta_3}{N^2}
        +\frac{11 \zeta_2 \zeta_3}{16}
        -\frac{41 \zeta_5}{32}
\Bigr) S_{-1}(N)
\nonumber\\&&
+\Big[
        \Li_4\Big(\frac{1}{2}\Big)
        +\frac{\ln(2)^4}{24}
        -\frac{1}{4} \ln(2)^2 \zeta_2
        -\frac{1}{8} \zeta_2^2
        -\frac{1}{8} \ln(2) \zeta_3
\Big] S_{-2}(N)
+\frac{S_{-2,1,1}(N)}{N^2}
\nonumber\\&&
+\Bigl[
        -2 \Li_4\Big(\frac{1}{2}\Big)
        -\frac{\ln(2)^4}{3}
        -\frac{\ln(2)^2}{4 N^2}
        -\frac{17 \ln(2) \zeta_3}{16}
\Bigr] \zeta_2
+\big(
        \frac{\ln(2)^2}{2}
        -\frac{1}{8 N^2}
\big) \zeta_2^2
\nonumber\\&&
-\frac{87}{280} \zeta_2^3
+\Big(
        \frac{7 \ln(2)
        ^3}{12}
        -\frac{\ln(2)}{8 N^2}
\Big) \zeta_3
+S_{2,-1}(N) \zeta_3
+\frac{105}{128} \zeta_3^2
-\frac{103}{32} \ln(2) \zeta_5 ~.
\end{eqnarray}
In this way, it is possible to evaluate the quantity $c_{2,q}^{(3)NS}$ at complex $N$ without resorting to the approximate formulas in \cite{Vermaseren:2005qc}. The accuracy of this evaluation for integer moments can be found in Table \ref{tab:accuracyWCs}.

\subsubsection{Structure of the massive OMEs}

The two-mass OMEs have been employed in our library to assemble the heavy quark contributions to the Wilson coefficients in the asymptotic regime. They have the following structure, which we repeat here from \cite{Ablinger:2017err} for clarity:
\begin{eqnarray}
	A_{qq,Q}^{NS} (N_F+2) &=& 1+a^2 A_{qq,Q}^{NS(2)}(N_F+2) + a^3 A_{qq,Q}^{NS(3)}(N_F+2) \\
[0.4cm]
	A_{qq,Q}^{NS(2)} (N_F+2) &=& A_{qq,Q}^{NS(2)}(m_c) + A_{qq,Q}^{NS(2)}(m_b) \\
	A_{qq,Q}^{NS(3)} (N_F+2) &=& A_{qq,Q}^{NS(3)}(m_c) + A_{qq,Q}^{NS(3)}(m_b) + \tilde A_{qq,Q}^{NS(3)}(m_c,m_b)
\end{eqnarray}
\vspace{5mm}
\begin{eqnarray}
	A_{Qq}^{PS} (N_F+2) &=& a^2 A_{Qq}^{PS(2)}(N_F+2) + a^3 A_{Qq}^{PS(3)}(N_F+2) \\
[0.4cm]
	A_{Qq}^{PS(2)} (N_F+2) &=& A_{Qq}^{PS(2)}(m_c) + A_{Qq}^{PS(2)}(m_b) \\
	\tilde{\tilde A}_{Qq}^{PS(3)} (N_F+2) &=& e_c^2 A_{Qq}^{PS(3)}(m_c) + e_b^2 A_{Qq}^{PS(3)}(m_b) + e_c^2 \bar A_{Qq}^{PS(3)}(m_c,m_b) + e_b^2 \bar A_{Qq}^{PS(3)}(m_b,m_c)
\nonumber\\
\end{eqnarray}
\vspace{5mm}
\begin{eqnarray}
	A_{qq,Q}^{PS} (N_F+2)    &=& a^3 A_{qq,Q}^{PS(3)} (N_F+2) \\
[0.4cm]
	A_{qq,Q}^{PS(3)} (N_F+2) &=& A_{qq,Q}^{PS(3)}(m_c) + A_{qq,Q}^{PS(3)}(m_b)
\end{eqnarray}
\vspace{5mm}
\begin{eqnarray}
	A_{Qg} (N_F+2) &=& a A_{Qg}^{(1)}(N_F+2) + a^2 A_{Qg}^{(2)}(N_F+2) + a^3 A_{Qg}^{(3)}(N_F+2) \\
[0.4cm]
	A_{Qg}^{(1)}(N_F+2) &=& A_{Qg}^{(1)}(m_c) + A_{Qg}^{(1)}(m_b) \\
	A_{Qg}^{(2)}(N_F+2) &=& A_{Qg}^{(2)}(m_c) + A_{Qg}^{(2)}(m_b) + \tilde A_{Qg}^{(2)}(m_b,m_c) \\
	\tilde{\tilde A}_{Qg}^{(3)}(N_F+2) &=& e_c^2 A_{Qg}^{(3)}(m_c) + e_b^2 A_{Qg}^{(3)}(m_b) + e_c^2 \bar A_{Qg}^{(3)}(m_c,m_b) + e_b^2 \bar A_{Qg}^{(3)}(m_b,m_c)
\end{eqnarray}
\vspace{5mm}
\begin{eqnarray}
	A_{qg,Q} (N_F+2) &=& a^2 A_{qg,Q}^{(2)}(N_F+2) + a^3 A_{qg,Q}^{(3)}(N_F+2) \\
[0.4cm]
	A_{qg,Q}^{(2)}(N_F+2) &=& A_{qg,Q}^{(2)}(m_c) + A_{qg,Q}^{(2)}(m_b) \\
	A_{qg,Q}^{(3)}(N_F+2) &=& A_{qg,Q}^{(3)}(m_c) + A_{qg,Q}^{(3)}(m_b) + \tilde A_{qg,Q}^{(3)}(m_c,m_b)
\end{eqnarray}
\vspace{5mm}
\begin{eqnarray}
	A_{gg,Q}(N_F+2) &=& 1 + a A_{gg,Q}^{(1)}(N_F+2) + a^2 A_{gg,Q}^{(2)}(N_F+2) + a^3 A_{gg,Q}^{(3)}(N_F+2) \\
[0.4cm]
	A_{gg,Q}^{(1)}(N_F+2) &=& A_{gg,Q}^{(1)}(m_c) + A_{gg,Q}^{(1)}(m_b) \\
	A_{gg,Q}^{(2)}(N_F+2) &=& A_{gg,Q}^{(2)}(m_c) + A_{gg,Q}^{(2)}(m_b) + \tilde A_{gg,Q}^{(2)}(m_c,m_b) \\
	A_{gg,Q}^{(3)}(N_F+2) &=& A_{gg,Q}^{(3)}(m_c) + A_{gg,Q}^{(3)}(m_b) + \tilde A_{gg,Q}^{(3)}(m_c,m_b) ~.
\end{eqnarray}

The two-mass unpolarized OMEs have been calculated in \cite{Ablinger:2017err,Ablinger:2017xml,Ablinger:2018brx}. The single-mass three-loop OMEs have been presented in \cite{Ablinger:2010ty,Ablinger:2014lka,Behring:2014eya,Ablinger:2014vwa,Ablinger:2014nga}.
The objects $\tilde{\tilde A}_{Qq}^{PS(3)} (N_F+2)$ and $\tilde{\tilde A}_{Qg}^{(3)}(N_F+2)$ are set to zero in our library.

In the polarized case, the program considers contributions to $g_1$ due to one massive quark. The three-loop polarized OMEs have been presented in \cite{Behring:2015zaa,Ablinger:2019etw,Ablinger:2020snj,Behring:2021asx}. The program includes the $\mathcal O(a_s^3)$ logarithmic contributions to the one-mass Wilson coefficients as presented in \cite{Behring:2015zaa,Blumlein:2021xlc}.

\subsection{Structure of the library}

The library is composed of a set of main files called {\tt F2.f, FL.f, F3WMW.f, F3WPW.f, g1.f}. Each will compute and print on screen the respective structure function. These programs draws on a set of shared routines which are described in the next section.

\subsubsection{List of routines}

\vspace{2mm}
\noindent
\textbf{Function {\tt AS($\Lambda_{QCD}$, $Q^2$, $n_f$) }}\\
This function returns the value of $a_s(Q^2)$ computed with Eq. \eqref{eq:asSol} for $n_f$ decoupling flavours.

\vspace{2mm}
\noindent
\textbf{Subroutine {\tt DISTTESTSPL} } \\
Prints diagnostic messages consisting in the difference between the inverse Mellin transform of certain convolutions of splitting functions with a test function, and their expected value.

\vspace{2mm}
\noindent
\textbf{Subroutine {\tt EVOLVE}, {\tt EVOLVEM} and {\tt EVOLVEPOL} }\\
Apply the respective evolution operators to $v_{3,8}(N,Q_0^2)$ and $\Sigma(N,Q_0^2)$; the combinations $(u+\bar u),(d+\bar d),(s+\bar s)$ and polarized counterparts are formed.

\vspace{2mm}
\noindent
\textbf{Subroutines {\tt F2MASSLESS, F2HEAVY}}\\
The structure functions $F_2^{massless}(N,Q^2)$ and $F_2^{heavy}(N,Q^2)$ are assembled using \eqref{eq:F2L_Factorization} and \eqref{eq:F2heavy}.

\vspace{2mm}
\noindent
\textbf{Subroutines {\tt F3WMWMASSLESS, F3WMWHEAVY}}\\
The structure function $F_3^{W^+ - W^-}(N,Q^2)$, massless and charm quark contributions, as in \eqref{eq:F3WMW}.

\vspace{2mm}
\noindent
\textbf{Subroutines {\tt F3WMPMASSLESS, F3WPWHEAVY}}\\
The structure function $F_3^{W^+ + W^-}(N,Q^2)$, massless and charm quark contributions, as in \eqref{eq:3flOddComb3}.

\vspace{2mm}
\noindent
\textbf{Subroutines {\tt G1MASSLESS, G1HEAVY}}\\
The structure functions $g_1^{massless}(N,Q^2)$ and $g_1^{heavy}(N,Q^2)$ are assembled.

\vspace{2mm}
\noindent
\textbf{Subroutine {\tt INIT}}\\
Provides the initialization of mathematical constants, including parameters for the Gaussian integration.

\vspace{2mm}
\noindent
\textbf{Subroutine {\tt INVERT($dat(640)$, $x$)}}\\
Evaluates numerically at $x$ the inverse Mellin transform of a function whose values along the contour are given as input in the array $dat(640)$. The result is stored in the common block {\tt INV}. The code is derived from the program Ancont \cite{Blumlein:2000hw}.

\vspace{2mm}
\noindent
\textbf{Subroutines {\tt MOMCHECKSPL}, {\tt MOMCHECKSPLM} }\\
Prints diagnostic messages consisting in the difference between the evaluation of moments of splitting functions and known values for the moments.

\vspace{2mm}
\noindent
\textbf{Subroutines {\tt PRECOMP1}, {\tt PRECOMP1M} and {\tt PRECOMP1POL} }\\
Evaluate on the contour the quantities $\mathbf R_0, \mathbf U_{1,2}$ required for the evolution of the singlet PDF, as in \eqref{eq:R0}, \eqref{eq:Uk}, and the corresponding non-singlet quantities $R_0, U_{1,2}$, Eq. \eqref{eq:R0NS}, \eqref{eq:U1NS}, \eqref{eq:U2NS}. They are stored in common blocks.

\vspace{2mm}
\noindent
\textbf{Subroutines {\tt PRECOMP2}, {\tt PRECOMP2M} and {\tt PRECOMP2POL} }\\
In this routines the assembly of the evolution operators of the PDFs is completed in those parts which depend on $a_s$.

\vspace{2mm}
\noindent
\textbf{Subroutines {\tt PRECOMPINIT}, {\tt PRECOMPINITM} and {\tt PRECOMPINITPOL} } \\
Initialize the contour and evaluates the user-defined PDFs over the contour, by calling the functions {\tt UPUB}, {\tt DPDB}, {\tt SPSB}, {\tt GLU} and their polarized counterparts {\tt UPUBPOL}, {\tt DPDBPOL}, {\tt SPSBPOL}, {\tt GLUPOL} which are described below.

\vspace{2mm}
\noindent
\textbf{Subroutines {\tt PRECOMPWCF2}, {\tt PRECOMPWCF3WMW}, {\tt PRECOMPWCF3WPW}, {\tt PRECOMPWCG1}, {\tt PRECOMPWCFL} }\\
Evaluation of massless Wilson coefficients and of the logarithmic terms of the massive Wilson coefficients.

\vspace{2mm}
\noindent
\textbf{Subroutine {\tt SETAS($\Lambda_{QCD}$, $n_f$) }}\\
Fills the common block {\tt AA0} with $a=a_Q^2$ and $a_0=a(Q_0^2)$. The parameter $n_f$ refers to the number of decoupling flavours in the running of the coupling constant. The values of $a$ and $a_0$ are calculated in the function {\tt AS} using Eq. \eqref{eq:asSol} truncated according to the parameter {\tt ORDER}.

\vspace{2mm}
\noindent
\textbf{Subroutine {\tt SETMASSES($m_c^2$, $m_b^2$) }}\\
Fills in the common block {\tt MCONST} the values of the quark masses.

\vspace{2mm}
\noindent
\textbf{Subroutine {\tt SETPOINTSX}}\\
Here the user may choose the points in $x$ for which the output will be calculated.

\vspace{2mm}
\noindent
\textbf{Subroutine {\tt SETQ2($Q^2$)}}\\
Sets the virtuality to $Q^2$ in the common block {\tt PHYCONS}.

\vspace{2mm}
\noindent
\textbf{Functions {\tt UMUB, DMDB, SMSB} }\\
User-defined input distributions $(u-\bar u),(d-\bar d),(s-\bar s)$ in $N$-space at the scale $Q_0^2$.

\vspace{2mm}
\noindent
\textbf{Functions {\tt UPUB, DPDB, SPSB, GLU, UPUBPOL, DPDBPOL, SPSBPOL, GLUPOL}}\\
User-defined input distributions $(u+\bar u),(d+\bar d),(s+\bar s)$ and the gluon density in $N$-space at the scale $Q_0^2$, and their polarized counterparts.

\vspace{2mm}
\noindent
\textbf{Subroutine {\tt USERINIT}}\\
Initializes user options including output control, the order of the perturbative truncation, and the physical constants $Q_0^2, \Lambda_{QCD}$, and the values of the quark charges.

\vspace{2mm}
\noindent
\textbf{Subroutine {\tt WRITEUPF2}, {\tt WRITEUPF3WMW}, {\tt WRITEUPF3WPW}, {\tt WRITEUPFL}, {\tt WRITEUPG1} }\\
Prints as output the values of $Q_0^2, Q^2, \Lambda_{QCD}, a,a_0$ and the PDFs together with the respective structure functions. These routines also create a file containing moment and convolution checks.

\vspace{2mm}
\noindent
\textbf{Functions for the analytic continuation of sums}\\
The functions required for the analytic continuation of harmonic sums are defined in the file {\tt allfuncs.f}. The weight-6 functions are defined in {\tt w6func.f}.

\subsubsection{User options}

In the source code file {\tt userinit.f}, the following switches can be modified:

\vspace{2mm}
\noindent
\textbf{\tt \underline{ORDER}=0,1,2,3}: chooses the order of the calculation. The possible values and their effects are explained in Table \ref{tab:orders}.
\begin{table}[H]
\centering
\begin{tabular}{ |c|c|c|c| } 
 \hline
{\tt ORDER}	& \eqref{eq:solAP}	& \eqref{eq:asSol} 		& Wilson coefficients \\ 
\hline
 0 			& LO 				& LO 					& $\mathcal O(a_s^0)$\\ 
 1			& NLO 				& NLO 					& $\mathcal O(a_s)$\\ 
 2 			& NNLO 				& NNLO 					& $\mathcal O(a_s^2)$\\ 
 3 			& NNLO 				& $\text{N}^3\text{LO}$ & $\mathcal O(a_s^3)$, see \protect\footnotemark.\\ 
 \hline
\end{tabular}
\caption{\label{tab:orders}Truncations applied according to the value of the switch {\tt ORDER}.}
\end{table}
\footnotetext{The OMEs $\tilde{\tilde A}_{Qq}^{PS(3)}$ and $\tilde{\tilde A}_{Qg}^{(3)}$ are set to zero, and so are the $\mathcal O(a_s^3)$ polarized massless Wilson coefficients. For the unpolarized ones, the approximate representations given in \cite{Vermaseren:2005qc} are used after Mellin transforming them to $N$-space.}
\vspace{2mm}
\noindent
{\tt \underline{MOMCHK}=0,1}: chooses whether or not to calculate and save numerical self-checks on the program, consisting on the evaluation of fixed moments and convolutions, compared with expected values hard-coded in the program. The differences to the expected values are saved in files called {\tt F2\_checks.txt, FL\_ckecks.txt}, etc. If the program is running correctly, these differences do not significantly depart from zero. The precision expected from the programs is discussed more in what follows.

\subsubsection{User initialization}
The programs accept as input the following parameters and settings:

\begin{itemize}
\item For $F_2$, $F_L$, $xF_3^{W^+ + W^-}$ and $g_1$, parametrizations of the PDFs $(u+\bar u)(N,Q_0^2)$, $(d+\bar d)(N,Q_0^2)$, $(s+\bar s)(N,Q_0^2)$ and $g(N,Q_0^2)$ in the complex plane, as well as of the corresponding polarized PDFs. These parametrizations must be defined in the functions {\tt UPUB,DPDB,SPSB,GLU}; the polarized ones in {\tt UPUBPOL,DPDBPOL,SPSBPOL,GLUPOL}. 
\item For $xF_3^{W^+ - W^-}$, parametrizations of the valence distributions $(u-\bar u)(N,Q_0^2)$, $(d-\bar d)(N,Q_0^2)$, $(s-\bar s)(N,Q_0^2)$ must be defined in the functions {\tt UMUB,DMDB,SMSB}.
\item Values of the virtualities $Q_0^2$ and $Q^2$ must be set through a subroutine call to {\tt SETQ02} and to {\tt SETQ2}.
\item The values of the OMS masses $m_c^2$ and $m_b^2$ are set through a call to the routine {\tt SETMASSES}. The bottom quark mass is only used for the computation of $F_2$.
\item The values of $\Lambda_{QCD}$ and of the number of decoupling quarks in the running of $a_s$ are set by calling the subroutine {\tt SETAS}. 
\item The user can choose the order of the evolution by setting the switch {\tt ORDER} in the source code.
\item The values of $x$ to be computed are programmed by the user through the function {\tt \mbox{SETPOINTSX}}.
\end{itemize}

\subsubsection{Output}
The program prints the values of $Q_0^2$, $Q^2$, $a(Q^2)$, $a_0(Q_0^2)$, $m_c^2$ and $m_b^2$, and tabulates the values of $x$ and the corresponding values of the PDFs $x(q\pm\bar q)(x,Q^2)$ and $xg(x,Q^2)$, and separately the values of the structure functions. Optionally, the moment test and convolution test are also printed.

\subsubsection{Estimates of the numerical accuracy}

We reproduce in the following a short sample of the checks produced by the programs, to illustrate the accuracy in the numerical inverse Mellin transformation. The evaluation of harmonic polylogarithms was obtained using the code \cite{Gehrmann:2001pz}.

In Tables \ref{tab:accuracy1}-\ref{tab:accuracy5}  we reproduce the results of the inverse Mellin transform of individual functions, and compare to a direct evaluation in $x$-space.
\begin{table}[H]
\centering
\resizebox{\textwidth}{!}{%
\texttt{\small%
\begin{tabular}{ |c|c|c|c| } 
 \hline
$x$	               & $a=1/(x-1)$         & $b=M^{-1}[S_1(N-1)](x) $& $(b-a)/a$ \\ 
\hline
 1.000000000000E-04 & -1.000100010001E+00 & -1.000099987789E+00 & -2.220935986079E-08 \\ 
 1.584893192461E-04 & -1.000158514442E+00 & -1.000158502934E+00 & -1.150539496217E-08 \\
 2.511886431509E-04 & -1.000251251754E+00 & -1.000251245838E+00 & -5.914269026505E-09 \\
 3.981071705534E-04 & -1.000398265722E+00 & -1.000398262715E+00 & -3.006285850799E-09 \\
 6.309573444801E-04 & -1.000631355702E+00 & -1.000631354196E+00 & -1.505841968705E-09 \\
 1.000000000000E-03 & -1.001001001001E+00 & -1.001001000260E+00 & -7.397988681656E-10 \\
 1.584893192461E-03 & -1.001587409066E+00 & -1.001587408710E+00 & -3.549958361238E-10 \\
 2.511886431509E-03 & -1.002518211893E+00 & -1.002518211727E+00 & -1.658781496371E-10 \\
 3.981071705534E-03 & -1.003996983985E+00 & -1.003996983909E+00 & -7.550888326096E-11 \\
 6.309573444801E-03 & -1.006349636945E+00 & -1.006349636911E+00 & -3.372868363612E-11 \\
 1.000000000000E-02 & -1.010101010101E+00 & -1.010101010085E+00 & -1.497991730659E-11 \\
 1.584893192461E-02 & -1.016104165751E+00 & -1.016104165744E+00 & -6.611049988000E-12 \\
 2.511886431509E-02 & -1.025766078956E+00 & -1.025766078953E+00 & -2.718826733572E-12 \\
 3.981071705534E-02 & -1.041461322014E+00 & -1.041461322014E+00 & -7.850202576436E-13 \\
 6.309573444801E-02 & -1.067344911073E+00 & -1.067344911073E+00 & -1.474964870854E-13 \\
 1.000000000000E-01 & -1.111111111111E+00 & -1.111111111111E+00 & -5.381695089567E-13 \\
 1.584893192461E-01 & -1.188339046515E+00 & -1.188339046516E+00 & -7.150860737426E-13 \\
 2.511886431509E-01 & -1.335449831060E+00 & -1.335449831061E+00 & -7.992575917488E-13 \\
 3.981071705534E-01 & -1.661425341982E+00 & -1.661425341983E+00 & -8.347595087882E-13 \\
 6.309573444801E-01 & -2.709713863811E+00 & -2.709713863814E+00 & -8.473002428732E-13 \\
 9.000000000000E-01 & -1.000000000000E+01 & -1.000000000000E+01 & -8.512301974405E-13 \\
 9.499999999999E-01 & -1.999999999999E+01 & -2.000000000001E+01 & -8.521183758603E-13 \\
 \hline
\end{tabular}
}
}
\caption{\label{tab:accuracy1} Accuracy in the inverse Mellin transform of $S_1(N-1)$. The program output is reproduced with 13 digits for reasons of space.}
\end{table}

\begin{table}[H]
\centering
\resizebox{\textwidth}{!}{%
\texttt{\small%
\begin{tabular}{ |c|c|c|c| } 
 \hline
$x$	               & $a=\frac{\ln x}{x+1}$         & $b=M^{-1}[\beta_1(N)-\frac{\zeta_2}{2}](x) $& $(b-a)/a$ \\ 
\hline
 1.000000000000E-04  & -9.209419430033E+00  & -9.209419470448E+00  & -4.388443722660E-09 \\ 
 1.584893192461E-04  & -8.748436819581E+00  & -8.748436839625E+00  & -2.291173719895E-09 \\
 2.511886431509E-04  & -8.287224678056E+00  & -8.287224687898E+00  & -1.187672521548E-09 \\
 3.981071705534E-04  & -7.825673859301E+00  & -7.825673864078E+00  & -6.103743606524E-10 \\
 6.309573444801E-04  & -7.363626163571E+00  & -7.363626165861E+00  & -3.110277938206E-10 \\
 1.000000000000E-03  & -6.900854424557E+00  & -6.900854425642E+00  & -1.572016298233E-10 \\
 1.584893192461E-03  & -6.437036245458E+00  & -6.437036245967E+00  & -7.905129767632E-11 \\
 2.511886431509E-03  & -5.971720956940E+00  & -5.971720957177E+00  & -3.977833130825E-11 \\
 3.981071705534E-03  & -5.504291245051E+00  & -5.504291245162E+00  & -2.022241689528E-11 \\
 6.309573444801E-03  & -5.033925283296E+00  & -5.033925283349E+00  & -1.056461036735E-11 \\
 1.000000000000E-02  & -4.559574441572E+00  & -4.559574441598E+00  & -5.788308115063E-12 \\
 1.584893192461E-02  & -4.079989688562E+00  & -4.079989688576E+00  & -3.395767201996E-12 \\
 2.511886431509E-02  & -3.593862406631E+00  & -3.593862406638E+00  & -2.171968528102E-12 \\
 3.981071705534E-02  & -3.100198023848E+00  & -3.100198023853E+00  & -1.533299120142E-12 \\
 6.309573444801E-02  & -2.599109395380E+00  & -2.599109395383E+00  & -1.207994830544E-12 \\
 1.000000000000E-01  & -2.093259175449E+00  & -2.093259175451E+00  & -1.044424505925E-12 \\
 1.584893192461E-01  & -1.590060472541E+00  & -1.590060472543E+00  & -9.557330066054E-13 \\
 2.511886431509E-01  & -1.104190853520E+00  & -1.104190853521E+00  & -9.091396253330E-13 \\
 3.981071705534E-01  & -6.587721289155E-01  & -6.587721289161E-01  & -8.825871230364E-13 \\
 6.309573444801E-01  & -2.823599404100E-01  & -2.823599404103E-01  & -8.579498338942E-13 \\
 9.000000000000E-01  & -5.545290297780E-02  & -5.545290297784E-02  & -7.716847343866E-13 \\
 9.499999999999E-01  & -2.630425353207E-02  & -2.630425353209E-02  & -6.234761888181E-13 \\
 \hline
\end{tabular}
}
}
\caption{\label{tab:accuracy2} Accuracy in the inverse Mellin transform of $\beta_1$. }
\end{table}
\begin{table}[H]
\centering
\resizebox{\textwidth}{!}{%
\texttt{\small%
\begin{tabular}{ |c|c|c|c| } 
 \hline
$x$	              & $a=\frac{\Li_2(x)-\zeta_2}{1-x}$ & $b=M^{-1}[S_{2,1}(N-1)](x) $& $(b-a)/a$ \\ 
\hline
 1.000000000000E-04  &  -1.644998564204E+00  &  -1.644998540407E+00  &  -1.446647820702E-08 \\ 
 1.584893192461E-04  &  -1.645036291930E+00  &  -1.645036279635E+00  &  -7.474110813401E-09 \\
 2.511886431509E-04  &  -1.645096091884E+00  &  -1.645096085586E+00  &  -3.828541288504E-09 \\
 3.981071705534E-04  &  -1.645190882335E+00  &  -1.645190879148E+00  &  -1.937364086139E-09 \\
 6.309573444801E-04  &  -1.645341150031E+00  &  -1.645341148443E+00  &  -9.650914623935E-10 \\
 1.000000000000E-03  &  -1.645579396133E+00  &  -1.645579395357E+00  &  -4.711063973839E-10 \\
 1.584893192461E-03  &  -1.645957211621E+00  &  -1.645957211251E+00  &  -2.245911407573E-10 \\
 2.511886431509E-03  &  -1.646556564352E+00  &  -1.646556564180E+00  &  -1.042902219387E-10 \\
 3.981071705534E-03  &  -1.647507872860E+00  &  -1.647507872782E+00  &  -4.725261822036E-11 \\
 6.309573444801E-03  &  -1.649019119964E+00  &  -1.649019119929E+00  &  -2.103084084651E-11 \\
 1.000000000000E-02  &  -1.651423186977E+00  &  -1.651423186962E+00  &  -9.254507404824E-12 \\
 1.584893192461E-02  &  -1.655255929989E+00  &  -1.655255929983E+00  &  -3.943331715728E-12 \\
 2.511886431509E-02  &  -1.661387852969E+00  &  -1.661387852966E+00  &  -1.437673816447E-12 \\
 3.981071705534E-02  &  -1.671253765538E+00  &  -1.671253765537E+00  &  -1.953058093093E-13 \\
 6.309573444801E-02  &  -1.687273908713E+00  &  -1.687273908714E+00  &  -3.912456697356E-13 \\
 1.000000000000E-01  &  -1.713684750832E+00  &  -1.713684750833E+00  &  -6.404716396125E-13 \\
 1.584893192461E-01  &  -1.758360042961E+00  &  -1.758360042962E+00  &  -7.580550801870E-13 \\
 2.511886431509E-01  &  -1.837462925114E+00  &  -1.837462925115E+00  &  -8.120652257532E-13 \\
 3.981071705534E-01  &  -1.990498234443E+00  &  -1.990498234445E+00  &  -8.357496386142E-13 \\
 6.309573444801E-01  &  -2.355468624473E+00  &  -2.355468624475E+00  &  -8.459583217938E-13 \\
 9.000000000000E-01  &  -3.452193438432E+00  &  -3.452193438435E+00  &  -8.495367348769E-13 \\
 9.499999999999E-01  &  -4.086005397563E+00  &  -4.086005397567E+00  &  -8.512242038690E-13 \\
 \hline
\end{tabular}
}
}
\caption{\label{tab:accuracy3} Accuracy in the inverse Mellin transform of $S_{2,1}(N-1)$. }
\end{table}
\begin{table}[H]
\centering
\resizebox{\textwidth}{!}{%
\texttt{\small%
\begin{tabular}{ |c|c|c|c| } 
 \hline
$x$	              & $a=\frac{H_{0,1,1}}{x^2(1+x)}$ & $b=M^{-1}[f_7(N-2)](x) $& $(b-a)/a$ \\ 
\hline
 1.000000000000E-04 & 2.499916688546E-01  &   2.499916756757E-01   & 2.728536296967E-08 \\ 
 1.584893192461E-04 & 2.499867975171E-01  &   2.499868010011E-01   & 1.393687802374E-08 \\
 2.511886431509E-04 & 2.499790801242E-01  &   2.499790817952E-01   & 6.684638324199E-09 \\
 3.981071705534E-04 & 2.499668557469E-01  &   2.499668565807E-01   & 3.335627022470E-09 \\
 6.309573444801E-04 & 2.499474989831E-01  &   2.499474993735E-01   & 1.561576512285E-09 \\
 1.000000000000E-03 & 2.499168644706E-01  &   2.499168646514E-01   & 7.236653027069E-10 \\
 1.584893192461E-03 & 2.498684222556E-01  &   2.498684223412E-01   & 3.426906042173E-10 \\
 2.511886431509E-03 & 2.497919230923E-01  &   2.497919231306E-01   & 1.530682820801E-10 \\
 3.981071705534E-03 & 2.496713736071E-01  &   2.496713736252E-01   & 7.270504517586E-11 \\
 6.309573444801E-03 & 2.494820529162E-01  &   2.494820529246E-01   & 3.353604243581E-11 \\
 1.000000000000E-02 & 2.491863455175E-01  &   2.491863455214E-01   & 1.583557950207E-11 \\
 1.584893192461E-02 & 2.487285250337E-01  &   2.487285250356E-01   & 7.781940476849E-12 \\
 2.511886431509E-02 & 2.480298918862E-01  &   2.480298918872E-01   & 4.048020924753E-12 \\
 3.981071705534E-02 & 2.469893222921E-01  &   2.469893222926E-01   & 2.278190730003E-12 \\
 6.309573444801E-02 & 2.455038639210E-01  &   2.455038639214E-01   & 1.489280827377E-12 \\
 1.000000000000E-01 & 2.435479273673E-01  &   2.435479273676E-01   & 1.137355787167E-12 \\
 1.584893192461E-01 & 2.414097389914E-01  &   2.414097389916E-01   & 9.760048719883E-13 \\
 2.511886431509E-01 & 2.403559269819E-01  &   2.403559269822E-01   & 9.044156761646E-13 \\
 3.981071705534E-01 & 2.446909830076E-01  &   2.446909830079E-01   & 8.748943344455E-13 \\
 6.309573444801E-01 & 2.711423442143E-01  &   2.711423442145E-01   & 8.586403903031E-13 \\
 9.000000000000E-01 & 3.802317499521E-01  &   3.802317499524E-01   & 8.577085251947E-13 \\
 9.499999999999E-01 & 4.374596695453E-01  &   4.374596695457E-01   & 8.550139887979E-13 \\
\hline
\end{tabular}
}
}
\caption{\label{tab:accuracy4} Accuracy in the inverse Mellin transform of $f_7$. }
\end{table}
\begin{table}[H]
\centering
\resizebox{\textwidth}{!}{%
\texttt{\small%
\begin{tabular}{ |c|c|c|c| } 
 \hline
$x$	              & $a=\frac{H_{0,1,1,1}(x)}{1-x}-\frac{2 \zeta_2^2}{5(1-x)}$ & $b=M^{-1}[S_{2,1,1,1}(N-1)](x) $& $(b-a)/a$ \\ 
\hline
 1.000000000000E-04 &  5.555638071193E-02 & 5.555625334407E-02 &  2.292587369082E-06 \\ 
 1.584893192461E-04 &  5.555664409123E-02 & 5.555665887067E-02 &  2.660248001242E-07 \\
 2.511886431509E-04 &  5.555729293858E-02 & 5.555730385240E-02 &  1.964426350397E-07 \\
 3.981071705534E-04 &  5.555832437271E-02 & 5.555832866686E-02 &  7.729080447635E-08 \\
 6.309573444801E-04 &  5.555995611372E-02 & 5.555995782154E-02 &  3.073829955091E-08 \\
 1.000000000000E-03 &  5.556255083106E-02 & 5.556255146465E-02 &  1.140314245553E-08 \\
 1.584893192461E-03 &  5.556669059384E-02 & 5.556669087928E-02 &  5.136982282049E-09 \\
 2.511886431509E-03 &  5.557332337352E-02 & 5.557332348222E-02 &  1.955951471582E-09 \\
 3.981071705534E-03 &  5.558401636196E-02 & 5.558401640846E-02 &  8.366584091823E-10 \\
 6.309573444801E-03 &  5.560141809671E-02 & 5.560141811601E-02 &  3.470683809650E-10 \\
 1.000000000000E-02 &  5.563014003017E-02 & 5.563014003790E-02 &  1.390042926803E-10 \\
 1.584893192461E-02 &  5.567853150687E-02 & 5.567853151009E-02 &  5.796285222075E-11 \\
 2.511886431509E-02 &  5.576244532849E-02 & 5.576244532979E-02 &  2.345220891379E-11 \\
 3.981071705534E-02 &  5.591362216565E-02 & 5.591362216618E-02 &  9.432983307131E-12 \\
 6.309573444801E-02 &  5.619919413788E-02 & 5.619919413809E-02 &  3.791382036271E-12 \\
 1.000000000000E-01 &  5.676911112229E-02 & 5.676911112238E-02 &  1.609525904497E-12 \\
 1.584893192461E-01 &  5.797855415690E-02 & 5.797855415696E-02 &  9.753948876302E-13 \\
 2.511886431509E-01 &  6.073950831360E-02 & 6.073950831365E-02 &  9.513924303335E-13 \\
 3.981071705534E-01 &  6.777446683517E-02 & 6.777446683523E-02 &  8.819196207566E-13 \\
 6.309573444801E-01 &  9.071028395085E-02 & 9.071028395093E-02 &  8.464949793590E-13 \\
 9.000000000000E-01 &  1.891923641417E-01 & 1.891923641419E-01 &  8.547067122406E-13 \\
 9.499999999999E-01 &  2.538238785232E-01 & 2.538238785235E-01 &  8.544588835467E-13 \\
\hline
\end{tabular}
}
}
\caption{\label{tab:accuracy5} Accuracy in the inverse Mellin transform of $S_{2,1,1,1}(N-1)$.}
\end{table}

We compared the numerical evaluation in $N$-space of the massless Wilson coefficients to known results. In Tables \ref{tab:accuracyWCsOdd} and \ref{tab:accuracyWCs} we quantify the numerical discrepancies,

\begin{equation}
	\Delta_a = \left|\frac{a_{\text{this program}}-a_{\text{exact}}}{a_{\text{exact}}}\right| ~.
\end{equation}

\begin{table}[H]
\centering
\resizebox{\textwidth}{!}{%
\texttt{\footnotesize%
\begin{tabular}{ |l|c|c|c|c|c|c| } 
 \hline
 \multicolumn{1}{|c|}{\small$a$}				& \multicolumn{6}{c|}{\small$\Delta_a$} \\
 \hline
 	 				& {\small$N=2$}			& {\small$N=4$} 		& {\small$N=6$} 		& {\small$N=8$} 		& {\small$N=10$} 		& {\small$N=12$} \\ 
 \hline
{\small$c_{2,q}^{(1)}$}		& 6.4948E-15	& 5.85612E-16	& 1.58934E-16	& 2.28766E-16	& 1.84073E-16	& 1.57002E-16	\\[2mm]
{\small$c_{2,g}^{(1)}$}		& 4.44089E-16	& 0.			& 0.			& 1.14914E-16	& 0.			& 0.		\\[1mm]
{\small$c_{2,q}^{(2)NS}$}	& 2.37352E-5		& 9.33086E-16	& 3.9048E-16	& 6.36842E-16	& 1.14E-16			& 1.77746E-16		\\[2mm]
{\small$c_{2,q}^{(2)PS}$}	& 6.76151E-16	& 3.04694E-16	& 1.52019E-16	& 0.			& 1.21499E-16	& 1.50975E-16	\\[2mm]
{\small$c_{2,g}^{(2)}$}		& 4.12294E-6		& 3.42287E-16	& 1.06955E-16	& 1.27729E-17	& 8.36773E-17	& 7.10658E-16	\\[1mm]
{\small$c_{2,q}^{(3)NS,appr.}$}&3.93616E-2  & 1.11349E-4	& 6.75312E-5	& 5.36697E-5	& 4.67686E-5	& 4.25631E-5 \\[2mm]
{\small$c_{2,q}^{(3)NS}$}	& 2.47144E-5		& 2.58686E-10	& 1.13248E-9	& 3.39733E-10	& 	& 1.6304E-11	\\[2mm]
{\small$c_{2,q}^{(3)PS}$}	& 2.18454E-5	& 9.60885E-5	& 5.23834E-5	& 8.06659E-5	& 9.16942E-5	& 9.66326E-5	\\[2mm]
{\small$c_{2,g}^{(3)}$}		& 4.53521E-5	& 1.81324E-4	& 4.07667E-5	& 2.0697E-6		& 1.40843E-5	& 2.10654E-5	\\[2mm]
\hline
{\small$c_{L,q}^{(1)}$}		& 0.			& 0.			& 0.			& 0.			& 0.			& 0.		\\[1mm]
{\small$c_{L,g}^{(1)}$}		& 0.			& 0.			& 1.29526E-16	& 0.			& 0.			& 0.		\\[2mm]
{\small$c_{L,q}^{(2)NS}$}	& 1.6779E-7		& 9.41017E-17	& 4.04832E-17	& 1.06525E-16	& 1.18235E-16	& 1.48074E-16	\\[1mm]
{\small$c_{L,q}^{(2)PS}$}	& 0.			& 0.			& 0.			& 1.51994E-16	& 0.			& 0.		\\[1mm]
{\small$c_{L,g}^{(2)}$}		& 2.1386E-7		& 3.42845E-16	& 1.15412E-16	& 0.			& 0.			& 2.4515E-16	\\[1mm]
{\small$c_{L,q}^{(3)NS}$}	& 7.83837E-7	& 9.91557E-6	& 1.49838E-5	& 1.59852E-5	& 1.36995E-5	& 8.8628E-6	\\[2mm]
{\small$c_{L,q}^{(3)PS}$}	& 5.11562E-5	& 3.35633E-5	& 3.05032E-5	& 4.14035E-5	& 5.79093E-5	& 7.5186E-5	\\[2mm]
{\small$c_{L,g}^{(3)}$}		& 1.27371E-5	& 9.99753E-5	& 1.00776E-4	& 9.06315E-5	& 7.55299E-5	& 5.71978E-5	\\[2mm]
\hline
{\small$c_{3,q}^{W^+\pm W^-,(1)}$}		& 1.3739E-15	& 5.47507E-16	& 6.86427E-16	& 4.76866E-16	& 1.89059E-16	& 3.20049E-16 \\[2mm] 
{\small$c_{3,q}^{W^++W^-,NS,(2)}$}		& 2.39981E-15	& 7.05237E-16	& 3.87358E-16	& 1.74118E-16	& 6.03027E-16	& 5.54728E-16 \\[2mm]
\hline
\end{tabular}
}
}
\caption{\label{tab:accuracyWCs} Accuracy in the library's evaluation of integer moments of massless Wilson coefficients. The evaluations refer to $N_F=3$ light flavours. The quantities $c_{2,q}^{(3)NS,appr.}$, $c_{2,q}^{(3)PS}$, $c_{2,g}^{(3)}$ refer to the approximate formulas given in \cite{Vermaseren:2005qc} whereas $c_{2,q}^{(3)NS}$ refers to the analytic continuation of the exact formula.\\
The second moment of $a=c_{2,q}^{(2)NS},c_{2,g}^{(2)},c_{L,q}^{(2)NS},c_{L,g}^{(2)}$ is evaluated numerically as 
\newline
\leavevmode\\\begin{minipage}{\linewidth}
  \begin{equation*}
\frac{1}{2}[a(N=2+\ep) + a(N=2-\ep)] , \qquad \ep = 10^{-9}.
  \end{equation*}
  \end{minipage}
For $c_{2,q}^{(3)NS}$, the accuracy is shown for $\ep=10^{-5}$. 
}
\end{table}

\begin{table}[H]
\centering
\resizebox{\textwidth}{!}{%
\texttt{\footnotesize%
\begin{tabular}{ |l|c|c|c|c|c|c| } 
 \hline
 \multicolumn{1}{|c|}{\small$a$}				& \multicolumn{6}{c|}{\small$\Delta_a$} \\
 \hline
 	 				& {\small$N=1$}			& {\small$N=3$} 		& {\small$N=5$} 		& {\small$N=7$} 		& {\small$N=9$} 		& {\small$N=11$} \\ 
\hline 	 				
{\small$c_{3,q}^{W^+\pm W^-,(1)}$}	& 1.77636E-15	& 5.32907E-16	& 2.29262E-16	& 8.3774E-16	& 2.1003E-16	& 1.72896E-16 \\[2mm]
{\small$c_{3,q}^{W^+-W^-,NS,(2)}$}	& 4.0864E-7		& 1.18963E-14 	& 1.23781E-15 	& 2.23423E-16 	& 1.42485E-16 			& 6.27689E-16 \\[2mm]
{\small$c_{3,q}^{W^+-W^-,NS,(3)}$}	& 4.69528E-5	& 2.05149E-4	& 2.06337E-4	& 7.25476E-5	& 4.1283E-5	& 2.7121E-5	\\[2mm]
\hline
{\small$c_{g_1,q}^{(1)}$}			& 5.70972E-16 	& 2.71212E-15 	& 3.60433E-16 	& 7.08728E-16 	& 0. 		& 1.73578E-16 \\[2mm]
{\small$c_{g_1,g}^{(1)}$}			& 4.51589E-16	& 0. 			& 0. 			& 0. 			& 1.11982E-16 	& 2.48665E-16  \\[2mm]
{\small$c_{g_1,q}^{(2),NS}$}		&1.07975E-8		& 6.11274E-15 	& 1.54687E-15 	& 3.96442E-16 	& 0. 			& 7.35235E-16 \\[2mm]
{\small$c_{g_1,q}^{(2),PS}$}		&1.07975E-8		& 0. 		& 0. 			& 1.53761E-16 & 2.20764E-16 & 2.97226E-16 \\[2mm]
{\small$c_{g_1,g}^{(2)}$}			&9.21074E-7		& 6.99694E-17 & 1.55394E-16 & 8.18172E-17 & 7.9721E-17 & 1.59193E-16 \\[1mm]
\hline
\end{tabular}
}
}
\caption{\label{tab:accuracyWCsOdd} Accuracy in the library's evaluation of integer moments of massless Wilson coefficients. The first moment of $c_{3,q}^{W^+-W^-,NS,(2)},c_{g_1,g}^{(1)},c_{g_1,q}^{(2),PS},c_{g_1,g}^{(2)}$ is evaluated numerically as 
\leavevmode\\\begin{minipage}{\linewidth}
  \begin{equation*}
    \frac{1}{2}[a(N=1+10^{-9}) + a(N=1-10^{-9})] .
  \end{equation*}
  \end{minipage}
In the case of $c_{g_1,g}^{(1)},c_{g_1,q}^{(2),PS},c_{g_1,g}^{(2)}$, we show the absolute error.}
\end{table}

\begin{table}[H]
\centering
\resizebox{\textwidth}{!}{%
\texttt{\footnotesize%
\begin{tabular}{ |l|c|c|c|c|c|c| }
 \hline
 	 				& {\small$N=2$}			& {\small$N=4$} 		& {\small$N=6$} 		& {\small$N=8$} 		& {\small$N=10$} 		& {\small$N=12$} \\ 
\hline 	 				
$\Delta_{gg}^{(2),C_A}$            &	-8.57019E-02 & -8.06314E-03 & -1.96624E-03 & -7.03654E-04 & -3.12204E-04 & -1.59213E-04 \\ [1mm] 
$\Delta_{gg}^{(2),C_F}$            &  	 1.08208E+00 &  9.35395E-01 &  7.36498E-01 &  5.82599E-01 &  4.70983E-01 &  3.89033E-01 \\ [1mm] 
$\Delta_{qg}^{(2),C_A}$            & 	-1.81706E+01 & -2.72475E+01 & -3.06467E+01 & -3.20367E+01 & -3.25043E+01 & -3.25071E+01 \\ [1mm] 
$\Delta_{qg}^{(2),C_F}$            & 	-2.55081E+01 & -3.67969E+01 & -4.15168E+01 & -4.34496E+01 & -4.40779E+01 & -4.40482E+01 \\ [1mm] 
$\Delta_{qq,C\bar E}^{(2)}$        &   	 2.24711E-03 & -1.57089E-02 & -9.44382E-03 & -5.86092E-03 & -3.92770E-03 & -2.80004E-03 \\ [1mm] 
$\Delta_{qq,C\bar F}^{(2)}$        &   	-1.12471E-01 & -9.03905E-03 & -2.00897E-03 & -6.79961E-04 & -2.90890E-04 & -1.44607E-04 \\ [1mm] 
$\Delta_{q\bar q}^{(1)}$           &   	 3.70178E+01 &  5.94904E+01 &  7.74200E+01 &  9.22970E+01 &  1.05079E+02 &  1.16334E+02 \\ [1mm] 
$\Delta_{q\bar q,A\bar A}^{(2)}$   &   	-1.71367E+02 & -1.35350E+02 & -1.13349E+02 & -9.85219E+01 & -8.77308E+01 & -7.94559E+01 \\ [1mm] 
$\Delta_{q\bar q,A\bar C}^{(2)}$   &   	-3.90245E+00 & -2.54535E+00 & -1.91964E+00 & -1.55866E+00 & -1.32124E+00 & -1.15196E+00 \\ [1mm] 
$\Delta_{q\bar q,B\bar B}^{(2)}$   &   	 4.55339E-02 &  1.60061E-03 &  1.96688E-04 &  4.21689E-05 &  1.24460E-05 &  4.52546E-06  \\ [1mm] 
$\Delta_{q\bar q,B\bar C}^{(2)}$   &   	 6.35907E-02 &  4.10797E-03 &  8.04686E-04 &  2.50991E-04 &  1.01396E-04 &  4.83097E-05 \\ [1mm] 
$\Delta_{q\bar q}^{(2),C_A}$       &   	 6.38582E+02 &  5.65000E+02 &  4.93668E+02 &  4.39274E+02 &  3.97207E+02 &  3.63728E+02 \\ [1mm] 
$\Delta_{q\bar q,C\bar C}^{(2)}$   &   	 2.09305E+00 &  1.17181E+00 &  7.80317E-01 &  5.59664E-01 &  4.24532E-01 &  3.35383E-01 \\ [1mm] 
$\Delta_{q\bar q,C\bar D}^{(2)}$   &   	 2.95342E+00 &  2.66503E-01 &  6.74218E-02 &  2.59248E-02 &  1.24901E-02 &  6.92420E-03 \\ [1mm] 
$\Delta_{q\bar q}^{(2),C_F}$       &   	 7.33957E+02 &  9.55661E+02 &  9.91477E+02 &  9.82852E+02 &  9.60469E+02 &  9.33912E+02 \\ [1mm] 
$\Delta_{q\bar q}^{(2),S+V}$       &   	 1.61679E+02 &  1.65599E+03 &  3.41258E+03 &  5.18771E+03 &  6.92816E+03 &  8.61957E+03 \\ [1mm]
\hline
\end{tabular}
}
}
\caption{\label{tab:mellinDY} Mellin moments of the Wilson coefficients for the Drell-Yan process.}
\end{table}

\begin{table}[H]
\centering
\resizebox{\textwidth}{!}{%
\texttt{\footnotesize%
\begin{tabular}{ |l|c|c|c|c|c|c| } 
 \hline
 	 				& {\small$N=2$}			& {\small$N=4$} 		& {\small$N=6$} 		& {\small$N=8$} 		& {\small$N=10$} 		& {\small$N=12$} \\ 
\hline 
$\Delta_{gg}^{(1)}$               	&	 1.52650E+02 &  1.99742E+02 &  2.37478E+02 &  2.69033E+02 &  2.96335E+02 &  3.20510E+02 \\ [1mm]
$\Delta_{gg}^{(2),C_A^2}$        	&  	 1.65181E+03 &  2.77754E+03 &  3.85776E+03 &  4.89074E+03 &  5.88033E+03 &  6.83078E+03 \\ [1mm]
$\Delta_{gg}^{(2),C_AT_FN_F}$     	&  	-2.47950E+02 & -3.38074E+02 & -4.13326E+02 & -4.79228E+02 & -5.38449E+02 & -5.92551E+02 \\ [1mm]
$\Delta_{gg}^{(2),C_FT_FN_F}$     	&  	 6.82658E+00 &  1.64545E+00 & -9.12071E-01 & -2.35346E+00 & -3.25173E+00 & -3.85374E+00 \\ [1mm]
$\Delta_{gq}^{(1)}$              	& 	-3.01851E+00 & -2.29037E+00 & -1.82594E+00 & -1.53336E+00 & -1.33113E+00 & -1.18186E+00 \\ [1mm]
$\Delta_{q_1q_2}^{(2)C_F^2}$     	&  	 6.82972E+00 &  3.11709E+00 &  1.92019E+00 &  1.34053E+00 &  1.00564E+00 &  7.90772E-01 \\ [1mm]
$\Delta_{qg}^{(2),C_AC_F}$       	&  	-1.47342E+02 & -1.23953E+02 & -1.09978E+02 & -1.00549E+02 & -9.35410E+01 & -8.80073E+01 \\ [1mm]
$\Delta_{qg}^{(2),C_F^2}$        	&  	-8.56482E+00 & -1.63375E+01 & -1.94112E+01 & -2.07630E+01 & -2.13430E+01 & -2.15314E+01 \\ [1mm]
$\Delta_{qg}^{(2),C_FT_FN_F}$    	&  	 1.65789E+01 &  1.11063E+01 &  8.69434E+00 &  7.27149E+00 &  6.31066E+00 &  5.60875E+00 \\ [1mm]
$\Delta_{qq}^{(2),C_AC_F^2}$     	&  	-6.23101E-01 & -8.16917E-02 & -2.35849E-02 & -9.66799E-03 & -4.83050E-03 & -2.73995E-03 \\ [1mm]
$\Delta_{qq}^{(2),C_F^3}$        	&    1.24620E+00 &  1.63383E-01 &  4.71698E-02 &  1.93359E-02 &  9.66100E-03 &  5.47991E-03 \\ [1mm]
$\Delta_{q\bar q}^{(1)}$           	&    2.37037E-01 &  3.38624E-02 &  9.40623E-03 &  3.59147E-03 &  1.65760E-03 &  8.68267E-04 \\ [1mm]
$\Delta_{q\bar q}^{(2),C_AC_F^2}$  	&    8.39792E-01 &  3.76668E-01 &  1.54603E-01 &  7.48312E-02 &  4.09579E-02 &  2.45098E-02 \\ [1mm]
$\Delta_{q\bar q}^{(2),C_F^2T_FN_F}$ &	-4.62222E-01 & -1.32003E-01 & -4.58497E-02 & -1.97860E-02 & -9.90172E-03 & -5.50229E-03 \\ [1mm]
$\Delta_{q\bar q}^{(2),C_F^3}$       &	 6.56292E+00 &  1.13663E+00 &  3.55301E-01 &  1.47112E-01 &  7.20582E-02 &  3.95033E-02 \\ [1mm]
\hline
\end{tabular}
}
}
\caption{\label{tab:mellinH} Mellin moments of the Wilson coefficients for Higgs production.}
\end{table}

\begin{table}[H]
\centering
\resizebox{\textwidth}{!}{%
\texttt{\footnotesize%
\begin{tabular}{ |l|c|c|c|c|c|c| } 
 \hline
 	 				& {\small$N=1$}			& {\small$N=3$} 		& {\small$N=5$} 		& {\small$N=7$} 		& {\small$N=9$} 		& {\small$N=11$} \\ 
\hline 
$\delta\Delta_{gg}^{(2),C_A}$          &  	 7.45059E-01 &  1.16898E-02 &  2.21010E-03 &  7.42325E-04 &  3.21754E-04 &  1.62305E-04 \\ [1mm] 
$\delta\Delta_{gg}^{(2),C_F}$          &	-2.70548E+00 & -6.91115E-01 & -7.02238E-01 & -5.94443E-01 & -4.91636E-01 & -4.09331E-01 \\ [1mm] 
$\delta\Delta_{qg}^{(2),C_A}$          &  	 3.86088E+01 &  2.34024E+01 &  2.81792E+01 &  3.06162E+01 &  3.16928E+01 &  3.20546E+01 \\ [1mm] 
$\delta\Delta_{qg}^{(2),C_F}$          &  	-2.15459E+01 &  2.17852E+01 &  3.49012E+01 &  4.00584E+01 &  4.22081E+01 &  4.29881E+01 \\ [1mm] 
$\delta\Delta_{q\bar q,C\bar C}^{(2)}$ &   	 4.05315E+00 & -2.21268E+00 & -1.33162E+00 & -8.96581E-01 & -6.51721E-01 & -4.99097E-01 \\ [1mm] 
$\delta\Delta_{q\bar q,C\bar D}^{(2)}$ &   	 5.37496E+00 & -2.49894E-02 & -2.84472E-02 & -1.46864E-02 & -8.12490E-03 & -4.89008E-03 \\ [1mm]
\hline
\end{tabular}
}
}
\caption{\label{tab:mellinDYpol} Mellin moments of the Wilson coefficients for the polarized Drell-Yan process.}
\end{table}

\begin{table}[H]
\centering
\resizebox{\textwidth}{!}{%
\texttt{\footnotesize%
\begin{tabular}{ |l|c|c|c|c|c|c| } 
 \hline
 	 				& {\small$N=1$}			& {\small$N=3$} 		& {\small$N=5$} 		& {\small$N=7$} 		& {\small$N=9$} 		& {\small$N=11$} \\ 
\hline 
$\Delta_{gg,A-H}^{(1)}$              		&  2.40000E+01 &  2.40000E+01 &  2.40000E+01 &  2.40000E+01 &  2.40000E+01 &  2.40000E+01 \\ [1mm] 
$\Delta_{gg,A-H}^{(2)C_A^2}$           		&  4.10884E+02 &  5.44791E+02 &  6.57038E+02 &  7.48636E+02 &  8.26657E+02 &  8.95012E+02 \\ [1mm] 
$\Delta_{gg,A-H}^{(2)C_AT_FN_F}$        	& -4.88888E+00 & -6.35555E+00 & -6.53121E+00 & -6.58994E+00 & -6.61710E+00 & -6.63196E+00 \\ [1mm] 
$\Delta_{gg,A-H}^{(2)C_FT_FN_F}$        	& -5.13333E+01 & -5.00333E+01 & -5.00042E+01 & -5.00009E+01 & -5.00003E+01 & -5.00001E+01 \\ [1mm] 
$\Delta_{q_1q_2,A-H}^{(2)C_F^2}$         	& -5.33333E+00 & -1.22222E-01 & -1.86243E-02 & -5.24376E-03 & -2.01122E-03 & -9.30046E-04 \\ [1mm] 
$\Delta_{qg,A-H}^{(2)C_AC_F}$          		& -2.04444E+01 & -1.64511E+01 & -1.24188E+01 & -1.01209E+01 & -8.62627E+00 & -7.56492E+00 \\ [1mm] 
$\Delta_{qg,A-H}^{(2)C_F^2}$           		&  0.00000E+00 &  4.16666E-01 &  2.07407E-01 &  1.20535E-01 &  7.82222E-02 &  5.47138E-02 \\ [1mm] 
$\Delta_{qq,A-H}^{(2)C_AC_F^2}$         	&  4.00000E+00 &  1.66666E-01 &  2.96296E-02 &  8.92857E-03 &  3.55555E-03 &  1.68350E-03 \\ [1mm] 
$\Delta_{qq,A-H}^{(2)C_F^3}$           		& -8.00000E+00 & -3.33333E-01 & -5.92592E-02 & -1.78571E-02 & -7.11111E-03 & -3.36700E-03 \\ [1mm] 
$\Delta_{q\bar q,A-H}^{(2)C_AC_F^2}$       	&  8.88888E-01 & -1.00000E-01 & -5.92592E-02 & -3.24735E-02 & -1.91784E-02 & -1.21545E-02 \\ [1mm] 
$\Delta_{q\bar q,A-H}^{(2)C_F^2T_FN_F}$    	& -1.77777E+00 & -8.88888E-02 & -1.69312E-02 & -5.29100E-03 & -2.15488E-03 & -1.03600E-03 \\ [1mm] 
$\Delta_{q\bar q,A-H}^{(2)C_F^3}$         	&  4.00000E+00 &  4.33333E-01 &  1.31216E-01 &  5.65476E-02 &  2.94141E-02 &  1.72235E-02 \\ [1mm]
\hline
\end{tabular}
}
}
\caption{\label{tab:mellinPSH} Mellin moments of the Wilson coefficients for pseudoscalar Higgs boson production.}
\end{table}

\subsubsection{List of mathematical functions}

In our Fortran code, we implemented the special functions and Mellin transforms as listed in Table \ref{tab:fortransums}. Further special functions whose implementation is in part taken from \cite{Blumlein:2000hw} are listed in Table \ref{tab:fortranfunc}. Among these are the classical polylogarithms,
\begin{equation}
	\Li_{k+1}(z) = \int_0^z \frac{\Li_k(t)}{t} dt, \qquad \Li_1(z) = -\ln(1-z) ,
\end{equation}
and Euler's $\Gamma$-function,
\begin{equation}
	\Gamma(z) = \int_0^\infty e^{-t} \ t^{z-1}
\end{equation}
for $\text{Re}(z)>0$, and Euler's Beta function
\begin{equation}
	B(z,w) = \int_0^1 t^{z-1} (1-t)^{w-1} dt = \frac{\Gamma(z)\Gamma(w)}{\Gamma(z+w)} .
\end{equation}

\begin{table}[H]
\centering
\begin{tabular}{ |l|c||l|c| } 
\hline
 Name 			& Definition			& Name							& Definition		\\
\hline
{\tt BETA}		& Euler's $B(z,w)$		& {\tt GAMMAL}					& $\ln\Gamma(z)$  \\
{\tt FLI}\it i	& $\Li_i(x), i=2,3,4$	& {\tt SUM}\it i, $i=1,2,3,4$	& Harmonic sums of integer argument and depth $i$	\\
\hline
\end{tabular}
\caption{\label{tab:fortranfunc} Special functions whose code is partly lifted from \cite{Blumlein:2000hw}. }
\end{table}

\begin{table}[H]
\centering
\begin{tabular}{ |l|l|c||l|l|c| } 
\hline
 Name 	& Name,  	& Definition			& Name	& Name,				& Definition		\\
 		& $|z|>15$	&						&		& $|z|>15$			& 				\\
\hline
{\tt S1} 	& {\tt ASYS1} 	& $S_1(z)$  	& {\tt S31}		& {\tt ASYS31}		& $S_{3,1}(z)$ 		\\
{\tt S2}	&{\tt ASYS2} 	& $S_2(z)$	 	& {\tt S41}		& {\tt ASYS41}		& $S_{4,1}(z)$ 		\\     
{\tt S3}	&{\tt ASYS3} 	& $S_3(z)$	 	& {\tt S23}		& {\tt ASYS23}		& $S_{2,3}(z)$ 		\\     
{\tt S4}	&{\tt ASYS4} 	& $S_4(z)$	 	& {\tt S211}	& {\tt ASYS211}		& $S_{2,1,1}(z)$	\\   
{\tt S5}	&{\tt ASYS5} 	& $S_5(z)$	 	& {\tt S221}	& {\tt ASYS221}		& $S_{2,2,1}(z)$	\\   
{\tt S6}	&{\tt ASYS6} 	& $S_6(z)$	 	& {\tt S311}	& {\tt ASYS311}		& $S_{3,1,1}(z)$	\\   
{\tt S21}	&{\tt ASYS21} 	& $S_{2,1}(z)$ 	& {\tt S2111}	& {\tt ASYS2111}	& $S_{2,1,1,1}(z)$  \\
\hline
{\tt B}			&{\tt ASYB}		& $\beta(z)$	& {\tt MM1H0M100}	&{\tt ASYMM1H0M100}	& $f_6(z)$		\\
{\tt B1}		&{\tt ASYB1}	& $\beta_1(z)$	& {\tt MM1H011}		&{\tt ASYMM1H011}	& $f_7(z)$		\\
{\tt B2}		&{\tt ASYB2	}	& $\beta_2(z)$	& {\tt MM1H0101}	&{\tt ASYMM1H0101}	& $f_8(z)$		\\
{\tt B3}		&{\tt ASYB3	}	& $\beta_3(z)$	& {\tt MM1H0M1M10}	&{\tt ASYMM1H0M1M10}& $f_9(z)$		\\
{\tt B4}		&{\tt ASYB4	}	& $\beta_4(z)$	& {\tt MM1H0111}	&{\tt ASYMM1H0111}	& $f_{10}(z)$	\\
{\tt MM1H01}	&{\tt ASYMM1H01}	& $f_1(z)$	& {\tt MM1H0011}	& {\tt ASYMM1H0011}	& $f_{11}(z)$	\\
{\tt MM1H0M1}	&{\tt ASYMM1H0M1}	& $f_2(z)$	& {\tt MM1H0001}	& {\tt ASYMM1H0001}	& $f_{12}(z)$	\\
{\tt MM1H001}	&{\tt ASYMM1H001}	& $f_3(z)$	& {\tt MM1H0100}	& {\tt ASYMM1H0100}	& $f_{13}(z)$	\\
{\tt MM1H010}	&{\tt ASYMM1H010}	& $f_4(z)$	& {\tt M1F}			& {\tt ASYM1F}		& $f_{14}(z)$	\\
{\tt MM1H0010}	&{\tt ASYMM1H0010}	& $f_5(z)$	& 					&					&				\\
\hline
\end{tabular}
\caption{\label{tab:fortransums} Special functions implemented in the Fortran code covering harmonic sums up to weight 5. Their interface is: {\tt COMPLEX*16} {\it name}{\tt (COMPLEX*16 }{\it z}{\tt )}. A larger set of about 50 more functions has been coded, targeting the weight-6 sums appearing in $c_{2,q}^{(3)NS}$. These follow the same naming convention and are not listed here.}
\end{table}

\subsubsection{Comparison to the code {\tt Pegasus}}

We perform the evolution of the input distributions $u+\bar u$, $d+\bar d$, $s+\bar s$ and $g$ with $N_F=3$ active flavours and with the same input as in the software {\tt Pegasus} \cite{Vogt:2004ns}. For the purpose of comparison, we import the same routine for the running of $a_s$, which, differing from our library, is based on a numerical solution of the renormalization group equation rather than a perturbative solution. Within these assumptions, our code delivers the same results largely within $10^{-5}$. We show in Table \ref{tab:accuracyPDFs} the discrepancies at LO, NLO and NNLO,
\begin{equation}
\begin{split}
	\Delta_a &= \max_x \left|\frac{a_{\text{this program}}-a_{\text{Pegasus}}}{a_{\text{this program}}}\right|, \\[2mm]
	x&\in\{10^{-7},10^{-6},10^{-5},10^{-4},10^{-3},10^{-2},10^{-1},0.3,0.5,0.7,0.9\} .
	\end{split}
\end{equation}

\begin{table}[h]
\centering
\texttt{\small%
\begin{tabular}{ |c|c|c|c|c|c|c| } 
 \hline
 $Q^2$	& $\Delta_{u+\bar u}$ & $\Delta_{d+\bar d}$ & $\Delta_{s+\bar s}$ & $\Delta_g$ & $\Delta_{v_3^+}$ & $\Delta_{v_8^+}$ \\ 
\hline
 \multicolumn{7}{|c|}{\normalfont LO} \\
 \hline
$10$   & 1.47881E-5 & 1.48382E-5 & 2.18277E-5 & 2.67289E-5 & 3.38616E-5 & 4.34145E-5  \\
$10^2$ & 1.52455E-5 & 1.5381E-5  & 3.75231E-5 & 1.01455E-5 & 9.12542E-5 & 8.97385E-6  \\ 
$10^3$ & 1.52071E-5 & 8.00595E-6 & 3.0937E-5  & 3.44221E-5 & 1.56461E-5 & 8.29479E-5  \\
$10^4$ & 2.17135E-5 & 2.17248E-5 & 1.67569E-5 & 7.7087E-6  & 6.02844E-5 & 3.55746E-4  \\
\hline
\multicolumn{7}{|c|}{\normalfont NLO} \\
\hline
$10$    & 1.02625E-5 & 1.26863E-5 & 2.03265E-5 & 3.52649E-5 & 2.35638E-5 & 4.16867E-4 \\
$10^2$  & 2.0161E-5  & 2.85297E-5 & 2.76987E-5 & 2.16919E-5 & 6.44566E-5 & 6.64705E-4 \\
$10^3$  & 8.31654E-6 & 5.64951E-6 & 2.13323E-5 & 2.36291E-5 & 5.47708E-5 & 2.49781E-4 \\
$10^4$  & 2.4118E-5  & 2.10801E-5 & 2.7311E-5  & 2.25101E-5 & 7.49564E-5 & 4.19059E-5 \\
\hline
\multicolumn{7}{|c|}{\normalfont NNLO} \\
\hline
$10$   & 4.83695E-6 & 3.21303E-5 & 1.32566E-5 & 3.80698E-5 & 8.5179E-5  & 5.08875E-5 \\ 
$10^2$ & 1.68509E-5 & 1.73187E-5 & 3.68565E-5 & 1.53346E-5 & 1.60866E-5 & 4.45352E-4 \\
$10^3$ & 2.5088E-5  & 2.51587E-5 & 2.82473E-5 & 2.96935E-5 & 8.42049E-5 & 8.62838E-4 \\
$10^4$ & 3.66124E-5 & 2.93829E-5 & 3.27043E-5 & 1.57975E-5 & 5.42801E-5 & 6.54498E-4 \\
\hline
\end{tabular}
}
\caption{\label{tab:accuracyPDFs} Relative errors in the evolution of the PDFs with respect to Pegasus. The input used is the default input of Pegasus, with $N_F=3$.}
\end{table}

\begin{table}[h]
\centering
\texttt{\small%
\begin{tabular}{ |c|c|c|c|c|c|c| } 
 \hline
 $Q^2$	& $\Delta_{\Delta u+\Delta \bar u}$ & $\Delta_{\Delta d+\Delta \bar d}$ & $\Delta_{\Delta s+\Delta \bar s}$ & $\Delta_{\Delta g}$ & $\Delta_{\Delta v_3^+}$ & $\Delta_{\Delta v_8^+}$ \\ 
\hline
 \multicolumn{7}{|c|}{\normalfont LO} \\
 \hline
$10$	& 2.81877E-4 & 1.08887E-4 & 9.39064E-5 & 3.11749E-4 & 2.4857E-4  & 4.38255E-4 \\
$10^2$	& 4.26776E-3 & 1.6991E-4  & 2.44907E-4 & 1.42952E-4 & 2.96073E-4 & 7.90195E-4 \\
$10^3$	& 9.36861E-3 & 3.57655E-4 & 1.81351E-4 & 1.25137E-4 & 1.29966E-4 & 1.43862E-3 \\
$10^4$	& 3.10387E-4 & 2.39955E-4 & 3.15498E-4 & 3.10184E-4 & 1.18058E-4 & 3.17004E-3 \\
\hline
\multicolumn{7}{|c|}{\normalfont NLO} \\
\hline 
$10$	& 3.15904E-3 & 2.29253E-4 & 1.84147E-4 & 2.69914E-4 & 1.93248E-4 & 1.53125E-3 \\  
$10^2$	& 1.20987E-3 & 1.8644E-4  & 2.63862E-4 & 3.1633E-4  & 2.31416E-4 & 1.28598E-3 \\
$10^3$	& 1.48338E-3 & 2.67103E-4 & 5.67934E-4 & 2.38033E-4 & 1.41118E-4 & 1.34927E-3 \\
$10^4$	& 4.21179E-4 & 3.92808E-4 & 9.55412E-4 & 2.21378E-4 & 1.52547E-4 & 1.02015E-2 \\
 \hline
\end{tabular}
}
\caption{\label{tab:accuracyPolPDFs} Relative errors in the evolution of the polarized PDFs as compared to Pegasus. The input used is the default input of Pegasus, with $N_F=3$.}
\end{table}

\subsubsection{Numerical results}

In the following we plot the evolution of polarized and unpolarized input PDFs obtained using our numerical evolution library to the scales $Q^2=\{10,10^2,10^3,10^4\}~\GeV^2$. For the unpolarized PDFs, we use, after Mellin-transforming it, the input \cite{Vogt:2004ns}
\begin{equation}
\begin{split}
	x u_v(x,Q_0^2)	&= 5.107200 \ x^{0.8}  \ (1-x)^3, \\
	x d_v(x,Q_0^2)	&= 3.064320 \ x^{0.8}  \ (1-x)^4, \\
	xg   (x,Q_0^2)	&= 1.7      \ x^{-0.1} \ (1-x)^5, \\
	x\bar d(x,Q_0^2)&= 0.1939875\ x^{-0.1} \ (1-x)^6, \\
	x\bar u(x,Q_0^2)&= (1-x) \ x\bar d(x,Q_0^2), \\
	xs(x,Q_0^2)		&= x\bar s(x,Q_0^2) = 0.2 \ x(\bar u +\bar d)(x,Q_0^2)
\label{eq:unpol-input-pdf}
\end{split}
\end{equation}
at $Q_0^2=2~\GeV^2$ and $\Lambda_{QCD}=0.226~\GeV$ with three light flavours. In producing Figures \ref{figpdf1a}-\ref{figpdf7} we kept the same functional form for $a_s(Q^2)$, namely obtained from Eq.\ \eqref{eq:asSol} truncated to NNLO. 
In Fig.\ \ref{figpdf1a}-\ref{figpdf1d} we show the lowest order result of the evolution, obtained by truncating Eqs.\ \eqref{eq:solAP} and \eqref{eq:solAPNS} to lowest order. 
In Fig.\ \ref{figpdf2a}-\ref{figpdf2d} we show the relative size of the NLO corrections to the LO evolved PDFs. For $v_3^+$, their size is of $\mathcal O$(10-15\%) at $x=10^{-4}$ and decreases at larger values of $x$, but become sizeable again at very large $x$. For $v_8^+$, the NLO corrections are appreciable at large $x$, reaching $\mathcal O$(5\%). For $\Sigma$, the NLO corrections are  of $\mathcal O$(30\%) at $x=10^{-4}$, and decrease with increasing $x$, but are again sizeable at very large $x$. A similar pattern is visible for the gluon density, with corrections of $\mathcal O$(10\%) at $x=10^{-4}$ and $\mathcal O$(20\%) at large $x$. 
In Fig.\ \ref{figpdf3a}-\ref{figpdf3d} the relative size of the NNLO corrections to the NLO evolved PDFs is shown. These corrections are within $\mathcal O$(0.5\%) for $v_3^+$ and $v_8^+$ in the range $x\in(10^{-4},1)$, and within $\mathcal O$(5\%) for $\Sigma$ and $g$, with a larger impact in the small-$x$ region than at moderate $x$.

For the polarized input PDFs we use \cite{Blumlein:2010rn}
\begin{equation}
\begin{split}
	x \Delta u_v(x,Q_0^2)	&=  0.130669  \ x^{0.239} \ (1-x)^{3.031} (1+27.64 x), \\
	x \Delta d_v(x,Q_0^2)	&=	-0.0270518 \ x^{0.128} \ (1-x)^{4.055} (1+44.26 x), \\
	x \Delta \bar u(x,Q_0^2)&= x\Delta \bar d(x,Q_0^2) = x\Delta \bar s(x,Q_0^2) = x\Delta s(x,Q_0^2) \\
							&= -0.059801  \ x^{0.365} \ (1-x)^{8.08}, \\
	x \Delta g (x,Q_0^2)	&=  7.08988   \ x^{1.365} \ (1-x)^{5.61} 
\label{eq:pol-input-pdf}
\end{split}
\end{equation}
at $Q_0^2=4~\GeV^2$ and $\Lambda_{QCD}=0.226~\GeV$ with three light flavours. We kept the same functional form for $a_s(Q^2)$ as for the unpolarized case. In Fig.\ \ref{figpdf4a}-\ref{figpdf6d}, the same plots as for the unpolarized case are presented. In Fig.\ \ref{figpdf4a}-\ref{figpdf4d} we plot the LO evolved PDFs. 
In Fig.\ \ref{figpdf5a}-\ref{figpdf5d}, where the NLO corrections are shown, we can see that in the region $x\in(10^{-4},1)$ they are within 10\% for $\Delta v_3^+$ and $\Delta v_8^+$, within 5\% for $\Delta \Sigma$ (except where it vanishes) and can exceed 80\% for $\Delta g$ at small and at large $x$. 
The NNLO corrections (Fig.\ \ref{figpdf6a}-\ref{figpdf6d}) are within 1\% for $\Delta v_3^+$ and $\Delta v_8^+$, with the largest impact at large $x$; within 5\% for $\Delta\Sigma$ and of $\mathcal O$(10\%) for $\Delta g$. 
In Fig. \ref{figpdf7} we show the difference of the NNLO and LO evolved gluon PDF normalized to the LO-evolved one.

\begin{figure}[H]
        \centering
        \includegraphics[width=0.7\textwidth]{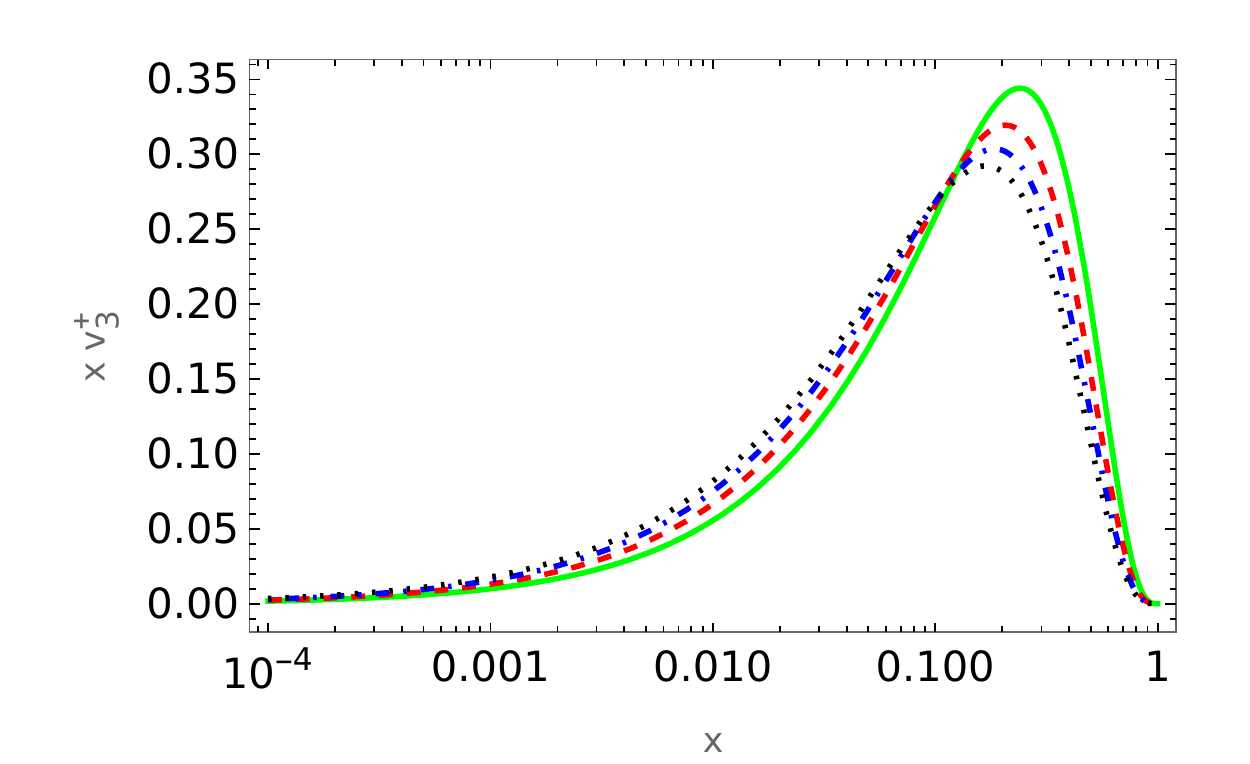}
\vspace{-0.2cm}
\caption{\sf The unpolarized PDF $v_3^+(x,Q^2)$ resulting from the LO evolution of the input \eqref{eq:unpol-input-pdf}. Solid lines: $Q^2=10~\GeV^2$. Dashed lines: $Q^2 = 10^2~\GeV^2$, Dash-dotted lines: $Q^2 = 10^3~\GeV^2$; Dotted lines: $Q^2 = 10^4~\GeV^2$. }
\label{figpdf1a}
\end{figure}
\begin{figure}[H]
        \centering
        \includegraphics[width=0.7\textwidth]{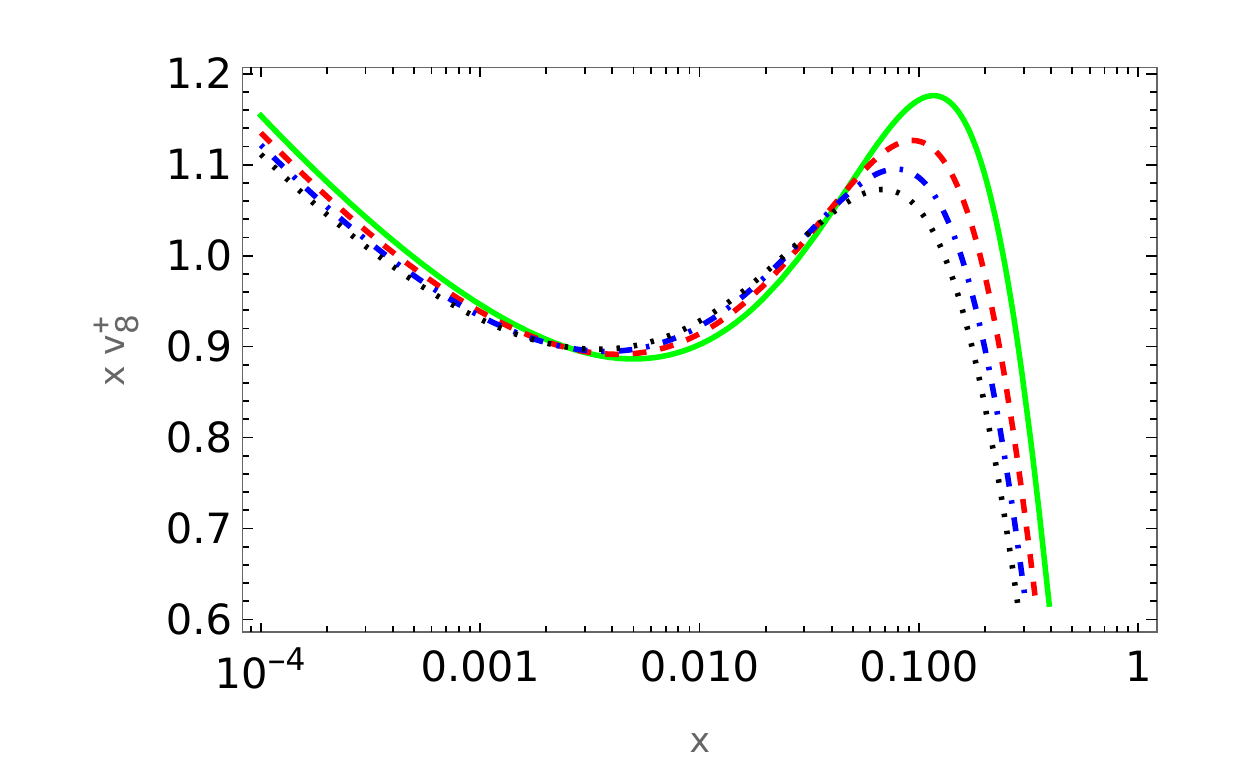}
\vspace{-0.2cm}
\caption{\sf The unpolarized PDF $v_8^+(x,Q^2)$ resulting from the LO evolution of the input \eqref{eq:unpol-input-pdf}. Solid lines: $Q^2=10~\GeV^2$. Dashed lines: $Q^2 = 10^2~\GeV^2$, Dash-dotted lines: $Q^2 = 10^3~\GeV^2$; Dotted lines: $Q^2 = 10^4~\GeV^2$. }
\label{figpdf1b}
\end{figure}
\begin{figure}[H]
        \centering
        \includegraphics[width=0.7\textwidth]{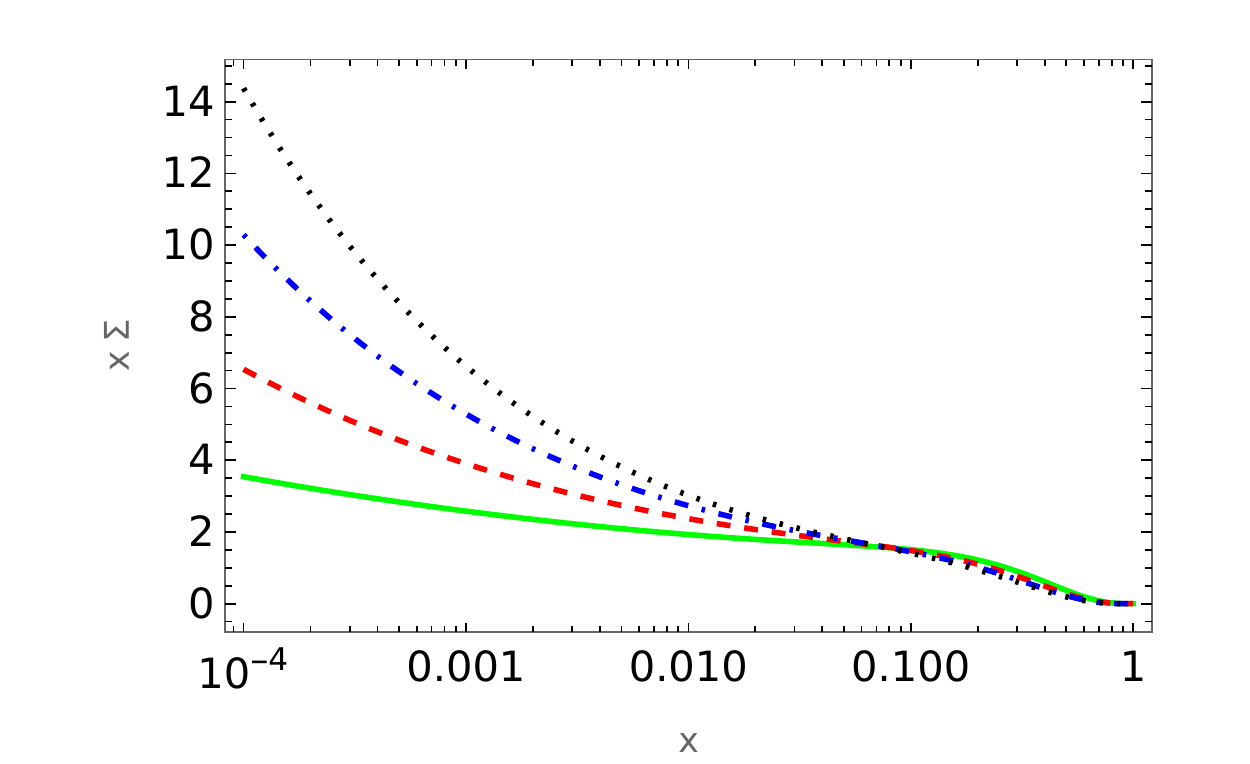}
\vspace{-0.2cm}
\caption{\sf The unpolarized PDF $\Sigma(x,Q^2)$ resulting from the LO evolution of the input \eqref{eq:unpol-input-pdf}. Solid lines: $Q^2=10~\GeV^2$. Dashed lines: $Q^2 = 10^2~\GeV^2$, Dash-dotted lines: $Q^2 = 10^3~\GeV^2$; Dotted lines: $Q^2 = 10^4~\GeV^2$. }
\label{figpdf1c}
\end{figure}
\begin{figure}[H]
        \centering
        \includegraphics[width=0.7\textwidth]{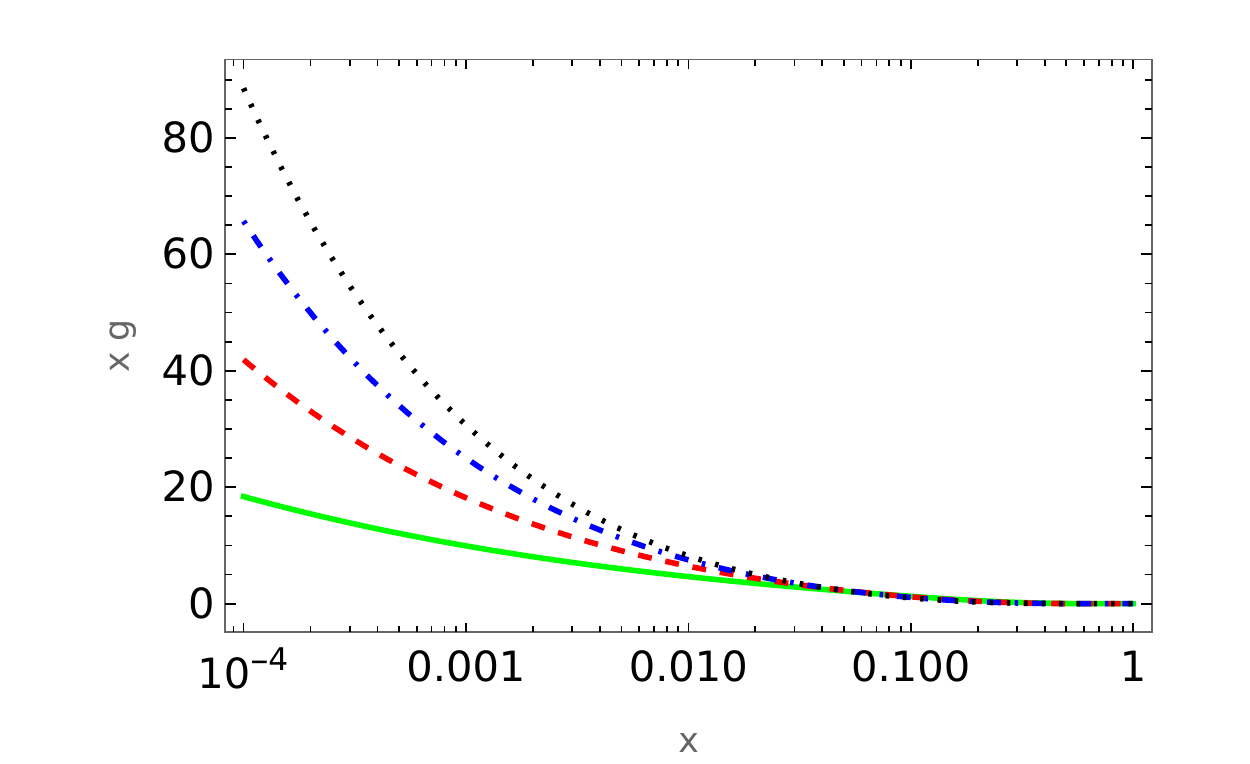}
\vspace{-0.2cm}
\caption{\sf The unpolarized PDF $g(x,Q^2)$ resulting from the LO evolution of the input \eqref{eq:unpol-input-pdf}. Solid lines: $Q^2=10~\GeV^2$. Dashed lines: $Q^2 = 10^2~\GeV^2$, Dash-dotted lines: $Q^2 = 10^3~\GeV^2$; Dotted lines: $Q^2 = 10^4~\GeV^2$. }
\label{figpdf1d}
\end{figure}

\begin{figure}[H]
        \centering
        \includegraphics[width=0.7\textwidth]{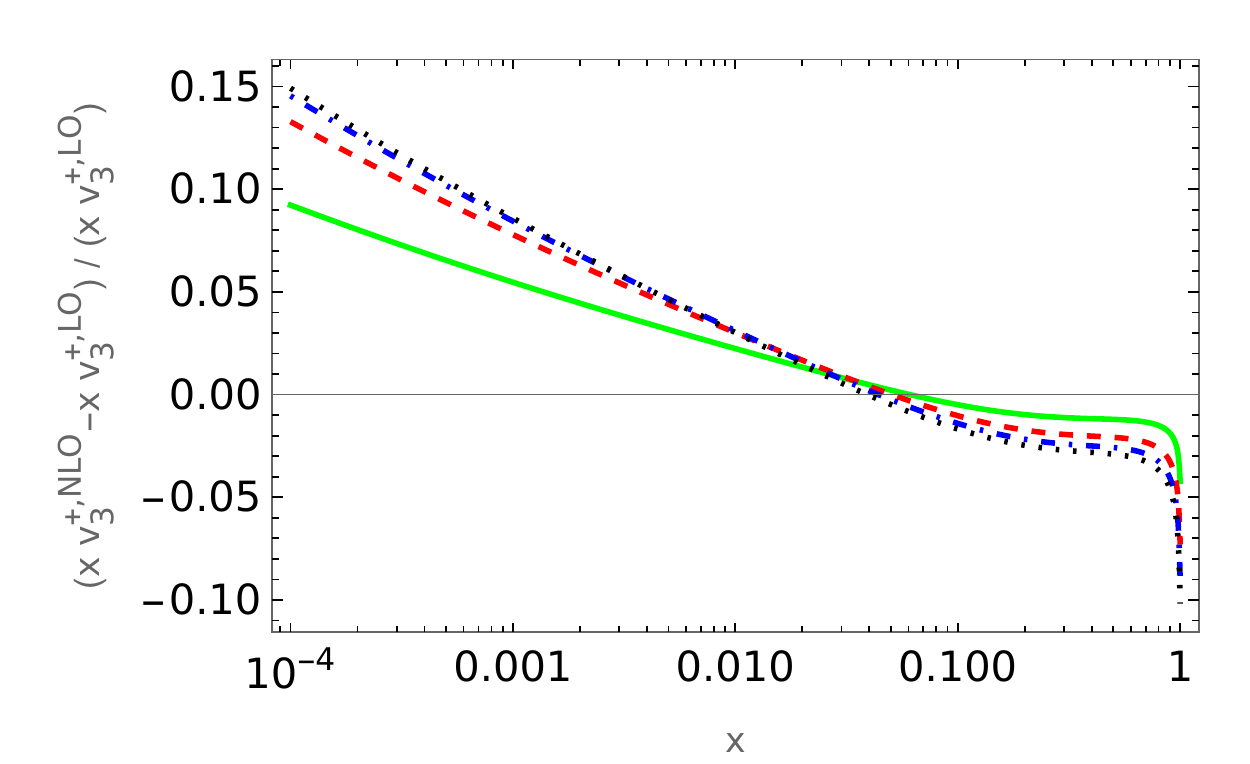}
\caption{\sf Relative size of the NLO corrections to the PDF evolution for the unpolarized PDF $v_3^+(x,Q^2)$. Solid lines: $Q^2=10~\GeV^2$. Dashed lines: $Q^2 = 10^2~\GeV^2$, Dash-dotted lines: $Q^2 = 10^3~\GeV^2$; Dotted lines: $Q^2 = 10^4~\GeV^2$. }
\label{figpdf2a}
\end{figure}
\begin{figure}[H]
        \centering
        \includegraphics[width=0.7\textwidth]{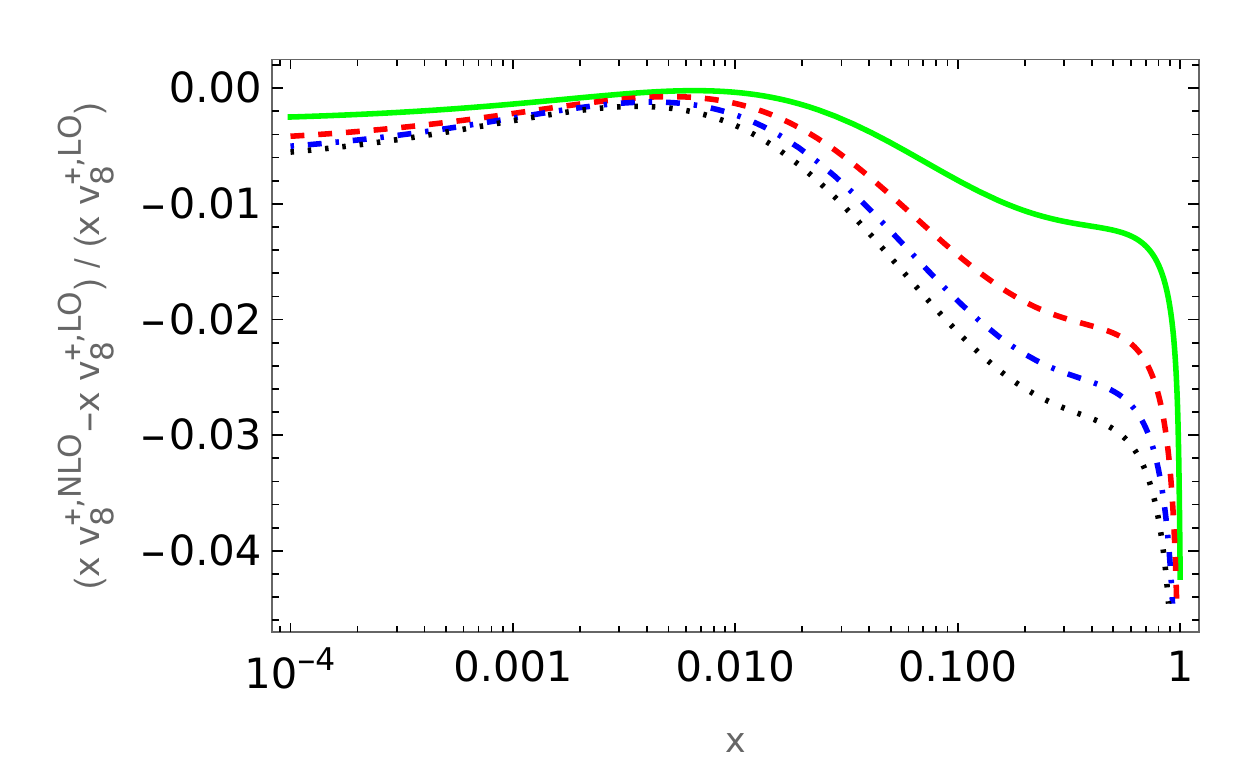}
\caption{\sf Relative size of the NLO corrections to the evolution for the unpolarized PDF $v_8^+(x,Q^2)$. Solid lines: $Q^2=10~\GeV^2$. Dashed lines: $Q^2 = 10^2~\GeV^2$, Dash-dotted lines: $Q^2 = 10^3~\GeV^2$; Dotted lines: $Q^2 = 10^4~\GeV^2$. }
\label{figpdf2b}
\end{figure}
\begin{figure}[H]
        \centering
        \includegraphics[width=0.7\textwidth]{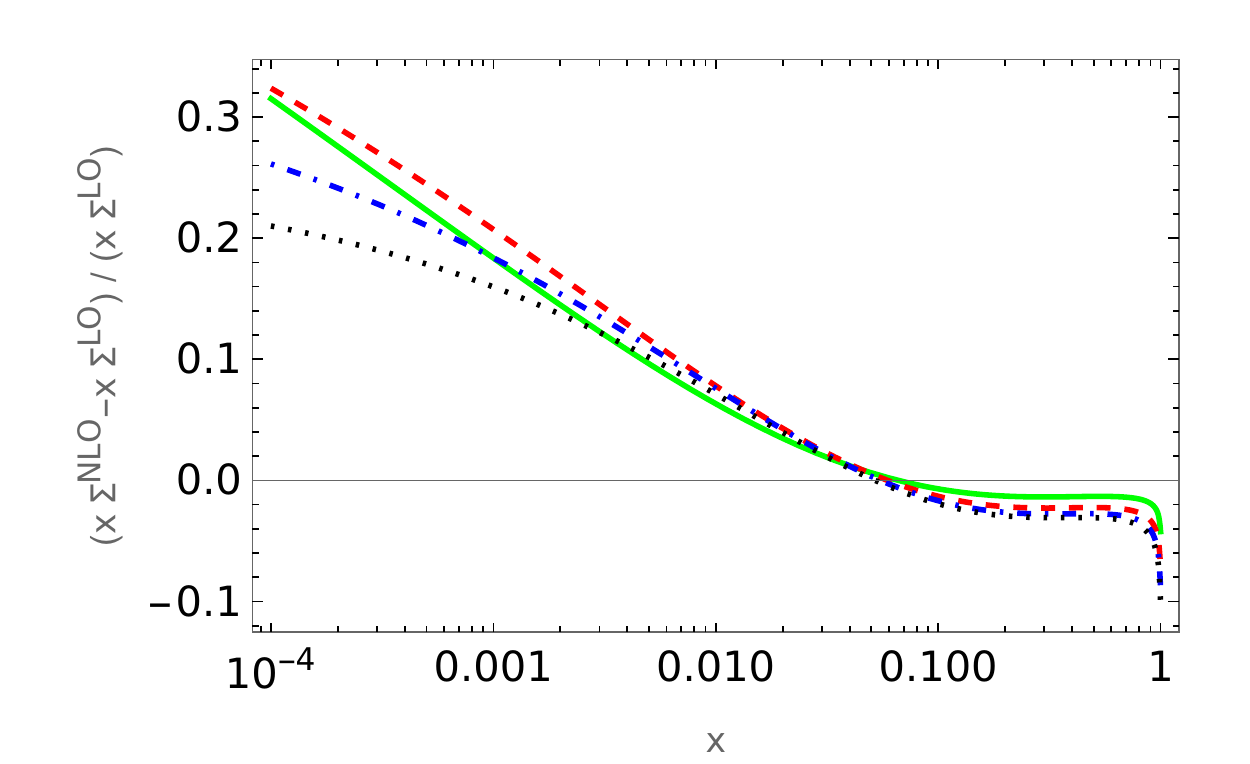}
\caption{\sf Relative size of the NLO corrections to the evolution for the unpolarized PDF $\Sigma(x,Q^2)$. Solid lines: $Q^2=10~\GeV^2$. Dashed lines: $Q^2 = 10^2~\GeV^2$, Dash-dotted lines: $Q^2 = 10^3~\GeV^2$; Dotted lines: $Q^2 = 10^4~\GeV^2$. }
\label{figpdf2c}
\end{figure}
\begin{figure}[H]
        \centering
        \includegraphics[width=0.7\textwidth]{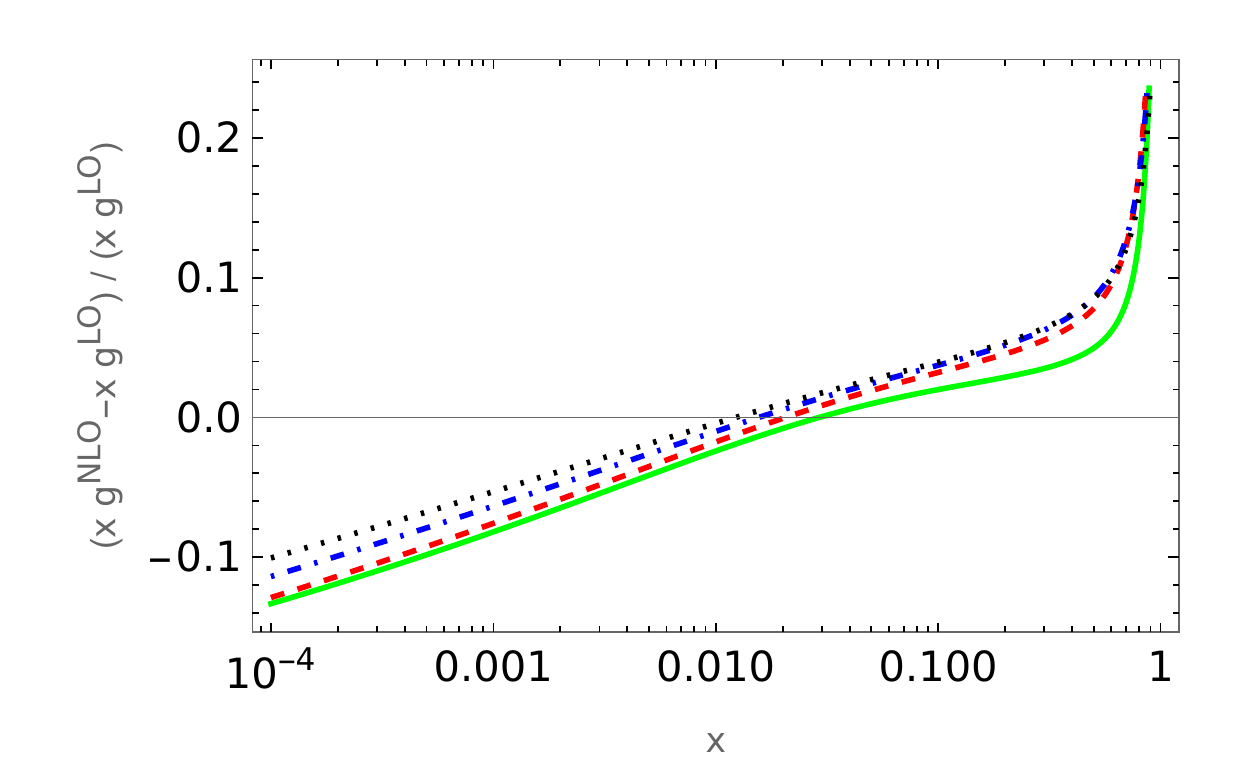}
\caption{\sf Relative size of the NLO corrections to the evolution for the unpolarized PDF $g(x,Q^2)$. Solid lines: $Q^2=10~\GeV^2$. Dashed lines: $Q^2 = 10^2~\GeV^2$, Dash-dotted lines: $Q^2 = 10^3~\GeV^2$; Dotted lines: $Q^2 = 10^4~\GeV^2$. }
\label{figpdf2d}
\end{figure}

\begin{figure}[H]
        \centering
        \includegraphics[width=0.7\textwidth]{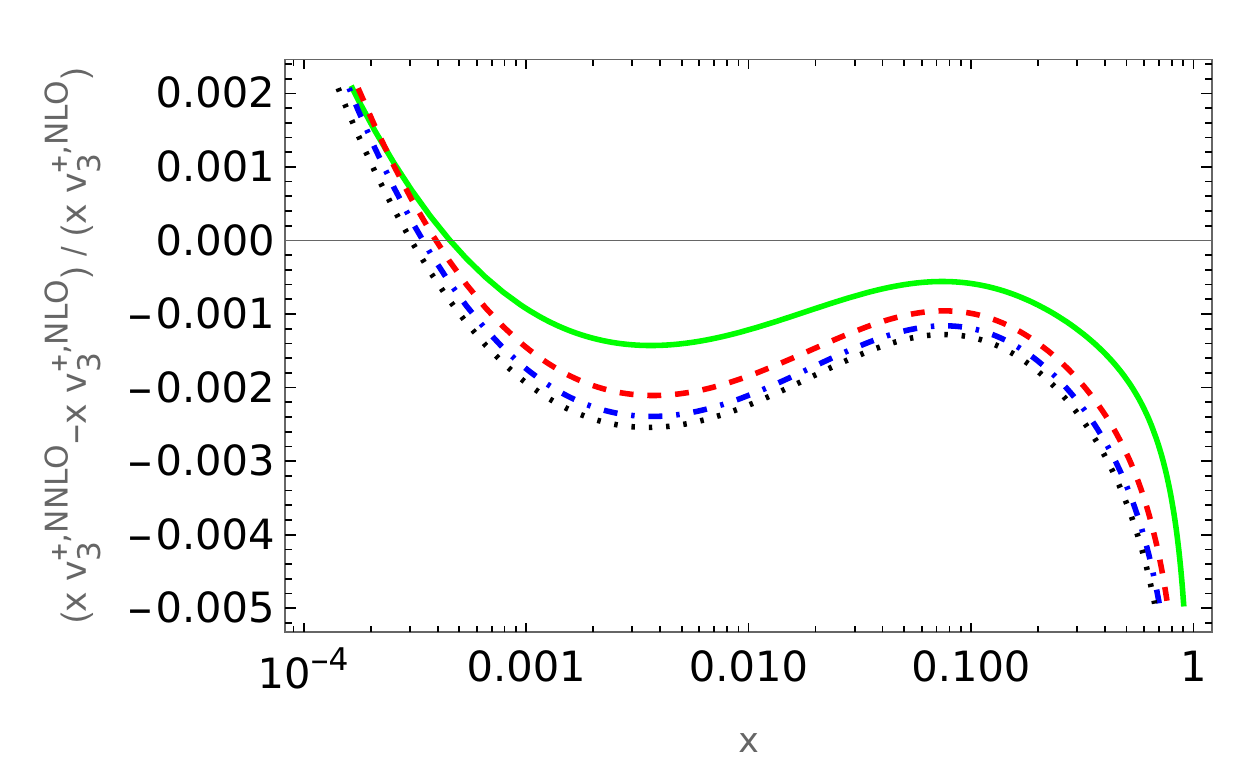}
\caption{\sf Relative size of the NNLO corrections to the evolution for the unpolarized PDF $v_3(x,Q^2)$. Solid lines: $Q^2=10~\GeV^2$. Dashed lines: $Q^2 = 10^2~\GeV^2$, Dash-dotted lines: $Q^2 = 10^3~\GeV^2$; Dotted lines: $Q^2 = 10^4~\GeV^2$. }
\label{figpdf3a}
\end{figure}
\begin{figure}[H]
        \centering
        \includegraphics[width=0.7\textwidth]{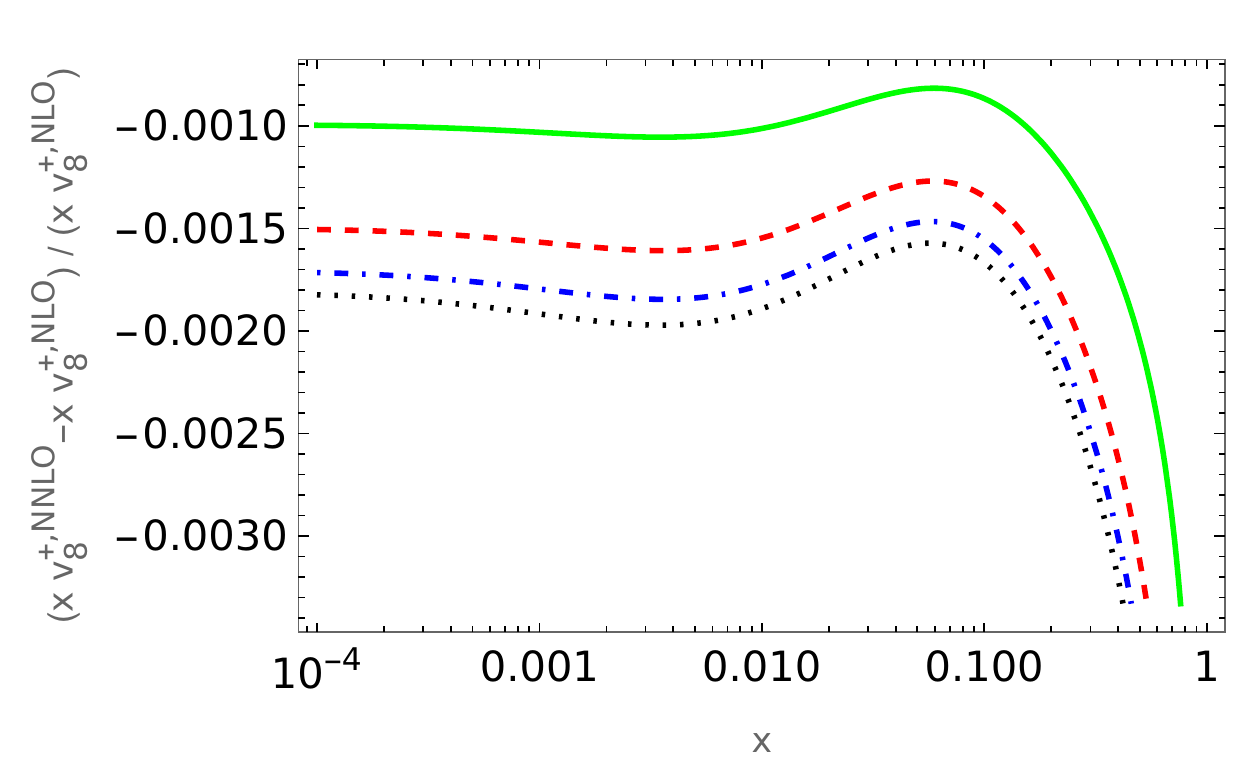}
\caption{\sf Relative size of the NNLO corrections to the evolution for the unpolarized PDF $v_8(x,Q^2)$. Solid lines: $Q^2=10~\GeV^2$. Dashed lines: $Q^2 = 10^2~\GeV^2$, Dash-dotted lines: $Q^2 = 10^3~\GeV^2$; Dotted lines: $Q^2 = 10^4~\GeV^2$. }
\label{figpdf3b}
\end{figure}
\begin{figure}[H]
        \centering
        \includegraphics[width=0.7\textwidth]{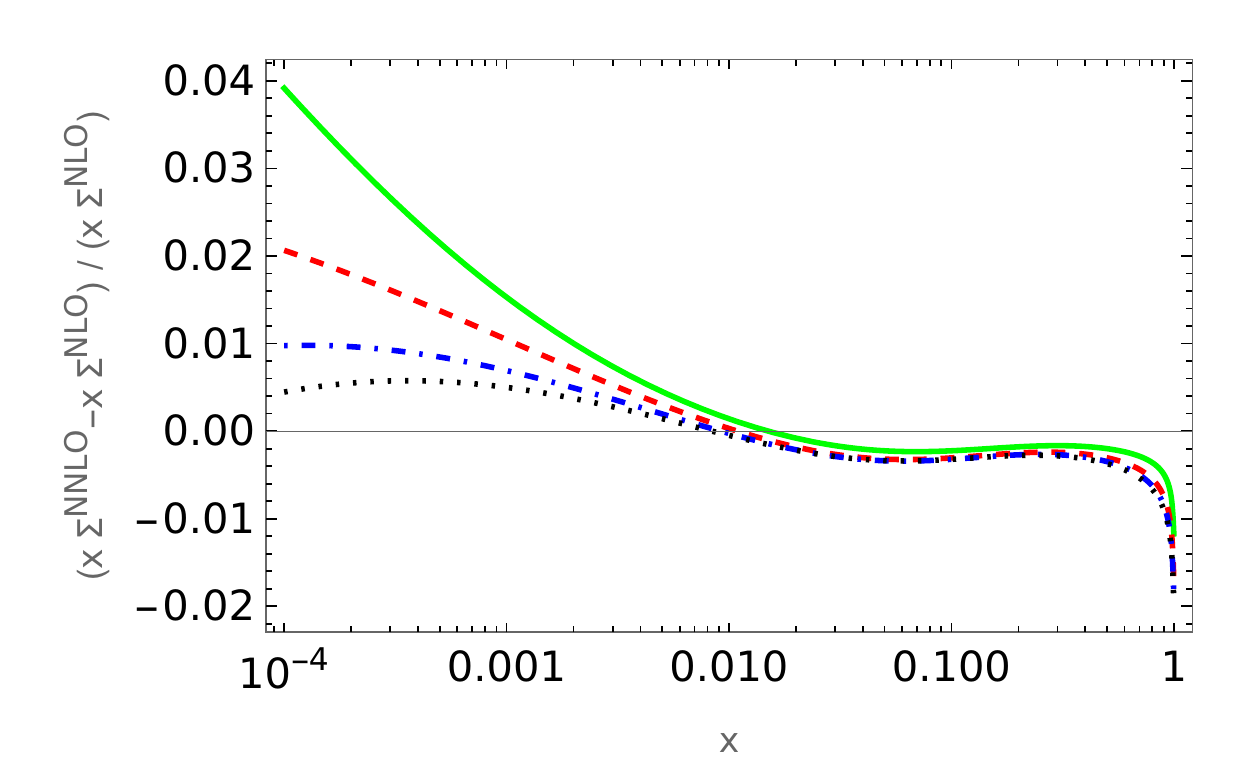}
\caption{\sf Relative size of the NNLO corrections to the evolution for the unpolarized PDF $\Sigma(x,Q^2)$. Solid lines: $Q^2=10~\GeV^2$. Dashed lines: $Q^2 = 10^2~\GeV^2$, Dash-dotted lines: $Q^2 = 10^3~\GeV^2$; Dotted lines: $Q^2 = 10^4~\GeV^2$. }
\label{figpdf3c}
\end{figure}
\begin{figure}[H]
        \centering
        \includegraphics[width=0.7\textwidth]{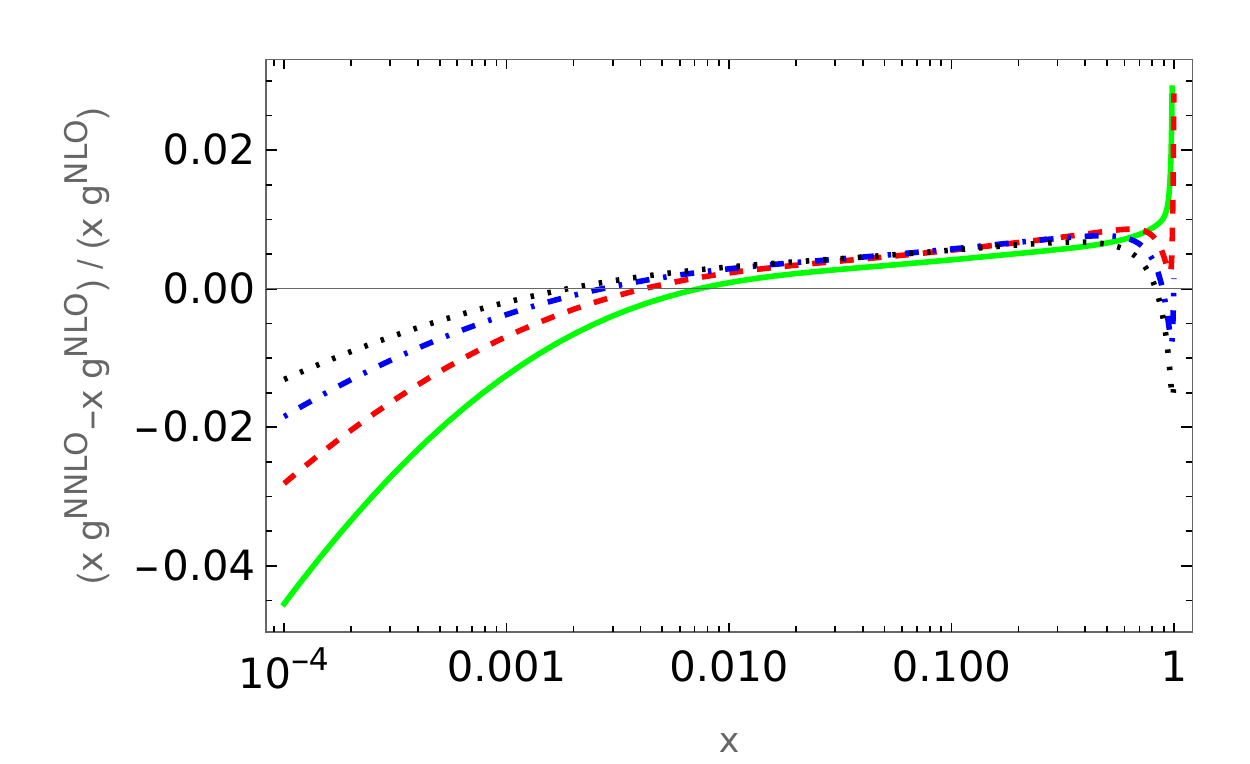}
\caption{\sf Relative size of the NNLO corrections to the evolution for the unpolarized PDF $g(x,Q^2)$. Solid lines: $Q^2=10~\GeV^2$. Dashed lines: $Q^2 = 10^2~\GeV^2$, Dash-dotted lines: $Q^2 = 10^3~\GeV^2$; Dotted lines: $Q^2 = 10^4~\GeV^2$. }
\label{figpdf3d}
\end{figure}

\begin{figure}[H]
        \centering
        \includegraphics[width=0.7\textwidth]{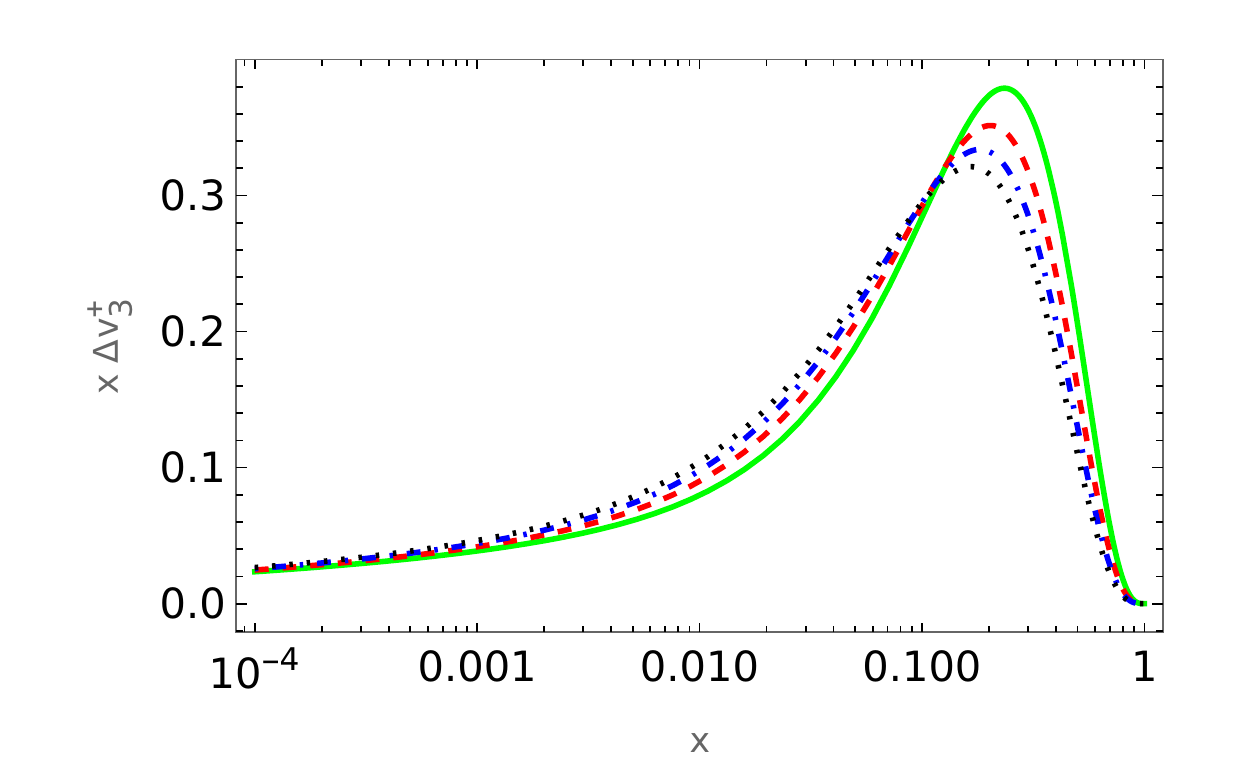}
\caption{\sf The polarized PDF $\Delta v_3^+(x,Q^2)$ resulting from the LO evolution of the input  \eqref{eq:pol-input-pdf}. Solid lines: $Q^2=10~\GeV^2$. Dashed lines: $Q^2 = 10^2~\GeV^2$, Dash-dotted lines: $Q^2 = 10^3~\GeV^2$; Dotted lines: $Q^2 = 10^4~\GeV^2$. }
\label{figpdf4a}
\end{figure}    
\begin{figure}[H]
        \centering
        \includegraphics[width=0.7\textwidth]{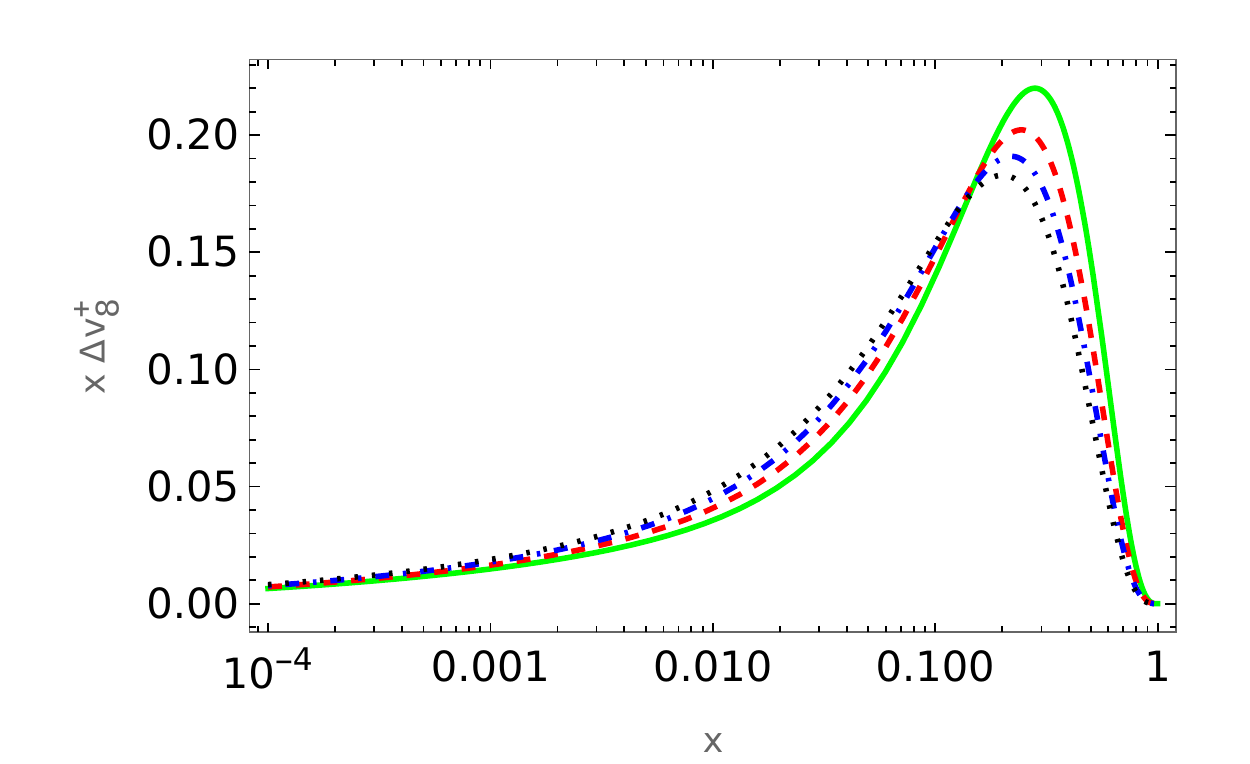}
\caption{\sf The polarized PDF $\Delta v_8^+(x,Q^2)$ resulting from the LO evolution of the input  \eqref{eq:pol-input-pdf}. Solid lines: $Q^2=10~\GeV^2$. Dashed lines: $Q^2 = 10^2~\GeV^2$, Dash-dotted lines: $Q^2 = 10^3~\GeV^2$; Dotted lines: $Q^2 = 10^4~\GeV^2$. }
\label{figpdf4b}
\end{figure}    
\begin{figure}[H]
        \centering
        \includegraphics[width=0.7\textwidth]{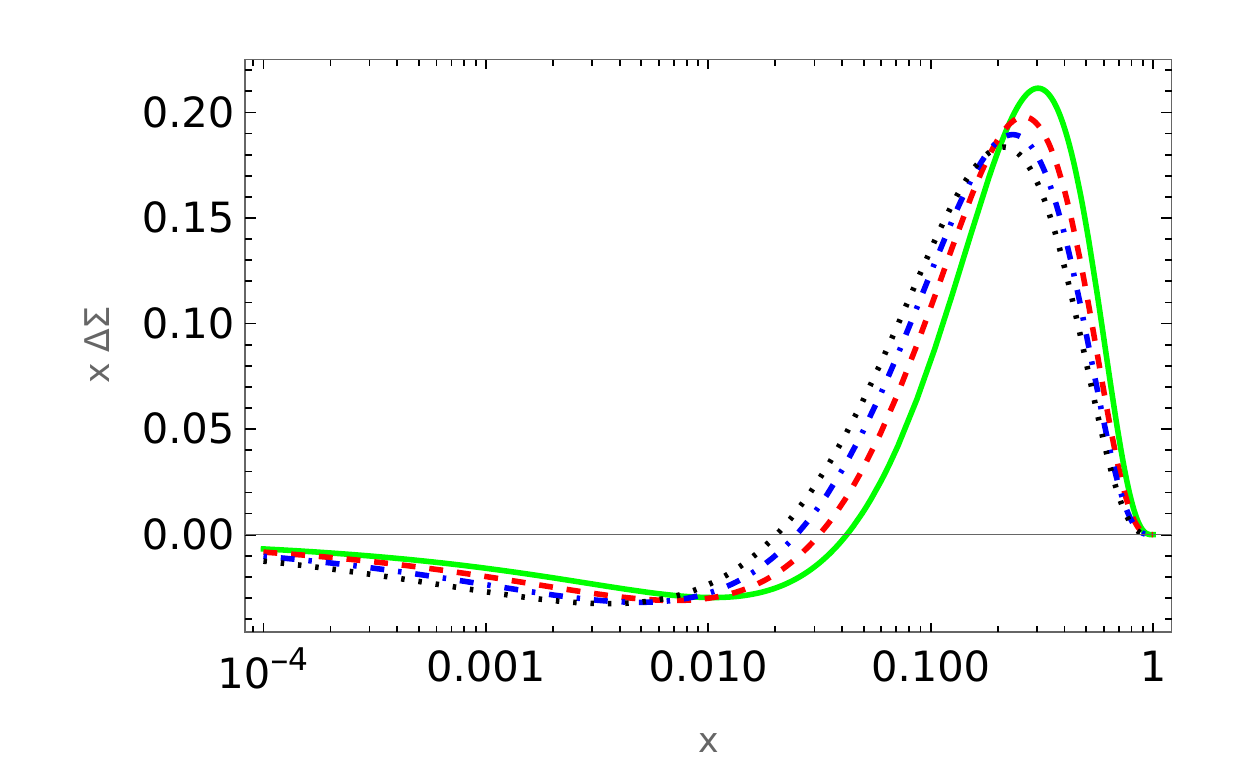}
\caption{\sf The polarized PDF $\Delta \Sigma(x,Q^2)$ resulting from the LO evolution of the input  \eqref{eq:pol-input-pdf}. Solid lines: $Q^2=10~\GeV^2$. Dashed lines: $Q^2 = 10^2~\GeV^2$, Dash-dotted lines: $Q^2 = 10^3~\GeV^2$; Dotted lines: $Q^2 = 10^4~\GeV^2$. }
\label{figpdf4c}
\end{figure}    
\begin{figure}[H]
        \centering
        \includegraphics[width=0.7\textwidth]{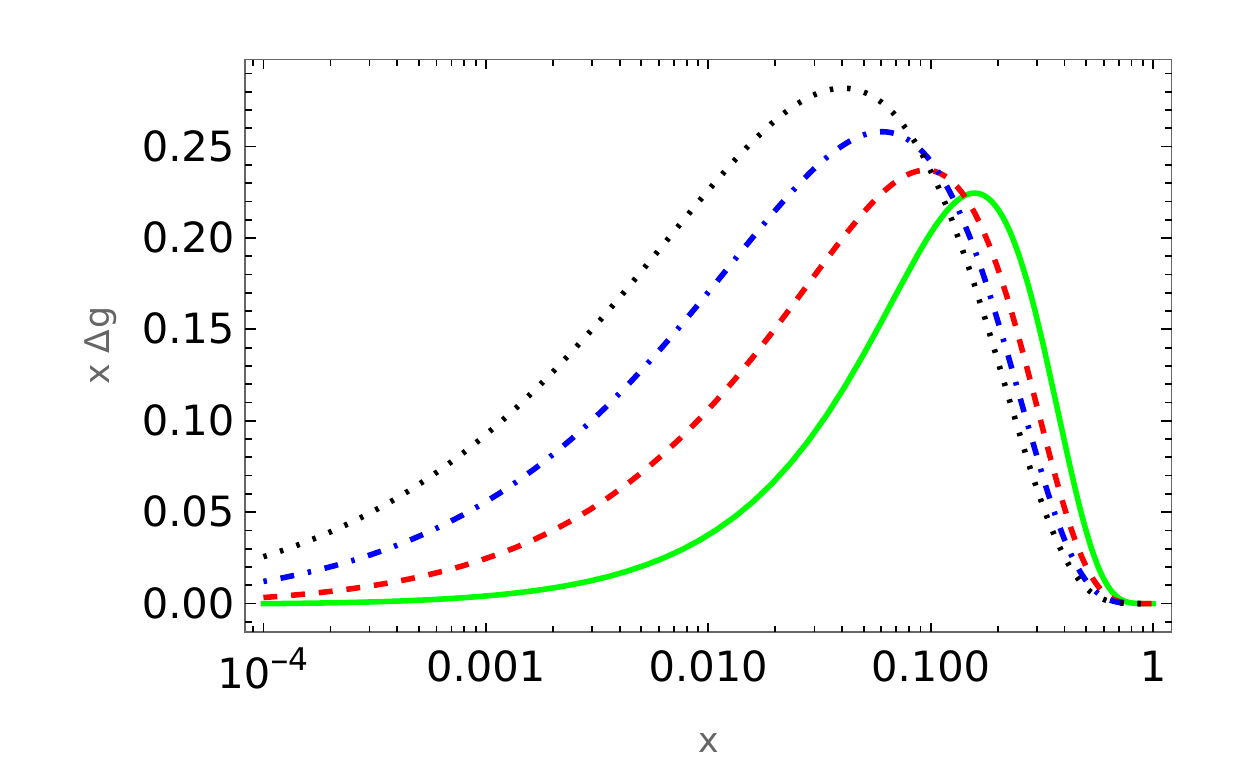}
\caption{\sf The polarized PDF $\Delta g(x,Q^2)$ resulting from the LO evolution of the input  \eqref{eq:pol-input-pdf}. Solid lines: $Q^2=10~\GeV^2$. Dashed lines: $Q^2 = 10^2~\GeV^2$, Dash-dotted lines: $Q^2 = 10^3~\GeV^2$; Dotted lines: $Q^2 = 10^4~\GeV^2$. }
\label{figpdf4d}
\end{figure}    

\begin{figure}[H]
        \centering
        \includegraphics[width=0.7\textwidth]{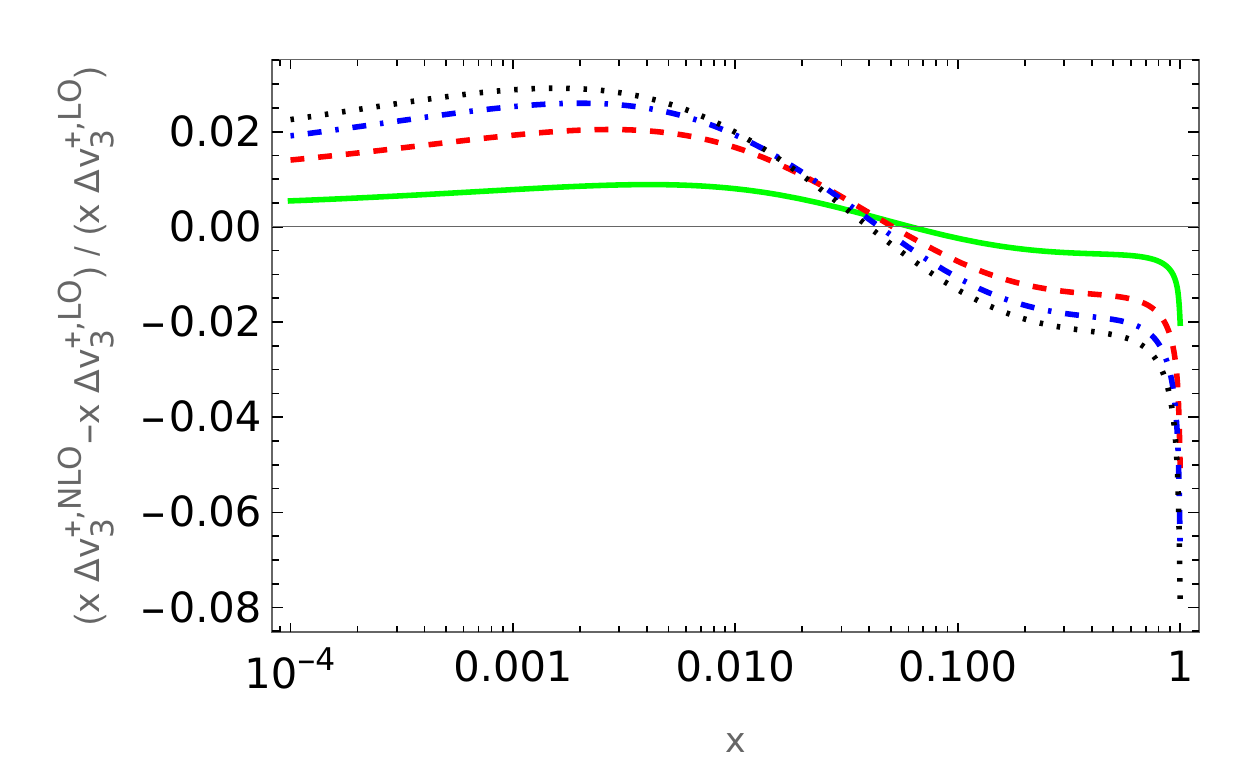}
\caption{\sf Relative size of the NLO corrections to the evolution of the polarized PDF $\Delta v_3^+(x,Q^2)$. Solid lines: $Q^2=10~\GeV^2$. Dashed lines: $Q^2 = 10^2~\GeV^2$, Dash-dotted lines: $Q^2 = 10^3~\GeV^2$; Dotted lines: $Q^2 = 10^4~\GeV^2$. }
\label{figpdf5a}
\end{figure}
\begin{figure}[H]
        \centering
        \includegraphics[width=0.7\textwidth]{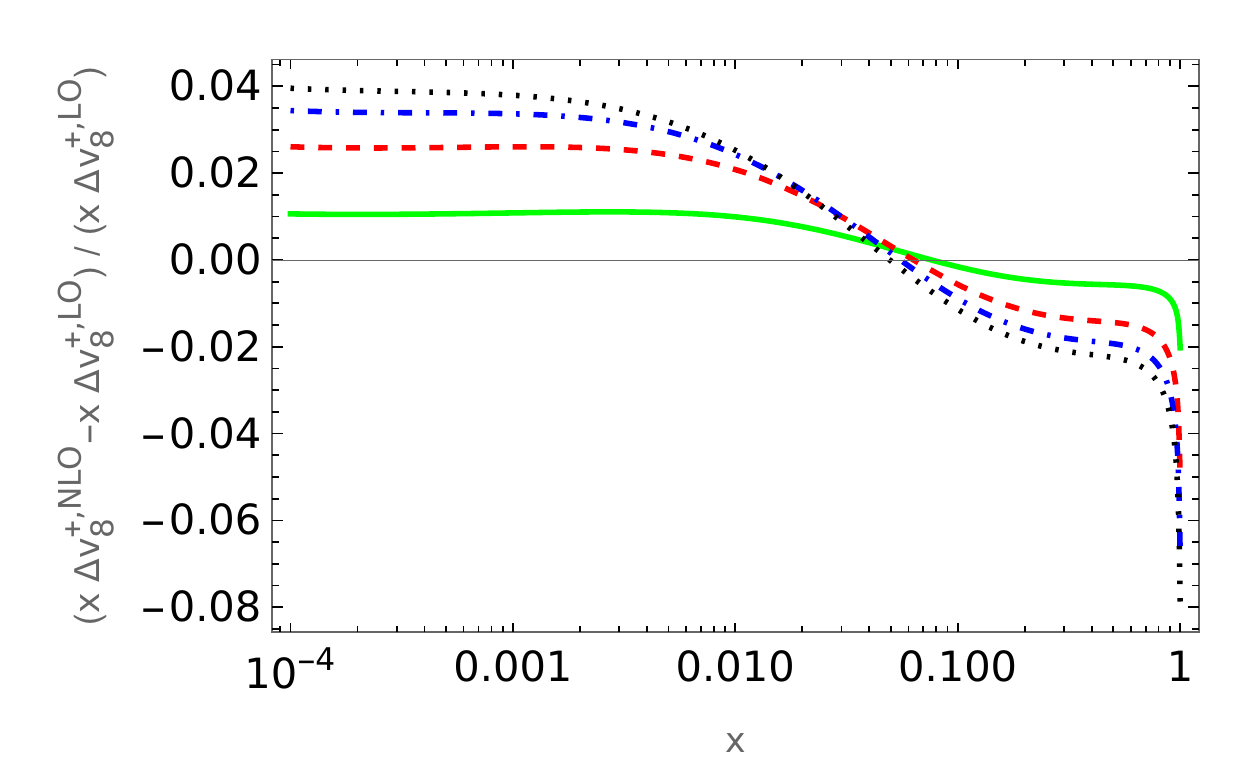}
\caption{\sf Relative size of the NLO corrections to the evolution of the polarized PDF $\Delta v_8^+(x,Q^2)$. Solid lines: $Q^2=10~\GeV^2$. Dashed lines: $Q^2 = 10^2~\GeV^2$, Dash-dotted lines: $Q^2 = 10^3~\GeV^2$; Dotted lines: $Q^2 = 10^4~\GeV^2$. }
\label{figpdf5b}
\end{figure}
\begin{figure}[H]
        \centering
        \includegraphics[width=0.7\textwidth]{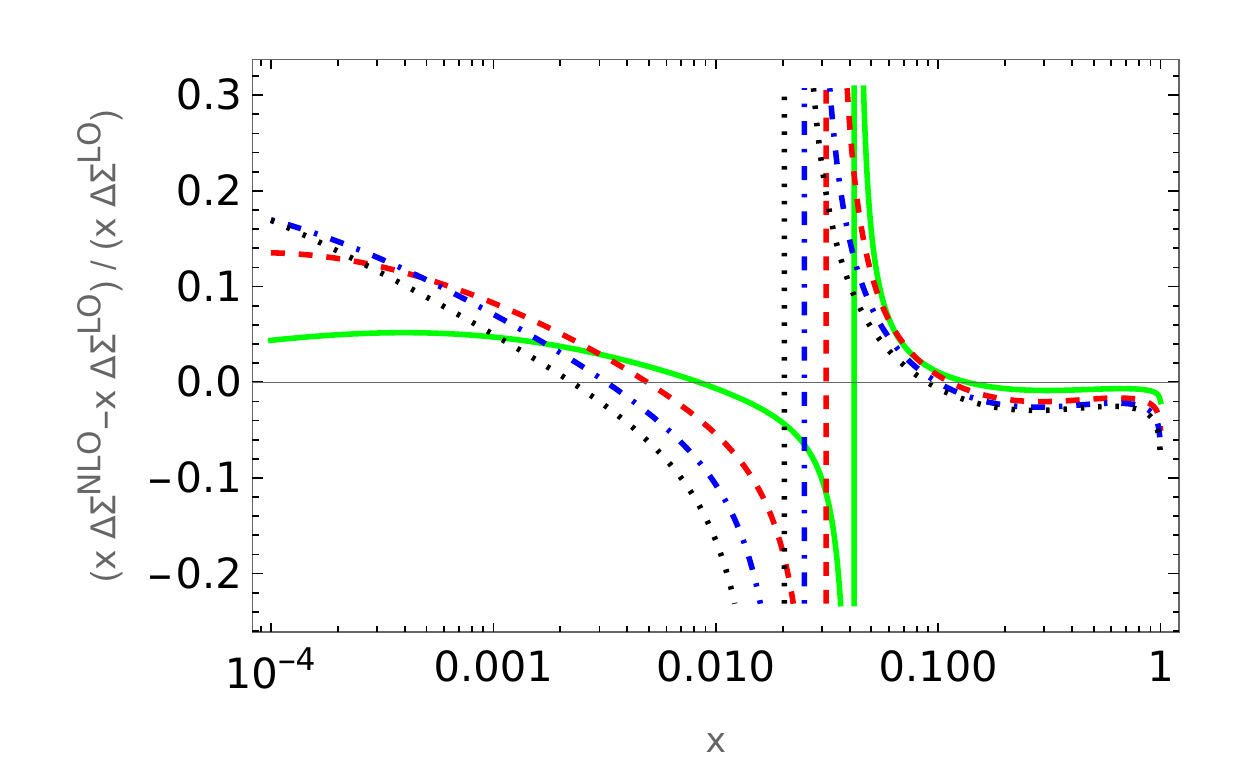}
\caption{\sf Relative size of the NLO corrections to the evolution of the polarized PDF $\Delta \Sigma(x,Q^2)$. Solid lines: $Q^2=10~\GeV^2$. Dashed lines: $Q^2 = 10^2~\GeV^2$, Dash-dotted lines: $Q^2 = 10^3~\GeV^2$; Dotted lines: $Q^2 = 10^4~\GeV^2$. }
\label{figpdf5c}
\end{figure}
\begin{figure}[H]
        \centering
        \includegraphics[width=0.7\textwidth]{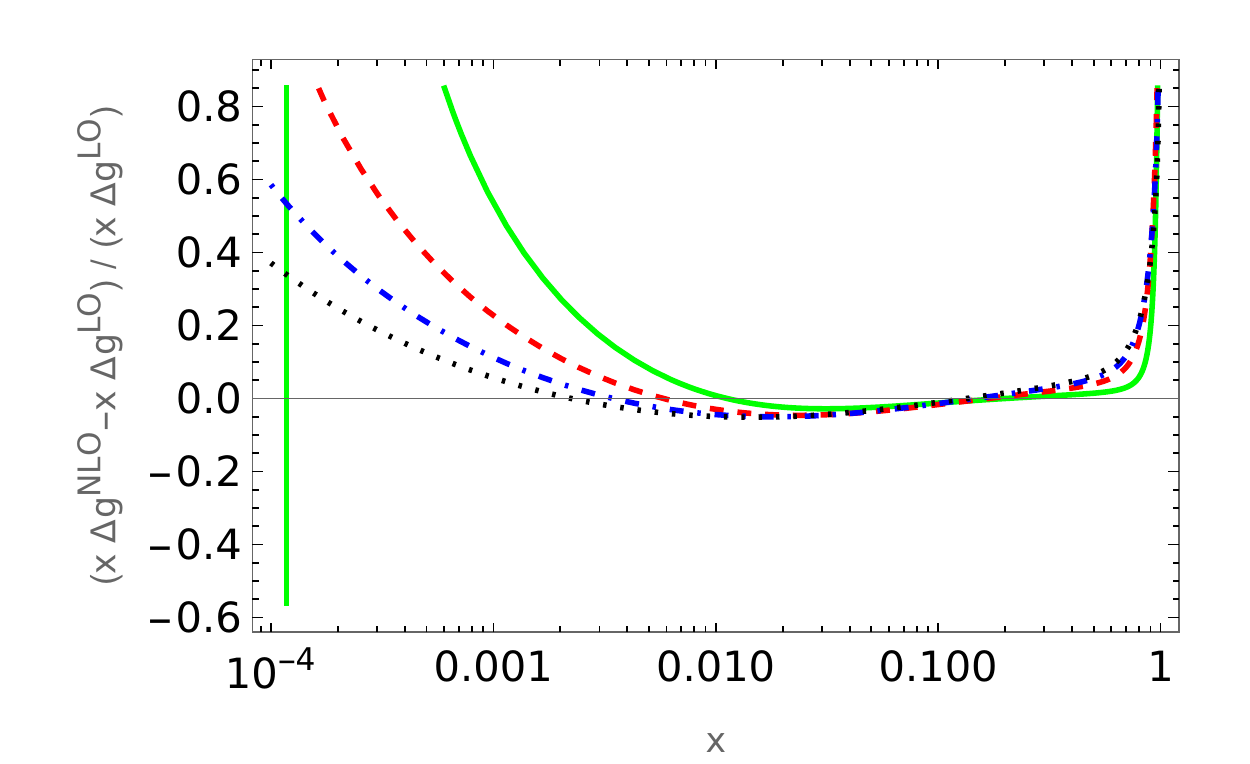}
\caption{\sf Relative size of the NLO corrections to the evolution of the polarized PDF $\Delta g(x,Q^2)$. Solid lines: $Q^2=10~\GeV^2$. Dashed lines: $Q^2 = 10^2~\GeV^2$, Dash-dotted lines: $Q^2 = 10^3~\GeV^2$; Dotted lines: $Q^2 = 10^4~\GeV^2$. }
\label{figpdf5d}
\end{figure}

\begin{figure}[H]
        \centering
        \includegraphics[width=0.7\textwidth]{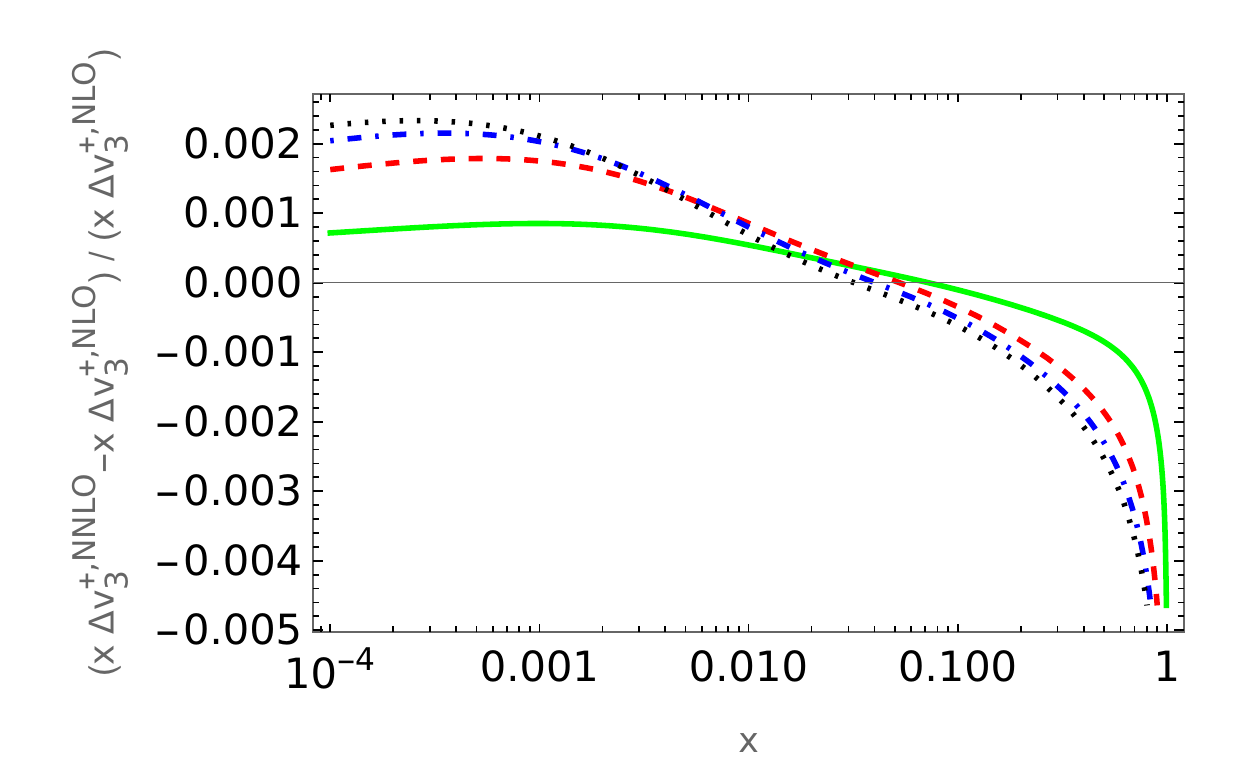}
\caption{\sf Relative size of the NNLO corrections to the evolution of the polarized PDF  $\Delta v_3^+(x,Q^2)$. Solid lines: $Q^2=10~\GeV^2$. Dashed lines: $Q^2 = 10^2~\GeV^2$, Dash-dotted lines: $Q^2 = 10^3~\GeV^2$; Dotted lines: $Q^2 = 10^4~\GeV^2$. }
\label{figpdf6a}
\end{figure}
\begin{figure}[H]
        \centering
        \includegraphics[width=0.7\textwidth]{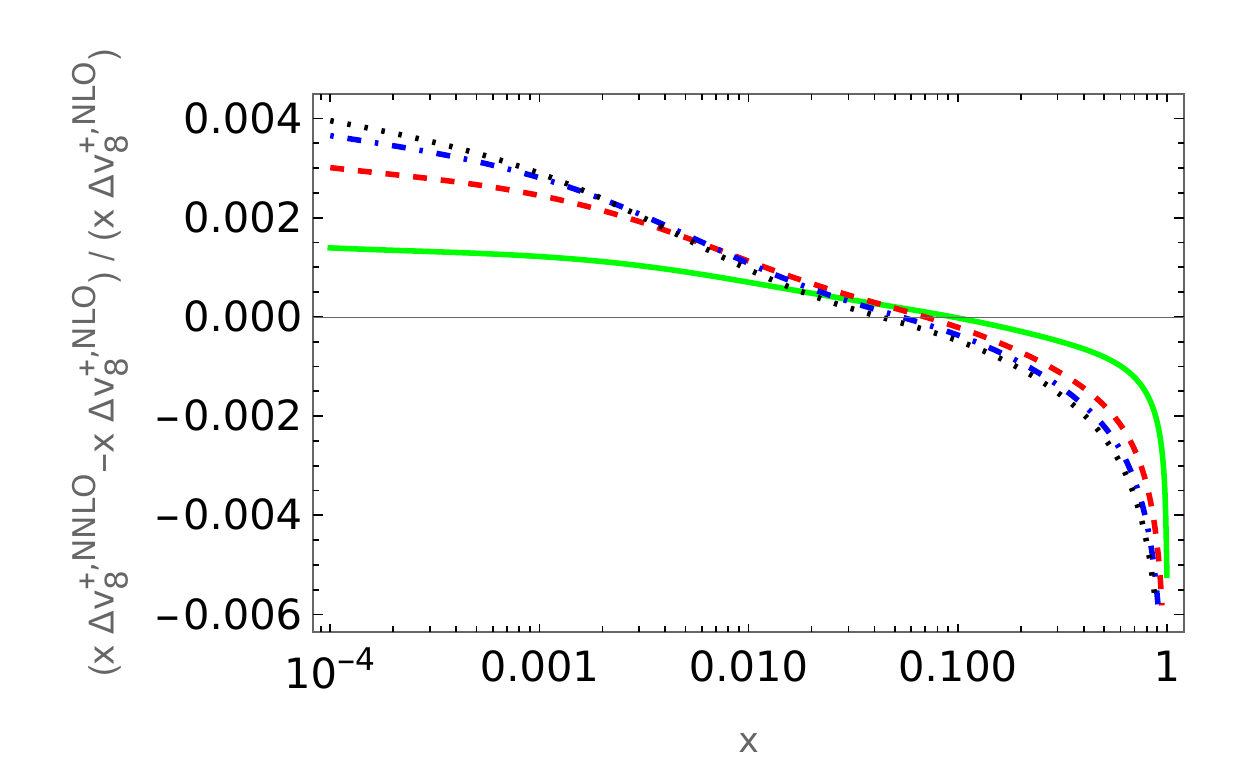}
\caption{\sf Relative size of the NNLO corrections to the evolution of the polarized PDF  $\Delta v_8^+(x,Q^2)$. Solid lines: $Q^2=10~\GeV^2$. Dashed lines: $Q^2 = 10^2~\GeV^2$, Dash-dotted lines: $Q^2 = 10^3~\GeV^2$; Dotted lines: $Q^2 = 10^4~\GeV^2$. }
\label{figpdf6b}
\end{figure}
\begin{figure}[H]
        \centering
        \includegraphics[width=0.7\textwidth]{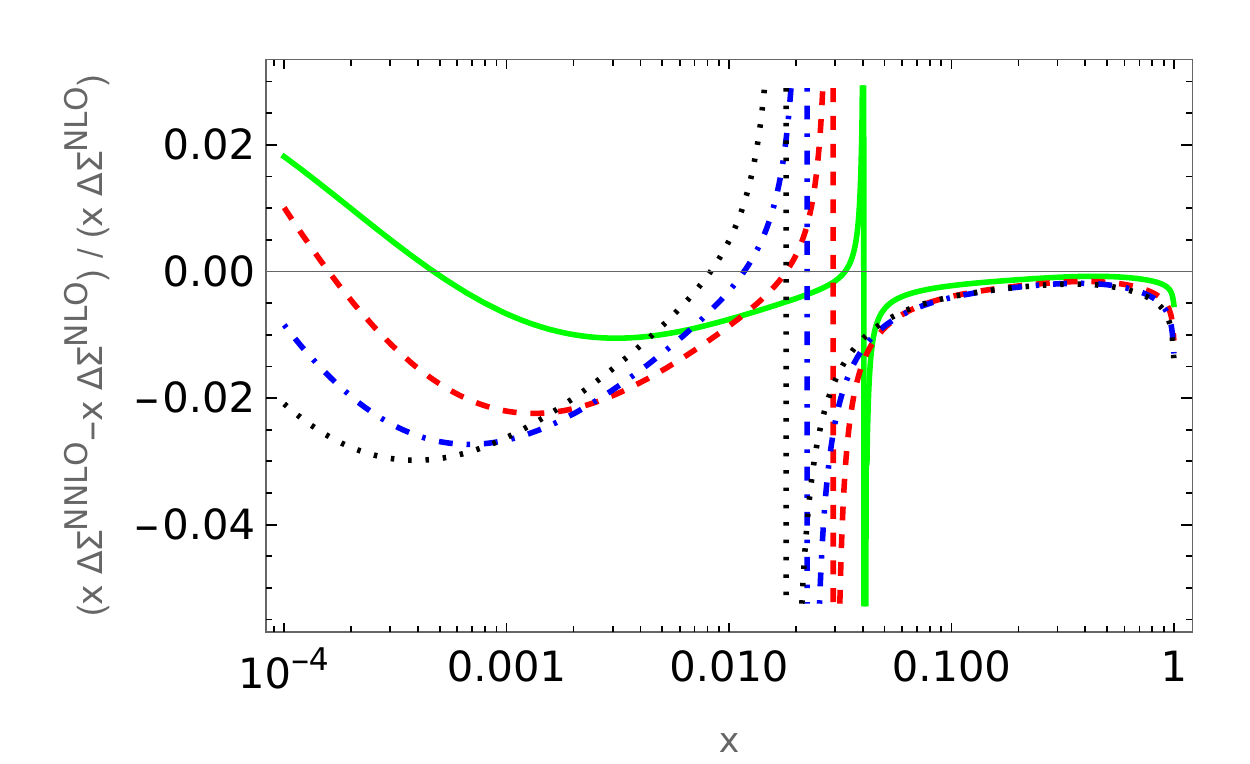}
\caption{\sf Relative size of the NNLO corrections to the evolution of the polarized PDF  $\Delta \Sigma(x,Q^2)$. Solid lines: $Q^2=10~\GeV^2$. Dashed lines: $Q^2 = 10^2~\GeV^2$, Dash-dotted lines: $Q^2 = 10^3~\GeV^2$; Dotted lines: $Q^2 = 10^4~\GeV^2$. }
\label{figpdf6c}
\end{figure}
\begin{figure}[H]
        \centering
        \includegraphics[width=0.7\textwidth]{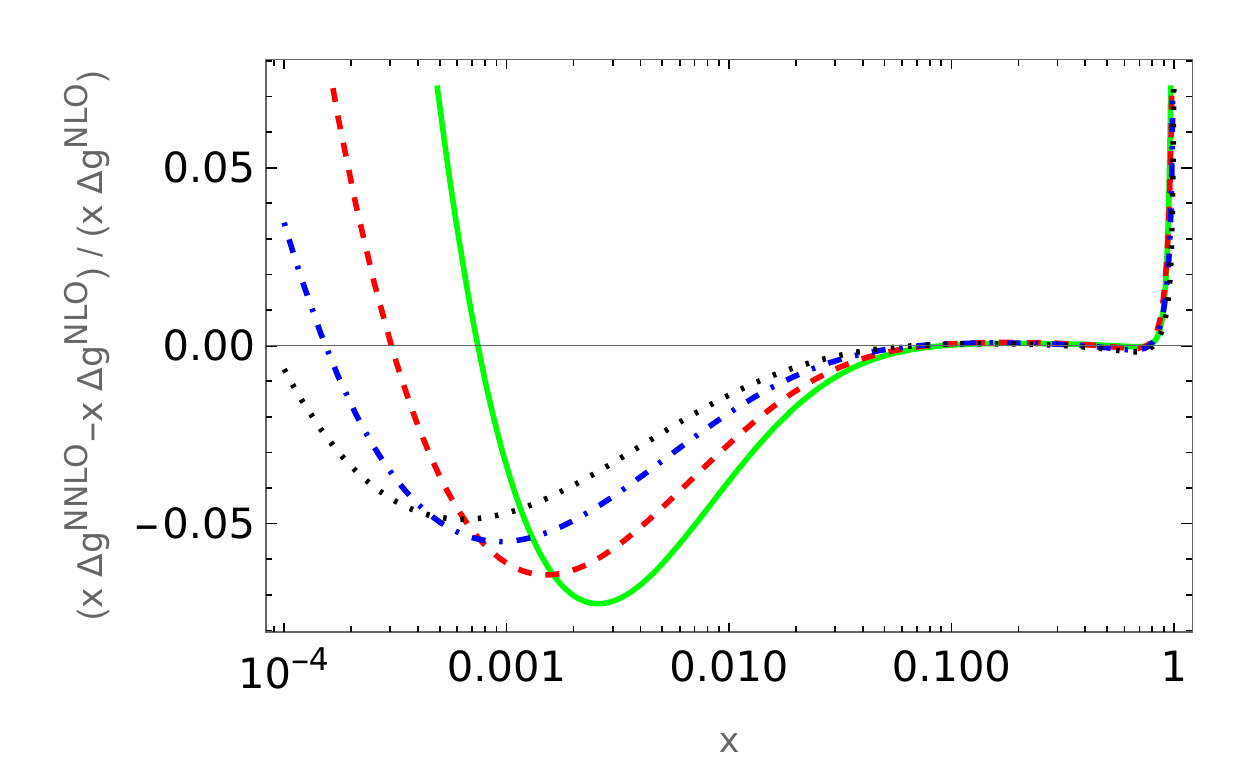}
\caption{\sf Relative size of the NNLO corrections to the evolution of the polarized PDF  $\Delta g(x,Q^2)$. Solid lines: $Q^2=10~\GeV^2$. Dashed lines: $Q^2 = 10^2~\GeV^2$, Dash-dotted lines: $Q^2 = 10^3~\GeV^2$; Dotted lines: $Q^2 = 10^4~\GeV^2$. }
\label{figpdf6d}
\end{figure}

\begin{figure}[H]
        \centering
        \includegraphics[width=0.7\textwidth]{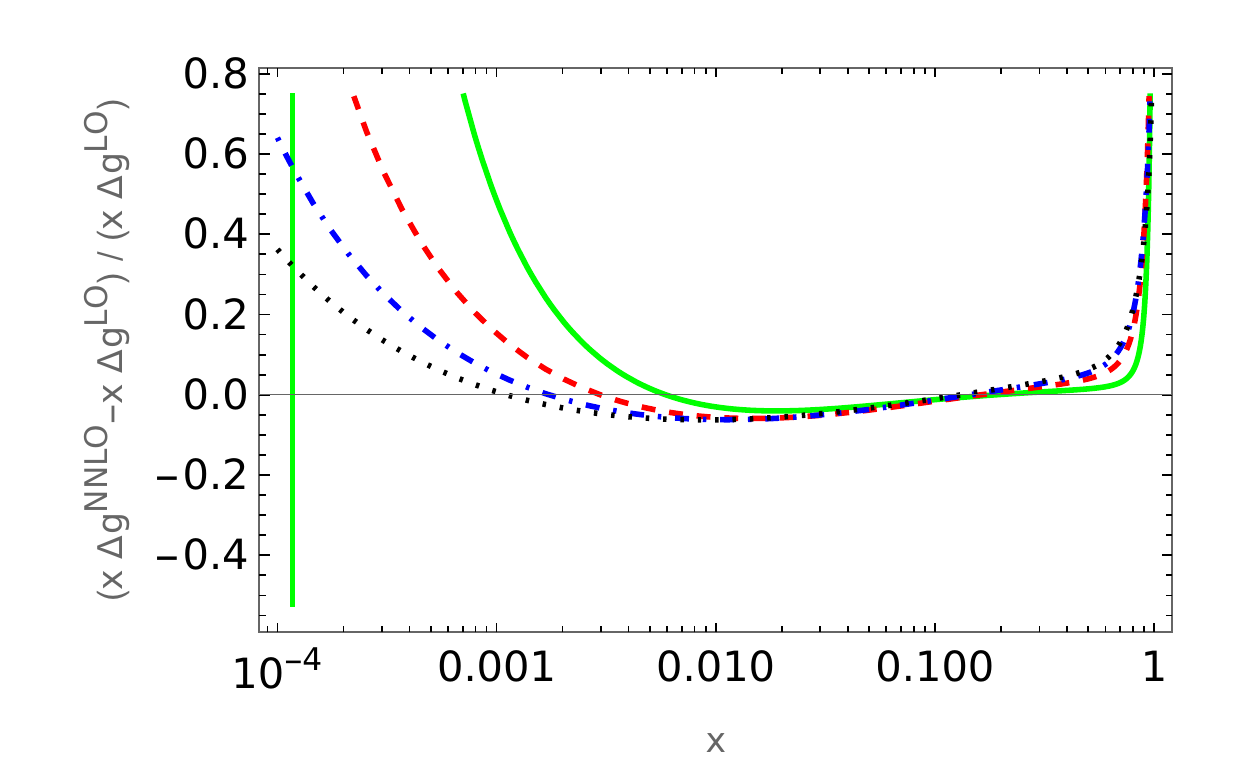}
\caption{\sf Comparison of the NNLO-evolved polarized gluon PDF to the LO-evolved one. The relative difference is shown. Solid lines: $Q^2=10~\GeV^2$. Dashed lines: $Q^2 = 10^2~\GeV^2$, Dash-dotted lines: $Q^2 = 10^3~\GeV^2$; Dotted lines: $Q^2 = 10^4~\GeV^2$. }
\label{figpdf7}
\end{figure}

\clearpage
\thispagestyle{empty}
~

\clearpage
\section{Conclusions}

In order to match the increasing experimental precision delivered by colliders, the precise determination of the parameters of the Standard Model, such as the strong coupling constant, the masses of the heavy quarks, as well as the parton distributions functions, is of great interest. Important experimental input comes from the study of structure functions in deep-inelastic scattering. The variable flavour number scheme, which constitutes a practical operational framework for experimental fits, requires the calculation of universal OMEs, which have been known for a long time at $\mathcal O(a_s^2)$, and which, starting to $\mathcal O(a_s^3)$, acquire contributions due to diagrams with two quark lines. Because the ratio of the masses of charm and bottom quarks, $m_c^2/m_b^2\sim0.1$, is not negligible, and in order to be fully consistent on theoretical grounds, it is desirable to consider a VNFS where the two quarks are decoupled together, not one at a time. This requires the calculation of two-mass contributions to the OMEs analytically in the quark mass ratio.

In this thesis, two such two-mass contributions to the polarized OMEs $\Delta A_{gg,Q}^{(3)}$ and $\Delta A_{Qq}^{PS(3)}$ have been calculated at $\text{N}^3\text{LO}$ in analytic and semi-analytic form in Chapter \ref{sec:PolDIS}, using Mellin-Barnes integrals and with the help of computer-algebraic packages for summation theory, namely {\tt Sigma} \cite{Schneider:2001,SIG1,SIG2}, {\tt HarmonicSums} \cite{HARMONICSUMS}, {\tt EvaluateMultiSums} and {\tt SumProduction} \cite{EMSSP}, mirroring previous work on the unpolarized case. In $N$-space, the result for $\Delta A_{gg,Q}^{(3)}$ is given in analytic form involving binomial sums. In the case of $\Delta A_{Qq}^{PS(3)}$ it was not possible with our methods to derive the $N$-space OME, and only a momentum-fraction $z$-space semi-analytic result is presented. For both OMEs the result is given in terms of iterated integrals over root-valued alphabets. We conclude that the size of the two-mass corrections to these OMEs is not negligible if compared to the single-mass corrections, hence they should be taken into account in the definition of the VFNS at $\mathcal O(a_s^3)$. Only the OME $A_{Qg}$ remains to be computed both in the polarized and in the unpolarized case to be able to complete all the OMEs required to define the VFNS at $\mathcal O(a_s^3)$. It is, however, known by now that the OME $A_{Qg}$ as well as the $N$-space OME $\Delta A_{Qq}^{PS(3)}$ satisfy difference equations which do not factorize to first order and therefore belong to a wider class of functions than the one considered in this thesis, possibly leading to elliptic integrals or to other classes of functions, of a form to date unknown.

The factorization of massive Wilson coefficients into universal OMEs and massless Wilson coefficients, which holds for $Q^2\gg m^2$ with $m$ the quark mass, was employed in Chapter \ref{sec:logWC} to write the asymptotic logarithmic charm contributions to the single-mass Wilson coefficients for the polarized structure function $g_1$, following earlier work for the unpolarized case, to $\mathcal O(a_s^3)$. Such logarithmic terms are determined by the renormalization structure of the theory and can be reconstructed from the knowledge of lower-order OMEs and anomalous dimensions. Contrary to the logarithmic terms, the constant part of the Wilson coefficients is still out of reach, because the corresponding massless Wilson coefficients are still unknown, as is the polarized $\Delta A_{Qg}^{(3)}$.

Fits of DIS structure functions are historically performed by choosing the parametrization of some functional form for the PDFs at an initial scale $Q_0^2$, evolving such parametrizations to the relevant scale $Q^2$, and applying a convolution with the relevant Wilson coefficients. An error minimization procedure simultaneously delivers the maximum-likelihood values of the PDF parameters and of the physical constants such as $a_s(M_Z^2)$. Such a procedure inevitably incurs in sources of theoretical uncertainty, among which, at higher perturbative order, a dependence of the PDF evolution as well as of the Wilson coefficients on the factorization scale and scheme. By contrast, a scheme-invariant evolution procedure takes as input the observable structure function and evolves it to a different scale: in this way all dependence on the factorization scheme is cancelled. In Chapter \ref{sec:schemeinv} we have extended such an evolution procedure for $F_2(x,Q^2)$ and $g_1(x,Q^2)$ to include the heavy flavour contributions and given some numerical illustrations to $\text N^3\text{LO}$, using Pad\'e approximants, of the effects of the (unknown) four-loop anomalous dimensions.

Feynman integrals form a fertile ground for mathematical exploration into integration theory, and novel insight into integration techniques and into unexplored classes of functions has historically been gained from their study. New classes of functions have been systematized in order to deal with the results of Feynman integration, for example, in the classes of harmonic polylogarithms and iterated integrals. Feynman integrals which evaluate to hypergeometric series are also known. 
In Chapter \ref{sec:hyperg} we considered the properties of hypergeometric series in several variables, among which the classical functions examined by Horn, Appell, Lauricella and Exton, and we examine a class of systems of differential equations obeyed by them, and describe an algorithm to solve such systems. The solutions appear as series having a nested hypergeometric product as summand. It is well-known that classical hypergeometric series have appeared in particle physics calculations, and, additionally, series containing hypergeometric (nested) products as summand arise in certain algorithms for the solutions of GKZ systems, which have been examined in the literature in the context of Feynman integrals. We give examples of the series expansion of such functions using {\tt Sigma} and show some classes of functions which arise in these expansions.

In Chapter \ref{sec:PLDEsolver} we described the problem of solving partial linear difference equations in several variables. These problems arise in particle physics, among others, when the Laporta algorithm is applied to Feynman integrals. In the univariate case, which is well studied, many algorithms and implementations are available; it is ubiquitous in the case study of deep-inelastic scattering, where the difference equations are in the variable $N$. In the multivariate case, we restricted our focus to the solution space of rational functions, possibly containing harmonic sums and Pochhammer symbols in the numerator. We implemented in a Mathematica package one algorithm to constrain the denominator of the solution, and a solver for the numerator based on heuristic methods and user-programmable anz\"atze, with an eye to delivering some solutions within an acceptable amount of computing time, rather than on the completeness of the solution space.

In Chapter \ref{sec:program} we describe a Mellin-space numerical library which encodes the splitting functions and asymptotic logarithmic corrections for a number of structure functions, namely $F_2$, $F_L$, $g_1$ under photon exchange, and $F_3^{W^+\pm W^-}$ under charged-current exchange. It also contains an implementation of the known Wilson coefficients for the Drell-Yan process and for Higgs production to order $a_s^2$.
The $N$-space evaluation is achieved by analytically continuing the harmonic sums in the complex plane: an asymptotic expansion is obtained in the large-$N$ limit for the Mellin transform of harmonic polylogarithms, which are related to harmonic sums, and recursion relations can be used to compute the analytic continuation of the harmonic sums for smaller $|N|$. The library encodes all the special functions necessary to calculate in $N$-space the two-loops Wilson coefficients for DIS and the 3-loop splitting functions, which are expressible using weight-5 harmonic sums, and several weight-6 harmonic sums, namely those contained in $c_{2,q}^{(3)NS}$. The other three-loop Wilson coefficients are encoded using the approximate representations given in the literature. The library can perform a fast numerical evolution of the PDFs and the numerical Mellin inversion to momentum fraction space, and may be suitable for experimental fits in future applications. It can accept a user-provided parametrization of the light-quark PDF combinations $u\pm\bar u$, $d\pm\bar d$, and $s\pm\bar s$, which are the relevant combinations for the calculation of the DIS structure functions in the asymptotic region $Q^2\gg m_{c,b}^2$ in the fixed flavour number scheme. The code can compute the running of $a_s(Q^2)$ from the solution of the renormalization-group equations in perturbation theory, with a fixed number of light flavours, from a value of $\Lambda_{QCD}$ chosen by the user. We provided some indications of the numerical precision attainable by the library by computing fixed Mellin moments of Wilson coefficients and of individual harmonic sums, and a comparison to the code Pegasus \cite{Vogt:2004ns}. We showed the results of the numerical evolution of an input set of PDFs to various scales under the fixed flavour number regime.

\clearpage
\appendix

\section{Representation of certain iterated integrals}
\label{sec:A}
\vspace*{1mm}
\noindent
In the following we present a series of integrals and relations which appeared in intermediate steps of the calculation of $A_{gg,Q}^{(3),\text{two-mass}}$
and which may be of further use in similar applications, extending the results given in 
\cite{Ablinger:2018brx} before. We obtained the following iterative integrals

\begin{eqnarray}
G_{44} &=& G\left(
        \left\{\frac{\sqrt{1-x}}{x}\right\},z\right) 
=
-\HA_{-1}\big(u_1\big)
-\HA_1\big(u_1\big)
+2 \ln(2)
-2
\nonumber \\ &&
+u_1 \Big[
\HA_0(z) + \HA_1\big(u_1\big) - \HA_{-1}\big(u_1\big) +2
\Big], \\
G_{45} &=& G\left(\left\{\frac{\sqrt{1-x}}{\sqrt{x} (\eta -x \eta +x)}\right\},z\right)
\nonumber \\ &=&
\frac{\pi}{\sqrt{\eta} + \eta} 
+ 2 \frac{{\rm arcsin}(\sqrt{1-z})}{1-\eta} 
- 2 \frac{{\rm arctan}\left(
\frac{\sqrt{\eta} \sqrt{1-z}}{\sqrt{z}}\right)}
{\sqrt{\eta}(1-\eta)} ,
\\
G_{46} &=& G\left(\left\{\frac{\sqrt{1-x}}{\sqrt{x} (x \eta -x+1)}\right\},z\right)
\nonumber \\ &=&
\frac{1}{\eta -1} \Biggl[\pi  \big(
        \sqrt{\eta }-1\big)
-2 \sqrt{\eta } \, {\rm arctan}\left(\frac{u_1}{\sqrt{\eta z}}\right)
+2 {\rm arctan}\left(\frac{u_1}{\sqrt{z}}\right)
\Biggr], \\
G_{47} &=& G\left(\left\{\frac{\sqrt{x}}{x \eta -x+1}\right\},z\right)
\nonumber \\ &=&
\frac{1}{v_1^3} \Big[
2 \HA_{-1}\big(v_1 \sqrt{z}\big)
+\HA_1(z-z \eta)
\Big]
-\frac{2 \sqrt{z}}{v_1^2}, \\
G_{48} &=& G\left(\left\{\sqrt{1-x} \sqrt{x \eta -x+1}\right\},z\right)
=
\frac{2-\eta}{4 (1-\eta)}
-\frac{u_1^3 u_2}{4}
\nonumber \\ &&
-\frac{u_1 u_2^3}{4 (1-\eta)}
+\frac{\eta^2}{4 v_1^3} \Biggl[
 {\rm arcsinh}\left(\frac{u_1 v_1}{\sqrt{\eta}}\right)
-{\rm arcsinh}\left(\frac{v_1}{\sqrt{\eta}}\right)
\Biggr], \\
G_{49} &=& G\left(\left\{\frac{\sqrt{x \eta -x+1}}{x}\right\},z\right)
=
-\HA_1\big(u_2\big)
-\HA_0(z-z \eta)
-\HA_{-1}\big(u_2\big)
\nonumber \\ &&
+\HA_0(z)
+2 \ln(2)
-2
+u_2 \Big[
         \HA_0(z-z \eta)
        +2
        +\HA_1\big(u_2\big)
        -\HA_{-1}\big(u_2\big)
\Big], \\ 
G_{50} &=& G\left(\left\{\frac{\sqrt{1-x}}{x},\frac{1}{x}\right\},z\right)
=
u_1 \biggl[
         \frac{1}{2} H^2_0(z)
        -4
\biggr]
+2 \ln^2(2)
\nonumber \\ &&
-4 \ln(2)
+4
+\big(1+u_1\big)
\biggl[2 \HA_{-1}\big(u_1\big)
        -\frac{1}{2} \HA_{-1}^2\big(u_1\big)
        +\HA_{-1,1}\big(u_1\big)
\biggr]
\nonumber \\ &&
+\big(1-u_1\big)
\biggl[2 \HA_1\big(u_1\big)
        -\HA_{1,-1}\big(u_1\big)
        +\frac{1}{2} \HA_1^2\big(u_1\big)
\biggr]
-\zeta_2, \\
G_{51} &=& G\left(\left\{\frac{\sqrt{x}}{\eta -x \eta +x},\frac{1}{x}\right\},z\right)
=
       \frac{2 \sqrt{z}}{v_1^2} \big(
                \HA_0(z)
                -2
        \big)
\nonumber \\ &&
+\frac{\sqrt{\eta }}{v_1^3} \Biggl[
        -2 \HA_0(z) \HA_{\{4,0\}}\left(v_1 \sqrt{\frac{z}{\eta}}\right)
        +4 \HA_{\{0,0\},\{4,0\}}\left(v_1 \sqrt{\frac{z}{\eta}}\right)
\Biggr], \\
G_{52} &=& G\left(\left\{\frac{\sqrt{x \eta -x+1}}{x},\frac{1}{x}\right\},z\right)
=
u_2 \big(2 \HA_0(z)-4\big)
+\frac{1}{2} H^2_0(z)
\nonumber \\ &&
+\big(1+u_2\big)
\biggl[-\HA_{-1}\big(u_2\big) \HA_0(z)
       +\HA_{-1,-1}\big(u_2\big)
       -\HA_{1,-1}\big(u_2\big)
\biggr]
\nonumber \\ &&
-2 \HA_0(z)
+\big(1-u_2\big)
\biggl[-\HA_0(z) \HA_1\big(u_2\big)
       -\HA_0(z) \HA_0(z-z \eta)
\nonumber \\ &&
       +\frac{1}{2} \HA_0^2(z-z \eta)
       +\HA_{-1,1}\big(u_2\big)
       -\frac{1}{2} \HA_1^2\big(u_2\big)
\biggr]
+4 \HA_{-1}\big(u_2\big)
-\zeta_2
\nonumber \\ &&
+2 \ln^2(2)
+2 \ln(2) \biggl[\HA_1\big(u_2\big) - \HA_{-1}\big(u_2\big)+\HA_0(z)\biggr]
-4 \ln(2)
+4, \\
G_{53} &=& G\left(\left\{\frac{\sqrt{1-x}}{x},\frac{\sqrt{1-x}}{x},\frac{1}{x}\right\},z\right)
=
z \HA_0(z)
-\frac{z}{2} H^2_0(z)
+\frac{1}{6} H^3_0(z)
\nonumber \\ &&
+2 u_1 \biggl[
        -\frac{1}{2} \HA_{-1}^2\big(u_1\big)
        +\HA_{-1,1}\big(u_1\big)
        -\HA_{1,-1}\big(u_1\big)
       +\frac{1}{2} \HA_1^2\big(u_1\big)
\biggr]
\nonumber \\ &&
-(z+2) \HA_1\big(u_1\big) \HA_{-1}\big(u_1\big)
+\frac{z-2}{2} \Big[\HA_{-1}^2\big(u_1\big)+\HA_1^2\big(u_1\big)\Big]
\nonumber \\ &&
+2 \HA_{-1,1,-1}\big(u_1\big)
-2 \HA_{-1,1,1}\big(u_1\big)
+2 \HA_{1,-1,-1}\big(u_1\big)
-2 \HA_{1,-1,1}\big(u_1\big)
\nonumber \\ &&
+\big(\zeta_2
        +4 \ln(2)
        -2 \ln^2(2)
        +4 u_1
\big)
\Big[
         \HA_{-1}\big(u_1\big)
        +\HA_1\big(u_1\big)
\Big]
-\frac{3}{2} \zeta_3
\nonumber \\ &&
-(3 z+4) \HA_{-1}\big(u_1\big)
+(3 z-4) \HA_1\big(u_1\big)
+\frac{4}{3} \ln^3(2)
\nonumber \\ &&
+\big(
        8
        -8 \ln(2)
        +4 \ln^2(2)
        -2 \zeta_2
\big) u_1
-4 \ln^2(2)
+8 \ln(2)
+6 z
-8, \\
G_{54} &=& G\left(\left\{\frac{1}{x \eta -x+1},\frac{1}{x},\frac{1}{x}\right\},z\right)
=
\frac{1}{1-\eta} \biggl[
\frac{1}{2} H^2_0(z) \HA_1(z-z \eta)
\nonumber \\ &&
-\HA_0(z) \HA_{0,1}(z-z \eta)
+\HA_{0,0,1}(z-z \eta)
\biggr], \\ 
G_{55} &=& G\left(\left\{\frac{\sqrt{x \eta -x+1}}{x},\frac{1}{x},\frac{1}{x}\right\},z\right)
=
-8
+
\frac{4 \ln^3(2)}{3}
\nonumber \\ &&
+\ln(2) \Big[
        8
        -4 \HA_1(\eta )
        -2 \zeta_2
        +\HA_1(\eta )^2
\Big]
\nonumber \\ &&
+u_2 \bigg\{
        8
        +\frac{1}{2} \Big[
                -\HA_1(u_2)
                +\HA_{-1}(u_2)
        \Big]^2 \Big[
                2-\HA_1(\eta )\Big]
\nonumber \\ &&
        -\frac{1}{2} \Big[
                -\HA_1(u_2)
                +\HA_{-1}(u_2)
        \Big]
\Big[-4+\HA_1(\eta )\Big] \HA_1(\eta )
        +\frac{1}{6} \HA_0(z)^3
        -4 \HA_1(\eta )
        -\frac{1}{6} \HA_1(\eta )^3
\nonumber \\ &&
        -\frac{1}{2} \HA_{-1}(u_2) \HA_1(u_2)^2
        +\frac{1}{6} \HA_1(u_2)^3
        -4 \HA_{-1}(u_2)
        -\frac{1}{6} \HA_{-1}(u_2)^3
        +\HA_1(\eta )^2
\nonumber \\ &&
        +\frac{1}{2} \HA_1(u_2) \Big[
                8+\HA_{-1}(u_2)^2\Big]
\bigg\}
+\bigg\{
        4
        +\frac{1}{2} \HA_1(u_2)^2
        +2 \HA_{-1}(u_2)
        -\frac{1}{2} \HA_{-1}(u_2)^2
        -\zeta_2
\nonumber \\ &&
        +\HA_1(u_2) \Big[
                2-\HA_{-1}(u_2)\Big]
\bigg\} \HA_1(\eta )
-\HA_1(\eta )^2
-\frac{1}{2} \Big[
        \HA_1(u_2)
        +\HA_{-1}(u_2)
\Big] \HA_1(\eta )^2
\nonumber \\ &&
+\frac{1}{6} \HA_1(\eta )^3
-\frac{1}{6} \HA_1(u_2)^3
-4 \HA_{-1}(u_2)
-\frac{1}{6} \HA_{-1}(u_2)^3
\nonumber \\ &&
+\Big[
        -4
        +2 \HA_1(\eta )
        +2 \HA_{-1}(u_2)
\Big] \HA_{-1,1}(u_2)
-2 \HA_{-1,1,1}(u_2)
-2 \HA_{-1,-1,1}(u_2)
\nonumber \\ &&
+2 \zeta_2
+2 \zeta_3
+2 \ln^2(2) \Big[
        -2+\HA_1(\eta )\Big]
+\frac{1}{2} \HA_1(u_2)^2 \Big[
        -2+\HA_{-1}(u_2)\Big]
+\HA_{-1}(u_2)^2
\nonumber \\ &&
+\frac{1}{2} \HA_1(u_2) \Big[
        -8+4 \HA_{-1}(u_2)-\HA_{-1}(u_2)^2\Big]
, \\
G_{56} &=& G\left(\left\{\frac{\sqrt{x \eta -x+1}}{x},\frac{1}{x \eta -x+1},\frac{1}{x}\right\},z\right) \nonumber \\
&=&
\frac{u_2}{1-\eta} \biggl\{
         4 \HA_{-1,0}(u_2)
        -4 \HA_{1,0}(u_2)
        -4 \HA_{-1,-1,0}(u_2)
        +4 \HA_{-1,1,0}(u_2)
\nonumber \\ &&
        +\Big[
                4
                -4 \HA_0(u_2)
                +2 \HA_{-1,0}(u_2)
                +\HA_{0,1}(z-z \eta)
                -2 \HA_{1,0}(u_2)
                -\zeta_2
        \Big] \HA_0(z)
\nonumber \\ &&
        -4 \HA_{1,1,0}(u_2)
        -2 \HA_{0,0,1}(z-z \eta)
        +4 \HA_{1,-1,0}(u_2)
        +2 \zeta_2 \HA_{-1}(u_2)
\nonumber \\ &&
        -2 \zeta_2 \HA_1(u_2)
        -2 \zeta_2
        +2 \zeta_3
        -8
\biggr\}
+\frac{1}{1-\eta} \biggl\{
-4 \HA_{-1,-1,0}(u_2)
\nonumber \\ &&
+\Big[
         2 \HA_{-1,0}(u_2)
        +2 \HA_{1,0}(u_2)
        +3 \zeta_2
        -4
\Big] \HA_0(z)
+8
-\frac{7}{2} \zeta_3
\nonumber \\ &&
+4 \HA_{1,1,0}(u_2)
-2 \big(
         \zeta_2
        -4
\big) \HA_{-1}(u_2)
+4 \HA_1(u_2) \zeta_2
-8 \ln(2)
\biggr\} ,
\end{eqnarray}
where
\begin{eqnarray}
u_1 &=& \sqrt{1-z}, \\
u_2 &=& \sqrt{1-z (1-\eta)}, \\
v_1 &=& \sqrt{1-\eta}.
\end{eqnarray}
We also encountered the following iterated integrals $K_i$, which evaluate to:


where $C$ is  Catalan's constant
\begin{equation}
C = \sum_{n=0}^\infty \frac{(-1)^n}{(2n+1)^2} \approx 0.915965594 ,
\end{equation}
and
\begin{eqnarray}
v_2 &=& 1-2 \sqrt{(1-\eta) \eta}, \\
v_3 &=& \frac{v_2}{1-2 \eta} ,
\end{eqnarray}
and
\begin{eqnarray}
G\left(\left\{\frac{\sqrt{1-x}}{x},\frac{\sqrt{x}}{1-x}\right\},\eta\right) &=& 
\pi^2 - 2 \sqrt{(1 - \eta) \eta} - 2 \arcsin(\sqrt{\eta}) 
\nonumber\\ &&
+
 4 \sqrt{1 - \eta} \arctanh(\sqrt{\eta}) 
+
 8 \arctan\left[-1 + \frac{\sqrt{1 - \eta}}{\sqrt{\eta}}\right] 
\nonumber\\ &&
-
 8 \arctanh\left[\frac{\sqrt{1 - \sqrt{\eta}}}{\sqrt{1 + \sqrt{\eta}}}\right] 
 \arctanh(\sqrt{\eta}) 
\nonumber\\ &&
+
 4 \Li_2\left( -\frac{\sqrt{1 - \sqrt{\eta}}}{\sqrt{1 + \sqrt{\eta}}} \right)
 -
 4 \Li_2\left( \frac{\sqrt{1 - \sqrt{\eta}}}{\sqrt{1 + \sqrt{\eta}}} \right).
\end{eqnarray}
The following constant is calculated numerically
\begin{eqnarray}
G\left(\left\{\frac{\sqrt{2-x}}{x},
\frac{\sqrt{1-x}}{x},
\frac{\sqrt{1-x}}{x}\right\},1\right) &=& 0.413734026910741614953~.
\end{eqnarray}

\clearpage
\section{Relations between certain functions}
\label{sec:appB}
We further present representations of a series of functions $g_i$ which are functions of $x$ and $\eta$. These functions emerged in the calculation of $A_{gg,Q}^{(3)\text{two-mass}}$.  
The symbol $y$, not to be confused with its meaning in the main text, is defined here as
\begin{equation}
y = \frac{1-2 \sqrt{1-x} \sqrt{x}}{1-2 x},
\end{equation}
and the formulas are valid for
\begin{equation}
\begin{gathered}
0 <\eta <1, \\
0 <x    <1,
\end{gathered} 
\end{equation}


\clearpage
\section{Polarized operator matrix elements}
\label{sec:polOMEs}

In the following we present all logarithmic single-mass contributions to the polarized operator matrix elements of twist-two operators in DIS, cf. Ref. \cite{Blumlein:2021xlc}, discussed in Section \ref{sec:logWC}.
We use the abbreviations
\begin{eqnarray}
	L_Q &=& \ln\frac{Q^2}{\mu^2} ~, \\
	L_M &=& \ln\frac{m^2}{\mu^2} ~,
\end{eqnarray}
where $\mu^2$ refers to the renormalization and factorization scale.

In $z$-space, the OME $A_{gg,Q}^S$ is distribution-valued, while the other OMEs presented here contain only regular contributions. For $A_{gg,Q}^S(z)$ we present separately three terms, namely
\begin{eqnarray}
	 A_{gg,Q}^S(N) = \int_0^1 dz z^{N-1} \big[ A_{gg,Q}^{reg} + \delta(1-z) A_{gg,Q}^{(\delta)} ] + \int_0^1 dz (z^{N-1}-1) A_{gg,Q}^{(+)}
\end{eqnarray}

\subsection{$A_{qq,Q}^{PS(3)}$ in $N$ space}
\begin{eqnarray}
A_{qq,Q}^{PS(3)} &=& 
	a_s^3 \Biggl\{
	a_{qq,Q}^{PS(3)}
	+ \textcolor{blue}{C_F N_F T_F^2} \biggl\{
                -\frac{32  L_M ^3 (N-1) (2+N)}{9 N^2 (1+N)^2}
\nonumber\\&&		
                +\frac{32 (N-1) (2+N) \big(
                        98+369 N+408 N^2+164 N^3\big)}{81 N^2 
(1+N)^5}
\nonumber\\&&
		+ L_M ^2 \biggl[
                        -\frac{32 (2+N) \big(
                                3+4 N-3 N^2+8 N^3\big)}{9 N^3 
(1+N)^3}
                        +\frac{32 (N-1) (2+N)}{3 N^2 (1+N)^2} S_1 
		\biggr]
\nonumber\\&&		
		+ L_M  \biggl[
                        -\frac{32 (2+N) P_1}{27 N^4 (1+N)^4}
                        +\frac{64 (2+N) \big(
                                3+4 N-3 N^2+8 N^3\big)}{9 N^3 
(1+N)^3} S_1 
\nonumber\\&&
                        -\frac{32 (N-1) (2+N)}{3 N^2 
(1+N)^2} S_1 ^2
                        -\frac{32 (N-1) (2+N)}{3 N^2 (1+N)^2} S_2 
		\biggr]
\nonumber\\&&		
		+\biggl[
                        -\frac{32 (N-1) (2+N) \big(
                                22+41 N+28 N^2\big)}{27 N^2 (1+N)^4}
                        -\frac{16 (N-1) (2+N) S_2 }{3 N^2 (1+N)^2}
		\biggr] S_1 
\nonumber\\&&		
                +\frac{16 (N-1) (2+N) (2+5 N)}{9 N^2 (1+N)^3} 
	(S_1 ^2 + S_2)
		-\frac{16 (N-1) (2+N)}{9 N^2 (1+N)^2} (S_1 ^3 + 2 S_3)
\nonumber\\&&		
		+\biggl[
                        \frac{16 (2+N) \big(
                                3+2 N-6 N^2+13 N^3\big)}{9 N^3 (1+N)^3}
                        -\frac{32 (N-1) (2+N) S_1 }{3 N^2 (1+N)^2}
		\biggr] \zeta_2
\nonumber\\&&		
                +\frac{32 (N-1) (2+N)}{9 N^2 (1+N)^2} \zeta_3
	\biggr\}
	\Biggr\} ~,
\end{eqnarray}

\begin{eqnarray}
P_1=86 N^5+38 N^4+40 N^3-8 N^2-15 N-9
\end{eqnarray}

\subsection{$A_{qg,Q}^{S(3)}$ in $N$ space}
\begin{eqnarray}
A_{qg,Q} &=&
	a_s^3 \Biggl\{
	a_{qg,Q}^{(3)}
	+ \textcolor{blue}{C_A N_F T_F^2} \biggl\{
                -
                \frac{8 (N-1) P_8}{81 N^5 (1+N)^5}
		+ L_M ^3 \biggl[
                        -\frac{64 (N-1)}{9 N^2 (1+N)^2}
                        +\frac{32 (N-1)}{9 N (1+N)} S_1
		\biggr]
\nonumber\\&&		
		+ L_M ^2 \biggl[
                        \frac{8 P_4}{9 N^3 (1+N)^3}
                        +\frac{32 \big(
                                1+5 N^2\big)}{9 N (1+N)^2} S_1
			-\frac{16 (N-1)}{3 N (1+N)} (S_1^2 + S_2)		
                        -\frac{32 (N-1)}{3 N (1+N)} S_{-2}		
		\biggr]
\nonumber\\&&
		+ L_M  \Biggl[
                        \frac{16 P_7}{27 N^4 (1+N)^4}
			+\biggl[
                                \frac{16 \big(
                                        -1+44 N+67 N^2+94 
N^3\big)}{27 N (1+N)^3}
                                -\frac{16 (N-1) S_2}{3 N (1+N)}
			\biggr] S_1
\nonumber\\&&			
                        -\frac{32 \big(
                                -2+5 N^2\big)}{9 N (1+N)^2} S_1^2
                        +\frac{16 (N-1)}{9 N (1+N)} S_1^3
                        -\frac{32 \big(
                                -2+6 N+5 N^2\big)}{9 N (1+N)^2} S_2
                        +\frac{32 (N-1)}{9 N (1+N)} S_3
\nonumber\\&&
                        -\frac{64 (-2+5 N)}{9 N (1+N)} S_{-2}
                        +\frac{64 (N-1)}{3 N (1+N)} S_{-3}
                        +\frac{64 (N-1)}{3 N (1+N)} S_{2,1}
		\Biggr]
\nonumber\\&&		
                -\frac{16 (N-1) \big(
                        283+584 N+328 N^2\big)}{81 N (1+N)^3} S_1
                -\frac{8 (N-1)}{3 N (1+N)^2} S_1^2
                +\frac{8 (N-1) (1+2 N)}{3 N (1+N)^2} S_2
\nonumber\\&&		
		+\biggl[
                        -\frac{8 P_3}{9 N^3 (1+N)^3}
                        -\frac{32 \big(
                                -2+5 N^2\big)}{9 N (1+N)^2} S_1
                        +\frac{8 (N-1) S_1^2}
                        {3 N (1+N)}
\nonumber\\&&			
                        +
                        \frac{8 (N-1) S_2}{3 N (1+N)}
                        +\frac{16 (N-1) S_{-2}}{3 N (1+N)}
		\biggr] \zeta_2
	+\biggl[
                        \frac{64 (N-1)}{9 N^2 (1+N)^2}
                        -\frac{32 (N-1) S_1}{9 N (1+N)}
		\biggr] \zeta_3
	\biggr\}
\nonumber\\&&	
	+ \textcolor{blue}{C_F N_F T_F^2} \biggl\{
                -\frac{(N-1) P_{10}}{81 N^6 (1+N)^6}
		+ L_M ^3 \biggl[
                        \frac{8 (N-1) P_2}{9 N^3 (1+N)^3}
                        -\frac{32 (N-1)}{9 N (1+N)} S_1
		\biggr]
\nonumber\\&&		
		+ L_M ^2 \biggl[
                        \frac{4 (N-1) P_6}{9 N^4 (1+N)^4}
                        -\frac{32 (N-1) (3+5 N)}{9 N^2 (1+N)} S_1
			+\frac{16 (N-1)}{3 N (1+N)} (S_1^2 + S_2)
		\biggr]
\nonumber\\&&		
		+ L_M  \Biggl[
                        \frac{4 P_9}{27 N^5 (1+N)^5}
			+\biggl[
                                -\frac{16 \big(
                                        -24-52 N+103 N^2\big)}{27 N^2 
(1+N)}
                                -\frac{16 (N-1)}{3 N (1+N)} S_2
			\biggr] S_1
\nonumber\\&&			
                        +\frac{16 (N-1) (3+10 N)}{9 N^2 (1+N)} S_1^2
                        -\frac{16 (N-1)}{9 N (1+N)} S_1^3
                        +\frac{16 \big(
                                -3+5 N+10 N^2\big)}{9 N^2 (1+N)} S_2
\nonumber\\&&				
                        +\frac{64 (N-1)}{9 N (1+N)} S_3
		\Biggr]
                +\frac{5248 (N-1) S_1}{81 N (1+N)}
                -\frac{896 (N-1) S_2}{27 N (1+N)}
                +\frac{160 (N-1) S_3}{9 N (1+N)}
\nonumber\\&&		
                -\frac{32 (N-1)}{3 N (1+N)} S_4
		+\biggl[
                        -\frac{4 (N-1) P_5}{9 N^4 (1+N)^4}
                        +\frac{16 (N-1) (3+10 N)}
                        {9 N^2 (1+N)} S_1
\nonumber\\&&			
                        -\frac{8 (N-1) S_1^2}{3 N (1+N)}
                        -\frac{8 (N-1) S_2}{N (1+N)}
		\biggr] \zeta_2
		+\biggl[
                        -\frac{8 (N-1) P_2}{9 N^3 (1+N)^3}
                        +\frac{32 (N-1) S_1}{9 N (1+N)}
		\biggr] \zeta_3
	\biggr\}
\Biggr\}
\end{eqnarray}

\begin{eqnarray}
P_2&=&3 N^4+6 N^3-N^2-4 N+12,\\
P_3&=&6 N^5+6 N^4-67 N^3+6 N^2+43 N-18, \\
P_4&=&9 N^5+9 N^4-79 N^3+15 N^2+22 N-24,\\
P_5&=&18 N^6+54 N^5+5 N^4-20 N^3+95 N^2-132 N-108,\\
P_6&=&33 N^6+99 N^5+41 N^4-11 N^3+86 N^2-216 N-144,\\
P_7&=&99 N^7+198 N^6-410 N^5-344 N^4+128 N^3-130 N^2-39 N+90, \\
P_8&=&255 N^8+1020 N^7-532 N^6-4536 N^5-4344 N^4-1138 N^3+3 N^2+36 N
\nonumber\\&&
+108,\\
P_9&=&159 N^9+477 N^8-220 N^7-710 N^6+117 N^5-1081 N^4+2536 N^3
\nonumber\\&&
+1026 N^2-1800 N-1080,\\
P_{10}&=&1551 N^{10}+7755 N^9+10982 N^8+1910 N^7+2427 N^6+14975 N^5
\nonumber\\&&
+13952 N^4-1488 N^3-7488 N^2-6912 N-2592
\end{eqnarray}

\subsection{$A_{Qg}^{S}$ in $N$ space}

with the polynomials
\begin{eqnarray}
P_{11}&=&-3 N^4-54 N^3-95 N^2-12 N+36,\\
P_{12}&=&N^4-94 N^3-256 N^2-161 
N+78,\\
P_{13}&=&N^4+2 N^3-5 N^2-12 N+2,\\
P_{14}&=&N^4+4 N^3-N^2-10 
N+2, \\
P_{15}&=&N^4+10 N^3+27 N^2+30 N+4,\\
P_{16}&=&N^4+17 N^3+43 N^2+33 
N+2, \\
P_{17}&=&2 N^4-4 N^3-3 N^2+20 N+12,\\
P_{18}&=&2 N^4+3 N^3-12 
N^2-23 N+6, \\
P_{19}&=&2 N^4+39 N^3+100 N^2+73 N+2,\\
P_{20}&=&3 N^4+6 N^3-N^2-4 N+12,\\
P_{21}&=&3 N^4+30 N^3+47 N^2+4 N-20,\\
P_{22}&=&3 N^4+48 N^3+123 N^2+98 N+8,\\
P_{23}&=&5 N^4+10 N^3+8 N^2+7 N+2, \\
P_{24}&=&5 N^4+13 N^3+14 N^2+16 N+6,\\
P_{25}&=&5 N^4+37 N^3+82 N^2+41 
N-48, \\
P_{26}&=&6 N^4+11 N^3-6 N^2-N-2,\\
P_{27}&=&6 N^4+12 N^3+7 
N^2+N+6, \\
P_{28}&=&9 N^4+102 N^3+245 N^2+192 N+12,\\
P_{29}&=&10 N^4+53 
N^3+92 N^2+37 N-48,\\
P_{30}&=&11 N^4-68 N^3-263 N^2-184 N+72, \\
P_{31}&=&11 N^4-26 N^3-227 N^2-286 N+48,\\
P_{32}&=&11 N^4+4 N^3-239 
N^2-304 N+240,\\
P_{33}&=&13 N^4+23 N^3+4 N^2-14 N-5,\\
P_{34}&=&13 N^4+140 N^3+365 N^2+190 N-276,\\
P_{35}&=&17 N^4+34 N^3+82 N^2+161 
N-78, \\
P_{36}&=&29 N^4+60 N^3+149 N^2+336 N+74,\\
P_{37}&=&55 N^4+86 
N^3-343 N^2-422 N+384,\\
P_{38}&=&55 N^4+182 N^3-175 N^2-542 N+240, \\
P_{39}&=&76 N^4+183 N^3+196 N^2+267 N-38,\\
P_{40}&=&97 N^4+494 N^3+1079 
N^2+898 N-408,\\
P_{41}&=&153 N^4+306 N^3+165 N^2+12 N+4,\\
P_{42}&=&183 N^4+366 N^3+305 N^2+122 N+96,\\
P_{43}&=&N^5+N^4-4 N^3+3 N^2-7 N-2,\\ 
P_{44}&=&2 N^5+6 N^4+3 N^3+11 N+2,\\
P_{45}&=&2 N^5+10 N^4+29 N^3+64 
N^2+67 N+8,\\
P_{46}&=&3 N^5+8 N^4+6 N^3+10 N^2+7 N+2,\\
P_{47}&=&8 N^5+7 
N^4-9 N^3+7 N^2+13 N+6,\\
P_{48}&=&9 N^5+9 N^4-79 N^3+15 N^2+22 N-24, \\
P_{49}&=&15 N^5+15 N^4-103 N^3+33 N^2-20 N-36, \\
P_{50}&=&18 N^5-15 
N^4-198 N^3-381 N^2-216 N+4,\\
P_{51}&=&18 N^5+47 N^4-35 N^3-141 N^2-5 
N-120, \\
P_{52}&=&40 N^5+73 N^4-142 N^3-163 N^2-150 N+54,\\
P_{53}&=&45 
N^5+45 N^4-47 N^3+27 N^2-190 N-24,\\
P_{54}&=&51 N^5+89 N^4+6 N^3-66 
N^2-104 N-72,\\
 P_{55}&=&66 N^5+336 N^4+627 N^3+415 N^2-10 N-194, \\
P_{56}&=&69 N^5+69 N^4-55 N^3+51 N^2-338 N-36,\\
 P_{57}&=&85 N^5+85 
N^4-73 N^3+197 N^2-342 N-108,\\
 P_{58}&=&103 N^5+103 N^4-79 N^3+317 
N^2-612 N-144,\\
 P_{59}&=&337 N^5+403 N^4-541 N^3-583 N^2-300 N+108, \\
P_{60}&=&436 N^5+1780 N^4+2689 N^3+2782 N^2+2167 N-134,\\
 P_{61}&=&489 
N^5+489 N^4-1187 N^3-57 N^2-742 N-144,\\
 P_{62}&=&N^6+18 N^5+63 N^4+84 
N^3+30 N^2-64 N-16,\\
 P_{63}&=&N^6+23 N^5+73 N^4+85 N^3+58 N^2+24 
N-24, \\
P_{64}&=&3 N^6+30 N^5+107 N^4+124 N^3+48 N^2+20 N+8,\\
 P_{65}&=&3 
N^6+51 N^5+153 N^4+185 N^3+160 N^2+80 N-72,\\
 P_{66}&=&4 N^6+41 N^5+126 
N^4+163 N^3+58 N^2-128 N-32,\\
 P_{67}&=&5 N^6+26 N^5+77 N^4+168 N^3+159 
N^2+19 N-22,\\
 P_{68}&=&6 N^6+75 N^5+345 N^4+719 N^3+323 N^2-696 N-96, \\
P_{69}&=&18 N^6+87 N^5+199 N^4+185 N^3+63 N^2+44 N+20, \\
P_{70}&=&23 
N^6+39 N^5-89 N^4-219 N^3-172 N^2-130 N-28,\\
 P_{71}&=&25 N^6-118 
N^5-662 N^4-500 N^3+421 N^2+186 N-264, \\
P_{72}&=&33 N^6+99 N^5+41 
N^4-11 N^3+86 N^2-216 N-144, \\
P_{73}&=&33 N^6+99 N^5+137 N^4+157 
N^3+62 N^2+8 N-16,\\
 P_{74}&=&36 N^6+48 N^5-297 N^4-977 N^3-976 N^2-362 
N+24,\\
 P_{75}&=&37 N^6+207 N^5+753 N^4+1771 N^3+1598 N^2-118 N+48, \\
P_{76}&=&57 N^6+297 N^5+567 N^4+615 N^3+468 N^2+220 N+16,\\
 P_{77}&=&80 
N^6+201 N^5-775 N^4-3495 N^3-4405 N^2-2238 N-72,\\
 P_{78}&=&94 N^6+282 
N^5+79 N^4+42 N^3+286 N^2-585 N-18, \\
P_{79}&=&129 N^6+387 N^5+509 
N^4+349 N^3+50 N^2+240 N+144,\\
 P_{80}&=&170 N^6+543 N^5+221 N^4-15 
N^3+425 N^2-864 N-288, \\
P_{81}&=&170 N^6+873 N^5+1547 N^4+951 N^3-1717 
N^2-2976 N+864,\\
P_{82}&=&243 N^6+729 N^5+923 N^4+583 N^3+14 N^2+300 
N+216,\\
P_{83}&=&321 N^6+1353 N^5+1521 N^4-713 N^3-2842 N^2-2216 
N+96,\\
P_{84}&=&333 N^6+999 N^5+1075 N^4+389 N^3-68 N^2+384 N+216, \\
P_{85}&=&633 N^6+1899 N^5+1967 N^4+697 N^3-4 N^2-48 N+8, \\
P_{86}&=&-891 
N^7-1782 N^6-3712 N^5-3058 N^4+6775 N^3
\nonumber\\&&
+7144 N^2+276 N-144, \\
P_{87}&=&9 N^7+18 N^6-124 N^5-109 N^4+199 N^3-191 N^2+138 N+72, \\
P_{88}&=&69 N^7+138 N^6-667 N^5-541 N^4+952 N^3-1277 N^2+990 N+432, \\
P_{89}&=&95 N^7+378 N^6+853 N^5+1832 N^4+2190 N^3+364 N^2-780 N-432,\\
P_{90}&=&251 N^7+1335 N^6+1745 N^5+243 N^4-529 N^3-4161 N^2-5400 
N-756,\\
P_{91}&=&N^8+427 N^7+2161 N^6+4081 N^5+3554 N^4
\nonumber\\&&
+1404 N^3+228 N^2+64 N+16,\\
P_{92}&=&2 N^8+10 N^7+22 N^6+36 N^5+29 N^4+4 N^3+33 
N^2+12 N+4,\\
P_{93}&=&2 N^8+29 N^7+135 N^6+297 N^5+333 N^4+204 N^3+28 
N^2-44 N-24,\\
P_{94}&=&3 N^8+33 N^7+149 N^6+267 N^5+196 N^4+104 N^3+64 
N^2-88 N-48,\\
P_{95}&=&12 N^8+52 N^7+60 N^6-25 N^4-2 N^3+3 N^2+8 N+4, \\
P_{96}&=&36 N^8+348 N^7+1210 N^6+2229 N^5+2168 N^4+505 N^3-424 N^2
\nonumber\\&&
-68 
N+48,\\
P_{97}&=&111 N^8+480 N^7+286 N^6-468 N^5+82 N^4+246 N^3+295 
N^2
\nonumber\\&&
+228 N+252,\\
P_{98}&=&201 N^8+840 N^7+565 N^6-699 N^5-344 N^4-645 
N^3-314 N^2
\nonumber\\&&
+324 N+360, \\
P_{99}&=&296 N^8+1184 N^7+2744 N^6+5900 
N^5+4088 N^4+476 N^3+9477 N^2
\nonumber\\&&
+4725 N+702,\\
P_{100}&=&-7299 N^{10}-39375 N^9-79900 N^8-85198 N^7-17323 N^6+129917 N^5
\nonumber\\&&
+137090 N^4+25904 N^3+12072 N^2+30672 N+8640, \\
P_{101}&=&-149 N^{10}-793 
N^9-1404 N^8-1170 N^7-1341 N^6-1221 N^5+1710 N^4
\nonumber\\&&
+2800 N^3+2256 N^2+368 N-32, \\
P_{102}&=&4 N^{10}+22 N^9+45 N^8+36 N^7-11 N^6-15 
N^5+25 N^4-41 N^3-21 N^2
\nonumber\\&&
-16 N-4, \\
P_{103}&=&10 N^{10}+62 N^9+403 
N^8+1523 N^7+2997 N^6+3197 N^5
\nonumber\\&&
+1812 N^4+478 N^3+46 N^2+24 N+8, \\
P_{104}&=&26 N^{10}+132 N^9+159 N^8-351 N^7-877 N^6+531 N^5+1820 
N^4-300 N^3
\nonumber\\&&
-252 N^2-192 N-48, \\
P_{105}&=&28 N^{10}+139 N^9+444 N^8+803 
N^7+451 N^6+3 N^5+490 N^4+219 N^3
\nonumber\\&&
+51 N^2-60 N-12, \\
P_{106}&=&435 N^{10}+2391 N^9+6946 N^8+11512 N^7+4822 N^6-7016 N^5-5369 N^4
\nonumber\\&&
-6743 N^3-2406 N^2-1764 N-216, \\
P_{107}&=&531 N^{10}+2799 N^9+4124 N^8+446 
N^7-3445 N^6-5245 N^5+4358 N^4
\nonumber\\&&
+18128 N^3-1968 N^2-10800 N-4320, \\
P_{108}&=&939 N^{10}+4893 N^9+5386 N^8-5198 N^7-10400 N^6-17636 
N^5-18137 N^4
\nonumber\\&&
+7177 N^3-21672 N^2-14112 N-4104, \\
P_{109}&=&1773 
N^{10}+9153 N^9+14204 N^8+2930 N^7-9151 N^6-8431 N^5-250 N^4
\nonumber\\&&
+13772 N^3+1920 N^2-8928 N-4320, \\
P_{110}&=&-23 N^{11}-92 N^{10}-53 N^9+322 
N^8+465 N^7-348 N^6-929 N^5-384 N^4
\nonumber\\&&
+132 N^3+102 N^2+32 N+8, \\
P_{111}&=&87 N^{12}+490 N^{11}+949 N^{10}+368 N^9-1285 N^8-2214 
N^7-1591 N^6
\nonumber\\&&
-126 N^5+644 N^4-86 N^3-268 N^2-184 N-48, \\
P_{112}&=&385 N^{12}+2182 N^{11}+4181 N^{10}+1458 N^9-5589 N^8-8414 N^7-5041 
N^6
\nonumber\\&&
-1754 N^5-760 N^4-176 N^3+152 N^2+224 N+96, \\
P_{113}&=&1623 
N^{12}+9602 N^{11}+20093 N^{10}+15520 N^9-3305 N^8-13494 N^7-5099 
N^6
\nonumber\\&&
+9414 N^5+10456 N^4+5270 N^3+1624 N^2+40 N-96 .
\end{eqnarray}

\subsection{$A_{gg,Q}^{S(3)}$ in $N$ space}
\begin{eqnarray}
A_{gg,Q} &=&
	\frac{4  a_s   L_M   \textcolor{blue}{T_F} }{3}
+ a_s ^2 \Biggl\{
	\frac{16  L_M ^2  \textcolor{blue}{T_F ^2} }{9}
	+ \textcolor{blue}{C_F T_F} \biggl[
                \frac{4  L_M  P_{147}}{N^3 (1+N)^3}
                +\frac{P_{166}}{N^4 (1+N)^4}
\nonumber\\&&		
                +\frac{4  L_M ^2 (N-1) (2+N)}{N^2 (1+N)^2}
        \biggr]
+ \textcolor{blue}{C_A T_F} \biggl\{
                \frac{2 P_{155}}{27 N^3 (1+N)^3}
                + L_M  \biggl[
                        \frac{16 P_{134}}{9 N^2 (1+N)^2}
                        -\frac{80}{9} S_ 1
                \biggr]
\nonumber\\&&		
                + L_M ^2 \biggl[
                        \frac{16}{3 N (1+N)}
                        -\frac{8}{3} S_ 1
                \biggr]
                -\frac{4 (47+56 N) S_ 1}{27 (1+N)}
	\biggr\}
	\Biggr\}
+ a_s ^3 \Biggl\{
	\frac{64  L_M ^3  }{27} \textcolor{blue}{T_F ^3}
\nonumber\\&&	
	+ \textcolor{blue}{C_F T_F^2} \biggl\{
                \frac{2 P_{177}}{9 N^5 (1+N)^5}
                +\frac{80  L_M ^3 (N-1) (2+N)}{9 N^2 (1+N)^2}
                + L_M ^2 \biggl[
                        \frac{8 P_{153}}{9 N^3 (1+N)^3}
\nonumber\\&&			
                        +\frac{32 (N-1) (2+N) S_ 1}{3 N^2 (1+N)^2}
                \biggr]
                + L_M  \biggl[
                        -\frac{8 P_{173}}{27 N^4 (1+N)^4}
\nonumber\\&&			
                        +\frac{32 (N-1) (2+N) \big(
                                -6-8 N+N^2\big)}{9 N^3 (1+N)^3} S_ 1
                        +\frac{16 (N-1) (2+N) S_ 1^2}{3 N^2 (1+N)^2}
\nonumber\\&&			
                        -\frac{16 (N-1) (2+N) S_ 2}{N^2 (1+N)^2}
                \biggr]
                +\biggl[
                        -\frac{8 P_{154}}{9 N^3 (1+N)^3}
                        -
                        \frac{16 (N-1) (2+N) S_ 1}{3 N^2 (1+N)^2}
                \biggr] \zeta_2
\nonumber\\&&		
                -\frac{80 (N-1) (2+N) \zeta_3}{9 N^2 (1+N)^2}
	\biggr\}
+ \textcolor{blue}{C_F N_F T_F^2} \biggl\{
                \frac{2 P_{178}}{81 N^5 (1+N)^5}
                +\frac{64  L_M ^3 (N-1) (2+N)}{9 N^2 (1+N)^2}
\nonumber\\&&		
                + L_M  \biggl[
                        -\frac{4 P_{171}}{9 N^4 (1+N)^4}
                        -\frac{32 (N-1) (2+N) \big(
                                4+6 N+N^2\big)}{3 N^3 (1+N)^3} S_ 1
\nonumber\\&&				
                        +\frac{16 (N-1) (2+N) S_ 1^2}{N^2 (1+N)^2}
                        -\frac{80 (N-1) (2+N) S_ 2}{3 N^2 (1+N)^2}
                \biggr]
\nonumber\\&&		
                +\biggl[
                        \frac{32 (N-1) (2+N) \big(
                                22+41 N+28 N^2\big)}{27 N^2 (1+N)^4}
                        +\frac{16 (N-1) (2+N) S_ 2}{3 N^2 (1+N)^2}
                \biggr] S_ 1
\nonumber\\&&		
                -\frac{16 (N-1) (2+N) (2+5 N)}{9 N^2 (1+N)^3} S_ 1^2
                +\frac{16 (N-1) (2+N) S_ 1^3}{9 N^2 (1+N)^2}
\nonumber\\&&		
                -\frac{16 (N-1) (2+N) (2+5 N)}{9 N^2 (1+N)^3} S_ 2
                +\frac{32 (N-1) (2+N) S_ 3}{9 N^2 (1+N)^2}
                +\biggl[
                        \frac{4 P_{162}}{9 N^3 (1+N)^3}
\nonumber\\&&			
                        +\frac{16 (N-1) (2+N) S_ 1}{3 N^2 (1+N)^2}
                \biggr] \zeta_2
                -\frac{64 (N-1) (2+N) \zeta_3}{9 N^2 (1+N)^2}
	\biggr\}
+ \textcolor{blue}{C_A^2 T_F}  \biggl\{
                -\frac{4 P_{174}}{243 N^4 (1+N)^4}
\nonumber\\&&		
                + L_M ^3 \biggl[
                        -\frac{352}{27 N (1+N)}
                        +\frac{176}
                        {27} S_ 1
                \biggr]
                + L_M ^2 \Biggl[
                        -
                        \frac{2 P_{149}}{9 N^3 (1+N)^3}
                        +\biggl[
                                -\frac{8 P_{140}}{9 N^2 (1+N)^2}
\nonumber\\&&				
                                +\frac{64}{3} S_ 2
                        \biggr] S_ 1
                        -\frac{128 S_ 2}{3 N (1+N)}
                        +\frac{32}{3} S_ 3
                        +\biggl[
                                -\frac{128}{3 N (1+N)}
                                +\frac{64}{3} S_ 1
                        \biggr] S_{-2}
                        +\frac{32}{3} S_{-3}
\nonumber\\&&			
                        -\frac{64}{3} S_ {-2,1}
                \Biggr]
                + L_M  \Biggl[
                        \frac{16 S_ 2 P_{133}}{9 N^2 (1+N)^2}
                        +\frac{16 S_{-3} P_{139}}{9 N^2 (1+N)^2}
                        -\frac{32 S_{-2,1} P_{139}}{9 N^2 (1+N)^2}
                        +\frac{8 S_ 3 P_{142}}{9 N^2 (1+N)^2}
\nonumber\\&&			
                        +\frac{P_{183}}{81 (N-1) N^5 (1+N)^5 (2+N)}
                        +\biggl[
                                -\frac{4 P_{176}}{81 (N-1) N^4 
(1+N)^4 (2+N)}
\nonumber\\&&
                                +\frac{640}{9} S_ 2
                                -\frac{32}{3} S_ 3
                        \biggr] S_ 1
                        +\biggl[
                                -\frac{16 P_{163}}{9 (N-1) N^3 
(1+N)^3 (2+N)}
\nonumber\\&&
                                +\frac{32 P_{157}}{9 (N-1) N^2 
(1+N)^2 (2+N)} S_ 1
                        \biggr] S_{-2}
                        +\frac{32}{3} S_ {-2}^2
\nonumber\\&&			
                        +\biggl[
                                \frac{64 \big(
                                        -3+2 N+2 N^2\big)}{N^2 (1+N)^2}
                                -64 S_ 1
                        \biggr] \zeta_3
                \Biggr]
                -\frac{8 \big(
                        2339+4876 N+2834 N^2\big)}{243 (1+N)^2} S_ 1
                -\frac{44 S_ 1^2}{9 (1+N)}
\nonumber\\&&		
                +\frac{44 (1+2 N) S_ 2}{9 (1+N)}
                +\Biggl[
                        \frac{4 P_{160}}{27 N^3 (1+N)^3}
                        +\big(
                                \frac{16 \big(
                                        36+72 N+N^2+2 
N^3+N^4\big)}{27 N^2 (1+N)^2}
                                -
                                \frac{32}{3} S_ 2
                        \big) S_ 1
\nonumber\\&&			
                        +\frac{64 S_ 2}{3 N (1+N)}
                        -\frac{16}{3} S_ 3
                        +\biggl[
                                \frac{64}{3 N (1+N)}
                                -\frac{32}{3} S_ 1
                        \biggr] S_{-2}
                        -\frac{16}{3} S_{-3}
                        +\frac{32}{3} S_ {-2,1}
                \Biggr] \zeta_2
\nonumber\\&&		
                +\biggl[
                        \frac{352}{27 N (1+N)}
                        -\frac{176}{27} S_ 1
                \biggr] \zeta_3
	\biggr\}
	+ \textcolor{blue}{C_A N_F T_F^2} \biggl\{
                \frac{16 P_{172}}{243 N^4 (1+N)^4}
                + L_M  \biggl[
                        -\frac{16 S_ 1 P_{145}}{81 N^2 (1+N)^2}
\nonumber\\&&			
                        -\frac{4 P_{164}}{81 N^3 (1+N)^3}
                \biggr]
                + L_M ^3 \biggl[
                        \frac{128}{27 N (1+N)}
                        -\frac{64}{27} S_ 1
                \biggr]
                +\frac{32 \big(
                        283+584 N+328 N^2\big)}{243 (1+N)^2} S_ 1
\nonumber\\&&			
                +\frac{16 S_ 1^2}{9 (1+N)}
                -\frac{16 (1+2 N) S_ 2}{9 (1+N)}
                +\biggl[
                        -\frac{4 P_{137}}{27 N^2 (1+N)^2}
                        +\frac{160}{27} S_ 1
                \biggr] \zeta_2
\nonumber\\&&		
                +\biggl[
                        -\frac{128}{27 N (1+N)}
                        +\frac{64}{27} S_ 1
                \biggr] \zeta_3
	\biggr\}
+ \textcolor{blue}{C_A T_F^2} \biggl\{
                -\frac{8 P_{167}}{81 N^4 (1+N)^4}
                + L_M  \biggl[
                        -\frac{8 S_ 1 P_{141}}{9 N^2 (1+N)^2}
\nonumber\\&&			
                        -\frac{2 P_{159}}{27 N^3 (1+N)^3}
                \biggr]
                + L_M ^2 \biggl[
                        \frac{8 P_{143}}{27 N^2 (1+N)^2}
                        -\frac{640}{27} S_ 1
                \biggr]
                + L_M ^3 \biggl[
                        \frac{448}{27 N (1+N)}
                        -
                        \frac{224}{27} S_ 1
                \biggr]
\nonumber\\&&		
                +\frac{16 \big(
                        283+584 N+328 N^2\big)}{81 (1+N)^2} S_ 1
                +\frac{8 S_ 1^2}{3 (1+N)}
                -\frac{8 (1+2 N) S_ 2}{3 (1+N)}
\nonumber\\&&		
                +\biggl[
                        -\frac{4 P_{144}}{27 N^2 (1+N)^2}
                        +\frac{560}{27} S_ 1
                \biggr] \zeta_2
                +\biggl[
                        -\frac{448}{27 N (1+N)}
                        +\frac{224}{27} S_ 1
                \biggr] \zeta_3
	\biggr\}
\nonumber\\&&	
	+ \textcolor{blue}{C_F^2 T_F}  \biggl\{
                \frac{8 S_ 3 P_{136}}{3 N^3 (1+N)^3}
                +\frac{4 S_ 2 P_{151}}{N^4 (1+N)^4}
                +\frac{P_{180}}{N^6 (1+N)^6}
\nonumber\\&&		
                + L_M ^3 \biggl[
                        -\frac{4 (N-1) (2+N) \big(
                                2+3 N+3 N^2\big)}{3 N^3 (1+N)^3}
                        +\frac{16 (N-1) (2+N) S_ 1}{3 N^2 (1+N)^2}
                \biggr]
\nonumber\\&&		
                + L_M ^2 \biggl[
                        -\frac{8 (N-1) (2+N) \big(
                                -2-3 N+N^3+2 N^4\big)}{N^4 (1+N)^4}
                        +\frac{8 (N-1)^2 (2+N) (2+3 N)}{N^3 (1+N)^3} 
S_ 1
\nonumber\\&&
                        -\frac{16 (N-1) (2+N) S_ 2}{N^2 (1+N)^2}
                \biggr]
                + L_M  \Biggl[
                        -\frac{4 S_ 2 P_{138}}{N^3 (1+N)^3}
                        -\frac{2 P_{181}}{(N-1) N^5 (1+N)^5 (2+N)}
\nonumber\\&&			
                        +\biggl[
                                -\frac{8 P_{152}}{N^4 (1+N)^4}
                                +\frac{24 (N-1) (2+N) S_ 2}{N^2 
(1+N)^2}
                        \biggr] S_ 1
                        +\frac{4 \big(
                                -6-13 N+3 N^3\big)}{N^2 (1+N)^3} S_ 1^2
\nonumber\\&&				
                        -\frac{8 (N-1) (2+N) S_ 1^3}{3 N^2 (1+N)^2}
                        +\frac{16 \big(
                                14+5 N+5 N^2\big)}
                        {3 N^2 (1+N)^2} S_ 3
                        +\biggl[
                                -
                                \frac{32 \big(
                                        10+N+N^2\big)}{(N-1) N (1+N) 
(2+N)}
\nonumber\\&&
                                +\frac{256 S_ 1}{N^2 (1+N)^2}
                        \biggr] S_{-2}
                        +\frac{128 S_{-3}}{N^2 (1+N)^2}
                        -\frac{32 (N-1) (2+N) S_{2,1}}{N^2 (1+N)^2}
                        -\frac{256 S_{-2,1}}{N^2 (1+N)^2}
\nonumber\\&&			
                        -\frac{96 \big(
                                2+N+N^2\big) \zeta_3}{N^2 (1+N)^2}
                \Biggr]
                +\biggl[
                        -\frac{8 P_{132}}{N^3 (1+N)^3}
                        +\frac{8 \big(
                                -2+3 N+3 N^2\big)}{N^3 (1+N)^2} S_ 2
\nonumber\\&&				
                        -\frac{16 (N-1) (2+N) S_ 3}{3 N^2 (1+N)^2}
                        -\frac{32 (N-1) (2+N) S_{2,1}}{N^2 (1+N)^2}
                \biggr] S_ 1
\nonumber\\&&		
                +\biggl[
                        \frac{4 \big(
                                -36-22 N-6 N^2+N^3\big)}{N^3 (1+N)^2}
                        -\frac{4 (N-1) (2+N) S_ 2}{N^2 (1+N)^2}
                \biggr] S_ 1^2
\nonumber\\&&		
                +\frac{8 \big(
                        -2+3 N+3 N^2\big)}{3 N^3 (1+N)^2} S_ 1^3
                -\frac{2 (N-1) (2+N) S_ 1^4}{3 N^2 (1+N)^2}
                -\frac{2 (N-1) (2+N) S_ 2^2}{N^2 (1+N)^2}
\nonumber\\&&		
                +\frac{12 (N-1) (2+N) S_ 4}{N^2 (1+N)^2}
                -\frac{32 (2+N) S_{2,1}}{N^3 (1+N)}
                -\frac{32 (N-1) (2+N) S_{3,1}}{N^2 (1+N)^2}
\nonumber\\&&		
                +\frac{64 (N-1) (2+N) S_{2,1,1}}{N^2 (1+N)^2}
                +\biggl[
			128 \ln(2)
                        -\frac{2 P_{170}}{N^4 (1+N)^4}
\nonumber\\&&			
                        -\frac{4 (N-1) (2+N) \big(
                                -4-3 N+3 N^2\big)}{N^3 (1+N)^3} S_ 1
                        -\frac{4 (N-1) (2+N) S_ 1^2}{N^2 (1+N)^2}
\nonumber\\&&			
                        +
                        \frac{12 (N-1) (2+N) S_ 2}{N^2 (1+N)^2}
                \biggr] \zeta_2
                +\biggl[
                        -\frac{4 P_{158}}{3 N^3 (1+N)^3}
                        -\frac{16 (N-1) (2+N) S_ 1}{3 N^2 (1+N)^2}
                \biggr] \zeta_3
	\biggr\}
\nonumber\\&&	
	+\textcolor{blue}{C_A C_F T_F} \biggl\{
                -\frac{4 S_ 2 P_{148}}{N^4 (1+N)^4}
                +\frac{P_{179}}{18 N^6 (1+N)^6}
\nonumber\\&&		
                + L_M ^3 \biggl[
                        -\frac{8 (N-1) (2+N) \big(
                                -12+11 N+11 N^2\big)}{9 N^3 (1+N)^3}
                        -\frac{16 (N-1) (2+N) S_ 1}{3 N^2 (1+N)^2}
                \biggr]
\nonumber\\&&		
                + L_M ^2 \biggl[
                        -\frac{8 S_ 1 P_{150}}{3 N^3 (1+N)^3}
                        -\frac{2 P_{169}}{9 N^4 (1+N)^4}
                        -\frac{16 (N-1) (2+N) S_ 2}{N^2 (1+N)^2}
\nonumber\\&&			
                        -\frac{32 (N-1) (2+N) S_{-2}}{N^2 (1+N)^2}
                \biggr]
                + L_M  \Biggl[
                        \frac{4 S_ 2 P_{135}}{N^3 (1+N)^3}
                        +\frac{8 P_{182}}{27 (N-1) N^5 (1+N)^5 
(2+N)}
\nonumber\\&&
                        +\biggl[
                                -\frac{8 P_{175}}{9 (N-1) N^4 
(1+N)^4 (2+N)}
                                -\frac{40 (N-1) (2+N) S_ 2}{N^2 
(1+N)^2}
                        \biggr] S_ 1
\nonumber\\&&			
                        -\frac{4 \big(
                                -12-16 N+5 N^2+11 N^3\big)}{3 N^3 
(1+N)^2} S_ 1^2
                        +\frac{8 (N-1) (2+N) S_ 1^3}{3 N^2 (1+N)^2}
                        -\frac{16 \big(
                                26+5 N+5 N^2\big)}{3 N^2 (1+N)^2} S_3
\nonumber\\&&				
                        +\biggl[
                                -\frac{16 P_{146}}{(N-1) N^2 (1+N)^3 
(2+N)}
                                +\frac{32 P_{130}}{(N-1) N^2 (1+N)^2 
(2+N)}
                                 S_ 1
                        \biggr] S_{-2}
                        +
\nonumber\\&&			
                        \frac{16 \big(
                                -22+5 N+5 N^2\big)}{N^2 (1+N)^2} S_{-3}
                        +\frac{32 (N-1) (2+N) S_{2,1}}{N^2 (1+N)^2}
                        -\frac{32 \big(
                                -14+N+N^2\big) S_{-2,1}}{N^2 (1+N)^2}
\nonumber\\&&				
                        +\biggl[
                                -\frac{32 (-3+N) (4+N)}{N^2 (1+N)^2}
                                +64 S_ 1
                        \biggr] \zeta_3
                \Biggr]
                +\biggl[
                        -\frac{2 P_{165}}{9 N^2 (1+N)^5}
                        +\frac{8 \big(
                                -13+3 N^2\big)}{N^2 (1+N)^3} S_ 2
\nonumber\\&&				
                        +\frac{160 (N-1) (2+N) S_ 3}{3 N^2 (1+N)^2}
                        -\frac{64 (N-1) (2+N) S_{-2,1}}{N^2 (1+N)^2}
                \biggr] S_ 1
                +\biggl[
                        -\frac{4 P_{131}}{N^2 (1+N)^4}
\nonumber\\&&			
                        +\frac{20 (N-1) (2+N) S_ 2}{N^2 (1+N)^2}
                \biggr] S_ 1^2
                -\frac{8 \big(
                        5+4 N+N^2\big)}{3 N^2 (1+N)^3} S_ 1^3
                +\frac{2 (N-1) (2+N) S_ 1^4}{3 N^2 (1+N)^2}
\nonumber\\&&		
                +\frac{2 (N-1) (2+N) S_ 2^2}{N^2 (1+N)^2}
                -\frac{64 \big(
                        -6+8 N+7 N^2+N^3\big)}{3 N^3 (1+N)^3} S_ 3
                +\frac{36 (N-1) (2+N) S_ 4}{N^2 (1+N)^2}
\nonumber\\&&		
                +\biggl[
                        -\frac{32 (2+N) \big(
                                3+N^2\big)}{N^2 (1+N)^4}
                        +\frac{64 (N-1) (2+N) S_ 1}{N^2 (1+N)^3}
                        +\frac{32 (N-1) (2+N) S_ 1^2}{N^2 (1+N)^2}
\nonumber\\&&			
                        +\frac{32 (N-1) (2+N) S_ 2}{N^2 (1+N)^2}
                \biggr] S_{-2}
                +\biggl[
                        \frac{32 (N-1) (2+N)}{N^2 (1+N)^3}
                        +\frac{32 (N-1) (2+N) S_ 1}{N^2 (1+N)^2}
                \biggr] S_{-3}
\nonumber\\&&		
                +\frac{16 (N-1) (2+N) S_{-4}}{N^2 (1+N)^2}
                -\frac{16 (N-1) (2+N) S_{3,1}}{N^2 (1+N)^2}
                -\frac{64 (N-1) (2+N) S_{-2,1}}{N^2 (1+N)^3}
\nonumber\\&&		
                -\frac{32 (N-1) (2+N) S_{-2,2}}{N^2 (1+N)^2}
                -\frac{32 (N-1) (2+N) S_{-3,1}}{N^2 (1+N)^2}
                -\frac{16 (N-1) (2+N) S_{2,1,1}}{N^2 (1+N)^2}
\nonumber\\&&		
                +\frac{64 (N-1) (2+N) S_{-2,1,1}}{N^2 (1+N)^2}
                +\biggl[
			-64 \ln(2)
                        -\frac{4 S_ 1 P_{161}}{3 N^3 (1+N)^3}
                        +\frac{4 P_{168}}{9 N^4 (1+N)^4}
\nonumber\\&&			
                        +\frac{4 (N-1) (2+N) S_ 1^2}{N^2 (1+N)^2}
                        +\frac{12 (N-1) (2+N) S_ 2}{N^2 (1+N)^2}
                        +\frac{24 (N-1) (2+N) S_{-2}}{N^2 (1+N)^2}
                \biggr] \zeta_2
\nonumber\\&&		
                +\biggl[
                        \frac{8 P_{156}}{9 N^3 (1+N)^3}
                        +\frac{16 (N-1) (2+N) S_ 1}{3 N^2 (1+N)^2}
                \biggr] \zeta_3
	\biggr\}
	-\frac{64}{27}  \textcolor{blue}{T_F ^3} \zeta_3
+a_{gg,Q}^{(3)}
\Biggr\} ~,
\end{eqnarray}
where the polynomials $P_i$ read
\begin{eqnarray}
P_{130} &=& N^4+2 N^3-7 N^2-8 N+28, \\ 
P_{131} &=& N^4+2 N^3-5 N^2-12 N+2, \\ 
P_{132} &=& 2 N^4-4 N^3-3 N^2+20 N+12, \\ 
P_{133} &=& 3 N^4+6 N^3-89 N^2-92 N+12, \\ 
P_{134} &=& 3 N^4+6 N^3+16 N^2+13 N-3, \\ 
P_{135} &=& 3 N^4+32 N^3+65 N^2-16 N-60, \\ 
P_{136} &=& 3 N^4+48 N^3+123 N^2+98 N+8, \\ 
P_{137} &=& 9 N^4+18 N^3+113 N^2+104 N-24, \\ 
P_{138} &=& 11 N^4+36 N^3+43 N^2+46 N+8, \\ 
P_{139} &=& 20 N^4+40 N^3+11 N^2-9 N+54, \\ 
P_{140} &=& 23 N^4+46 N^3+23 N^2+96 N+48, \\ 
P_{141} &=& 40 N^4+74 N^3+25 N^2-9 N+16, \\ 
P_{142} &=& 40 N^4+80 N^3+73 N^2+33 N+54, \\ 
P_{143} &=& 63 N^4+126 N^3+271 N^2+208 N-48, \\
P_{144} &=& 99 N^4+198 N^3+463 N^2+364 N-84, \\
P_{145} &=& 136 N^4+254 N^3+37 N^2-81 N+144, \\
P_{146} &=& 3 N^5+7 N^4-29 N^3-51 N^2-2 N-8, \\ 
P_{147} &=& N^6+3 N^5+5 N^4+N^3-8 N^2+2 N+4, \\ 
P_{148} &=& N^6+18 N^5+63 N^4+84 N^3+30 N^2-64 N-16, \\ 
P_{149} &=& 3 N^6+9 N^5-163 N^4-341 N^3+164 N^2-432 N-192, \\ 
P_{150} &=& 3 N^6+9 N^5+20 N^4+25 N^3-11 N^2-46 N-12, \\ 
P_{151} &=& 3 N^6+30 N^5+107 N^4+124 N^3+48 N^2+20 N+8, \\ 
P_{152} &=& 6 N^6+23 N^5-14 N^4-121 N^3-114 N^2-20 N+8, \\ 
P_{153} &=& 15 N^6+45 N^5+49 N^4-13 N^3-64 N^2+40 N+48, \\
P_{154} &=& 15 N^6+45 N^5+56 N^4+N^3-68 N^2+29 N+42, \\ 
P_{155} &=& 15 N^6+45 N^5+374 N^4+601 N^3+161 N^2-24 N+36, \\ 
P_{156} &=& 18 N^6+54 N^5+65 N^4+40 N^3-23 N^2-34 N+24, \\ 
P_{157} &=& 20 N^6+60 N^5+11 N^4-78 N^3-13 N^2+36 N-108, \\ 
P_{158} &=& 24 N^6+72 N^5+69 N^4+18 N^3+N^2+4 N+4, \\ 
P_{159} &=& 27 N^6+81 N^5-1247 N^4-2341 N^3-720 N^2+32 N-240, \\ 
P_{160} &=& 27 N^6+81 N^5+148 N^4+161 N^3+253 N^2-390 N-144, \\ 
P_{161} &=& 30 N^6+90 N^5+79 N^4+8 N^3+23 N^2+70 N+12, \\ 
P_{162} &=& 63 N^6+189 N^5+157 N^4+35 N^3+80 N^2+4 N-24, \\ 
P_{163} &=& 95 N^6+285 N^5+92 N^4-291 N^3-97 N^2+96 N-36, \\ 
P_{164} &=& 297 N^6+891 N^5-461 N^4-2119 N^3-872 N^2-96 N-432, \\ 
P_{165} &=& 233 N^7+1093 N^6+1970 N^5+1538 N^4-167 N^3
\nonumber\\&&
-2143 N^2-2412 N-288, \\ 
P_{166} &=& -15 N^8-60 N^7-82 N^6-44 N^5-15 N^4-4 N^2-12 N-8, \\ 
P_{167} &=& 3 N^8+12 N^7+2080 N^6+5568 N^5+4602 N^4+1138 N^3
\nonumber\\&&
-3 N^2-36 N-108, \\ 
P_{168} &=& 15 N^8+60 N^7+242 N^6+417 N^5+344 N^4+285 N^3
\nonumber\\&&
+185 N^2+456 N+108, \\ 
P_{169} &=& 33 N^8+132 N^7-82 N^6-840 N^5-571 N^4+564 N^3+308 N^2
\nonumber\\&&
+984 N+288, \\
P_{170} &=& 40 N^8+160 N^7+205 N^6+61 N^5+18 N^4+113 N^3+75 N^2
\nonumber\\&&
-20 N-12, \\ 
P_{171} &=& 67 N^8+268 N^7+194 N^6-508 N^5-533 N^4+480 N^3
\nonumber\\&&
+616 N^2+344 N+144, \\ 
P_{172} &=& 126 N^8+504 N^7-1306 N^6-5052 N^5-4473 N^4-1138 N^3
\nonumber\\&&
+3 N^2+36 N+108, \\ 
P_{173} &=& 219 N^8+876 N^7+1142 N^6+288 N^5-217 N^4+240 N^3
\nonumber\\&&
+410 N^2+366 N+180, \\ 
P_{174} &=& 1386 N^8+5544 N^7-11270 N^6-46284 N^5-39915 N^4
\nonumber\\&&
-9422 N^3+33 N^2+396 N+1188, \\ 
P_{175} &=& 15 N^{10}+75 N^9+2 N^8-469 N^7-506 N^6+524 N^5
\nonumber\\&&
+781 N^4+26 N^3-1192 N^2-1128 N-432, \\ 
P_{176} &=& 310 N^{10}+1748 N^9+4811 N^8+14192 N^7+24974 N^6+3194 N^5
\nonumber\\&&
-29393 N^4-16866 N^3+8694 N^2+7128 N+1944, \\ 
P_{177} &=& 391 N^{10}+1955 N^9+3622 N^8+3046 N^7+1595 N^6+1327 N^5
\nonumber\\&&
+1152 N^4+216 N^3-288 N^2-360 N-144, \\ 
P_{178} &=& 1593 N^{10}+7965 N^9+11578 N^8+1594 N^7-1379 N^6+12793 N^5
\nonumber\\&&
+17152 N^4+4432 N^3-1728 N^2-2160 N-864, \\ 
P_{179} &=& -3135 N^{12}-18810 N^{11}-42713 N^{10}-44692 N^9-22145 N^8
\nonumber\\&&
-9290 N^7-8167 N^6-4136 N^5-960 N^4+11232 N^3
\nonumber\\&&
+6720 N^2+3360 N+576, \\ 
P_{180} &=& -39 N^{12}-234 N^{11}-521 N^{10}-492 N^9-85 N^8-42 N^7-883 N^6
\nonumber\\&&
-1660 N^5-1324 N^4-492 N^3-52 N^2+48 N+16, \\ 
P_{181} &=& N^{12}+6 N^{11}-3 N^{10}-58 N^9-21 N^8+222 N^7+609 N^6
\nonumber\\&&
+1144 N^5+1122 N^4+142 N^3-180 N^2+40 N+48, \\ 
P_{182} &=& 276 N^{12}+1656 N^{11}+3334 N^{10}+869 N^9-6591 N^8-7395 N^7
\nonumber\\&&
+7452 N^6+13479 N^5+3167 N^4+7303 N^3+1110 N^2-5004 N
\nonumber\\&&
-2376, \\ 
P_{183} &=& 2493 N^{12}+14958 N^{11}+42317 N^{10}+75910 N^9+45511 N^8-60782 N^7
\nonumber\\&&
-29777 N^6+17194 N^5-130384 N^4-115536 N^3+25776 N^2
\nonumber\\&&
+24192 N+5184 .
\end{eqnarray}

\subsection{$A_{qq,Q}^{PS(3)}$ in $z$ space}
\begin{eqnarray}
	\lefteqn{A_{qq,Q}^{PS}(x) =}
\nonumber\\&&	
	a_s^3 \Biggl\{
		 \textcolor{blue}{C_F N_F T_F^2} \biggl\{
                 L_M ^3 \biggl[
                        \frac{160}{9} (x-1)
                        -\frac{64}{9} (1+x) \HA_0
                \biggr]
\nonumber\\&&		
                + L_M ^2 \biggl[
                        -\frac{32}{9} (x-1) \big(
                                -1+15 \HA_1\big)
                        +(1+x) \Bigl(
                                -\frac{32}{3} \HA_0^2
                                +\frac{64}{3} \HA_{0,1}
                                -\frac{64}{3} \zeta_2
                        \Bigr)
\nonumber\\&&			
                        -\frac{32}{9} (7+x) \HA_0
                \biggr]
                + L_M  \Biggl[
                        (1+x) \Bigl(
                                -\frac{176}{9} \HA_0^2
                                -\frac{32}{9} \HA_0^3
                                +\frac{128}{3} \HA_{0,0,1}
                                -\frac{128}{3} \HA_{0,1,1}
                        \Bigr)
\nonumber\\&&			
                        -\frac{32}{27} (55+31 x) \HA_0
                        +(x-1) \Bigl(
                                \frac{5312}{27}-\frac{64}{9} 
\HA_1+\frac{160}{3} \HA_1^2\Bigr)
                        +\frac{64}{9} (7+x) \HA_{0,1}
\nonumber\\&&			
                        +\biggl[
                                -\frac{64}{9} (7+x)
                                -\frac{128}{3} (1+x) \HA_0
                        \biggr] \zeta_2
                \Biggr]
                +(1+x) \Bigl(
                        -\frac{64}{3} \HA_{0,1,1,1}
                        +\frac{448}{15} \zeta_2^2
                \Bigr)
\nonumber\\&&		
                +\frac{128}{81} (49+82 x) \HA_0
                +(x-1) \Bigl(
                        -\frac{10880}{81}+\frac{3712}{27} 
\HA_1-\frac{64}{9} \HA_1^2+\frac{80}{9} \HA_1^3\Bigr)
                -\frac{128}{27} (11+14 x) \HA_{0,1}
\nonumber\\&&		
                +\frac{64}{9} (2+5 x) \HA_{0,1,1}
                +\biggl[
                        \frac{16}{27} (103+97 x)
                        +(1+x) \Bigl(
                                \frac{176}{9} \HA_0
                                +\frac{16}{3} \HA_0^2
                                -\frac{64}{3} \HA_{0,1}
                        \Bigr)
\nonumber\\&&			
                        +\frac{160}{3} (x-1) \HA_1
                \biggr] \zeta_2
                +\biggl[
                        -\frac{32}{9} (-1+15 x)
                        +\frac{64}{9} (1+x) \HA_0
                \biggr] \zeta_3
	\biggr\}
	+a_{qq,Q}^{PS(3)}
\Biggr\} ~.
\end{eqnarray}

\subsection{$A_{qg,Q}^{S(3)}$ in $z$ space}
\begin{eqnarray}
	\lefteqn{A_{qg,Q} =}
\nonumber\\&&	
	a_s^3 \Biggl\{
	 \textcolor{blue}{ C_F N_F T_F^2} \biggl\{
                \frac{1}{27} (58501-59018 x)
                + L_M ^3 \biggl[
                        \frac{8}{3} (-41+42 x)
\nonumber\\&&			
                        +(-1+2 x) \Bigl(
                                \frac{16}{3} \HA_0^2
                                -\frac{32}{9} \HA_1
                        \Bigr)
                        -\frac{32}{9} (13+19 x) \HA_0
                \biggr]
                + L_M ^2 \biggl[
                        \frac{4}{3} (-285+296 x)
\nonumber\\&&			
                        +(-1+2 x) \Bigl(
                                \frac{32}{3} \HA_0^3
                                +\frac{16}{3} \HA_1^2
                                +\frac{32}{3} \HA_{0,1}
                                -\frac{32}{3} \zeta_2
                        \Bigr)
                        -\frac{8}{9} (425+2 x) \HA_0
                        -16 (7+6 x) \HA_0^2
\nonumber\\&&			
                        -\frac{32}{9} (1+4 x) \HA_1
                \biggr]
                + L_M  \Biggl[
                        \frac{212}{9} (-69+70 x)
                        +(-1+2 x) \Bigl(
                                \frac{20}{3} \HA_0^4
                                -\frac{16}{9} \HA_1^3
                                +\frac{32}{3} \HA_{0,0,1}
\nonumber\\&&				
                                -\frac{32}{3} \HA_{0,1,1}
                                +\frac{32}{3} \zeta_3
                        \Bigr)
                        +\biggl[
                                -\frac{8}{27} (4103+1565 x)
                                +\frac{64}{3} (x-1) \HA_1
                                -\frac{32}{3} (-1+2 x) \HA_{0,1}
                        \biggr] \HA_0
\nonumber\\&&			
                        +\biggl[
                                \frac{4}{3} (-343+10 x)
                                +\frac{16}{3} (-1+2 x) \HA_1
                        \biggr] \HA_0^2
                        -\frac{16}{9} (50+23 x) \HA_0^3
                        -\frac{16}{27} (-28+131 x) \HA_1
\nonumber\\&&			
                        +\frac{32}{9} (-2+7 x) \HA_1^2
                        +\frac{64}{9} (1+4 x) \HA_{0,1}
                        -\frac{64}{9} (-2+7 x) \zeta_2
                \Biggr]
                +(-1+2 x) \Bigl(
                        -\frac{4}{15} \HA_0^5
\nonumber\\&&			
                        +\frac{5248}{81} \HA_1
                        -\frac{896}{27} \HA_{0,1}
                        +\frac{160}{9} \HA_{0,0,1}
                        -\frac{32}{3} \HA_{0,0,0,1}
                        +
                        \frac{16}{3} \zeta_2^2
                \Bigr)
                +\biggl[
                        \frac{16}{81} (7333+4195 x)
\nonumber\\&&			
                        +(-1+2 x) \Bigl(
                                \frac{896}{27} \HA_1
                                -\frac{160}{9} \HA_{0,1}
                                +\frac{32}{3} \HA_{0,0,1}
                        \Bigr)
                \biggr] \HA_0
                +\biggl[
                        \frac{8}{27} (1331+140 x)
\nonumber\\&&			
                        +(-1+2 x) \Bigl(
                                \frac{80}{9} \HA_1
                                -\frac{16}{3} \HA_{0,1}
                        \Bigr)
                \biggr] \HA_0^2
                +\biggl[
                        -\frac{16}{27} (-101+28 x)
                        +\frac{16}{9} (-1+2 x) \HA_1
                \biggr] \HA_0^3
\nonumber\\&&		
                +\frac{4}{9} (13+7 x) \HA_0^4
                +\Biggl[
                        -\frac{4}{3} (-299+305 x)
                        +(-1+2 x) \Bigl(
                                -8 \HA_0^3
                                -\frac{8}{3} \HA_1^2
                                -\frac{32}{3} \HA_{0,1}
                        \Bigr)
\nonumber\\&&			
                        +\biggl[
                                \frac{4}{9} (757+178 x)
                                +\frac{16}{3} (-1+2 x) \HA_1
                        \biggr] \HA_0
                        +\frac{8}{3} (34+25 x) \HA_0^2
                        +\frac{32}{9} (-2+7 x) \HA_1
                \Biggr] \zeta_2
\nonumber\\&&		
                +\biggl[
                        -\frac{8}{3} (-41+42 x)
                        +(-1+2 x) \Bigl(
                                -\frac{16}{3} \HA_0^2
                                +\frac{32}{9} \HA_1
                        \Bigr)
                        +\frac{32}{9} (13+19 x) \HA_0
                \biggr] \zeta_3
	\biggr\}
\nonumber\\&&	
	+ \textcolor{blue}{C_A N_F T_F^2} \biggl\{
                -\frac{8}{27} (-4028+4113 x)
                + L_M ^3 \biggl[
                        \frac{64}{3} (x-1)
                        -\frac{64}{9} (1+x) \HA_0
\nonumber\\&&			
                        +\frac{32}{9} (-1+2 x) \HA_1
                \biggr]
                + L_M ^2 \Biggl[
                        \frac{8}{3} (-65+68 x)
                        +\biggl[
                                -\frac{16}{9} (47+38 x)
                                -\frac{32}{3} (1+2 x) \HA_{-1}
                        \biggr] \HA_0
\nonumber\\&&			
                        -\frac{32}
                        {3} \HA_0^2
                        +
                        \frac{32}{9} (1+4 x) \HA_1
                        -\frac{16}{3} (-1+2 x) \HA_1^2
                        +\frac{32}{3} (1+2 x) \HA_{0,-1}
                        -\frac{32}{3} \zeta_2
                \Biggr]
\nonumber\\&&		
                + L_M  \Biggl[
                        \frac{16}{9} (-514+547 x)
                        +(-1+2 x) \Bigl(
                                \frac{16}{9} \HA_1^3
                                +\frac{32}{3} \HA_{0,0,1}
                                +32 \HA_{0,1,1}
                        \Bigr)
\nonumber\\&&			
                        +\biggl[
                                -\frac{32}{27} (463+211 x)
                                +(-1+2 x) \Bigl(
                                        \frac{16}{3} \HA_1^2
                                        -\frac{32}{3} \HA_{0,1}
                                \Bigr)
                                -\frac{64}{9} (2+7 x) \HA_{-1}
\nonumber\\&&				
                                +\frac{64}{3} (1+2 x) \HA_{0,-1}
                        \biggr] \HA_0
                        +\biggl[
                                -\frac{8}{9} (133+72 x)
                                +\frac{16}{3} (-1+2 x) \HA_1
                                -\frac{32}{3} (1+2 x) \HA_{-1}
                        \biggr] \HA_0^2
\nonumber\\&&			
                        +\frac{16}{9} (-5+6 x) \HA_0^3
                        +\biggl[
                                \frac{16}{27} (-1+95 x)
                                -\frac{64}{3} (-1+2 x) \HA_{0,1}
                        \biggr] \HA_1
                        -\frac{32}{9} (-2+7 x) \HA_1^2
\nonumber\\&&			
                        +\frac{64}{3} x \HA_{0,1}
                        +\frac{64}{9} (2+7 x) \HA_{0,-1}
                        -\frac{64}{3} (1+2 x) \HA_{0,0,-1}
                        +\biggl[
                                -\frac{64}{9} (2+3 x)
\nonumber\\&&				
                                +\frac{32}{3} (-1+2 x) \HA_1
                        \biggr] \zeta_2
                        +32 \zeta_3
                \Biggr]
                +\biggl[
                        \frac{8}{81} (5836+4771 x)
                        -\frac{16}{3} (-1+2 x) \HA_1
                \biggr] \HA_0
\nonumber\\&&		
                -\frac{4}{27} (-649+128 x) \HA_0^2
                +\frac{16}{27} (17+2 x) \HA_0^3
                -\frac{4}{9} (-1+2 x) \HA_0^4
                -\frac{16}{81} (-283+611 x) \HA_1
\nonumber\\&&		
                -\frac{8}{3} (x-1) \HA_1^2
                -\frac{16}
                {9} (3+10 x) \HA_{0,1}
                +\frac{32}{3} x \HA_{0,1,1}
                +\Biggl[
                        -\frac{8}{9} (-231+205 x)
\nonumber\\&&			
                        +\biggl[
                                \frac{8}{9} (97+100 x)
                                +\frac{16}{3} (1+2 x) \HA_{-1}
                        \biggr] \HA_0
                        -\frac{8}{3} (-3+2 x) \HA_0^2
                        -\frac{32}{9} (-2+7 x) \HA_1
\nonumber\\&&			
                        +\frac{8}{3} (-1+2 x) \HA_1^2
                        -\frac{16}{3} (1+2 x) \HA_{0,-1}
                \Biggr] \zeta_2
                +\frac{16}{3} \zeta_2^2
                +\biggl[
                        -\frac{32}{3} (-2+3 x)
                        +\frac{64}{9} (1+x) \HA_0
\nonumber\\&&			
                        -\frac{32}{9} (-1+2 x) \HA_1
                \biggr] \zeta_3
	\biggr\}
	+a_{qg,Q}^{(3)}
\Biggr\}
\end{eqnarray}

\subsection{$A_{Qg}^{S}$ in $z$ space}


\subsection{$A_{gg,Q}^{S(3)}$ in $z$ space}
\begin{eqnarray}
	\lefteqn{A_{gg,Q}^{(\delta)} =}
\nonumber\\&&
	a_s \frac{4 L_M }{3} \textcolor{blue}{T_F}
+a_s^2 \biggl[
	\textcolor{blue}{C_F T_F} (-15+4  L_M )  
	+ \textcolor{blue}{C_A T_F} \frac{2}{9}   (5+24  L_M )  
	+\frac{16  L_M ^2  }{9} \textcolor{blue}{T_F ^2}
\biggr]
\nonumber\\&&
	+a_s^3 \Biggl\{
	\frac{64  L_M ^3  }{27} \textcolor{blue}{T_F ^3}
	+ \textcolor{blue}{C_A N_F T_F^2} \biggl[
                \frac{224}{27}
                -\frac{44  L_M }{3}
                -\frac{4 \zeta_2}{3}
        \biggr]
+ \textcolor{blue}{ C_F N_F T_F^2} \biggl[
                \frac{118}{3}
\nonumber\\&&		
                -\frac{268  L_M }{9}
                +28 \zeta_2
        \biggr]
+ \textcolor{blue}{C_A T_F^2} \biggl[
                -\frac{8}{27}
                -2  L_M 
                +\frac{56  L_M ^2}{3}
                -\frac{44}{3} \zeta_2
        \biggr]
+ \textcolor{blue}{C_F T_F^2} \biggl[
                \frac{782}{9}
\nonumber\\&&		
                -\frac{584  L_M }{9}
                +\frac{40  L_M ^2}{3}
                -\frac{40}{3} \zeta_2
        \biggr]
+ \textcolor{blue}{C_A^2 T_F}  \biggl[
                -\frac{616}{27}
                + L_M  \Bigl(
                        \frac{277}{9}
                        +\frac{16}{3} \zeta_2^2
                        +\frac{160}{9} \zeta_3
                \Bigr)
\nonumber\\&&		
                + L_M ^2 \Bigl(
                        -\frac{2}{3}
                        +\frac{16 \zeta_3}{3}
                \Bigr)
                +\Bigl(
                        4
                        -\frac{8 \zeta_3}{3}
                \Bigr) \zeta_2
        \biggr]
+ \textcolor{blue}{C_F^2 T_F}  \biggl[
                -39
                -2  L_M 
		+16 \Bigl[-5+8 \ln(2) \Bigr] \zeta_2
\nonumber\\&&		
                -32 \zeta_3
        \biggr]
+ \textcolor{blue}{ C_A C_F T_F} \biggl[
                -\frac{1045}{6}
                +\frac{736  L_M }{9}
                -\frac{22  L_M ^2}{3}
		-\frac{4}{3} \Bigl[-5+48 \ln(2) \Bigr] \zeta_2
                +16 \zeta_3
        \biggr]
\nonumber\\&&	
        -\frac{64}{27}  T_F ^3 \zeta_3
	+a_{gg,Q}^{(3),\delta}
\Biggr\} ~, 
\end{eqnarray}

\begin{eqnarray}
	\lefteqn{A_{gg,Q}^{(+)} =}
\nonumber\\&&	
-a_s^2 \frac{8   \big(
	28+30  L_M+9  L_M^2\big) }{27 (x-1)} \textcolor{blue}{C_A T_F}
	+a_s^3 \Biggl\{
		\textcolor{blue}{C_A N_F T_F^2} \Biggl[
		\frac{1}{x-1}\bigg[-\frac{2176  L_M}{81}
                -\frac{64  L_M^3}{27}
\nonumber\\&&		
                -\frac{4}{243} \big(
                        -2624
                        -441 \zeta_2
                        +81 x \zeta_2
                        -144 \zeta_3
                \big)
                +\frac{32}{9} \HA_0
		\bigg]
                +\frac{4}{3} \zeta_2
        \Biggr]	
+ \textcolor{blue}{C_A   T_F ^2} \Biggl[
		\frac{1}{x-1}\bigg[-\frac{320  L_M}{9}
\nonumber\\&&		
                -\frac{640  L_M^2}{27}
                -\frac{224  L_M^3}{27}
                -\frac{4}{81} \big(
                        -1312
                        -717 \zeta_2
                        +297 x \zeta_2
                        -168 \zeta_3
                \big)
                +\frac{16}{3} \HA_0
		\bigg]
                +\frac{44}{3} \zeta_2
        \Biggr]
\nonumber\\&&	
	+ \textcolor{blue}{C_A ^2  T_F}  \biggl\{
		\frac{1}{x-1}\biggl[\frac{176  L_M^3}{27}
                -\frac{4}{243} \big(
                        5668
                        +207 \zeta_2
                        -243 x \zeta_2
                        +324 \zeta_2^2
                        +396 \zeta_3
                        -162 \zeta_2 \zeta_3
\nonumber\\&&			
                        +162 x \zeta_2 \zeta_3
                \big)
                +\Bigl(
                        -\frac{88}{9}
                        +\frac{32 \HA_1 \zeta_2}{3}
                \Bigr) \HA_0
                +\frac{8}{3} \HA_0^2 \zeta_2
		\biggr]
                + L_M \Biggl[
			\frac{1}{x-1}\bigg[-\frac{16}{9} \HA_0^2 \big(
                                10+3 \HA_1\big)
\nonumber\\&&				
                        +\frac{8}{81} \big(
                                -155
                                +360 \zeta_2
                                -54 \zeta_2^2
                                +54 x \zeta_2^2
                                -1044 \zeta_3
                                +180 x \zeta_3
                        \big)
                        +\Bigl(
                                -\frac{16}
                                {3}
                                -\frac{640}{9} \HA_1
\nonumber\\&&				
                                +\frac{32}{3} \HA_{0,1}
                                -\frac{64}{3} \HA_{0,-1}
                        \Bigr) \HA_0
                        -\frac{64}{3} \HA_{0,0,1}
                        +\frac{128}{3} \HA_{0,0,-1}
			\bigg]
                        -\frac{16}{3} \zeta_2^2
                        -\frac{160}{9} \zeta_3
                \Biggr]
\nonumber\\&&		
                + L_M^2 \Biggl[
			\frac{1}{x-1}\biggl[\frac{8}{9} \bigl(
                                -23
                                +12 \zeta_2
                                -6 \zeta_3
                                +6 x \zeta_3
                        \bigr)
                        -\frac{16}{3} \HA_0^2
                        -\frac{64}{3} \HA_0 \HA_1
			\biggr]
                        -\frac{16}{3} \zeta_3
                \Biggr]
\nonumber\\&&	
                +\Bigl(
                        -4
                        +\frac{8 \zeta_3}{3}
                \Bigr) \zeta_2
	\biggr\}
	+\textcolor{blue}{ C_A   C_F   T_F } \biggl\{
		\frac{1}{x-1}\Biggl[-\frac{2}{9} \big(
                        233
                        +210 \zeta_2
                        -288 \ln(2)  \zeta_2
\nonumber\\&&			
                        -30 x \zeta_2
                        +288 \ln(2)  x \zeta_2
                        +72 \zeta_3
                        -72 x \zeta_3
                \big)
                +\frac{8}{3}  L_M \big(
                        -5
                        +24 \zeta_3
                \big)
		-8 L_M^2
		\Biggr]
\nonumber\\&&		
		+\frac{4}{3} \Bigl[-5+48 \ln(2) \Bigr] \zeta_2
                -16 \zeta_3
	\biggr\}
+ \textcolor{blue}{C_F ^2  T_F}  \biggl\{
                16 \Bigl[
                        -5 \zeta_2
                        +8 \ln(2)  \zeta_2
                        -2 \zeta_3
                \Bigr]
\nonumber\\&&	
	-16 \Bigl[-5+8 \ln(2) \Bigr] \zeta_2
                +32 \zeta_3
	\biggr\}
+a_{gg,Q}^{(3),(+)}
\Biggr\} ~,
\end{eqnarray}



\clearpage
\section{Polarized massive Wilson coefficients}
\label{sec:polWCs}

In this section we present the polarized heavy Wilson coefficients for the structure function $g_1$ in the asymptotic region $Q^2\gg m^2$. 

\subsection{$L_q^{PS}$ in $N$ space}
\begin{eqnarray}
\lefteqn{L_q^{PS} = \frac{1}{2}\left[1-(-1)^N\right] }
\nonumber\\&&
	\times\Bigg\{
	a_s^3 \Biggl\{
	 \textcolor{blue}{ C_F   N_F   T_F ^2} \biggl\{
                -\frac{32 (N-1)^2 (2+N) \big(
                        22+41 N+28 N^2\big)}{27 N^3 (1+N)^4}
\nonumber\\&&
                + L_M  \Biggl[
                        -\frac{64 (N-1)^2 (2+N) (2+5 N)}{9 N^3 
(1+N)^3}
                        +(N-1) \biggl[
                                -\frac{64 (2+N) \big(
                                        3+2 N+2 N^2\big)}{9 N^3 
(1+N)^3} S_1
\nonumber\\&&
                                +\frac{64 (2+N) S_1^2}{3 N^2 (1+N)^2}
                        \biggr]
                \Biggr]
                + L_M ^2 \biggl[
                        -\frac{32 (N-1)^2 (2+N)}{3 N^3 (1+N)^2}
                        -\frac{32 (N-1) (2+N) S_1}{3 N^2 (1+N)^2}
                \biggr]
\nonumber\\&&
                +(N-1) \Biggl[
                         L_Q  \Biggl[
                                \frac{32  L_M ^2 (2+N)}{3 N^2 
(1+N)^2}
                                +\frac{32 (2+N) \big(
                                        22+41 N+28 N^2\big)}{27 N^2 
(1+N)^4}
                                + L_M  \biggl[
                                        \frac{64 (2+N) (2+5 N)}{9 N^2 
(1+N)^3}
\nonumber\\&&
                                        -\frac{64 (2+N) S_1}{3 N^2 
(1+N)^2}
                                \biggr]
                                -\frac{32 (2+N) (2+5 N)}{9 N^2 
(1+N)^3} S_1
                                +\frac{16 (2+N) S_1^2}{3 N^2 (1+N)^2}
                                +\frac{16 (2+N) S_2}{3 N^2 (1+N)^2}
                        \Biggr]
\nonumber\\&&
                        +\biggl[
                                -\frac{32 (2+N) \big(
                                        6+37 N+35 N^2+13 N^3\big)}{27 
N^3 (1+N)^4}
                                -\frac{16 (2+N) S_2}{3 N^2 (1+N)^2}
                        \biggr] S_1
\nonumber\\&&
                        +\frac{16 (2+N) \big(
                                3+4 N+7 N^2\big)}{9 N^3 (1+N)^3} 
S_1^2
                        -\frac{16 (2+N) S_1^3}{3 N^2 (1+N)^2}
                \Biggr]
                -\frac{16 (N-1)^2 (2+N)}{3 N^3 (1+N)^2} S_2
	\biggr\}
\nonumber\\&&
+A_{qq,Q}^{PS(3)}
+ \textcolor{blue}{N_F} \hat{\tilde{C}}_q^{PS(3)}\left(L_Q,N_F\right)
\Biggr\}
\Bigg\} ~.
\end{eqnarray}

\subsection{$L_g^S$ in $N$ space}
\begin{eqnarray}
\lefteqn{ L_g^S = \frac{1}{2}\left[1-(-1)^N\right] }
\nonumber\\&&
	\times\Bigg\{
	a_s^2  \textcolor{blue}{N_F   T_F ^2} \Biggl\{
        \frac{16  L_M   L_Q  (N-1)}{3 N (1+N)}
        + L_M  \biggl[
                -\frac{16 (N-1)^2}{3 N^2 (1+N)}
                -\frac{16 (N-1) S_1}{3 N (1+N)}
        \biggr]
	\Biggr\}
\nonumber\\&&
	+a_s^3 \Biggl\{
         \textcolor{blue}{N_F   T_F ^3} \biggl\{
                \frac{64  L_M ^2  L_Q  (N-1)}{9 N (1+N)}
                + L_M ^2 \biggl[
                        -\frac{64 (N-1)^2}{9 N^2 (1+N)}
                        -\frac{64 (N-1) S_1}{9 N (1+N)}
                \biggr]
	\biggr\}
\nonumber\\&&
	+ \textcolor{blue}{C_A   N_F   T_F ^2} \biggl\{
                -\frac{8 (N-1)^2 Q_9}{27 N^5 (1+N)^4}
                + L_Q  \Biggl[
                        (N-1) \Biggl[
                                \frac{8 Q_9}{27 N^4 (1+N)^4}
                                + L_M ^2 \biggl[
                                        \frac{64}{3 N^2 (1+N)^2}
\nonumber\\&&
                                        -\frac{32 S_1}{3 N (1+N)}
                                \biggr]
                                -\frac{16 (47+56 N) S_1}{27 N 
(1+N)^2}
                        \Biggr]
                        + L_M  \Biggl[
                                \frac{32 Q_5}{9 N^3 (1+N)^3}
                                +(N-1) \biggl[
                                        \frac{32 S_1^2}{3 N (1+N)}
\nonumber\\&&
                                        -\frac{32 S_2}{3 N (1+N)}
                                        -\frac{64 S_{-2}}{3 N (1+N)}
                                \biggr]
                                -\frac{64 \big(
                                        -9-2 N+3 N^2+2 N^3\big)}{9 
N^2 (1+N)^2} S_1
                        \Biggr]
                \Biggr]
                + L_M ^2 \Biggl[
                        -\frac{64 (N-1)^2}{3 N^3 (1+N)^2}
\nonumber\\&&
                        +(N-1) \biggl[
                                \frac{32 \big(
                                        -3+N^2\big) S_1}{3 N^2 (1+N)^2}
                                +\frac{32 S_1^2}{3 N (1+N)}
                        \biggr]
                \Biggr]
                +(N-1) \Biggl[
                        \frac{8 S_1 Q_{10}
                        }{27 N^4 (1+N)^4}
\nonumber\\&&
                        + L_M   L_Q ^2 \biggl[
                                \frac{64}{3 N^2 (1+N)^2}
                                -\frac{32 S_1}{3 N (1+N)}
                        \biggr]
                        +\frac{16 (47+56 N) S_1^2}{27 N (1+N)^2}
                \Biggr]
\nonumber\\&&
                + L_M  \Biggl[
                        -\frac{16 Q_{13}}{9 (N-1) N^4 (1+N)^4 (2+N)^2}
                        +(N-1) \biggl[
                                -\frac{16 S_1^3}{9 N (1+N)}
                                -\frac{64 S_{2,1}}{3 N (1+N)}
                        \biggr]
\nonumber\\&&
                        +\biggl[
                                -\frac{16 Q_8}{9 N^3 (1+N)^3 (2+N)}
                                +\frac{80 (N-1) S_2}{3 N (1+N)}
                        \biggr] S_1
                        +\frac{32 \big(
                                -9-10 N+3 N^2+7 N^3\big)}{9 N^2 
(1+N)^2} S_1^2
\nonumber\\&&
                        +\frac{32 \big(
                                3-2 N+N^2+N^3\big)}{3 N^2 (1+N)^2} 
S_2
                        +\frac{32 \big(
                                2+5 N+5 N^2\big)}{9 N (1+N) (2+N)} 
S_3
\nonumber\\&&
                        +\biggl[
                                \frac{64 Q_3}{3 (N-1) N (1+N)^2 
(2+N)^2}
                                +\frac{64 \big(
                                        2+N+N^2\big)}{3 N (1+N) 
(2+N)} S_1
                        \biggr] S_{-2}
                        -\frac{64 \big(
                                -4+N+N^2\big)}{3 N (1+N) (2+N)} 
S_{-3}
\nonumber\\&&
                        -\frac{256 S_{-2,1}}{3 N (1+N) (2+N)}
                        -\frac{32 \big(
                                2+N+N^2\big)}{N (1+N) (2+N)} \zeta_3
                \Biggr]
	\biggr\}
	+ \textcolor{blue}{C_F  N_F   T_F ^2} \biggl\{
                \frac{4 (N-1)^2 Q_{12}}{N^6 (1+N)^5}
\nonumber\\&&
                +(N-1) \Biggl[
                        \frac{4 S_1 Q_{12}}{N^5 (1+N)^5}
                        + L_M   L_Q ^2 \biggl[
                                \frac{8 Q_2}{3 N^3 (1+N)^3}
                                -\frac{32 S_1}{3 N (1+N)}
                        \biggr]
                \Biggr]
\nonumber\\&&
                + L_Q  \Biggl[
                        \frac{16  L_M 
                        ^2 (N-1)^2 (2+N)}{N^3 (1+N)^3}
                        +(N-1) \Biggl[
                                -\frac{4 Q_{12}}{N^5 (1+N)^5}
                                + L_M  \biggl[
                                        \frac{16 S_1 Q_1}{3 N^3 
(1+N)^3}
\nonumber\\&&
                                        -\frac{16 Q_6}{3 N^4 (1+N)^4}
                                        +\frac{64 S_1^2}{3 N (1+N)}
                                        -\frac{64 S_2}{3 N (1+N)}
                                \biggr]
                        \Biggr]
                \Biggr]
                + L_M ^2 \biggl[
                        -\frac{16 (N-1)^3 (2+N)}{N^4 (1+N)^3}
\nonumber\\&&
                        -\frac{16 (N-1)^2 (2+N)}{N^3 (1+N)^3} S_1
                \biggr]
                + L_M  \Biggl[
                        \frac{8 S_2 Q_4}{3 N^3 (1+N)^3}
                        -\frac{8 Q_{14}}{3 (N-1) N^5 (1+N)^5 
(2+N)^2}
\nonumber\\&&
                        +(N-1) \biggl[
                                -\frac{8 \big(
                                        -4+2 N+3 N^2
                                \big)
\big(3+2 N+3 N^2\big)}{3 N^3 (1+N)^3} S_1^2
                                -\frac{80 S_1^3}{9 N (1+N)}
                                +\frac{64 S_{2,1}}{3 N (1+N)}
                        \biggr]
\nonumber\\&&
                        +\biggl[
                                -\frac{32 Q_{11}}{3 N^4 (1+N)^4 
(2+N)}
                                +\frac{16 (N-1) S_2}{N (1+N)}
                        \biggr] S_1
                        -\frac{256 \big(
                                1+N+N^2\big)}{9 N (1+N) (2+N)} S_3
\nonumber\\&&
                        +\biggl[
                                \frac{64 Q_7}{3 (N-1) N^2 (1+N)^2 
(2+N)^2}
                                -\frac{512 S_1}{3 N (1+N) (2+N)}
                        \biggr] S_{-2}
                        -\frac{256 S_{-3}}{3 N (1+N) (2+N)}
\nonumber\\&&
                        +\frac{512 S_{-2,1}}{3 N (1+N) (2+N)}
                        +\frac{64 \big(
                                2+N+N^2\big)}{N (1+N) (2+N)} \zeta_3
                \Biggr]
	\biggr\}
+A_{qg,Q}^{(3)}
+ \textcolor{blue}{N_F} \hat{\tilde{C}}_g^{S(3)}\left(L_Q,N_F\right)   
\Biggr\}
\Bigg\} ~,
\end{eqnarray}
with
\begin{eqnarray}
Q_1&=&3 N^4+2 N^3-N^2-12, \\
Q_2&=&3 N^4+6 N^3-N^2-4 N+12, \\
Q_3&=&N^5+3 N^4-3 N^3-9 N^2-8 N-8, \\
Q_4&=&9 N^5-N^4-23 N^3-15 N^2-14 N+12, \\
Q_5&=&9 N^5+9 N^4-4 N^3+15 N^2-41 N-12, \\
Q_6&=&N^6-19 N^4-22 N^3+22 N^2-34 N-36, \\
Q_7&=&N^6+3 N^5+13 N^4+21 N^3+22 N^2+12 N-24, \\
Q_8&=&13 N^6+44 N^5+71 N^4+94 N^3-90 N^2-288 N-72, \\
Q_9&=&15 N^6+45 N^5+374 N^4+601 N^3+161 N^2-24 N+36, \\
Q_{10}&=&97 N^6+161 N^5-392 N^4-807 N^3-255 N^2+24 N-36, \\
Q_{11}&=&N^8+3 N^7+11 N^6+20 N^5-15 N^4-17 N^3+49 N^2-20 N-36 
\\
Q_{12}&=&15 N^8+60 N^7+82 N^6+44 N^5+15 N^4+4 N^2+12 N+8, \\
Q_{13}&=&24 N^{10}+102 N^9+58 N^8-210 N^7-209 N^6+23 N^5+529 N^4
\nonumber\\&&
+1109 N^3+234 N^2-388 N-120, \\
Q_{14}&=&8 N^{12}+50 N^{11}+122 N^{10}+98 N^9-457 N^8-1398 N^7-1232 N^6
\nonumber\\&&
-634 N^5-793 N^4+388 N^3+1128 N^2-16 N-336 .
\end{eqnarray}

\subsection{$H_q^{PS}$ in $N$ space}
\begin{eqnarray}
	\lefteqn{ H_q^{PS} = \frac{1}{2} \left[ 1-(-1)^N \right] }
\nonumber\\&&
	\times \Bigg\{
	a_s^2  \textcolor{blue}{C_F   T_F}  \Biggl\{
        \frac{8 Q_{16}}{(N-1) N^4 (1+N)^4 (2+N)}
        +(2+N) \Biggl[
                \frac{8  L_M  \big(
                        1+2 N+N^3\big)}{N^3 (1+N)^3}
\nonumber\\&&			
                + L_Q  \biggl[
                        -\frac{8 \big(
                                2+N-N^2+2 N^3\big)}{N^3 (1+N)^3}
                        -\frac{8 (N-1) S_1}{N^2 (1+N)^2}
                \biggr]
                +(N-1) \biggl[
                        -\frac{4  L_M ^2}{N^2 (1+N)^2}
                        +\frac{4  L_Q ^2}{N^2 (1+N)^2}
\nonumber\\&&
                        +\frac{4 S_1^2}{N^2 (1+N)^2}
                        -\frac{12 S_2}{N^2 (1+N)^2}
                \biggr]
                +\frac{8 \big(
                        2+N-N^2+2 N^3\big)}{N^3 (1+N)^3} S_1
        \Biggr]
\nonumber\\&&
        -\frac{64}{(N-1) N (1+N) (2+N)} S_{-2}
	\Biggr\}
+a_s^3 \Biggl\{
		 \textcolor{blue}{C_F ^2  T_F} (2+N) \biggl\{
                        \frac{8 S_2 Q_{15}}{N^4 (1+N)^4}
\nonumber\\&&
                        +\frac{4 \big(
                                -1-3 N-4 N^2+4 N^3
                        \big)
\big(-2+5 N+6 N^2+9 N^3\big)}{N^6 (1+N)^5}
\nonumber\\&&
                        + L_Q  \Biggl[
                                -\frac{4 \big(
                                        2+3 N+3 N^2
                                \big)
\big(-1-3 N-4 N^2+4 N^3\big)}{N^5 (1+N)^5}
\nonumber\\&&
                                + L_M  \biggl[
                                        \frac{8 \big(
                                                2+3 N+3 N^2
                                        \big)
\big(1+2 N+N^3\big)}{N^4 (1+N)^4}
                                        -\frac{32 \big(
                                                1+2 N+N^3\big)}{N^3 
(1+N)^3} S_1
                                \biggr]
\nonumber\\&&
                                +(N-1) \Biggl[
                                         L_M ^2 \biggl[
                                                -\frac{4 \big(
                                                        2+3 N+3 
N^2\big)}{N^3 (1+N)^3}
                                                +\frac{16 S_1}{N^2 
(1+N)^2}
                                        \biggr]
                                        -\frac{8 \big(
                                                2+3 N+3 N^2\big)}
                                        {N^3 (1+N)^3} S_2
                                \Biggr]
\nonumber\\&&
                                +\biggl[
                                        \frac{16 \big(
                                                -1-3 N-4 N^2+4 
N^3\big)}{N^4 (1+N)^4}
                                        +\frac{32 (N-1) S_2}{N^2 
(1+N)^2}
                                \biggr] S_1
                        \Biggr]
\nonumber\\&&
                        + L_M  \biggl[
                                -\frac{8 \big(
                                        1+2 N+N^3
                                \big)
\big(-2+5 N+6 N^2+9 N^3\big)}{N^5 (1+N)^4}
\nonumber\\&&
                                +\frac{8 \big(
                                        -2+3 N+3 N^2
                                \big)
\big(1+2 N+N^3\big)}{N^4 (1+N)^4} S_1
                                +\frac{16 \big(
                                        1+2 N+N^3\big)}{N^3 (1+N)^3} 
S_1^2
                                -\frac{16 \big(
                                        1+2 N+N^3\big)}{N^3 (1+N)^3} 
S_2
                        \biggr]
\nonumber\\&&
                        +(N-1) \Biggl[
                                 L_M ^2 \biggl[
                                        \frac{4 \big(
                                                -2+5 N+6 N^2+9 
N^3\big)}{N^4 (1+N)^3}
                                        -\frac{4 \big(
                                                -2+3 N+3 
N^2\big)}{N^3 (1+N)^3} S_1
                                        -\frac{8 S_1^2}{N^2 (1+N)^2}
\nonumber\\&&
                                        +\frac{8 S_2}{N^2 (1+N)^2}
                                \biggr]
                                +\frac{16 S_2^2}{N^2 (1+N)^2}
                        \Biggr]
                        +\biggl[
                                -\frac{4 \big(
                                        -2+3 N+3 N^2
                                \big)
\big(-1-3 N-4 N^2+4 N^3\big)}{N^5 (1+N)^5}
\nonumber\\&&
                                -\frac{8 (N-1) \big(
                                        -2+3 N+3 N^2\big)}{N^3 
(1+N)^3} S_2
                        \biggr] S_1
                        +\biggl[
                                -\frac{8 \big(
                                        -1-3 N-4 N^2+4 N^3\big)}{N^4 
(1+N)^4}
                                -\frac{16 (N-1) S_2}{N^2 (1+N)^2}
                        \biggr] S_1^2
		\biggr\}
\nonumber\\&&
		+ \textcolor{blue}{C_F   T_F ^2} (2+N) \biggl\{
                        -\frac{32 (N-1)^2 \big(
                                22+41 N+28 N^2\big)}{27 N^3 (1+N)^4}
                        + L_M  \Biggl[
                                -\frac{64 (N-1)^2 (2+5 N)}{9 N^3 
(1+N)^3}
\nonumber\\&&
                                +(N-1) \biggl[
                                        -\frac{64 \big(
                                                3+2 N+2 N^2\big)}
                                        {9 N^3 (1+N)^3} S_1
                                        +\frac{64 S_1^2}{3 N^2 (1+N)^2}
                                \biggr]
                        \Biggr]
                        + L_M ^2 \biggl[
                                -\frac{32 (N-1)^2}{3 N^3 (1+N)^2}
\nonumber\\&&
                                -\frac{32 (N-1) S_1}{3 N^2 (1+N)^2}
                        \biggr]
                        +(N-1) \Biggl[
                                 L_Q  \Biggl[
                                        \frac{32  L_M ^2}{3 N^2 
(1+N)^2}
                                        +\frac{32 \big(
                                                22+41 N+28 
N^2\big)}{27 N^2 (1+N)^4}
\nonumber\\&&
                                        + L_M  \biggl[
                                                \frac{64 (2+5 N)}{9 
N^2 (1+N)^3}
                                                -\frac{64 S_1}{3 N^2 
(1+N)^2}
                                        \biggr]
                                        -\frac{32 (2+5 N) S_1}{9 N^2 
(1+N)^3}
                                        +\frac{16 S_1^2}{3 N^2 (1+N)^2}
                                        +\frac{16 S_2}{3 N^2 (1+N)^2}
                                \Biggr]
\nonumber\\&&
                                +\biggl[
                                        -\frac{32 \big(
                                                6+37 N+35 N^2+13 
N^3\big)}{27 N^3 (1+N)^4}
                                        -\frac{16 S_2}{3 N^2 (1+N)^2}
                                \biggr] S_1
                                +\frac{16 \big(
                                        3+4 N+7 N^2\big)}{9 N^3 
(1+N)^3} S_1^2
\nonumber\\&&
                                -\frac{16 S_1^3}{3 N^2 (1+N)^2}
                        \Biggr]
                        -\frac{16 (N-1)^2 S_2}{3 N^3 (1+N)^2}
		\biggr\}
+A_{Qq}^{PS(3)}
+\tilde{C}_q^{PS(3)}\left(L_Q,N_F+1\right)
\Biggr\}
\Bigg\} ~,
\end{eqnarray}
where
\begin{eqnarray}
Q_{15}&=&9 N^5+6 N^4-12 N^2-8 N+1, \\
Q_{16}&=&3 N^8+10 N^7-N^6-22 N^5-14 N^4-18 N^3-30 N^2+8 .
\end{eqnarray}

\subsection{$H_g^S$ in $N$ space}

with
\begin{eqnarray}
Q_{17}&=&N^4-22 N^3-79 N^2-72 N-4, \\
Q_{18}&=&N^4+3 N^3-4 N^2-8 N-4, \\
Q_{19}&=&N^4+3 N^3-2 N^2+3 N+3, \\
Q_{20}&=&N^4+17 N^3+43 N^2+33 N+2, \\
Q_{21}&=&2 N^4-N^3-24 N^2-17 N+28, \\
Q_{22}&=&3 N^4+6 N^3+2 N^2-N+6, \\
Q_{23}&=&3 N^4+6 N^3+4 N^2+N-6, \\
Q_{24}&=&3 N^4+6 N^3+11 N^2+8 N-12, \\
Q_{25}&=&3 N^4+18 N^3+47 N^2+56 N-4, \\
Q_{26}&=&3 N^4+42 N^3+71 N^2+8 N-28, \\
Q_{27}&=&4 N^4+5 N^3+3 N^2-4 N-4, \\
Q_{28}&=&7 N^4+74 N^3+171 N^2+128 N+4, \\
Q_{29}&=&9 N^4+6 N^3-55 N^2-44 N+44, \\
Q_{30}&=&9 N^4+6 N^3-35 N^2-16 N+20, \\
Q_{31}&=&9 N^4+12 N^3+8 N^2+5 N-2, \\
Q_{32}&=&9 N^4+15 N^3+17 N^2+9 N-6, \\
Q_{33}&=&9 N^4+46 N^3+93 N^2+48 N-76, \\
Q_{34}&=&11 N^4+42 N^3+47 N^2+32 N+12, \\
Q_{35}&=&25 N^4+26 N^3-3 N^2-52 N-36, \\
Q_{36}&=&29 N^4+58 N^3+5 N^2-24 N+36, \\
Q_{37}&=&43 N^4+86 N^3+107 N^2-8 N-56, \\
Q_{38}&=&45 N^4+78 N^3+53 N^2+12 N-12, \\
Q_{39}&=&85 N^4+170 N^3+61 N^2-24 N+36, \\
Q_{40}&=&111 N^4+198 N^3+135 N^2+24 N-32, \\
Q_{41}&=&763 N^4+1418 N^3+985 N^2+978 N+72, \\
Q_{42}&=&N^5+N^4-4 N^3+3 N^2-7 N-2, \\
Q_{43}&=&N^5+3 N^4-3 N^3-9 N^2-8 N-8, \\
Q_{44}&=&2 N^5+6 N^4+N^3-6 N^2+11 N+10, \\
Q_{45}&=&4 N^5-13 N^4-114 N^3-241 N^2-144 N+4, \\
Q_{46}&=&4 N^5+13 N^4-19 N^3-69 N^2-15 N-10, \\
Q_{47}&=&5 N^5+39 N^4+118 N^3+171 N^2+93 N+2, \\
Q_{48}&=&9 N^5+9 N^4-4 N^3+15 N^2-41 N-12, \\
Q_{49}&=&9 N^5+15 N^4+8 N^3+3 N^2+7 N-6, \\
Q_{50}&=&15 N^5+36 N^4-97 N^3-366 N^2-264 N+16, \\
Q_{51}&=&439 N^5+439 N^4-937 N^3-367 N^2-582 N-144, \\
Q_{52}&=&815 N^5+923 N^4-1349 N^3-491 N^2-258 N-216, \\
Q_{53}&=&N^6-7 N^4-4 N^3+16 N^2-10 N-12, \\
Q_{54}&=&2 N^6+5 N^5-22 N^4-95 N^3-114 N^2-24 N+16, \\
Q_{55}&=&2 N^6+5 N^5-3 N^4-7 N^3+2 N^2-11 N-8, \\
Q_{56}&=&11 N^6+30 N^5+9 N^4-22 N^3-10 N^2+2 N+4, \\
Q_{57}&=&13 N^6+44 N^5+71 N^4+94 N^3-90 N^2-288 N-72, \\
Q_{58}&=&15 N^6+45 N^5+374 N^4+601 N^3+161 N^2-24 N+36, \\
Q_{59}&=&15 N^6+57 N^5+47 N^4-N^3+12 N^2-38 N+4, \\
Q_{60}&=&21 N^6+45 N^5+55 N^4-13 N^3-12 N^2-8 N+8, \\
Q_{61}&=&25 N^6+85 N^5+119 N^4+75 N^3-118 N^2-102 N+44, \\
Q_{62}&=&57 N^6+153 N^5+233 N^4+163 N^3-70 N^2+120 N+144, \\
Q_{63}&=&97 N^6+161 N^5-392 N^4-807 N^3-255 N^2+24 N-36, \\
Q_{64}&=&247 N^6+795 N^5+555 N^4-71 N^3+210 N^2-360 N-432, \\
Q_{65}&=&15 N^7+28 N^6-116 N^5-453 N^4-575 N^3-221 N^2+114 N+88, \\
Q_{66}&=&21 N^7-52 N^6-86 N^5-60 N^4+17 N^3+176 N^2+24 N-40, \\
Q_{67}&=&21 N^7+220 N^6+713 N^5+1132 N^4+1010 N^3+486 N^2+38 N-52, \\
Q_{68}&=&753 N^7+1308 N^6-44 N^5-1118 N^4-1549 N^3-766 N^2-600 N-288, \\
Q_{69}&=&3479 N^7+7444 N^6-5160 N^5-13414 N^4-8111 N^3-5478 N^2
\nonumber\\&&
+1368 N+864, \\
Q_{70}&=&N^8+3 N^7+5 N^6+5 N^5-9 N^4-8 N^3+19 N^2-8 N-12, \\
Q_{71}&=&N^8+14 N^7+88 N^6+229 N^5+202 N^4+47 N^3+77 N^2+14 N+8, \\
Q_{72}&=&2 N^8+10 N^7+22 N^6+36 N^5+29 N^4+4 N^3+33 N^2+12 N+4, \\
Q_{73}&=&4 N^8-43 N^7-277 N^6-500 N^5-308 N^4+25 N^3+245 N^2+102 N-24, \\
Q_{74}&=&11 N^8+55 N^7+99 N^6+119 N^5+90 N^4+2 N^3+136 N^2+48 N+16, \\
Q_{75}&=&12 N^8+52 N^7+60 N^6-25 N^4-2 N^3+3 N^2+8 N+4, \\
Q_{76}&=&15 N^8+60 N^7+82 N^6+44 N^5+15 N^4+4 N^2+12 N+8, \\
Q_{77}&=&21 N^8+101 N^7+193 N^6+321 N^5+528 N^4+550 N^3+302 N^2+88 N+8, \\
Q_{78}&=&N^{10}+3 N^9-15 N^8-56 N^7-8 N^6+90 N^5+60 N^4+67 N^3+86 N^2
\nonumber\\&&
-12 N-24, \\
Q_{79}&=&5 N^{10}+23 N^9+31 N^8-N^7+54 N^6+268 N^5+342 N^4+98 N^3-60 N^2
\nonumber\\&&
-8 N+16, \\
Q_{80}&=&15 N^{10}+87 N^9+154 N^8+96 N^7+18 N^6-64 N^5+47 N^4+153 N^3-22 N^2
\nonumber\\&&
-12 N-8, \\
Q_{81}&=&18 N^{10}+162 N^9+740 N^8+2296 N^7+4511 N^6+5341 N^5+3593 N^4
\nonumber\\&&
+1065 N^3-130 N^2-148 N-8, \\
Q_{82}&=&24 N^{10}+102 N^9+58 N^8-210 N^7-209 N^6+23 N^5+529 N^4+1109 N^3
\nonumber\\&&
+234 N^2-388 N-120, \\
Q_{83}&=&29 N^{10}+229 N^9+620 N^8+1434 N^7+2173 N^6+505 N^5-86 N^4+712 N^3
\nonumber\\&&
+704 N^2-16 N-160, \\
Q_{84}&=&1371 N^{10}+6171 N^9+10220 N^8+5678 N^7-9493 N^6-17113 N^5-9154 N^4
\nonumber\\&&
-10864 N^3-10656 N^2+1872 N+4320, \\
Q_{85}&=&20283 N^{10}+92379 N^9+127804 N^8+11278 N^7-154181 N^6-222809 N^5
\nonumber\\&&
-170666 N^4-111392 N^3-62568 N^2+5040 N+8640, \\
Q_{86}&=&8 N^{11}+42 N^{10}+44 N^9-102 N^8-415 N^7-539 N^6-195 N^5-241 N^4
\nonumber\\&&
-414 N^3+268 N^2+104 N-96 .
\end{eqnarray}

\subsection{$L_q^{PS}$ in $z$ space}
\begin{eqnarray}
\lefteqn{ L_q^{PS}(z) = }
\nonumber\\&&
	a_s^3  C_F   N_F   T_F ^2 \Biggl\{
         L_M ^2 \biggl[
                \frac{32}{3} (z-1) \big(
                        12+5 \HA_1\big)
                +(1+z) \Bigl(
                        -\frac{32}{3} \HA_0^2
                        -\frac{64}{3} \HA_{0,1}
                        +\frac{64}{3} \zeta_2
                \Bigr)
\nonumber\\&&		
                +\frac{32}{3} (-7+z) \HA_0
        \biggr]
        +(z-1) \biggl[
                \frac{3296}{9}
                -\frac{32}{27} \HA_1 \big(
                        -125+12 \HA_0\big)
                +\frac{16}{9} \HA_1^2 \big(
                        28+15 \HA_0\big)
                +\frac{80}{3} \HA_1^3
        \biggr]
\nonumber\\&&
        + L_Q  \Biggl[
                (1+z) \Bigl(
                        \frac{64}{3} \HA_{0,1,1}
                        -\frac{64 \zeta_3}{3}
                \Bigr)
                + L_M ^2 \biggl[
                        -\frac{160}{3} (z-1)
                        +\frac{64}{3} (1+z) \HA_0
                \biggr]
\nonumber\\&&
                + L_M  \biggl[
                        \frac{64}{9} (z-1) \big(
                                -4+15 \HA_1\big)
                        +(1+z) \Bigl(
                                -\frac{128 \HA_{0,1}}{3}
                                +\frac{128 \zeta_2}{3}
                        \Bigr)
                        +\frac{128}{9} (2+5 z) \HA_0
                \biggr]
\nonumber\\&&
                +\frac{128}{27} (11+14 z) \HA_0
                +(z-1) \Bigl(
                        -\frac{3712}{27}+\frac{128}{9} 
\HA_1-\frac{80}{3} \HA_1^2\Bigr)
                -\frac{64}{9} (2+5 z) \HA_{0,1}
\nonumber\\&&
                +\frac{64}{9} (2+5 z) \zeta_2
        \Biggr]
        + L_M  \Biggl[
                (z-1) \biggl[
                        -\frac{64}{3}
                        -\frac{64}{9} \HA_1 \big(
                                32+15 \HA_0\big)
                        -\frac{320}{3} \HA_1^2
                \biggr]
\nonumber\\&&
                +(1+z) \biggl[
                        \frac{128}{9} \HA_{0,1} \big(
                                1+3 \HA_0\big)
                        -\frac{128}{3} \HA_{0,0,1}
                        +\frac{256}{3} \HA_{0,1,1}
                        -\frac{128}{3} \zeta_3
                \biggr]
                -\frac{512}{9} (1+2 z) \HA_0
\nonumber\\&&
                -\frac{64}
                {9} (2+5 z) \HA_0^2
                +\biggl[
                        \frac{64}{9} (-17+13 z)
                        -\frac{128}{3} (1+z) \HA_0
                \biggr] \zeta_2
        \Biggr]
        +(1+z) \Bigl(
                \frac{64}{3} \HA_{0,0,1,1}
                -64 \HA_{0,1,1,1}
\nonumber\\&&
                +\frac{32}{15} \zeta_2^2
        \Bigr)
        +\frac{128}{27} (-40+z) \HA_0
        -\frac{64}{27} (11+14 z) \HA_0^2
        +\big(
                \frac{128}{27} (-8+z)
                +\frac{64}{9} (2+5 z) \HA_0
        \big) \HA_{0,1}
\nonumber\\&&
        -\frac{64}{9} (2+5 z) \HA_{0,0,1}
        +\biggl[
                \frac{64}{9} (1+4 z)
                -\frac{64}{3} (1+z) \HA_0
        \biggr] \HA_{0,1,1}
        +\biggl[
                \frac{128}{27} (5+2 z)
                -\frac{64}{9} (2+5 z) \HA_0
\nonumber\\&&
                -\frac{160}{3} (z-1) \HA_1
                +\frac{64}{3} (1+z) \HA_{0,1}
        \biggr] \zeta_2
        +\biggl[
                -\frac{32}{9} (-17+13 z)
                +\frac{64}{3} (1+z) \HA_0
        \biggr] \zeta_3
\nonumber\\&&
+ A_{qq,Q}^{PS(3)}(z)
+ N_F \hat{\tilde{C}}_q^{PS(3)}(L_Q,N_F,z)
\Biggr\} .
\end{eqnarray}

\subsection{$L_g^S$ in $z$ space}
\begin{eqnarray}
\lefteqn{ L_g^S(z) = }
\nonumber\\&&
	a_s^2  \textcolor{blue}{N_F   T_F ^2} \Biggl\{
        \frac{16}{3}  L_M   L_Q  (2z-1)
        + L_M  \biggl[
                -\frac{16}{3} (-3+4 z)
                +(2z-1) \Bigl(
                        -\frac{16}{3} \HA_0
                        -\frac{16}{3} \HA_1
                \Bigr)
        \biggr]
\Biggr\}
\nonumber\\&&
	+a_s^3 \Biggl\{
		\textcolor{blue}{ N_F   T_F ^3} \biggl\{
                \frac{64}{9}  L_M ^2  L_Q  (2z-1)
                + L_M ^2 \biggl[
                        -\frac{64}{9} (-3+4 z)
                        +(2z-1) \Bigl(
                                -\frac{64}{9} \HA_0
                                -\frac{64}{9} \HA_1
                        \Bigr)
                \biggr]
	\biggr\}
\nonumber\\&&
	+ \textcolor{blue}{C_A   N_F   T_F ^2} \biggl\{
                \frac{8}{27} (-3943+3928 z)
                + L_M   L_Q ^2 \biggl[
                        -64 (z-1)
                        +\frac{64}{3} (1+z) \HA_0
                        -\frac{32}{3} (2z-1) \HA_1
                \biggr]
\nonumber\\&&
                + L_M ^2 \Biggl[
                        \frac{512}{3} (z-1)
                        +(1+z) \Bigl(
                                -\frac{32}{3} \HA_0^2
                                -\frac{64}{3} \HA_{0,1}
                        \Bigr)
                        +\biggl[
                                \frac{64}{3} (-4+z)
                                +\frac{32}{3} (2z-1) \HA_1
                        \biggr] \HA_0
\nonumber\\&&
                        +\frac{32}{3} (-9+10 z) \HA_1
                        +\frac{32}{3} (2z-1) \HA_1^2
                        +32 \zeta_2
                \Biggr]
                + L_Q  \Biggl[
                        -\frac{40}{9} (-108+107 z)
\nonumber\\&&
                        + L_M ^2 \biggl[
                                -64 (z-1)
                                +\frac{64}{3} (1+z) \HA_0
                                -\frac{32}{3} (2z-1) \HA_1
                        \biggr]
                        + L_M  \Biggl[
                                -\frac{32}{3} (-22+19 z)
\nonumber\\&&
                                +\biggl[
                                        \frac{32}{9} (5+98 z)
                                        +\frac{64}{3} (2z-1) \HA_1
                                \biggr] \HA_0
                                -\frac{64}{3} (1+2 z) \HA_{-1}
                                 \HA_0
                                -
                                \frac{64}{3} (1+3 z) \HA_0^2
\nonumber\\&&
                                +\frac{128}{9} (-8+7 z) \HA_1
                                +\frac{32}{3} (2z-1) \HA_1^2
                                -\frac{128}{3} (1+z) \HA_{0,1}
                                +\frac{64}{3} (1+2 z) \HA_{0,-1}
                                +\frac{128}{3} \zeta_2
                        \Biggr]
\nonumber\\&&
                        +\frac{8}{27} (785+644 z) \HA_0
                        +\frac{16}{9} (17+2 z) \HA_0^2
                        -\frac{16}{9} (2z-1) \HA_0^3
                        -\frac{16}{27} (-47+103 z) \HA_1
                        -\frac{32}{3} z \HA_{0,1}
\nonumber\\&&
                        +\frac{32}{3} z \zeta_2
                \Biggr]
                +(2z-1) \Bigl(
                        \frac{4}{9} \HA_0^4
                        +\frac{32}{3} \HA_{0,0,0,1}
                        -\frac{64}{15} \zeta_2^2
                \Bigr)
                + L_M  \Biggl[
                        \frac{16}{9} (-325+301 z)
\nonumber\\&&
                        +\biggl[
                                \frac{32}{9} \big(
                                        -58-42 z+3 z^2\big)
                                -\frac{128}{9} (-8+7 z) \HA_1
                                -16 (2z-1) \HA_1^2
                                -\frac{32}{3} \big(
                                        -3-6 z+2 z^2\big) \HA_{0,1}
\nonumber\\&&
                                +\frac{64}{3} \big(
                                        -1-2 z+z^2\big) \HA_{0,-1}
                        \biggr] \HA_0
                        +\frac{32}{3} \big(
                                1+2 z+2 z^2\big) \HA_{-1}^2 \HA_0
                        +\biggl[
                                \frac{8}{9} \big(
                                        23-124 z+44 z^2\big)
\nonumber\\&&
                                +\frac{16}{3} \big(
                                        3-6 z+2 z^2\big) \HA_1
                        \biggr] \HA_0^2
                        +\frac{16}{9} (5+14 z) \HA_0^3
                        +\biggl[
                                \frac{16}{9} \big(
                                        -180+161 z+6 z^2\big)
\nonumber\\&&
                                +\frac{64}{3} (2z-1) \HA_{0,1}
                        \biggr] \HA_1
                        -\frac{32}{9} (-8+z) \HA_1^2
                        -\frac{16}{9} (2z-1) \HA_1^3
\nonumber\\&&
                        +\biggl[
                                -\frac{64 \big(
                                        2+6 z+12 z^2+11 z^3\big)}
                                {9 z} \HA_0
                                -
                                \frac{32}{3} \big(
                                        -1-2 z+z^2\big) \HA_0^2
                                +\frac{64}{3} (1+2 z) \HA_{0,1}
\nonumber\\&&
                                -\frac{64}{3} \big(
                                        1+2 z+2 z^2\big) \HA_{0,-1}
                        \biggr] \HA_{-1}
                        -\frac{32}{9} (41+62 z) \HA_{0,1}
                        +\frac{64 \big(
                                2+6 z+12 z^2+11 z^3\big)}{9 z} \HA_{0,-1}
\nonumber\\&&
                        +\frac{32}{3} \big(
                                -3+2 z+2 z^2\big) \HA_{0,0,1}
                        -\frac{64}{3} \big(
                                -1-2 z+z^2\big) \HA_{0,0,-1}
                        -\frac{32}{3} (-7+2 z) \HA_{0,1,1}
\nonumber\\&&
                        -\frac{64}{3} (1+2 z) \HA_{0,1,-1}
                        -\frac{64}{3} (1+2 z) \HA_{0,-1,1}
                        +\frac{64}{3} \big(
                                1+2 z+2 z^2\big) \HA_{0,-1,-1}
\nonumber\\&&
                        +\biggl[
                                -\frac{32}{9} \big(
                                        3-90 z+22 z^2\big)
                                -\frac{64}{3} (3+4 z) \HA_0
                                -\frac{64}{3} (z-1)^2 \HA_1
                                +\frac{32}{3} \big(
                                        -1-2 z+2 z^2\big) \HA_{-1}
                        \biggr] \zeta_2
\nonumber\\&&
                        -\frac{64}{3} \big(
                                2+z+2 z^2\big) \zeta_3
                \Biggr]
                +\biggl[
                        \frac{8}{27} (-2405+413 z)
                        +\frac{16}{27} (-47+103 z) \HA_1
                        +\frac{32}{3} z \HA_{0,1}
                \biggr] \HA_0
\nonumber\\&&
                -\frac{4}{27} (989+764 z) \HA_0^2
                +\frac{160}{27} (-2+z) \HA_0^3
                +\frac{8}{27} (-1884+1981 z) \HA_1
                +\frac{16}{27} (-47+103 z) \HA_1^2
\nonumber\\&&
                -\frac{8}{27} (785+572 z) \HA_{0,1}
                -\frac{32}{9} (17+5 z) \HA_{0,0,1}
                +\frac{64}{3} z \HA_{0,1,1}
                +\biggl[
                        \frac{8}{9} (293+122 z)
                        -\frac{32}
                        {9} (-17+z) \HA_0
\nonumber\\&&
                        -
                        \frac{16}{3} (2z-1) \HA_0^2
                \biggr] \zeta_2
                +\biggl[
                        -\frac{32}{9} (-17+z)
                        -\frac{32}{3} (2z-1) \HA_0
                \biggr] \zeta_3
	\biggr\}
	+ \textcolor{blue}{C_F   N_F   T_F ^2} \biggl\{
                4 (-889+904 z)
\nonumber\\&&
                + L_M   L_Q ^2 \biggl[
                        8 (-41+42 z)
                        +(2z-1) \Bigl(
                                16 \HA_0^2
                                -\frac{32}{3} \HA_1
                        \Bigr)
                        -\frac{32}{3} (13+19 z) \HA_0
                \biggr]
\nonumber\\&&
                + L_Q  \Biggl[
                        -20 (-67+70 z)
			+ L_M ^2 \Bigl[
                                -336 (z-1)
                                +48 (3+4 z) \HA_0
                                -16 (2z-1) \HA_0^2
			\Bigr]
\nonumber\\&&
                        + L_M  \Biggl[
                                -\frac{16}{3} (-472+473 z)
                                +(2z-1) \Bigl(
                                        -32 \HA_0^3
                                        +\frac{64}{3} \HA_1^2
                                        -64 \HA_{0,0,1}
                                        +64 \zeta_3
                                \Bigr)
\nonumber\\&&
                                +\biggl[
                                        \frac{32}{3} (156+25 z)
                                        +\frac{128}{3} (2z-1) \HA_1
                                \biggr] \HA_0
                                +\frac{16}{3} (73+52 z) \HA_0^2
                                -\frac{16}{3} (-109+106 z) \HA_1
\nonumber\\&&
                                +\frac{64}{3} (14+17 z) \HA_{0,1}
				+\Bigl[
                                        -64 (4+7 z)
                                        +64 (2z-1) \HA_0
				\Bigr] \zeta_2
                        \Biggr]
                        +32 (23+17 z) \HA_0
\nonumber\\&&
                        -8 (-23+9 z) \HA_0^2
                        +\frac{8}{3} (9+4 z) \HA_0^3
                        -\frac{4}{3} (2z-1) \HA_0^4
                \Biggr]
                + L_M ^2 \biggl[
                        928 (z-1)
\nonumber\\&&
                        +(2z-1) \Bigl(
                                \frac{16}{3} \HA_0^3
                                +32 \HA_{0,0,1}
                                -32 \zeta_3
                        \Bigr)
                        -16 (30+7 z) \HA_0
                        -8 (11+4 z) \HA_0^2
\nonumber\\&&
                        +336 (z-1) \HA_1
                        -48 (3+4 z) \HA_{0,1}
			+\Bigl[
                                48 (3+4 z)
                                -32 (2z-1) \HA_0
			\Bigr] \zeta_2
                \biggr]
\nonumber\\&&
                + L_M 
                 \Biggl[
                        \frac{8}{9} (-8533+8509 z)
                        +(1+z)^2 \Bigl(
                                -\frac{128}{3} \HA_{-1}^2 \HA_0
                                -\frac{256}{3} \HA_{0,-1,-1}
                        \Bigr)
                        +(2z-1) \Bigl(
                                \frac{28}{3} \HA_0^4
\nonumber\\&&
                                -\frac{80}{9} \HA_1^3
                                +64 \HA_{0,0,1,1}
                                -\frac{32}{5} \zeta_2^2
                        \Bigr)
                        +\biggl[
                                -\frac{8}{9} \big(
                                        5606+209 z+24 z^2\big)
                                +(2z-1) \Bigl(
                                        -\frac{64}{3} \HA_1^2
\nonumber\\&&
                                        +64 \HA_{0,0,1}
                                \Bigr)
                                +48 (-11+10 z) \HA_1
                                +\frac{32}{3} \big(
                                        -25-40 z+4 z^2\big) \HA_{0,1}
\nonumber\\&&
                                -\frac{128}{3} \big(
                                        -2+4 z+z^2\big) \HA_{0,-1}
                        \biggr] \HA_0
                        +\biggl[
                                -\frac{4}{9} \big(
                                        3063+462 z+32 z^2\big)
                                -\frac{16}{3} \big(
                                        -1+2 z+4 z^2\big) \HA_1
                        \biggr] \HA_0^2
\nonumber\\&&
                        -\frac{16}{9} (104+17 z) \HA_0^3
                        +\biggl[
                                -\frac{32}{3} \big(
                                        229-230 z+2 z^2\big)
                                -\frac{64}{3} (2z-1) \HA_{0,1}
                        \biggr] \HA_1
                        +24 (-11+10 z) \HA_1^2
\nonumber\\&&
                        +\biggl[
                                (1+z)^2 \Bigl(
                                        \frac{64}{3} \HA_0^2
                                        +\frac{256}{3} \HA_{0,-1}
                                \Bigr)
                                +\frac{64 \big(
                                        4+45 z+48 z^2+4 z^3\big)}{9 
z} \HA_0
                        \biggr] \HA_{-1}
                        -\frac{16}{3} (203+156 z) \HA_{0,1}
\nonumber\\&&
                        -\frac{64 \big(
                                4+45 z+48 z^2+4 z^3\big)}{9 z} \HA_{0,-1}
                        -\frac{64}{3} (z-1) (-11+2 z) \HA_{0,0,1}
\nonumber\\&&
                        +\frac{128}{3} \big(
                                -5+6 z+z^2\big) \HA_{0,0,-1}
                        -\frac{32}{3} (31+28 z) \HA_{0,1,1}
                        +\biggl[
                                \frac{32}
                                {9} \big(
                                        543+99 z+8 z^2\big)
\nonumber\\&&
                                +\frac{32}{3} (83+56 z) \HA_0
                                -96 (2z-1) \HA_0^2
                                +\frac{64}{3} \big(
                                        -1+2 z+2 z^2\big) \HA_1
                                -\frac{128}{3} (1+z)^2 \HA_{-1}
                        \biggr] \zeta_2
\nonumber\\&&
                        +\biggl[
                                \frac{64}{3} \big(
                                        33-4 z+4 z^2\big)
                                -128 (2z-1) \HA_0
                        \biggr] \zeta_3
                \Biggr]
                +(2z-1) \Bigl(
                        \frac{4}{15} \HA_0^5
                        +32 \HA_{0,0,0,0,1}
                        -32 \zeta_5
                \Bigr)
\nonumber\\&&
                +4 (-519+10 z) \HA_0
                -8 (69+8 z) \HA_0^2
                -\frac{8}{3} (32+3 z) \HA_0^3
                +\frac{2}{3} (-11+4 z) \HA_0^4
                +20 (-67+70 z) \HA_1
\nonumber\\&&
                -32 (23+17 z) \HA_{0,1}
                +16 (-23+9 z) \HA_{0,0,1}
                -16 (9+4 z) \HA_{0,0,0,1}
                +\biggl[
                        32 (23+17 z)
\nonumber\\&&
                        -16 (-23+9 z) \HA_0
                        +8 (9+4 z) \HA_0^2
                        -\frac{16}{3} (2z-1) \HA_0^3
                \biggr] \zeta_2
                +\biggl[
                        \frac{32}{5} (9+4 z)
                        -\frac{64}{5} (2z-1) \HA_0
                \biggr] \zeta_2^2
\nonumber\\&&
		+\Bigl[
                        -16 (-23+9 z)
                        +16 (9+4 z) \HA_0
                        -16 (2z-1) \HA_0^2
		\Bigr] \zeta_3
	\biggr\}
\nonumber\\&&	
+A_{qg,Q}^{(3)}
+N_F \hat{\tilde{C}}_g^{S(3)}\left(L_Q,N_F,z\right)
\Biggr\} ~.
\end{eqnarray}

\subsection{$H_q^{PS}$ in $z$ space}
\begin{eqnarray}
\lefteqn{ H_q^{PS} = }
\nonumber\\&&
	a_s^2  \textcolor{blue}{C_F   T_F } \Biggl\{
        -
        \frac{4}{3} (z-1) \big(
                148+66 \HA_1+15 \HA_1^2\big)
	+ L_Q  \Bigl[
                8 (z-1) \big(
                        11+5 \HA_1\big)
\nonumber\\&&			
                +(1+z) \big(
                        -16 \HA_0^2
                        -16 \HA_{0,1}
                        +16 \zeta_2
                \big)
                +32 (-2+z) \HA_0
	\Bigr]
	+ L_M ^2 \Bigl[
                20 (z-1)
                -8 (1+z) \HA_0
	\Bigr]
\nonumber\\&&
	+ L_Q ^2 \Bigl[
                -20 (z-1)
                +8 (1+z) \HA_0
	\Bigr]
	+ L_M  \Bigl[
                8 (z-1)
                -8 (-1+3 z) \HA_0
                +8 (1+z) \HA_0^2
	\Bigr]
\nonumber\\&&
        +(1+z) \Bigl(
                \frac{16}{3} \HA_0^3
                -32 \HA_{0,0,1}
                +16 \HA_{0,1,1}
                +16 \zeta_3
        \Bigr)
        +\biggl[
                -\frac{256}{3} (-2+z)
                -80 (z-1) \HA_1
\nonumber\\&&
                -\frac{32 (1+z)^3 \HA_{-1}}{3 z}
                +32 (1+z) \HA_{0,1}
        \biggr] \HA_0
        +\frac{8}{3} \big(
                21+2 z^2\big) \HA_0^2
        +16 (-1+3 z) \HA_{0,1}
\nonumber\\&&
        +\frac{32 (1+z)^3 \HA_{0,-1}}{3 z}
        +\biggl[
                -\frac{32}{3} \big(
                        9-3 z+z^2\big)
                -32 (1+z) \HA_0
        \biggr] \zeta_2
	\Biggr\}
\nonumber\\&&
	+a_s^3 \Biggl\{
		\textcolor{blue}{C_F   T_F ^2} \biggl\{
                 L_M ^2 \biggl[
                        \frac{32}{3} (z-1) \big(
                                12+5 \HA_1\big)
                        +(1+z) \Bigl(
                                -\frac{32}{3} \HA_0^2
                                -\frac{64}{3} \HA_{0,1}
                                +\frac{64}{3} \zeta_2
                        \Bigr)
\nonumber\\&&
                        +\frac{32}{3} (-7+z) \HA_0
                \biggr]
                + L_Q  \Biggl[
                        (1+z) \Bigl(
                                \frac{64}{3} \HA_{0,1,1}
                                -\frac{64 \zeta_3}{3}
                        \Bigr)
                        + L_M ^2 \biggl[
                                -\frac{160}{3} (z-1)
                                +\frac{64}
                                {3} (1+z) \HA_0
                        \biggr]
\nonumber\\&&
                        + L_M  \biggl[
                                \frac{64}{9} (z-1) \big(
                                        -4+15 \HA_1\big)
                                +(1+z) \Bigl(
                                        -\frac{128 \HA_{0,1}}{3}
                                        +\frac{128 \zeta_2}{3}
                                \Bigr)
                                +\frac{128}{9} (2+5 z) \HA_0
                        \biggr]
\nonumber\\&&
                        +\frac{128}{27} (11+14 z) \HA_0
                        +(z-1) \Bigl(
                                -\frac{3712}{27}+\frac{128}{9} 
\HA_1-\frac{80}{3} \HA_1^2\Bigr)
                        -\frac{64}{9} (2+5 z) \HA_{0,1}
\nonumber\\&&
                        +\frac{64}{9} (2+5 z) \zeta_2
                \Biggr]
                + L_M  \Biggl[
                        (1+z) \Bigl(
                                \frac{128}{9} \HA_{0,1}
                                -\frac{128}{3} \HA_{0,0,1}
                                +\frac{256}{3} \HA_{0,1,1}
                                -\frac{128}{3} \zeta_3
                        \Bigr)
\nonumber\\&&
                        +\biggl[
                                -\frac{512}{9} (1+2 z)
                                -\frac{320}{3} (z-1) \HA_1
                                +\frac{128}{3} (1+z) \HA_{0,1}
                        \biggr] \HA_0
                        -\frac{64}{9} (2+5 z) \HA_0^2
\nonumber\\&&
                        +(z-1) \Bigl(
                                -\frac{64}{3}-\frac{2048}{9} 
\HA_1-\frac{320}{3} \HA_1^2\Bigr)
                        +\biggl[
                                \frac{64}{9} (-17+13 z)
                                -\frac{128}{3} (1+z) \HA_0
                        \biggr] \zeta_2
                \Biggr]
\nonumber\\&&
                +(1+z) \Bigl(
                        \frac{64}{3} \HA_{0,0,1,1}
                        -64 \HA_{0,1,1,1}
                        +\frac{32}{15} \zeta_2^2
                \Bigr)
                +\biggl[
                        \frac{128}{27} (-40+z)
                        +\frac{16}{9} (z-1) \HA_1 \big(
                                -8+15 \HA_1\big)
\nonumber\\&&
                        +\frac{64}{9} (2+5 z) \HA_{0,1}
                        -\frac{64}{3} (1+z) \HA_{0,1,1}
                \biggr] \HA_0
                -\frac{64}{27} (11+14 z) \HA_0^2
                +(z-1) \Bigl(
                        \frac{3296}{9}+\frac{4000}{27} \HA_1
\nonumber\\&&
			+\frac{448}{9} \HA_1^2+
                        \frac{80}{3} \HA_1^3\Bigr)
                +\frac{128}{27} (-8+z) \HA_{0,1}
                -\frac{64}{9} (2+5 z) \HA_{0,0,1}
                +\frac{64}{9} (1+4 z) \HA_{0,1,1}
\nonumber\\&&
                +\biggl[
                        \frac{128}{27} (5+2 z)
                        -\frac{64}{9} (2+5 z) \HA_0
                        -\frac{160}{3} (z-1) \HA_1
                        +\frac{64}{3} (1+z) \HA_{0,1}
                \biggr] \zeta_2
\nonumber\\&&
                +\biggl[
                        -\frac{32}{9} (-17+13 z)
                        +\frac{64}{3} (1+z) \HA_0
                \biggr] \zeta_3
	\biggr\}
	+ \textcolor{blue}{C_F ^2  T_F}  \biggl\{
		(z-1) \Bigl[
                        2656
                        +8 \HA_1 \big(
                                103+40 \HA_{0,0,1}\big)
\nonumber\\&&				
                        +\frac{8}{3} \HA_0^4
                        +144 \HA_1^2
                        -80 \HA_{0,1}^2
                        +160 \HA_{0,1,1,1}
		\Bigr]
                + L_M ^2 \biggl[
                        4 (z-1) \big(
                                92+43 \HA_1+10 \HA_1^2\big)
\nonumber\\&&
                        +(1+z) \Bigl(
                                -\frac{8}{3} \HA_0^3
                                +48 \HA_{0,0,1}
                                -32 \HA_{0,1,1}
                                -16 \zeta_3
                        \Bigr)
			+\Bigl[
                                -4 (51+7 z)
                                +80 (z-1) \HA_1
\nonumber\\&&
                                -32 (1+z) \HA_{0,1}
			\Bigr] \HA_0
                        +8 (-6+5 z) \HA_0^2
                        -80 z \HA_{0,1}
			+\Bigl[
                                80
                                +16 (1+z) \HA_0
			\Bigr] \zeta_2
                \biggr]
\nonumber\\&&
                + L_Q  \Biggl[
			 L_M ^2 \Bigl[
                                -4 (z-1) \big(
                                        13+20 \HA_1\big)
                                +(1+z) \big(
                                        8 \HA_0^2
                                        +32 \HA_{0,1}
                                        -32 \zeta_2
                                \big)
                                -16 (-2+3 z) \HA_0
			\Bigr]
\nonumber\\&&
                        +(z-1) \big(
                                -392
                                -288 \HA_1
                                -160 \HA_{0,1,1}
                        \big)
                        + L_M  \biggl[
                                -8 (z-1) \big(
                                        5+4 \HA_1\big)
\nonumber\\&&
                                +(1+z) \Bigl(
                                        -8 \HA_0
                                        -\frac{16}{3} \HA_0^3
                                        -64 \HA_{0,0,1}
                                        +64 \zeta_3
                                \Bigr)
                                +32 z \HA_0^2
                                +32 (-1+3 z) \HA_{0,1}
\nonumber\\&&
				+\Bigl[
                                        -32 (-1+3 z)
                                        +64 (1+z) \HA_0
				\Bigr] \zeta_2
                        \biggr]
                        +(1+z) \Bigl(
                                \frac{2}{3} \HA_0^4
                                -64 \HA_{0,0,0,1}
                                +128 \HA_{0,0,1,1}
                                -\frac{256}{5} \zeta_2^2
                        \Bigr)
\nonumber\\&&
			+\Bigl[
                                -52 (-3+z)
                                +8 (z-1) \HA_1 \big(
                                        13+10 \HA_1\big)
                                +(1+z) \big(
                                        96 \HA_{0,0,1}
                                        -64 \HA_{0,1,1}
                                \big)
\nonumber\\&&
                                -32 (-3+2 z) \HA_{0,1}
			\Bigr] \HA_0
			+\Bigl[
                                6 (3+19 z)
                                +80 (z-1) \HA_1
                                -32 (1+z) \HA_{0,1}
			\Bigr] \HA_0^2
                        -\frac{16}{3} z \HA_0^3
\nonumber\\&&
                        +8 (19+17 z) \HA_{0,1}
                        -16 (1+5 z) \HA_{0,0,1}
			+\Bigl[
                                -48 (1+5 z)
                                +(1+z) \big(
                                        -16 \HA_0^2
                                        +64 \HA_{0,1}
                                \big)
\nonumber\\&&
                                +16 (-1+3 z) \HA_0
                                -160 (z-1) \HA_1
			\Bigr] \zeta_2
			+\Bigl[
                                48 (-3+5 z)
                                -96 (1+z) \HA_0
			\Bigr] \zeta_3
                \Biggr]
\nonumber\\&&
                + L_M  \biggl[
                        (z-1) \Bigl(
                                312
                                -\frac{64}{3} \HA_0^3
                                +88 \HA_1
                                +16 \HA_1^2
                        \Bigr)
                        +(1+z) \Bigl(
                                \frac{4}{3} \HA_0^4
                                -160 \HA_{0,0,0,1}
                                +64 \HA_{0,0,1,1}
\nonumber\\&&
                                +\frac{288}{5} \zeta_2^2
                        \Bigr)
			+\Bigl[
                                -8 (-2+21 z)
                                +32 (z-1) \HA_1
                                -32 (-1+3 z) \HA_{0,1}
                                +64 (1+z) \HA_{0,0,1}
			\Bigr] \HA_0
\nonumber\\&&
                        +4 (7+13 z) \HA_0^2
                        -8 (-11+21 z) \HA_{0,1}
                        +32 (1+7 z) \HA_{0,0,1}
                        -32 (-1+3 z) \HA_{0,1,1}
\nonumber\\&&
			+\Bigl[
                                8 (-7+17 z)
                                -32 (3+z) \HA_0
                                -16 (1+z) \HA_0^2
			\Bigr] \zeta_2
			+\Bigl[
                                -64 (1+2 z)
                                +32 (1+z) \HA_0
			\Bigr] \zeta_3
                \biggr]
\nonumber\\&&
                +(1+z) \Bigl(
                        -\frac{2}{15} \HA_0^5
                        -80 \HA_{0,0,0,0,1}
                        +832 \HA_{0,0,0,1,1}
                        +256 \HA_{0,0,1,0,1}
                        -128 \HA_{0,0,1,1,1}
                        +80 \zeta_5
                \Bigr)
\nonumber\\&&
                +\Biggl[
                        -4 (263+195 z)
                        +(z-1) \biggl[
                                -32 \HA_1 \big(
                                        14+5 \HA_{0,1}\big)
                                -172 \HA_1^2
                                -\frac{80}{3} \HA_1^3
                        \biggr]
\nonumber\\&&
                        +(1+z) \big(
                                32 \HA_{0,1}^2
                                +96 \HA_{0,0,0,1}
                                -224 \HA_{0,0,1,1}
                                +64 \HA_{0,1,1,1}
                        \big)
                        +16 (1+10 z) \HA_{0,1}
\nonumber\\&&
                        -16 (-1+17 z) \HA_{0,0,1}
                        +160 (-2+3 z) \HA_{0,1,1}
                \Biggr] \HA_0
                +\biggl[
                        -6 (26+15 z)
                        -4 (z-1) \HA_1 \big(
                                43+10 \HA_1\big)
\nonumber\\&&
                        +(1+z) \big(
                                -48 \HA_{0,0,1}
                                +32 \HA_{0,1,1}
                        \big)
                        +80 z \HA_{0,1}
                \biggr] \HA_0^2
                +\biggl[
                        -2 (5+23 z)
                        -\frac{80}{3} (z-1) \HA_1
\nonumber\\&&
                        +\frac{32}{3} (1+z) \HA_{0,1}
                \biggr] \HA_0^3
		+\Bigl[
                        4 (-169+35 z)
                        -128 (1+z) \HA_{0,0,1}
		\Bigr] \HA_{0,1}
                -4 (109+33 z) \HA_{0,0,1}
\nonumber\\&&
                +8 (-49+13 z) \HA_{0,1,1}
                +32 (-8+15 z) \HA_{0,0,0,1}
                -144 (-1+3 z) \HA_{0,0,1,1}
                +\biggl[
                        4 (57+77 z)
\nonumber\\&&
                        +8 (z-1) \HA_1 \big(
                                43+10 \HA_1\big)
                        +(1+z) \Bigl(
                                \frac{8}{3} \HA_0^3
                                +96 \HA_{0,0,1}
                                -64 \HA_{0,1,1}
                                -96 \zeta_3
                        \Bigr)
			+\Bigl[
                                12 (5+13 z)
\nonumber\\&&
                                +160 (z-1) \HA_1
                                -64 (1+z) \HA_{0,1}
			\Bigr] \HA_0
                        +8 (3+z) \HA_0^2
                        -160 z \HA_{0,1}
                \biggr] \zeta_2
                +\biggl[
                        -\frac{8}{5} (-145+83 z)
\nonumber\\&&
                        +\frac{48}{5} (1+z) \HA_0
                \biggr] \zeta_2^2
		+\Bigl[
                        4 (207+7 z)
                        +(1+z) \big(
                                24 \HA_0^2
                                +128 \HA_{0,1}
                        \big)
                        -32 (-13+8 z) \HA_0
\nonumber\\&&
                        -320 (z-1) \HA_1
		\Bigr] \zeta_3
	\biggr\}
+A_{Qq}^{PS(3)}
+\tilde{C}^{PS(3)}(L_Q,N_F+1)
\Biggr\} ~.
\end{eqnarray}

\subsection{$H_g^S$ in $z$ space}


\clearpage
\section{Definition of certain iterated integrals}
\label{sec:ex2GL}

We collect in the following the definitions of the functions $G_i$ appearing in Chapter \ref{sec:ex2}.
%

\clearpage
\section*{Acknowledgements}
I would like to thank Prof. Johannes Bl\"umlein for accepting to patiently train me in this project and for his constant involvement as well as for building an international research environment in field theory. I also thank Kay Sch\"onwald for much help and support. I thank Prof. Carsten Schneider and Jakob Ablinger for their help and hospitality at RISC, and all the scholars at DESY whom I have met, particularly Abilio De Freitas. I also thank Prof. Gabriele Travaglini and all the organizers and lecturers in the SAGEX network, for the many training events which have been offered.

This project has received funding from the European Union’s Horizon 2020 research and innovation programme under the Marie Sklodowska-Curie grant agreement No. 764850, SAGEX.

\clearpage
\thispagestyle{empty}
\vspace*{-1cm}
\hspace*{-2.5cm}
\parbox[c][\textheight]{\textwidth}{
\includegraphics[width=209mm,page=1]{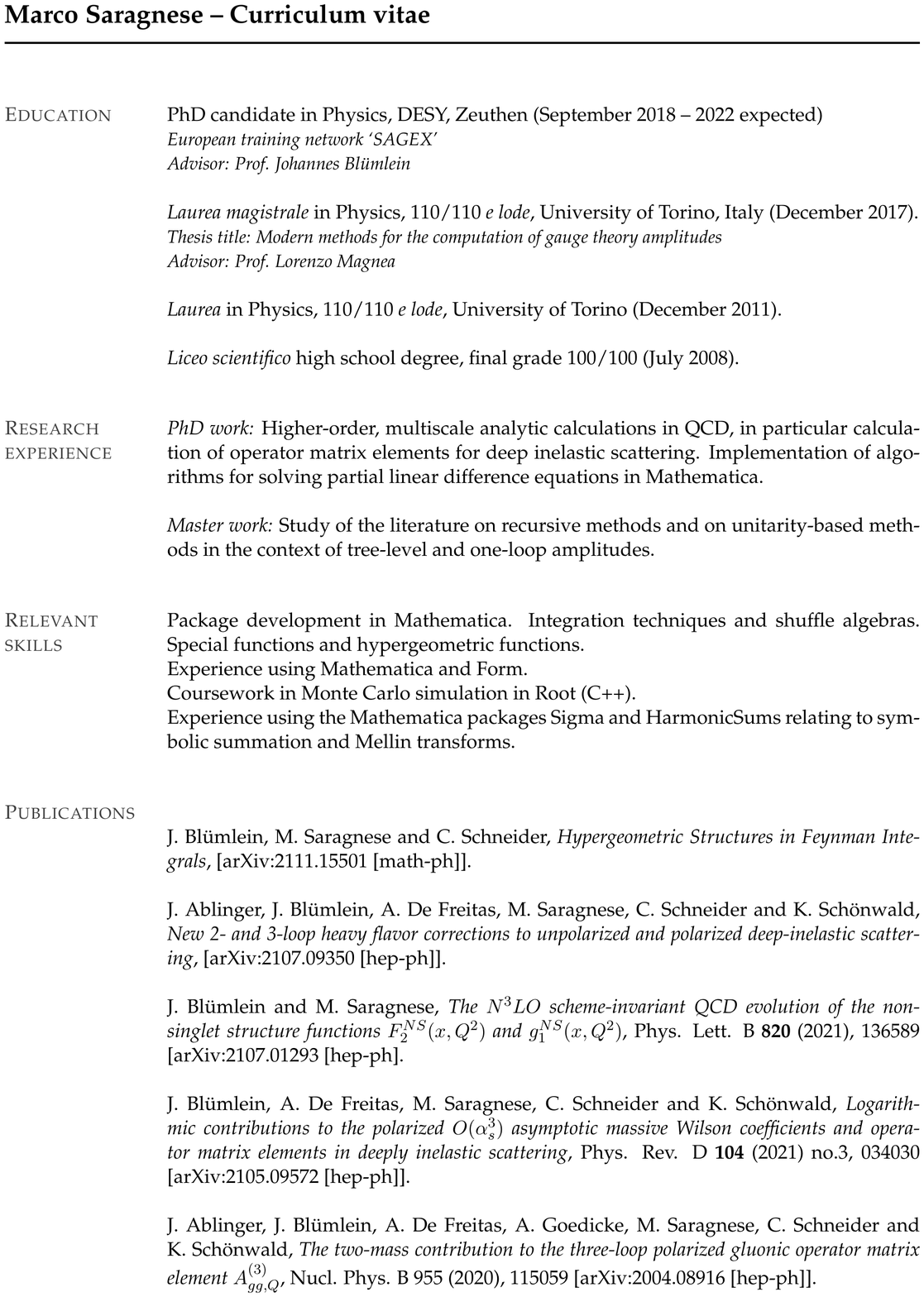}
}

\clearpage
\thispagestyle{empty}
\thispagestyle{empty}
\vspace*{-1cm}
\hspace*{-2.5cm}
\parbox[c][\textheight]{\textwidth}{
\includegraphics[width=209mm,page=2]{figs/cv_with_publ.pdf}
}

\clearpage

\end{document}